\documentclass[a4paper,12pt,twoside,openright]{book}
\usepackage{color}
\usepackage[utf8]{inputenc}
\usepackage{lmodern}
\usepackage{textgreek}
\usepackage[hang,flushmargin]{footmisc} 

\usepackage{bm}
\usepackage[skip=10pt,font=footnotesize,bf]{caption}
\usepackage{url}
\usepackage{amsmath,amssymb}
\usepackage[breaklinks,colorlinks=false,urlcolor=blue,citecolor=blue,linkcolor=red]{hyperref}

\usepackage{epsfig,graphicx,verbatim,xspace,multirow,mathtools}
\usepackage{booktabs}

\usepackage[english]{babel}
\usepackage{epigraph}
\usepackage{fancyhdr}
\usepackage{indentfirst}

\usepackage{braket}
\usepackage{slashed}
\usepackage{subcaption}
\usepackage{color}
\usepackage{xfrac}
\usepackage{cite}
\usepackage{mathtools}

\newcommand{\mychi}{\raisebox{1pt}[1ex][0.6ex]{$\chi$}}

\usepackage[final]{pdfpages}

\def\eqref#1{{(\ref{#1})}}

\usepackage[hyperpageref]{backref}

\renewcommand*{\backref}[1]{}  
\renewcommand*{\backrefalt}[4]{
\ifcase #1 
No cited.
\or
{Cited on page} #2.
\else
{Cited on page} #2.
\fi}

\makeatletter
\def \cleardoublepage {\clearpage \if@twoside
\ifodd \c@page
\else
\null\thispagestyle{empty}\clearpage
\fi
\fi}
\makeatother

\usepackage{fancyhdr}
\renewcommand{\headrulewidth}{0.0pt}

\fancypagestyle{plain}{%
\fancyhead{} 
\fancyfoot[C]{\thepage}
\renewcommand{\headrulewidth}{0pt} 
}

\newcommand{\kdf}{\mathcal{K}_{\text{df},3} }

\newcommand{\tr}{\text{tr } }
\definecolor{orange}{rgb}{0.93, 0.57, 0.13}
\definecolor{purple}{rgb}{0.75, 0.0, 1.0}
\newcommand{\red}[1]{\textcolor{red}{#1}}
\newcommand{\orange}[1]{\textcolor{orange}{#1}}
\newcommand{\yellow}[1]{\textcolor{yellow}{#1}}
\newcommand{\green}[1]{\textcolor{green}{#1}}
\newcommand{\purple}[1]{\textcolor{purple}{#1}}
\newcommand{\cyan}[1]{\textcolor{cyan}{#1}}
\newcommand{\blue}[1]{\textcolor{blue}{#1}}
\newcommand{\largeNc}{\red{L}\orange{a}\yellow{r}\green{g}\cyan{e} $\blue{\text{N}}_{\purple{\text{c}}}$ }

\newcommand{\cY}[0]{\mathcal Y}
\newcommand{\cM}[0]{\mathcal M}
\newcommand{\cD}[0]{\mathcal D}
\newcommand{\cL}[0]{\mathcal L}
\newcommand{\cR}[0]{\mathcal R}
\newcommand{\cK}[0]{\mathcal K}
\newcommand{\cS}[0]{\mathcal S}
\newcommand{\df}[0]{\text{df}}
\newcommand{\mc}[1]{\mathcal{#1}}

\begin{document}


\setlength{\unitlength}{1cm} 
\thispagestyle{empty}
\begin{center}

\begin{center}
\includegraphics[scale=0.08]{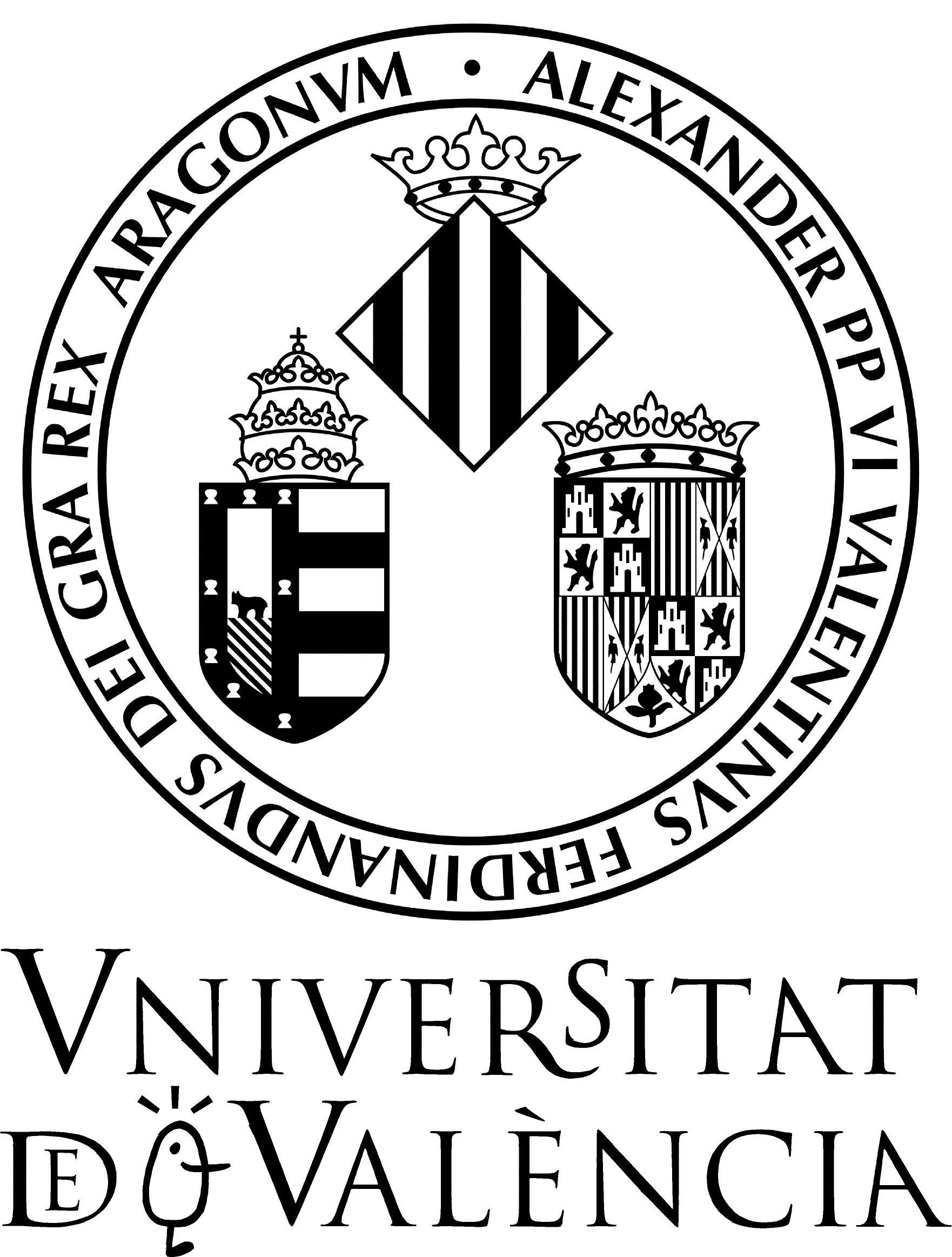}
\end{center}

\vspace{1cm}

\textbf{\Large{Kaon decays and other hadronic processes\\ in lattice QCD }\\ \vspace{0.5cm}
{\Large PhD Thesis}}

\vspace{2.5cm}

{\Large \bf{Fernando Romero L\'opez}} \\[2ex]
{IFIC - Universitat de Val\`{e}ncia - CSIC}\\
{Departament de F\'isica Te\`{o}rica}\\
{Programa de Doctorat en F\'isica }\\[4ex]

\textbf{Under the supervision of}\\[3ex]

 {\large \textbf{Pilar Hern\'andez Gamazo} \\[1ex] \large and \\[1.1ex] \large \textbf{Stephen R. Sharpe} } \\

\vspace{1.8cm}

{\large \bf{Val\`{e}ncia, July 2021}}

\end{center}


\begin{titlepage}
\cleardoublepage
\thispagestyle{empty}

\thispagestyle{empty}
\vspace*{3cm}

\noindent \textbf{Pilar Hern\'andez Gamazo}, catedr\'atica del Departamento de F\'isica Te\'orica
de la Universitat de Val\`{e}ncia, y \vspace{0.2cm}

\noindent \textbf{Stephen R. Sharpe}, profesor del Departamento de Física de la Universidad de Washington (Seattle, EE. UU.), \vspace{0.2cm}

\noindent \textbf{certifican}:\\[2ex]
\noindent que la presente memoria, ``Kaon decays and other hadronic processes in lattice QCD'', ha sido realizada bajo su direcci\'on en el Instituto de F\'isica Corpuscular, centro mixto de la Universitat de Val\`{e}ncia y del CSIC, por Fernando Romero L\'opez, y constituye su tesis para optar al grado de Doctor en Ciencias F\'isicas.\\[2ex]
\noindent Y para que as\'i conste, en cumplimiento de la legislaci\'on vigente, presentan al Departamento de F\'isica Te\'orica de la Universitat de Val\`{e}ncia la referida Tesis Doctoral, y firman el presente certificado.\\[4ex]

\noindent Val\`{e}ncia, Junio de 2021,\\[12ex]

\noindent \hspace{1cm} Pilar Hern\'andez Gamazo  \hspace{2cm}  Stephen R. Sharpe

\newpage
\thispagestyle{empty}
$$ $$
\newpage
\thispagestyle{empty}

\end{titlepage}


\frontmatter
\fancyfoot[C]{\thepage}

\chapter*{List of Publications}
\addcontentsline{toc}{chapter}{List of Publications}  

This PhD thesis is based on the following publications: 
\begin{itemize}

 \item \emph{Implementing the three-particle quantization condition including higher partial waves}~\cite{Blanton:2019igq}\\
 T. D. Blanton, F. Romero-L\'opez,  and S. R. Sharpe. \\
 \href{http://dx.doi.org/10.1007/JHEP03(2019)106}{\emph{JHEP} \textbf{03} (2019), 106} [\href{http://arxiv.org/abs/1901.07095}{{\tt 1901.07095}}].
 
 \item \emph{Large $N_c$ scaling of meson masses and decay constants} \cite{Hernandez:2019qed}\\
 P.~Hern\'andez, C.~Pena and F.~Romero-L\'opez. \\
\href{https://doi.org/10.1140/epjc/s10052-019-7395-y}{\emph{Eur. Phys. J. C} \textbf{79} (2019) no.10, 865}
[\href{http://arxiv.org/abs/1907.11511}{{\tt 1907.11511}}].

\item \emph{$I=3$ Three-Pion Scattering Amplitude from Lattice QCD}~\cite{Blanton:2019vdk} \\
T.~D.~Blanton, F.~Romero-L\'opez and S.~R.~Sharpe. \\
\href{http://dx.doi.org/10.1103/PhysRevLett.124.032001}{\emph{Phys. Rev. Lett.} \textbf{124} (2020) no.3, 032001} [\href{http://arxiv.org/abs/1909.02973}{{\tt 1909.02973}}].

\item \emph{Dissecting the $\Delta I= 1/2$ rule at large $N_c$}~\cite{Donini:2020qfu}\\
A.~Donini, P.~Hern\'andez, C.~Pena and F.~Romero-L\'opez. \\
\href{http://dx.doi.org/10.1140/epjc/s10052-020-8192-3}{\emph{Eur. Phys. J. C} \textbf{80} (2020) no.7, 638}
[\href{http://arxiv.org/abs/2003.10293}{{\tt 2003.10293}}].

\item \emph{Generalizing the relativistic quantization condition to include all three-pion isospin channels}~\cite{Hansen:2020zhy}\\
M.~T.~Hansen, F.~Romero-L\'opez and S.~R.~Sharpe. \\
\href{http://dx.doi.org/10.1007/JHEP07(2020)047}{\emph{JHEP} \textbf{20} (2020), 047} 
[\href{http://arxiv.org/abs/2003.10974}{{\tt 2003.10974}}]  
[erratum: \href{https://doi.org/10.1007/JHEP02(2021)014}{\emph{JHEP} \textbf{02} (2021), 014}]

\item \emph{Decay amplitudes to three hadrons from finite-volume matrix elements}~\cite{Hansen:2021ofl}\\
M.~T.~Hansen, F.~Romero-L\'opez and S.~R.~Sharpe. \\
\href{http://dx.doi.org/10.1007/JHEP04(2021)113}{\emph{JHEP} \textbf{04} (2021), 113}
[\href{http://arxiv.org/abs/2101.10246}{{\tt 2101.10246}}].
\end{itemize}

Authors are ordered alphabetically, implying an equal contribution to the work. 

\clearpage

The following are additional articles I worked on during my doctorate. They are however not summarized in this document, and will be quoted as ordinary references when necessary.

\begin{itemize}

\item \emph{Nonleptonic kaon decays at large $N_c$}~\cite{Donini:2016lwz} \\
 A.~Donini, P.~Hern\'andez, C.~Pena and F.~Romero-L\'opez. \\
\href{http://dx.doi.org/10.1103/PhysRevD.94.114511}{Phys.\ Rev.\ D {\bf 94} (2016) no.11,  114511}
[\href{http://arxiv.org/abs/1607.03262}{{\tt 1607.03262}}].

\item \emph{{Two- and three-body interactions in $\varphi^4$ theory from lattice simulations}}~\cite{Romero-Lopez:2018rcb}\\
F.~Romero-L\'opez, A.~Rusetsky and C.~Urbach.\\  
 \href{http://dx.doi.org/10.1140/epjc/s10052-018-6325-8}{\emph{Eur. Phys. J. C}  {\bf 78} (2018) no.10, 846}, [\href{http://arxiv.org/abs/1806.02367}{{\tt 1806.02367}}]. 

\item \emph{Numerical exploration of three relativistic particles in a finite volume including two-particle resonances and bound states}~\cite{Romero-Lopez:2019qrt}\\
F.~Romero-L\'opez, \textit{et al}.\\
\href{http://dx.doi.org/10.1007/JHEP10(2019)007}{\emph{JHEP} \textbf{10} (2019), 007} [\href{http://arxiv.org/abs/1908.02411}{{\tt 1908.02411}}].

\item \emph{Scattering of two and three physical pions at maximal isospin from lattice QCD}~\cite{Fischer:2020jzp} \\
M. Fischer, \textit{et al}.\\
\href{https://doi.org/10.1140/epjc/s10052-021-09206-5}{\emph{Eur. Phys. J. C} \textbf{81} (2021) no.5, 436}~ [\href{http://arxiv.org/abs/2008.03035}{{\tt 2008.03035}}]. 

\item \emph{{Relativistic $N$-particle energy shift in finite volume}}~\cite{Romero-Lopez:2020rdq}\\
F.~Romero-L\'opez, A.~Rusetsky, N.~Schlage and C.~Urbach.\\  
 \href{http://dx.doi.org/10.1007/JHEP02(2021)060}{\emph{JHEP} \textbf{02} (2021), 060}, [\href{http://arxiv.org/abs/2010.11715}{{\tt 2010.11715}}].

\item \emph{The Large $N_c$ limit of QCD on the lattice}~\cite{Hernandez:2020tbc}\\
P.~Hern\'andez and F.~Romero-L\'opez. \\
 \href{http://dx.doi.org/10.1140/epja/s10050-021-00374-2}{\emph{Eur. Phys. J. A} \textbf{57} (2021), 52}, [\href{http://arxiv.org/abs/2012.03331}{{\tt 2012.03331}}]. 

\end{itemize}



\chapter*{Preface}
\addcontentsline{toc}{chapter}{Preface} 
\vspace{-0.3cm}

This doctoral thesis deals with the study of properties and interactions of light mesons. More specifically, we focus on hadronic decay and scattering processes, which  are dominated by effects of the strong interaction in the low-energy regime. Concrete examples that will be addressed are the weak decay of a kaon into two pions, and the scattering of three pions.

A peculiarity of the strong interaction is that perturbative expansions fail at hadronic energy scales.  For this reason, genuine nonperturbative tools are essential to obtain first-principles predictions. The central technique employed in this work is Lattice Field Theory, which uses a discretized spacetime to stochastically estimate physical quantities in a quantum field theory. We will also make use of Effective Field Theories, as they provide a complementary description to the dynamics of light hadrons. The mathematical formulation of the strong interaction---Quantum Chromodynamics (QCD)---and the methods to resolve its dynamics will be addressed in Chapter~\ref{sec:LFT}.

The original research of this dissertation is divided in two parts, each with a dedicated chapter.  Chapter~\ref{sec:largeN} describes our study of the 't Hooft limit of QCD using lattice simulations, while in Chapter~\ref{sec:multiparticle} we consider processes that involve multiparticle states.

The 't Hooft limit provides a simplification of nonabelian gauge theories that leads to precise nonperturbative predictions. We will analyze the scaling with the number of colours of various observables, such as meson masses, decay constants and weak transition matrix elements. An important question we address is the origin of the long-standing puzzle of the $\Delta I=1/2$ rule, that is, the large hierarchy in the isospin amplitudes of the $K \to \pi\pi$ weak decay. This is an example in which the 't Hooft limit seems to fail.

Regarding multiparticle processes, we will discuss generalizations of the well-established L\"uscher formalism to explore three-particle processes from lattice simulations. The focus will be on the highlights of our contribution, such as our implementation of the finite-volume formalism that includes higher partial waves, and the first application of the formalism to a full lattice QCD spectrum. We will also comment on the extension of the approach to generic three-pion systems. These will enable lattice explorations of scattering processes in some resonant channels, as well as phenomenologically interesting decays to three pions.

A detailed summary in Spanish of the motivations, methodology, results and achievements of this thesis will be given in Chapter~\ref{sec:resumen}. The final part of the thesis (Part \ref{sec:papers}) includes the peer-reviewed publications that constitute the body of this dissertation. Their original published form has been kept.


\hypersetup{urlcolor=black}
\fancyfoot[C]{\thepage}

\chapter*{Agradecimientos}
\addcontentsline{toc}{chapter}{Agradecimientos}

This has been a long and exciting journey, full of colours and flavours. I have met many people and discovered many places. With these sincere words, I would like to thank all of you for helping me grow as a scientist, and as a person.

A mi directora, Pilar. Me has transmitido una visión creativa y divertida de la física, y me has enseñado a ser inconformista y a cuestionarlo todo (también en lo que va más allá de la física). Siempre me has animado a empujar mis límites, y gracias a eso he llegado más lejos de lo que jamás había imaginado.  Me has dado la máxima libertad, pero siempre has estado ahí para aconsejarme en el camino. El científico que soy y seré te lo debo en gran parte a ti. 

To Steve. When I first arrived in Seattle, I could not even imagine how much that would change me. I really appreciate how welcomed I felt in each visit, and I am genuinely grateful I have the chance to work with you. Solving problems together has been an extremely enjoyable challenge, and what I have learnt from you is simply immeasurable. I have very much enjoyed your company around the world, and I am very looking forward to more of that.

A Carlos. Las visitas a Madrid siempre han sido productivas y muy divertidas. Gracias por tantas conversaciones llenas de contenido de las cuales he aprendido tanto. Te agradezco, además, todos los consejos y el apoyo durante estos años. 

Max, your vision of physics has been tremendously inspiring. Thank you so much for great discussions, a very fruitful collaboration, interesting conversations, and good advice. 

To Akaki, and Carsten. The way I carry out research has your imprint. From you I learnt the basics of many topics I use now every day. Working with you, and learning from you, has been a great pleasure. I am beyond grateful for the support you provided me during various stages of my career. 

I am also thankful to my collaborators in Plymouth. The time I spent there was very fun, and I felt truly welcomed. 

To everyone else I have worked with: Andrea, Raúl, Tyler and the members of the Bonn lattice group. Working with you has been an enriching experience. 

A todos los miembros presentes y pasados de mi grupo, SOM. He disfrutado de vuestra compañía dentro y fuera del IFIC. Especial mención a los latticeros: ¡ha sido un placer trabajar con vosotros!

I cannot forget those, whose support in this journey has been essential. For our everlasting friendship that extends through continents, and for our crazy adventures.  Por descubrir lugares lejanos mano a mano escuchando nuestra playlist. Por surcar los siete mares. Por las risas y sonrisas con los de siempre. Por noches reversibles en cualquier otra parte. Por hacer que el peor año, pueda ser el mejor si estamos juntos. Gracias.

A mi familia. Este doctorado es tan mío como vuestro. Si he llegado hasta aquí, ha sido por vuestro apoyo incondicional a lo largo de los años. Por muy lejos que vaya, siempre estáis conmigo. 

This PhD thesis has been possible thanks to funding from the European Union Horizon 2020 research and innovation program under the Marie Skłodowska-Curie grant agreement No. 713673 and “La Caixa” Foundation (ID 100010434).
\hypersetup{urlcolor=blue}
\fancyfoot[C]{} 
\hypersetup{linkcolor=black} 
\tableofcontents
\definecolor{linkcolour}{rgb}{0.85,0.15,0.15}
\hypersetup{linkcolor=linkcolour}

\mainmatter 
\renewcommand{\headrulewidth}{0.5pt} 



\lhead[{\bfseries \thepage}]{ \rightmark}
\rhead[ Chapter \thechapter. \leftmark]{\bfseries \thepage}
\chapter{Resolving the dynamics of the strong interaction}
\label{sec:LFT}

The strong interaction is one of the fundamental forces known in Nature. Its name originates from the fact that at the femtometer scale it is much stronger than the other three interactions: electromagnetism, the weak force and gravitation. Historically, the study of the strong interaction is tightly linked to nuclear physics. In fact, a well-known manifestation of the strong force is that it holds nucleons (protons and neutrons) together in atomic nuclei. Its strength is such that it overcomes the electromagnetic repulsion of the positively charged protons.

Nowadays, we know that quarks and gluons are the fundamental particles that carry the colour charge responsible for the strong force. Yet, what we observe in experiments are colourless bound states thereof---what we call hadrons. This phenomenon is called confinement, and it will be addressed later in this thesis, along with the mathematical theory behind the strong interaction---Quantum Chromodynamics (QCD). It is interesting to point out that most of the mass of nucleons is the energy of the strong force that binds the constituent quarks. The largest fraction of the mass of the visible Universe has therefore its origin in this interaction.

Whilst QCD is well established, obtaining predictions from first principles is a challenging endeavour. More specifically, methods that compute physical observables by means of perturbative expansions fail to converge in the low-energy regime. The formulation of QCD on a spacetime lattice---lattice QCD---is the state-of-the-art \textit{ab-initio} treatment. It is a numerical approach in which physical observables are obtained from stochastically estimated correlation functions. Lattice QCD has flourished in the last decades achieving a precision matching or exceeding that of experimental measurements in many observables of interest. In addition,  Effective Field Theories (EFTs) provide a complementary tool, based on symmetry relations, which enable the extraction of physical information in an efficient way.

In this first introductory chapter, I will present the mathematical formulation of QCD, along with its peculiarities in comparison to other theories, specifically its low-energy behaviour. Then, I will turn to the discussion of existing methods to solve it. The concept of Effective Field Theories will be introduced in Section~\ref{sec:EFT}, and more specifically, the paradigmatic Chiral Perturbation Theory. The final part of this chapter---Section~\ref{sec:LQCD}---will be dedicated to Lattice QCD.

\section{Quantum Chromodynamics}

The Standard Model (SM) of particle physics is the theory that successfully describes all known
phenomena in the subatomic domain. It is a quantum field theory based on the following gauge symmetry group:
\begin{equation}
 SU(3)_c \otimes SU(2)_L \otimes U(1)_Y,
\end{equation}
which explains the strong and electroweak force between three families of elementary fermions (quarks and leptons). In addition, a scalar sector describes the Higgs force, giving different masses to all the elementary particles. We refer to Quantum Chromodynamics (QCD) as the subset of elementary fields that are charged under the $SU(3)_c$ subgroup.

The matter content in QCD includes the gauge fields, or gluons, and the fermionic fields, or quarks. There are six flavours\footnote{Each quark flavour is abbreviated to the first letter of its name, e.g., $u$ for up.} of the latter (up, down, charm, strange, top and bottom), organized in three families:
\begin{equation}
\begin{pmatrix}
u \\ d
\end{pmatrix}
\begin{pmatrix}
c \\ s
\end{pmatrix}
\begin{pmatrix}
t \\ b
\end{pmatrix}. \label{eq:quarks}
\end{equation}
Each family contains two quarks with different electric charge. The quarks in the upper row of Eq.~(\ref{eq:quarks}) are positively charged ($Q=+2/3$), and the ones in the lower row are negatively charged ($Q=-1/3$). As will be seen in Chapter~\ref{sec:largeN}, electroweak interactions that involve quarks from different families will be a central topic of this thesis. 

The charge of the strong interaction is called colour. The name is an analogy to red, green and blue, as it can take three different values in QCD. More rigorously, (anti)quarks transform under the (anti)fundamental irreducible representation (irrep) of the $SU(3)_c$ colour group. In the absence of interactions, the quark Lagrangian would be 
\begin{equation}
\mathcal{L}_{free} = \sum_{f}\bar q_f (i \gamma_\mu \partial^\mu -m_f )  q_f, \label{eq:freeL}
\end{equation}
where each quark field is really a colour triplet $q_f \equiv (q_f^{(r)}, q_f^{(g)}, q_f^{(b)})$, and $r,g,b$ label the three possible colours. It is the easy to see that $\mathcal{L}_{free}$ is invariant under \textit{global} $SU(3)_c$ transformations. As we will see in Chapter~\ref{sec:largeN}, it will be useful to leave the number of colours in the gauge group, $N_c$, and the number of active flavours, $N_f$, as parameters that one can vary.

The QCD Lagrangian~\cite{Fritzsch:1973pi} follows from imposing the principle of gauge invariance to the Lagrangian in Eq.~(\ref{eq:freeL}). In other words, we promote $SU(3)_c$ to be a {\it local} (gauge) symmetry. This simply means that the colour convention can be chosen locally, without altering the physical outcome. The corresponding gauge transformation of the quark fields is
\begin{equation}
q_f \to U(x) q_f, \text{ with } U(x)= e^{i t_a \theta^a(x)} \in SU(3) \label{eq:gaugeq}
\end{equation}
where $t_a$ are the $SU(3)$ generators (Gell-Mann matrices) and $\theta^a$ are real and scalar functions of the spacetime position. The consequence of this is the need for an additional vector field---the gluon field---that transforms under the adjoint irrep of the gauge group:
\begin{equation}
A_\mu \to U A_\mu  U^\dagger + \frac{i}{g} (\partial_\mu U) U^\dagger.  \label{eq:gaugeA}
\end{equation}
Note that there are 8 gluons, one per generator: $A_\mu = A_\mu^a t_a$.

The most general renormalizable CP-conserving\footnote{C is charge conjugation and P is parity.  CP is the composition of both transformations.} Lagrangian that is invariant under the simultaneous action of the two transformations in Eqs.~(\ref{eq:gaugeq}) and (\ref{eq:gaugeA}) is
\begin{equation}
\mathcal{L}_{QCD} =  \sum_{f }\bar q_f (i \gamma_\mu D^\mu -m_f )  q_f -\frac{1}{2} \tr F_{\mu\nu} F^{\mu\nu} , \label{eq:LQCD}
\end{equation} 
with
\begin{equation}
 D_\mu = \partial_\mu + i g_s t_a A^a_\mu \ \text{ and } \ F_{\mu\nu} = \frac{-i}{g_s} [D_\mu, D_\nu],
\end{equation}
and $g_s$ being the QCD coupling. This simple expression is the Lagrangian of Quantum Chromodynamics. Interactions between quarks and gluons are encoded in the covariant derivative, $D_\mu$. In addition, note that the second term in Eq.~(\ref{eq:LQCD}) is a kinetic term for the gluons, and also includes gluonic self-interactions as $SU(3)$ is nonabelian~\cite{Yang:1954ek}.

A further term that is allowed by gauge invariance is the $\theta$-term:
\begin{equation}
\mathcal{L}_\theta =-\theta\,  N_f  \,\frac{ \alpha_s}{8 \pi}  \tr F_{\mu \nu} \widetilde F^{\mu\nu}, \label{eq:thetaterm}
\end{equation}
where $F^{\mu\nu} = \epsilon^{\mu\nu\rho\sigma} F_{\rho \sigma}$. This term is interesting for various reasons. First, it is a total derivative, and yet its integral is a topological invariant that takes integer values: the topological charge. Second, it violates CP. Since no CP-violation has been found in the strong interactions, the coupling $\theta$ is generally set to zero. It will be however relevant for part of the discussion in Chapter~\ref{sec:largeN}.

While Eq.~(\ref{eq:LQCD}) is rather simple, there remains the question on how to use it for predictions of physical quantities. One would be tempted to use perturbation theory and Feynman diagrams, as is customary for, e.g., Quantum Electrodynamics (QED). However, this will turn out to be useful only in the high-energy regime.

%

\subsection{Asymptotic freedom and confinement}

In contrast to QED, the magnitude of the strong coupling decreases with growing energy, such that $g_s(\mu) \to 0$ when $\mu \to \infty$. This is known as asymptotic freedom.  The understanding of this behaviour has played a crucial role in the development of QCD, as recognized by the 2004 Nobel prize to the discoverers: Gross, Politzer and Wilczek~\cite{Gross:1973id,Politzer:1973fx}. The other side of the coin is that the interactions become strong at lower energies (long distances). This leads to a failure of perturbative expansions, but also to the confinement of quarks and gluons within composite states. These are called hadrons, and they are the asymptotic states of QCD. 

In the framework of perturbative QCD, all quantities can be computed as an expansion in the coupling, $\alpha_s = {g_s^2}/({4 \pi})$.  When considering higher orders in the loop expansion, divergences appear and need to be reabsorbed in a redefinition (renormalization) of the bare gauge coupling and bare quark masses. The regularization procedure introduces an arbitrary energy scale, at which the renormalization condition is set. The fact that observables do not depend on this arbitrary scale leads to a scale dependence of the renormalized coupling. The physical interpretation is that this is the effective coupling at the center-of-mass energy of the process of interest.

 In perturbation theory, the scale dependence of the coupling is described via the beta function:
\begin{equation}
\frac{d \alpha_s}{d \log \mu^2} = \beta(\alpha_s).
\label{eq:beta}
\end{equation}
At one loop~\cite{Gross:1973id,Politzer:1973fx}, it takes the form\footnote{It must be noted that the beta function has been computed up to five loops~\cite{vanRitbergen:1997va,Luthe:2017ttg}.}
\begin{equation}
\beta(\alpha_s) = -\frac{\alpha_s^2}{4\pi} \beta_0  \left[ 1 + \mathcal{O}(\alpha_s) \right], \ \text{ with }  \ \beta_0 = \frac{11}{3} N_c - \frac{2}{3}N_f.
\label{eq:beta1}
\end{equation}
Note that with $N_c=3$ and $N_f\leq 6$, one has $\beta_0 >0$, which ensures a decreasing coupling with increasing energy, {\it ergo}, asymptotic freedom. Combining Eqs.~(\ref{eq:beta}) and (\ref{eq:beta1}), we obtain the one-loop expression for the running coupling:
\begin{equation}
\frac{1}{\alpha_s(\mu^2)}= \beta_0 \log \frac{\mu^2}{\Lambda_{QCD}^2}, \label{eq:alphasQ}
\end{equation}
where $\Lambda_{QCD}$ is an integration constant that fixes the coupling. It has the physical interpretation of a dynamically generated scale that defines the nonperturbative regime, $\alpha_s( \Lambda_{QCD}) \to \infty$. Experimentally, one finds $\Lambda_{QCD} \simeq 300$ MeV. Perturbation theory breaks down around and below that energy scale, and other tools such as effective theories and lattice QCD are essential to study the dynamics of the strong interaction. This will be addressed below in Sections~\ref{sec:EFT} and \ref{sec:LQCD}.

\begin{figure}[h!]
\centering
\includegraphics[width=8cm]{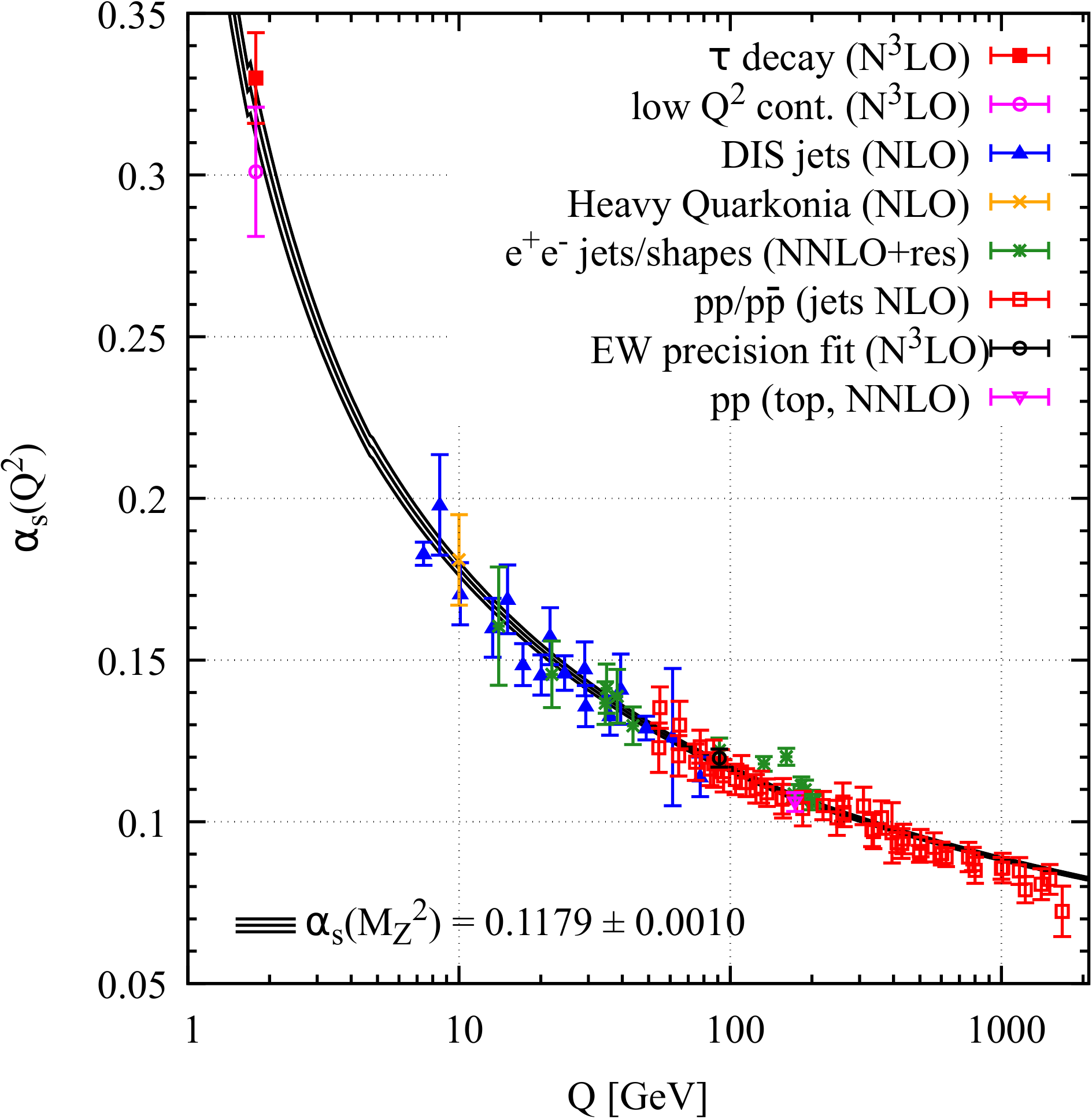}
\caption{Summary of determinations of $\alpha_s$ as a function of the energy scale $Q$. Source: PDG \cite{Zyla:2020zbs}.}
\label{fig:alphas}
\end{figure}

Over the years, experimentalists have collected a plethora of data of the running coupling, along with convincing evidence for asymptotic freedom.  This is summarized in Fig.~\ref{fig:alphas}. 

\subsection{Symmetries in QCD} \label{subsec:QCDsym} 

Symmetries (and symmetry breaking) play a crucial role in the strong interaction. As already mentioned before, the relevant degrees of freedom at low energies are the hadrons. In fact,  the accidental and/or approximate symmetries of the QCD Lagrangian determine to a large extent the properties of hadrons and their interactions.

According to the Noether's theorem \cite{Noether:1918zz}, each continuous symmetry transformation implies a conserved charge. The most obvious example in QCD is a global phase transformation of all quark fields, $q_f \to \exp \left(i \theta\right) q_f$, which leads to baryon number conservation. Since a phase is an element of the group $U(1)$, we will say that this is a symmetry group. In addition, a similar transformation can be applied to each quark independently
\begin{equation}
q_f \to \exp \left(i \theta_f\right) q_f,
\end{equation}
leading to individual quark flavour conservation, e.g., strangeness and charmness conservation. 

Chiral symmetry is the most important one in the description of the low-energy spectrum of QCD. To see this, let us first consider the Lagrangian in Eq.~(\ref{eq:LQCD}) in the massless limit. If we decompose the quark fields in their chiral components:
\begin{equation}
q = \frac{1-\gamma_5}{2} q +  \frac{1+\gamma_5}{2} q =P_L q + P_R q = q_L + q_R  ,
\end{equation}
the Lagrangian takes the form:
\begin{equation}
\lim_{m_f \to 0} \mathcal{L}_{QCD} \supset  \sum_{f} \bar q_{f,R}  (i \gamma_\mu D^\mu  ) q_{f,R} +  \sum_{f} \bar q_{f,L}  (i \gamma_\mu D^\mu  ) q_{f,L}, \label{eq:LQCDm0}
\end{equation}
which means that the two chiralities decouple in the massless limit.  Since a phase transformation can be applied to each flavour and chiral component independently, it is clear that the global symmetry group is
\begin{equation}
 G = U(1)_R \otimes SU(N_f)_R \otimes U(1)_L \otimes SU(N_f)_L. \label{eq:symgroups}
\end{equation}
It will be convenient to take linear combinations of the transformations: vector transformations rephase both chiralities in the same way, while axial transformations do it in opposite directions.

The dynamics of the strong interaction results in a nonvanishing quark condensate,
\begin{equation}
\Sigma = \braket{0 | \bar q q | 0} = \braket{0 | \bar q_L q_R +  \bar q_R q_L | 0} \neq 0,
\end{equation}
which is not invariant under the action of axial transformations. Therefore, the symmetry group is spontaneously broken to the vector subgroup:
\begin{equation}
 G \longrightarrow U(1)_V \otimes SU(N_f)_V,
\end{equation}
where the subscript $V$ indicates vector transformations. It turns out that  the $U(1)_V$ symmetry is just baryon number. Moreover, in the case of only up and down quarks, the $SU(2)_V$ group is related to famous isospin quantum number. Its associated conserved charges are thus the total isospin, and its third component, $I$ and $I_3$.

It is well known that a spontaneously broken global symmetry leads to massless particles, known as Nambu-Goldston bosons (NGB)~\cite{Nambu:1960tm,Goldstone:1961eq,Goldstone:1962es}. The Goldstone theorem states that there are as many massless excitations as broken generators. They have the same quantum numbers as the associated Noether charge, i.e., they are pseudoscalars (spin zero, but negative parity).

The previous discussion is however only valid for QCD with massless quarks. In the real world, the mass term mixes left and right components, and thus the axial symmetries are also explicitly broken. This causes the would-be NGB to obtain a nonzero mass---they become pseudo-Nambu-Goldstone bosons (pNGB). The pNGB can be identified with the three pions ($\pi^\pm$, $\pi^0$), since they are the lightest hadrons in  the QCD spectrum. In the next section, flavour symmetries will be used to classify the hadronic states.

An important point that has been omitted so far is related to the axial $U(1)_A$ symmetry. While at the classical level it is conserved, it is broken at the quantum level by the chiral anomaly~\cite{Adler:1969gk,Bell:1969ts}. One can see this in the fact that the divergence of the conserved current is nonvanishing, and couples to the topological term of QCD:
\begin{equation}
\partial_\mu J^\mu_A = N_f \, \frac{\alpha_s}{8 \pi} F^{\mu\nu} \widetilde {F}^{\mu\nu}, \ \text{ with } \ J^\mu_A = \sum_f \bar q_f \gamma_\mu \gamma_5 q_f. \label{eq:anomaly}
\end{equation}
An elegant explanation for this is that the measure of the path integral is not invariant under axial transformations~\cite{Fujikawa:1979ay}. The chiral anomaly also explains why the $\eta'$ meson it is not a light hadron, i.e., it is heavier than pions, kaons and the eta meson~\cite{tHooft:1976snw,tHooft:1976rip,Callan:1976je,Jackiw:1976pf}. We will come back to the properties of the $\eta'$ meson in Chapter~\ref{sec:largeN}.

\subsection{Low-energy hadron spectrum}

In the early days of the study of the strong interaction, more and more experimental evidence for hadronic states appeared. It then became  clear that a classification scheme ought to be developed. This is the origin of the so-called quark model~\cite{GellMann:1964nj,Zweig:1981pd,Zweig:1964jf}, which in fact precedes the development of QCD. Our present understanding is that hadrons are strongly-interacting particles made up of quarks and gluons. The quark model, at least in its original form, assumes that all the quantum numbers are carried by the quarks within hadrons. The hadrons are thus colourless objects (singlets) of the gauge group, that is, colour is permanently confined.

There are various ways to build up colourless objects with quarks. First, a colour singlet can be made up of a quark-antiquark pair. In the language of group theory, one object in the fundamental irrep and one in the antifundamental irrep may be combined into a singlet: $  \boldsymbol{3_c} \otimes  {\boldsymbol{\overline{3}_c}}   \supset \boldsymbol{1_c} $. The resulting state---a meson---will have an integer spin, and will carry no baryon number. Similarly, three quarks can be combined into colourless state, since $\boldsymbol{3_c} \otimes \boldsymbol{3_c} \otimes \boldsymbol{3_c} \supset \boldsymbol{1_c}$. The composite fermions are called baryons, and they carry one unit of baryon number. Antibaryons can also be built from antiquarks. We will not cover more exotic states such as tetraquarks or pentaquarks, whose existence is under debate.

Let us discuss the case of mesons, which is the main focus of this thesis. A $\bar q q$ state can have total spin $s=0$ and $1$. In the case of zero relative angular momentum, this results into pseudoscalar ($J^P=0^-$) and vector ($1^-$) states. With higher $\ell$, scalar, axial and tensor states can also be constructed. We now consider only states built from $u$, $d$ and $s$ quarks. Thus, we will assume an approximate flavour $SU(3)$ symmetry. A single (anti)quark transforms under the (anti)fundamental irrep of the flavour group. Thus, a single meson state will have either octet or singlet flavour quantum numbers:
\begin{equation}
\boldsymbol{3_f} \otimes \boldsymbol{\overline{3}_f}  \to \boldsymbol{8_f} \oplus \boldsymbol{1_f}.
\end{equation}
Note that the pseudoscalar octet includes the lightest particles, as they are the pNGB of the spontaneously broken axial symmetries. This is confirmed experimentally in the masses of $\pi$, $K$ and $\eta$ mesons. The mass of the pseudoscalar singlet, the $\eta'$, is found to be much heavier than the octet due to the anomaly. As expected, the vector resonances, such as  $\rho(770)$ and $K^*(892)$, are also heavier because they are not pNGB.

\begin{figure}[h!]
\centering
\includegraphics[width=8cm]{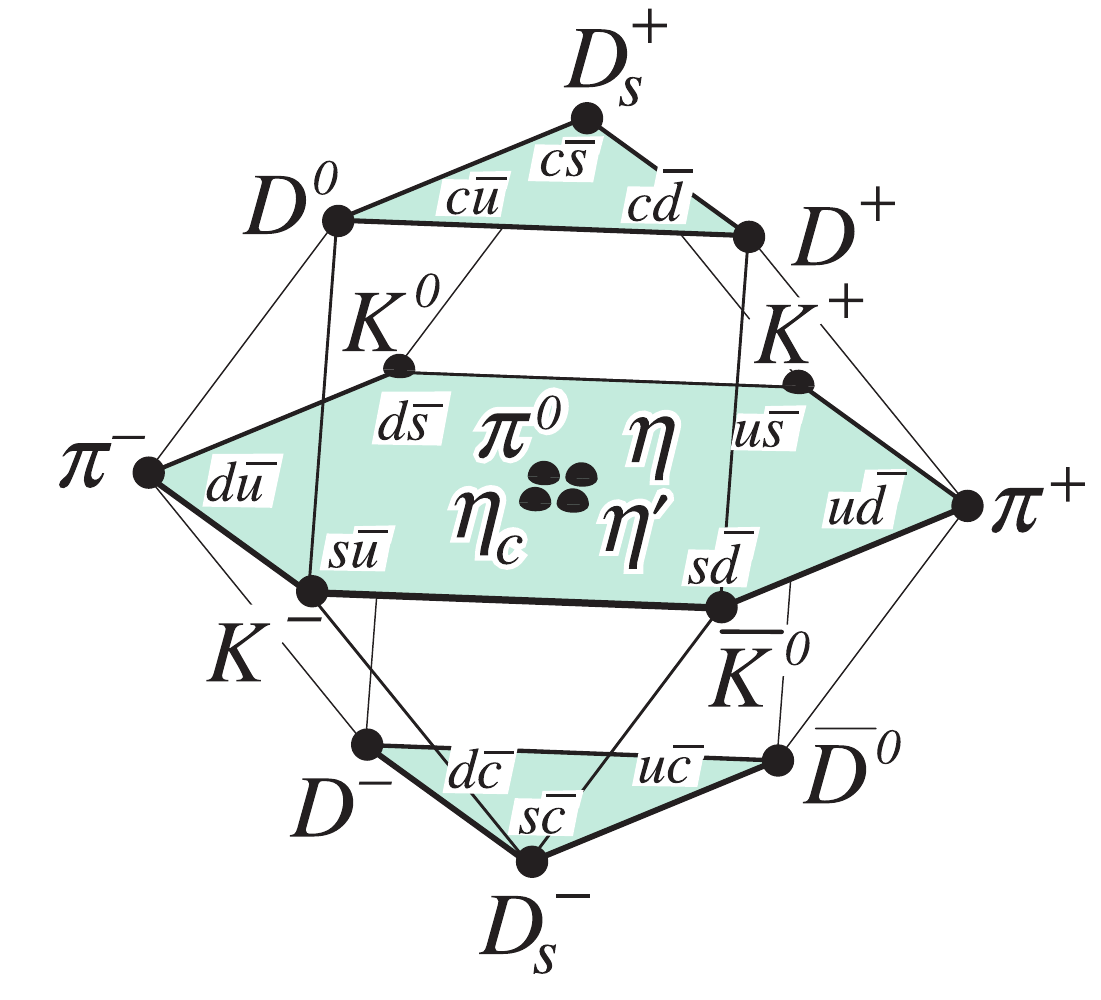}
\caption{Lightest pseudoscalar mesons, and their quark content in the quark model picture. Source: PDG \cite{Zyla:2020zbs}.}
\label{fig:mesons}
\end{figure}

For reasons that will become clear in the next chapter, it is useful to include the charm quark in this analysis ($N_f=4)$. Then, one would have a singlet and a $\boldsymbol{15_f}$ multiplet in quark-antiquark states:
\begin{equation}
\boldsymbol{4_f} \otimes \boldsymbol{\overline{4}_f}  \to \boldsymbol{15_f} \oplus \boldsymbol{1_f}.
\end{equation}
This is illustrated in Fig.~\ref{fig:mesons}, where the $D$, $D_s$ and $\eta_c$ mesons are included. Note that the middle layer corresponds to charmless mesons ($C=0$), which is the case discussed in the previous paragraph (ignoring the $\eta_c$ meson).

A similar classification can be done for baryons states, with the additional difficulty of Fermi statistics. One then concludes that in the $N_f=3$ case, there is a baryon octet (which includes the proton and neutron), and a decuplet (with the $\Delta$ baryons). This is nicely reviewed in the PDG booklet~\cite{Zyla:2020zbs}.

The study of the interactions of the pseudoscalar mesons is the central topic of this thesis. In the following two sections, I will introduce the state-of-the-art techniques for this purpose.

\clearpage
\section{Effective Field Theories \label{sec:EFT}}

Effective Field Theories are a powerful tool to describe the dynamics of a system, without precise knowledge of its high-energy behaviour. Specifically, EFTs incorporate the active degrees of freedom assuming the most general interactions constrained by symmetries. Their range of validity is restricted to energy scales below some cutoff $\Lambda$. At that energy, additional degrees of freedom may become active, or the substructure of existing ones can be resolved. Our modern understanding of EFTs is based upon the unproved, yet unquestioned, theorem of Weinberg~\cite{Weinberg:1978kz}: 
\begin{center}
``\textit{if one writes down the most general possible Lagrangian, including all terms consistent with assumed symmetry principles, and then calculates matrix elements with this Lagrangian to any given order of perturbation theory, the result will simply be the most general possible S-matrix consistent with analyticity, perturbative unitarity, cluster decomposition and the assumed symmetry principles}''.
\end{center}

Before turning to EFTs for QCD, we will discuss the classic example of an effective theory: the Fermi theory. This will be useful to introduce some basic concepts.

\subsection{From the Fermi theory to the strong interaction}

\begin{figure}[h!]
\begin{subfigure}{.5\textwidth}
  \centering
  \includegraphics[width=.8\linewidth]{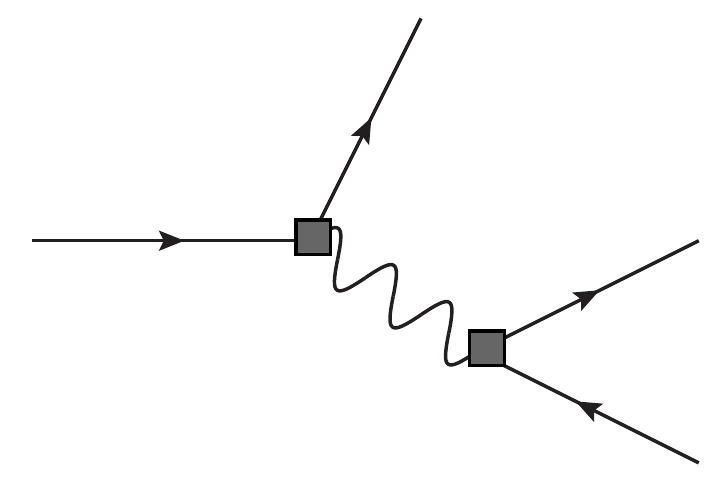}  
  \caption{} \label{fig:fermia}
\end{subfigure}
\begin{subfigure}{.5\textwidth}
  \centering
  \includegraphics[width=.8\linewidth]{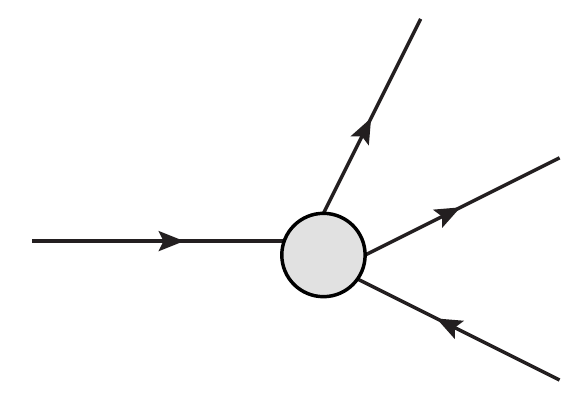}  
  \caption{\label{fig:fermib}}
\end{subfigure}
\caption{Feynman diagrams explaining beta decay in the fundamental electroweak theory (left), and in the effective Fermi theory (right). Solid straight lines are fermions, while wavy lines represent the $W$ boson. }
\label{fig:fermi}
\end{figure}

In the 1930s, Enrico Fermi developed a theory to explain beta decay~\cite{Fermi:1934hr}. His great success was to write down a simple Hamiltonian with four-fermion interactions that could explain the observed beta spectrum. In fact, his proposal preceded the development of the electroweak theory by decades. Nowadays we know that there exists a heavy particle, the $W$ boson with mass $M_W$, whose exchange mediates beta decays, among other processes. At hadronic energy scales, the $W$ boson is much heavier than the typical momentum transfer, and so, the interaction can be approximated by a four-fermion local interaction:
\begin{equation}
\mathcal{L}_\text{Fermi} = G_F \left[ \bar u \gamma_\mu (1 - \gamma_5) d \right]  \left[ \bar e \gamma_\mu (1 - \gamma_5) \nu_e \right].
\end{equation}
In Fig.~\ref{fig:fermi}, both the fundamental (left) and effective (right) interactions are shown.

An important notion in the context of EFTs is the so-called power counting. Thus means that every effective theory has a small expansion parameter, $\delta$. In the case of the Fermi theory, we have  $\delta \sim q^2/M_W^2$, with $q^2$ being the (maximal) momentum transfer in the decay. Thus, the picture of Fig.~\ref{fig:fermib} is only valid up to relative $\mathcal{O}(q^2/M_W^2)$ corrections. 

The connection between the two theories is what we call ``matching''. In this case,  it can be carried out in perturbation theory. The idea is to calculate the same process in the fundamental, and in the Fermi theory using the diagrams in Figs.~\ref{fig:fermia} and \ref{fig:fermib}, respectively. Then, one can relate the respective couplings by equating the amplitudes. This gives:
\begin{equation}
G_F = \frac{g_W^2}{4 \sqrt{2} M_W^2},
\end{equation}
which is the relation between the Fermi constant, $G_F$, and the weak coupling, $g_W$. 

EFTs are also a central subject in QCD. While we have a very successful theory at high-energies with a ``simple'' Lagrangian [see Eq.~(\ref{eq:LQCD})], we also know that the relevant states at low-energies are the hadrons. Due to confinement, the low- and high-energy regime of QCD cannot be matched in perturbation theory, and yet, an EFT description of hadronic interactions is still possible. The hadronic EFT for QCD is Chiral Perturbation Theory, which describes the interactions of pseudoscalar mesons in a consistent power counting at sufficiently low momenta. As this EFT will be particularly important for the dissertation, it will be discussed in detail in the next section.

\subsection{Chiral Perturbation Theory} \label{sec:ChPT}

As explained in Section~\ref{subsec:QCDsym}, the nonsinglet pseudoscalar mesons are the (pseudo-)Nambu-Goldstone bosons that result from the breaking of chiral symmetry. In fact, their Goldstone nature implies strong constraints on their interactions. This can be incorporated into a low-energy EFT description: Chiral Perturbation Theory (ChPT). Early ChPT-like calculations of pion scattering go back to Weinberg in the 1960s \cite{Weinberg:1966kf}, however, a more modern version of ChPT was systematized about a decade later by Weinberg~\cite{Weinberg:1978kz}, as well as Gasser and Leutwyler~\cite{Gasser:1983yg}.

For simplicity, we will first focus on the case of pions ($N_f=2$). As we have seen, we know that the QCD Lagrangian is invariant under the symmetry group $G=SU(2)_L \otimes SU(2)_R$, which is spontaneously broken to $H= SU(2)_V$. This gives rise to three broken generators, and hence, to three pions. Since these fields live in the coset space, that is, $G/H \cong SU(2)$,  their transformation properties are fixed, except for the freedom in the choice of coordinates on $SU(2)$. The standard choice is to use $U(x) \in SU(2)$ with
\begin{equation}
U(x) = \exp \left[i \frac{\phi(x)}{F} \right] \text{, and }  \
 \phi(x) 
 = \begin{pmatrix}
\pi^0(x) & \sqrt{2} \pi^+(x)  \\
\sqrt{2} \pi^-(x)  & -\pi^0(x) 
\end{pmatrix}, \label{eq:pionfields}
\end{equation}
where $F$ is a constant with units of energy that will be defined below. This object transforms under the action of the group $G$ as
\begin{equation}
U'(x) = R U(x) L^\dagger,  \label{eq:piontrafo}
\end{equation}
with $R \in SU(2)_R$, and similarly for $L$.

Following Weinberg's rule, we should write down the most general Lagrangian using the object in Eq.~(\ref{eq:pionfields}) that is consistent with chiral symmetry. Since we aim at describing the low-momentum regime, this Lagrangian will be organized in (even) powers of momentum, or equivalently, derivatives.  The only allowed term with no derivatives is a meaningless constant  in the Lagrangian, because $U^\dagger U = 1$. Thus, the lowest order Lagrangian has two derivatives:
\begin{align}
&\mathcal{L}_2 = \frac{F^2}{4} \tr \left[ \partial_\mu U  \partial^\mu U^\dagger  \right],
\end{align} 
and will be given in terms of an unknown coupling, $F$. This quantity will be very important throughout this work, because it is the pion decay constant\footnote{We use the $F\simeq 87$ MeV normalization throughout this work.} in the chiral limit. Note that a transformation like that in Eq.~(\ref{eq:piontrafo}) leaves $\mathcal{L}_2$ unchanged.

While the previous Lagrangian describes the dynamics of massless pions at low energies, we also know that chiral symmetry is explicitly broken by the mass term. The way to incorporate this is to treat the mass as an external source. For this, we introduce a spurion field, $\chi$, that transforms as $\chi' \to  R \chi L^\dagger$, and whose expectation value is related to the quark mass. This way, an additional operator is invariant under chiral symmetry:
\begin{equation}
\tr \left[ U \chi^\dagger  + U^\dagger \chi \right].
\end{equation}
Therefore, the most general Lagrangian at this order becomes:
\begin{align}
&\mathcal{L}_2 = \frac{F^2}{4} \tr \left[ \partial_\mu U  \partial^\mu U^\dagger  \right]
+ \frac{ B F^2}{2 } \tr \left[ U \chi^\dagger  + U^\dagger \chi \right], \label{eq:fullL2}
\end{align} 
where $B$ is an additional effective coupling related to the quark condensate. In isospin-symmetric QCD, we have $\chi =  \text{ diag } (m,  m)$, where $m$ is the quark mass.
Expanding to $\mathcal{O}(\phi^2)$, we have
\begin{equation}
\mathcal{L}_2 = \frac{1}{4} \tr \partial_\mu \phi  \partial^\mu \phi  - \frac{2 B m}{4} \tr \phi^2  + \mathcal{O}(\phi^4) \supset \partial_\mu \pi^+ \partial^\mu \pi^-  -\, 2 B m \pi^+ \pi^-,
\end{equation}
which means that $M^2 = 2 B m$, with $M$ being the tree-level mass of the pions. The beauty of Eq.~(\ref{eq:fullL2}) is that it describes the QCD dynamics at low energies in terms of only two unknown couplings, $F$ and $2Bm$, which may be fixed by experimental input.

The previous discussion is also valid when the strange quark is included. This is called $N_f=3$ ChPT, for which the Goldstone fields looks like:
\begin{equation}
 \phi
 = \begin{pmatrix}
\pi^0  + \frac{1}{\sqrt{3}} \eta& \sqrt{2} \pi^+ & \sqrt{2}K^+  \\
\sqrt{2} \pi^-  & -\pi^0 + \frac{1}{\sqrt{3}} \eta& \sqrt{2}K^0 \\
\sqrt{2}K^-& \sqrt{2}\bar K^0& - \frac{2}{\sqrt{3}} \eta
\end{pmatrix}.
\end{equation}
The Lagrangian is formally\footnote{We also use the same name for the effective couplings, although their values depend implicitly in $N_f$.} identical to that of Eq.~(\ref{eq:freeL}), although including the strange quark mass, $m_s$. Therefore, one has $\chi = \text{ diag } (m,  m, m_s)$.

At this point, it will be useful to discuss in more detail the power counting in ChPT, and its range of validity. As we have seen, at leading order an operator with two derivatives appears together with the mass term. This way, we should have $O(p^2) \sim O(m) \sim O(M^2)$  in the low-momentum expansion. We also expect that the expansion parameter is
\begin{equation}
\delta \sim \frac{M^2}{\Lambda_\chi^2} \sim \frac{p^2}{\Lambda_\chi^2}, 
\end{equation}
where $\Lambda_\chi$ should correspond to the high-energy scale at which the chiral expansion breaks down. Thus, $\Lambda_\chi$ must be of the order of the mass of lightest resonance in the QCD spectrum. A standard choice is $\Lambda_\chi = 4 \pi F_\pi$, as it naturally appears in perturbative calculations in ChPT~\cite{Georgi:1985kw}. Numerically, $4 \pi F_\pi$ is of the order of 1 GeV, and it is not far from the mass of the $\rho$ resonance. 

Although $\mathcal L_2$ is very predictive, higher-order corrections are to be expected, and could be significant in some observables. To improve on this, one would like to construct the next-to-leading-order (NLO) Lagrangian
\begin{equation}
\mathcal{L}_4= \sum_i L_i \mathcal{O}_i,
\end{equation} 
which in the chiral power counting is $O(p^4)$. The operators $\mathcal{O}_i$ will be Lorentz-invariant and chirally-symmetric combinations of  $\partial_\mu U$ and $\chi$, such as:
\begin{equation}
\mathcal{O}_5 = \tr\left[ \partial_\mu U^\dagger   \partial^\mu \left( U \chi^\dagger + U^\dagger \chi   \right)  \right]. \label{eq:O5}
\end{equation}
While for $SU(N_f)$ ChPT there are 11 linearly independent terms, some relations exist in the case of $SU(3)$ and $SU(2)$, reducing the number of independent operators to 10 and 8, respectively. The arbitrary couplings that multiply the operators in the Lagrangian, $L_i$, are called Low Energy Constants (LECs). The full list of the operators can be found in these reviews~\cite{Pich:1995bw,Scherer:2002tk}.

An important point concerns renormalization in ChPT. When calculating observables in this EFT, one can see that the tree-level diagrams from $\mathcal{L}_4$, and the one-loop contributions from $\mathcal{L}_2$ have the same power of $\delta$ in the momentum expansion. As usual, loop diagrams can be divergent, requiring a renormalization procedure. In ChPT the solution is to absorb the infinities of loops from $\mathcal{L}_2$ by an appropriate renormalization of the NLO LECs that appear in $\mathcal{L}_4$~\cite{Gasser:1984gg}. Thus, we say that ChPT is renormalizable order by order.

During the present dissertation, we will make use of various ChPT predictions. The results in Refs.~\cite{Bijnens:2009qm,Bijnens:2013yca,Bijnens:2011fm} will be of special importance, as they include ChPT calculations for generic $N_f$ theories. Specifically, the $N_f=4$ results will be used in Chapter~\ref{sec:largeN}, while ChPT predictions for pion scattering will be needed in Chapter~\ref{sec:multiparticle}.

\clearpage
\section{Lattice QCD\label{sec:LQCD}}

The formulation of QCD on the lattice is due to the work of Kenneth Wilson in the 1970s~\cite{Wilson:1974sk} (see also~\cite{Wilson:1975id}). 
Today, lattice QCD (LQCD) is a well-established {\it ab-initio} approach to solve the dynamics of the strong interaction in the nonperturbative regime.

Lattice calculations rely on high-performance computing. In recent decades, technological and algorithmic advances have enabled enormous progress in LQCD. In fact, the uncertainty achieved in lattice results is comparable to the experimental one in many relevant quantities, e.g., the violation of CP in kaons ($\epsilon'/\epsilon$). An additional example---very important in this thesis---are three-particle scattering quantities. While LQCD calculations already exist, they are difficult to access experimentally.

Another interesting point about LQCD is the following. In real-world measurements we are limited to a specific value of quark masses, number of flavours, and number of colours. In contrast, we can pick our simulation parameters on the lattice, and so it is an excellent tool to experiment with QCD, and explore various nonabelian gauge theories.

In this section, we will review the formulation of QCD on the lattice. Part of the discussion will be based on existing reviews~\cite{Gupta:1997nd,Luscher:1998pe,Hernandez:2009zz}. 

\subsection{Preliminaries}

The key feature of LQCD is that the theory can be treated as a statistical system.  Here, we will introduce the relevant concepts and definitions using the simplest case of a scalar theory.

Let us start with a complex scalar theory with a $U(1)$ symmetry, whose Lagrangian is
\begin{equation}
\mathcal{L} = \partial_\mu \phi^\dagger \partial^\mu \phi - V(|\phi|).
\end{equation}
In the path integral formulation of a quantum field theory\footnote{Based on Feynman's path integral formulation of quantum mechanics~\cite{Feynman:1948ur}.}, the partition function takes the form:
\begin{equation}
\mathcal{Z} = \int D\phi \, e^{i S[\phi]}, \text{ with } S[\phi] = \int d^4x\, \mathcal{L},
\end{equation}
where $S[\phi]$ is the action, and the integral is over all possible field configurations, that is, all possible values of the field $\phi(x)$. As can be seen, $\mathcal{Z}$ is complex and does not allow for a simple statistical treatment. However, this can be solved by performing a Wick rotation to the so-called Euclidean time ($x^0 \to -i x_0^E$). This way, the action becomes:
\begin{equation}
S = \int d^4 x\, \mathcal{L} \longrightarrow i S_E = i \int d^4 x\, \mathcal{L}_E,
\end{equation}
with
\begin{equation}
\mathcal{L}_E = \partial_\mu \phi^\dagger \partial_\mu \phi + V(|\phi|).
\end{equation}
Note that the double subscript implies Euclidean metric. It is now clear that the partition function is strictly real:
\begin{equation}
\mathcal{Z} = \int \, D \phi e^{- S_E[\phi]}, \text{ with } S_E[\phi] = \int d^4x \, \mathcal{L}_E, \label{eq:Zeur}
\end{equation}
and it has now statistical meaning\footnote{Assuming that the potential is bounded from below.}, since the exponential may be interpreted as a
Boltzmann weight factor. Hence, the dynamics of this theory will be the consequence of a statistical average over all possible field configurations with weight $\exp \left(-S_E\right)$. The configurations contributing the most are the ones near the minimum of the action (its classical solutions). 

All the physical information of the theory is contained in the Euclidean correlation functions. These are defined as the expectation value of a product of local fields. For instance, the two-point function in the scalar theory is:
\begin{equation}
C(x-y) = \langle \phi(x) \phi(y) \rangle = \frac{1}{\mathcal{Z}}  \int D \phi \,\, \phi(x) \phi(y)    e^{- S_E[\phi]} . \label{eq:corrphi}
\end{equation}
As we will see later, from correlation functions we can extract energy levels---the spectrum---or the $S$-matrix elements.

The Euclidean continuum theory needs to be discretized, so that it can be solved by numerical methods. We define the physical fields on a lattice with $T$ points in the time direction, and $L$ points in each of the three spatial directions. For the scalar theory, the discretization is achieved by replacing derivatives by forward differences:
\begin{equation}
 \partial_\mu \phi(x)  \to \hat \partial_\mu \phi(x)   = \frac{1}{a} \left[  \phi(x+a \hat \mu) - \phi(x) \right],
\end{equation}
where $a$ is the lattice spacing and $\hat \mu$ is a unit vector in the direction $\mu$. One must also choose the boundary conditions, typically, periodic boundary conditions are considered. 

The final ingredient is a numerical method to compute correlation functions, which involves a multidimensional integral over $T \times L^3$ complex variables in the complex scalar theory.
To do so, Monte Carlo methods are combined with importance sampling techniques. The main idea is to generate field configurations, $\{\phi_i\}$ , distributed according to the probability distribution:
\begin{equation}
p[\{\phi_i\}] = \frac{1}{\mathcal{Z}} e^{-S_E[\phi]}.
\end{equation} 
Then, the expectation value of any observable can be calculated as:
\begin{equation}
\langle \mathcal{O} \rangle = \frac{1}{\mathcal{Z}}  \int D \phi\, \mathcal{O}(\phi) \, e^{- S_E[\phi]} \simeq \frac{1}{N_\text{conf}} \sum_{i=1}^{N_\text{conf}} \mathcal{O}(\{\phi_i\}) + O\left( \frac{1}{\sqrt{N_\text{conf}}}\right),
\end{equation}
that is, an average over the field configurations. In order to obtain a sequence of configurations with the appropriate distribution, one can use Markov-chain Monte-Carlo methods. Modern lattice QCD calculations use the Hybrid Monte Carlo (HMC) algorithm~\cite{Duane:1987de}, which combines molecular dynamics with a Metropolis accept-reject step~\cite{Metropolis:1953am,Hastings:1970aa}.

Observables calculated on the lattice suffer from discretization effects. In order to get rid of them, one must perform a continuum extrapolation by simulating at different values of the lattice spacing. In addition, quantities on the lattice are affected by finite-volume effects. These can be avoided if $L$ and $T$ are much larger than the longest correlation length in the theory, which is the inverse of the mass of the lightest particle in the spectrum. However, as we will see in Chapter~\ref{sec:multiparticle}, some finite-volume effects can be used in our favour to study scattering processes.

While the scalar theory is useful to introduce some concepts, it does not have two complications present in QCD:  fermions and gauge symmetry. These will be addressed in the subsequent sections.

\subsection{Fermions in lattice QCD}

Unlike for scalars, the naive discretization of fermions is not enough, due to the problem of fermion doubling. We discuss the origin of this, and how it can be cured.

Let us first consider free fermions. We recall that the Euclidean continuum Lagrangian can be written as 
\begin{equation}
\mathcal L (\psi, \bar \psi) = \frac{1}{2} \left( \bar \psi \gamma_\mu \partial_\mu \psi  - \partial_\mu \bar \psi \gamma_\mu \psi  \right) + m_0 \bar \psi \psi.
\end{equation}
In the previous equation, we can pick the chiral representation of the $\gamma_\mu$ matrices:
\begin{equation}
\gamma_0 = \begin{pmatrix}
0 & -I \\ -I & 0
\end{pmatrix}, \text{ and } \gamma_k = \begin{pmatrix}
0 & -i \sigma_k \\ i \sigma^\dagger_k & 0
\end{pmatrix},
\end{equation}
where $I$ is the $2 \times 2$ identity, and $\sigma_k$ are the Pauli matrices. The discretization can be achieved replacing derivatives with finite differences. The resulting action can be written in a compact manner:
\begin{align}
\begin{split}
S[ \psi, \bar \psi ] &= a^4 \sum_x \bar \psi(x)\left[ \frac{1}{2} \gamma_\mu(\hat \partial_\mu + \hat \partial^*_\mu)    + m_0 \right]\psi(x) \\ & \equiv a^4 \sum_{x,y} \bar \psi_\alpha(x) D^{\alpha \beta}_{xy} \psi_\beta(y),
\end{split}
\end{align}
where $\hat \partial_\mu$($\hat \partial^*_\mu$) is the forward(backward) difference operator, and the discrete Dirac operator is
\begin{equation}
D^{\alpha \beta}_{xy} =  \sum_\mu \frac{1}{2a} (\gamma_\mu)_{\alpha \beta} \left[ \delta_{y, x+a\hat \mu} + \delta_{y, x-a\hat \mu} \right] + m_0 \delta_{\alpha \beta} \delta_{x,y}. \label{eq:Dnaive}
\end{equation}
In momentum space, the previous equation takes the form:
\begin{equation}
D^{\alpha \beta}_{pk} =   (2 \pi a)^4 \delta( p + k) \left( \sum_\mu \frac{i}{a} (\gamma_\mu)_{\alpha\beta} \sin (a p_\mu) + \delta_{\alpha\beta} m_0  \right),
\end{equation}
and so, the Fermi propagator becomes
\begin{equation}
\langle \psi(x) \bar \psi(y) \rangle_F = \int_\text{BZ} \frac{d^4 k}{(2 \pi)^4} \frac{e^{ik(x-y)}}{ m_0 + \sum_\mu i \gamma_\mu \frac{\sin k_\mu a}{a} }, \label{eq:propF}
\end{equation}
where the integral runs over the Brillouin zone, i.e., $p_\mu \in [-\pi/a, +\pi/a]$.

By exploring Eq.~(\ref{eq:propF}), we can understand the particle content of this discretized theory. One-particle states correspond to poles in the Fermi propagator. As can be seen in Eq.~(\ref{eq:propF}), there is one at $k_\mu \sim 0$, but also more at the end of the Brillouin zone in each direction, that is, when $k_\mu \sim \pi/a$. In total, one has $2^d$ poles, where $d$ is the number of space-time dimensions. The interpretation behind this fact is that this discretization really describes $2^d$ continuum fermions, that is, 16 mass-degenerate quarks in QCD. This undesirable situation is usually referred to as fermion doubling~\cite{Wilson:1975id,Susskind:1976jm}. It is in fact a general result for all discretizations of the Dirac operator under very general assumptions: the Nielsen-Ninomiya no-go theorem~\cite{Nielsen:1981hk}. The statement is that any local, hermitian, fermionic lattice action, that has chiral symmetry and translational invariance, will necessarily have fermion doubling. 

Let us now discuss Wilson's solution to fermion doubling---the so-called Wilson fermions~\cite{Wilson:1975id}. His proposal was to give up chiral symmetry by adding the following term (``Wilson term'') to the action
\begin{equation}
\Delta S_W = - \frac{r}{2} a^5 \sum_x \bar \psi(x) \hat \partial^*_\mu \hat \partial_\mu \psi(x) ,
\end{equation}
where $r=1$ was Wilson's choice. Note that the corresponding Dirac operator maintains C, P and T invariance\footnote{T is the time-reversal transformation.} as well as, $\gamma_5$-hermiticity: 
\begin{equation}
D^\dagger = \gamma_5 D \gamma_5.
\end{equation}
The Feynman propagator then becomes:
\begin{equation}
\langle \psi(x) \bar \psi(y) \rangle_F = \int_\text{BZ} \frac{d^4 k}{(2 \pi)^4} \frac{e^{ik(x-y)}}{ m_0 + \sum_\mu i \gamma_\mu \frac{\sin k_\mu a}{a}  + \frac{r}{a} \sum_\mu (1 - \cos a k_\mu) }. \label{eq:propFW}
\end{equation}
As can be seen, in the $a \to 0$ limit, $k_\mu \sim 0$ yields the correct continuum denominator. However, around $k_\mu \sim \pi/a$ the last term becomes a  $O(a^{-1})$ contribution to the mass of the doublers.
 Consequently, they decouple in the continuum limit, as they become infinitely heavy. In practice, there is a price to pay for a broken chiral symmetry: (i) low-momentum modes are affected by discretization effects of $O(a)$, as opposed to $O(a^2)$ if chiral symmetry is preserved, and (ii) some quantities, such as the quark mass, get both additive and multiplicative renormalization
\begin{equation}
m_R = Z_m(m_0-m_c),
\end{equation}
where $m_c$ is the so-called critical mass. Since $m_0$ and $m_c$ are linearly divergent in the cutoff, some fine tuning will be needed to take the continuum limit at fixed renormalized mass.

\subsection{Gauge symmetry on the lattice}

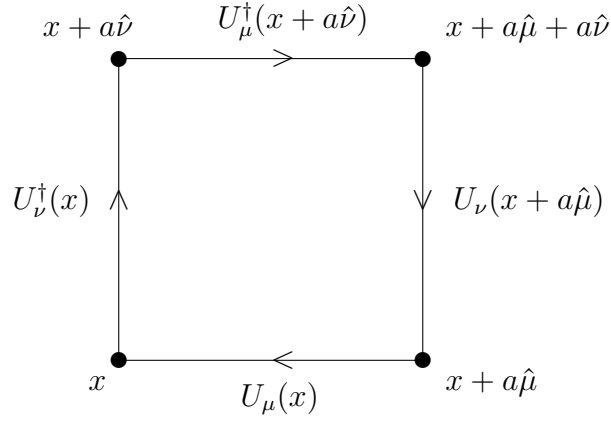
\begin{figure}
\setlength{\unitlength}{0.1cm}
\begin{center}
\begin{picture}(40,60)
\put (0,10){\line(1,0){40}}
\put (40,10){\line(0,1){40}}
\put (40,50){\line(-1,0){40}}
\put (0,50){\line(0,-1){40}}
\put(20,9.){$<$}
\put(16,4){$U_\mu(x)$}
\put(38.6,30){$\vee$}
\put(44,30){$U_\nu(x+a \hat{\mu})$}
\put(-1.4,30){$\wedge$}
\put(-14,30){$U^\dagger_\nu(x)$}
\put(20,49){$>$}
\put(13,54){$U^\dagger_\mu(x+a\hat{\nu})$}
\put(0,10){\circle*{2}}
\put(-4,6){$x$}
\put(40,10){\circle*{2}}
\put(43,6){$x+a \hat{\mu}$}
\put(40,50){\circle*{2}}
\put(43,53){$x+a \hat{\mu} +a \hat{\nu}$}
\put(0,50){\circle*{2}}
\put(-10,53){$x+a \hat{\nu}$}
\end{picture}
\caption{Representation of a plaquette, $U^\text{plaq}_{\mu\nu}$.}
\label{fig:plaquette}
\end{center}
\end{figure}

The treatment of gauge symmetries on the lattice also goes back to the \textit{magnum opus} of Wilson~\cite{Wilson:1974sk}. While the continuum gauge fields belong to the algebra of the gauge group, in the Wilsonian formulation, the gauge field is represented by an element of the gauge group, i.e., $SU(3)$ for QCD. If the discretized fields are assigned to the lattice sites, the gauge fields are assigned to the links between two neighbouring sites. A link is characterized by a position, $x$, and a direction $\mu$, $U_\mu (x)$. This way, we have:
\begin{equation}
U_\mu (x)  = e^{i a g_0  A_\mu(x) }, \ \text{ with } \ A_\mu = t_a A^a_\mu, \label{eq:Adisc}
\end{equation}
and gauge transformations act as:
\begin{equation}
U_\mu (x) \to \Omega(x) U_\mu (x) \Omega^\dagger(x+a\hat \mu),\ \text{ with } \ \Omega \in SU(3).
\end{equation}
Note that the gauge link transforms as a parallel transporter between two adjacent points, $x$ and $x + a\hat \mu$. The smallest, and most local, combination of links that is gauge invariant is the plaquette:
\begin{equation}
\tr U^\text{plaq}_{\mu\nu} = \tr \left( U_\mu (x) U_\nu (x + a \hat \mu) U^\dagger_\mu (x+a \hat \nu) U^\dagger_\nu (x) \right).
\end{equation}
A graphical representation of a plaquette is shown in Fig.~\ref{fig:plaquette}. In the naive continuum limit, the plaquette is related to the field strength tensor as 
\begin{equation}
U^\text{plaq}_{\mu\nu} = e^{- i a^2 g_0 F_{\mu\nu} + O(a^3)}.
\end{equation}
Therefore, the lattice action
\begin{equation}
S^\text{plaq}_\text{YM}[U] =\frac{\beta}{2N_c} \sum_{\mu\nu} \sum_x \text{Re }\tr\left(1-U^\text{plaq}_{\mu\nu}\right),
\end{equation}
with $\beta={2 N_c}/{g_0^2}$, becomes the Euclidean action of a pure Yang-Mills theory in the continuum limit:
\begin{equation}
S^\text{plaq}_\text{YM}[U]= \int d^4 x \frac{1}{2} \tr F_{\mu\nu} F_{\mu\nu}+ \mathcal{O}(a^2).
\end{equation}

We can also add fermions in the fundamental representation of the gauge group, which transform as $\psi(x) \to \Omega(x) \psi(x)$. Then, the coupling of these fermions to the gauge fields can be incorporated in a gauge invariant way by replacing the discrete derivatives with a discrete analogue of the covariant derivative:
\begin{align}
\hat \partial_\mu \psi &\to \nabla_\mu \psi = \frac{1}{a} \left[ U_\mu (x) \psi(x+a\hat \mu) - \psi(x)  \right], \\
\hat \partial^*_\mu \psi &\to \nabla^*_\mu \psi = \frac{1}{a} \left[\psi(x)  -  U^\dagger_\mu (x -a\hat \mu) \psi(x-a\hat \mu)  \right].
\end{align}
Note that using Eq.~(\ref{eq:Adisc}), one has $\nabla_\mu \psi  = (\partial_\mu + ig_0A_\mu) \psi + O(a)$. It can be easily seen that the combination $\bar \psi(x) \nabla_\mu \psi(x)$ is gauge invariant. 

\subsection{The Lattice QCD action(s)}

Let us now consider QCD. There is not a unique discretization of theory, but as long as the degrees of freedom and symmetries are preserved, all versions should lead to the same continuum results. We  briefly describe the LQCD actions that have been used in our work, indicating the advantages of each choice. Improved actions will be of special relevance, as they suffer from less cutoff effects.

The standard Wilson formulation of QCD is given by the following Euclidean action:
\begin{equation}
S_{LQCD} = S^\text{plaq}_\text{YM}[U]  + a^4 \sum_f \sum_{x,y} \bar \psi_f(x) D^W_{xy}  \psi_f(y), \label{eq:SWLQCD}
\end{equation}
with
\begin{align}
\begin{split}
D^\text{W} _{xy} = \delta_{xy} - \kappa_f \bigg [ \sum_\mu& (1- \gamma_\mu) U_\mu(x) \delta_{x,y-a \hat \mu} \\ +&(1+ \gamma_\mu) U^\dagger_\mu(x- a \hat \mu) \delta_{x,y+a \hat \mu} \bigg ],
\end{split}
\end{align}
where $\kappa_f = (2 a m_f + 8)^{-1} $, and the fermion fields have been rescaled with respect to those in Eq.~(\ref{eq:Dnaive}) as $\psi_f \to \psi_f/\sqrt{2 \kappa_f}$. As a consequence of the breaking of chiral symmetry, the action in Eq.~(\ref{eq:SWLQCD}) leads to $O(a)$ corrections to physical quantities. While this is acceptable in principle, the cutoff effects can be sizeable at the typical values of the lattice spacing that can be simulated. Thus, a reliable continuum extrapolation becomes computationally expensive.

Alternative fermionic discretizations are also available, e.g., staggered fermions~\cite{Kogut:1974ag}, or domain-wall fermions~\cite{Kaplan:1992bt}. We will not discuss them further as they are not used in this dissertation.

\subsubsection{Twisted-mass fermions}

A variation of Wilson fermions that we have used are twisted-mass Wilson fermions~\cite{Frezzotti:2000nk} (see Ref.~\cite{Shindler:2007vp} for review). It uses a Dirac operator with a chirally-rotated Wilson term:
\begin{equation}
D = \frac{1}{2} \big\{ \gamma_\mu(\hat \nabla_\mu + \hat \nabla^*_\mu) \tau_0 - a e^{-i  \omega \gamma_5 \tau_3}\nabla_\mu \nabla^*_\mu \big\}  + m \tau_0,
\end{equation}
which acts on a flavour doublet of quark fields, $\psi$. In the previous equation, $\omega$ is the so-called twist angle. Moreover, $\tau_3$ and $\tau_0$ are matrices in flavour space---the third Pauli matrix and the identity, respectively. Upon the following change of variables:
\begin{equation}
\psi = e^{i \frac{\omega}{2} \gamma_5 \tau_3} \chi, \quad \bar \psi =\bar \chi e^{i \frac{\omega}{2}  \gamma_5 \tau_3},
\end{equation}
the operator becomes
\begin{equation}
D = \frac{1}{2} \big\{ \gamma_\mu(\hat \nabla_\mu + \hat \nabla^*_\mu)  - a \nabla_\mu \nabla^*_\mu \big\} \tau_0  +  m e^{i  \omega \gamma_5 \tau_3} .
\end{equation}
The new $\chi$ variables usually receive the name of twisted basis. In this basis, the mass term can be written as:
\begin{equation}
m e^{ i \omega \gamma_5 \tau_3}   = m \left( \tau_0  \cos \omega + i \tau_3 \sin \omega  \right).
\end{equation}
A favourable situation is achieved at maximal twist ($\omega=\pi/2$), for which the mass term becomes purely imaginary. In this case, the action also has an exact flavoured chiral symmetry in the physical basis:
\begin{equation}
\psi \to e^{ i \theta \gamma_5 \tau_{k}} \psi, \quad \text{ with } k=1,2.
\end{equation} 

A subtlety here is the renormalization. The imaginary part of the mass renormalizes multiplicatively, while the real part additively. Therefore, one requires some fine tuning to achieve maximal twist in a nonperturbative way. In practice, the bare twisted-mass lattice action is
\begin{equation}
S^\text{TM}= a^4 \sum_x \bar \chi\left[ \frac{1}{2} \big\{ \gamma_\mu(\hat \nabla_\mu + \hat \nabla^*_\mu)  - a\nabla_\mu \nabla^*_\mu \big\} \tau_0  + m_0 \tau_0 + i \mu_0 \gamma_5 \tau_3\right]\chi,
\end{equation}
where $m_0$ and $\mu_0$ are now bare parameters, and the latter is called the bare twisted mass. Maximal twist is ensured if $m_0$ is tuned to its critical value. 

There are important advantages of twisted-mass QCD at maximal twist: (i) $\mu_0$ plays the role of the bare quark mass that renormalizes multiplicatively, (ii) the axial current associated with the exact chiral symmetry does not requiere renormalization, and (iii) physical observables are only affected by $O(a^2)$ effects, i.e, there is automatic $O(a)$-improvement~\cite{Frezzotti:2003ni}. A clear disadvantage is that isospin symmetry and parity are broken by cutoff effects, which implies for instance that charged and neutral pions are nondegenerate. Although this is an $O(a^2)$ effect, it is found to be numerically significant.

\subsubsection{Improved actions} \label{sec:improvedact}

Improved actions are discretizations with a better scaling to the continuum\footnote{A discussion about this can be found in Ref.~\cite{Sommer:1997jg}.}. They are especially useful in the case of Wilson fermions, since they eliminate the leading $O(a)$ cutoff effects.

The improvement procedure is also referred to as Symanzik improvement~\cite{Symanzik:1983pq,Symanzik:1983dc}.  The key point is that close to the continuum limit the lattice theory may be described in terms of a local EFT:
\begin{equation}
\mathcal{L}_\text{eff} = \mathcal{L}_0 + a \mathcal{L}_1 + a^2 \mathcal{L}_2 + \hdots,
\end{equation}
where $\mathcal{L}_0$ is the continuum Lagrangian, and $\mathcal{L}_1, \mathcal{L}_2,$ etc., are linear combinations of local, gauge-invariant operators:
\begin{equation}
\mathcal{L}_k =  \sum_i \, c_i \mathcal{O}^k_i(x).
\end{equation}
Here, the operators $\mathcal{O}^k_i(x)$ have dimension $4+k$, and they respect the symmetries of the lattice theory. For the case of Wilson fermions, it can be seen that the only relevant operator at dimension 5 is:
\begin{equation}
\mathcal{O}^1 = i\bar \psi \sigma_{\mu\nu} F_{\mu\nu} \psi.
\end{equation}
Hence, the proposal by Sheikholeslami and Wohlert~\cite{Sheikholeslami:1985ij} is to add a term to the Dirac operator:
\begin{equation}
D^\text{imp} = D^\text{W} + \frac{ia}{4 } c_{sw}   \, \sigma_{\mu\nu} F_{\mu\nu},
\end{equation}
and choose the coefficient $c_{sw}$ to cancel $O(a)$ effects\footnote{An alternative version with the $c_{sw}$ term in an exponential has been proposed in Ref.~\cite{Francis:2019muy}.}. Using lattice perturbation theory, one can see that $c_{sw} = 1 + O(g_0^2)$. Setting $c_{sw}=1$ is called tree-level Symanzik improvement. While one loop expressions are also available~\cite{Luscher:1996vw}, a complete $O(a)$ improvement needs a nonperturbative determination of $c_{sw}$~\cite{Luscher:1996ug,Luscher:1996sc}. Although twisted-mass fermions already have automatic $O(a)$-improvement, the $c_{sw}$ term can also be included in the action. This will alter only the $O(a^2)$ effects, but reduces in practice\footnote{This statement may depend on the specific choice of gauge action.} to reduce isospin-breaking effects~\cite{Becirevic:2006ii}.

By means of the improvement of the action, on-shell quantities (particle masses, scattering amplitudes) approach the continuum as $O(a^2)$ (up to logarithms). However, the improvement of correlation functions requires also the improvement of the fields, which involves additional counterterms for the unimproved fields. A particular example is the axial operator, whose cutoff effects can be parametrized\footnote{This valid for degenerate quarks.} as~\cite{Bhattacharya:2005rb}:
\begin{equation}
A^a_\mu(x) = Z_A(1 + b_A a m_q) \left[ A_\mu^a + a c_A \partial_\mu P^a  \right],
\end{equation}
where $Z_A$ is the renormalization constant, and $b_A, c_A$ are improvement coefficients. An appropriate tuning of the latter is needed to ensure full $O(a)$-improvement.

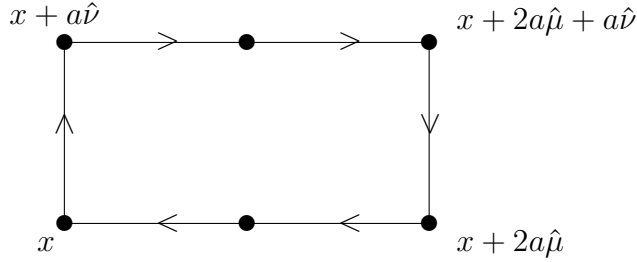
\begin{figure}[h!]
\setlength{\unitlength}{0.06cm}
\begin{center}
\begin{picture}(80,60)
\put (0,10){\line(1,0){80}}
\put (80,10){\line(0,1){40}}
\put (40,50){\line(-1,0){40}}
\put (0,50){\line(0,-1){40}}
\put (40,50){\line(1,0){40}}
\put(20,8.4){$<$}
\put(60,8.4){$<$}
\put(77.9,30){$\vee$}
\put(-2.3,30){$\wedge$}
\put(20,48.5){$>$}
\put(60,48.5){$>$}
\put(0,10){\circle*{3.33}}
\put(-6,4){$x$}
\put(40,10){\circle*{3.33}}
\put(86,4){$x+2a \hat{\mu}$}
\put(40,50){\circle*{3.33}}
\put(86,53){$x+2a \hat{\mu} +a \hat{\nu}$}
\put(0,50){\circle*{3.33}}
\put(80,50){\circle*{3.33}}
\put(80,10){\circle*{3.33}}
\put(-12,54){$x+a \hat{\nu}$}
\end{picture}
\caption{Representation of a rectangle Wilson loop, $U^\text{rec}_{\mu\nu}(x)$.}
\label{fig:rec}
\end{center}
\end{figure}

To conclude, we comment on the improvement of the gauge part of the action. Although the plaquette action suffers only from $O(a^2)$ discretization effects, Symanzik improvement can also be applied to reduce them. As proposed by L\"uscher and Weisz~\cite{Luscher:1984xn}, this can be achieved by including more complicated Wilson loops in the action. The most common choice is to add rectangular Wilson loops\footnote{Other parallelograms can also be included, but are less common in actual simulations.}---as shown in Fig.~\ref{fig:rec}---to the action:
\begin{equation}
S[U] = \frac{\beta}{2N_c} \sum_{\mu \nu} \sum_x \left[ c_0 \text{Re }\tr\left(1-U^\text{plaq}_{\mu\nu}\right) +  c_1 \text{Re }\tr\left(1-U^\text{rec}_{\mu\nu}\right)  \right].
\end{equation}
Note that an appropriate continuum limit constrains the relation between the two coefficients: $c_0+8c_1=1$. The choice $c_1=-1/12$, based on tree-level improvement, is called the L\"uscher-Weisz action \cite{Luscher:1984xn}. Another common choice, based on empirical evidence, is $c_1=-0.331$, and is referred to as the Iwasaki action~\cite{Iwasaki:2011np}.

\subsection{Euclidean Correlation functions in QCD}

In this section, we will discuss how to interpret correlation functions that we will compute from lattice QCD. In particular, we will focus on the extraction of the spectrum. 

Let us start with an example. Consider a field with the quantum numbers of a single positively charged pion ($J^P=0^-$ and $I, I_3 = 1,1$). An example of such operator is $\hat \pi^+(x) \equiv \bar d(x) \gamma_5 u (x)$. Its Fourier transform at zero momentum is:
\begin{equation}
\hat \pi^+(t) = \sum_{\boldsymbol x} \hat \pi^+(\boldsymbol x,t).
\end{equation}
We now consider the following correlation function at zero momentum:
\begin{equation}
C_\pi(t) = \langle  \hat \pi^+(t)  \hat \pi^-(0) \rangle  = \braket{0 | e^{\hat H t}  \hat \pi^+(0) e^{-\hat H t} \hat \pi^+(0)  | 0},  \label{eq:corrpi}
\end{equation}
where the time evolution of the operator in terms of the Hamiltonian has been used in the last step. Inserting a complete set of states, we reach the spectral decomposition of the correlation function:
\begin{equation}
C_\pi(t) = \frac{1}{L^3} \sum_n \frac{| \braket{0 | \hat \pi^+ |  n}|^2}{2 E_n} e^{-E_n t}, \label{eq:corrOBC}
\end{equation}
where the relativistic normalization of the states has been used, and the energy of the vacuum is taken to be zero. In the previous equation, the sum runs over all states with the same quantum numbers: $\pi^+$, but also $\pi^+ \pi^0 \pi^0$, and many more.  A particularly useful limit is $E_n t \gg E_0 t > 1$, as it provides a clean way to measure the mass of the ground state:
\begin{equation}
C_\pi(t)  \xrightarrow[\ \ \ t/a \gg 1 \ \ \ ]{} \frac{1}{L^3} \frac{| \braket{0 | \hat \pi^+ |  \pi^+}|^2}{2 M_\pi} e^{-M_\pi t}.
\end{equation}

\begin{figure}[h!]
\centering
\includegraphics[width=0.8\linewidth]{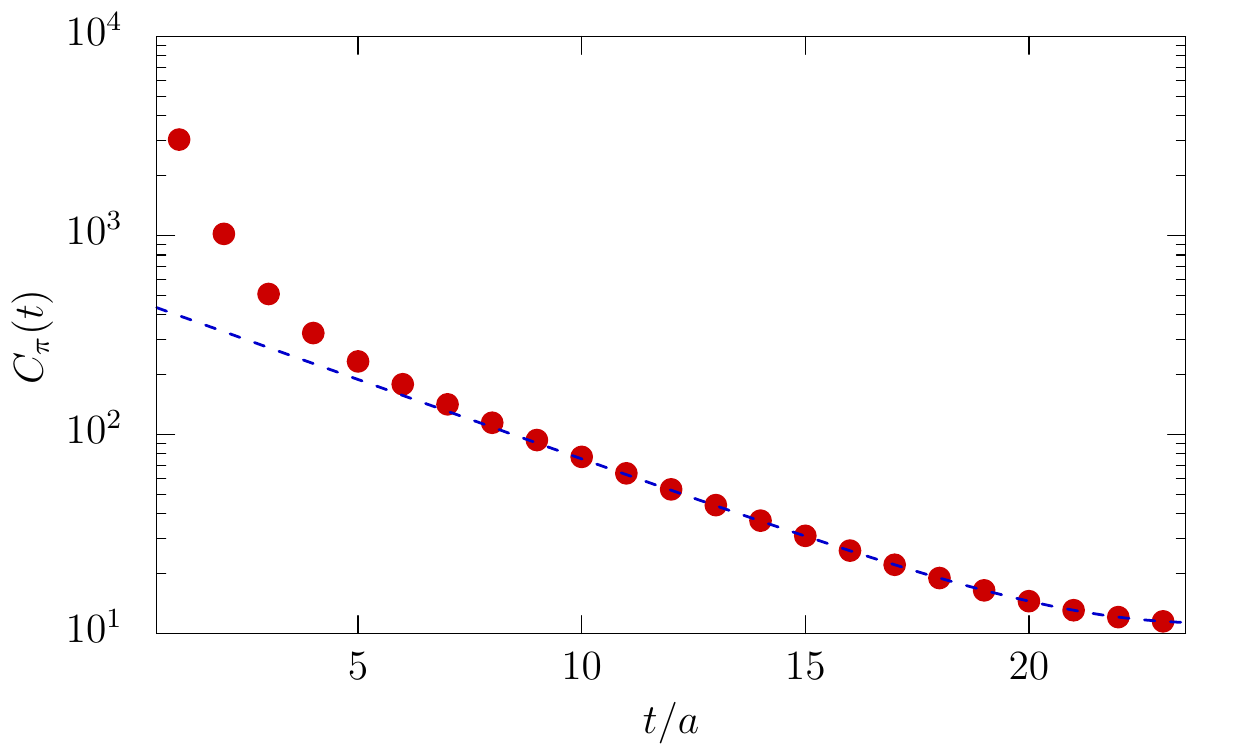}
\caption{Euclidean correlator of a pion, see Eq.~(\ref{eq:corrpi}). Statistical errors are too small to be seen, and the $y$-axis has an unimportant overall normalization. The dashed blue line is a fit to the last few time slices. The lattice action is $N_f=4$ $O(a)$-improved Wilson fermions. The lattice spacing is $a \simeq 0.075$ fm and the pion mass is $M_\pi \simeq 480$ MeV. For more technical details see Ensemble 3A20 in Ref.~\cite{Hernandez:2019qed}.}
\label{fig:corrpi}
\end{figure}

In practice, many simulations are carried out using periodic boundary conditions (PBC) in time. In this setup, the particle can also propagate backwards in time, and so Eq.~(\ref{eq:corrOBC}) becomes:
\begin{equation}
C_\pi(t) = \frac{1}{L^3} \frac{1}{Z_T}\sum_{n,m} \frac{| \braket{m | \hat \pi^+ |  n}|^2}{2 E_n} e^{-E_n t} e^{-E_m(T-t)}, \label{eq:corrPBC}
\end{equation}
with $Z_T = \tr \left( e^{-\hat H T} \right)$. Note that this implies that the ground state has the following asymptotic dependence:
\begin{equation}
 C_\pi(t)  \xrightarrow[\ \ T/a \gg   t/a \gg 1 \ \ ]{}   \frac{1}{L^3} \frac{| \braket{0 | \hat \pi^+ | \pi}|^2}{2 M_\pi \sinh{ M_\pi T/2}}   \cosh { M_\pi(t - T/2)}. \label{eq:corrPBC0}
\end{equation}
In Fig.~\ref{fig:corrpi}, we show an example for the pion correlator extracted from a lattice simulation with PBC. The dashed blue line is a fit of the last few time slices to Eq.~(\ref{eq:corrPBC0}). As can be seen, the mass of the pion can be measured to a high accuracy. Moreover, one can clearly see how excited states fall off faster than the ground state, and they are irrelevant in this case for $t/a>10$.

We will see in Chapter~\ref{sec:multiparticle} that one needs many levels in each channel to study multiparticle interactions on the lattice. The usual approach involves solving a generalized eigenvalue problem (GEVP). This consists on measuring a $N \times N$ matrix of correlation functions:
\begin{equation}
C_{ij} = \langle  \hat O_i(t)  \hat O_j^\dagger(0) \rangle,
\end{equation}
where $O_i$ are distinct operators with the same quantum numbers. Then, one can solve the eigenvalue equation:
\begin{equation}
C(t) v_n(t,t_0) = \lambda_n(t,t_0) C(t_0) v_n(t,t_0),
\end{equation}
where $t$ and $t_0$ are different Euclidean times, with $t>t_0$. $N$ energy levels can be extracted from the time dependence of each eigenvalues $ \lambda_n(t,t_0)$~\cite{Luscher:1990ck}. The method relies on the fact that the coupling of each operator to each state is different.
 
\clearpage
\chapter[Kaon decays and the large ${\text{N}}_{{\text{c}}}$ limit of QCD]{Kaon decays and the \largeNc limit of QCD}
\label{sec:largeN}

This chapter is focused on the study of the 't Hooft~\cite{tHooft:1973alw} (or large $N_c$) limit of QCD using lattice methods.  This limit is a well-known and useful simplification of $SU(N_c)$ gauge theories, with and without matter content. Despite the increased number of degrees of freedom as $N_c$ grows, the theory simplifies to the extent that exact nonperturbative predictions can be made. In fact, a long-term aspiration has been to solve the theory analytically in this limit.  Our main goal here is to address an open problem in QCD related to kaon decays. 

Even if we solve the theory in the 't Hooft limit, and it provides a good approximation to $N_c=3$ for some observables, the description of hadron decays and interactions involves $1/N_c$ corrections. Lattice QCD can provide a quantitative, first-principles determination of the subleading ${\mathcal O}(1/N_c)$ corrections to the 't Hooft limit by directly simulating $SU(N_c)$ theories at different values of the number of colours~\cite{Teper:1998te,Lucini:2001ej,Bali:2013kia}. 

We will study a famous failure of large $N_c$ in the $K \to \pi \pi$ weak decay. Experimentally, one observes a large ratio of decay amplitudes in the two possible isospin channels, while large $N_c$ arguments predict no such hierarchy. This is known as the puzzle of the ``$\Delta I=1/2$ rule'' in kaon decays, and indicates the relevance of at least some of the subleading $1/N_c$ corrections. We will use lattice simulations to dissect the large $N_c$ behaviour of the amplitudes. We will also see that the large $N_c$ predictions work reasonably well, e.g., for meson masses and decay constants. 

This chapter is organized as follows. First the 't Hooft limit will be introduced, together with its nonperturbative predictions. The $U(1)_A$ problem at large $N_c$ will also be discussed---another example in which the naive $N_c$ counting seemed to fail. Next, we will address the ChPT description of large-$N_c$ QCD, as well as the ``$\Delta I=1/2$ rule'' in the context of large $N_c$. Then, we will discuss some technical aspects of simulating large-$N_c$ QCD on the lattice. 
After that, we will summarize the main results of two of the articles included in the thesis:
(i) the $N_c$ scaling of meson masses and decay constants~\cite{Hernandez:2019qed}, and (ii) the exploration of weak decay amplitudes related to the ``$\Delta I=1/2$ rule''~\cite{Donini:2020qfu}. We will end with some remarks.

\section{The 't Hooft limit} \label{sec:thooft}

We will now address the mathematical formulation and properties of the 't Hooft limit. We use ``large $N_c$ limit'' and ``'t Hooft limit'' interchangeably. Part of this discussion is based on Ref.~\cite{Manohar:1998xv}, and our recent review~\cite{Hernandez:2020tbc}.

The precise definition of the 't Hooft limit is
\begin{equation}
N_c\rightarrow\infty, \quad \lambda=g_s^2 N_c ={\rm fixed} , \quad N_f={\rm fixed},
\end{equation}
where $g_s$ is the standard QCD coupling, and $\lambda$ is the so-called 't Hooft coupling. 
The renormalization group equation for $\lambda$ at large $N_c$,
\begin{eqnarray}
\mu {d \lambda \over d \mu} = - {11\over 3} {\lambda^2\over  8 \pi^2} +{\mathcal O}(\lambda^3),
\end{eqnarray}
indicates that asymptotic freedom survives, and that the limit is nontrivial since the coupling becomes strong at low energies. As in QCD, we expect that a nonperturbative scale is generated dynamically, as well as colour confinement, and the spontaneous breaking of chiral symmetry.
Hence, the large $N_c$ limit captures the most relevant nonperturbative phenomena of the strong interaction.

\begin{figure}[h!]
  \centering
  \includegraphics[width=0.9\linewidth]{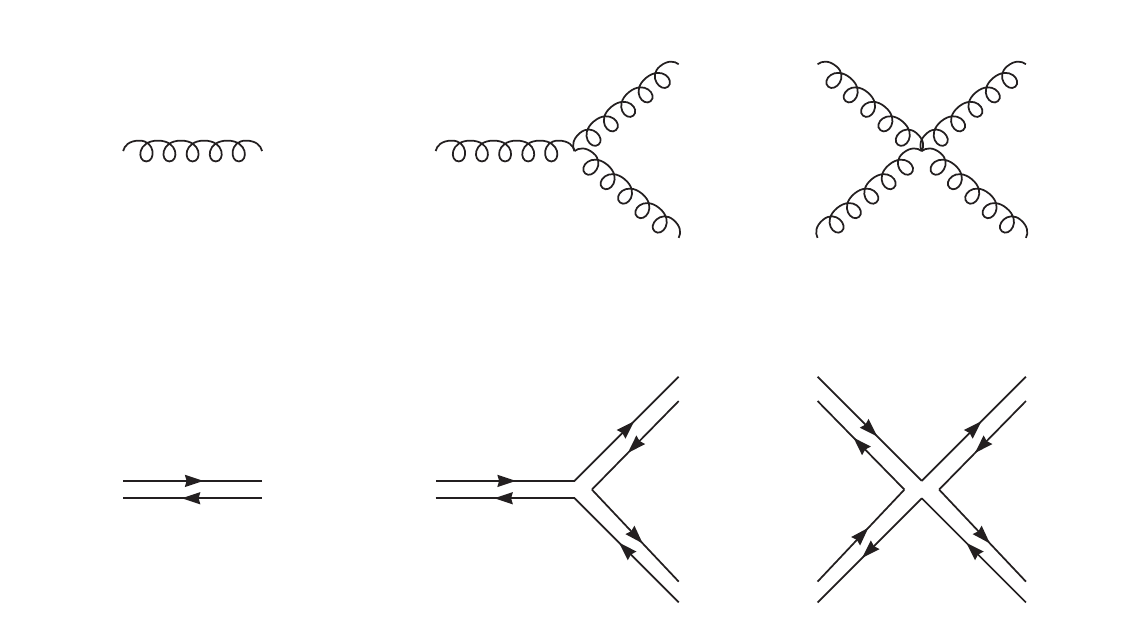}  
  \caption{'t Hooft double-line notation for gluon lines. Source: Ref.~\cite{Hernandez:2020tbc}.}
  \label{fig:doubleline}
\end{figure}

The main predictions in the large $N_c$ limit originate from counting powers of $N_c$ in correlation functions calculated to all orders in perturbation theory~\cite{tHooft:1973alw}. An important point is that (anti)quarks are in the (anti)fundamental irrep of $SU(N_c)$, while gluons live in the adjoint. Thus, the former have a single colour index, whereas the latter are represented by traceless matrices with two colour indices. In order to incorporate this, the usual notation for gluons in Feynman diagrams becomes the double-line 't Hooft notation, depicted in Fig.~\ref{fig:doubleline}. 
Each diagram can then be assigned a power of $N_c$ by simply counting closed loops, and using the fact that QCD vertices scale as $ g_s \sim 1/\sqrt{N_c} $.  The power of $N_c$ in each diagram is also related to the topology of the surface and its Euler characteristic.
In the following subsection, we will see some applications of this to obtain predictions at large $N_c$.

\subsection{Nonperturbative predictions at large $\text{N}_\text{c}$}

Let us first address the predictions for mesons at large $N_c$. For this, we consider hermitian operators with the quantum numbers of a meson, such as:
\begin{equation}
\mathcal{O}_\Gamma(x) = \frac{1}{\sqrt{N_c}} \bar q_i (x) \Gamma q_j (x),  \label{eq:opNc}
\end{equation}
where $\Gamma$ is a gamma matrix or product thereof, and for simplicity the quark fields have different flavours, $i \neq j$. In the previous equation, the normalization 
$1/\sqrt{N_c}$ ensures that the operator creates mesons with $O(N_c^0)$ amplitudes.

\begin{figure}[h!]
 \begin{center}
\[\begin{array}{ccccc}
\multicolumn{5}{c}{\includegraphics[scale=0.27]{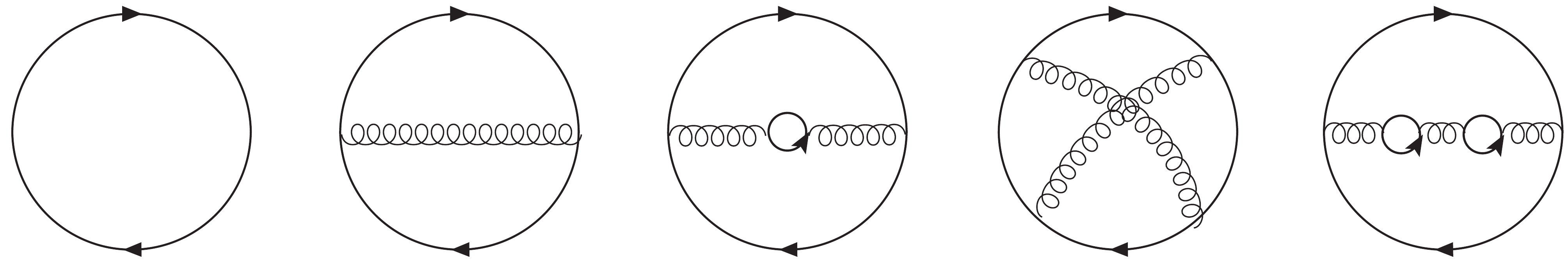}} \\ \vspace{-0.2cm} \\
\multicolumn{5}{c}{\includegraphics[scale=0.27]{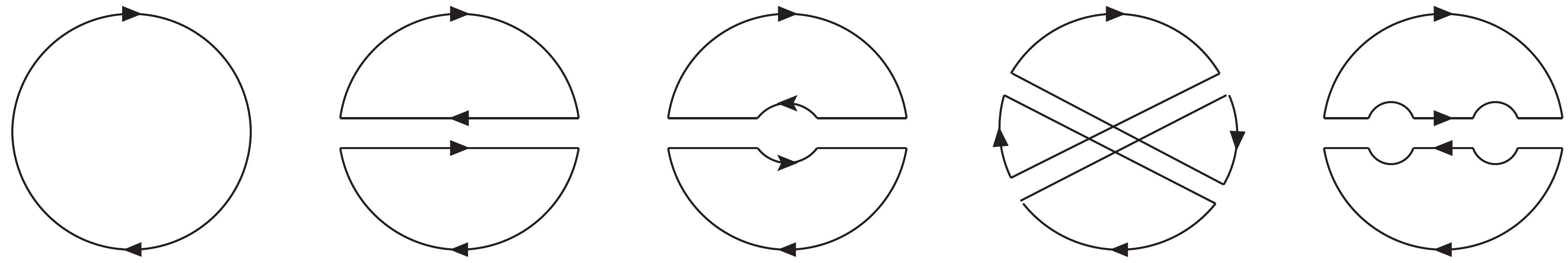}}  \\
\quad \ \, \, (a) \ \quad \quad &\quad\, (b) \quad\quad &\quad\,\,\, (c) \quad\quad& \quad\,\,\, (d) \quad &\,\,\,\ (e) \\  \\
N_c^0 \ \ & N_c^0 \ \ & N_c^{-1} N_f \ &\   N_c^{-2} &\ \ N_c^{-2} N_f^2 
\end{array}\]
\caption{Various diagrams contributing to the correlation function of two meson operators with the Feynman notation (top), and the 't Hooft double-line notation (bottom). The power of $N_c$ and $N_f$ associated to each diagram is also given.  \label{fig:diag2pt}  }
\end{center}
\end{figure}

A simple case to explore is that of the two point function
\begin{equation}
C_{2,\Gamma} = \langle \mathcal{O}_\Gamma(x_1)  \mathcal{O}_\Gamma(x_2)  \rangle.
\end{equation}

By inspecting all contributing diagrams, one can gain insight into the $N_c$ dependence. Note that the normalization in Eq.~(\ref{eq:opNc}) adds a factor $1/N_c$ to each diagram. Let us comment on the examples shown in Fig.~\ref{fig:diag2pt}. It is trivial to see that the dominant one [diagram (a)] has an overall scaling of $N_c^0$. Introducing one gluon, as in diagram (b), does not alter the counting: there are two closed loops, and a $g_s^2 \sim 1/N_c$ factor. More generally, diagrams with any number of gluons that do not cross are called planar diagrams, and have the same power as the diagram without gluons. An example of a nonplanar diagram is given in (d), since the two gluons cross. Diagrams (c) and (e) are two examples in which quark loops are included. Each quark loop reduces a power $N_c$ while including a factor of the number of flavours, $N_f$.

If the operator in the two-point correlation function is chosen to have axial quantum numbers, $\Gamma=\gamma_0 \gamma_5$, it is dominated at large time separation by the pion contribution. The matrix element is then related to the pion decay constant, $C_{2,\Gamma}  \propto F^2_\pi/N_c$. Based on the expansion in Fig.~\ref{fig:diag2pt}, a simple prediction can then be derived 
\begin{equation}
\frac{F_\pi^2}{N_c} = \left( A + B \frac{N_f}{N_c} + \cdots \right), \label{eq:Fpiexpansion}
\end{equation}
with $A$ and $B$ being constants that do not depend upon $N_c$ and $N_f$. This can be used to relate the value of $F_\pi$ across gauge theories with different matter content. 

Similarly, one can consider four-point functions in order to study scattering processes. In particular, the dispersive properties are contained in the connected part of the correlation functions. For instance, the $s$-wave scattering length\footnote{The scattering length is proportional to the two-particle $s$-wave scattering amplitude at threshold.} is just
\begin{eqnarray}
a_0 &\propto &{ \langle O_\Gamma O_\Gamma O_\Gamma O_\Gamma \rangle_c \over  |\langle 0| O_\Gamma |\pi\rangle|^4} \propto N_c^{-1},
\label{eq:aM}
\end{eqnarray}
and so it decreases with growing $N_c$. When inspecting three-point functions, one can see that similar arguments hold for decay processes. Hence, mesons in large-$N_c$ QCD neither scatter nor decay, and QCD at large $N_c$ is a theory of free and infinitely narrow states~\cite{tHooft:1973alw,Witten:1979kh,Coleman:1980nk}.

\subsection{The Witten-Veneziano equation}

In this section, we will comment on the so-called $U(1)_A$ problem in the context of large $N_c$. We will see that a naive counting of powers of $N_c$ in correlation functions seems to be in conflict with phenomenology regarding the expected pNGB associated to the singlet axial current---the $\eta'$. The resolution of this problem has brought new insights into QCD and the chiral anomaly~\cite{Witten:1979vv,Veneziano:1979ec}.

Consider the following gluonic correlation function in QCD:
\begin{equation}
C(k) = \int d^4 x e^{ikx} \langle q(x) q(0)  \rangle, \label{eq:Cketa}
\end{equation}
where the topological charge operator is
 \begin{equation}
q(x) \equiv {\lambda \over 32 \pi^2 N_c} {\rm Tr}[F_{\mu\nu}(x) {\tilde F}^{\mu\nu}(x)],
\end{equation}
and its four-dimensional integral is equal to the topological charge. Formally, the correlation function at zero momentum can be related to the partition function in the path integral formulation with a $\theta$-term [see Eq.~(\ref{eq:thetaterm})]:
\begin{equation}
\frac{\partial^2 \mathcal Z}{\partial \theta^2} \bigg \rvert_{\theta=0} \propto C(0). \label{eq:thetaC}
\end{equation}
Furthermore, the topological susceptibility is just the correlation function in Eq.~(\ref{eq:Cketa}) at zero momentum, $\chi = C(0)$.

A diagrammatic analysis of this two-point functions yields a $O(N_c^0)$ scaling, since it is a closed gluon loop with a normalization $1/N_c^2$. In the previous section, we have argued that the contributions of increasing number of quark loops are suppressed by the corresponding powers of $N_c$:
\begin{equation}
C(k) = C_0(k) + C_1(k) + \hdots, \label{eq:Cexpansion}
\end{equation}
where $C_0$ is the sum of all planar diagrams with zero quark loops,  $C_1$ with a single quark loop, and so on. Note that their $N_c$ scaling is $C_0 \propto N_c^0$, and $C_1 \propto N_c^{-1}$.

In the case of massless quarks, $C(0)$ must vanish. This is because the $\theta$-term can be reabsorbed by a chiral rotation. Therefore, there cannot be a dependence with $\theta$, or equivalently, all derivatives with respect to $\theta$ are zero. In the pure gauge theory, this is not the case and $C_0(k)$ is in general nonzero. This way, there is an apparent contradiction in Eq.~(\ref{eq:Cexpansion}) at zero momentum: how can the full correlation function vanish, if the term with the leading $N_c$ power does not? In order to answer this, let us write the spectral decomposition of the correlation function as sums over one-particle poles:
\begin{equation}
C(k) = \sum_\text{glueballs} \frac{a_n}{k^2 - m_g^2}  + \sum_\text{mesons} \frac{b_n/N_c}{k^2 - M_h^2}, \label{eq:Ck}
\end{equation} 
where $a_n$ and $b_n$ are $O(N_c^0)$ coefficients. The sum over glueballs\footnote{Bound states of gluons.} determines the correlation function in the pure gauge theory, $C_0(k)$.
Inspecting Eq.~(\ref{eq:Ck}), one can deduce that the only way that a cancellation at $k=0$ can occur is if there is a meson, such that, $M_h^2 \propto 1/N_c$. From the quantum numbers, one can deduce that this hadron is the $\eta'$ meson---see Eq.~(\ref{eq:anomaly}) in the previous chapter.

This is an example where the diagrammatic analysis leads to a wrong conclusion: the leading $N_c$ scaling of the correlation function is cancelled by what naively looks like a subleading one. The consequence of this is the well-known Witten-Veneziano equation, which connects the mass of the $\eta'$ meson to the topological susceptibility of the pure gauge theory, $\mychi_{YM}$:
\begin{equation}
M_{\eta'}^2= {2 N_f \over F^2_{\eta'}} \mychi_{YM} \equiv {2 N_f \over F^2_{\eta'}}  \int d^4 x \langle q(x) q(0)\rangle_{YM},
\label{eq:WV}
\end{equation}
where $F_{\eta'}$ is the decay constant of the $\eta'$. As written, Eq.~(\ref{eq:WV}) is valid for the case of massless quarks. If quarks are massive and degenerate, then 
\begin{equation}
M_{\eta'}^2 = M_\pi^2 + {2 N_f \over F^2_{\eta'}} \mychi_{YM}.  \label{eq:WVmpi}
\end{equation}
Note that $F_{\eta'} = F_\pi$ at large $N_c$. While $\mychi_{YM}$ cannot be measured experimentally, it has been determined using lattice QCD~\cite{Ce:2015qha,Ce:2016awn}.

\subsection{Chiral Perturbation Theory at large $\text{N}_\text{c}$} \label{sec:chptlargenc}

As suggested by the running of the 't Hooft coupling, spontaneous chiral symmetry breaking survives at large $N_c$~\cite{Coleman:1980mx}. This means that the lightest particles in the large $N_c$ spectrum are also the pseudoscalar mesons. At leading order in the quark mass, the pion mass is $M^2_\pi = 2 \Sigma m_q/F_\pi^2 $, and thus of order $N_c^0$---see Section~\ref{sec:ChPT}. One would therefore expect that the ChPT description of the pseudoscalar states is still valid. 

A subtlety of the chiral EFT in the large $N_c$ limit is the treatment of the $\eta'$. From Eq.~(\ref{eq:WVmpi}), it is clear that the $\eta'$ becomes a pNGB\footnote{This assumes that $N_f$ is kept fixed. If however $N_f/N_c = \text{ const}$, then the singlet remains heavy (Veneziano limit).} at large $N_c$, and hence, it must be included in the EFT as a relevant degree of freedom~\cite{DiVecchia:1980yfw,Rosenzweig:1979ay,Witten:1980sp,Kawarabayashi:1980dp,Gasser:1984gg,Leutwyler:1996sa,HerreraSiklody:1996pm,Kaiser:2000gs}. Specifically, the matrix of pseudoscalar fields must be modified as ($N_f=3$ is assumed)
\begin{equation}
  \phi
 = \begin{pmatrix}
\pi^0   + \frac{1}{\sqrt{3}} ( \sqrt{2}\eta'  + \eta)& \sqrt{2} \pi^+ & \sqrt{2}K^+  \\
\sqrt{2} \pi^-  & -\pi^0 + \frac{1}{\sqrt{3}} (\sqrt{2}\eta' + \eta) & \sqrt{2}K^+ \\
\sqrt{2}K^-& \sqrt{2}\bar K^0&  \frac{1}{\sqrt{3}} (\sqrt{2}\eta'-2 \eta )
\end{pmatrix}, \label{eq:pionfieldseta}
\end{equation}
with $U = \exp \left( i \phi/F \right)$. The LO chiral Lagrangian then becomes
\begin{equation}
\mathcal{L}_2 = \frac{F^2}{4} \tr \left[ \partial_\mu U  \partial^\mu U^\dagger  \right]
+ \frac{B F^2}{2 } \tr \left[ U \chi^\dagger + \chi U^\dagger  \right] - {N_f}  \frac{\tau}{F^2} (\eta' - \theta)^2, 
\end{equation}
where $\chi=  \text{diag }(m,m,m_s)$, and the new coupling $\tau$ is the topological susceptibility at leading order. We have also included the vacuum angle, $\theta$. Expanding, one can see that the quadratic terms in $\eta'$ are
\begin{equation}
\mathcal{L}_2 \supset \frac{1}{2} \partial_\mu \eta'  \partial^\mu \eta'  - \frac{1}{2}(\eta')^2 \left[ \frac{2}{3}B(2m+m_s)  +2 N_f \frac{\tau}{F^2} \right],
\end{equation}
which means
\begin{equation}
M_{\eta'}^2 =  \frac{1}{3}  (2M_K^2+M_\pi^2) + \frac{2 N_f  \tau}{F^2},
\end{equation}
and coincides with the Witten-Veneziano equation at this order, $\tau = \mychi_{YM}$, for $M_K=M_\pi$.

Beyond leading order, we must revisit the power counting of this EFT. A consistent choice for the expansion parameter in large-$N_c$ ChPT is~\cite{Kaiser:2000gs}
\begin{eqnarray}
 \delta  \sim  \left(M_\pi\over 4 \pi F_\pi\right)^2 \sim  \left(p \over 4 \pi F_\pi\right)^2  \sim  {1\over N_c}. \label{eq:powercounting}
\end{eqnarray}
Even if $\delta$ becomes smaller and smaller with growing $N_c$, the range of validity of the chiral effective theory does not increase. This is because the failure of the chiral expansion will be abrupt when the energy scale reaches the mass of the lightest resonances, $\Lambda_\chi$.  This mass is expected to scale as $O(N_c^0)$, and so, it remains constant at large $N_c$.  Typically, one considers $\Lambda_\chi \sim M_\rho$. However, loop corrections in the form of logarithms are suppressed, and they become irrelevant as $N_c \to \infty$.

An additional simplification of ChPT at large $N_c$ is related to the scaling of the NLO low-energy constants with the number of colours. Based on general rules, one can show that only a subset thereof is leading in $N_c$, i.e., $L_i \propto {\mathcal O}(N_c)$. They are the ones that correspond to operators with a single flavour trace. A particular example is $L_5$, whose operator is given in Eq.~(\ref{eq:O5}). The operators with two flavour traces correspond diagramatically to at least two fermion loops, and thus are suppressed by $1/N_c$. In the case of $N_f=3$, one has~\cite{Gasser:1984gg,Peris:1994dh}:
\begin{align}
\begin{split}
L_1,L_2,L_3,L_5,L_8,L_9,L_{10} &\propto  {\mathcal O}(N_c), \\
2 L_1-L_2, L_4,L_6,L_7 &\propto  {\mathcal O}(1).
\end{split}
\end{align}
Phenomenological approaches have estimated the leading $N_c$ behaviour of these LECs by assuming that ChPT can be matched onto a theory that includes heavier resonances with other $J^P$ quantum numbers---the resonant chiral theory~\cite{Ecker:1988te}. The values for the LECs result from the exchange of these resonances, and they can be extracted in terms of the measured spectrum, simple large $N_c$ arguments, and imposing the correct behaviour at large $p$ of certain correlation functions. Alternatively, we will measure them on the lattice.

\subsection[The elusive $\Delta \text{I=1/2}$ rule]{The elusive $\Delta \text{I=1/2}$ rule\footnote{Part of this discussion is based on the review in Ref.~\cite{Cirigliano:2011ny}}}

The weak decay of a kaon into two pions is a very appealing process in the context of the $1/N_c$ expansion. An exact nonperturbative prediction can be obtained in the 't Hooft limit, but this prediction is in conflict with experimental results. While for many years it has been a benchmark process for both phenomenological and lattice calculations, it still remains a challenging one.

In the limit of approximate isospin symmetry, the $K \to \pi \pi$ weak decay has two different decay channels: the two pions in the final state can either have total isospin of $I=2$ or $I=0$. Thus, the relevant matrix elements are:
\begin{equation}
 i A_I e^{i \delta_I} = \braket{(\pi\pi)_I | \mathcal{H}_w | K}, 
\end{equation}
where $\mathcal{H}_w$ is the electroweak Hamiltonian, and $\delta_I$ are the strong scattering phases. Experimentally, it has been known for quite some time that the $A_0$ amplitude is strongly enhanced with respect to $A_2$~\cite{Zyla:2020zbs}
\begin{equation}
{\bigg \rvert} \frac{A_0}{A_2 }{\bigg \rvert} = 22.45(6). \label{eq:deltaIexp}
\end{equation}
This fact is referred to as the ``$\Delta I=1/2$ rule'', since the transition that dominates is the one where the isospin quantum number changes by half a unit.

 \begin{figure}[h!]
 \begin{center}
\[\begin{array}{cc}
\raisebox{-.45\height}{\includegraphics[scale=1]{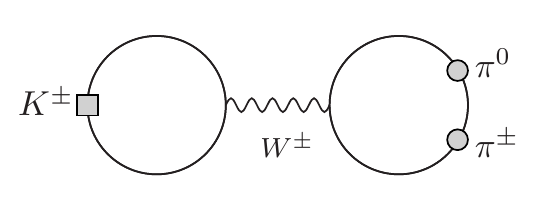}} & \hspace{0.3cm} { \propto  N_c^{1/2}} \\ \\
\raisebox{-.43\height}{\includegraphics[scale=1]{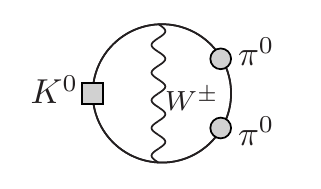}} &\hspace{0.3cm} \propto  N_c^{-1/2}\\ 
\end{array}\]
\caption{\label{fig:deltaI} Leading diagrams in $N_c$ for the decays of charged kaons (top), and neutral kaons (bottom). Source: Ref.~\cite{Hernandez:2020tbc}.}
\end{center}
\end{figure}

In order to derive the large $N_c$ prediction, let us consider the following physical decay amplitudes:
\begin{align}
& \mathcal{T} \left[ K^0\, \to \pi^0 \,\pi^0 \right] = \sqrt{\frac{2}{3}} A_2 e^{i \delta_2} -\frac{1}{\sqrt{3}} A_0 e^{i \delta_0}, \label{eq:K0decay} \\ 
& \mathcal{T} \left[ K^+ \to \pi^+ \pi^0 \right] = \frac{\sqrt{3}}{2}  A_2 e^{i \delta_2},
\end{align}
where on the right-hand side we have used the isospin decompositions of the states using the standard Clebsch-Gordan coefficients. In Fig.~\ref{fig:deltaI}, the leading diagrams for each of the amplitudes are shown, including their $N_c$ counting, as explained in the previous section. From this scaling, one can infer that the neutral kaon does not decay at large $N_c$. By means of the isospin decomposition in Eq.~(\ref{eq:K0decay}), the following prediction can then be derived:
\begin{equation}
\text{Re }\frac{A_0}{A_2 }{\bigg \rvert}_{N_c \to \infty} = \sqrt{2} + O(N_c^{-1}).
\end{equation}
This is over an order of magnitude smaller than the measured value, indicating large $1/N_c$ corrections, or a breakdown of the large $N_c$ expansion for this observable.  It seems unlikely that beyond-the-standard-model (BSM) physics can explain the discrepancy. Since this enhancement enters in the SM prediction for direct CP violation in kaons (the famous $\epsilon'/\epsilon$), a good handle on the real part of the amplitude is of great phenomenological interest.

Several explanations have been proposed over the years. First, the multiscale dynamics ($M_W \gg m_c \gg M_K$) may produce corrections that are parametrically large but subleading in $1/N_c$---large logarithms~\cite{Shifman:1975tn}. Second, rescattering effects from the pions in the final state have also been proposed as a source of enhancement~\cite{Pallante:1999qf,Pallante:2000hk}. Finally, it is possible that the enhancement may be  largely dominated by intrinsic QCD effects, which could be understood in an EFT picture. 

A few years ago, the RBC-UKQCD collaboration~\cite{Boyle:2012ys} analysed the various contributions to $K \to \pi\pi$. Their results suggested that the main source of the enhancement comes from a strong cancellation in $A_2$. More specifically, there is a negative relative sign between a colour-connected contraction and a colour-disconnected one, which have different $N_c$ scaling but comparable magnitude. A lattice exploration of the $N_c$ scaling of the amplitudes involved in this process may have the potential to shed light on the origin of this enhancement. In this manner, one should be able to disentangle the two contributions rigorously. This has been studied in Refs. \cite{Donini:2016lwz,Donini:2020qfu}, and will be addressed below in Section~\ref{sec:dissecting}.

\clearpage

\section{Lattice QCD with varying $\text{N}_\text{c}$}

In this section, we will address our study of the large $N_c$ limit of QCD on the lattice.  We will present some technical details of the simulations that we have carried out. Then, we will discuss two of the articles~\cite{Hernandez:2019qed,Donini:2020qfu} that are included in this thesis.

\subsection{Technical aspects} \label{sec:technical}

The lattice simulations for this project have been carried out using HiRep~\cite{DelDebbio:2008zf,DelDebbio:2009fd}, which is a state-of-the-art lattice QCD code that allows for simulations with different gauge groups, matter content and fermionic representations.

The choice for the gauge action in our simulations is the Iwasaki gauge action, introduced in Section~\ref{sec:improvedact}. For $N_c=3$, we use the same value of $\beta$ as the ETM Collaboration~\cite{Alexandrou:2018egz}. For the other values of $N_c$, $\beta$ is tuned such that the lattice spacing is as close as possible. Two additional ensembles with finer lattice spacing are also included. Our simulations have $N_f=4$ active quarks. This will be important to study the amplitudes related to the $\Delta I=1/2$ rule, for which we need an active light charm quark.  Furthermore, we use $O(a)$-improved Wilson fermions. For $N_c=3$, we take the one-loop value~\cite{Aoki:2003sj}
\begin{equation}
c_{sw} = 1 + \frac{g_0^2}{P} c^{(1)}_{sw}, \quad \text{ with } \quad  c^{(1)}_{sw} = 0.113,
\end{equation}
where we use the bare coupling boosted by the average plaquette. For $N_c>3$, the complete result cannot be easily reproduced from Ref.~\cite{Aoki:2003sj}. Instead, we use the fact that $c^{(1)}_{sw}$ is dominated by the tadpole contribution\footnote{The tadpole diagram is shown in Fig. 4(d) of Ref. \cite{Aoki:2003sj}.}, which is of order $N_c$, according to Eq.~(58) in Ref. \cite{Aoki:2003sj}. This means that $c_{sw}$ is constant in $N_c$, up to effects $O(a^2/N_c)$. 

A summary of the simulation parameters is given in Table \ref{tab:ensembles}. The naming scheme for the ensembles is the following. The first number indicates the value of $N_c$. The letter in the second position refers to the lattice spacing: ``A'' for the coarsest. In the third position, there is a number that indicates the pion mass: 1 for the heaviest. The final position is used to differentiate two ensembles that only differ in the volume.

\begin{table}[h!]
\centering
\begin{tabular}{c|c|c|c|c}
  Ensemble& $L^3 \times T$ &$\beta$ & $am^s$ & $aM^s_\pi$  \\ \hline \hline
3A10 & $20^3 \times 36$ &\multirow{ 5}{*}{1.778}&-0.4040 & 0.2204(21)   \\ \cline{1-2} \cline{4-5} 
3A11 & $24^3 \times 48$ & &-0.4040 & 0.2147(18)  \\ \cline{1-2} \cline{4-5} 
3A20 &$24^3 \times 48$ &&-0.4060 &  0.1845(14)   \\ \cline{1-2} \cline{4-5} 
3A30 &$24^3 \times 48$ &&-0.4070 & 0.1613(16)   \\ \cline{1-2} \cline{4-5} 
3A40 &$32^3 \times 60$ &&-0.4080 &  0.1429(12)   \\ \hline   \hline 

3B10 & $24^3 \times 48$ &\multirow{ 2}{*}{1.820}& -0.3915& 0.1755(15)  \\ \cline{1-2} \cline{4-5} 
3B20 & $32^3 \times 60$ & & -0.3946 & 0.1191(9)      \\ \hline   \hline 

4A10 &$20^3 \times 36$ &\multirow{ 4}{*}{3.570} &-0.3725& 0.2035(14)  \\ \cline{1-2} \cline{4-5} 
4A20 &$24^3 \times 48$& &-0.3752&0.1805(7)  \\ \cline{1-2} \cline{4-5} 
4A30 &$24^3 \times 48$& &-0.3760& 0.1714(8) \\\cline{1-2} \cline{4-5} 
4A40 &$32^3 \times 60$& &-0.3780& 0.1397(8)\\ \hline  \hline 
5A10 &$20^3 \times 36$& \multirow{ 4}{*}{5.969} &-0.3458& 0.2128(9)   \\ \cline{1-2} \cline{4-5} 
5A20 &$24^3 \times 48$& &-0.3490& 0.1802(6)   \\ \cline{1-2} \cline{4-5} 
5A30 &$24^3 \times 48$& &-0.3500& 0.1712(6) \\ \cline{1-2} \cline{4-5} 
5A40 &$32^3 \times 60$& &-0.3530&  0.1331(7)  \\ \hline  \hline 
6A10 &$20^3 \times 36$ &\multirow{ 4}{*}{8.974} &-0.3260& 0.2150(7)  \\ \cline{1-2} \cline{4-5} 
6A20 &$24^3 \times 48$& &-0.3300&0.1801(5)  \\\cline{1-2} \cline{4-5} 
6A30 &$24^3 \times 48$& &-0.3311& 0.1689(7) \\ \cline{1-2} \cline{4-5} 
6A40 &$32^3 \times 60$& &-0.3340& 0.1351(6) \\ \hline 
\end{tabular}
\caption{  Summary of ensembles used in this dissertation: $\beta$,  sea quark bare mass parameter, $m^s$, and sea pion mass $M^s_\pi$ .  We keep $c_{sw}=1.69$ in the ``A'' ensembles, and $c_{sw}=1.66$ in the ``B''. In this simulations, $N_f=4$.}
\label{tab:ensembles}
\end{table}

We employ maximally twisted quarks \cite{Frezzotti:2000nk} for the valence Dirac operator, i.e., a mixed-action setup~\cite{Bar:2002nr}. Maximal twist is ensured by tuning the untwisted bare valence mass $m_0^{\rm v}$ to the critical value for which the valence PCAC mass is zero:
\begin{equation}
\lim_{m^{\rm v} \rightarrow m_{\rm cr}} m^{\rm v}_{\rm pcac} \equiv \lim_{m^{\rm v} \rightarrow m_{\rm cr}} \frac{\partial_0\braket{A_0(x) P^\dagger(y)}}{2 \braket{P(x) P^\dagger(y)}}= 0,
\end{equation}
with $A_0 = \bar u \gamma_0 \gamma_5 d$, and  $P = \bar u \gamma_5 d$.
The bare twisted-mass, $\mu_0$, is tuned such that the pion mass in the valence and sea sectors match, $M_\pi^{\rm v}= M_\pi^s$.

This choice has some advantages. First, we achieve automatic $O(a)$ improvement\footnote{Up to residual sea quark mass effects~\cite{Bussone:2018ljj}.}~\cite{Frezzotti:2003ni} regardless of the value of $c_{sw}$. We observed in Ref.~\cite{Donini:2020qfu} that, for our gauge action, the choice $c_{sw}=0$ in the twisted-mass valence sector minimizes the isospin breaking effects and leads to smaller statistical errors. Moreover, the renormalized pion decay constant, $F_\pi$, can be obtained with no need for a renormalization constant~\cite{Shindler:2007vp}:
\begin{equation}
F_\pi = \frac{\sqrt{2} \mu_0  \braket{0 | P | \pi}_{\text{bare}} }{M_\pi^2}. \label{eq:Fpitm}
\end{equation}
This fact will be a central point in Ref.~\cite{Hernandez:2019qed}. Finally, it avoids the mixing of different-chirality operators for weak matrix elements, which will be essential for Ref.~\cite{Donini:2020qfu}.

\subsubsection{Scale setting}

The procedure of computing the lattice spacing, $a$, in physical units receives the name of scale setting.  Having this conversion is crucial for lattice calculations, since their outcomes are always given in terms of the lattice spacing. The main idea is to compute some observable on the lattice with a very high accuracy, and then use its known value from experiment to fix the lattice spacing. The scale setting of the ensembles in Table~\ref{tab:ensembles} has been carried out in Ref.~\cite{Hernandez:2019qed}, and revisited in Ref.~\cite{Hernandez:2020tbc}. In this section, we will summarize the key points.

The gradient flow~\cite{Luscher:2010iy} is nowadays a standard tool for setting the scale on the lattice~\cite{Bruno:2013gha,Sommer:2014mea}. It consists on a differential equation that evolves the gauge fields in a fictitious dimension $t$, the flow time. In the continuum, the flow equation is
\begin{equation}
\frac{dB_\mu(x,t)}{dt} = D_\nu G_{\nu \mu}(x,t),
\end{equation}
where 
\begin{equation}
G_{\nu \mu} = \partial_\nu B_\mu - \partial_\mu B_\nu + [B_\nu, B_\mu].
\end{equation}
Here, $B_\mu(x,t)$ are the flowed gauge fields, with boundary conditions:
\begin{equation}
B_\mu(x, t=0) = A_\mu(x),
\end{equation}
and $A_\mu(x)$ are simply the gluon fields of the QCD Lagrangian.

\begin{figure}[h!]
  \centering
  \includegraphics[width=0.9\linewidth]{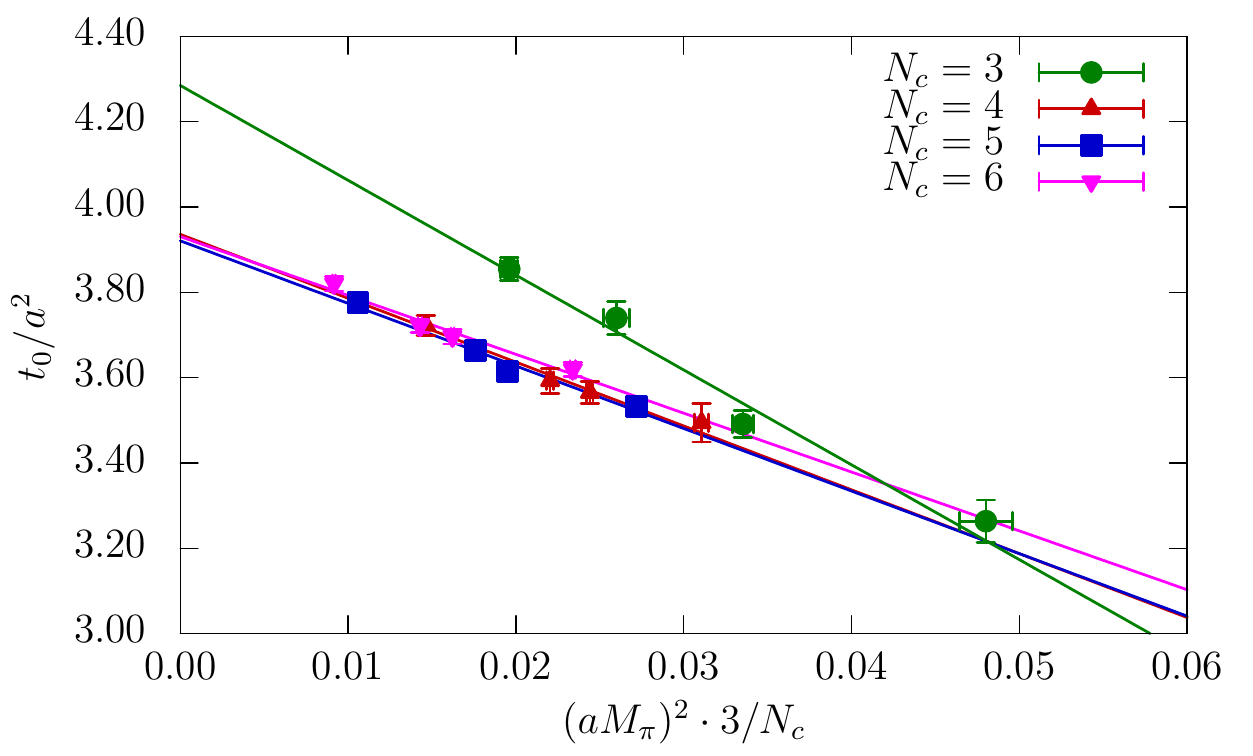}  
  \caption{Chiral dependence of $t_0$. Source: Refs.~\cite{Hernandez:2019qed,Hernandez:2020tbc}.}
  \label{fig:t0}
\end{figure}

The main advantage of the gradient flow is that it allows for a simple definition of a renormalized coupling. In particular, the energy density can be related to the 't Hooft coupling in the gradient flow (GF) scheme:
\begin{equation}
\langle  E(t) \rangle \equiv \frac{1}{4} \langle G_{\mu \nu }^a G^{\mu \nu}_a \rangle    = \frac{3}{138 \pi^2 t^2} \frac{N_c^2 -1}{N_c} \lambda_{GF}(\mu),
\end{equation}
where $ \lambda_{GF}(\mu)$ is defined at the scale $\mu=1/\sqrt{8 t}$. The two-loop matching between the GF and $\overline{\text{MS}}$ schemes is known~\cite{Harlander:2016vzb}. A conventional scale $t_0$ is defined in the literature via the implicit equation
\begin{equation}
 t^2 \langle  E(t) \rangle \big \rvert_{t=t_0} = 0.3. \label{eq:t0}
\end{equation}
While $t_0$ cannot be measured experimentally, it is an observable quantity that can be determined from lattice simulations~\cite{Bruno:2013gha,Sommer:2014mea,Bruno:2016plf}. For our simulations with $N_c>3$, we generalize the definition in Eq.~(\ref{eq:t0}) as in Ref. \cite{Ce:2016awn}:
\begin{equation}
 t^2 \langle  E(t) \rangle \big \rvert_{t=t_0} = 0.3 \times \frac{3}{8} \frac{N_c^2-1}{N_c}. \label{eq:t0Nc}
\end{equation}
From previous results~\cite{Bruno:2013gha,Sommer:2014mea,Bruno:2016plf}, one can infer that 
\begin{equation}
\sqrt{t_0} \bigg \rvert^{N_f=4}_{M_\pi=420 \text{ MeV}}  = 0.1450(39) \text{ fm}. 
\end{equation}
Then, our scale setting condition becomes
\begin{equation}
\left( M_\pi \sqrt{t_0} \right) \bigg \rvert^{N_f=4}_{M_\pi=420 \text{ MeV}} = 0.3091(83). \label{eq:condition}
\end{equation}
In practise, this is how the procedure works. First, we measure $t_0/a^2$ and the pion mass in each ensemble. Then, we fit to the Chiral Perturbation Theory prediction for $t_0$~\cite{Bar:2013ora}:
\begin{equation}
t_0(M_\pi) = t_0^\chi \left(1 + \frac{\widetilde k}{N_c} M_\pi^2 \right) + O(M_\pi^4),
\end{equation}
with $t_0^\chi, \widetilde k$ being low-energy constants. Note that the mass dependence of $t_0$ is suppressed with $N_c$.  Finally, for each value of $N_c$ we look for the point in which the condition in Eq.~(\ref{eq:condition}) is met. In Fig.~\ref{fig:t0} we show the chiral fits for $t_0$ in the ``A'' ensembles of Table \ref{tab:ensembles}. The results for the lattice spacing is summarized in Table~\ref{tab:scalesetting}.

\begin{table}[h!]
\centering
\begin{tabular}{c|c}
 Ensembles & $a$ ($\times 10^{-2}$ fm) \\ \hline\hline
3A     &$7.5(2)$    \\ \hline
3B    &   $6.5(2)$   \\ \hline
4A      &$7.6(2)$  \\ \hline 
5A     &$7.5(2)$ \\ \hline
6A   &  $7.5(2)$  \\ 
\end{tabular}
\caption{Results for the lattice spacing in the various sets of ensembles used in this work. The error is dominated by that of Eq.~(\ref{eq:condition}).}
\label{tab:scalesetting}
\end{table}

\subsection{Large $\text{N}_\text{c}$ scaling of meson masses and decay constants}

Ref.~\cite{Hernandez:2019qed} contains a study of the $N_c$ scaling of meson masses and decay constants. The results allow us to confront the expected $N_c$ scaling of the LECs of the chiral Lagrangian with results from lattice simulations.  Our work goes beyond previous explorations in the literature. The most extensive one is Ref.~\cite{Bali:2013kia}, which is a thorough study carried out in the quenched approximation. While this limit captures the correct large $N_c$ result, it modifies subleading effects in an uncontrolled way. Furthermore, in Ref.~\cite{DeGrand:2016pur} the same quantities were explored with $N_f = 2$ dynamical fermions, but at larger pion masses, and no chiral fits were performed. 

The lattice setup of this work is the one described in the previous section: four dynamical fermions, and $N_c=3-6$. We extract the pion mass and decay constant from the pseudoscalar two-point function. For the latter, we use Eq.~(\ref{eq:Fpitm}). Furthermore, we only included the ``A'' ensembles in Table~\ref{tab:ensembles}. 

\begin{figure}[h!]
\centering
\begin{subfigure}{0.8\textwidth}
  \centering
  \includegraphics[width=.999\linewidth]{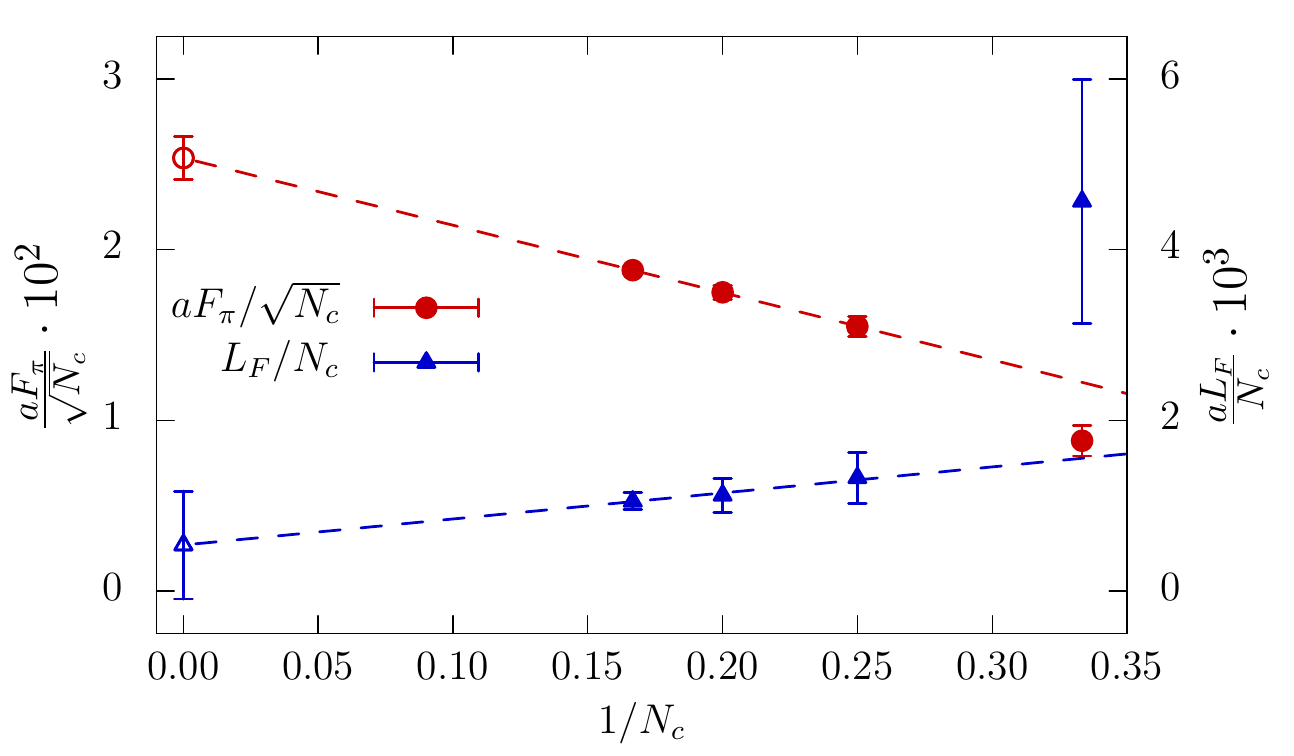}  
  \caption{LECs of the decay constant.}
  \label{fig:fpiscaling}
\end{subfigure} 
\begin{subfigure}{0.8\textwidth}
  \centering
  \includegraphics[width=.999\linewidth]{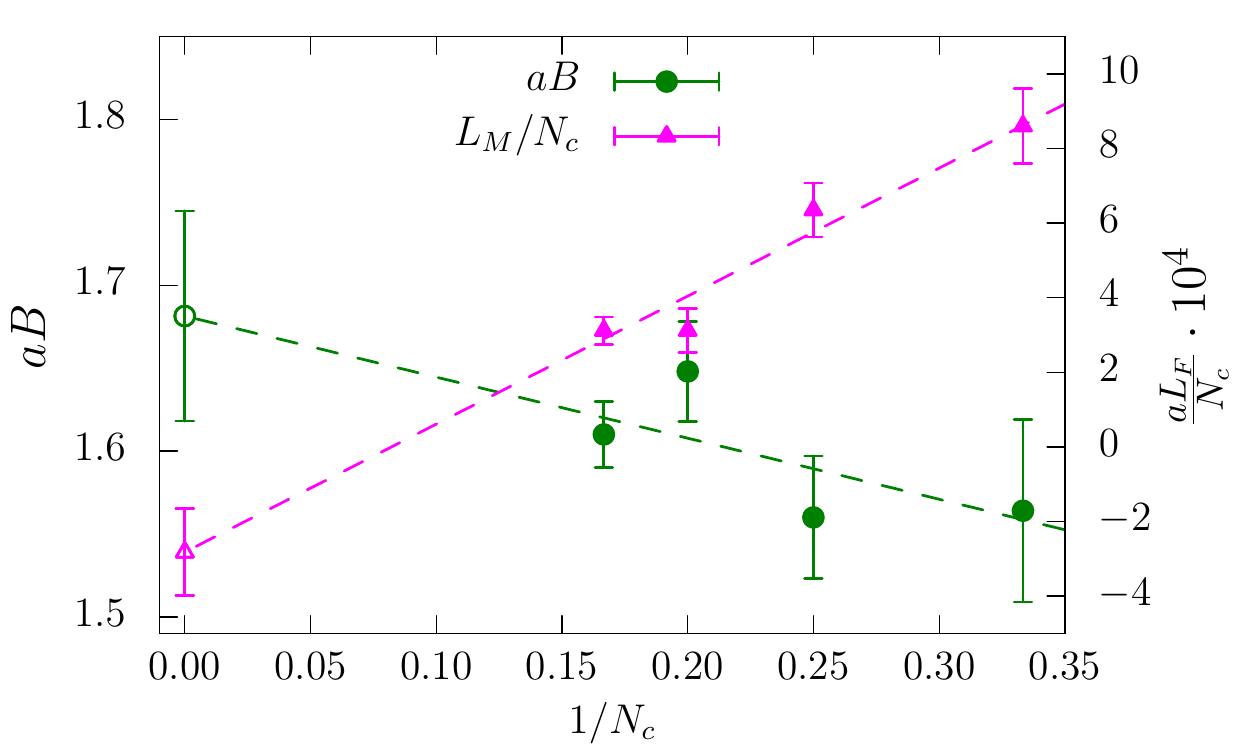}  
  \caption{LECs of the meson mass.}
  \label{fig:massscaling}
\end{subfigure}
\caption{$N_c$ dependence of the LO and NLO LECs extracted from fits to Eqs. ~(\ref{eq:Fpi1}) and (\ref{eq:Mpi1}). The figure is taken from Ref.~\cite{Hernandez:2020tbc}, but it uses data from the original article~\cite{Hernandez:2019qed}.}  
\label{fig:chiral1}
\end{figure}

First, the ensembles at fixed value of the number of colours are considered separately, and compared to the $SU(N_f)$ NLO ChPT predictions for $F_\pi$ and $M_\pi$: 
\begin{align}
F_\pi &= {F} \Bigg[1   - {\frac{N_f}{2} \frac{M_\pi^2}{(4 \pi F_\pi)^2}\log \frac{M_\pi^2}{\mu^2}}     +4 \frac{M_\pi^2}{F_\pi^2} L_F\Bigg],  
\label{eq:Fpi1}
 \\
\frac{M^2_\pi}{m} &= 2B \Bigg[1  + {\frac{1}{N_f} \frac{M_\pi^2}{(4 \pi F_\pi)^2}\log \frac{M_\pi^2}{\mu^2}}  +8\frac{M_\pi^2}{F_\pi^2} L_M\Bigg]. \label{eq:Mpi1} 
\end{align} 
We employ here the same notation as in Section \ref{sec:ChPT}. Note that if valence twisted-mass fermions are used, the quark mass is $m = \mu_0/Z_P$, where $Z_P$ is the pseudoscalar renormalization constant. Moreover, $L_M$, $L_F$ are combinations of renormalized LECs:
\begin{equation}
L_F = L_5^r  +  N_f L_4^r , \ \ \ L_M = 2L_8^r - L_5^r + N_f (2 L_6^r  - L_4^r ).
\end{equation}
As explained before, $F^2, L_5$ and $L_8$ are $O(N_c)$ and $B, L_4$ and $L_6$ are $O(N_c^0)$. 

\begin{figure}[h!]
\centering
\begin{subfigure}{.8\textwidth}
  \centering
  \includegraphics[width=.999\linewidth]{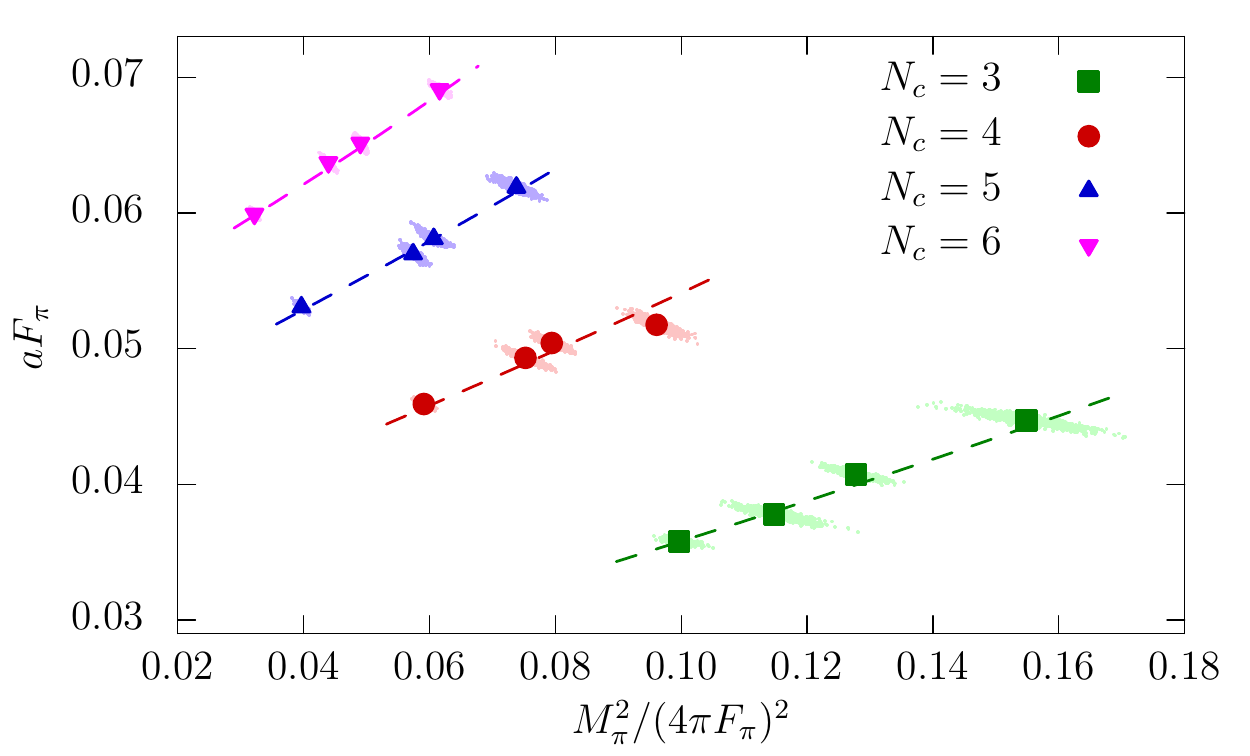}  
  \caption{}
  \label{fig:fpichiral}
\end{subfigure} 
\begin{subfigure}{.8\textwidth}
  \centering
  \includegraphics[width=.999\linewidth]{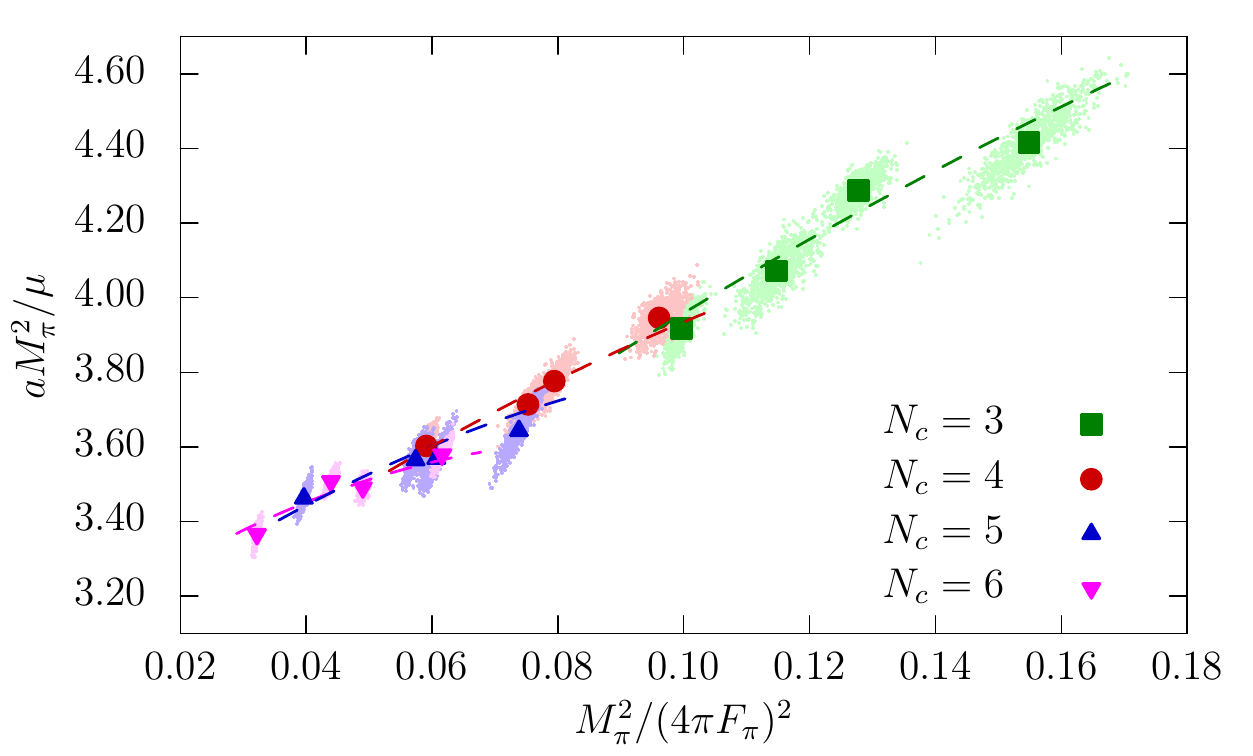}  
  \caption{}
  \label{fig:mpichiral}
\end{subfigure}
\caption{Simultaneous chiral and $N_c$ fits for $F_\pi$ (top) and $M_\pi$ (bottom). Bootstrap samples are depicted as shaded areas around the corresponding central value. The figure is taken from Ref.~\cite{Hernandez:2020tbc}, but it uses data from the original article~\cite{Hernandez:2019qed}.  } 
\label{fig:chiral2}
\end{figure}

The results of the fits to Eqs.~(\ref{eq:Fpi1}) and (\ref{eq:Mpi1}) are shown in Figs.~\ref{fig:fpiscaling} and \ref{fig:massscaling}, respectively. We also show a fit of the LECs to a leading and subleading coefficient in the $1/N_c$ expansion:
\begin{align}
{L_{F,M}} = L_{F,M}^{(0)} N_c &+ L_{F,M}^{(1)} ,\\
F = \sqrt{N_c} \left(F_0 + F_1 \frac{1}{N_c}\right) &, \quad
B = B_0 + B_1 \frac{1}{N_c}. \label{eq:expansionNc}
\end{align}
As can be seen, the scaling for $N_c=4-6$ is well described by Eq.~(\ref{eq:expansionNc}), while $1/N_c^2$ corrections are significant for $F_\pi$ with $N_c=3$. Also note that the extracted $B$ is bare, due to the use of the unrenormalized twisted mass.

In Section \ref{sec:chptlargenc}, we have discussed how the chiral Lagrangian, and its power counting is modified to incorporate the $\eta'$ meson---see Eq.~(\ref{eq:powercounting}). In this case, the NNLO predictions [$\mathcal O (\delta^2)$] for the pion mass and decay constant are~\cite{Guo:2015xva}:
\begin{align}
\begin{split}
&  F_\pi =\sqrt{N_c} \left( {F}_0+ {{F}_1\over N_c} + {{ F}_2 \over N_c^2} \right) \Bigg[1    - {\frac{N_f}{2} \frac{M_\pi^2}{(4 \pi F_\pi)^2}\log \frac{M_\pi^2}{\mu^2}}  \\
&+4 \frac{M_\pi^2}{F_\pi^2} \Big( N_c {L}_F^{(0)} + {L}_F^{(1)} \Big) + N_c^2 K_F^{(0)} \left({M_\pi^2\over F_\pi^2}\right)^2 + \ {\mathcal O}(\delta^3)  \Bigg] , 
\label{eq:Fpi}
\end{split}
\end{align}
and 
\begin{align}
\begin{split}
\frac{M^2_\pi}{2m} =&  \left( { B}_0+ {{ B}_1\over N_c} + {{B}_2 \over N_c^2} \right)  \Bigg[
1+ {\frac{1}{N_f} \frac{M_\pi^2}{(4 \pi F_\pi)^2}\log \frac{M_\pi^2}{\mu^2}}  \\-& {\frac{1}{N_f} \frac{M_{\eta'}^2}{(4 \pi F_\pi)^2}\log \frac{M_{\eta'}^2}{\mu^2}} +8\frac{M_\pi^2}{F_\pi^2} \Big( N_c {L}_M^{(0)} + {L}_M^{(1)} \Big) \\+& N_c^2 K_M^{(0)} \left({M_\pi^2\over F_\pi^2}\right)^2 +{\mathcal O}(\delta^3) \Bigg]   .\label{eq:Mpi}
\end{split}
\end{align}
In the previous equations, $F_i, B_i, L_M^{(i)}$ and $L_F^{(i)}$ are the coefficients of the $1/N_c$ expansion of the corresponding couplings---see Eq.~(\ref{eq:expansionNc}). Furthermore, $K_{F,M}$ are complicated combinations of LECs that contribute at the next order in the chiral expansion: $\mathcal{O}(M_\pi^4)$. Since the mass of the $\eta'$ meson is not measured directly, the Witten-Veneziano equation is assumed. Another technical point is that we choose $\mu^2 = \frac{3}{N_c} (4 \pi F_\pi)^2 $ for the renormalization scale, in order to cancel the leading $N_c$ dependence. The chiral dependence for $M_\pi$ and $F_\pi$, along with a global chiral and $N_c$ fit to Eqs.~(\ref{eq:Mpi}) and~(\ref{eq:Fpi}) are shown in Fig. \ref{fig:chiral2}. As can be seen, the chiral predictions seem to describe data well, with $\chi^2/\text{dof} < 1$ for $F_\pi$ and $\chi^2/\text{dof} \sim 2$ for $M_\pi$---see Tables VI and VII in Ref.~\cite{Hernandez:2019qed}. An interesting observation is that the subleading contribution to some of the LECs is larger than the leading one at $N_c=3$, as shown in Table VIII in Ref.~\cite{Hernandez:2019qed}.

Another result that was exploited in Ref. \cite{Hernandez:2019qed} is that by studying the first subleading term in the $1/N_c$ expansion, one can derive the values of certain observables in theories with different number of flavours. This was discussed explicitly for the decay constant in Eq.~(\ref{eq:Fpiexpansion}), where the leading correction goes as $N_f/N_c$. This way, we can infer:
\begin{align}
\begin{split}
 F^{N_c=3, N_f=2} = 81(7)  \text{ MeV},  \\  F^{N_c=3, N_f=3} = 68(7)  \text{ MeV}.\label{eq:FNf2}
\end{split}
\end{align}
These numbers are in good agreement with various determinations---see the FLAG report~\cite{Aoki:2019cca} for a summary. 

\subsection{Dissecting the $\Delta \text{I=1/2}$ rule at large $\text{N}_\text{c}$} \label{sec:dissecting}

The goal of Ref.~\cite{Donini:2020qfu} is to understand the origin of the large $1/N_c$ corrections to the $K \to \pi\pi$ amplitudes. For this, we studied for the first time the $N_c$ scaling of weak matrix elements relevant to the $\Delta I=1/2$ rule. An earlier exploratory study in the quenched approximation was presented by us in Ref.~\cite{Donini:2016lwz}.

A direct computation of the $K \to \pi\pi$ amplitudes from lattice simulations is possible---the Lellouch-L\"uscher formalism~\cite{Lellouch:2000pv}. It is however a complex calculation with large uncertainties, as evidenced by the recent work of the RBC-UKQCD collaborations~\cite{Abbott:2020hxn}. We follow an indirect path, based on earlier work on this subject~\cite{Giusti:2004an,Giusti:2006mh}, that exploits ChPT and involves the evaluation of simpler $K \to \pi$ amplitudes.

The lattice setup is again the one described in Section~\ref{sec:technical}. We use both, the ``A''  and ``B'' ensembles of Table~\ref{tab:ensembles}. The ``B'' ensembles have a finer lattice spacing, and they are used to estimate discretization effects.

Let us now discuss our strategy. In Section~\ref{sec:EFT}, we argued that at energies below $M_W$, the electroweak gauge bosons can be integrated out. The weak interactions can then be represented by four-fermion operators. This is in fact a necessary step to study weak interactions on the lattice, due to the large separation of scales: $M_W \gg \frac{1}{a} \gg \Lambda_{QCD}$.  For the case of CP-conserving transitions with variation of strangeness of one unit,  $\Delta S=1$, the Hamiltonian takes the simple form~\cite{Gaillard:1974nj}:
\begin{eqnarray}
\label{eq:heffs1}
{\mathcal H}^{N_f=4}_{\Delta S=1} =\sqrt{2}G_{\rm F}V_{us}^*V_{ud}   ( k^+ \, \bar{Q}^+(x)+k^- \, \bar{Q}^-(x) )\,, 
\label{eq:hw4}
\end{eqnarray}
with
\begin{align}
\begin{split}
\bar{Q}^\pm &= Z_Q^\pm\, Q^\pm  \\&
=Z_Q^\pm \, \bigg(  (\bar{s}_L \gamma_\mu u_L)(\bar{u}_L \gamma^\mu d_L) \pm (\bar{s}_L \gamma_\mu d_L)(\bar{u}_L \gamma^\mu u_L)  -[u\leftrightarrow c]  \bigg).  \label{eq:currento}
\end{split}
\end{align}
The flavour symmetry group is ${SU}(4)_{\rm L} \otimes SU(4)_{\rm R}$. $Q^+$ transforms under the $(84,1)$ irrep, while $Q^-$ under the $(20,1)$. Whereas both operators contribute to $A_0$, $\bar{Q}^+$ fully determines $A_2$. Thus, the hierarchy of the amplitudes must be translated into a hierarchy of the matrix elements of the operators. In addition, $k^\pm$ are the Wilson coefficients, and $ Z_Q^\pm$ are the renormalization constants of the bare operator in some regularization scheme. The Hamiltonian in Eq.~(\ref{eq:hw4}) is valid above the charm mass, $m_c$. An interesting observation is that the separation of scales $M_W \gg m_c$ induces large logarithms that enhance the ratio of Wilson coefficients~\cite{Gaillard:1974nj,Altarelli:1974exa}:  $k^-(m_c)/k^+(m_c) \sim 2$. This is clearly not enough, and suggests that the main source of enhancement lies elsewhere. 

The conventional approach in the literature is to integrate out the charm quark. The resulting $N_f=3$ effective weak Hamiltonian~\cite{Buchalla:1995vs} has ten different operators, including the famous penguin operators. In fact, it was proposed that the latter could be responsible for the $\Delta I=1/2$ rule~\cite{Shifman:1975tn}. However, as seen by the RBC-UKQCD collaboration~\cite{Bai:2015nea,Blum:2015ywa,Boyle:2012ys}, the contribution from penguin diagrams is not dominant. The effect of the charm can then be disentangled by considering the so-called GIM limit, i.e.,  $m_u=m_c$~\cite{Giusti:2004an,Giusti:2006mh}. In this limit there is no charm threshold, and the weak Hamiltonian keeps the same structure with just two current-current operators, after the renormalization-group running. If the $\Delta I=1/2$ enhancement still occurs in this limit, one can conclude that it is a low-energy non-perturbative phenomenon, unrelated to the charm threshold. From the technical point of view, the GIM limit is also advantageous because no closed quark propagator contributes to the amplitudes. This explains the choice $N_f=4$ for our lattice simulations.

At hadronic scales, a further simplification is possible. This consists of matching the effective Hamiltonian in Eq.~(\ref{eq:hw4}) to ChPT. At leading order, only two chiral structures appear with the same transformation properties as the operators in Eq.~(\ref{eq:currento}).  Correspondingly, there are two weak LECs, $g^\pm$, that need to be determined nonperturvatively.  This way, the chiral weak hamiltonian is \cite{Giusti:2004an,Giusti:2006mh}
\begin{eqnarray}
{\mathcal H}^{N_f=4}_{\rm ChPT} =\sqrt{2}G_{\rm F}V_{us}^*V_{ud}   ( g^+ \, {\mathcal Q}^++g^- \, {\mathcal Q}^-)\,, 
\label{eq:chpt4}
\end{eqnarray}
with
\begin{eqnarray}
{\mathcal Q}^\pm &=& {F_\pi^4\over 4} \left[(U \partial_\mu U^\dagger)_{us} (U \partial_\mu U^\dagger)_{du}\pm (U \partial_\mu U^\dagger)_{ds} (U \partial_\mu U^\dagger)_{uu} \right. \nonumber\\
&-&\left. (u\rightarrow c)\right] .
\label{eq:currentochpt}
\end{eqnarray}
At this order in ChPT, the ratio of $K \to \pi\pi$ isospin amplitudes is given in terms of the ratio of LECs:
\begin{eqnarray}
{A_0\over A_2} ={1\over 2 \sqrt{2}} \left(1+ 3 {g^-\over g^+}\right). \label{eq:A0overA2}
\end{eqnarray}
It is now clear that in this approximation an enhancement in $g^-/g^+$ could explain the  $\Delta I=1/2$ rule.
The couplings can be extracted from the appropriate matrix elements obtained from Euclidean correlation function on the lattice. In particular, the $K \to \pi$ amplitudes correspond to $g^\pm$ in the chiral limit:
\begin{eqnarray}
A^\pm = \braket{K| k^\pm \bar{Q}^\pm  |\pi}, \quad
 \lim_{M_\pi \rightarrow 0} A^\pm = g^\pm. 
\end{eqnarray}
More concretely, $A^\pm$ can be obtained from the following ratio (up to Wilson coefficients and renormalization constants):
\begin{eqnarray}
R^\pm  = \kern-1.0em
\lim_{ \substack{z_0-x_0\to\infty \\ y_0-z_0\to \infty}}
\frac{\sum_{{\mathbf x},{\mathbf y}} \langle P(y) Q^\pm(z) P(x) \rangle}
{\sum_{{\mathbf x,\mathbf y}}  \langle P(y) A_0(z) \rangle  \langle P(x) A_0(z) \rangle}, \label{eq:ratios}
\end{eqnarray}
where $A_0$ and $P$ are nonsinglet axial and pseudoscalar currents with appropriate flavour content.

 \begin{figure}[h!]
 \begin{center}
\renewcommand{\arraystretch}{2}
\[\begin{array}{lll}
 \raisebox{-.5\height}{\includegraphics[scale=1.2]{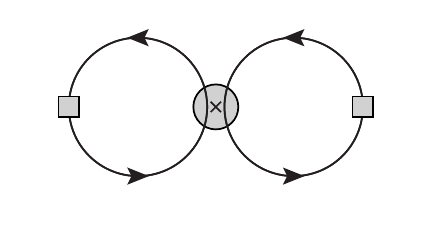}} & \mp &\raisebox{-.5\height}{\includegraphics[scale=1.2]{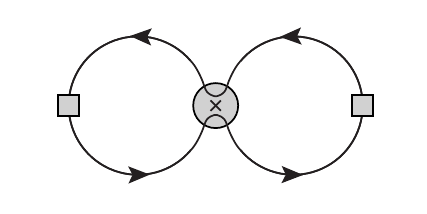}}\\
 \end{array}\]
 \caption{\label{fig:condis} Colour-disconnected (left) and colour-connected (right) contributions to the three-point function in $R^\pm$. Source: Ref.~\cite{Hernandez:2020tbc}. }
\end{center}
\end{figure}

It turns out that the three-point function in the numerator of Eq.~(\ref{eq:ratios}) gets contributions from two separate contractions that scale differently with $N_c$. More specifically, there is a colour-connected contraction that is suppressed with $1/N_c$ with respect to the colour-disconnected one, and changes sign for $R^\pm$---see Fig.~\ref{fig:condis}. Therefore, in the strict large $N_c$ limit, one has $A^+=A^-$, and so Eq.~(\ref{eq:A0overA2}) recovers the large $N_c$ result of $\sqrt{2}$.

\begin{figure}[h!]
\centering
\begin{subfigure}{.8\textwidth}
  \centering
  \includegraphics[width=.999\linewidth]{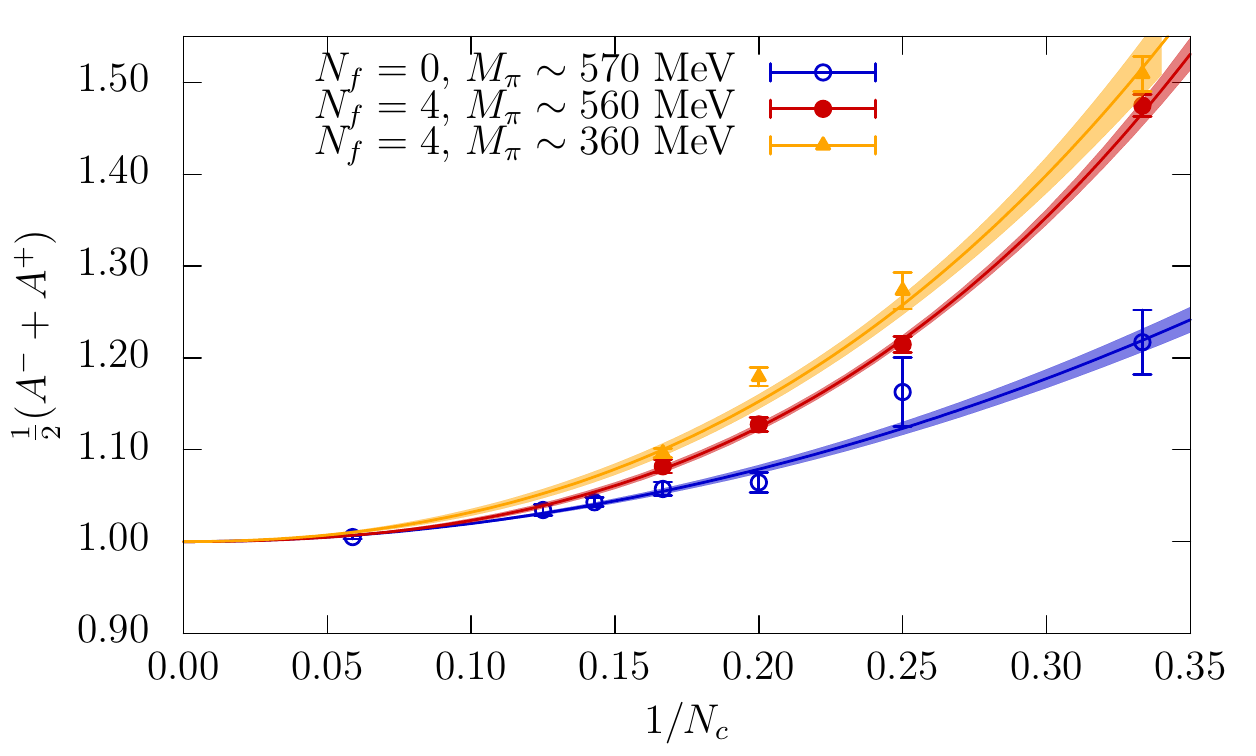}  
  \caption{}
  \label{fig:ammap}
\end{subfigure}
\begin{subfigure}{.8\textwidth}
  \centering
  \includegraphics[width=.999\linewidth]{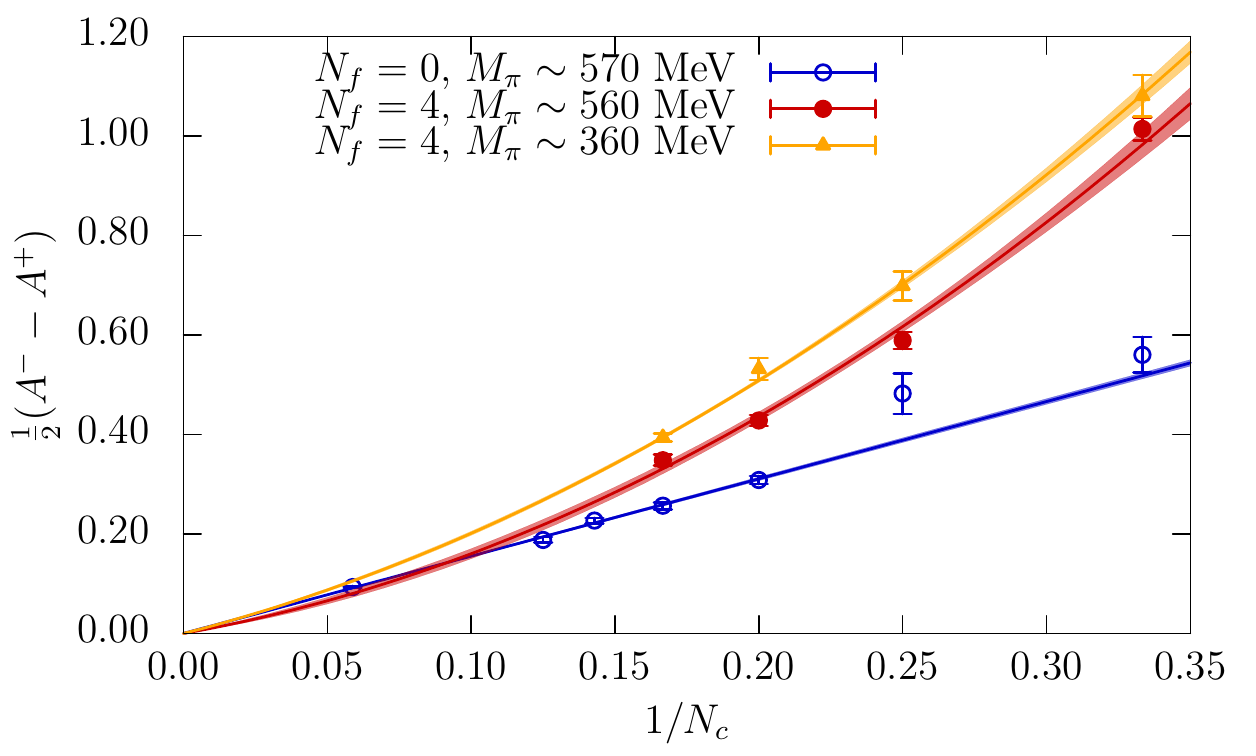}  
  \caption{}
  \label{fig:ampap}
\end{subfigure}
\caption{Half-sum and half-difference of $A^\pm$ as a function of $1/N_c$. Three different cases are shown: (i) quenched in blue, (ii) dynamical at a pion similar to the quenched case (red), and (iii) dynamical at lower $M_\pi$ (orange). Errors are only statistical. The figure is taken from Ref.~\cite{Hernandez:2020tbc}, but it uses data from the original article~\cite{Donini:2020qfu}. } 
\label{fig:ktopi1}
\end{figure}

A careful analysis of the subleading contributions in $1/N_c$ to the amplitudes $A^\pm$ was carried out in Ref.~\cite{Donini:2020qfu}. This is indeed very similar to the one for $F_\pi$ in Eq.~({\ref{eq:Fpiexpansion}}). The result is that the amplitudes can be expanded as
\begin{eqnarray}
A^\pm = 1 \pm \tilde a {1 \over N_c}\pm \tilde b {N_f \over N^2_c}+\tilde c {1 \over N^2_c}+ \tilde d {N_f \over N^3_c}+\cdots, \label{eq:rnc}
\end{eqnarray}
with coefficients $\tilde a- \tilde d$ that are 
independent of $N_c$ and $N_f$, but can depend on the pseudoscalar mass. A {\it natural} expectation for their magnitude is  ${\mathcal O}(1)$. It will be convenient to study the linear combinations
\begin{align}
\frac{A^- + A^+}{2} &= 1 + \tilde c \frac{1}{N_c^2}  + \tilde d \frac{N_f}{N_c^3} , \\ \frac{A^- - A^+}{2} &=  \tilde a \frac{1}{N_c}+\tilde b \frac{N_f}{N_c^2}, \label{eq:halfcomb}
\end{align}
as they isolate the (anti)correlated coefficients. In our work, we have studied them in three different situations: (i) quenched simulations ($N_f=0$) with $M_\pi \sim 570$ MeV~\cite{Donini:2016lwz}, (ii) $N_f=4$ simulations with $M_\pi \sim 560$ MeV, and (iii) $N_f=4$ with lighter pions: $M_\pi \sim 360$ MeV.  The dependence on $N_c$ of the half-sum and half-difference of the amplitudes are shown\footnote{See also Table V in Ref.~\cite{Donini:2020qfu}} in Fig.~\ref{fig:ktopi1}.  A fit to the forms in Eq.~(\ref{eq:halfcomb}) is also shown as the colour band. Interestingly, all coefficients are found to be of the natural size. In addition, $\tilde a$ and  $\tilde b$ are both negative. This reduces $A^+$, while enhancing $A^-$  in a correlated way. Because of the coefficient $\tilde b$, fermion loops are a signifcant contribution to the enhancement. Regarding the mass dependence in the dynamical simulations, it seems that it affects mostly the coefficient $\tilde a$, and increases the ratio $A^-/A^+$ towards the chiral limit.

\begin{figure}[t!]
\centering
\begin{subfigure}{.8\textwidth}
  \centering
  \includegraphics[width=.999\linewidth]{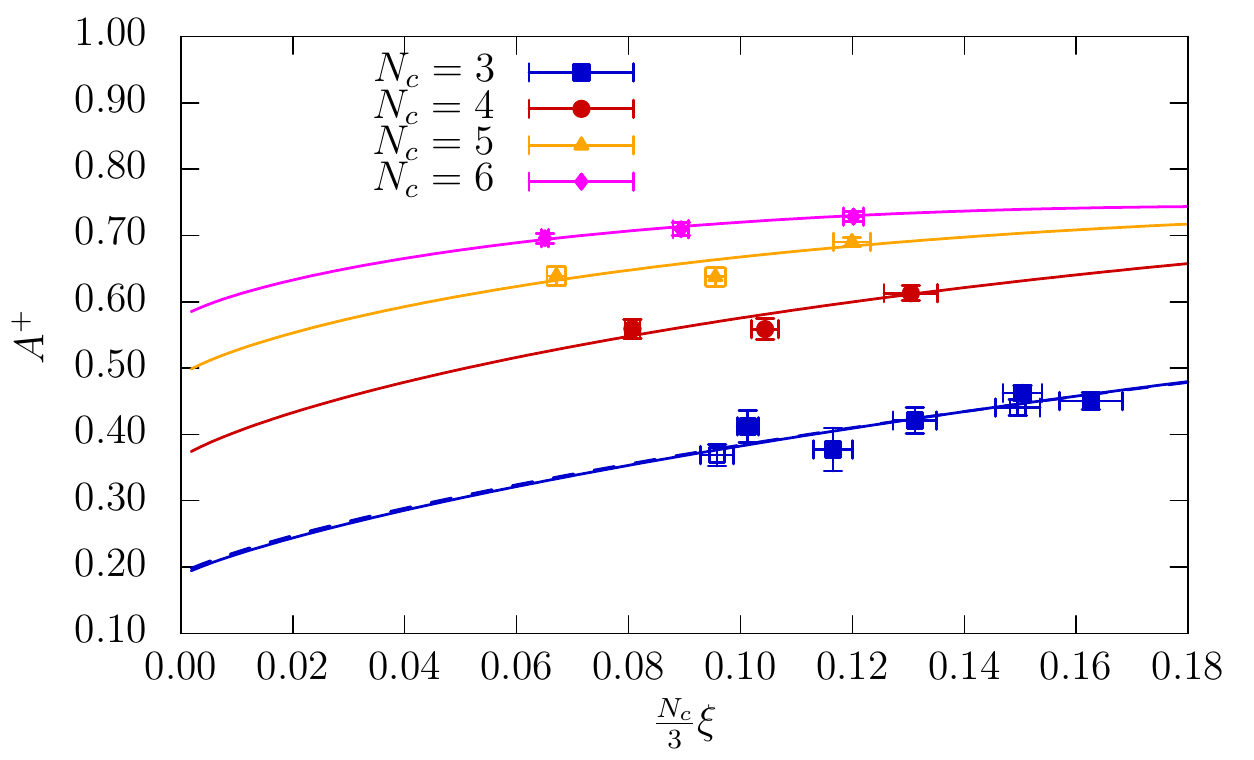}  
  \caption{}
  \label{fig:plotplus}
\end{subfigure}
\begin{subfigure}{.8\textwidth}
  \centering
  \includegraphics[width=.999\linewidth]{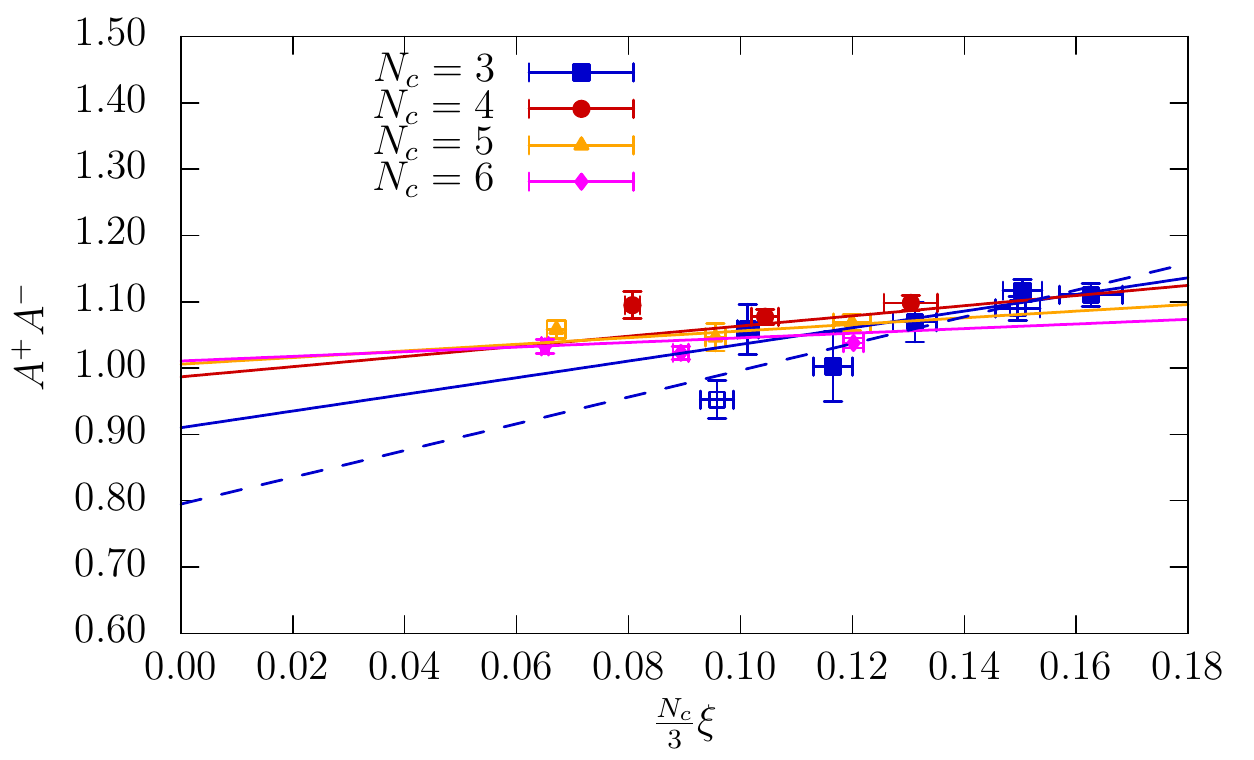}  
  \caption{}
  \label{fig:plotprod}
\end{subfigure}
\caption{Chiral extrapolation of $A^+$ and $A^+A^-$. We use $\xi = \left[M_\pi/(4\pi F_\pi)\right]^2$ in the x-axis. The data points come from the ensembles in Table~\ref{tab:ensembles}, and we use empty squares for the ``B'' ensembles (finer lattice spacing). Solid lines indicate a simultaneous chiral and $N_c$ fit. Dashed lines correspond to the chiral extrapolation at $N_c = 3$.  The figure is taken from Ref.~\cite{Hernandez:2020tbc}, but it uses data from the original article~\cite{Donini:2020qfu}. } 
\label{fig:ktopi2}
\end{figure}

In order to extract $g^\pm$, we need to perform a chiral extrapolation. Alternatively, we incorporate the mass corrections in ChPT. At NLO, the chiral dependence of $A^\pm$~\cite{Hernandez:2006kz,Kambor:1989tz} is given by
\begin{equation}
A^\pm = g^\pm \left[ 1 \mp 3 \left( \frac{M_\pi}{4 \pi F_\pi}\right)^2\log \frac{M_\pi^2}{\Lambda_\pm^2} \right]. 
\end{equation}
The result of the chiral fit for $A^+$ to this function is shown in Fig. \ref{fig:plotplus}, and for the product $A^+ A^-$  in Fig. \ref{fig:plotprod}. With these results, the ratio of couplings is found to be:
\begin{equation}
\frac{g^-}{g^+} \Bigg \rvert_{N_c=3}= 22(5),
\end{equation}
where the error is only statistical. Finally, using the LO ChPT formula in Eq.~(\ref{eq:A0overA2}), as well as the NLO correction derived in Ref. \cite{Donini:2020qfu}, an indirect estimate for the ratio of isospin amplitudes is:
\begin{equation}
\frac{A_0}{A_2} \Bigg \rvert_{N_f=4, N_c=3} = 24(5)_\text{stat}(7)_\text{sys}, \label{eq:A0A2}
\end{equation}
which is valid in the theory with a light charm quark.

We end by stating the main conclusions of this work. First, the large enhancement observed in the $\Delta I=1/2$ rule seems consistent with coefficients in the $1/N_c$ expansion that are of the natural size, i.e., $O(1)$. It must be mentioned that a sizeable contribution to the hierarchy originates in quark loops, that is, $N_f/N_c$ effects.  Second, the result in Eq.~(\ref{eq:A0A2}) suggests that the enhancement may indeed be largely dominated by intrinsic QCD effects, instead of rescattering effects or the crossing of the charm threshold. In fact, even in the simplified setup, our results are consistent with the recent RBC/UKQCD update at the physical point~\cite{Abbott:2020hxn}, which appeared after our work.

\subsection{Concluding remarks}

Lattice Field Theory offers the possibility of exploring the parameter space of nonabelian gauge theories: different number of colours, flavours and even fermionic representations. We have used this possibility to study of QCD in the large $N_c$ limit. Our main motivation has been to understand the origin of the large $1/N_c$ corrections in the ratio of isospin amplitudes of the $K \to \pi \pi$ weak decay. To this end, we have tested the scaling of various observables with the number of colours: meson masses and decay constants~\cite{Hernandez:2019qed}, as well as weak matrix elements~\cite{Donini:2020qfu}.  

We have observed that all the explored quantities have a $1/N_c$ expansion with coefficients of $O(1)$. For the case of pion masses and decay constants, we have been able to disentangle the leading and subleading terms, and even found that some subleading contributions are non-negligible. In addition, a milestone in our work has been to reconcile this with the observed $\Delta I=1/2$ rule. 

Further insight can be gained by exploring other observables using lattice QCD. A nonperturbative test of the Witten-Veneziano equation at large $N_c$ would also be of interest, in other words, properties of the $\eta'$ meson at large $N_c$. Another compelling direction is the exploration of scattering observables with growing $N_c$. In fact, some preliminary results on a two-$\pi^+$ system were presented by us in Ref.~\cite{Romero-Lopez:2019gqt}. More attractive are resonant channels---while we know that resonances become stable at $N_c \to \infty$, subleading corrections may show surprising features. A related question is if exotics, such as tetraquarks, survive at large $N_c$, and whether this can be explored on the lattice. We expect to pursue this line of research in the future.


\chapter{Multiparticle processes on the lattice}
\label{sec:multiparticle}

The extraction of scattering and decay amplitudes from lattice QCD simulations has become a hot topic for the lattice QCD community. The case of two-particle scattering is by now well established for generic $2 \to 2$ processes, with many applications to different systems, e.g., two baryons or coupled-channel scattering. In this context, the present frontier has become the determination of three-particle scattering amplitudes and related decays. Interestingly, lattice QCD can already offer access to three-particle scattering processes that are hard to determine experimentally.

Compared to collider experiments, the study of hadronic interactions is intrinsically different in lattice QCD. The reason for this is simple: multiple particles in a box can never be pulled apart, and thus one cannot define asymptotic states. Therefore, scattering quantities must be extracted in some other way. A solution to this was developed by M. L\"uscher in the 1980s. He realized that the energy levels of the theory in finite volume (and their volume dependence) contain information about the interactions. The so-called L\"uscher formalism is nothing else than a mapping between the two-particle spectrum and the two-particle scattering amplitude~\cite{Luscher:1986pf,Luscher:1990ux}. The existing generalizations to three particles follow the same lines, although with technical complications that will be address below.

This chapter is organized as follows. In the first section, we will introduce some relevant concepts to understand scattering processes in infinite volume. Subsequently, we will revisit the main ideas behind the finite-volume two-particle formalism for scattering processes and decays. We will then turn to processes involving three particles in Section~\ref{sec:3particles}. After a brief review of the formalism, we will discuss four of the papers included in this thesis: (i) implementing the three-particle quantization condition including $d$-wave interactions~\cite{Blanton:2019igq}, (ii) the first application of the three-particle formalism to analyze a full three-$\pi^+$ spectrum~\cite{Blanton:2019vdk}, (iii) generalizing the three-particle formalism to a generic three-pion system~\cite{Hansen:2020zhy}, and (iv) the formalism for three-pion decays, such as $K \to 3\pi$~\cite{Hansen:2021ofl}. We will conclude with some remarks.

\section{Scattering quantities from lattice QCD}

In this section, we will cover important concepts of scattering in infinite volume, and the related finite-volume formalism. First, we will introduce the $S$-matrix, the scattering amplitude and the phase shifts, as well as the notion of a resonance. After that, we will present the L\"uscher method, i.e., the relation between the finite-volume spectrum and the two-particle interactions. We will end by commenting on the Lellouch-L\"uscher formalism, used to study two-particle decays from finite-volume matrix elements. This section will serve as a warm up for the next section, where we will deal with three-particle processes.

\subsection{Scattering in infinite volume}

The scattering matrix, or $S$-matrix, is an operator that contains information about all the interactions in a given quantum field theory, including the presence of resonances. Its matrix elements can be obtained from\footnote{For simplicity, we focus on two identical particles.} 
\begin{equation}
S_{f,i} = \braket{\text{out} | \hat S | \text{in}},
\end{equation}
where the incoming state is $\ket{\text{in}} \equiv \ket{\boldsymbol p_1, \boldsymbol p_2}$, and  $\ket{\text{out}}\equiv \ket{\boldsymbol k_1, \boldsymbol k_2}$ is the outgoing one. Note that both are considered to be free asymptotic states. The scattering amplitude is defined as the connected part of this matrix element:
\begin{equation}
\braket{\text{out} |\,i \hat T\, | \text{in}} = (2 \pi)^4 \delta^{(4)}( P_\text{in} - P_\text{out}) i \mathcal{M}(\boldsymbol k_1, \boldsymbol k_2; \boldsymbol p_1, \boldsymbol p_2),
\end{equation}
with $\hat S = 1+i \hat T$.  

The fact that the $S$-matrix is unitary, $\hat S^\dagger \hat S = 1$, implies the following constraint for the amplitude of elastic scattering:
\begin{align}
\begin{split}
\mathcal{M}_2(\boldsymbol k_1, \boldsymbol k_2; &\boldsymbol p_1, \boldsymbol p_2) - \mathcal{M}^*_2(\boldsymbol p_1, \boldsymbol p_2; \boldsymbol k_1, \boldsymbol k_2) =\\\frac{i}{2}& \int \frac{d^3q_1\, d^3q_2}{(2\pi)^6 4 \omega(q_1) \omega(q_2)}  \mathcal{M}_2(\boldsymbol k_1, \boldsymbol k_2; \boldsymbol q_1, \boldsymbol q_2) \mathcal{M}_2^*(\boldsymbol p_1, \boldsymbol p_2; \boldsymbol q_1, \boldsymbol q_2) \\ &\times (2\pi)^4 \delta^{(4)}( k_1+k_2-q_1-q_2), \label{eq:unitarity}
\end{split}
\end{align}
with $\omega(q) = \sqrt{m^2 + q^2}$, and the factor $1/2$ arises because of having identical particles. This relation is known as two-particle unitarity. It can be seen that the following expression satisfies the $s$-wave projection of the unitarity condition:
\begin{equation}
\mathcal{M}^s_2 = \frac{16 \pi \sqrt{s} }{k \cot \delta_0 - i k}, \label{eq:M2s}
\end{equation}
where $\delta_0$ is the $s$-wave phase shift, and $s = 4(M^2+k^2)$. The $K$-matrix is closely related to the scattering amplitude:
\begin{equation}
\frac{1}{\mathcal{M}^s_2} = \frac{1}{\mathcal{K}^s_2} - i \rho,
\end{equation}
where $\rho = k/(16 \pi \sqrt{s}) $ is the two-particle phase space. Therefore,
\begin{equation}
\mathcal{K}^s_2 = \frac{16 \pi \sqrt{s} }{k \cot \delta_0},  \label{eq:K2s}
\end{equation}
which is strictly real.  A standard parametrization for $\delta_0$ is given by a momentum expansion, the so-called effective range expansion (ERE):
\begin{equation}
k \cot \delta_0 = -\frac{1}{a_0} + \frac{1}{2} r_0 k^2 + O(k^4). \label{eq:EREs}
\end{equation}
This defines $a_0$ as the $\ell=0$ scattering length, and $r_0$ as the effective range.

\begin{figure}[h!]
  \centering
  \includegraphics[width=0.8\linewidth]{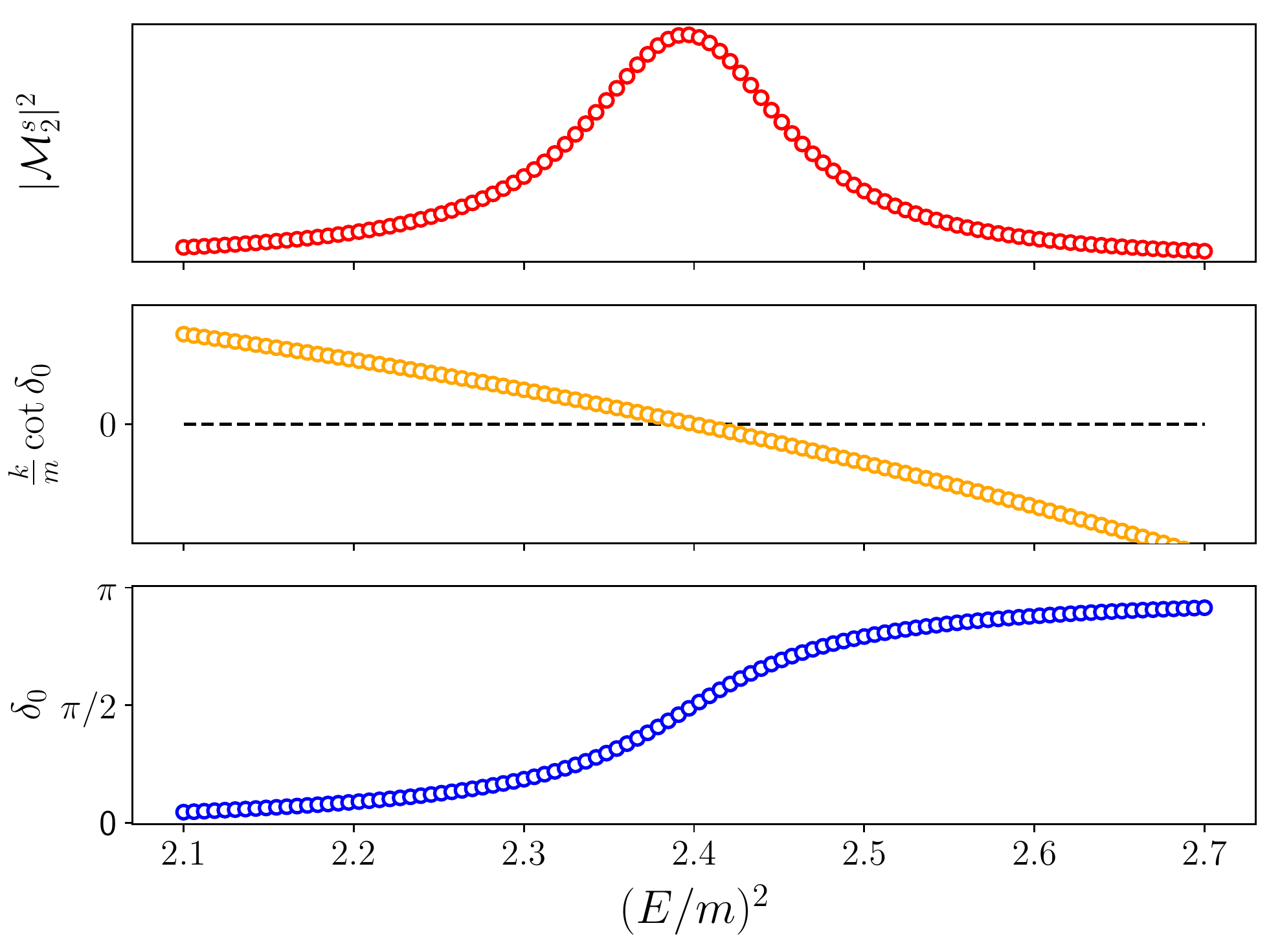}  
  \caption{Toy example of a narrow resonance with $M_R \sim 2.4 m$, $\Gamma_R \sim 0.15 m$. The upper panel shows the squared magnitude of the scattering amplitude as function of the energy. The middle one is the behaviour of the phase shift in the form $k \cot \delta_0$. The lower plot corresponds to the phase shift growing from zero to $\pi$. Units are arbitrary. }
  \label{fig:resonance}
\end{figure}

An interesting outcome of particle scattering is the appearance of resonances. The experimental signatures of a resonances is a bump in the cross-section ($\sigma$), which is proportional to the squared magnitude of the scattering amplitude, $\sigma \propto |\mathcal{M}_2|^2$. Mathematically, resonances correspond to poles of $\mathcal M_2$ in the complex plane at $\sqrt{s} = M_R - i \Gamma_R/2$, where $M_R$ is the mass, and $\Gamma_R$ its width. The behaviour of an idealized toy resonance is depicted in Fig.~\ref{fig:resonance}. In this example, it can be seen that the bump in the cross-section translates into a zero crossing from above in $k \cot \delta_0$. Equivalently, we see that the phase shift grows from 0 to $\pi$ as the energy crosses  $M_R$.

\subsection{The L\"uscher formalism}

Since lattice calculations are performed in a finite box, scattering amplitudes cannot be obtained in the same manner as in experiments or perturbative calculations. A relevant perspective on this challenge came from the work of Maiani and Testa~\cite{Maiani:1990ca}. They showed that one cannot in general obtain on-shell amplitudes from matrix elements of Euclidean correlation functions\footnote{A recent proposal tries to overcome this in a different way~\cite{Bruno:2020kyl}.}. 
An ingenious alternative strategy is to exploit the finite-size scaling:
restricting particles to a finite volume shifts their energy in a way that depends on their interactions. Early work by Huang and Yuan showed this for the case of hard spheres~\cite{Huang:1957im}, but the quantum field theory formalism for two-particle scattering was pioneered by L\"uscher~\cite{Luscher:1986pf,Luscher:1990ux}. In the subsequent discussion, we will assume periodic boundary conditions in the spatial directions, and an infinite time extent. In addition, discretization effects will be neglected.

Let us consider the simplest case of a state of two identical particles at rest with mass $m$ in a box of size $L$. Lüscher showed~\cite{Luscher:1986pf} that the energy of the ground state differs from that of the one-particle states by a correction that can be expanded in powers of $1/L$---the so-called threshold expansion:
\begin{equation}
\Delta E_2 = E_2 -2 m =\frac{4\pi a_0}{m L^3}\,\biggl\{1+c_1\biggl(\frac{a_0}{L}\biggr)
+c_2\biggl(\frac{a_0}{L}\biggr)^2\biggr\}
+\mathcal{O}(L^{-6})\, ,  
\label{eq:thresholdE2}
\end{equation}
where $c_1 \simeq 2.837$, and   $ c_2 \simeq  6.375$. To the given order in $L$, this corresponds to a one-to-one mapping between the energy shift of the two-particle ground state and the $s$-wave scattering length, $a_0$. Because of its perturbative nature, Eq.~(\ref{eq:thresholdE2}) is only valid for big enough boxes, $a_0/L \ll 1$. In practice, it is only useful for weak enough interactions $m a_0 \ll 1$, i.e., in the absence of resonances or bound states. A physical system for which Eq.~(\ref{eq:thresholdE2}) has been successfully applied is isospin-2 $\pi\pi$ scattering ($ 2 \pi^+$ system). Some examples are Refs.~\cite{Feng:2009ij,Helmes:2015gla}, where results for $a_0$ at heavier-than-physical pion masses were combined with a chiral extrapolation to reach the physical point. It must also be mentioned that perturbative expansions in $1/L$ have been extended to three and more particles, as well as excited states~\cite{Beane:2007es,Hansen:2015zta,Sharpe:2017jej,Pang:2019dfe,Romero-Lopez:2020rdq,Muller:2020vtt}.

The nonperturbative mapping between the two-particle spectrum (up to inelastic thresholds) and the scattering amplitude was derived first by L\"uscher in his seminal work for identical scalars in an $s$ wave. Several generalizations have followed~\cite{Luscher:1991cf,Luscher:1990ux,Rummukainen:1995vs, Kim:2005gf, He:2005ey, Bernard:2010fp, Briceno:2012yi, Briceno:2014oea, Romero-Lopez:2018zyy,Luu:2011ep,Gockeler:2012yj}, and the formalism is currently able to treat any two-to-two system: multichannel scattering of nonidentical particles with spin. In fact, the formalism has been successfully applied to many systems---see the following review \cite{Briceno:2017max}.

We now turn to the description of the formalism. We will use the notation of Ref.~\cite{Kim:2005gf}, as it will be convenient in the three-particle case. The two-particle quantization condition (QC2) is a determinant equation whose solutions are the finite-volume energy levels in the presence of interactions. It has the form
\begin{equation}
\det \left[ F^{-1} (E, \boldsymbol P, L)+ \mathcal{K}_2(E^*) \right] =0, \label{eq:QC2}
\end{equation}
where $F$ and $\mathcal K_2$ are matrices with angular momentum indices: ${\ell\, m, \, \ell' m'}$. 
The definition of $F$ is:
\begin{equation}
F = \frac{1}{2} \left[  \frac{1}{L^3}\sum_{\boldsymbol k} - \text{PV} \int\frac{d^3k}{(2\pi)^3} \right] \frac{4 \pi Y_{\ell m}(\boldsymbol{\hat k}^*) Y^*_{\ell' m'}(\boldsymbol{\hat k}^*)  }{2 \omega_k 2 \omega_{Pk} (E- \omega_k- \omega_{Pk})} \left( \frac{k^*}{q^*}\right)^{\ell+\ell'}, \label{eq:F2}
\end{equation}
where $Y_{\ell m}$ are the usual spherical harmonics, $\boldsymbol{k}^*$ is the vector $\boldsymbol{k}$ boosted to the center-of-mass (CM) frame, and 
\begin{equation}
\omega_k = \sqrt{m^2+k^2}, \quad \omega_{Pk} = \sqrt{m^2+(\boldsymbol P - \boldsymbol k)^2}.
\end{equation}
Furthermore, $q^*$ is the back-to-back momentum in the CM frame, defined via
\begin{equation}
E^* = \sqrt{E^2 - \boldsymbol P^2} = 2\omega_{q^*} =2 \sqrt{m^2+(q^*)^2}.
\end{equation}

The pole in the integral in Eq.~(\ref{eq:F2}) is regulated using the principal value (PV) prescription. Further details and an efficient way to evaluate $F$ numerically are given in Ref.~\cite{Kim:2005gf}. Moreover, the partial-wave expansion of $\mathcal{K}_2$ in the CM frame reads 
\begin{equation}
\mathcal{K}_2( P, \boldsymbol q^* ,\boldsymbol q^{\prime\,*}) = Y_{\ell m}(\boldsymbol{\hat q}^*) \left(\mathcal{K}_2\right)_{\ell m, \ell'm'}(E^*)  Y^*_{\ell' m'}(\boldsymbol{\hat q}^{\prime\,*}),
\end{equation}
with
\begin{equation}
\left(\mathcal{K}_2\right)_{\ell m, \ell'm'} = \mathcal{K}_2^{\ell} \delta_{\ell, \ell'} \delta_{m,m'} .
\end{equation}
Note that the $\ell=0$ component is the same as in Eq.~(\ref{eq:K2s}). At this point, additional comments to this formalism are in order. First, the QC2 can be derived by noticing that finite-volume spectrum is given by poles in the finite-volume correlation function of two-particle operators in momentum space. Second, all power-law dependence of energy levels in $1/L$, such as the one in Eq.~(\ref{eq:thresholdE2}), is included in the quantization condition. However, effects that fall off like $e^{-mL}$ or faster are neglected.

\begin{figure}[h!]
  \centering
  \includegraphics[width=0.7\linewidth]{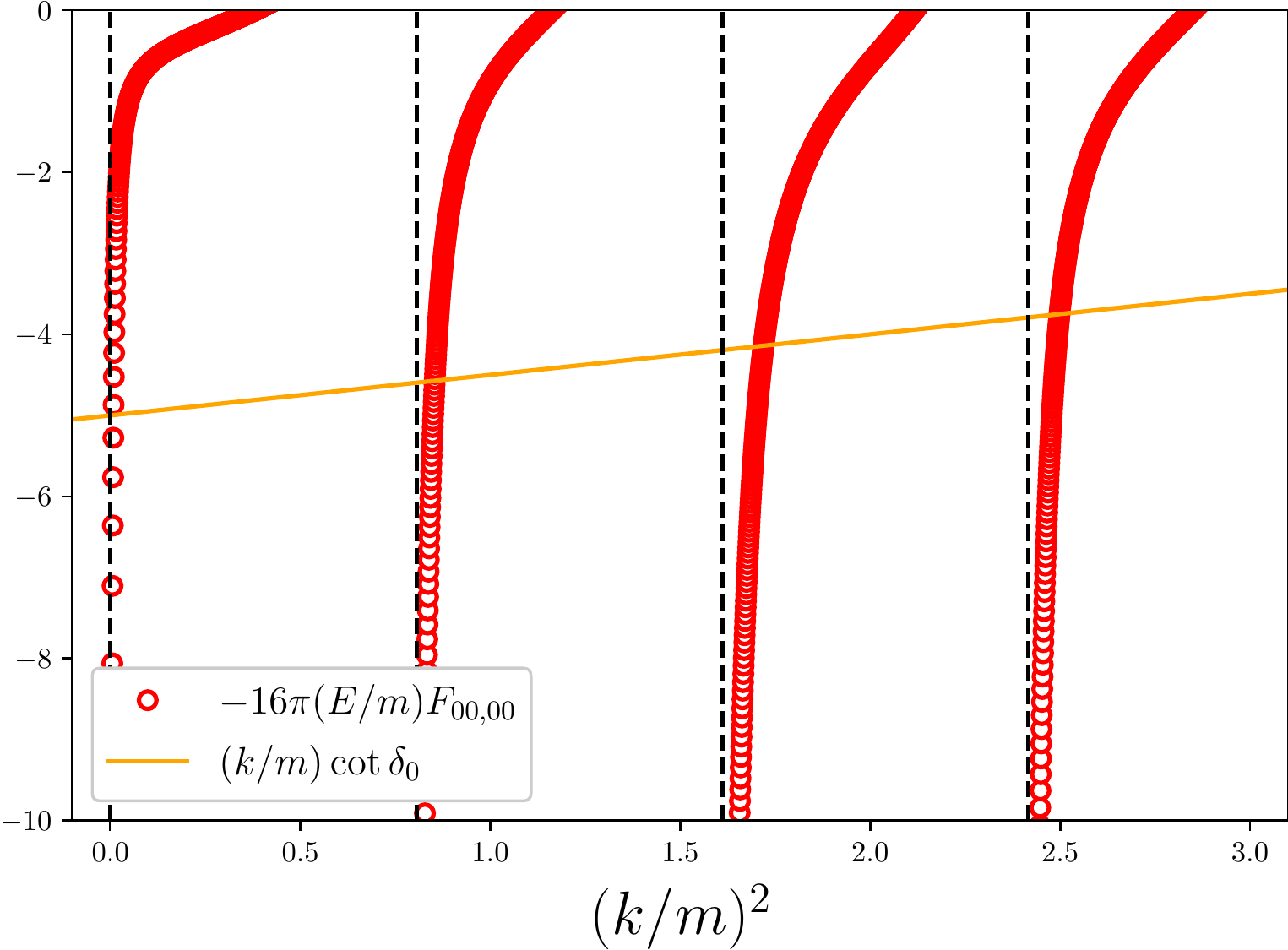}  
  \caption{Graphical representation of the QC2 in the case of two identical scalars with only $s$-wave  interactions. Further details are found in the main text.}
  \label{fig:luscher}
\end{figure}

In principle, the matrices in Eq.~(\ref{eq:F2}) are infinite dimensional, and all partial waves contribute\footnote{Only even  $\ell $ for identical particles.}. To render the quantization condition tractable, a truncation in $\ell$ must be applied. This is generally justified, since the scattering amplitude of higher partial waves is suppressed around the two-particle threshold: $\mathcal{K}_2^\ell \propto (q^*)^{2\ell}$. The simplest truncation is given by keeping only $\ell=0$ interactions, such that the QC2 becomes the algebraic relation
\begin{equation}
\frac{1}{\mathcal{K}_2^s} = - F_{00,00}.
\end{equation} 
Using Eq.~(\ref{eq:K2s}), it can be brought to the form:
\begin{align}
\begin{split}
q^* \cot &\delta_{0}(q^*)= \\& -8 \pi E^*  \left[  \frac{1}{L^3}\sum_{\boldsymbol k} - \text{PV} \int\frac{d^3k}{(2\pi)^3} \right] \frac{1 }{2 \omega_k 2 \omega_{Pk} (E- \omega_k- \omega_{Pk})}.
\end{split} \label{eq:QC2s}
\end{align}
A visualization of this equation is provided in Fig.~\ref{fig:luscher}. The yellow line corresponds to the $s$-wave phase shift in the form $(k/m) \cot \delta_0$ following an ERE parametrization [Eq.~(\ref{eq:EREs})] with $m a_0=0.2$ and $m r_0=1$. The red unfilled markers are the right-hand side of Eq.~(\ref{eq:QC2s}) with $m L=7$, and in the CM frame, i.e., $\boldsymbol P=0$. The points in which the two curves intersect correspond to the finite-volume energy levels. In addition,  $F_{00,00}$ diverges for the ``free'' finite-volume energies, that is, solutions when $a_0 \to 0$. These are plotted as vertical dashed lines, and they appear at
\begin{equation}
\bigg(\frac{k}{m}\bigg)^2 = \boldsymbol n^2 \bigg(\frac{2\pi}{m L} \bigg)^2, \quad \text{ with } \quad  \boldsymbol n \in \mathbb Z^3.
\end{equation}
Note that for this example the finite-volume energies are slightly shifted to the right with respect to the noninteracting ones, indicating mildly repulsive interactions.

It will also be useful to discuss the role of spatial symmetries in the L\"uscher method. Notice that because of the finite volume itself, full rotation invariance---the $SO(3)$ symmetry group---is reduced to a discrete subset of transformations that leave a cube unchanged---the octahedral group\footnote{$\boldsymbol P=0$ is implied. If $\boldsymbol P \neq 0$, the symmetry group is further reduced to subgroups of $O_h$.} ($O_h$). This leads to angular momentum nonconservation, which can be seen in the fact that $F$ in Eq.~(\ref{eq:QC2}) is not diagonal in $\ell$. The finite-volume energies are then shifted by interactions in multiple partial waves at the same time. Fortunately, this can be used in our favour. In the same way that $\ell\,m$ are the labels of irreducible representations of $SO(3)$, the finite-volume symmetry group has several irreps, labelled by $\Lambda\,\mu$, which correspond to good finite-volume quantum numbers\footnote{A summary of irreps can be found, e.g., in Appendix A of Ref.~\cite{Bernard:2008ax}.}. Thus, one can measure the spectrum in a particular irrep, $E_n^\Lambda(\boldsymbol P, L)$. Besides, the QC2 can be brought to a block-diagonal form, where each block corresponds to a particular choice of $\Lambda\, \mu$. In consequence, Eq.~(\ref{eq:QC2}) will factorize as:
\begin{equation}
\prod_{\Lambda \mu} \det _{\Lambda \mu} \left(   \mathbb P_{\Lambda \mu } \left[ F^{-1} (E, \boldsymbol P, L)+ \mathcal{K}_2(E^*) \right]   \mathbb P_{\Lambda \mu }  \right) = 0,  \label{eq:QC2irreps}
\end{equation}
where $ \mathbb P_{\Lambda \mu }$ are projectors to a given block, and the determinant runs over that same block. In other words, one has a separate quantization conditions for each irrep. This can be used to gain access to the phase shift of higher partial waves. For instance, the leading partial wave in the $E^+$ irrep is $d$-wave. Likewise, Eq.~(\ref{eq:QC2s}) corresponds to the $A_1^+$ QC2, which in the CM frame gets corrections from $\ell=4$ interactions that one usually neglects.

\subsection{Two-particle decays in finite volume} \label{sec:LL}

The decay of one particle into two other also gets distorted in a finite box due to the rescattering of the particles in the final state. The problem was first addressed for the $K \to \pi \pi$ weak decay by Lellouch and L\"uscher in Ref.~\cite{Lellouch:2000pv}. They found a way to correct these distortions, and provided a relation between a finite-volume matrix element and the infinite-volume decay amplitude. In later work, the relation has been generalized to multichannel decays~\cite{Hansen:2012tf}. In addition, it was further realized that a vacuum-to-two ($\gamma^* \to \pi\pi$) transition can be treated in a formally identical way~\cite{Briceno:2014uqa,Briceno:2015csa}.

The Lellouch-L\"uscher formalism works at leading order in the insertion of a external operator (such as $\mathcal H_w$), and to all orders in the strong interactions.  The transition amplitude of interest is that with a single insertion of the operator. In the case of $K \to \pi\pi$, it would be
\begin{equation}
\mathcal T_{\ell m} = \braket{ (\pi\pi)_{\ell m} | \mathcal H_w| K},
\end{equation}
where the kaon and two-pion states are understood to be asymptotic infinite-volume states. We have also included the partial wave projection of the amplitude. Note that angular momentum conservation ensures that only the $s$-wave amplitude is nonvanishing for $K \to \pi\pi$, but this may be different in other processes. From the lattice perspective, one would measure the following finite-volume matrix elements using the appropriate correlation functions:
\begin{equation}
M = \braket{E_n, \boldsymbol P, \Lambda \mu, L | \mathcal H_w(0) | K,\boldsymbol P, L}.
\end{equation}

To establish the relation between $\mathcal T$ and $M$, we assume that the two-pion system has an energy that matches that of the kaon, $E_K(\boldsymbol P,L) =E_{n}^\Lambda(\boldsymbol P, L) $. This way, the relation\footnote{We use the notation of Ref.~\cite{Briceno:2015csa}, as it will be more convenient below.} reads:
\begin{equation}
| M |^2 = \frac{1}{2 E_K(\boldsymbol P,L) L^6 } \mathcal T^\dagger_{\ell\, m}  \left[ \mathcal R_{\Lambda \mu} (E_n^\Lambda, \boldsymbol P, L) \right]_{\ell\, m, \ell' m'} \mathcal T_{\ell' m'}, \label{eq:LLres}
\end{equation}
where $\mathcal R_{\Lambda \mu}$ is the residue of the QC2 at the finite-volume energies
\begin{equation}
\mathcal R_{\Lambda \mu} (E_n^\Lambda, \boldsymbol P, L)  = \lim_{P_4 \to i E^\Lambda_n} -(E^\Lambda_n + i P_4) \mathbb P_{\Lambda\mu} \frac{1}{F_{i\epsilon}^{-1} + \mathcal M_2}  \mathbb P_{\Lambda\mu},
\end{equation}
and $\mathcal{T}$ has to be understood as a column vector in angular-momentum space. 
Note that this version of the QC2 differs from that in Eq.~(\ref{eq:QC2}). This one uses an $i \epsilon$ regularization for the sum minus integral difference ($F_{i \epsilon} = F + i \rho$), and we replace $\mathcal{K}_2$ by the scattering amplitude. Both versions lead to an identical finite-volume spectrum. 

In the CM frame, and neglecting the contribution from higher partial waves, Eq.~(\ref{eq:LLres}) can be brought to the original form by Lellouch and L\"uscher~\cite{Lellouch:2000pv}:
\begin{equation}
|\mathcal{T}|^2 = 8 \pi \left[ \eta \frac{\partial \phi(\eta)}{\partial \eta}  + k \frac{\partial \delta_0(k)}{\partial k}\right]_{k=k_\pi} \left( \frac{M_K}{k_\pi} \right)^3  |M|^2,  \label{eq:LLs}
\end{equation} 
with 
\begin{equation}
k_\pi = \sqrt{\frac{M_K^2}{4}- M_\pi^2}, \quad \eta=\frac{L k}{2\pi},
\end{equation}
 and
\begin{equation}
\tan \phi(\eta) = \frac{k}{16 \pi E} F_{00,00}^{-1}.
\end{equation}
An interpretation of Eq.~(\ref{eq:LLs}) is that the finite-volume matrix element and the infinite-volume decay amplitude differ only by a volume-dependent normalization factor. 

\clearpage

\section{Three-particle scattering in finite volume} \label{sec:3particles}

In the last few years, considerable theoretical effort has been devoted to generalizations of the two-particle L\"uscher formalism for more-than-two-particle systems. In fact, applications to simple systems (three charged mesons) have been successfully undertaken only very recently. In the present section, we will discuss how to deal with three particles in a finite volume, and review the contributions to the field achieved in this thesis.

The three-particle formalism has been derived following three different approaches: (i) a generic relativistic effective field theory (RFT)~\cite{Hansen:2014eka,Hansen:2015zga,Briceno:2017tce,Briceno:2018mlh,Briceno:2018aml,Blanton:2019igq,Romero-Lopez:2019qrt,Blanton:2019vdk,Hansen:2020zhy,Blanton:2020gha,Blanton:2020jnm,Blanton:2020gmf}, 
(ii) a nonrelativistic effective field theory (NREFT)~\cite{Hammer:2017uqm,Hammer:2017kms,Doring:2018xxx,Pang:2019dfe,Pang:2020pkl}, and (iii) the (relativistic) finite volume unitarity (FVU) approach~\cite{Mai:2017bge,Mai:2018djl,Mai:2019fba}. Recent reviews of the three approaches can be found in Refs.~\cite{Hansen:2019nir,Mai:2021lwb}. While the three versions should be completely equivalent, the connection is not easy to establish---see Ref.~\cite{Blanton:2020jnm} for FVU and RFT. A key point that differs is the precise definition of a scheme-dependent intermediate three-particle scattering quantity.

Before turning to details, it is worth commenting on the different status of the three methods. Only the RFT formalism has been explicitly worked out including higher partial waves~\cite{Blanton:2019igq}, although it should be possible in the other two cases. On top of that, formalisms for nonidentical scalars exist in the RFT~\cite{Hansen:2020zhy,Blanton:2020gmf}, as well as in the NREFT approach~\cite{Muller:2020vtt,Pang:2020pkl}. Moreover, both the RFT and FVU formalisms have been confronted with lattice QCD\footnote{See also similar work in $\varphi^4$ theory~\cite{Romero-Lopez:2018rcb,Romero-Lopez:2020rdq}.} data~\cite{Blanton:2019vdk,Fischer:2020jzp,Hansen:2020otl,Mai:2019fba,Culver:2019vvu,Alexandru:2020xqf,Brett:2021wyd}. Finally, a three-particle generalization of the Lellouch-L\"uscher formalism exists in two of the approaches: NREFT~\cite{Muller:2020wjo} and RFT~\cite{Hansen:2021ofl}.

In the remainder, we will focus on the RFT formalism. After a short summary of the approach, we will summarize the main results of four articles included in this thesis. We will close the chapter with some remarks.

\subsection{Relativistic finite-volume formalism}

The relativistic three-particle finite-volume formalism was first derived by Hansen and Sharpe in Refs.~\cite{Hansen:2014eka,Hansen:2015zga} for the case of identical scalars with a $\mathbb Z_2$ symmetry. Although extensions to more complex systems are available, we will concentrate on the original version for now. A physical system for which it is applicable---and has been applied---corresponds to three charged pions.

A complication of the three-particle formalism is the fact that three-particle scattering amplitudes have physical divergences. This is because it is possible for two particles to scatter, and then travel arbitrarily far before one of them scatters again off the third particle. The subtraction of these divergences will introduce a scheme dependence. This treatment can be identified in quantities labelled by the subscript ``df'', which stands for ``divergence-free''.

While in the two-particle case the quantization condition provides direct access to the scattering amplitude, for three particles it becomes a two-step process. First, the three-particle quantization condition relates the spectrum to an intermediate quasi-local three-particle scattering quantity, $\kdf$, and to the two-particle $K$-matrix, $\mathcal{K}_2$~\cite{Hansen:2014eka}. Even if $\kdf$ is a useful quantity to parametrize three-body interactions, it is scheme dependent and hence, unphysical. The second step is then necessary to get rid of the scheme dependence. It consists of a set of integral equations that map $\kdf$ and $\mathcal{K}_2$ into the three-particle scattering amplitude, $\mathcal M_3$.

\subsubsection{The three-particle quantization condition}

Let us start with the first step. This uses three-particle energies, obtained from correlation functions with three-particle quantum numbers, to access the three-particle $K$-matrix. The central element of the formalism is the three-particle quantization condition (QC3), which for identical, spinless particles with a $\mathbb Z_2$ symmetry reads\footnote{Up to exponentially-suppressed corrections.}:
\begin{equation}
\det \left[ F_3(E, \boldsymbol P, L )^{-1}   + \kdf(E^*) \right]=0. \label{eq:QC3}
\end{equation}
Even though this looks formally identical to the two-particle quantization condition in Eq.~(\ref{eq:QC2}), there are several differences. First, $\kdf$ and $F_3$ are matrices in a space that characterizes three on-shell particles in finite volume. Their indices are angular momentum of the interacting pair, $\ell\,m$, and the finite-volume momentum of the spectator particle, $\boldsymbol k$. We will refer to this as the $(k\ell m)$ space. In practice, a finite dimensionality is ensured by neglecting interactions in $\ell > \ell_\text{max}$, and using a cutoff function that truncates values of $|\boldsymbol k| > k_\text{max}$. In fact, the scheme in $\kdf$ is linked to the particular choice of cutoff function for $\boldsymbol k$. Finally, $F_3$ is not purely kinematical, but it also depends on two-particle interactions via $\mathcal K_2$. Qualitatively, this means that pairwise scattering is incorporated into $F_3$. It also implies that two-particle interactions must be under control before studying three particles in a finite volume. In addition, an analytic continuation of $\mathcal K_2$ below the two-particle threshold is needed.

\begin{figure}[h!]
  \centering
  \includegraphics[width=0.9\linewidth]{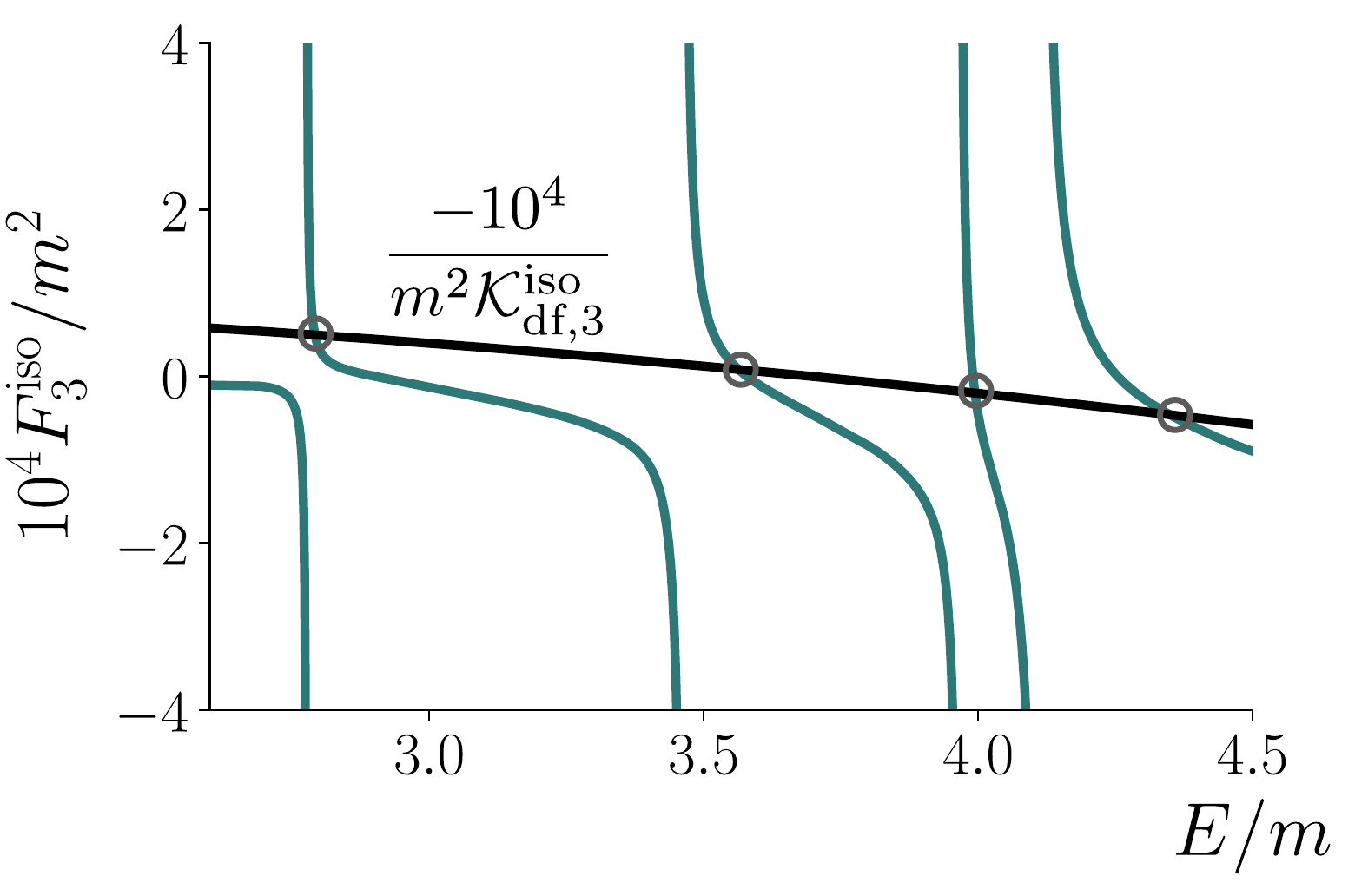}  
  \caption{Example three-particle quantization condition in the isotropic approximation. The blue line corresponds to $F_3^\text{iso}$, while the black line to $-1/\kdf^\text{iso}$. Here $mL = 6$, and the $s$-wave phase shift includes only the scattering length, $m a_0=-10$. The intersections of the two curves, marked by open circles, indicate finite-volume energy levels. Source: Ref.\cite{Briceno:2018mlh}.}
  \label{fig:qciso}
\end{figure}

A simplification of the QC3 is achieved within the so-called isotropic approximation. This involves three ingredients: (i) only $s$-wave interactions are considered for the pair, and so, only the $\ell=0$ component of the matrices in the QC3 is included; (ii) $\kdf$ is chosen to be independent of the spectator momentum, and it is only a function of the total energy; (iii) $F_3$ is projected onto the isotropic vector, $\ket{1}$, which has a one in each allowed entry. Because of this last step, solutions of the QC3 in the isotropic approximation live in the $A_1^+$ irrep for $\boldsymbol P=0$. The isotropic three-particle quantization condition becomes:
\begin{equation}
F_3^\text{iso}(E) = \braket{1 | F_3 |1} = -\frac{1}{\kdf^\text{iso}(E)}, \label{eq:qc3iso}
\end{equation}
and does not involve determinants anymore. One can understand this equation as follows. If one knows the two-particle interactions that enter in $F_3^\text{iso}$, and given an energy level from the lattice, one can determine the value of $\kdf$ at the given energy. It can be considered as the three-particle analogue of Eq.~(\ref{eq:QC2s}). An example of this is given in Fig.~\ref{fig:qciso}. A numerical exploration of the QC3 in this approximation was carried out in Ref.\cite{Briceno:2018mlh}.

\pagebreak
For completeness, we present now the definitions\footnote{These can also be found in, e.g., Appendix A of Ref~\cite{Blanton:2019igq}.} of the various objects involved. We choose the definition of the spherical harmonics, $\mc{Y}_{\ell m}$, as in Ref.~\cite{Blanton:2019igq}.  We begin with the cutoff function, which needs to be smooth in order to avoid spurious finite volume effects. Our choice is
\begin{align}
    H(\boldsymbol{k}) = J\left( \frac{E_{2,k}^{*2}}{4m^2}\right)\,, \quad
        \label{eq:H_k}
    J(z) =
    \begin{cases}
        0, & z\leq 0 \\
        \exp{\left( -\frac{1}{z}\exp{\left[ -\frac{1}{1-z} \right]} \right)}, & 0<z<1 \\
        1, & 1\leq z
    \end{cases}
\end{align}
with $E_{2,k}^{*\,2} = (P-k)^2$. The matrix $F_3$ is given by
\begin{equation}
F_3 = \frac{1}{2\omega L^3} \left[ \frac{ F}{3} -  F \frac{1}{\widetilde{\mathcal K}_2^{-1} + F +  G  }  F \right],
\end{equation}
where $\omega$ is a diagonal matrix with entries $\omega_k = (m^2 + \boldsymbol k^2)^{1/2}$. The other building blocks are yet to be defined. Qualitatively, $\widetilde{\mathcal K}_2$ accounts for two-particle interactions, $ G$ corresponds to finite-volume effects stemming from one-particle exchange diagrams, and $ F$ includes the sum-minus-integral difference from loops. More precisely, $\widetilde{\mathcal K}_2$ is a modified version of the two-particle $K$-matrix:
\begin{align}
\begin{split}
\big(\widetilde{ \mathcal K}_2&\big)^{-1}_{k\, \ell\,m,p\, \ell' m'} = \\  &\frac{\delta_{\ell,\ell'} \delta_{m,m'} \delta_{kp}}{16 \pi E_{2,k}^*  (q_{2,k}^*)^{2\ell}}  \left( (q_{2,k}^*)^{2\ell+1} \cot \delta_\ell + |q_{2,k}^* |^{2\ell+1}  \left[1-H(\boldsymbol k) \right]\right),
\end{split}
\end{align}
with $q_{2,k}^* = \sqrt{E_{2,k}^{*\,2}/4 - M^2}$. Next,
\begin{equation}
    {G}_{p\,\ell'm', k\,\ell\, m} = \frac{1}{L^3}  \frac{H(\boldsymbol{p}) H(\boldsymbol{k})}{b^2-m^2}
    \frac{4\pi \mc{Y}_{\ell'm'}(\boldsymbol{k}^*) \mc{Y}_{\ell m}(\boldsymbol{p}^{\,*}) }{q_{2,p}^{*\ell'}\,q_{2,k}^{*\ell} } \frac{1}{2\omega_k}\,, 
    \label{eq:defG}
\end{equation}
where $b= P-p - k$ is the momentum of the exchanged particle, $\boldsymbol p^{\,*}$ is the result of boosting $\boldsymbol p$ to the CM frame of the dimer for which $\boldsymbol k$ is the spectator momentum, and {\em vice versa} for $\boldsymbol k^{\,*}$ . Finally, 
\begin{align}
\begin{split}
 F_{k\,\ell' m',\, p\,\ell m} &= \\
\bigg[\frac{1}{L^3} \sum_{\boldsymbol a} - &\text{PV}\int \frac{d^3a}{(2\pi)^3} \bigg]
\frac{\delta_{pk} H(\boldsymbol k)}{(q_{2,k}^{*})^{\ell'+\ell}}
\frac{H(\boldsymbol a) H(\boldsymbol b') 4\pi \cY_{\ell' m'}(\boldsymbol a^{\,*}) \cY_{\ell m}(\boldsymbol a^{\,*})}
{8 \omega_a \omega _b(E-\omega_k-\omega_a -\omega_b)}
\,,
\label{eq:Ft0}
\end{split}
\end{align}
where $b'=P-k-a $, and $\boldsymbol a^*$ is the result of boosting $\boldsymbol a$ to the dimer rest frame, with spectator momentum $\boldsymbol k$. It is generally convenient to choose the real harmonics.

To conclude, we comment on an extension of the formalism proposed in Ref.~\cite{Romero-Lopez:2019qrt}. This lifts up a technical limitation of the original QC3, that prevented the inclusion of resonances or bound states in $\cK_2$. The solution is to use a modified principal value prescription to regulate the poles in the $F$ matrix, and requires the following changes:
\begin{align}
[ F]_{k\ell' m';p\ell m} &\to [ F]_{k\ell' m';p \ell m} 
+ \delta_{kp} \delta_{\ell'\ell}\delta_{m' m}
 H(\boldsymbol k)  \frac{I_{\ell}(q_{2,k}^{\star2})}{32\pi}\,,
\label{eq:Fellshift}
\\
\big [  ( \widetilde \cK_2)^{-1} \big]_{k\ell' m'; p \ell m} &\to\
\big [   (\widetilde \cK_2)^{-1} \big]_{k\ell' m'; p \ell m} 
- \delta_{kp} \delta_{\ell'\ell}\delta_{m' m}
 H(\boldsymbol k)  \frac{I_{\ell}(q_{2,k}^{\star2})}{32\pi}\,,
\label{eq:Kellshift}
\end{align}
where $I_{\ell}$ is a smooth function. This will be used below when the $\rho$ resonance is considered. 

\subsubsection{Relation to the three-particle scattering amplitude}

The relation between the two- and three-particle $K$-matrices and the scattering amplitude, $\mathcal{M}_3$, was initially derived in Ref.~\cite{Hansen:2015zga}. The authors found a way to define a finite-volume version of the three-particle scattering amplitude, $\mathcal{M}_{3,L}$, which turns into the desired object in the appropriate infinite-volume limit. 

The finite-volume amplitude is given by
\begin{equation}
\cM_{3,L} = \cS\left\{ \cM_{3,L}^{(u,u)}\right\}\,,
\end{equation}
where $\cS$ stands for the symmetrization operation, and $\cM_{3,L}^{(u,u)}$ is an unsymmetrized version of the amplitude. The later means that one of the incoming and of the outgoing particles is fixed to be the spectator. More details about the symmetrization procedure are discussed in Ref.~\cite{Hansen:2020zhy}. Furthermore, the unsymmetrized amplitude is given by:
\begin{align}
\cM_{3,L}^{(u,u)} &= \cD^{(u,u)} + \cM_{\df,3,L}^{(u,u)}, 
\label{eq:M3Luu}
\end{align}
where the different objects are defined as:
\begin{align}
\cD^{(u,u)} &= - \frac1{1+\cM_{2,L} G} \cM_{2,L} G \cM_{2,L} 2\omega L^3 \,,
\label{eq:Duu}
\\
\cM_{\df,3,L}^{(u,u)} &= \cL^{(u)}_L \frac1{1+\cK_{\df,3} F_3} \cK_{\df,3} \cR^{(u)}_L\,, 
\label{eq:M3dfL}
\\
\cL^{(u)}_L &= \left(\frac{F}{2\omega L^3}\right)^{-1} F_3
= \frac13 - \frac1{1+\cM_{2,L} G} \cM_{2,L} F\,,
\label{eq:LLu}
\\
\cR^{(u)}_L &= F_3\left(\frac{F}{2\omega L^3}\right)^{-1}
= \frac13 - F \cM_{2,L}\frac1{1+G \cM_{2,L} }\,,
\label{eq:RLu}
\end{align}
with $\cM_{2,L}^{-1} = \cK_2^{-1} + F$. We note that $\cD^{(u,u)}$ represents the sum over all possible pair-wise interactions mediated by one-particle exchanges, and $ \cM_{\df,3,L}^{(u,u)}$ can be understood as the short-distance contribution to the amplitude.

Finally, $\cM_3$ will be obtained from $\cM_{3,L}$ by taking the $L \to \infty$ limit in which poles in $F$ and $G$ are regulated by an $i \epsilon$ prescription.  Note that the infinite-volume limit of $\cD^{(u,u)}$ contains the kinematical singularities of the three-particle scattering amplitude. In contrast, the infinite-volume limit of the symmetrized version of Eq.~(\ref{eq:M3dfL}), $\cM_{\df,3}$, is regular. Examples of solutions to these equations are given in Refs.~\cite{Briceno:2018mlh,Hansen:2020otl,Jackura:2020bsk}

\subsection{Implementing the three-particle quantization condition including higher partial waves }

The RFT approach is the only one that has been explicitly studied including higher partial waves. This was carried out in Ref.~\cite{Blanton:2019igq}, which is one of the articles included as a part of this thesis.  In that paper, we include $d$-wave interactions to the three-body formalism, both in the two- and three-particle sectors.

\begin{figure}[h!]
  \centering
  \includegraphics[width=0.7\linewidth]{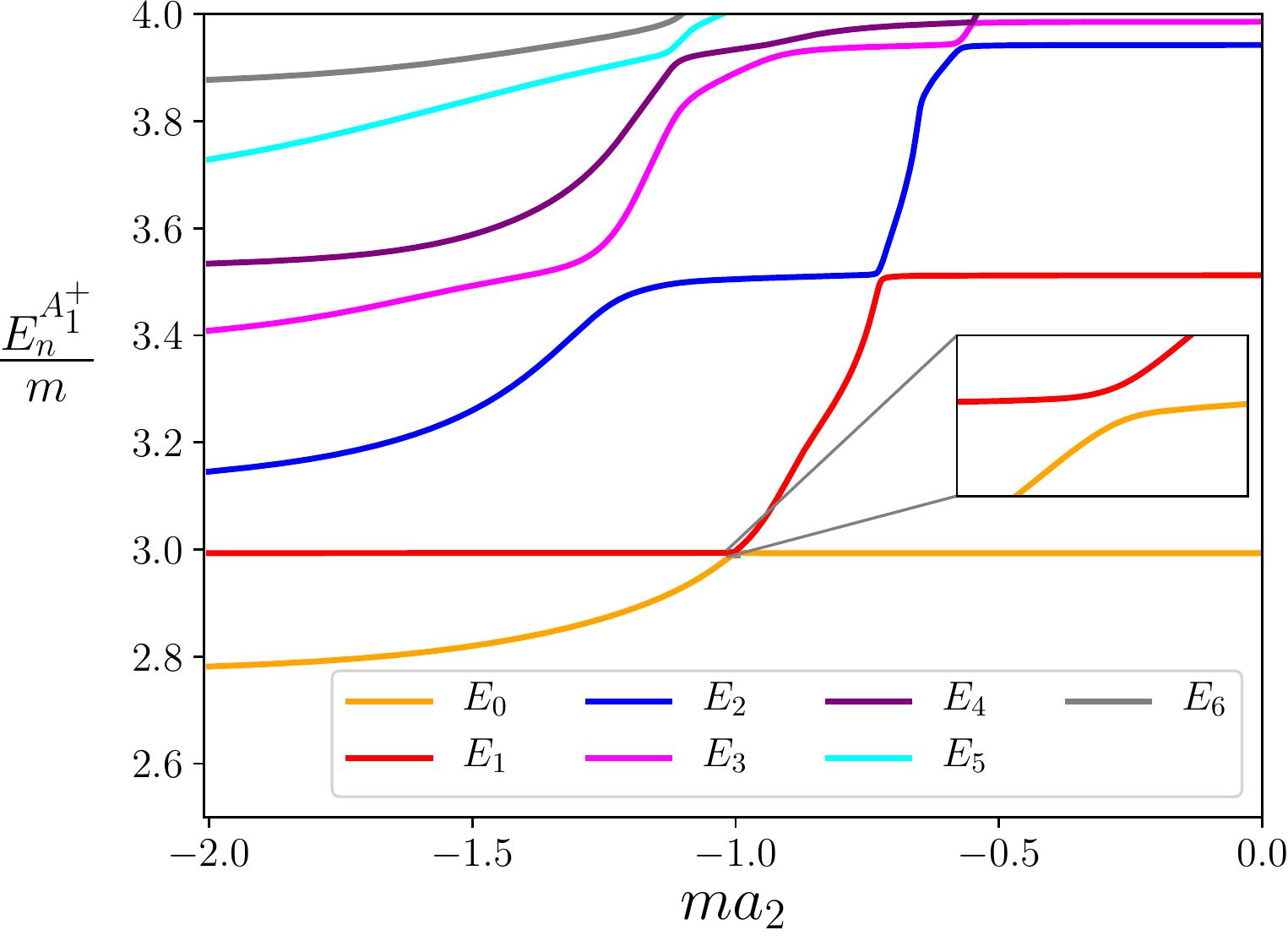}  
  \caption{Finite-volume spectrum in the $A_1^+$ irrep as a function of $ma_2$ in the region $E<4m$ with $mL=8.1$. The other parameters are: $m a_0 = -0.1$, $r_0=P_0=\kdf=0$. Source: Ref.\cite{Blanton:2019igq}.}
  \label{fig:a2spectrum}
\end{figure}

We first study the impact of two-particle $d$-wave interactions, and focus on the case of $\kdf=0$. We consider that the phase shifts in the two lowest partial waves are given as:
\begin{equation}
(q_{2,k}^*) \cot \delta_0 = -\frac{1}{a_0} + \frac{1}{2} r_0 (q_{2,k}^*)^2 + r_0^3 P_0 (q_{2,k}^*)^4,
\end{equation}
and
\begin{equation}
(q_{2,k}^*)^5 \cot \delta_2 = -\frac{1}{(a_2)^5},
\end{equation}
and neglect all $\ell>2$ interactions. An example of our numerical explorations is given in Fig.~\ref{fig:a2spectrum}. There we fix the $s$-wave interactions to be weakly repulsive ($m a_0 = -0.1$), and inspect the spectrum when varying the strength of $d$-wave interactions at fixed box size. As can be seen, the effect of $d$-wave interactions is small when $|m a_2| \ll 1$. However, the spectrum is significantly shifted when $|m a_2| \gtrsim 1$, and there is even a state well below threshold. As argued in the article, this appears to be a three-particle Efimov-like\footnote{A three-particle bound state produced by nearly-resonant two-body interactions.} bound state~\cite{Efimov:1970zz}, since it survives in the $L \to \infty$ limit.

Another important point we address is the expansion of $\kdf$ around the three-particle threshold. In particular, we consider how this can be done consistently, and at which order higher partial waves play a role. Since $\kdf$ is expected to be real and smooth in some region around threshold, one can expand it in a Taylor series in terms of Lorentz-invariant quantities. One can further use the symmetries of the theory---C, P, T and particle exchange---to constrain the expansion parameters. This way, the expansion to quadratic order worked out in Ref.~\cite{Blanton:2019igq} reads:
\begin{align}
    m^2 \kdf &= \cK^\text{iso}
    + \kdf^{(2,A)} \Delta^{(2)}_A + \kdf^{(2,B)}  \Delta^{(2)}_B + \mc{O}(\Delta^3)\,, 
    \label{eq:termsKdf}
    \\
    \cK^\text{iso} &= \kdf^\text{iso} + \cK^\text{iso,1} \Delta + \cK^\text{iso,2} \Delta^2,
    \label{eq:Kisoterm}
    \end{align}
where $\kdf^\text{iso} , \cK^\text{iso,1} , \cK^\text{iso,2},\kdf^{(2,A)}  $  and $\kdf^{(2,B)}$ are real coefficients, and 
  \begin{align}
    \Delta^{(2)}_A = \sum_{i=1}^3 (\Delta_{i}^2 + \Delta_{i}'^{\,2}) - \Delta^2, &\qquad		
    \Delta^{(2)}_B = \sum_{i,j=1}^3 \widetilde{t}_{ij}^{\;2} - \Delta^2 \,,
\end{align}
are relativistic invariants with
\begin{equation}
    \Delta \equiv \frac{s - 9m^2}{9m^2}\,, \ \
    \Delta_i \equiv \frac{s_{jk} - 4m^2}{9m^2} \,, \ \
     \quad \Delta_i' \equiv \frac{s_{jk}' - 4m^2}{9m^2}\,, \ \
    \widetilde{t}_{ij} \equiv \frac{t_{ij}}{9m^2}\,.
\end{equation}
Note that this result implies that, at quadratic order, only five constants account for three-body interactions of identical particles. An interesting observation is that only $ \kdf^{(2,A)} $ and $ \kdf^{(2,B)} $ depend on angular variables.

\begin{figure}[h!]
\centering
\begin{subfigure}{.72\textwidth}
  \centering
  \includegraphics[width=.999\linewidth]{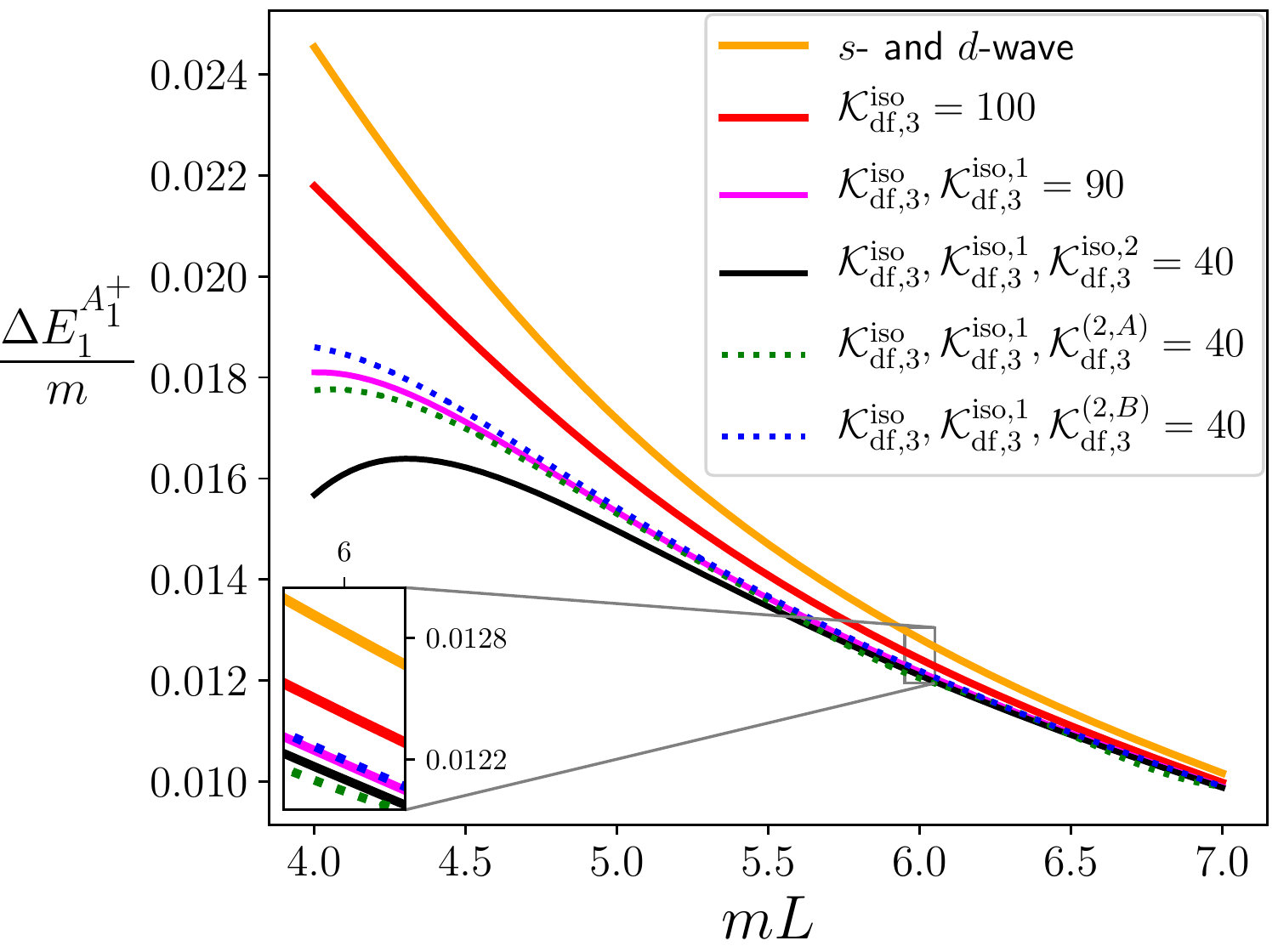}  
  \caption{}
  \label{fig:plotA1}
\end{subfigure}
\begin{subfigure}{.72\textwidth}
  \centering
  \includegraphics[width=.999\linewidth]{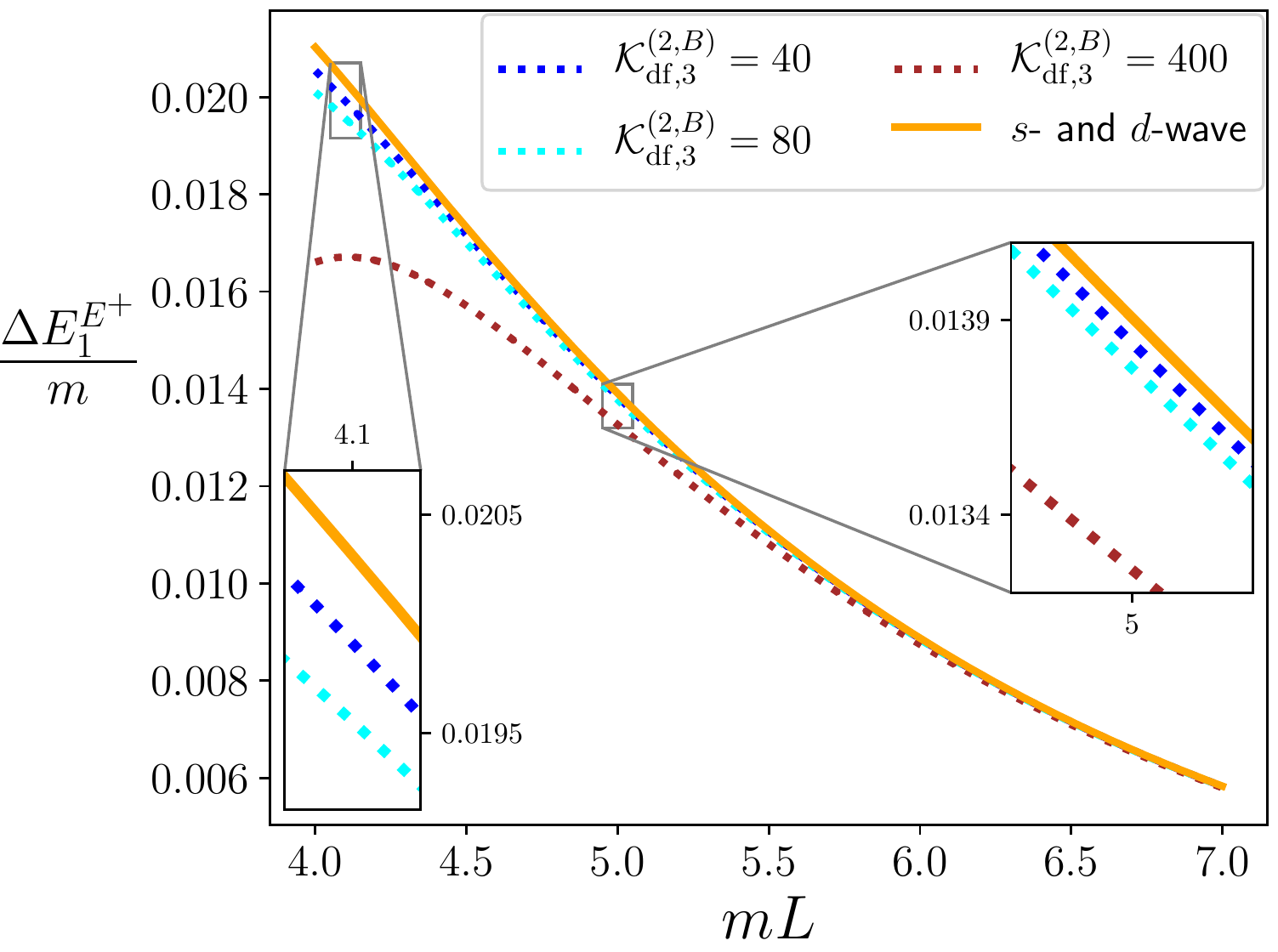}  
  \caption{}
  \label{fig:plotE}
\end{subfigure}
\caption{Energy shift of the first excited state in the $A^+_1$ irrep (top) and $E_+$ irrep (bottom)
with various choices of the parameters in $\kdf$. The two-particle interactions are set as in Eq.~(\ref{eq:paramdwave}). The parameters in $\kdf$ are explained by the legend, with the convention that a parameter value not given explicitly is set to the value given earlier in the legend.
Source: Ref.~\cite{Blanton:2019igq}. } 
\label{fig:toyspectrumdwave}
\end{figure}

To gain further insight on how the different terms of $\kdf$ affect the spectrum, we use a toy model in which the two-particle parameters are tuned to those of a physical $3\pi^+$ system~\cite{Yndurain:2002ud}:
\begin{equation}
m  a_0 = 0.0422, \ \, m r_0 = 56.21, \ \,  P_0 = -3.08 \cdot 10^{-4}, \ \, m a_2=-0.1867.
\label{eq:paramdwave}
\end{equation}
We then explore the shifts in the finite-volume energies produced by some choices of the terms in $\kdf$. An example of this is given in Fig.~\ref{fig:toyspectrumdwave} for the first excited state in two irreps\footnote{In Ref.~\cite{Blanton:2019igq} we explain how to project the QC3 to the finite-volume irreps. This is analogous to the two-particle case [Eq.~(\ref{eq:QC2irreps})]. }. One can notice that all terms shift the energies in the $A_1^+$ irrep, with a stronger sensitivity to the isotropic parameters. Interestingly, only $\kdf^{(2,B)}$ couples to the $E^+$ irrep. Using information from these and similar plots, we lay out a strategy to constrain the different terms in $\kdf$ from lattice QCD simulations.

Finally, in Ref.~\cite{Blanton:2019igq} we also explore the circumstances under which the quantization condition has unphysical solutions---solutions that are artefacts of the QC3. We concluded that this  unresolved issue will require further investigation.

\subsection{The $I=3$ three-pion scattering amplitude}

The relativistic three-particle formalism took a qualitative step forward with its first application to a full lattice QCD finite-volume spectrum. This was carried out in one of the articles of this dissertation, Ref.~\cite{Blanton:2019vdk}. There, we analysed the $2\pi^+$ and $3\pi^+$ energy levels in several irreps and moving frames measure by Hörz and Hanlon in Ref.~\cite{Horz:2019rrn} keeping only $s$-wave interactions. We found some statistical significance for the first two parameters in the expansion of $\kdf$, explained in Eq.~(\ref{eq:termsKdf}).

\begin{figure}[h!]
  \centering
  \includegraphics[width=0.85\linewidth]{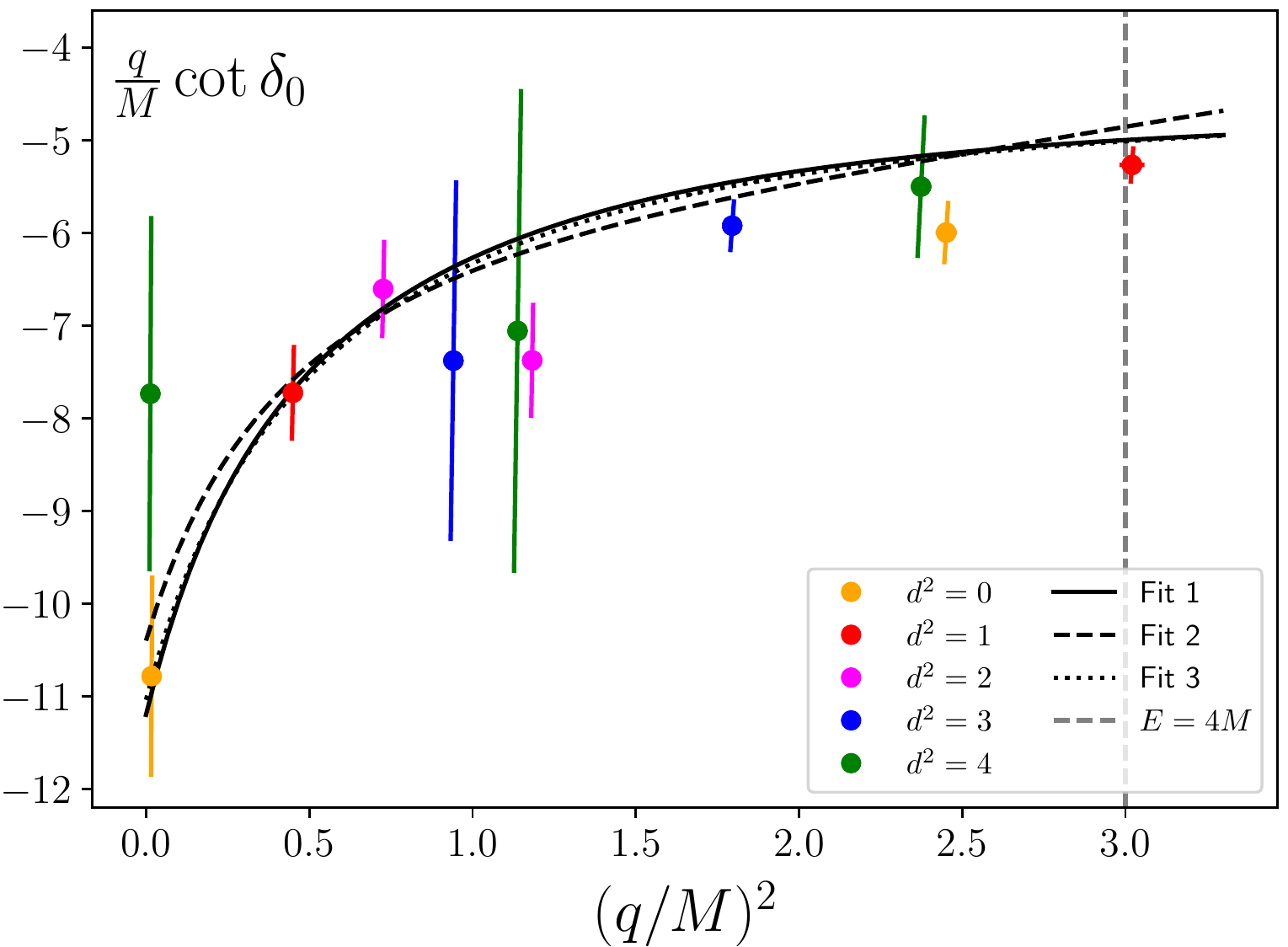}  
  \caption{Phase shift obtained from the $2\pi^+$ spectrum of Ref.~\cite{Horz:2019rrn} using the QC2. $d^2$ labels the moving frame from which each data point is obtained. Different fits are included. Fit 1, corresponds to the form in Eq.~(\ref{eq:Adlerzero}) with $B_2=0$ and $z_2^2=M^2$. Fit 2 is the same but with $B_2 \neq 0$.  Fit 3 has $B_2=0$, but we let $z_2^2$ free. Source: Ref.\cite{Blanton:2019vdk}.}
  \label{fig:kcotI2}
\end{figure}

As explained above, in order to study three-particle interactions, one must have the two-particle sector under control. For this, we study different parametrizations of the $s$-wave phase shift. An interesting observation is that the spectrum is better fit when incorporating the Adler zero~\cite{Adler:1964um}, which is a zero of the scattering amplitude below threshold required by chiral symmetry. Our proposed parametrization is:
\begin{equation}
\frac{q}{M} \cot \delta_0 = \frac{E_2^* M }{E_2^{*2} - 2z_2^2} \left( B_0 + B_1 \frac{q^2}{M^2} + B_2 \frac{q^4}{M^4} + \hdots \right). \label{eq:Adlerzero}
\end{equation}
Note that this diverges below threshold when $ E_2^{*2} = 2z_2^2 $, which limits the radius of convergence of polynomial expansions. The data for the two-particle phase shift is shown in Fig.~\ref{fig:kcotI2}, along with three different fits to Eq.~(\ref{eq:Adlerzero})---more details are given in the caption. We find a reasonable description when fixing $z_2^2$ to its LO ChPT result, $z_2^2=M^2$.

\begin{figure}[h!]
  \centering
  \includegraphics[width=0.85\linewidth]{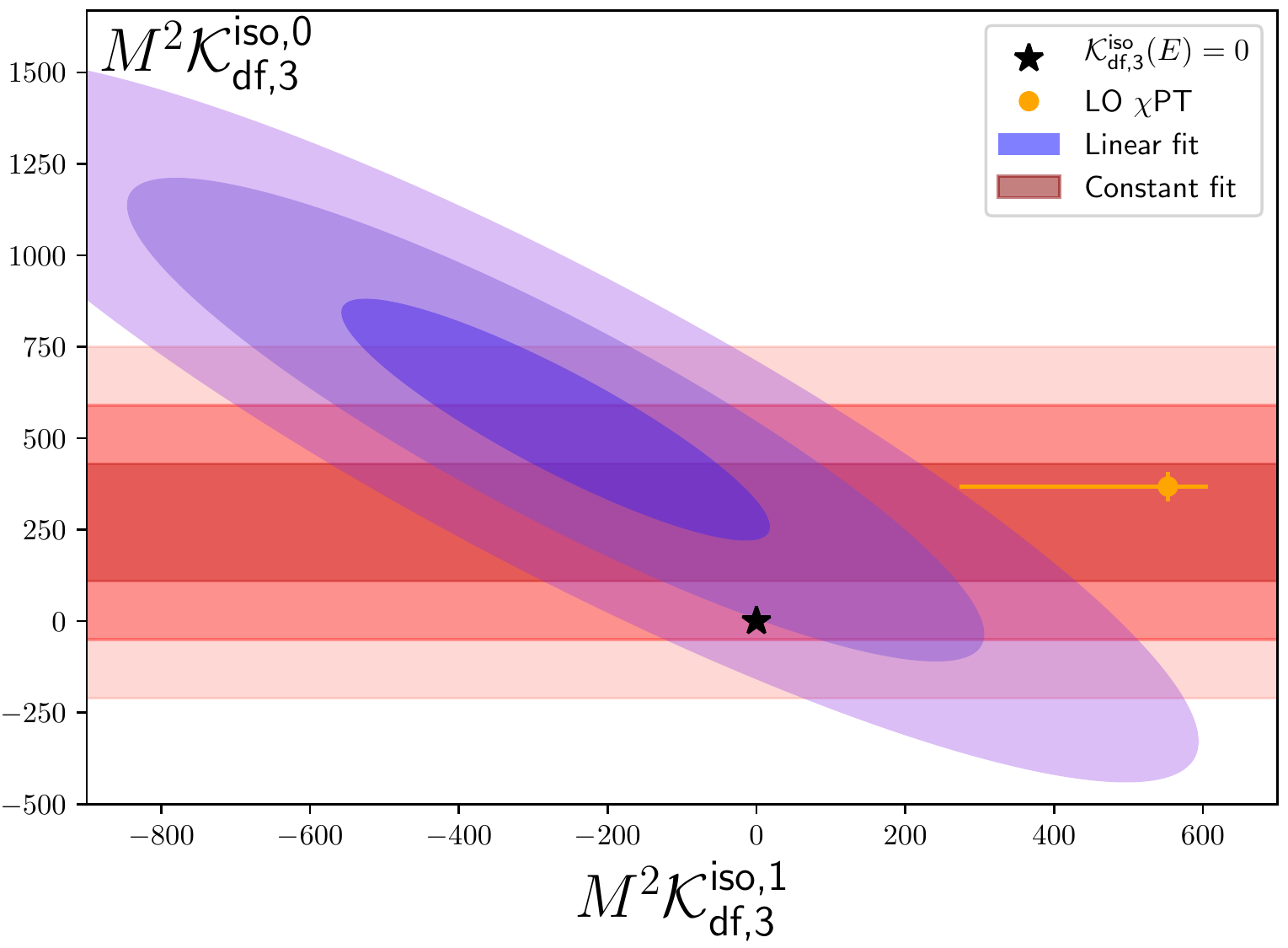}  
  \caption{1, 2 and 3$\sigma$ confidence intervals for $M^2 \kdf^\text{iso,0}$ for different two- and three-particle fits. The ``constant fit'' sets $\kdf^\text{iso,1}=0$ (fit 4 in the article), while the linear term leaves it free (fit 5 in the article). Source: Ref.\cite{Blanton:2019vdk}.}
  \label{fig:confK}
\end{figure}

Once we have a suitable model for the two-pion sector, we turn to the three-particle sector. For this, we perform a global two- and three-particle fit using simultaneously the QC2 and QC3. For $\kdf$, we use the following parametrization
\begin{equation}
\kdf = \kdf^\text{iso,0} + \kdf^\text{iso,1}  \Delta,
\end{equation}
which is consistent with keeping only $s$-wave interactions. The central results of these fits is given in Fig.~\ref{fig:confK}, where we show the confidence intervals of the parameters of $\kdf$ projected to the $(\kdf^\text{iso,0},\kdf^\text{iso,1})$ plane. As can be seen, the scenario $\kdf=0$ is disfavoured by $2\sigma$.

An additional result presented in Ref.~\cite{Blanton:2019vdk} is the leading order ChPT prediction of $\kdf$. For this, we use the fact that the relation between $\kdf$ and $\cM_\text{df,3}$ is trivial at this order of the chiral expansion, 
\begin{equation}
\kdf =\cM_\text{df,3} \left[ 1 + O(M^2/F^2) \right],
\end{equation}
which can be deduced from Eq.~(\ref{eq:M3dfL}). This way, the result is
\begin{equation}
M^2 \kdf = \frac{M^4}{F^4 } (18 + 27 \Delta) = (16 \pi M a_0)^2 (18 + 27 \Delta),
\end{equation} 
which is also indicated in Fig.~\ref{fig:confK}. Interestingly, the constant term seems to be reasonably well describe by LO ChPT, whereas there is a significant tension in the linear term. This behavious has been confirmed by later work~\cite{Fischer:2020jzp}, although there is no satisfactory explanation yet.

\subsection{A generic three-pion system in finite volume}

In its original form, the three-particle formalism is only valid for identical (pseudo)scalars. This limits its applicability to three charged mesons at maximal isospin, such as $3 \pi^+$ or $3K^-$. Even if they are satisfactory benchmark systems, they are weakly interacting, nonresonant channels. Motivated by this, in another paper of this thesis (Ref.~\cite{Hansen:2020zhy}) we provide a generalization of the RFT formalism to include 
nonidentical, mass-degenerate (pseudo)scalar particles. More precise, we focus on a generic three-pion system with exact G parity. To illustrate the physical relevance of such  extension, we summarize in Table~\ref{tab:resonances} the lowest-lying resonances with quantum numbers of three pions.

\begin{table}[h!]
\centering
\begin{tabular}{ccccc}
Resonance & $I_{\pi\pi\pi}$ &$J^P$ & Irrep ($\boldsymbol P =0$)  \\ \hline\hline
$\omega(782)$ & 0  & $1^-$ & $T_1^-$    \\ \hline
$h_1(1170)$ & 0  & $1^+$ &  $T_1^+$   \\ \hline
$\omega_3(1670)$ & 0  & $3^-$ & $A_2^-$   \\ \hline \hline
  $\pi(1300)$ & 1  & $0^-$ &  $A_1^-$   \\ \hline
 $a_1(1260)$ & 1  & $1^+$ &  $T_1^+$  \\ \hline
 $\pi_1(1400)$ & 1  & $1^-$ &  $T_1^-$   \\ \hline
 $\pi_2(1670)$ & 1  & $2^-$ &  $E^-$ and $T_2^-$  \\ \hline
 $a_2(1320)$ & 1  & $2^+$ &  $E^+$ and $T_2^+$  \\ \hline
 $a_4(1970)$ & 1  & $4^+$ &  $A_1^+$ \\ \hline
\end{tabular}
\caption{Lowest lying resonances with negative G-parity, and which couple to three pions,
in the different isospin ($I_{\pi\pi\pi}$) and $J^P$ channels. The fourth column shows the cubic group irreps that are subduced from the rotation group irreps in the CM frame ($\boldsymbol P=0$).
}
\label{tab:resonances}
\end{table}

Before turning to the derivation, it is useful to comment on three-pion states from the point of view of three objects with isospin 1. Their combination leads to seven irreps:
\begin{equation}
\bm 1 \otimes \bm 1 \otimes \bm 1 = (\bm0 \oplus \bm1 \oplus \bm2) \otimes \bm1
= (\bm 1) \oplus (\bm0  \oplus \bm1 \oplus \bm2) \oplus
(\bm1 \oplus \bm2 \oplus \bm3)\,,
\end{equation}
which means that total three-pion isospin will have values $I_{\pi \pi \pi} = 0, 1, 2, 3$,  with respective multiplicities $1,3,2,1$. The value of the multiplicity is given by the number of two-pion subchannels, each labelled by the two-pion isospin $I_{\pi\pi}$. We then have  $I_{\pi\pi}=0,1,2$ if  $I_{\pi\pi\pi}=1$, $I_{\pi\pi}=1,2$ for $I_{\pi\pi\pi}=2$, and only one value each for $I_{\pi\pi\pi}=0$ and $3$, namely $I_{\pi\pi}=1$ and $2$, respectively.

The starting point of the derivation is the finite-volume correlation function:
\begin{equation}
C_{L;jk}(P) \equiv \int d x^0 \int_{L^3} d^3 x \ e^{- i \boldsymbol P \cdot \boldsymbol x + i E t} \ \langle \text{T} \mathcal O_{j}(x) \mathcal O^\dagger_{k}(0) \rangle_L \,,
\end{equation}
where $\mathcal O_{j}$ are operators that annihilate three-pion states. It will be more convenient to use operators in momentum space\footnote{We use the notation $\int_k \equiv \int d k^0/(2 \pi) \sum_{\bm k}$, with
$\bm k$ being the finite-volume spectator momentum for $\bm P$. 
Also, the factor of $1/L^3$ accompanying each sum is left implicit.}, related to $\mathcal O_{j}$ as:
\begin{equation}
\label{eq:opdefB}
\mathcal O_j(x) \equiv \int_{a,b,k} \, f(a,b,k) \, e^{- i (a+b+k) \cdot x} \, \widetilde { \mathcal O}_j(a,b,k) \,,
\end{equation}
where $f(a,b,k)$ is a smooth function that specifies the detailed form of the operator. Because of isospin symmetry, all the relevant information can be obtained from the three-pion sector with zero
electric charge. Hence, we focus on the space of the seven neutral operators:
\begin{equation}
\label{eq:opdefA}
\widetilde { \mathcal O}(a,b,k) \equiv  \left(
\begin{array}{c}
 \widetilde{\pi}_{-}( a) \  \widetilde{\pi}_{0}( b) \  \widetilde{\pi}_{+}( k)  \\[5pt]
 \widetilde{\pi}_{0}( a) \  \widetilde{\pi}_{-}( b) \  \widetilde{\pi}_{+}( k)  \\[5pt]
 \widetilde{\pi}_{-}( a) \  \widetilde{\pi}_{+}( b) \  \widetilde{\pi}_{0}( k)  \\[5pt]
 \widetilde{\pi}_{0}( a) \  \widetilde{\pi}_{0}( b) \  \widetilde{\pi}_{0}( k)  \\[5pt]
 \widetilde{\pi}_{+}( a) \  \widetilde{\pi}_{-}( b) \  \widetilde{\pi}_{0}( k)  \\[5pt]
 \widetilde{\pi}_{0}( a) \  \widetilde{\pi}_{+}( b) \  \widetilde{\pi}_{-}( k)  \\[5pt]
 \widetilde{\pi}_{+}( a) \  \widetilde{\pi}_{0}( b) \  \widetilde{\pi}_{-}( k) 
\end{array}
\right) \,.
\end{equation}
As the previous equation suggests, all the objects appearing in the three-pion formalism will have an additional flavour index, running over this seven-dimensional space.

The detailed derivation is given in Ref.~\cite{Hansen:2020zhy}. Here we will just state the result, and comment on its structure. The three-pion quantization condition reads
\begin{equation}
\label{eq:genIsoQC}
\text{det}_{k,\ell,m, \textbf f} \big [1 - \textbf K_{\df,3}(E^\star) \, \textbf  F_3(E, \boldsymbol P, L) \big ] =0 \,,
\end{equation}
where the determinant runs over the $(k\ell m)$ space and the additional flavour index. The quantities $\textbf K_{\df,3}$ and $ \textbf  F_3$ are defined as their analogous for identical particles, but they have been promoted to matrices in flavour space---see Section 2.1 of Ref.~\cite{Hansen:2020zhy}.  Moreover, the generalized relation to the three-pion scattering amplitude is established in Section 2.3 of the same reference. 

The main result of Ref.~\cite{Hansen:2020zhy} is given in Sections 2.4 and 2.5. It corresponds to projecting the quantization condition of Eq.~(\ref{eq:genIsoQC}) to definite two- and three-pion isospin. By doing so, one in fact recovers four independent quantization conditions:
\begin{equation}
{\ \ \ 
\text{det} \big [1 - \textbf K^{[I]}_{\df,3}(E^\star) \, \textbf  F^{[I]}_3(E, \boldsymbol P, L) \big ] =0 \ \ \ 
},
\end{equation}
where the superscript $[I]$ accounts for the fixed three-pion isospin. All necessary definitions are given in Table 1 of the corresponding article. Similarly, one can bring the generalized relation to the three-pion scattering amplitude to a block-diagonal form. It is also important to note that in the same paper we also discuss the generalized threshold expansion of $\textbf K_{\df,3}$, as well as parametrizations for the three-pion resonances of Table~\ref{tab:resonances}.

\begin{figure}[h!]
  \centering
  \includegraphics[width=0.8\linewidth]{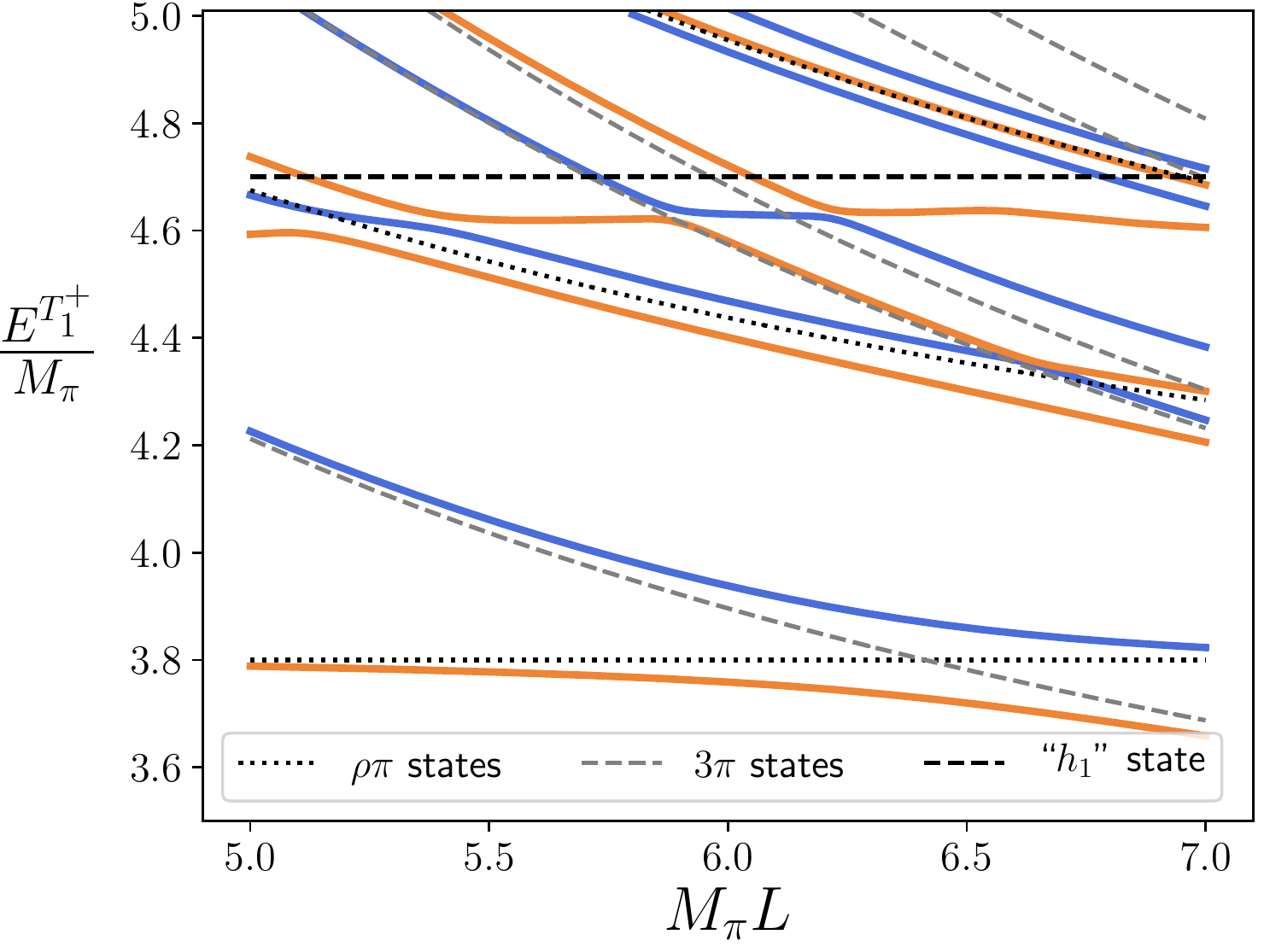}  
  \caption{ Example of finite-volume spectrum for three pions with $I_{\pi\pi\pi} = 0$ and irreps $T_1^+$. The interacting energies are depicted with solid lines with alternating colors. Dashed and dotted grey lines represent the noninteracting levels. More details about the parameters can be found in the paper. Source: Ref.\cite{Hansen:2020zhy}.}
  \label{fig:h1spec}
\end{figure}

We conclude with an example of the utility of this formalism. In Fig.~\ref{fig:h1spec}, we present a toy implementation\footnote{The various parametrizations used here do not correspond to the physical ones, and are chosen for illustrative purposes---see Section 4 of Ref.~\cite{Hansen:2020zhy}.} of the quantization condition with total $I_{\pi\pi\pi}=0$ in the $T_1^+$ irrep. This corresponds to the channel of the $h_1$ resonance, and it provides an example in which a complication of cascading resonant decays happens: $h_1 \to \rho \pi \to 3 \pi$. Along with the interacting energies, the free $3\pi$, $\rho \pi$ and $h_1$ energies are included for comparison. As can be seen, the actual spectral lines are significantly shifted with respect to the noninteracting levels. We also see the usual pattern of avoided level crossings. In addition, the finite-volume state related to the toy $h_1$ is well below the position of the pole in $\kdf$. Understanding this and other features will require further numerical and theoretical investigations.

\subsection{Three-particle decays}

The final article of this thesis, Ref.~\cite{Hansen:2021ofl}, deals with the generalization of the Lellouch-L\"uscher formalism, explained in Section~\ref{sec:LL}, to three-particle decays using the RFT approach. A physical process for which this is useful is the CP-violating $K \to 3\pi$ weak decay. Thus, it nicely connects to Chapter~\ref{sec:largeN}, where another nonleptonic kaon decay was studied: $K \to 2\pi$. Other transitions that can be treated with the formalism of that work are the isospin-violating $\eta \to 3\pi$ strong decay, or the electromagnetic $\gamma^* \to 3\pi$ amplitude that enters the calculation of the muonic $g-2$.

The article is divided in two parts. First, the formalism for identical scalars is presented. For this, we make use of the original form of the QC3 of Refs.~\cite{Hansen:2014eka,Hansen:2015zga}. This is helpful to understand the main features, even though it does not apply to any system in QCD. In the second part of the paper, the extension to generic three-pion decays is discussed. This requires the three-pion formalism of Ref.~\cite{Hansen:2020zhy}, introduced in the previous section. Here we will comment only on the first part, and refer the reader to the original reference for the second.

\begin{figure}[h!]
  \centering
  \includegraphics[width=1\linewidth]{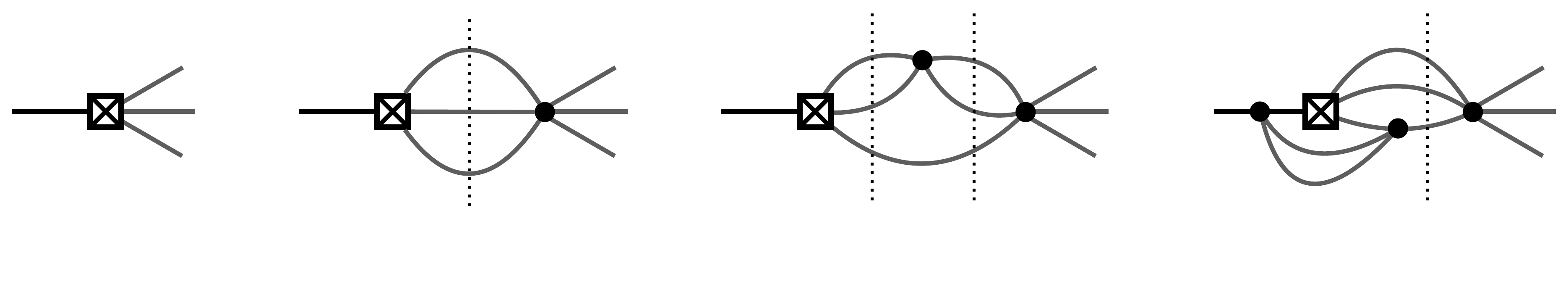}  
  \caption{Four examples of underlying diagrams contributing to $K \to 3\pi$, and the
corresponding finite-volume matrix element. The leftmost diagram is a local one-to-three transition, whose exponentially-suppressed finite-volume effects we neglect. By contrast, the middle two diagrams have power-like $1/L$ effects because of the on-shell intermediate states. This is indicated by vertical dashed lines. Finally, the rightmost diagram depicts a QCD induced dressing to the weak vertex. Our formalism includes all such interactions and dressing of the vertices. Source: Ref.\cite{Hansen:2021ofl}.}
  \label{fig:kto3pi}
\end{figure}

As in the two-body case, power-law finite-volume effects appear in decays to three particles. This is because final-state interactions are mangled in a finite box. Our goal is therefore to derive expressions that correct for this distortions (up to exponentially-suppressed corrections). To exemplify the origin of these effects, we show in Fig.~\ref{fig:kto3pi} three diagrams that produce them, and one that does not. Since we work in a generic relativistic EFT to all orders, all contributions are automatically incorporated.

A shared trait of three-particle formalisms is that they are two-step processes. This also extends to the three-body decay formalism. In the first part, the finite-volume matrix element---obtained from lattice QCD---is related to an intermediate quasilocal scheme-dependent quantity ($A_{K3\pi}^\text{PV}$):
\begin{equation}
\sqrt{2 E_K(\boldsymbol P) }L^3 \langle E_n, \boldsymbol P, \Lambda \mu,L |\mathcal H_W(0)| K,\boldsymbol P,L \rangle =
v^\dagger A_{K3\pi}^\text{PV}
\,,
\label{eq:MEfinal2}
\end{equation}
where $v$ is a vector, whose outer product defines the residue of the three-particle quantization condition in a given irrep:
\begin{equation}
\cR_{\Lambda \mu}(E^{\Lambda}_n,\boldsymbol P,L) = v(E^{\Lambda}_n,\boldsymbol P,\Lambda \mu,L) v^\dagger(E^{\Lambda}_n,\boldsymbol P,\Lambda \mu,L)\,.
\label{eq:residueres}
\end{equation}
In fact, $A_{K3\pi}^\text{PV}$ plays an analogous role to that of $\kdf$ in $3 \to 3$ scattering. For practical purposes, it will be convenient to parametrize it using the threshold expansion---see Section 2.3 in Ref.~\cite{Hansen:2021ofl}. The second step involves integral equations. In particular, one can define a finite-volume quantity:
\begin{align}
T_{K3\pi,L}^{(u)}
&= \cL_L^{(u)} \frac1{1+\cK_{\df,3} F_3} A_{K3\pi}^\text{PV}\,,
\label{eq:Tufinal}
\end{align}
whose infinite-volume limit taken in the appropriate way equates the infinite-volume decay amplitude:
\begin{equation}
T^{(u)}_{K3\pi}(\boldsymbol k)_{\ell m} = \lim_{\epsilon\to 0^+} \lim_{L\to \infty}
T^{(u)}_{K3\pi,L}(\boldsymbol k)_{\ell m}\bigg|_{E \to E + i\epsilon}\,.
\label{eq:limTu}
\end{equation}

A simplification of the expressions can be achieved in the isotropic approximation, that is, considering that $A_{K3\pi}^\text{PV} = A^\text{iso}$, with $A^\text{iso} \equiv \text{const}$. Explicit equations for this are given in Section 2.5 of the paper. It is expected that this is equivalent to the three-particle decay formalism derived in the NREFT approach in Ref.~\cite{Muller:2020wjo}, when the nonrelativistic limit of our result is taken.

\subsection{Concluding remarks}

The four articles discussed in this chapter have boosted the applicability of the three-particle formalism in many ways. We have implemented the formalism including $d$-wave interactions\cite{Blanton:2019igq}, as well as irrep projection~\cite{Blanton:2019igq,Blanton:2019vdk} and moving frames~\cite{Blanton:2019vdk}. We have established the threshold expansion of the three-particle $K$-matrix~\cite{Blanton:2019vdk}, and developed a strategy to constrain the different terms from lattice simulations~\cite{Blanton:2019igq,Blanton:2019vdk}. We have been able to constrain with statistical significance the leading two terms in $\kdf$, and tested useful parametrizations of the $s$-wave phase shift of two pions at maximal isospin~\cite{Blanton:2019vdk}. Moreover, we have extended the formalism to deal with degenerate nonidentical pions~\cite{Hansen:2020zhy}, which enables the study of some QCD resonances, such as the $\omega$ or $h_1$ resonances. Finally, we have presented the generalization of the Lellouch-L\"uscher formalism for three particles~\cite{Hansen:2021ofl}, and so, one can now study from lattice simulations some phenomenologically interesting decays: $K \to 3\pi$, $\eta \to 3\pi$ and $\gamma^* \to 3\pi$.

We have {\it de facto} entered an era of three-particle spectroscopy. We expect to see a blossoming of generalizations and applications of this formalism, some of which are already under way.  Compelling examples will be the extraction of resonance parameters from lattice simulations, and explorations of three-particle systems that include particles with spin. The latter is relevant for the Roper resonance, as well as studies of the three-nucleon force.

The long-term aspiration of hadron spectroscopy on the lattice is to deal with processes involving more than three hadrons. Future techniques might come in the form of $N$-particle quantization conditions, or possibly involve a shift of paradigm in the way finite-volume quantities are treated. In this manner, one hopes to obtain {\it ab-initio} studies of, e.g., the charmonium and bottomonium spectra. Weak decays of heavier hadrons also pose an interesting problem. An important example is the decay of $D$ mesons, where CP violation has been recently confirmed~\cite{Aaij:2019kcg}.

\addtocontents{toc}{\protect\clearpage\protect}

\renewcommand{\chaptername}{Capítulo}
\chapter{Resumen de tesis}
\label{sec:resumen}

En esta tesis doctoral se estudian las propiedades e interacciones de mesones ligeros. En particular, nos centramos en procesos hadrónicos de decaimiento y dispersión, como la desintegración débil de un kaón a dos piones y la dispersión de tres piones cargados. La predicción de estos procesos requiere resolver la teoría que describe las interacciones fuertes.

La formulación matemática de la interacción fuerte es la cromodinámica cuántica (QCD, por sus siglas en inglés). Un peculiaridad de esta teoría es que las expansiones perturbativas fallan en escalas de energía hadrónicas. Por esto, se necesitan herramientas no perturbativas para obtener predicciones de primeros principios. El principal método usado en esta tesis es la formulación de teorías cuánticas en el retículo. También emplearemos teorías efectivas, ya que proporcionan un punto de vista complementario para entender la dinámica de hadrones ligeros. En la Sección~\ref{sec:LFTspanish}, presentamos un resumen de estos métodos.

Los temas de investigación de esta disertación están divididos en dos apartados. El primero trata del estudio del límite del gran número de colores (límite de 't Hooft) usando simulaciones numéricas en el retículo. El objetivo pricipial es abordar el origen de la regla $\Delta I=1/2$ en la desintegración de los kaones, que es un problema abierto clásico en QCD.  El segundo se centra en el estudio de procesos multipartícula en volumen finito, que nos permitirá predecir la dispersión de tres piones a partir de simulaciones en el retículo. Estos temas han sido tratados en los Capítulos~\ref{sec:largeN} y \ref{sec:multiparticle}, respectivamente, y se resumen en las Secciones~\ref{sec:largeNspanish} y \ref{sec:multispanish}.

Por último, las publicaciones revisadas por pares que constituyen el cuerpo de esta tesis se pueden encontrar en el compendio de la Parte \ref{sec:papers}. Hemos mantenido la versión original de la revista.

\section{Resolviendo la dinámica de la interacción fuerte} \label{sec:LFTspanish}

La interacción fuerte es una de las fuerzas conocidas en la naturaleza. Su nombre se debe a que a distancias del orden de femtómetro su magnitud es mayor que la de las otras tres interacciones: electromagnetismo, la fuerza débil y la gravitación. Esta interacción es la responsable de la estructura y propiedades del núcleo atómico.  En esta breve sección, presentaremos la formulación matemática de la cromodinámica cuántica, así como algunas características clave. Asimismo, discutiremos los métodos existentes para resolver la teoría: las teorías efectivas y la formulación en el retículo. 

El Modelo Estándar de la física de partículas es una teoría que logra describir con éxito los fenómenos subatómicos. Es una teoría cuántica de campos que incluye las interacciones electrodébil y fuerte de tres familias de fermiones fundamentales (quarks y leptones). Además, el Modelo Estándar incluye un sector escalar, el bosón de Higgs, responsable de dar masa a las diferentes partículas elementales. Llamamos QCD al conjunto de campos fundamentales que interaccionan mediante la fuerza fuerte.

La carga de la interacción fuerte se denomina ``color". Las partículas fundamentales con carga de color son los seis quarks y los campos gauge (gluones). El lagrangiano~\cite{Fritzsch:1973pi} correspondiente viene dado por
\begin{equation}
\mathcal{L}_{QCD} =  \sum_{f }\bar q_f (i \gamma_\mu D^\mu -m )  q_f -\frac{1}{2} F_{\mu\nu} F^{\mu\nu} , \label{eq:LQCDsp}
\end{equation} 
donde $f$ es un índice de sabor con posibles valores $(u,d,c,s,t,b)$. Además,
\begin{equation}
 D_\mu = \partial_\mu + i g_s t_a A^a_\mu \ \text{ y } \ F_{\mu\nu} = \frac{-i}{g_s} [D_\mu, D_\nu],
\end{equation}
con $A^a_\mu$ siendo el campo gluónico, $t_a$ las matrices del Gell-Mann y $g_s$ el acoplo de la interacción. Este lagrangiano se deriva imponiendo la simetría gauge, o sea, invariancia local bajo transformaciones de $SU(N_c)$. En QCD, hay tres colores, o sea, $N_c=3$.

En el régimen de altas energías, QCD exhibe una característica que la diferencia de otras teorías de campos, como la electrodinámica cuántica. Esta es la libertad asintótica~\cite{Gross:1973id,Politzer:1973fx}, es decir, el hecho de que la constante de acoplo decrece al incrementarse la energía. Asimismo, en el marco de teoría de perturbaciones, el acoplo diverge si la energía se aproxima a una escala generada dinámicamente, $\Lambda_{QCD} \sim 300$ MeV. Esto indica una ruptura de la expansión perturbativa a bajas energías.

Una manifestación no perturbativa de la interacción fuerte es el confinamiento de los quarks y gluones dentro de estados compuestos (hadrones). Ello conlleva que no se puedan detectar quarks y gluones en libertad, sino que los hadrones son las únicas partículas observables. Sabemos que hay dos tipos de hadrones: mesones y bariones. Los primeros son bosones, y se pueden interpretar como estados ligados de un quark y un antiquark. Los segundos, generalmente más pesados, son fermiones y se asocian a estados de tres quarks. El modelo quark es una forma de sistematizar todos estos estados basándose en teoría de grupos~\cite{GellMann:1964nj,Zweig:1981pd,Zweig:1964jf}.

Durante este trabajo nos hemos centrado en las propiedades de los mesones más ligeros. Estos son el octete de mesones pseudoescalares (espín 0 y paridad negativa): los piones ($\pi^\pm, \pi^0$), los kaones ($K^\pm, K^0, \bar{K}^0$) y la eta ($\eta$). Su baja masa se debe a que se pueden interpretar como bosones de Goldstone~\cite{Nambu:1960tm,Goldstone:1961eq,Goldstone:1962es}, originados por la ruptura espontánea de la simetría quiral. Especial mención merece el mesón pseudoescalar que tiene números cuánticos de singlete de sabor, la eta prima ($\eta'$). Esta partícula es mucho más pesada que los otros mesones pseudoescalares, ya que recibe una contribución a su masa de origen topológico debido a la anomalía quiral~\cite{Adler:1969gk,Bell:1969ts}.

\subsection{Teoría de perturbaciones quiral}

Las teorías efectivas se basan en las ideas de Weinberg~\cite{Weinberg:1978kz}. Estas dicen que los distintos observables en una teoría se pueden calcular usando el lagrangiano más genérico que incluye los grados de libertad activos, y que es compatible con las simetrías existentes. El ejemplo más famoso de teoría efectiva es la teoría de Fermi~\cite{Fermi:1934hr}, que sirvió para calcular procesos de decaimiento electrodébiles mucho antes de descubrir los bosones $W$. La teoría efectiva más importante para este trabajo es la teoría de perturbaciones quirales (ChPT, por sus siglas en inglés)~\cite{Weinberg:1978kz,Gasser:1983yg}.

Las simetrías de sabor de QCD y la naturaleza de bosón de Goldstone imponen restricciones muy fuertes en las interacciones de los mesones pseudoescalares. ChPT es, por tanto, una teoría efectiva que describe las interacciones de estos mesones en la región de momento pequeño. En concreto, en esta teoría se organizan los diferentes operadores de acuerdo al siguiente contaje:
\begin{equation}
\delta \sim O(p^2) \sim O(M_\pi^2) \sim O(m),
\end{equation}
donde $M_\pi$ y $m$ son las masas del pion y del quark, respectivamente. El objeto principal es la matriz de campos mesónicos\footnote{Estas expresiones corresponden a ChPT con tres sabores: $u,d$ y $s$.},
\begin{equation}
 \phi(x) 
 = \begin{pmatrix}
\pi^0  + \frac{1}{\sqrt{3}} \eta& \sqrt{2} \pi^+ & \sqrt{2}K^+  \\
\sqrt{2} \pi^-  & -\pi^0 + \frac{1}{\sqrt{3}} \eta& \sqrt{2}K^0 \\
\sqrt{2}K^-& \sqrt{2}\bar K^0& - \frac{2}{\sqrt{3}} \eta
\end{pmatrix},
\end{equation}
que entra en el Lagragiano de esta manera:
\begin{equation}
U(x) = \exp \left[i \frac{\phi(x)}{F} \right],
\end{equation}
donde $F$ es una constante con dimensiones de energía cuya interpretación describiremos más adelante.

Mediante $U(x)$, e imponiendo las simetrías de sabor adecuadas, podemos escribir el lagrangiano de orden más bajo:
\begin{equation}
\mathcal{L}_2 = \frac{F^2}{4} \tr \left[ \partial_\mu U  \partial^\mu U^\dagger  \right]
+ \frac{ 2 B m F^2}{4 } \tr \left[ U  + U^\dagger  \right]. 
\end{equation}
Como se puede ver, hay dos operadores que aparecen con sus respectivos acoplos, cuyo valor no estará constreñido por las simetrías, pero que se podrían fijar con datos experimentales. De hecho, a este orden, $F$ es la constante de decaimiento\footnote{En este trabajo usamos la normalización $F_\pi  \simeq 92 $ MeV.} del pion, y $B$ está relacionado con el condensado quiral.

Típicamente se necesita ir más allá de primer orden. Para ello necesitaríamos el lagrangiano de segundo orden, $\mathcal{L}_4$. Este incluye operadores con cuatro derivadas, o con contaje equivalente. Estos aparecen multiplicados por unos acoplos genéricos de baja energía, $L_i$, que se  abrevian como LECs, por sus siglas en inglés. A lo largo de esta tesis se han usado varios resultados de ChPT. En concreto, en la Sección~\ref{sec:largeNspanish} hemos usado las predicciones de las Refs.~\cite{Bijnens:2009qm,Bijnens:2013yca,Bijnens:2011fm} para la constante de decaimiento del pion y las masas de los mesones. Además, en la Sección~\ref{sec:multispanish} hemos calculado amplitudes de dispersión de tres piones en ChPT.

\subsection{Teorías de campos en el retículo}

La formulación de la cromodinámica cuántica en el retículo (LQCD, por sus siglas en inglés) es un método numérico que permite resolver la dinámica de la interacción fuerte en el régimen no perturbativo. Se basa en el trabajo de Wilson en los setenta~\cite{Wilson:1974sk}. Mediante LQCD, se ha llegado a calcular observables con una precisión que compite o iguala a la experimental. 

El primer punto clave en LQCD es que la teoría de campos se puede tratar como un sistema estadístico. Para ello, es imprescindible realizar una rotación de Wick, de tal manera que trabajemos en el denominado tiempo euclídeo ($x^0 \to -i x_0^E$). Así pues, la función de partición toma la forma un significado probabilístico:
\begin{equation}
\mathcal{Z} = \int D \phi e^{- S_E[\phi]}, \label{eq:Zeursp}
\end{equation}
con $ S_E[\phi] = \int d^4x \mathcal{L}_E(\phi)$, donde $\mathcal{L}_E(\phi)$ es el lagrangiano euclídeo en función de un campo genérico. 

El tratamiento numérico de una teoría de campos requiere la discretización de la misma. En teorías escalares, es suficiente hacerlo de forma naíf: sustituir derivadas por diferencias finitas. En teorías gauge con fermiones hay varias sutilezas técnicas que discutiremos más abajo. El siguiente paso de una simulación de LQCD es generar configuraciones de campos que sigan la distribución de probabilidad marcada por la acción. Para ello existen una serie de algoritmos estandarizados. Uno de los más sencillos es el de Metropolis-Hastings~\cite{Metropolis:1953am,Hastings:1970aa}. Sin embargo, las simulaciones actuales utilizan el algoritmo Híbrido de Monte Carlo (HMC)~\cite{Duane:1987de}.

La obtención de predicciones físicas se consigue tras tomar el límite al continuo, es decir, el límite en el cual el espaciado del retículo, $a$, se va a cero. Asimismo, el tamaño del retículo ha de ser lo suficientemente grande para que no haya efectos apreciables por el volume finito.

\subsubsection{La acción discreta de QCD}

El proceso de discretización de QCD presenta dos complicaciones técnicas que requieren mención adicional. En esta subsección los describiremos de forma cualitativa.

El primero tiene que ver con la presencia de fermiones: una discretización naíf de la acción fermiónica tiene como consecuencia el problema de duplicación de fermiones~\cite{Wilson:1975id,Susskind:1976jm}. Esto significa que el límite al continuo de esta discretización no produce un solo campo fermiónico, sino $2^d$, donde $d$ es el número de dimensiones. Wilson propuso una solución pionera para este problema. Esta consiste en añadir un término a la acción con dimensión 5 y que rompe la simetría quiral (el llamado término de Wilson). Esto tiene como consecuencia que los fermiones adicionales adquieren una masa de orden $1/a$, y por tanto se desacoplan en el continuo. El precio a pagar es que todas las cantidades escalan como $O(a)$ al continuo, y no $O(a^2)$. A esto se le denomina fermiones de Wilson~\cite{Wilson:1975id}. 

Una alternativa que usaremos en esta tesis son los fermiones de {\it twisted mass}~\cite{Frezzotti:2000nk}. Esta formulación consiste en añadir un término de Wilson al que se le aplica una rotación quiral. Si el ángulo de esta rotación es $\omega=\pi/2$, tuneado de una manera no perturbativa, la teoría se aproxima al continuo como $O(a^2)$. Además, la renormalización de ciertos observables, como la constante de decaimiento del pion, se vuelve más fácil. Sin embargo, una desventaja es que esta discretización rompe las simetrías de isospín y paridad. Esto conlleva, por ejemplo, que el pion neutro y el cargado no tengan la misma masa. Aunque esto es un efecto de orden $O(a^2)$, habitualmente es numéricamente significativo.

El segundo asunto a tratar es la inclusión de los campos gauge en el retículo. Convencionalmente, los gluones viven en el álgebra del grupo gauge. Sin embargo, en la formulación wilsoniana los campos gauge se representan mediante elementos del grupo, los denominados enlaces gauge, $U_\mu(x)$. En el caso de QCD, estos son matrices de $SU(3)$ que se relacionan con los campos gluónicos como, 
\begin{equation}
U_\mu (x)  = e^{i a g_0  A_\mu(x) }.
\end{equation}
Sobre ellos, las transformaciones gauge actúan de la siguiente manera:
\begin{equation}
U_\mu (x) \to \Omega(x) U_\mu (x) \Omega^\dagger(x+a\hat \mu),\ \text{ donde } \ \Omega \in SU(3).
\end{equation}
Mediante la combinación de varios enlaces gauge en posiciones contiguas, se puede construir un invariante gauge denominado plaqueta,
\begin{equation}
\tr U^\text{plaq}_{\mu\nu} = \tr \left( U_\mu (x) U_\nu (x + a \hat \mu) U^\dagger_\mu (x+a \hat \nu) U^\dagger_\nu (x) \right),
\end{equation}
que está relacionado con el tensor del campo gluónico,
\begin{equation}
U^\text{plaq}_{\mu\nu} = e^{- i a^2 g_0 F_{\mu\nu} + O(a^3)}.
\end{equation}
Por tanto, la acción en el retículo
\begin{equation}
S^\text{plaq}_\text{YM}[U] =\frac{\beta}{2N_c} \sum_{\mu\nu} \sum_x \text{Re }\tr\left(1-U^\text{plaq}_{\mu\nu}\right),
\end{equation}
con $\beta={2 N_c}/{g_0^2}$, tiene como límite al continuo la acción de una teoría Yang-Mills. Asimismo, la versión discreta de la derivada covariante es:
\begin{align}
\nabla_\mu \psi = \frac{1}{a} \left[ U_\mu (x) \psi(x+a\hat \mu) - \psi(x)  \right].
\end{align}
Es fácil ver que el producto $\bar \psi \nabla_\mu \psi$ es un invariante gauge.

\subsubsection{Funciones de correlación}

Toda la información de una teoría de campos está contenida en las funciones de correlación. En concreto, lo que nos interesa para este trabajo son los niveles de energía y los elementos de matriz.  

Considérese la función de correlación  a dos puntos, 
\begin{equation}
C(t) = \langle  \hat O(t)  \hat O(0) \rangle,
\end{equation}
donde $\hat O$ es un operador hermítico con ciertos números cuánticos, por ejemplo, de un pion cargado. La descomposición espectral de $C(t)$ tiene la siguiente forma:
\begin{equation}
C(t) = \frac{1}{L^3} \sum_n \frac{| \braket{0 | \hat O |  n}|^2}{2 E_n} e^{-E_n t}, \label{eq:corrOBCsp}
\end{equation}
donde $n$ son todos los estados de la teoría con los mismos números cuánticos. La Ec.~(\ref{eq:corrOBCsp}) es, por tanto, una combinación lineal de exponenciales que decaen con el tiempo euclídeo. 

A partir de la Ec.~(\ref{eq:corrOBCsp}), se deducir ver que la extracción del estado fundamental es particularmente sencilla. Es se debe a que a tiempos grades, $t \gg 1$, $C(t)$ está dominada por la exponencial que decae más despacio:
\begin{equation}
C(t) \longrightarrow A_0 e^{-E_0 t}, \label{eq:corrOBCspground}
\end{equation}
donde $E_0$ es la energía del estado fundamental. Nótese que en presencia de condiciones de contorno periódicas, las expresiones anteriores adquieren correcciones por efectos del borde. Por ejemplo, la exponencial en la Ec.~\ref{eq:corrOBCspground} se convierte en un $\cosh$.

Como veremos más adelante, para estudios de dispersión en volumen finito es necesario determinar muchos niveles de energía (el espectro en un cierto canal). Esto se puede lograr usando tantos operadores con los mismos números cuánticos como niveles a determinar. Para ello, es necesario resolver el problema generalizado de los autovalores (GEVP, por sus siglas en inglés)~\cite{Luscher:1990ck}.

\clearpage
\section{Desintegraciones de kaones y el límite de 't Hooft en el retículo} \label{sec:largeNspanish}

El límite del gran número de colores, o límite de 't Hooft~\cite{tHooft:1973alw}, es una simplificación muy útil de teorías gauge $SU(N_c)$.  Matemáticamente, este límite corresponde a
\begin{equation}
N_c\rightarrow\infty, \quad \lambda=g_s^2 N_c ={\rm constante} , \quad N_f={\rm fijo},
\end{equation}
donde $g_s$ es el acoplo gauge, $\lambda$ se denomina el acoplo de 't Hooft, y $N_f$ es el número de sabores. Pese a que el número de grados de libertad aumenta con $N_c$, la teoría se simplifica de tal modo que se pueden realizar predicciones no perturbativas. Además, este limite preserva la libertad asintótica, el confinamiento y la ruptura espontánea de la simetría quiral. Por tanto, mantiene las características más relevantes de la interacción fuerte.

Es de esperar que el límite de 't Hooft se aproxime razonablemente a QCD. Sin embargo, la descripción de procesos de dispersión y decaimiento necesita correcciones subdominantes en $1/N_c$. Afortunadamente, LQCD es un método cuantitativo que permite determinar la magnitud de estas. Esto se consigue mediante simulaciones en el retículo a distintos valores de $N_c$~\cite{Teper:1998te}.

Uno de los objetivos de esta tesis ha sido explorar la dependencia de varios observables con el número de colores. Nos hemos centrado en dos temas, incluidos como sendos artículos en la tesis: (i) la dependencia en $N_c$ de las masas y constante de decaimiento del pion~\cite{Hernandez:2019qed}, y (ii) el estudio de amplitudes de transiciones débiles relacionadas con el proceso $K \to \pi\pi$ y la regla de $\Delta I=1/2$~\cite{Donini:2020qfu}.

El resto de la sección se organiza de la siguiente manera. Primero, en la Sección~\ref{sec:predicciones}, discutiremos ciertas predicciones del límite de 't Hooft para observables relacionados con mesones ligeros. Especialmente, nos centraremos en la regla de $\Delta I=1/2$, que corresponde a uno de los fallos más famosos de las predicciones el límite de 't Hooft. En la segunda parte, la Sección \ref{sec:simulaciones}, resumiremos los puntos clave de los dos artículos.

\subsection{Predicciones en el límite de 't Hooft} \label{sec:predicciones}

Las principales predicciones en el límite de 't Hooft provienen de contar potencias de $N_c$ en diagramas calculados en teoría de perturbaciones a todos los órdenes. Para ello, es importante darse cuenta de que los quarks viven en la representación fundamental del grupo gauge, mientras que los gluones en la adjunta. Esto implica que un bucle fermiónico escala como $N_c$, mientras que uno gluónico como $N_c^2$. Una representación útil para incorporar esto es la notación de doble línea para los gluones, mostrada en la Figura~\ref{fig:doubleline}. Por último, para asignar la potencia de $N_c$ a un diagrama se ha de tener en cuenta que cada vértice añade un factor de $g_s \sim 1/\sqrt{N_c}$. 

A continuación, mostraremos algunos ejemplos de predicciones en este límite. La primera concierne la constante de decaimiento del pion. Esta se puede sacar de la función a dos puntos de operadores con números cuánticos de vector axial. En la Figura~\ref{fig:diag2pt} su muestran varios ejemplos de diagramas que contribuyen a tal correlador, así como su correspondiente potencia de $N_c$. Combinando todas las contribuciones, se puede ver que la dependencia dominante en $N_c$ y $N_f$ toma la siguiente forma:
\begin{equation}
\frac{F_\pi^2}{N_c} = \left( A + B \frac{N_f}{N_c} + \cdots \right), \label{eq:Fpisp}
\end{equation}
donde $A$ y $B$ son constantes con dimensión de enería que no dependen de $N_f$ ni $N_c$. Esta simple expresión nos permite comparar el valor de $F_\pi$ en diferentes teorías gauge. 

Conclusiones parecidas se pueden sacar para la longitud de dispersión en onda $s$, $a_0$. Esta se puede extraer de la parte conexa de la función de correlación a cuatro puntos:
\begin{eqnarray}
a_0 &\propto &{ \langle O_\Gamma O_\Gamma O_\Gamma O_\Gamma \rangle_c \over  |\langle 0| O_\Gamma |\pi\rangle|^4} \propto N_c^{-1},
\end{eqnarray}
donde $O_\Gamma$ es un operador genérico que crea un pion. Este resultado implica que los procesos de dispersión están suprimidos con $N_c$. Argumentos similares aplican en decaimientos de mesones. Por tanto, se puede decir que los mesones en el límite de 't Hooft no interaccionan, y QCD  se vuelve una teoría de resonancias infinitinamente estrechas~\cite{tHooft:1973alw,Witten:1979kh,Coleman:1980nk}.

Otro punto a tratar son las propiedades de la $\eta'$ en el límite 't Hooft. Un análisis naíf de las potencias de $N_c$ en las funciones de correlación parece entrar en conflicto con la esperada naturaleza de bosón de Goldstone de esta partícula. La resolución de este problema aumentó nuestro entendimiento sobre la interacción fuerte. Esto se plasma en la ecuación de Witten y Veneziano~\cite{Witten:1979vv,Veneziano:1979ec}, que relaciona la masa de este mesón con la susceptibilidad topológica de la teoría puramente gauge:  
\begin{equation}
M_{\eta'}^2 - M_\pi^2= {2 N_f \over F^2_{\eta'}} \mychi_{YM} \equiv {2 N_f \over F^2_{\eta'}}  \int d^4 x \langle q(x) q(0)\rangle_{YM},
\end{equation}
con el operador de la carga topológica definido como
 \begin{equation}
q(x) \equiv {\lambda \over 32 \pi^2 N_c} {\rm Tr}[F_{\mu\nu}(x) {\tilde F}^{\mu\nu}(x)],
\end{equation}
y donde $F_{\eta'}$ es la constante de decaimiento de la $\eta'$. En el límite de 't Hooft, $F_{\eta'}$ coincide con $F_\pi$. Aunque la susceptibilidad topológica no puede ser medida experimentalmente, ha podido ser determinada usando LQCD~\cite{Ce:2015qha,Ce:2016awn}.

Debido a que la ruptura espontánea de la simetría quiral se mantiene en el límite de 't Hooft, es de esperar que ChPT proporcione una descripción adecuada de las interacciones de mesones ligeros. Una observación relevante es que la $\eta'$ se vuelve ligera en el límite de 't Hooft. Por tanto, ha de ser incorporada en la teoría de perturbaciones quirales como un grado de libertad adicional~\cite{DiVecchia:1980yfw,Rosenzweig:1979ay,Witten:1980sp,Kawarabayashi:1980dp,Gasser:1984gg,Leutwyler:1996sa,HerreraSiklody:1996pm,Kaiser:2000gs}. Esto implica que hay que modificar el contaje de la siguiente manera:
\begin{eqnarray}
 \delta  \sim  \left(M_\pi\over 4 \pi F_\pi\right)^2 \sim  \left(p \over 4 \pi F_\pi\right)^2  \sim  {1\over N_c}. 
\end{eqnarray}
A este contaje modificado lo llamaremos contaje de Leutwyler. Además, la matriz de campos se amplía a
\begin{equation}
  \phi
 = \begin{pmatrix}
\pi^0   + \frac{1}{\sqrt{3}} ( \sqrt{2}\eta'  + \eta)& \sqrt{2} \pi^+ & \sqrt{2}K^+  \\
\sqrt{2} \pi^-  & -\pi^0 + \frac{1}{\sqrt{3}} (\sqrt{2}\eta' + \eta) & \sqrt{2}K^0 \\
\sqrt{2}K^-& \sqrt{2}\bar K^0&  \frac{1}{\sqrt{3}} (\sqrt{2}\eta'-2 \eta )
\end{pmatrix}. 
\end{equation}
Una simplificación adicional de ChPT en el límite de 't Hooft tiene que ver con la dependencia de los acoplos efectivos con el número de colores. En el caso de tres sabores activos, se puede ver que algunas son $O(N_c)$, mientras que otras son $O(1)$~\cite{Gasser:1984gg,Peris:1994dh}:
\begin{align}
\begin{split}
L_1,L_2,L_3,L_5,L_8,L_9,L_{10} &\propto  {\mathcal O}(N_c), \\
2 L_1-L_2, L_4,L_6,L_7 &\propto  {\mathcal O}(1).
\end{split}
\end{align}

La última predicción que discutiremos tiene que ver con la desintegración débil de un kaón a dos piones, que es un canal muy interesante en el cual se ha detectado violación de CP. En el límite de simetría de isospín esta transición tiene dos modos diferentes, en los cuales los piones del estado final tienen un isospín total de valor $0$ o $2$. Los elementos de matriz relevantes son:
\begin{equation}
 i A_I e^{i \delta_I} = \braket{(\pi\pi)_I | \mathcal{H}_w | K}, 
\end{equation}
donde $\mathcal{H}_w$ es el hamiltoniano electrodébil, y $\delta_I$ la fase de dispersión fuerte. Los resultados experimentales muestran  que el canal isoescalar ($I=0$) domina con respecto al otro~\cite{Zyla:2020zbs}:
\begin{equation}
{\bigg \rvert} \frac{A_0}{A_2 }{\bigg \rvert} = 22.45(6). 
\end{equation}
A esto se le domina la regla de $\Delta I=1/2$, ya que la transición relevante es aquella donde el isospín cambia en media unidad. Sorprendentemente, el límite de 't Hooft no predice ninguna jerarquía y se equivoca por un orden de magnitud:
\begin{equation}
\text{Re }\frac{A_0}{A_2 }{\bigg \rvert}_{N_c \to \infty} = \sqrt{2} + O(N_c^{-1}).
\end{equation}
Esto parece indiciar que las correcciones subdominantes en $1/N_c$ son anormalmente grandes, o que la expansión falla para este observable. A lo largo de los años, se han propuesto algunas explicaciones: efectos del quark encanto, de la dispersión de los piones del estado final, o efectos intrínsecos de QCD que se pueden parametrizar como acoplos efectivos. De hecho, esta ha sido la pregunta que hemos tratado en un artículo de este trabajo~\cite{Donini:2020qfu}.

\subsection{Simulaciones de QCD en el límite de 't Hooft} \label{sec:simulaciones}

A lo largo de esta tesis hemos llevado a cabo simulaciones en el retículo variando el número de colores, $N_c=3-6$. Para las simulaciones, se ha usado un código publico, HiRep~\cite{DelDebbio:2008zf,DelDebbio:2009fd}. Hemos tomado cuatro sabores degenerados, $N_f=4$. Además, se ha usado la acción de Iwasaki~\cite{Iwasaki:2011np} para la parte gauge, que es una acción gauge mejorada. Respecto a los quarks, hemos utilizado fermiones de Wilson mejorados\footnote{Esto se consigue añadiendo el término de Sheikholeslami y Wohlert a la acción~\cite{Sheikholeslami:1985ij}.} en el mar, y fermiones de {\it twisted mass} en la valencia. Un resumen de nuestras simulaciones y los correspondientes parámetros se encuentra en la Tabla~\ref{tab:ensembles}. Para determinar el valor del espaciado del retículo en unidades físicas, hemos utilizado el método del {\it gradient flow}~\cite{Luscher:2010iy}. El resultado de nuestras determinaciones se resume en la Tabla~\ref{tab:scalesetting}. Como se puede ver, tenemos un espaciado aproximadamente constante, $a \sim 0.075$ fm, para todos los valores de $N_c$. Asimismo, disponemos de dos simulaciones con un espaciado más fino a $N_c=3$, $a \sim 0.065$ fm, para evaluar efectos de discretización.

\subsubsection{Dependencia en $N_c$ de las masas y constantes de decaimientos del pion}

En el primer artículo de esta tesis sobre este tema, hemos estudiado la dependencia de las masas y las constantes de decaimiento con el número de colores~\cite{Hernandez:2019qed}. Para ello, hemos usado las predicciones de ChPT, con y sin incluir la $\eta'$ como grado de libertad activo. Mediante ajustes a estas expresiones, hemos sido capaces de extraer la dependencia dominante y subdominante en $N_c$ de los acoplos efectivos. 

En la primera parte, hemos realizado ajustes a expresiones de ChPT estándar incluyendo solo los puntos a $N_c$ fijo. El resultado se muestra en la Figura~\ref{fig:chiral1}, donde se puede ver que el comportamiento de los acoplos es en general compatible con un término dominante y otro subdominante en $N_c$. La única excepción son los acoplos para $F_\pi$ en $N_c=3$, donde se pueden apreciar contribuciones de orden más alto. Despues de esto, hemos realizado ajustes a expresiones de ChPT\footnote{Hemos asumido la ecuación de Witten y Veneziano para la masa del singlete, ya que no la medimos directamente.} con la $\eta'$, en los que incluimos la dependencia en $M_\pi$  y $N_c$ al mismo tiempo. Como se puede ver en la Figura~\ref{fig:chiral2}, se consigue una descripción razonable a orden $\delta^3$ en contaje de Leutwyler.

Concluimos el resumen de este artículo con una observación. Usando la Ec.~(\ref{eq:Fpisp}) y nuestros resultados de los ajustes con $N_f=4$, es posible extrapolar a otros valores del número de sabores. Por ejemplo, obtenemos:
\begin{align}
\begin{split}
 F^{N_c=3, N_f=2} = 81(7)  \text{ MeV},   \quad  F^{N_c=3, N_f=3} = 68(7)  \text{ MeV}.
\end{split}
\end{align}
Estos valores son consistentes con aquellos recopilados por FLAG~\cite{Aoki:2019cca}.

\subsubsection{Diseccionando la regla de $\Delta I=1/2$ en el límite de 't Hooft}

El objetivo de otro de los artículos de esta tesis~\cite{Donini:2020qfu} es entender el origen de las enormes correcciones en $1/N_c$ de la regla de $\Delta I= 1/2$.  Este artículo es una continuación de otro trabajo exploratorio previo, Ref.~\cite{Donini:2016lwz}, donde un estudio similar se llevó acabo despreciando efectos de bucles de quark (la denominada aproximación {\it quenched}).
 
 Aunque ya existen cálculos directos de las amplitudes de $K \to \pi \pi$ en el retículo, estos son complejos y presentan incertidumbres elevadas~\cite{Abbott:2020hxn}. Por consiguiente, hemos usado un camino indirecto, basado en la estrategia de las Refs.~\cite{Giusti:2004an,Giusti:2006mh}. La idea principal es usar ChPT y las amplitudes de transición $K \to \pi$, que son más sencillas de computar.

A continuación, resumiremos el procedimiento. Al desacoplar el bosón $W$, el hamiltoniano electrodébil que describe transiciones con un cambio de extrañeza de una unidad ($\Delta S=1$) se compone de dos operadores tipo corriente-corriente. Al contrario de otros estudios en el retículo, optamos por matener el quark $c$ ligero, y degenerado con quark $u$ (límite de GIM). Esto tiene dos ventajas principales: (i) separar el efecto de diagramas de pinguino, y (ii) no se necesita evaluar propagadores cerrados de quarks. Esto justifica, por tanto, la elección de $N_f=4$ en nuestras simulaciones.

Una simplificación adicional es posible usando ChPT. A primer orden, existen únicamente dos operadores con las mismas propiedades de transformación que los operadores a nivel quark. De esta manera, ChPT predice que el cociente amplitudes viene dado en términos de dos acoplos efectivos, $g^\pm$:
\begin{eqnarray}
{A_0\over A_2} ={1\over 2 \sqrt{2}} \left(1+ 3 {g^-\over g^+}\right).
\end{eqnarray}
Por tanto, es de esperar que la jerarquía en las amplitudes se traduzca en un gran cociente de acoplos $g^-/g^+$. Asimismo, los acoplos efectivos se pueden extraer de simulaciones de LQCD usando las amplitudes $K \to \pi$:
\begin{eqnarray}
A^\pm = \braket{K| {Q}^\pm  |\pi}, \quad
 \lim_{M_\pi \rightarrow 0} A^\pm = g^\pm,
\end{eqnarray}
donde $Q^\pm$ son los dos operadores del hamiltoniano electrodébil. En nuestro trabajo hemos explorado la dependencia en $N_c$ de $A^\pm$ y extraído $g^\pm$ mediante ajustes quirales.

En la primera parte del artículo hemos investigado la dependencia en $N_c$ de $A^\pm$ a masa fija. 
En base a en un análisis perturbativo de las contribuciones a las funciones de correlación, esta sería
\begin{eqnarray}
A^\pm = 1 \pm \tilde a {1 \over N_c}\pm \tilde b {N_f \over N^2_c}+\tilde c {1 \over N^2_c}+ \tilde d {N_f \over N^3_c}+\cdots,
\end{eqnarray}
donde $\tilde a- \tilde d$ son coeficientes numéricos. Mediante ajustes de las amplitudes a la ecuación anterior, hemos podido comprobar que los coeficientes tienen la magnitud esperable, es decir, $O(1)$. Del mismo modo, los coeficientes $\tilde a$ y $\tilde b$ son negativos, lo que implica un incremento considerable en el cociente $A^-/A^+$. Además, parece que cuando la masa se reduce, $\tilde a$ cambia en la dirección de aumentar el cociente. Esto se muestra en la Figura~\ref{fig:ktopi1} en el texto principal.

En la segunda parte del artículo, hemos ajustado la dependencia en $M_\pi$ de $A^\pm$ a la expresión correspondiente en ChPT para obtener los acoplos $g^\pm$. Con ello, podemos obtener un estimador indirecto del cociente de amplitudes de isospín:
\begin{equation}
\frac{A_0}{A_2} \Bigg \rvert_{N_f=4, N_c=3} = 24(5)_\text{est}(7)_\text{sist}, \label{eq:A0A2sp}
\end{equation}
donde el primer error es estadístico, y el segundo, sistemático. Nótese además que este resultado es solo válido en la teoría con un quark encanto ligero.

Finalizamos la sección con las conclusiones principales de este trabajo. En primer lugar, parece que el enorme cociente de amplitudes es consistente con una expansión en $1/N_c$ con coeficientes de $O(1)$. Asimismo, una contribución importante proviene de efectos de bucles de quark, o sea, términos $N_f/N_c$. Por último, el resultado en la Ec.~(\ref{eq:A0A2sp}) sugiere que la regla de $\Delta I= 1/2$ podría estar dominada por efectos intrínsecos de QCD, y no por contribuciones la dispersión de los piones, o por haber cruzado el umbral del quark encanto. 

\subsubsection{Comentario final}

Mediante simulaciones en el retículo se puede explorar el espacio de parámetros de las teorías gauge. En nuestro caso, nos hemos centrado en variar el número de colores del grupo gauge. Hemos calculado varios observables variando $N_c$, y constatado que las cantidades exploradas tienen coeficientes $O(1)$ en la expansión en $1/N_c$. Un gran logro de nuestro trabajo ha sido reconciliar esto con la regla de $\Delta I=1/2$.

Existen otros observables que sería interesante explorar. Un ejemplo sería realizar un test no perturbativo de la ecuación de Witten y Veneziano, midiendo la masa y constante de decaimiento de la $\eta'$. También estudiar la dispersión de mesones ligeros al variar el número de colores, posiblemente incluyendo canales con resonancias. Además, podría resultar interesante investigar si los estados exóticos, como tetraquarks, sobreviven en el límite de 't Hooft, y si esto es factible de calcular mediante simulaciones en el retículo.

\clearpage
\section{Procesos multipartícula en un volumen finito} \label{sec:multispanish}

La extracción de cantidades de dispersión y decaimiento en el retículo es un tema candente en la comunidad de LQCD. Desde hace tiempo, existe un formalismo sólido para describir sistemas de hasta dos partículas, que ha sido aplicado ya a muchos sistemas complejos. El límite del marco teórico actual reside en sistemas de tres partículas, que es el tema central de esta parte de la tesis.

El estudio de procesos hadrónicos de varias partículas en el retículo es intrínsecamente diferente al experimental. Esto se debe a que no se pueden definir estados asintóticos en volumen finito, ya que no es posible separar las partículas. En los ochenta, Lüscher ideó un método para sortear este problema, basado en que los niveles de energía en volumen finito contienen información sobre las interacciones. El método de Lüscher~\cite{Luscher:1986pf,Luscher:1990ux} es por tanto una correspondencia entre el espectro en volumen finito y la amplitud de dispersión.

El resto de la sección está dividida en dos partes. En la primera revisaremos conceptos básicos de dispersión en volumen infinito, y presentaremos el método de Lüscher. En la segunda, comentaremos el formalismo relativista para tres partículas en volumen finito, así como las cuatro publicaciones sobre este tema que componen este trabajo.

\subsection{Dispersión en volumen infinito y finito}

La matriz $\hat S$, o de dispersión, es un operador que contiene toda la información sobre las interacciones de la teoría, inclusive la existencia de resonancias. El hecho de que sea unitario impone fuertes restricciones en su comportamiento. Por ejemplo, en el caso de amplitudes de dispersión elástica de dos partículas, sus elementos de matriz se pueden parametrizar usando unos ángulos. A estos se les denomina desfasajes, y existe uno para cada onda parcial, $\delta_\ell$.

Una característica interesante de los procesos de dispersión es la aparición de resonancias. Experimentalmente, estas se manifiestan como picos en la sección eficaz. Desde un punto de vista teórico, su presencia se puede ver en el desfasaje: este varía de $0$ a $\pi$ cuando la energía en el sistema centro de masas cruza la masa de la resonancia. Un ejemplo de resonancia idealizada se muestra en la Figura~\ref{fig:resonance}.

El cálculo de amplitudes de dispersión (o desfasajes) en el retículo se realiza mediante el formalismo de Lüscher~\cite{Luscher:1986pf,Luscher:1990ux}, y sus correspondientes generalizaciones\cite{Luscher:1991cf,Luscher:1990ux,Rummukainen:1995vs, Kim:2005gf, He:2005ey, Bernard:2010fp, Briceno:2012yi, Briceno:2014oea, Romero-Lopez:2018zyy,Luu:2011ep,Gockeler:2012yj}. A la expresión central de este método se le llama condición de cuantización de dos partículas (QC2, por su nombre en inglés).  Es una ecuación en forma de determinante, cuyas soluciones corresponden a niveles de energía en presencia de interacciones en volumen finito:
\begin{equation}
\det \left[ F^{-1} (E, \boldsymbol P, L)+ \mathcal{K}_2(E^*) \right] =0. \label{eq:QCsp}
\end{equation}
Esta ecuación tiene dos componentes. El primero, $F$, es una función de naturaleza cinemática con información sobre el volumen finito. Su valor está fijado si se conocen los niveles de energía en volumen finito, obtenidos de funciones de correlación en el retículo. El segundo, $\cK_2$, es una cantidad de volumen infinito trivialmente relacionada con los desfasajes. Los índices matriciales de la Ec.~\ref{eq:QCsp} son simplemente los de las ondas parciales, $\ell\, m$.  Dado que existen infitos valores de $\ell$, es necesario despreciar las interacciones a partir un valor de $\ell > \ell_\text{max}$. Una referencia útil para entender cómo aplicar este método es la siguiente revisión bibliográfica~\cite{Briceno:2017max}.

Igual que ocurre en los procesos de dispersión, los decaimientos a estados de dos partículas también se ven alterados en el retículo. La solución a esto es el método de Lellouch y Lüscher~\cite{Lellouch:2000pv}, que se emplea para corregir la distorsión provocada por el volumen finito. Un proceso para el cual esta técnica se ha aplicado es el decaimiento débil $K \to \pi\pi$~\cite{Abbott:2020hxn}, que ya fue comentado con anterioridad. Es método tambien posibilita la extracción de la amplitud $\gamma^* \to \pi\pi$.

\subsection{Tres partículas en un volume finito}

En los últimos años la generalización a tres partículas del formalismo de Lüscher ha progresado significativamente, e incluso se ha llegado a aplicar a sistemas sencillos de tres mesones cargados. Existen tres versiones del mismo, basados en: (i) una teoría efectiva relativista genérica (RFT)~\cite{Hansen:2014eka,Hansen:2015zga,Briceno:2017tce,Briceno:2018mlh,Briceno:2018aml,Blanton:2019igq,Romero-Lopez:2019qrt,Blanton:2019vdk,Hansen:2020zhy,Blanton:2020gha,Blanton:2020jnm,Blanton:2020gmf}, (ii) una teoría efectiva no relativista (NREFT)~\cite{Hammer:2017uqm,Hammer:2017kms,Doring:2018xxx,Pang:2019dfe}, y (iii) la unitariedad del volumen finito (FVU)~\cite{Mai:2017bge,Mai:2018djl,Mai:2019fba}. Aunque los tres deberían ser equivalentes, la conexión precisa no es fácil de establecer. Uno de los puntos que difiere es la naturaleza de una cantidad intermedia que parametriza las interacciones de tres partículas. En este trabajo nos hemos centrado únicamente en el método RFT.

Una característica del formalismo de tres partículas, que no tiene el de dos, es que es un proceso con dos pasos. En el paso inicial, se utiliza la condición de cuantización de tres partículas (QC3, por su nombre en inglés)~\cite{Hansen:2014eka}. En el caso de partículas idénticas, y sin transiciones $2 \to 3$, la condición de cuantización es:
\begin{equation}
\det \left[ F_3(E, \boldsymbol P, L )^{-1}   + \kdf(E^*) \right]=0. \label{eq:QC3sp}
\end{equation}
Aunque formalmente se asemeja a la Ec.~(\ref{eq:QCsp}), hay algunos detalles técnicos distintivos. En primer lugar, $\kdf$ no es una cantidad física ya que depende de una función de {\it cutoff}. Aun así, es una cantidad muy útil para parametrizar las interacciones cuasilocales de tres partículas. Por otro lado, $F_3$ es una función cinemática que también incluye una dependencia en la amplitud de dispersión de dos partículas. Esto último implica que las interacciones de dos partículas son un prerrequisito para estudiar las de tres. Asimismo, los índices de la matriz son tales que caracterizan el espacio de fases de tres partículas.

La dependencia de $\kdf$ en la función de {\it cutoff} se elimina en el segundo paso~\cite{Hansen:2015zga}. Este consiste en una serie de ecuaciones integrales que conectan $\kdf$ y los desfasajes con la amplitud de dispersión elástica de tres partículas, $\cM_3$. Varios ejemplos de resolución de estas ecuaciones están disponibles en la literatura~\cite{Briceno:2018mlh,Hansen:2020otl,Jackura:2020bsk}.

\subsubsection{Contribuciones al formalismo de tres partículas}

En esta subsección, procedemos a resumir los cuatro artículos sobre el formalismo de tres partículas que forman parte de esta disertación.

El primer artículo, Ref.~\cite{Blanton:2019igq}, es un estudio de la QC3 en presencia de interacciones en onda $d$. De hecho, el formalismo RFT es el único que se ha implementado explícitamente incluyendo ondas parciales distintas de $\ell=0$. Como mostramos en la publicación, los efectos de interacciones con $\ell=2$ pueden llegar a tener un impacto significativo en el espectro de tres partículas. Un ejemplo concreto se puede ver en la Figura~\ref{fig:a2spectrum}, donde una longitud de dispersión atractiva en la onda $d$ modifica notablemente los niveles de energía. Otro punto importante que tratamos es la expansión de $\kdf$ alrededor del umbral de tres partículas, que se simplifica al usar las simetrías de la teoría. Probamos que a segundo orden en las variables de Mandelstam, $\kdf$ está compuesta por cinco cantidades independientes, y solo dos dependen de variables angulares. Del mismo modo, en este trabajo establecemos una estrategia para extraer los diferentes términos de la expansión de $\kdf$ mediante simulaciones de LQCD.

A continuación, en otra publicación~\cite{Blanton:2019vdk}, aplicamos las condiciones de cuantización a los niveles de energía de dos y tres piones cargados obtenidos previamente por Hörz y Hanlon en simulaciones en el retículo~\cite{Horz:2019rrn}. Mediante ajustes combinados a los dos espectros, podemos constreñir el valor de los dos primeros términos en la expasión de $\kdf$. El resultado de estos ajustes sugiere que $\kdf$ es distinto de cero, con una significancia estadística de $2\sigma$. Además, calculamos la predicción de $\kdf$ a primer orden en ChPT. En nuestros resultados se aprecia que el primer término de la expansión de $\kdf$ es consistente con ChPT a primer orden, pero que la tensión es elevada para el segundo término. Este patrón ha sido confirmado en estudios posteriores~\cite{Fischer:2020jzp}, y su interpretación es todavía una incógnita. En la Figura~\ref{fig:confK} se resumen los principales resultados de las determinaciones de $\kdf$.

Asimismo, en la Ref.~\cite{Hansen:2020zhy} extendemos el formalismo de tres partículas para el caso de un sistema genérico de tres piones degenerados pero distinguibles. Esto tiene un alto interés fenomenológico, ya que existen varias resonancias con modos de decaimiento a tres piones (ver la Tabla~\ref{tab:resonances}). La característica principal de esta generalización es que los objetos de la condición de cuantización adquieren un índice adicional de sabor. En este trabajo también presentamos la expansión de $\kdf$ en todos los canales de tres piones, y en presencia de resonancias. Por tanto, este trabajo pone a disposición todos los ingredientes necesarios en cálculos realistas de LQCD para tratar canales con resonancias (como la $\omega$ y la $h_1$). Un ejemplo de implementación el canal de tres piones con isospín 0 se muestra en la Figura~\ref{fig:h1spec}.

Por último, en la Ref.~\cite{Hansen:2021ofl}, generalizamos el formalismo de Lellouch y Lüscher al caso de decaimientos de tres partículas. Para ello, nos centramos primero en un sistema simplificado donde asumimos que las tres partículas son idénticas. Aunque esto no tiene un análogo claro en QCD, nos sirve para entender los rasgos generales del formalismo. Igual que en el caso de procesos de dispersión, este es un método de dos pasos, donde existe un cantidad intermedia que depende del {\it cutoff}. Finalmente, extendemos el método a un sistema genérico de tres piones. Para ello, usamos el formalismo desarrollado previamente en la Ref.~\cite{Hansen:2020zhy}. En resumen, este trabajo establece el fundamento teórico que permitirá a medio plazo estudiar varios procesos de relevancia fenomenológica mediante simulaciones en el retículo. Algunos ejemplos que consideramos son: el decaimiento débil $K\to 3\pi$, la transición electromagnética $\gamma^* \to 3\pi$, y la desintegración $\eta \to 3\pi$, que es un proceso mediado por la interacción fuerte donde no se conserva el isospín.

\subsubsection{Comentario final}

Concluimos esta sección con unas reflexiones finales. El trabajo de esta tesis ha supuesto un antes y un después en el formalismo de tres partículas en volumen finito. Lo hemos implementado eficientemente, y aplicado con éxito a sistemas físicos sencillos. También hemos propuesto generalizaciones para sistemas con mayor relevancia física: resonancias y desintegraciones que involucran tres piones genéricos. 

Es de esperar que en los próximos años veamos un número considerable de aplicaciones y generalizaciones, por ejemplo, para incluir bariones en el formalismo. Esto permitiría estudiar la resonancia de Roper y tratar la fuerza de tres nucleones a partir de primeros principios.

A largo plazo, sería deseable desarrollar técnicas para tratar sistemas de más de tres partículas. Estos avances podrían venir, por ejemplo, en forma de condición de cuantización de $N$ partículas. Una aplicación relevante sería el estudio de decaimientos de mesones $D$, ya que es un sistema donde se ha detectado violación de la simetría de CP.

\renewcommand{\chaptername}{Chapter}


\cleardoublepage

\phantomsection

\addcontentsline{toc}{chapter}{{Bibliography}}

\renewcommand{\headrulewidth}{0.5pt}

\lhead[{\bfseries \thepage}]{The Bibliography}
\rhead[{The Bibliography}]{\bfseries \thepage}

\bibliographystyle{jhep}
\bibliography{biblio}

\providecommand{\href}[2]{#2}\begingroup\raggedright\begin{thebibliography}{100}

\bibitem{Blanton:2019igq}
T.~D. Blanton, F.~Romero-L\'opez and S.~R. Sharpe, \emph{{Implementing the
  three-particle quantization condition including higher partial waves}},
  \href{http://dx.doi.org/10.1007/JHEP03(2019)106}{\emph{JHEP} {\bf 03} (2019)
  106}, [\href{http://arxiv.org/abs/1901.07095}{{\tt 1901.07095}}].

\bibitem{Hernandez:2019qed}
P.~Hern\'andez, C.~Pena and F.~Romero-L\'opez, \emph{{Large $N_c$ scaling of
  meson masses and decay constants}},
  \href{http://dx.doi.org/10.1140/epjc/s10052-019-7395-y}{\emph{Eur. Phys. J.
  C} {\bf 79} (2019) 865}, [\href{http://arxiv.org/abs/1907.11511}{{\tt
  1907.11511}}].

\bibitem{Blanton:2019vdk}
T.~D. Blanton, F.~Romero-L\'opez and S.~R. Sharpe, \emph{{$I=3$ Three-Pion
  Scattering Amplitude from Lattice QCD}},
  \href{http://dx.doi.org/10.1103/PhysRevLett.124.032001}{\emph{Phys. Rev.
  Lett.} {\bf 124} (2020) 032001}, [\href{http://arxiv.org/abs/1909.02973}{{\tt
  1909.02973}}].

\bibitem{Donini:2020qfu}
A.~Donini, P.~Hern\'andez, C.~Pena and F.~Romero-L\'opez, \emph{{Dissecting the
  $\Delta I= 1/2$ rule at large $N_c$}},
  \href{http://dx.doi.org/10.1140/epjc/s10052-020-8192-3}{\emph{Eur. Phys. J.
  C} {\bf 80} (2020) 638}, [\href{http://arxiv.org/abs/2003.10293}{{\tt
  2003.10293}}].

\bibitem{Hansen:2020zhy}
M.~T. Hansen, F.~Romero-L\'opez and S.~R. Sharpe, \emph{{Generalizing the
  relativistic quantization condition to include all three-pion isospin
  channels}}, \href{http://dx.doi.org/10.1007/JHEP07(2020)047}{\emph{JHEP} {\bf
  20} (2020) 047}, [\href{http://arxiv.org/abs/2003.10974}{{\tt 2003.10974}}].

\bibitem{Hansen:2021ofl}
M.~T. Hansen, F.~Romero-L\'opez and S.~R. Sharpe, \emph{{Decay amplitudes to
  three hadrons from finite-volume matrix elements}},
  \href{http://arxiv.org/abs/2101.10246}{{\tt 2101.10246}}.

\bibitem{Donini:2016lwz}
A.~Donini, P.~Hern\'andez, C.~Pena and F.~Romero-L\'opez, \emph{{Nonleptonic
  kaon decays at large $N_c$}},
  \href{http://dx.doi.org/10.1103/PhysRevD.94.114511}{\emph{Phys. Rev. D} {\bf
  94} (2016) 114511}, [\href{http://arxiv.org/abs/1607.03262}{{\tt
  1607.03262}}].

\bibitem{Romero-Lopez:2018rcb}
F.~Romero-L\'opez, A.~Rusetsky and C.~Urbach, \emph{{Two- and three-body
  interactions in $\varphi ^4$ theory from lattice simulations}},
  \href{http://dx.doi.org/10.1140/epjc/s10052-018-6325-8}{\emph{Eur. Phys. J.
  C} {\bf 78} (2018) 846}, [\href{http://arxiv.org/abs/1806.02367}{{\tt
  1806.02367}}].

\bibitem{Romero-Lopez:2019qrt}
F.~Romero-L\'opez, S.~R. Sharpe, T.~D. Blanton, R.~A. Brice\~no and M.~T.
  Hansen, \emph{{Numerical exploration of three relativistic particles in a
  finite volume including two-particle resonances and bound states}},
  \href{http://dx.doi.org/10.1007/JHEP10(2019)007}{\emph{JHEP} {\bf 10} (2019)
  007}, [\href{http://arxiv.org/abs/1908.02411}{{\tt 1908.02411}}].

\bibitem{Fischer:2020jzp}
M.~Fischer, B.~Kostrzewa, L.~Liu, F.~Romero-L\'opez, M.~Ueding and C.~Urbach,
  \emph{{Scattering of two and three physical pions at maximal isospin from
  lattice QCD}},  \href{http://arxiv.org/abs/2008.03035}{{\tt 2008.03035}}.

\bibitem{Romero-Lopez:2020rdq}
F.~Romero-L\'opez, A.~Rusetsky, N.~Schlage and C.~Urbach, \emph{{Relativistic
  $N$-particle energy shift in finite volume}},
  \href{http://dx.doi.org/10.1007/JHEP02(2021)060}{\emph{JHEP} {\bf 02} (2021)
  060}, [\href{http://arxiv.org/abs/2010.11715}{{\tt 2010.11715}}].

\bibitem{Hernandez:2020tbc}
P.~Hern\'andez and F.~Romero-L\'opez, \emph{{The Large $N_c$ limit of QCD on
  the lattice}},
  \href{http://dx.doi.org/10.1140/epja/s10050-021-00374-2}{\emph{Eur. Phys. J.
  A} {\bf 57} (2021) 52}, [\href{http://arxiv.org/abs/2012.03331}{{\tt
  2012.03331}}].

\bibitem{Fritzsch:1973pi}
H.~Fritzsch, M.~Gell-Mann and H.~Leutwyler, \emph{{Advantages of the Color
  Octet Gluon Picture}},
  \href{http://dx.doi.org/10.1016/0370-2693(73)90625-4}{\emph{Phys. Lett. B}
  {\bf 47} (1973) 365--368}.

\bibitem{Yang:1954ek}
C.-N. Yang and R.~L. Mills, \emph{{Conservation of Isotopic Spin and Isotopic
  Gauge Invariance}},
  \href{http://dx.doi.org/10.1103/PhysRev.96.191}{\emph{Phys. Rev.} {\bf 96}
  (1954) 191--195}.

\bibitem{Gross:1973id}
D.~J. Gross and F.~Wilczek, \emph{{Ultraviolet Behavior of Nonabelian Gauge
  Theories}}, \href{http://dx.doi.org/10.1103/PhysRevLett.30.1343}{\emph{Phys.
  Rev. Lett.} {\bf 30} (1973) 1343--1346}.

\bibitem{Politzer:1973fx}
H.~D. Politzer, \emph{{Reliable Perturbative Results for Strong
  Interactions?}},
  \href{http://dx.doi.org/10.1103/PhysRevLett.30.1346}{\emph{Phys. Rev. Lett.}
  {\bf 30} (1973) 1346--1349}.

\bibitem{vanRitbergen:1997va}
T.~van Ritbergen, J.~A.~M. Vermaseren and S.~A. Larin, \emph{{The Four loop
  beta function in quantum chromodynamics}},
  \href{http://dx.doi.org/10.1016/S0370-2693(97)00370-5}{\emph{Phys. Lett. B}
  {\bf 400} (1997) 379--384}, [\href{http://arxiv.org/abs/hep-ph/9701390}{{\tt
  hep-ph/9701390}}].

\bibitem{Luthe:2017ttg}
T.~Luthe, A.~Maier, P.~Marquard and Y.~Schroder, \emph{{The five-loop Beta
  function for a general gauge group and anomalous dimensions beyond Feynman
  gauge}}, \href{http://dx.doi.org/10.1007/JHEP10(2017)166}{\emph{JHEP} {\bf
  10} (2017) 166}, [\href{http://arxiv.org/abs/1709.07718}{{\tt 1709.07718}}].

\bibitem{Zyla:2020zbs}
{\scshape Particle Data Group} collaboration, P.~A. Zyla et~al., \emph{{Review
  of Particle Physics}},
  \href{http://dx.doi.org/10.1093/ptep/ptaa104}{\emph{PTEP} {\bf 2020} (2020)
  083C01}.

\bibitem{Noether:1918zz}
E.~Noether, \emph{{Invariant Variation Problems}},
  \href{http://dx.doi.org/10.1080/00411457108231446}{\emph{Gott. Nachr.} {\bf
  1918} (1918) 235--257}, [\href{http://arxiv.org/abs/physics/0503066}{{\tt
  physics/0503066}}].

\bibitem{Nambu:1960tm}
Y.~Nambu, \emph{{Quasiparticles and Gauge Invariance in the Theory of
  Superconductivity}},
  \href{http://dx.doi.org/10.1103/PhysRev.117.648}{\emph{Phys. Rev.} {\bf 117}
  (1960) 648--663}.

\bibitem{Goldstone:1961eq}
J.~Goldstone, \emph{{Field Theories with Superconductor Solutions}},
  \href{http://dx.doi.org/10.1007/BF02812722}{\emph{Nuovo Cim.} {\bf 19} (1961)
  154--164}.

\bibitem{Goldstone:1962es}
J.~Goldstone, A.~Salam and S.~Weinberg, \emph{{Broken Symmetries}},
  \href{http://dx.doi.org/10.1103/PhysRev.127.965}{\emph{Phys. Rev.} {\bf 127}
  (1962) 965--970}.

\bibitem{Adler:1969gk}
S.~L. Adler, \emph{{Axial vector vertex in spinor electrodynamics}},
  \href{http://dx.doi.org/10.1103/PhysRev.177.2426}{\emph{Phys. Rev.} {\bf 177}
  (1969) 2426--2438}.

\bibitem{Bell:1969ts}
J.~S. Bell and R.~Jackiw, \emph{{A PCAC puzzle: $\pi^0 \to \gamma \gamma$ in
  the $\sigma$ model}}, \href{http://dx.doi.org/10.1007/BF02823296}{\emph{Nuovo
  Cim. A} {\bf 60} (1969) 47--61}.

\bibitem{Fujikawa:1979ay}
K.~Fujikawa, \emph{{Path Integral Measure for Gauge Invariant Fermion
  Theories}}, \href{http://dx.doi.org/10.1103/PhysRevLett.42.1195}{\emph{Phys.
  Rev. Lett.} {\bf 42} (1979) 1195--1198}.

\bibitem{tHooft:1976snw}
G.~'t~Hooft, \emph{{Computation of the Quantum Effects Due to a
  Four-Dimensional Pseudoparticle}},
  \href{http://dx.doi.org/10.1103/PhysRevD.14.3432}{\emph{Phys. Rev. D} {\bf
  14} (1976) 3432--3450}.

\bibitem{tHooft:1976rip}
G.~'t~Hooft, \emph{{Symmetry Breaking Through Bell-Jackiw Anomalies}},
  \href{http://dx.doi.org/10.1103/PhysRevLett.37.8}{\emph{Phys. Rev. Lett.}
  {\bf 37} (1976) 8--11}.

\bibitem{Callan:1976je}
C.~G. Callan, Jr., R.~F. Dashen and D.~J. Gross, \emph{{The Structure of the
  Gauge Theory Vacuum}},
  \href{http://dx.doi.org/10.1016/0370-2693(76)90277-X}{\emph{Phys. Lett. B}
  {\bf 63} (1976) 334--340}.

\bibitem{Jackiw:1976pf}
R.~Jackiw and C.~Rebbi, \emph{{Vacuum Periodicity in a Yang-Mills Quantum
  Theory}}, \href{http://dx.doi.org/10.1103/PhysRevLett.37.172}{\emph{Phys.
  Rev. Lett.} {\bf 37} (1976) 172--175}.

\bibitem{GellMann:1964nj}
M.~Gell-Mann, \emph{{A Schematic Model of Baryons and Mesons}},
  \href{http://dx.doi.org/10.1016/S0031-9163(64)92001-3}{\emph{Phys. Lett.}
  {\bf 8} (1964) 214--215}.

\bibitem{Zweig:1981pd}
G.~Zweig, \emph{{An SU(3) model for strong interaction symmetry and its
  breaking. Version 1}}, .

\bibitem{Zweig:1964jf}
G.~Zweig, \emph{{An SU(3) model for strong interaction symmetry and its
  breaking. Version 2}}, pp.~22--101.
\newblock 2, 1964.

\bibitem{Weinberg:1978kz}
S.~Weinberg, \emph{{Phenomenological Lagrangians}},
  \href{http://dx.doi.org/10.1016/0378-4371(79)90223-1}{\emph{Physica A} {\bf
  96} (1979) 327--340}.

\bibitem{Fermi:1934hr}
E.~Fermi, \emph{{An attempt of a theory of beta radiation. 1.}},
  \href{http://dx.doi.org/10.1007/BF01351864}{\emph{Z. Phys.} {\bf 88} (1934)
  161--177}.

\bibitem{Weinberg:1966kf}
S.~Weinberg, \emph{{Pion scattering lengths}},
  \href{http://dx.doi.org/10.1103/PhysRevLett.17.616}{\emph{Phys. Rev. Lett.}
  {\bf 17} (1966) 616--621}.

\bibitem{Gasser:1983yg}
J.~Gasser and H.~Leutwyler, \emph{{Chiral Perturbation Theory to One Loop}},
  \href{http://dx.doi.org/10.1016/0003-4916(84)90242-2}{\emph{Annals Phys.}
  {\bf 158} (1984) 142}.

\bibitem{Georgi:1985kw}
H.~Georgi, \emph{{Weak Interactions and Modern Particle Theory}}.
\newblock 1984.

\bibitem{Pich:1995bw}
A.~Pich, \emph{{Chiral perturbation theory}},
  \href{http://dx.doi.org/10.1088/0034-4885/58/6/001}{\emph{Rept. Prog. Phys.}
  {\bf 58} (1995) 563--610}, [\href{http://arxiv.org/abs/hep-ph/9502366}{{\tt
  hep-ph/9502366}}].

\bibitem{Scherer:2002tk}
S.~Scherer, \emph{{Introduction to chiral perturbation theory}}, {\emph{Adv.
  Nucl. Phys.} {\bf 27} (2003) 277},
  [\href{http://arxiv.org/abs/hep-ph/0210398}{{\tt hep-ph/0210398}}].

\bibitem{Gasser:1984gg}
J.~Gasser and H.~Leutwyler, \emph{{Chiral Perturbation Theory: Expansions in
  the Mass of the Strange Quark}},
  \href{http://dx.doi.org/10.1016/0550-3213(85)90492-4}{\emph{Nucl. Phys. B}
  {\bf 250} (1985) 465--516}.

\bibitem{Bijnens:2009qm}
J.~Bijnens and J.~Lu, \emph{{Technicolor and other QCD-like theories at
  next-to-next-to-leading order}},
  \href{http://dx.doi.org/10.1088/1126-6708/2009/11/116}{\emph{JHEP} {\bf 11}
  (2009) 116}, [\href{http://arxiv.org/abs/0910.5424}{{\tt 0910.5424}}].

\bibitem{Bijnens:2013yca}
J.~Bijnens, K.~Kampf and S.~Lanz, \emph{{Leading logarithms in N-flavour
  mesonic Chiral Perturbation Theory}},
  \href{http://dx.doi.org/10.1016/j.nuclphysb.2013.04.012}{\emph{Nucl. Phys. B}
  {\bf 873} (2013) 137--164}, [\href{http://arxiv.org/abs/1303.3125}{{\tt
  1303.3125}}].

\bibitem{Bijnens:2011fm}
J.~Bijnens and J.~Lu, \emph{{Meson-meson Scattering in QCD-like Theories}},
  \href{http://dx.doi.org/10.1007/JHEP03(2011)028}{\emph{JHEP} {\bf 03} (2011)
  028}, [\href{http://arxiv.org/abs/1102.0172}{{\tt 1102.0172}}].

\bibitem{Wilson:1974sk}
K.~G. Wilson, \emph{{Confinement of Quarks}}, .

\bibitem{Wilson:1975id}
K.~G. Wilson, \emph{{Quarks and Strings on a Lattice}},  in \emph{{13th
  International School of Subnuclear Physics: New Phenomena in Subnuclear
  Physics}}, p.~99, 11, 1975.

\bibitem{Gupta:1997nd}
R.~Gupta, \emph{{Introduction to lattice QCD: Course}},  in \emph{{Les Houches
  Summer School in Theoretical Physics, Session 68: Probing the Standard Model
  of Particle Interactions}}, pp.~83--219, 7, 1997.
\newblock \href{http://arxiv.org/abs/hep-lat/9807028}{{\tt hep-lat/9807028}}.

\bibitem{Luscher:1998pe}
M.~Luscher, \emph{{Advanced lattice QCD}},  in \emph{{Les Houches Summer School
  in Theoretical Physics, Session 68: Probing the Standard Model of Particle
  Interactions}}, pp.~229--280, 2, 1998.
\newblock \href{http://arxiv.org/abs/hep-lat/9802029}{{\tt hep-lat/9802029}}.

\bibitem{Hernandez:2009zz}
M.~P. Hernandez, \emph{{Lattice field theory fundamentals}},  in \emph{{Les
  Houches Summer School: Session 93: Modern perspectives in lattice QCD:
  Quantum field theory and high performance computing}}, pp.~1--91, 8, 2009.

\bibitem{Feynman:1948ur}
R.~P. Feynman, \emph{{Space-time approach to nonrelativistic quantum
  mechanics}}, \href{http://dx.doi.org/10.1103/RevModPhys.20.367}{\emph{Rev.
  Mod. Phys.} {\bf 20} (1948) 367--387}.

\bibitem{Duane:1987de}
S.~Duane, A.~D. Kennedy, B.~J. Pendleton and D.~Roweth, \emph{{Hybrid Monte
  Carlo}}, \href{http://dx.doi.org/10.1016/0370-2693(87)91197-X}{\emph{Phys.
  Lett. B} {\bf 195} (1987) 216--222}.

\bibitem{Metropolis:1953am}
N.~Metropolis, A.~W. Rosenbluth, M.~N. Rosenbluth, A.~H. Teller and E.~Teller,
  \emph{{Equation of state calculations by fast computing machines}},
  \href{http://dx.doi.org/10.1063/1.1699114}{\emph{J. Chem. Phys.} {\bf 21}
  (1953) 1087--1092}.

\bibitem{Hastings:1970aa}
W.~K. Hastings, \emph{{Monte Carlo Sampling Methods Using Markov Chains and
  Their Applications}},
  \href{http://dx.doi.org/10.1093/biomet/57.1.97}{\emph{Biometrika} {\bf 57}
  (1970) 97--109}.

\bibitem{Susskind:1976jm}
L.~Susskind, \emph{{Lattice Fermions}},
  \href{http://dx.doi.org/10.1103/PhysRevD.16.3031}{\emph{Phys. Rev. D} {\bf
  16} (1977) 3031--3039}.

\bibitem{Nielsen:1981hk}
H.~B. Nielsen and M.~Ninomiya, \emph{{No Go Theorem for Regularizing Chiral
  Fermions}}, \href{http://dx.doi.org/10.1016/0370-2693(81)91026-1}{\emph{Phys.
  Lett. B} {\bf 105} (1981) 219--223}.

\bibitem{Kogut:1974ag}
J.~B. Kogut and L.~Susskind, \emph{{Hamiltonian Formulation of Wilson's Lattice
  Gauge Theories}},
  \href{http://dx.doi.org/10.1103/PhysRevD.11.395}{\emph{Phys. Rev. D} {\bf 11}
  (1975) 395--408}.

\bibitem{Kaplan:1992bt}
D.~B. Kaplan, \emph{{A Method for simulating chiral fermions on the lattice}},
  \href{http://dx.doi.org/10.1016/0370-2693(92)91112-M}{\emph{Phys. Lett. B}
  {\bf 288} (1992) 342--347}, [\href{http://arxiv.org/abs/hep-lat/9206013}{{\tt
  hep-lat/9206013}}].

\bibitem{Frezzotti:2000nk}
{\scshape Alpha} collaboration, R.~Frezzotti, P.~A. Grassi, S.~Sint and
  P.~Weisz, \emph{{Lattice QCD with a chirally twisted mass term}},
  {\emph{JHEP} {\bf 08} (2001) 058},
  [\href{http://arxiv.org/abs/hep-lat/0101001}{{\tt hep-lat/0101001}}].

\bibitem{Shindler:2007vp}
A.~Shindler, \emph{{Twisted mass lattice QCD}},
  \href{http://dx.doi.org/10.1016/j.physrep.2008.03.001}{\emph{Phys. Rept.}
  {\bf 461} (2008) 37--110}, [\href{http://arxiv.org/abs/0707.4093}{{\tt
  0707.4093}}].

\bibitem{Frezzotti:2003ni}
R.~Frezzotti and G.~C. Rossi, \emph{{Chirally improving Wilson fermions. 1.
  O(a) improvement}},
  \href{http://dx.doi.org/10.1088/1126-6708/2004/08/007}{\emph{JHEP} {\bf 08}
  (2004) 007}, [\href{http://arxiv.org/abs/hep-lat/0306014}{{\tt
  hep-lat/0306014}}].

\bibitem{Sommer:1997jg}
R.~Sommer, \emph{{O(a) improved lattice QCD}},
  \href{http://dx.doi.org/10.1016/S0920-5632(97)00490-8}{\emph{Nucl. Phys. B
  Proc. Suppl.} {\bf 60} (1998) 279--294},
  [\href{http://arxiv.org/abs/hep-lat/9705026}{{\tt hep-lat/9705026}}].

\bibitem{Symanzik:1983pq}
K.~Symanzik, \emph{{IMPROVED LATTICE ACTIONS FOR NONLINEAR SIGMA MODEL AND
  NONABELIAN GAUGE THEORY}},  in \emph{{Workshop on Non-perturbative Field
  Theory and QCD}}, pp.~61--72, 1983.

\bibitem{Symanzik:1983dc}
K.~Symanzik, \emph{{Continuum Limit and Improved Action in Lattice Theories. 1.
  Principles and phi**4 Theory}},
  \href{http://dx.doi.org/10.1016/0550-3213(83)90468-6}{\emph{Nucl. Phys. B}
  {\bf 226} (1983) 187--204}.

\bibitem{Sheikholeslami:1985ij}
B.~Sheikholeslami and R.~Wohlert, \emph{{Improved Continuum Limit Lattice
  Action for QCD with Wilson Fermions}},
  \href{http://dx.doi.org/10.1016/0550-3213(85)90002-1}{\emph{Nucl. Phys. B}
  {\bf 259} (1985) 572}.

\bibitem{Francis:2019muy}
A.~Francis, P.~Fritzsch, M.~L\"uscher and A.~Rago, \emph{{Master-field
  simulations of O($a$)-improved lattice QCD: Algorithms, stability and
  exactness}}, \href{http://dx.doi.org/10.1016/j.cpc.2020.107355}{\emph{Comput.
  Phys. Commun.} {\bf 255} (2020) 107355},
  [\href{http://arxiv.org/abs/1911.04533}{{\tt 1911.04533}}].

\bibitem{Luscher:1996vw}
M.~Luscher and P.~Weisz, \emph{{O(a) improvement of the axial current in
  lattice QCD to one loop order of perturbation theory}},
  \href{http://dx.doi.org/10.1016/0550-3213(96)00448-8}{\emph{Nucl. Phys. B}
  {\bf 479} (1996) 429--458}, [\href{http://arxiv.org/abs/hep-lat/9606016}{{\tt
  hep-lat/9606016}}].

\bibitem{Luscher:1996ug}
M.~Luscher, S.~Sint, R.~Sommer, P.~Weisz and U.~Wolff, \emph{{Nonperturbative
  O(a) improvement of lattice QCD}},
  \href{http://dx.doi.org/10.1016/S0550-3213(97)00080-1}{\emph{Nucl. Phys.}
  {\bf B491} (1997) 323--343},
  [\href{http://arxiv.org/abs/hep-lat/9609035}{{\tt hep-lat/9609035}}].

\bibitem{Luscher:1996sc}
M.~Luscher, S.~Sint, R.~Sommer and P.~Weisz, \emph{{Chiral symmetry and O(a)
  improvement in lattice QCD}},
  \href{http://dx.doi.org/10.1016/0550-3213(96)00378-1}{\emph{Nucl. Phys.} {\bf
  B478} (1996) 365--400}, [\href{http://arxiv.org/abs/hep-lat/9605038}{{\tt
  hep-lat/9605038}}].

\bibitem{Becirevic:2006ii}
D.~Becirevic, P.~Boucaud, V.~Lubicz, G.~Martinelli, F.~Mescia, S.~Simula
  et~al., \emph{{Exploring twisted mass lattice QCD with the Clover term}},
  \href{http://dx.doi.org/10.1103/PhysRevD.74.034501}{\emph{Phys. Rev. D} {\bf
  74} (2006) 034501}, [\href{http://arxiv.org/abs/hep-lat/0605006}{{\tt
  hep-lat/0605006}}].

\bibitem{Bhattacharya:2005rb}
T.~Bhattacharya, R.~Gupta, W.~Lee, S.~R. Sharpe and J.~M.~S. Wu,
  \emph{{Improved bilinears in lattice QCD with non-degenerate quarks}},
  \href{http://dx.doi.org/10.1103/PhysRevD.73.034504}{\emph{Phys. Rev. D} {\bf
  73} (2006) 034504}, [\href{http://arxiv.org/abs/hep-lat/0511014}{{\tt
  hep-lat/0511014}}].

\bibitem{Luscher:1984xn}
M.~Luscher and P.~Weisz, \emph{{On-Shell Improved Lattice Gauge Theories}},
  \href{http://dx.doi.org/10.1007/BF01206178}{\emph{Commun. Math. Phys.} {\bf
  97} (1985) 59}.

\bibitem{Iwasaki:2011np}
Y.~Iwasaki, \emph{{Renormalization Group Analysis of Lattice Theories and
  Improved Lattice Action. II. Four-dimensional non-Abelian SU(N) gauge
  model}},  \href{http://arxiv.org/abs/1111.7054}{{\tt 1111.7054}}.

\bibitem{Luscher:1990ck}
M.~Luscher and U.~Wolff, \emph{{How to Calculate the Elastic Scattering Matrix
  in Two-dimensional Quantum Field Theories by Numerical Simulation}},
  \href{http://dx.doi.org/10.1016/0550-3213(90)90540-T}{\emph{Nucl. Phys. B}
  {\bf 339} (1990) 222--252}.

\bibitem{tHooft:1973alw}
G.~'t~Hooft, \emph{{A Planar Diagram Theory for Strong Interactions}},
  \href{http://dx.doi.org/10.1016/0550-3213(74)90154-0}{\emph{Nucl. Phys. B}
  {\bf 72} (1974) 461}.

\bibitem{Teper:1998te}
M.~J. Teper, \emph{{SU(N) gauge theories in (2+1)-dimensions}},
  \href{http://dx.doi.org/10.1103/PhysRevD.59.014512}{\emph{Phys. Rev. D} {\bf
  59} (1999) 014512}, [\href{http://arxiv.org/abs/hep-lat/9804008}{{\tt
  hep-lat/9804008}}].

\bibitem{Lucini:2001ej}
B.~Lucini and M.~Teper, \emph{{SU(N) gauge theories in four-dimensions:
  Exploring the approach to N = infinity}},
  \href{http://dx.doi.org/10.1088/1126-6708/2001/06/050}{\emph{JHEP} {\bf 06}
  (2001) 050}, [\href{http://arxiv.org/abs/hep-lat/0103027}{{\tt
  hep-lat/0103027}}].

\bibitem{Bali:2013kia}
G.~S. Bali, F.~Bursa, L.~Castagnini, S.~Collins, L.~Del~Debbio, B.~Lucini
  et~al., \emph{{Mesons in large-N QCD}},
  \href{http://dx.doi.org/10.1007/JHEP06(2013)071}{\emph{JHEP} {\bf 06} (2013)
  071}, [\href{http://arxiv.org/abs/1304.4437}{{\tt 1304.4437}}].

\bibitem{Manohar:1998xv}
A.~V. Manohar, \emph{{Large N QCD}},  in \emph{{Les Houches Summer School in
  Theoretical Physics, Session 68: Probing the Standard Model of Particle
  Interactions}}, pp.~1091--1169, 2, 1998.
\newblock \href{http://arxiv.org/abs/hep-ph/9802419}{{\tt hep-ph/9802419}}.

\bibitem{Witten:1979kh}
E.~Witten, \emph{{Baryons in the 1/n Expansion}},
  \href{http://dx.doi.org/10.1016/0550-3213(79)90232-3}{\emph{Nucl. Phys. B}
  {\bf 160} (1979) 57--115}.

\bibitem{Coleman:1980nk}
S.~R. Coleman, \emph{{1/N}},  in \emph{{17th International School of Subnuclear
  Physics: Pointlike Structures Inside and Outside Hadrons}}, p.~0011, 3, 1980.

\bibitem{Witten:1979vv}
E.~Witten, \emph{{Current Algebra Theorems for the U(1) Goldstone Boson}},
  \href{http://dx.doi.org/10.1016/0550-3213(79)90031-2}{\emph{Nucl. Phys. B}
  {\bf 156} (1979) 269--283}.

\bibitem{Veneziano:1979ec}
G.~Veneziano, \emph{{U(1) Without Instantons}},
  \href{http://dx.doi.org/10.1016/0550-3213(79)90332-8}{\emph{Nucl. Phys. B}
  {\bf 159} (1979) 213--224}.

\bibitem{Ce:2015qha}
M.~C\`e, C.~Consonni, G.~P. Engel and L.~Giusti, \emph{{Non-Gaussianities in
  the topological charge distribution of the SU(3) Yang--Mills theory}},
  \href{http://dx.doi.org/10.1103/PhysRevD.92.074502}{\emph{Phys. Rev. D} {\bf
  92} (2015) 074502}, [\href{http://arxiv.org/abs/1506.06052}{{\tt
  1506.06052}}].

\bibitem{Ce:2016awn}
M.~C\`e, M.~Garc\'\i{}a~Vera, L.~Giusti and S.~Schaefer, \emph{{The topological
  susceptibility in the large-$N$ limit of SU($N$) Yang\textendash{}Mills
  theory}}, \href{http://dx.doi.org/10.1016/j.physletb.2016.09.029}{\emph{Phys.
  Lett. B} {\bf 762} (2016) 232--236},
  [\href{http://arxiv.org/abs/1607.05939}{{\tt 1607.05939}}].

\bibitem{Coleman:1980mx}
S.~R. Coleman and E.~Witten, \emph{{Chiral Symmetry Breakdown in Large N
  Chromodynamics}},
  \href{http://dx.doi.org/10.1103/PhysRevLett.45.100}{\emph{Phys. Rev. Lett.}
  {\bf 45} (1980) 100}.

\bibitem{DiVecchia:1980yfw}
P.~Di~Vecchia and G.~Veneziano, \emph{{Chiral Dynamics in the Large n Limit}},
  \href{http://dx.doi.org/10.1016/0550-3213(80)90370-3}{\emph{Nucl. Phys. B}
  {\bf 171} (1980) 253--272}.

\bibitem{Rosenzweig:1979ay}
C.~Rosenzweig, J.~Schechter and C.~G. Trahern, \emph{{Is the Effective
  Lagrangian for QCD a Sigma Model?}},
  \href{http://dx.doi.org/10.1103/PhysRevD.21.3388}{\emph{Phys. Rev. D} {\bf
  21} (1980) 3388}.

\bibitem{Witten:1980sp}
E.~Witten, \emph{{Large N Chiral Dynamics}},
  \href{http://dx.doi.org/10.1016/0003-4916(80)90325-5}{\emph{Annals Phys.}
  {\bf 128} (1980) 363}.

\bibitem{Kawarabayashi:1980dp}
K.~Kawarabayashi and N.~Ohta, \emph{{The Problem of $\eta$ in the Large $N$
  Limit: Effective Lagrangian Approach}},
  \href{http://dx.doi.org/10.1016/0550-3213(80)90024-3}{\emph{Nucl. Phys. B}
  {\bf 175} (1980) 477--492}.

\bibitem{Leutwyler:1996sa}
H.~Leutwyler, \emph{{Bounds on the light quark masses}},
  \href{http://dx.doi.org/10.1016/0370-2693(96)85876-X}{\emph{Phys. Lett. B}
  {\bf 374} (1996) 163--168}, [\href{http://arxiv.org/abs/hep-ph/9601234}{{\tt
  hep-ph/9601234}}].

\bibitem{HerreraSiklody:1996pm}
P.~Herrera-Siklody, J.~I. Latorre, P.~Pascual and J.~Taron, \emph{{Chiral
  effective Lagrangian in the large N(c) limit: The Nonet case}},
  \href{http://dx.doi.org/10.1016/S0550-3213(97)00260-5}{\emph{Nucl. Phys. B}
  {\bf 497} (1997) 345--386}, [\href{http://arxiv.org/abs/hep-ph/9610549}{{\tt
  hep-ph/9610549}}].

\bibitem{Kaiser:2000gs}
R.~Kaiser and H.~Leutwyler, \emph{{Large N(c) in chiral perturbation theory}},
  \href{http://dx.doi.org/10.1007/s100520000499}{\emph{Eur. Phys. J. C} {\bf
  17} (2000) 623--649}, [\href{http://arxiv.org/abs/hep-ph/0007101}{{\tt
  hep-ph/0007101}}].

\bibitem{Peris:1994dh}
S.~Peris and E.~de~Rafael, \emph{{On the large N(c) behavior of the L(7)
  coupling in chi(PT)}},
  \href{http://dx.doi.org/10.1016/0370-2693(95)00160-M}{\emph{Phys. Lett. B}
  {\bf 348} (1995) 539--542}, [\href{http://arxiv.org/abs/hep-ph/9412343}{{\tt
  hep-ph/9412343}}].

\bibitem{Ecker:1988te}
G.~Ecker, J.~Gasser, A.~Pich and E.~de~Rafael, \emph{{The Role of Resonances in
  Chiral Perturbation Theory}},
  \href{http://dx.doi.org/10.1016/0550-3213(89)90346-5}{\emph{Nucl. Phys. B}
  {\bf 321} (1989) 311--342}.

\bibitem{Cirigliano:2011ny}
V.~Cirigliano, G.~Ecker, H.~Neufeld, A.~Pich and J.~Portoles, \emph{{Kaon
  Decays in the Standard Model}},
  \href{http://dx.doi.org/10.1103/RevModPhys.84.399}{\emph{Rev. Mod. Phys.}
  {\bf 84} (2012) 399}, [\href{http://arxiv.org/abs/1107.6001}{{\tt
  1107.6001}}].

\bibitem{Shifman:1975tn}
M.~A. Shifman, A.~I. Vainshtein and V.~I. Zakharov, \emph{{Light Quarks and the
  Origin of the Delta I = 1/2 Rule in the Nonleptonic Decays of Strange
  Particles}},
  \href{http://dx.doi.org/10.1016/0550-3213(77)90046-3}{\emph{Nucl. Phys. B}
  {\bf 120} (1977) 316--324}.

\bibitem{Pallante:1999qf}
E.~Pallante and A.~Pich, \emph{{Strong enhancement of epsilon-prime / epsilon
  through final state interactions}},
  \href{http://dx.doi.org/10.1103/PhysRevLett.84.2568}{\emph{Phys. Rev. Lett.}
  {\bf 84} (2000) 2568--2571}, [\href{http://arxiv.org/abs/hep-ph/9911233}{{\tt
  hep-ph/9911233}}].

\bibitem{Pallante:2000hk}
E.~Pallante and A.~Pich, \emph{{Final state interactions in kaon decays}},
  \href{http://dx.doi.org/10.1016/S0550-3213(00)00601-5}{\emph{Nucl. Phys. B}
  {\bf 592} (2001) 294--320}, [\href{http://arxiv.org/abs/hep-ph/0007208}{{\tt
  hep-ph/0007208}}].

\bibitem{Boyle:2012ys}
{\scshape RBC, UKQCD} collaboration, P.~A. Boyle et~al., \emph{{Emerging
  understanding of the $\Delta I = 1/2$ Rule from Lattice QCD}},
  \href{http://dx.doi.org/10.1103/PhysRevLett.110.152001}{\emph{Phys. Rev.
  Lett.} {\bf 110} (2013) 152001}, [\href{http://arxiv.org/abs/1212.1474}{{\tt
  1212.1474}}].

\bibitem{DelDebbio:2008zf}
L.~Del~Debbio, A.~Patella and C.~Pica, \emph{{Higher representations on the
  lattice: Numerical simulations. SU(2) with adjoint fermions}},
  \href{http://dx.doi.org/10.1103/PhysRevD.81.094503}{\emph{Phys. Rev. D} {\bf
  81} (2010) 094503}, [\href{http://arxiv.org/abs/0805.2058}{{\tt 0805.2058}}].

\bibitem{DelDebbio:2009fd}
L.~Del~Debbio, B.~Lucini, A.~Patella, C.~Pica and A.~Rago, \emph{{Conformal
  versus confining scenario in SU(2) with adjoint fermions}},
  \href{http://dx.doi.org/10.1103/PhysRevD.80.074507}{\emph{Phys. Rev. D} {\bf
  80} (2009) 074507}, [\href{http://arxiv.org/abs/0907.3896}{{\tt 0907.3896}}].

\bibitem{Alexandrou:2018egz}
C.~Alexandrou et~al., \emph{{Simulating twisted mass fermions at physical
  light, strange and charm quark masses}},
  \href{http://dx.doi.org/10.1103/PhysRevD.98.054518}{\emph{Phys. Rev. D} {\bf
  98} (2018) 054518}, [\href{http://arxiv.org/abs/1807.00495}{{\tt
  1807.00495}}].

\bibitem{Aoki:2003sj}
S.~Aoki and Y.~Kuramashi, \emph{{Determination of the improvement coefficient
  c(SW) up to one loop order with the conventional perturbation theory}},
  \href{http://dx.doi.org/10.1103/PhysRevD.68.094019}{\emph{Phys. Rev. D} {\bf
  68} (2003) 094019}, [\href{http://arxiv.org/abs/hep-lat/0306015}{{\tt
  hep-lat/0306015}}].

\bibitem{Bar:2002nr}
O.~Bar, G.~Rupak and N.~Shoresh, \emph{{Simulations with different lattice
  Dirac operators for valence and sea quarks}},
  \href{http://dx.doi.org/10.1103/PhysRevD.67.114505}{\emph{Phys. Rev. D} {\bf
  67} (2003) 114505}, [\href{http://arxiv.org/abs/hep-lat/0210050}{{\tt
  hep-lat/0210050}}].

\bibitem{Bussone:2018ljj}
{\scshape ALPHA} collaboration, A.~Bussone, S.~Chaves, G.~Herdo\'\i{}za,
  C.~Pena, D.~Preti, J.~A. Romero et~al., \emph{{Heavy-quark physics with a
  tmQCD valence action}},
  \href{http://dx.doi.org/10.22323/1.334.0270}{\emph{PoS} {\bf LATTICE2018}
  (2019) 270}, [\href{http://arxiv.org/abs/1812.01474}{{\tt 1812.01474}}].

\bibitem{Luscher:2010iy}
M.~L\"uscher, \emph{{Properties and uses of the Wilson flow in lattice QCD}},
  \href{http://dx.doi.org/10.1007/JHEP08(2010)071}{\emph{JHEP} {\bf 08} (2010)
  071}, [\href{http://arxiv.org/abs/1006.4518}{{\tt 1006.4518}}].

\bibitem{Bruno:2013gha}
{\scshape ALPHA} collaboration, M.~Bruno and R.~Sommer, \emph{{On the
  $N_f$-dependence of gluonic observables}},
  \href{http://dx.doi.org/10.22323/1.187.0321}{\emph{PoS} {\bf LATTICE2013}
  (2014) 321}, [\href{http://arxiv.org/abs/1311.5585}{{\tt 1311.5585}}].

\bibitem{Sommer:2014mea}
R.~Sommer, \emph{{Scale setting in lattice QCD}},
  \href{http://dx.doi.org/10.22323/1.187.0015}{\emph{PoS} {\bf LATTICE2013}
  (2014) 015}, [\href{http://arxiv.org/abs/1401.3270}{{\tt 1401.3270}}].

\bibitem{Harlander:2016vzb}
R.~V. Harlander and T.~Neumann, \emph{{The perturbative QCD gradient flow to
  three loops}}, \href{http://dx.doi.org/10.1007/JHEP06(2016)161}{\emph{JHEP}
  {\bf 06} (2016) 161}, [\href{http://arxiv.org/abs/1606.03756}{{\tt
  1606.03756}}].

\bibitem{Bruno:2016plf}
M.~Bruno, T.~Korzec and S.~Schaefer, \emph{{Setting the scale for the CLS $2 +
  1$ flavor ensembles}},
  \href{http://dx.doi.org/10.1103/PhysRevD.95.074504}{\emph{Phys. Rev. D} {\bf
  95} (2017) 074504}, [\href{http://arxiv.org/abs/1608.08900}{{\tt
  1608.08900}}].

\bibitem{Bar:2013ora}
O.~Bar and M.~Golterman, \emph{{Chiral perturbation theory for gradient flow
  observables}},
  \href{http://dx.doi.org/10.1103/PhysRevD.89.034505}{\emph{Phys. Rev. D} {\bf
  89} (2014) 034505}, [\href{http://arxiv.org/abs/1312.4999}{{\tt 1312.4999}}].

\bibitem{DeGrand:2016pur}
T.~DeGrand and Y.~Liu, \emph{{Lattice study of large $N_c$ QCD}},
  \href{http://dx.doi.org/10.1103/PhysRevD.94.034506}{\emph{Phys. Rev. D} {\bf
  94} (2016) 034506}, [\href{http://arxiv.org/abs/1606.01277}{{\tt
  1606.01277}}].

\bibitem{Guo:2015xva}
X.-K. Guo, Z.-H. Guo, J.~A. Oller and J.~J. Sanz-Cillero, \emph{{Scrutinizing
  the $\eta$-$\eta'$ mixing, masses and pseudoscalar decay constants in the
  framework of U(3) chiral effective field theory}},
  \href{http://dx.doi.org/10.1007/JHEP06(2015)175}{\emph{JHEP} {\bf 06} (2015)
  175}, [\href{http://arxiv.org/abs/1503.02248}{{\tt 1503.02248}}].

\bibitem{Aoki:2019cca}
{\scshape Flavour Lattice Averaging Group} collaboration, S.~Aoki et~al.,
  \emph{{FLAG Review 2019: Flavour Lattice Averaging Group (FLAG)}},
  \href{http://dx.doi.org/10.1140/epjc/s10052-019-7354-7}{\emph{Eur. Phys. J.
  C} {\bf 80} (2020) 113}, [\href{http://arxiv.org/abs/1902.08191}{{\tt
  1902.08191}}].

\bibitem{Lellouch:2000pv}
L.~Lellouch and M.~Luscher, \emph{{Weak transition matrix elements from finite
  volume correlation functions}},
  \href{http://dx.doi.org/10.1007/s002200100410}{\emph{Commun. Math. Phys.}
  {\bf 219} (2001) 31--44}, [\href{http://arxiv.org/abs/hep-lat/0003023}{{\tt
  hep-lat/0003023}}].

\bibitem{Abbott:2020hxn}
{\scshape RBC, UKQCD} collaboration, R.~Abbott et~al., \emph{{Direct CP
  violation and the $\Delta I=1/2$ rule in $K\to\pi\pi$ decay from the standard
  model}}, \href{http://dx.doi.org/10.1103/PhysRevD.102.054509}{\emph{Phys.
  Rev. D} {\bf 102} (2020) 054509},
  [\href{http://arxiv.org/abs/2004.09440}{{\tt 2004.09440}}].

\bibitem{Giusti:2004an}
L.~Giusti, P.~Hernandez, M.~Laine, P.~Weisz and H.~Wittig, \emph{{A Strategy to
  study the role of the charm quark in explaining the Delta I = 1/2 rule}},
  \href{http://dx.doi.org/10.1088/1126-6708/2004/11/016}{\emph{JHEP} {\bf 11}
  (2004) 016}, [\href{http://arxiv.org/abs/hep-lat/0407007}{{\tt
  hep-lat/0407007}}].

\bibitem{Giusti:2006mh}
L.~Giusti, P.~Hernandez, M.~Laine, C.~Pena, J.~Wennekers and H.~Wittig,
  \emph{{On K ---\ensuremath{>} pi pi amplitudes with a light charm quark}},
  \href{http://dx.doi.org/10.1103/PhysRevLett.98.082003}{\emph{Phys. Rev.
  Lett.} {\bf 98} (2007) 082003},
  [\href{http://arxiv.org/abs/hep-ph/0607220}{{\tt hep-ph/0607220}}].

\bibitem{Gaillard:1974nj}
M.~K. Gaillard and B.~W. Lee, \emph{{$\Delta$ I = 1/2 Rule for Nonleptonic
  Decays in Asymptotically Free Field Theories}},
  \href{http://dx.doi.org/10.1103/PhysRevLett.33.108}{\emph{Phys. Rev. Lett.}
  {\bf 33} (1974) 108}.

\bibitem{Altarelli:1974exa}
G.~Altarelli and L.~Maiani, \emph{{Octet Enhancement of Nonleptonic Weak
  Interactions in Asymptotically Free Gauge Theories}},
  \href{http://dx.doi.org/10.1016/0370-2693(74)90060-4}{\emph{Phys. Lett. B}
  {\bf 52} (1974) 351--354}.

\bibitem{Buchalla:1995vs}
G.~Buchalla, A.~J. Buras and M.~E. Lautenbacher, \emph{{Weak decays beyond
  leading logarithms}},
  \href{http://dx.doi.org/10.1103/RevModPhys.68.1125}{\emph{Rev. Mod. Phys.}
  {\bf 68} (1996) 1125--1144}, [\href{http://arxiv.org/abs/hep-ph/9512380}{{\tt
  hep-ph/9512380}}].

\bibitem{Bai:2015nea}
{\scshape RBC, UKQCD} collaboration, Z.~Bai et~al., \emph{{Standard Model
  Prediction for Direct CP Violation in
  K\textrightarrow{}\ensuremath{\pi}\ensuremath{\pi} Decay}},
  \href{http://dx.doi.org/10.1103/PhysRevLett.115.212001}{\emph{Phys. Rev.
  Lett.} {\bf 115} (2015) 212001}, [\href{http://arxiv.org/abs/1505.07863}{{\tt
  1505.07863}}].

\bibitem{Blum:2015ywa}
T.~Blum et~al., \emph{{$K \rightarrow \pi\pi$ $\Delta I=3/2$ decay amplitude in
  the continuum limit}},
  \href{http://dx.doi.org/10.1103/PhysRevD.91.074502}{\emph{Phys. Rev. D} {\bf
  91} (2015) 074502}, [\href{http://arxiv.org/abs/1502.00263}{{\tt
  1502.00263}}].

\bibitem{Hernandez:2006kz}
P.~Hernandez and M.~Laine, \emph{{Probing the chiral weak Hamiltonian at finite
  volumes}}, \href{http://dx.doi.org/10.1088/1126-6708/2006/10/069}{\emph{JHEP}
  {\bf 10} (2006) 069}, [\href{http://arxiv.org/abs/hep-lat/0607027}{{\tt
  hep-lat/0607027}}].

\bibitem{Kambor:1989tz}
J.~Kambor, J.~H. Missimer and D.~Wyler, \emph{{The Chiral Loop Expansion of the
  Nonleptonic Weak Interactions of Mesons}},
  \href{http://dx.doi.org/10.1016/0550-3213(90)90236-7}{\emph{Nucl. Phys. B}
  {\bf 346} (1990) 17--64}.

\bibitem{Romero-Lopez:2019gqt}
F.~Romero-López, A.~Donini, P.~Hernández and C.~Pena, \emph{{Meson
  interactions at large $N_c$ from Lattice QCD}},
  \href{http://dx.doi.org/10.22323/1.363.0005}{\emph{PoS} {\bf LATTICE2019}
  (2019) 005}, [\href{http://arxiv.org/abs/1910.10418}{{\tt 1910.10418}}].

\bibitem{Luscher:1986pf}
M.~Luscher, \emph{{Volume Dependence of the Energy Spectrum in Massive Quantum
  Field Theories. 2. Scattering States}},
  \href{http://dx.doi.org/10.1007/BF01211097}{\emph{Commun. Math. Phys.} {\bf
  105} (1986) 153--188}.

\bibitem{Luscher:1990ux}
M.~Luscher, \emph{{Two particle states on a torus and their relation to the
  scattering matrix}},
  \href{http://dx.doi.org/10.1016/0550-3213(91)90366-6}{\emph{Nucl. Phys. B}
  {\bf 354} (1991) 531--578}.

\bibitem{Maiani:1990ca}
L.~Maiani and M.~Testa, \emph{{Final state interactions from Euclidean
  correlation functions}},
  \href{http://dx.doi.org/10.1016/0370-2693(90)90695-3}{\emph{Phys. Lett. B}
  {\bf 245} (1990) 585--590}.

\bibitem{Bruno:2020kyl}
M.~Bruno and M.~T. Hansen, \emph{{Variations on the Maiani-Testa approach and
  the inverse problem}},  \href{http://arxiv.org/abs/2012.11488}{{\tt
  2012.11488}}.

\bibitem{Huang:1957im}
K.~Huang and C.~N. Yang, \emph{{Quantum-mechanical many-body problem with
  hard-sphere interaction}},
  \href{http://dx.doi.org/10.1103/PhysRev.105.767}{\emph{Phys. Rev.} {\bf 105}
  (1957) 767--775}.

\bibitem{Feng:2009ij}
X.~Feng, K.~Jansen and D.~B. Renner, \emph{{The pi+ pi+ scattering length from
  maximally twisted mass lattice QCD}},
  \href{http://dx.doi.org/10.1016/j.physletb.2010.01.018}{\emph{Phys. Lett. B}
  {\bf 684} (2010) 268--274}, [\href{http://arxiv.org/abs/0909.3255}{{\tt
  0909.3255}}].

\bibitem{Helmes:2015gla}
{\scshape ETM} collaboration, C.~Helmes, C.~Jost, B.~Knippschild, C.~Liu,
  J.~Liu, L.~Liu et~al., \emph{{Hadron-hadron interactions from N$_{f}$ = 2 + 1
  + 1 lattice QCD: isospin-2 \ensuremath{\pi}\ensuremath{\pi} scattering
  length}}, \href{http://dx.doi.org/10.1007/JHEP09(2015)109}{\emph{JHEP} {\bf
  09} (2015) 109}, [\href{http://arxiv.org/abs/1506.00408}{{\tt 1506.00408}}].

\bibitem{Beane:2007es}
S.~R. Beane, W.~Detmold, T.~C. Luu, K.~Orginos, M.~J. Savage and A.~Torok,
  \emph{{Multi-Pion Systems in Lattice QCD and the Three-Pion Interaction}},
  \href{http://dx.doi.org/10.1103/PhysRevLett.100.082004}{\emph{Phys. Rev.
  Lett.} {\bf 100} (2008) 082004}, [\href{http://arxiv.org/abs/0710.1827}{{\tt
  0710.1827}}].

\bibitem{Hansen:2015zta}
M.~T. Hansen and S.~R. Sharpe, \emph{{Perturbative results for two and three
  particle threshold energies in finite volume}},
  \href{http://dx.doi.org/10.1103/PhysRevD.93.014506}{\emph{Phys. Rev. D} {\bf
  93} (2016) 014506}, [\href{http://arxiv.org/abs/1509.07929}{{\tt
  1509.07929}}].

\bibitem{Sharpe:2017jej}
S.~R. Sharpe, \emph{{Testing the threshold expansion for three-particle
  energies at fourth order in $\phi^4$ theory}},
  \href{http://dx.doi.org/10.1103/PhysRevD.96.054515}{\emph{Phys. Rev. D} {\bf
  96} (2017) 054515}, [\href{http://arxiv.org/abs/1707.04279}{{\tt
  1707.04279}}].

\bibitem{Pang:2019dfe}
J.-Y. Pang, J.-J. Wu, H.~W. Hammer, U.-G. Mei\ss{}ner and A.~Rusetsky,
  \emph{{Energy shift of the three-particle system in a finite volume}},
  \href{http://dx.doi.org/10.1103/PhysRevD.99.074513}{\emph{Phys. Rev. D} {\bf
  99} (2019) 074513}, [\href{http://arxiv.org/abs/1902.01111}{{\tt
  1902.01111}}].

\bibitem{Muller:2020vtt}
F.~M\"uller, T.~Yu and A.~Rusetsky, \emph{{Finite-volume energy shift of the
  three-pion ground state}},
  \href{http://dx.doi.org/10.1103/PhysRevD.103.054506}{\emph{Phys. Rev. D} {\bf
  103} (2021) 054506}, [\href{http://arxiv.org/abs/2011.14178}{{\tt
  2011.14178}}].

\bibitem{Luscher:1991cf}
M.~Luscher, \emph{{Signatures of unstable particles in finite volume}},
  \href{http://dx.doi.org/10.1016/0550-3213(91)90584-K}{\emph{Nucl. Phys. B}
  {\bf 364} (1991) 237--251}.

\bibitem{Rummukainen:1995vs}
K.~Rummukainen and S.~A. Gottlieb, \emph{{Resonance scattering phase shifts on
  a nonrest frame lattice}},
  \href{http://dx.doi.org/10.1016/0550-3213(95)00313-H}{\emph{Nucl. Phys. B}
  {\bf 450} (1995) 397--436}, [\href{http://arxiv.org/abs/hep-lat/9503028}{{\tt
  hep-lat/9503028}}].

\bibitem{Kim:2005gf}
C.~h. Kim, C.~T. Sachrajda and S.~R. Sharpe, \emph{{Finite-volume effects for
  two-hadron states in moving frames}},
  \href{http://dx.doi.org/10.1016/j.nuclphysb.2005.08.029}{\emph{Nucl. Phys. B}
  {\bf 727} (2005) 218--243}, [\href{http://arxiv.org/abs/hep-lat/0507006}{{\tt
  hep-lat/0507006}}].

\bibitem{He:2005ey}
S.~He, X.~Feng and C.~Liu, \emph{{Two particle states and the S-matrix elements
  in multi-channel scattering}},
  \href{http://dx.doi.org/10.1088/1126-6708/2005/07/011}{\emph{JHEP} {\bf 07}
  (2005) 011}, [\href{http://arxiv.org/abs/hep-lat/0504019}{{\tt
  hep-lat/0504019}}].

\bibitem{Bernard:2010fp}
V.~Bernard, M.~Lage, U.~G. Meissner and A.~Rusetsky, \emph{{Scalar mesons in a
  finite volume}}, \href{http://dx.doi.org/10.1007/JHEP01(2011)019}{\emph{JHEP}
  {\bf 01} (2011) 019}, [\href{http://arxiv.org/abs/1010.6018}{{\tt
  1010.6018}}].

\bibitem{Briceno:2012yi}
R.~A. Briceno and Z.~Davoudi, \emph{{Moving multichannel systems in a finite
  volume with application to proton-proton fusion}},
  \href{http://dx.doi.org/10.1103/PhysRevD.88.094507}{\emph{Phys. Rev. D} {\bf
  88} (2013) 094507}, [\href{http://arxiv.org/abs/1204.1110}{{\tt 1204.1110}}].

\bibitem{Briceno:2014oea}
R.~A. Briceno, \emph{{Two-particle multichannel systems in a finite volume with
  arbitrary spin}},
  \href{http://dx.doi.org/10.1103/PhysRevD.89.074507}{\emph{Phys. Rev. D} {\bf
  89} (2014) 074507}, [\href{http://arxiv.org/abs/1401.3312}{{\tt 1401.3312}}].

\bibitem{Romero-Lopez:2018zyy}
F.~Romero-L\'opez, A.~Rusetsky and C.~Urbach, \emph{{Vector particle scattering
  on the lattice}},
  \href{http://dx.doi.org/10.1103/PhysRevD.98.014503}{\emph{Phys. Rev. D} {\bf
  98} (2018) 014503}, [\href{http://arxiv.org/abs/1802.03458}{{\tt
  1802.03458}}].

\bibitem{Luu:2011ep}
T.~Luu and M.~J. Savage, \emph{{Extracting Scattering Phase-Shifts in Higher
  Partial-Waves from Lattice QCD Calculations}},
  \href{http://dx.doi.org/10.1103/PhysRevD.83.114508}{\emph{Phys. Rev. D} {\bf
  83} (2011) 114508}, [\href{http://arxiv.org/abs/1101.3347}{{\tt 1101.3347}}].

\bibitem{Gockeler:2012yj}
M.~Gockeler, R.~Horsley, M.~Lage, U.~G. Meissner, P.~E.~L. Rakow, A.~Rusetsky
  et~al., \emph{{Scattering phases for meson and baryon resonances on general
  moving-frame lattices}},
  \href{http://dx.doi.org/10.1103/PhysRevD.86.094513}{\emph{Phys. Rev. D} {\bf
  86} (2012) 094513}, [\href{http://arxiv.org/abs/1206.4141}{{\tt 1206.4141}}].

\bibitem{Briceno:2017max}
R.~A. Briceno, J.~J. Dudek and R.~D. Young, \emph{{Scattering processes and
  resonances from lattice QCD}},
  \href{http://dx.doi.org/10.1103/RevModPhys.90.025001}{\emph{Rev. Mod. Phys.}
  {\bf 90} (2018) 025001}, [\href{http://arxiv.org/abs/1706.06223}{{\tt
  1706.06223}}].

\bibitem{Bernard:2008ax}
V.~Bernard, M.~Lage, U.-G. Meissner and A.~Rusetsky, \emph{{Resonance
  properties from the finite-volume energy spectrum}},
  \href{http://dx.doi.org/10.1088/1126-6708/2008/08/024}{\emph{JHEP} {\bf 08}
  (2008) 024}, [\href{http://arxiv.org/abs/0806.4495}{{\tt 0806.4495}}].

\bibitem{Hansen:2012tf}
M.~T. Hansen and S.~R. Sharpe, \emph{{Multiple-channel generalization of
  Lellouch-Luscher formula}},
  \href{http://dx.doi.org/10.1103/PhysRevD.86.016007}{\emph{Phys. Rev. D} {\bf
  86} (2012) 016007}, [\href{http://arxiv.org/abs/1204.0826}{{\tt 1204.0826}}].

\bibitem{Briceno:2014uqa}
R.~A. Brice\~no, M.~T. Hansen and A.~Walker-Loud, \emph{{Multichannel 1
  $\rightarrow$ 2 transition amplitudes in a finite volume}},
  \href{http://dx.doi.org/10.1103/PhysRevD.91.034501}{\emph{Phys. Rev. D} {\bf
  91} (2015) 034501}, [\href{http://arxiv.org/abs/1406.5965}{{\tt 1406.5965}}].

\bibitem{Briceno:2015csa}
R.~A. Brice\~no and M.~T. Hansen, \emph{{Multichannel 0 $\to$ 2 and 1 $\to$ 2
  transition amplitudes for arbitrary spin particles in a finite volume}},
  \href{http://dx.doi.org/10.1103/PhysRevD.92.074509}{\emph{Phys. Rev. D} {\bf
  92} (2015) 074509}, [\href{http://arxiv.org/abs/1502.04314}{{\tt
  1502.04314}}].

\bibitem{Hansen:2014eka}
M.~T. Hansen and S.~R. Sharpe, \emph{{Relativistic, model-independent,
  three-particle quantization condition}},
  \href{http://dx.doi.org/10.1103/PhysRevD.90.116003}{\emph{Phys. Rev. D} {\bf
  90} (2014) 116003}, [\href{http://arxiv.org/abs/1408.5933}{{\tt 1408.5933}}].

\bibitem{Hansen:2015zga}
M.~T. Hansen and S.~R. Sharpe, \emph{{Expressing the three-particle
  finite-volume spectrum in terms of the three-to-three scattering amplitude}},
  \href{http://dx.doi.org/10.1103/PhysRevD.92.114509}{\emph{Phys. Rev. D} {\bf
  92} (2015) 114509}, [\href{http://arxiv.org/abs/1504.04248}{{\tt
  1504.04248}}].

\bibitem{Briceno:2017tce}
R.~A. Brice\~no, M.~T. Hansen and S.~R. Sharpe, \emph{{Relating the
  finite-volume spectrum and the two-and-three-particle $S$ matrix for
  relativistic systems of identical scalar particles}},
  \href{http://dx.doi.org/10.1103/PhysRevD.95.074510}{\emph{Phys. Rev. D} {\bf
  95} (2017) 074510}, [\href{http://arxiv.org/abs/1701.07465}{{\tt
  1701.07465}}].

\bibitem{Briceno:2018mlh}
R.~A. Brice\~no, M.~T. Hansen and S.~R. Sharpe, \emph{{Numerical study of the
  relativistic three-body quantization condition in the isotropic
  approximation}},
  \href{http://dx.doi.org/10.1103/PhysRevD.98.014506}{\emph{Phys. Rev. D} {\bf
  98} (2018) 014506}, [\href{http://arxiv.org/abs/1803.04169}{{\tt
  1803.04169}}].

\bibitem{Briceno:2018aml}
R.~A. Brice\~no, M.~T. Hansen and S.~R. Sharpe, \emph{{Three-particle systems
  with resonant subprocesses in a finite volume}},
  \href{http://dx.doi.org/10.1103/PhysRevD.99.014516}{\emph{Phys. Rev. D} {\bf
  99} (2019) 014516}, [\href{http://arxiv.org/abs/1810.01429}{{\tt
  1810.01429}}].

\bibitem{Blanton:2020gha}
T.~D. Blanton and S.~R. Sharpe, \emph{{Alternative derivation of the
  relativistic three-particle quantization condition}},
  \href{http://dx.doi.org/10.1103/PhysRevD.102.054520}{\emph{Phys. Rev. D} {\bf
  102} (2020) 054520}, [\href{http://arxiv.org/abs/2007.16188}{{\tt
  2007.16188}}].

\bibitem{Blanton:2020jnm}
T.~D. Blanton and S.~R. Sharpe, \emph{{Equivalence of relativistic
  three-particle quantization conditions}},
  \href{http://dx.doi.org/10.1103/PhysRevD.102.054515}{\emph{Phys. Rev. D} {\bf
  102} (2020) 054515}, [\href{http://arxiv.org/abs/2007.16190}{{\tt
  2007.16190}}].

\bibitem{Blanton:2020gmf}
T.~D. Blanton and S.~R. Sharpe, \emph{{Relativistic three-particle quantization
  condition for nondegenerate scalars}},
  \href{http://dx.doi.org/10.1103/PhysRevD.103.054503}{\emph{Phys. Rev. D} {\bf
  103} (2021) 054503}, [\href{http://arxiv.org/abs/2011.05520}{{\tt
  2011.05520}}].

\bibitem{Hammer:2017uqm}
H.-W. Hammer, J.-Y. Pang and A.~Rusetsky, \emph{{Three-particle quantization
  condition in a finite volume: 1. The role of the three-particle force}},
  \href{http://dx.doi.org/10.1007/JHEP09(2017)109}{\emph{JHEP} {\bf 09} (2017)
  109}, [\href{http://arxiv.org/abs/1706.07700}{{\tt 1706.07700}}].

\bibitem{Hammer:2017kms}
H.~W. Hammer, J.~Y. Pang and A.~Rusetsky, \emph{{Three particle quantization
  condition in a finite volume: 2. general formalism and the analysis of
  data}}, \href{http://dx.doi.org/10.1007/JHEP10(2017)115}{\emph{JHEP} {\bf 10}
  (2017) 115}, [\href{http://arxiv.org/abs/1707.02176}{{\tt 1707.02176}}].

\bibitem{Doring:2018xxx}
M.~D\"oring, H.~W. Hammer, M.~Mai, J.~Y. Pang, t.~A. Rusetsky and J.~Wu,
  \emph{{Three-body spectrum in a finite volume: the role of cubic symmetry}},
  \href{http://dx.doi.org/10.1103/PhysRevD.97.114508}{\emph{Phys. Rev. D} {\bf
  97} (2018) 114508}, [\href{http://arxiv.org/abs/1802.03362}{{\tt
  1802.03362}}].

\bibitem{Pang:2020pkl}
J.-Y. Pang, J.-J. Wu and L.-S. Geng, \emph{{$DDK$ system in finite volume}},
  \href{http://dx.doi.org/10.1103/PhysRevD.102.114515}{\emph{Phys. Rev. D} {\bf
  102} (2020) 114515}, [\href{http://arxiv.org/abs/2008.13014}{{\tt
  2008.13014}}].

\bibitem{Mai:2017bge}
M.~Mai and M.~D\"oring, \emph{{Three-body Unitarity in the Finite Volume}},
  \href{http://dx.doi.org/10.1140/epja/i2017-12440-1}{\emph{Eur. Phys. J. A}
  {\bf 53} (2017) 240}, [\href{http://arxiv.org/abs/1709.08222}{{\tt
  1709.08222}}].

\bibitem{Mai:2018djl}
M.~Mai and M.~Doring, \emph{{Finite-Volume Spectrum of $\pi^+\pi^+$ and
  $\pi^+\pi^+\pi^+$ Systems}},
  \href{http://dx.doi.org/10.1103/PhysRevLett.122.062503}{\emph{Phys. Rev.
  Lett.} {\bf 122} (2019) 062503}, [\href{http://arxiv.org/abs/1807.04746}{{\tt
  1807.04746}}].

\bibitem{Mai:2019fba}
M.~Mai, M.~D\"oring, C.~Culver and A.~Alexandru, \emph{{Three-body unitarity
  versus finite-volume $\pi^+\pi^+\pi^+$ spectrum from lattice QCD}},
  \href{http://dx.doi.org/10.1103/PhysRevD.101.054510}{\emph{Phys. Rev. D} {\bf
  101} (2020) 054510}, [\href{http://arxiv.org/abs/1909.05749}{{\tt
  1909.05749}}].

\bibitem{Hansen:2019nir}
M.~T. Hansen and S.~R. Sharpe, \emph{{Lattice QCD and Three-particle Decays of
  Resonances}},
  \href{http://dx.doi.org/10.1146/annurev-nucl-101918-023723}{\emph{Ann. Rev.
  Nucl. Part. Sci.} {\bf 69} (2019) 65--107},
  [\href{http://arxiv.org/abs/1901.00483}{{\tt 1901.00483}}].

\bibitem{Mai:2021lwb}
M.~Mai, M.~D\"oring and A.~Rusetsky, \emph{{Multi-particle systems on the
  lattice and chiral extrapolations: a brief review}},
  \href{http://arxiv.org/abs/2103.00577}{{\tt 2103.00577}}.

\bibitem{Hansen:2020otl}
{\scshape Hadron Spectrum} collaboration, M.~T. Hansen, R.~A. Brice\~no, R.~G.
  Edwards, C.~E. Thomas and D.~J. Wilson, \emph{{Energy-Dependent $\pi^+ \pi^+
  \pi^+$ Scattering Amplitude from QCD}},
  \href{http://dx.doi.org/10.1103/PhysRevLett.126.012001}{\emph{Phys. Rev.
  Lett.} {\bf 126} (2021) 012001}, [\href{http://arxiv.org/abs/2009.04931}{{\tt
  2009.04931}}].

\bibitem{Culver:2019vvu}
C.~Culver, M.~Mai, R.~Brett, A.~Alexandru and M.~D\"oring, \emph{{Three pion
  spectrum in the $I=3$ channel from lattice QCD}},
  \href{http://dx.doi.org/10.1103/PhysRevD.101.114507}{\emph{Phys. Rev. D} {\bf
  101} (2020) 114507}, [\href{http://arxiv.org/abs/1911.09047}{{\tt
  1911.09047}}].

\bibitem{Alexandru:2020xqf}
A.~Alexandru, R.~Brett, C.~Culver, M.~D\"oring, D.~Guo, F.~X. Lee et~al.,
  \emph{{Finite-volume energy spectrum of the $K^-K^-K^-$ system}},
  \href{http://dx.doi.org/10.1103/PhysRevD.102.114523}{\emph{Phys. Rev. D} {\bf
  102} (2020) 114523}, [\href{http://arxiv.org/abs/2009.12358}{{\tt
  2009.12358}}].

\bibitem{Brett:2021wyd}
R.~Brett, C.~Culver, M.~Mai, A.~Alexandru, M.~D\"oring and F.~X. Lee,
  \emph{{Three-body interactions from the finite-volume QCD spectrum}},
  \href{http://arxiv.org/abs/2101.06144}{{\tt 2101.06144}}.

\bibitem{Muller:2020wjo}
F.~M\"uller and A.~Rusetsky, \emph{{On the three-particle analog of the
  Lellouch-L\"uscher formula}},
  \href{http://dx.doi.org/10.1007/JHEP03(2021)152}{\emph{JHEP} {\bf 21} (2020)
  152}, [\href{http://arxiv.org/abs/2012.13957}{{\tt 2012.13957}}].

\bibitem{Jackura:2020bsk}
A.~W. Jackura, R.~A. Brice\~no, S.~M. Dawid, M.~H.~E. Islam and C.~McCarty,
  \emph{{Solving relativistic three-body integral equations in the presence of
  bound states}},  \href{http://arxiv.org/abs/2010.09820}{{\tt 2010.09820}}.

\bibitem{Efimov:1970zz}
V.~Efimov, \emph{{Energy levels arising form the resonant two-body forces in a
  three-body system}},
  \href{http://dx.doi.org/10.1016/0370-2693(70)90349-7}{\emph{Phys. Lett. B}
  {\bf 33} (1970) 563--564}.

\bibitem{Yndurain:2002ud}
F.~J. Yndurain, \emph{{Low-energy pion physics}},
  \href{http://arxiv.org/abs/hep-ph/0212282}{{\tt hep-ph/0212282}}.

\bibitem{Horz:2019rrn}
B.~H\"orz and A.~Hanlon, \emph{{Two- and three-pion finite-volume spectra at
  maximal isospin from lattice QCD}},
  \href{http://dx.doi.org/10.1103/PhysRevLett.123.142002}{\emph{Phys. Rev.
  Lett.} {\bf 123} (2019) 142002}, [\href{http://arxiv.org/abs/1905.04277}{{\tt
  1905.04277}}].

\bibitem{Adler:1964um}
S.~L. Adler, \emph{{Consistency conditions on the strong interactions implied
  by a partially conserved axial vector current}},
  \href{http://dx.doi.org/10.1103/PhysRev.137.B1022}{\emph{Phys. Rev.} {\bf
  137} (1965) B1022--B1033}.

\bibitem{Aaij:2019kcg}
{\scshape LHCb} collaboration, R.~Aaij et~al., \emph{{Observation of CP
  Violation in Charm Decays}},
  \href{http://dx.doi.org/10.1103/PhysRevLett.122.211803}{\emph{Phys. Rev.
  Lett.} {\bf 122} (2019) 211803}, [\href{http://arxiv.org/abs/1903.08726}{{\tt
  1903.08726}}].

\end{thebibliography}\endgroup


\setcounter{part}{1}

\part{Scientific Research\label{sec:papers}}\thispagestyle{empty}
\renewcommand{\headrulewidth}{0pt}



\section*{Supplementary material}

# Part II

# Scientific Research



# Implementing the three-particle quantization condition including higher partial waves

**Tyler D. Blanton,[a] Fernando Romero-López[b] and Stephen R. Sharpe[a]**

[a]*Department of Physics, University of Washington,*
*3910 15th Ave. NE, Seattle, WA 98195-1560, U.S.A.*

[b]*Instituto de Física Corpuscular, Universitat de València and CSIC,*
*Edificio Institutos Investigación, Catedrático José Beltrán 2, 46980 Paterna, Spain*

*E-mail:* blanton1@uw.edu, fernando.romero@uv.es, srsharpe@uw.edu

ABSTRACT: We present an implementation of the relativistic three-particle quantization condition including both $s$- and $d$-wave two-particle channels. For this, we develop a systematic expansion of the three-particle K matrix, $\mathcal{K}_{\mathrm{df},3}$, about threshold, which is the generalization of the effective range expansion of the two-particle K matrix, $\mathcal{K}_2$. Relativistic invariance plays an important role in this expansion. We find that $d$-wave two-particle channels enter first at quadratic order. We explain how to implement the resulting multichannel quantization condition, and present several examples of its application. We derive the leading dependence of the threshold three-particle state on the two-particle $d$-wave scattering amplitude, and use this to test our implementation. We show how strong two-particle $d$-wave interactions can lead to significant effects on the finite-volume three-particle spectrum, including the possibility of a generalized three-particle Efimov-like bound state. We also explore the application to the $3\pi^+$ system, which is accessible to lattice QCD simulations, where we study the sensitivity of the spectrum to the components of $\mathcal{K}_{\mathrm{df},3}$. Finally, we investigate the circumstances under which the quantization condition has unphysical solutions.

KEYWORDS: Lattice QCD, Lattice Quantum Field Theory, Scattering Amplitudes

ARXIV EPRINT: 1901.07095

## Contents

# 1 Introduction

There has been considerable recent progress developing the formalism necessary to extract the properties of resonances coupling to three-particle channels from simulations of lattice QCD, with three different approaches being followed [1–7]. For a recent review, see ref. [8]. The outputs of this work are quantization conditions, which relate the finite-volume spectrum with given quantum numbers to the infinite-volume two- and three-particle interactions. This development is timely since simulations now have extensive results for the finite-volume spectrum above the three-particle threshold; see, e.g., refs. [9–11] and the recent review in ref. [12]. Turning the formalism into a practical tool remains, however, a significant challenge. To date, this has been done only for the simplest case, in which all particles are spinless and identical, the total momentum vanishes, the two-particle interaction is purely $s$-wave, and three particles interact only via a momentum-independent contact interaction [4, 6, 13–15].[1] This is the analog in the three-particle system of the initial implementations of the two-particle quantization condition of Lüscher [16, 17], which assumed only $s$-wave interactions and vanishing total momentum.

In the two-particle case, such an approximation makes sense for levels close to the two-particle threshold, since higher partial waves are suppressed by powers of the relative momentum. In the meson sector it begins to fail for energies around 1 GeV. Indeed, recent applications of the two-particle quantization condition use multiple partial waves (see, e.g., refs. [18, 19]). Similar considerations apply for three particles, and we expect that for many resonances of interest one will need to include higher partial waves.

The aim of this paper is to take the first step in this direction by including the first higher partial wave that enters in the case of identical, spinless particles, namely the $d$ wave.[2] In the language of refs. [3, 4, 6], we include dimers (two-particle channels) with both $\ell = 0$ and $\ell = 2$. At the same time, for consistency, we make a corresponding extension of the three-particle interaction beyond its local (pure $s$-wave) form. We will explain how to implement the formalism in this generalized setting, and show examples for which the higher-order terms have a significant impact on the finite-volume spectrum.

Three-particle quantization conditions have been developed with three different approaches. These use, respectively, generic relativistic effective field theory analyzed diagrammatically to all orders in perturbation theory (the RFT approach) [1, 5, 7], non-relativistic effective field theory (NREFT) [3, 4], and unitarity constraints on the two- and three-particle S-matrix elements applied to finite-volume amplitudes (the finite-volume unitarity or FVU approach) [6]. To date, only in the RFT approach has the formalism been worked out explicitly with no limitations on the two-particle partial waves, whereas in the other two approaches the quantization condition has been written down only for $s$-wave dimers.[3] Therefore we adopt the RFT approach in this work. Specifically, we use the formalism of ref. [1], which applies to identical, spinless particles, with a $G$-parity-

---

[1]There is also an induced three-particle interaction due to the exchange of a virtual particle between a pair of two-particle interactions. This is included in all approaches.

[2]The $p$ wave is absent due to Bose symmetry.

[3]It is expected, however, that there is no barrier to extending to higher waves.

like $\mathbb{Z}_2$ symmetry that forbids $2 \leftrightarrow 3$ transitions. Another important feature of this approach is that it can be made relativistic [5], which turns out to simplify the expansion about threshold. Although we use the RFT approach, we expect that many of the technical considerations and general conclusions will apply to all three approaches to the quantization condition.

The formalism of ref. [1] is restricted to two-particle interactions that do not lead to poles in $\mathcal{K}_2$, the two-particle K matrix. If there are such poles, then one should use the generalized, and more complicated, formalism derived in ref. [7]. For simplicity, we consider here only examples in which there are no K-matrix poles.

Since our main goal is to show how the formalism works when including higher waves, our numerical examples are mainly chosen for illustrative purposes and do not represent physical systems. However, there is one case in nature for which our simplified setting applies, namely the $3\pi^+$ system. Thus, in one of our examples, we set the two-particle scattering parameters to those measured experimentally for two charged pions, and illustrate the dependence of the resulting three-pion spectrum on the three-particle scattering parameters. This is similar to the study made in ref. [15] using the FVU approach, except here we include $d$-wave dimers.

All three-particle quantization conditions involve an intermediate three-particle scattering quantity that is not physical, but that can be related, in a second step, to the infinite-volume scattering amplitude by solving integral equations. In the RFT formalism this quantity is called $\mathcal{K}_{\mathrm{df},3}$, and the second step is explained in ref. [2]. We do not discuss the implementation of this second step in the present work. Clearly, it will be important to do so in the future, but the methods required are quite different from those needed for the quantization condition.

This paper develops the ideas already sketched in section 4 of ref. [20]. It is organized as follows. In the next section we recall the quantization condition of ref. [1], and explain how one can consistently expand $\mathcal{K}_{\mathrm{df},3}$ about the three-particle threshold, with $d$-wave interactions entering at quadratic order. In section 3 we describe the implementation of the quantization condition including $d$-wave interactions, focusing on how to make use of the factorization into different irreducible representations (irreps) of the cubic group. Subsequently, in section 4 we show results illustrating the effect of $d$-wave interactions on the three-particle spectrum, including in section 4.3 the case of the $3\pi^+$ system with realistic interactions, which is a target for a potential lattice QCD study. In addition, in section 4.4, we address the issue of characterizing unphysical solutions to the quantization condition. We summarize and close the discussion in section 5.

We also include seven appendices describing technical details. Appendix A is a collection of relevant definitions, whereas appendices B and C provide further details concerning the topics of section 3. Appendix D describes the calculation of the leading contribution of $d$-wave scattering to the threshold expansion. Finally, the remaining appendices relate to the free solutions discussed in section 4.4.3: appendix E motivates the presence of these solutions in excited states, appendix F explains why they are absent in the isotropic approximation of refs. [1, 13], and appendix G explains in an example why removing the free solutions requires higher orders in the threshold expansion of $\mathcal{K}_{\mathrm{df},3}$.

## 2 Threshold expansion of the three-particle quantization condition

As noted above, we consider a theory of identical, scalar particles, with interactions constrained only by the imposition of a $\mathbb{Z}_2$ global symmetry that prevents odd-legged vertices. In such a theory, the spectrum of odd-particle-number states in a cubic box of length $L$, with periodic boundary conditions, is determined by solutions to the quantization condition [1]

$$\det \left[ F_3(E, L)^{-1} + \mathcal{K}_{\mathrm{df},3}(E) \right] = 0 \,. \tag{2.1}$$

This holds up to finite-volume corrections that are exponentially suppressed, i.e., which fall as $\exp(-mL)$ up to powers of $L$, where $m$ is the mass of the particle. In eq. (2.1), $F_3$ and $\mathcal{K}_{\mathrm{df},3}$ are matrices with index space $\{\vec{k}, \ell, m\}$, where $\vec{k} \in (2\pi/L)\mathbb{Z}^3$ is the finite-volume momentum assigned to one of the particles (the "spectator"), while $\ell$ and $m$ specify the angular momentum of the other two (the "dimer").[4] This matrix space will be truncated, as explained in section 3 below, so that the quantization condition (2.1) becomes tractable. The matrix $F_3$ is a complicated object given in eq. (3.1) below; all we need to know for now is that it depends on the two-particle K matrix, $\mathcal{K}_2$. Thus the infinite-volume quantities that enter into the quantization condition are $\mathcal{K}_2$ and the three-particle quasilocal interaction $\mathcal{K}_{\mathrm{df},3}$.[5]

The quantization condition (2.1) is valid only when the CM (center of momentum) energy lies in the range $m < E^* < 5m$, within which the only odd-particle-number states that can go on shell involve three particles (rather than one, five, seven, etc.). Here $E^* = \sqrt{E^2 - \vec{P}^2}$, with $(E, \vec{P})$ the total four-momentum of the state. As in the previous numerical studies [3, 6, 13, 14], we further restrict our considerations to the overall rest frame, with $\vec{P} = 0$, implying $E^* = E$ henceforth. We also recall that eq. (2.1) assumes that there are no poles in $\mathcal{K}_2$ in the kinematic regime of interest. We discuss the constraints that this places on the two-particle scattering parameters in section 3.

The aim of this section is to develop a systematic expansion of $\mathcal{K}_{\mathrm{df},3}$ about the three-particle threshold at $E = 3m$. To that end, we make use of the fact that, unlike the matrix $F_3$, $\mathcal{K}_{\mathrm{df},3}$ is an infinite-volume quantity, and so is defined for arbitrary choices of the three incoming and three outgoing on-shell momenta in the scattering process, and not just for finite-volume momenta. It is also important that it can be chosen to be relativistically invariant, if an appropriate choice of the kinematic function $\widetilde{G}$ entering $F_3$ is made [5] [see eq. (A.3)].

In the remainder of this section, we first recall the threshold expansion of $\mathcal{K}_2$ and its relation to the partial wave decomposition, and then describe the generalization of the threshold expansion to $\mathcal{K}_{\mathrm{df},3}$, extending an analysis given in ref. [13]. Finally, we show how the terms in this expansion are decomposed into the matrix form needed for eq. (2.1).

---

[4]Context determines which meaning of $m$ is intended.

[5]The subscript "df" stands for "divergence-free", indicating that a long distance one-particle exchange contribution that can diverge has been removed. For further details, see ref. [1].

## 2.1 Warm up: expanding $\mathcal{K}_2$ about threshold

To illustrate the method that we employ for $\mathcal{K}_{df,3}$, we first consider the simpler, and well-understood, case of the two-particle K matrix, $\mathcal{K}_2$. Since $\mathcal{K}_2$ is relativistically invariant, it depends only on the standard Mandelstam variables $s_2$, $t_2$ and $u_2 = 4m^2 - s_2 - t_2$. It is convenient to use dimensionless variables that vanish at threshold,

$$\widetilde{\Delta}_2 = \frac{s_2 - 4m^2}{4m^2} = \frac{q_2^{*2}}{m^2}, \;\; \widetilde{t}_2 = \frac{t_2}{4m^2} = -\frac{q_2^{*2}}{2m^2}(1 - c_\theta), \;\; \widetilde{u}_2 = \frac{u_2}{4m^2} = -\frac{q_2^{*2}}{2m^2}(1 + c_\theta), \quad (2.2)$$

where $q_2^*$ is the magnitude of the momentum of each particle in the CM frame, and $c_\theta$ is the cosine of the scattering angle. For physical scattering, $\widetilde{\Delta}_2$, $-\widetilde{t}_2$ and $-\widetilde{u}_2$ are all non-negative, and satisfy

$$\widetilde{\Delta}_2 = -\widetilde{t}_2 - \widetilde{u}_2, \quad (2.3)$$

implying that $-\widetilde{t}_2$ and $-\widetilde{u}_2$ are both bounded by $\widetilde{\Delta}_2$.

Since $\mathcal{K}_2$ is known to be analytic near threshold, we can expand it in powers of $\widetilde{\Delta}_2$, $\widetilde{t}_2$ and $\widetilde{u}_2$. The previous considerations imply that, for generic kinematics (i.e., $\theta \neq 0$ or $\pi$), all three quantities are of the same order. Bose symmetry implies that the expression must be symmetric under $\widetilde{t}_2 \leftrightarrow \widetilde{u}_2$. Thus, through quadratic order we have

$$\mathcal{K}_2 = \widetilde{c}_0 + \widetilde{c}_1 \widetilde{\Delta}_2 + \widetilde{c}_2 \widetilde{\Delta}_2^2 + \widetilde{c}_3 \left( \widetilde{t}_2^2 + \widetilde{u}_2^2 \right) + \mathcal{O}(\widetilde{\Delta}_2^3), \quad (2.4)$$

where the $\widetilde{c}_i$ are constants (which are real since $\mathcal{K}_2$ is real), and we have used the constraint (2.3) to reduce the number of independent terms. We now decompose this result into partial waves, using

$$\mathcal{K}_2 = \sum_{\ell=0}^{\infty} (2\ell + 1) \mathcal{K}_2^{(\ell)}(\widetilde{\Delta}_2) P_\ell(\cos\theta). \quad (2.5)$$

All odd partial waves vanish by Bose symmetry, while eq. (2.4) leads to

$$\mathcal{K}_2^{(0)} = \widetilde{c}_0 + \widetilde{c}_1 \widetilde{\Delta}_2 + \left( \widetilde{c}_2 + \frac{2}{3} \widetilde{c}_3 \right) \widetilde{\Delta}_2^2 + \mathcal{O}(\widetilde{\Delta}_2^3), \quad (2.6)$$

$$\mathcal{K}_2^{(2)} = \frac{1}{15} \widetilde{c}_3 \widetilde{\Delta}_2^2 + \mathcal{O}(\widetilde{\Delta}_2^3). \quad (2.7)$$

The first equation gives the first three terms in the effective range expansion for $\mathcal{K}_2$, while from the second equation we recover the well-known result that $\mathcal{K}_2^{(2)} \propto q_2^{*4}$ near threshold. By extending this analysis, one can show that $\mathcal{K}_2^{(\ell)}$ only enters when we include terms of $\mathcal{O}(\widetilde{\Delta}_2^\ell)$ in the threshold expansion [13].

The threshold expansion has a finite radius of convergence. In particular, we know that $\mathcal{K}_2$ has a left-hand cut at $\widetilde{\Delta}_2 = -1$, so that the radius of convergence cannot be greater than $|\widetilde{\Delta}_2| = 1$. In practice, we truncate the expansion at the order shown in eqs. (2.6) and (2.7) (and set $\mathcal{K}_2^{(\ell)} = 0$ for $\ell \geq 3$), use a cutoff function such that $\widetilde{\Delta}_2 > -1$, and restrict $E < 5m$ implying that $\widetilde{\Delta}_2 < 3$. We are thus assuming that the deviations from the truncated threshold expansion are small over this kinematic range.

– 4 –

## 2.2 Invariants for three-particle scattering

To extend the analysis to the three-particle amplitude $\mathcal{K}_{\mathrm{df},3}$, we begin by listing the generalized Mandelstam variables,

$$s \equiv E^2\,,\quad s_{ij} \equiv (p_i + p_j)^2 = s_{ji}\,,\quad s'_{ij} \equiv (p'_i + p'_j)^2 = s'_{ji}\,,\quad t_{ij} \equiv (p_i - p'_j)^2\,, \qquad (2.8)$$

where $p_i$ ($p'_i$), $i = 1$–$3$, are the incoming (outgoing) momenta. As in the two-particle case, it is convenient to use dimensionless quantities that vanish at threshold,

$$\Delta \equiv \frac{s - 9m^2}{9m^2}\,,\quad \Delta_i \equiv \frac{s_{jk} - 4m^2}{9m^2}\,,\qquad \Delta'_i \equiv \frac{s'_{jk} - 4m^2}{9m^2}\,,\quad \widetilde{t}_{ij} \equiv \frac{t_{ij}}{9m^2}\,, \qquad (2.9)$$

where in the definitions of $\Delta_i$ and $\Delta'_i$, $(i, j, k)$ form a cylic permutation of $(1, 2, 3)$. These sixteen quantities are constrained by the following eight independent relations,

$$\sum_{i=1}^{3} \Delta_i = \sum_{i=1}^{3} \Delta'_i = \Delta \qquad (2.10)$$

$$\sum_{j=1}^{3} \widetilde{t}_{ij} = \Delta_i - \Delta\,,\quad \sum_{j=1}^{3} \widetilde{t}_{ji} = \Delta'_i - \Delta\,. \qquad [i = 1, 2, 3]\,. \qquad (2.11)$$

Thus only eight are independent: the overall CM energy (parametrized here by $\Delta$) and seven "angular" degrees of freedom.[6] This counting is as expected: six on-shell momenta with total incoming and outgoing 4-momentum fixed have $3 \cdot 6 - 4 \cdot 2 = 10$ degrees of freedom, which is reduced to 7 by overall rotation invariance.

For physical scattering, it is straightforward to show that $\Delta_i$, $\Delta'_i$, $-\widetilde{t}_{ij}$ are all non-negative, and the constraint equations then lead to the inequality

$$0 \le \Delta_i, \Delta'_i, -\widetilde{t}_{ij} \le \Delta\,. \qquad (2.12)$$

Thus all the variables $\{\Delta, \Delta_i, \Delta'_i, \widetilde{t}_{ij}\}$ can be treated as being of the same order in an expansion about threshold.

## 2.3 Expanding $\mathcal{K}_{\mathrm{df},3}$ about threshold

By construction, $\mathcal{K}_{\mathrm{df},3}$ is a smooth function for some region around threshold.[7] Thus it can be expanded in a Taylor series in the variables $\{\Delta, \Delta_i, \Delta'_i, \widetilde{t}_{ij}\}$, which are all treated as being of $\mathcal{O}(\Delta)$. Since $\mathcal{K}_{\mathrm{df},3}$ is real, the coefficients in this expansion must also be real. The expansion must also respect the symmetries of $\mathcal{K}_{\mathrm{df},3}$, which is invariant under [5]:[8]

- Interchange of any two incoming particles: $p_i \leftrightarrow p_j \Rightarrow \Delta_i \leftrightarrow \Delta_j$ and $\widetilde{t}_{ik} \leftrightarrow \widetilde{t}_{jk}$

- Interchange of any two outgoing particles: $p'_i \leftrightarrow p'_j \Rightarrow \Delta'_i \leftrightarrow \Delta'_j$ and $\widetilde{t}_{ki} \leftrightarrow \widetilde{t}_{kj}$

- Time reversal: $p_i \leftrightarrow p'_i$ ($\forall i$) $\Rightarrow \Delta_i \leftrightarrow \Delta'_i$ and $\widetilde{t}_{ij} \leftrightarrow \widetilde{t}_{ji}$ ($\forall ij$)

---

[6]We call these variables angular since they span a compact space.

[7]More precisely, what is shown in ref. [1] is that $\mathcal{K}_{\mathrm{df},3}$ has no kinematic singularities at threshold, a result that is checked by the explicit perturbative calculations of refs. [21, 22]. There can be dynamical singularities due to a three-particle resonance, but, generically, this will lie away from threshold.

[8]The first two symmetries hold because we are considering identical bosons. They would not hold in the more general case of nonidentical particles, allowing additional terms to be present in $\mathcal{K}_{\mathrm{df},3}$.

It is then a tedious but straightforward exercise to write down the allowed terms at each order in $\Delta$, and simplify them using the constraints (2.10)–(2.11). Through quadratic order we find

$$m^2 \mathcal{K}_{\mathrm{df},3} = \mathcal{K}^{\mathrm{iso}} + \mathcal{K}_{\mathrm{df},3}^{(2,A)} \Delta_A^{(2)} + \mathcal{K}_{\mathrm{df},3}^{(2,B)} \Delta_B^{(2)} + \mathcal{O}(\Delta^3), \tag{2.13}$$

$$\mathcal{K}^{\mathrm{iso}} = \mathcal{K}_{\mathrm{df},3}^{\mathrm{iso}} + \mathcal{K}_{\mathrm{df},3}^{\mathrm{iso},1} \Delta + \mathcal{K}_{\mathrm{df},3}^{\mathrm{iso},2} \Delta^2 \tag{2.14}$$

$$\Delta_A^{(2)} = \sum_{i=1}^{3} (\Delta_i^2 + \Delta_i'^2) - \Delta^2, \tag{2.15}$$

$$\Delta_B^{(2)} = \sum_{i,j=1}^{3} \widetilde{t}_{ij}^2 - \Delta^2, \tag{2.16}$$

where $\mathcal{K}_{\mathrm{df},3}^{\mathrm{iso}}$, $\mathcal{K}_{\mathrm{df},3}^{\mathrm{iso},1}$, $\mathcal{K}_{\mathrm{df},3}^{\mathrm{iso},2}$, $\mathcal{K}_{\mathrm{df},3}^{(2,A)}$ and $\mathcal{K}_{\mathrm{df},3}^{(2,B)}$ are real, dimensionless constants. We thus see that there is a single term both at leading (zeroth) order and at first order, while there are three independent terms at quadratic order. The particular linear combinations of the quadratic terms that appear in eqs. (2.15) and (2.16) (and in particular the subtraction of $\Delta^2$ in $\Delta_A^{(2)}$ and $\Delta_B^{(2)}$) are chosen based on our numerical experiments described below in order to ensure that their contributions to the finite-volume spectrum are distinct.

As noted in ref. [13], the leading order contribution to $\mathcal{K}_{\mathrm{df},3}$ in eq. (2.13) is independent of momenta $p_i$ and $p_j'$. This shows that the isotropic approximation to $\mathcal{K}_{\mathrm{df},3}$, defined as independence of the seven angular variables, arises naturally in the same way as the $s$-wave approximation to $\mathcal{K}_2$. What we add here is the result that $\mathcal{K}_{\mathrm{df},3}$ remains isotropic at $\mathcal{O}(\Delta)$, having only an overall linear dependence on $s$. Furthermore, at quadratic order, we find only two terms that depend on angular variables ($\Delta_A^{(2)}$ and $\Delta_B^{(2)}$), compared to the seven angular variables that are needed to fully characterize three-particle scattering. Thus, if it is a good approximation to truncate the threshold expansion at $\mathcal{O}(\Delta^2)$, the number of parameters needed to describe $\mathcal{K}_{\mathrm{df},3}$ is smaller than one might naively have expected.

For most of our numerical investigations, we have restricted ourselves to quadratic order in the expansion of $\mathcal{K}_{\mathrm{df},3}$. It is interesting, however, to push the classification to higher order for at least three reasons. First, in order to know how rapidly the number of parameters grows; second, to see which dimer partial waves enter; and, third, to investigate the issue of solutions to the quantization condition with energies given by those of three noninteracting particles (see section 4.4.3). Thus we have classified all terms of cubic order. We find eight independent terms: three that are just $\Delta$ times each of the terms of quadratic order, plus five new angular terms,

$$\Delta_A^{(3)} = \sum_i \left( \Delta_i^3 + \Delta_i'^3 \right), \qquad \Delta_B^{(3)} = \sum_{i,j} \widetilde{t}_{ij}^3 \tag{2.17}$$

$$\Delta_C^{(3)} = \sum_{i,j} \Delta_i \widetilde{t}_{ij} \Delta_j', \qquad \Delta_D^{(3)} = \sum_{i,j} \widetilde{t}_{ij}^2 \left( \Delta_i + \Delta_j' \right) \tag{2.18}$$

$$\Delta_E^{(3)} = \sum_{\sigma \in S_3} \widetilde{t}_{1\sigma(1)} \widetilde{t}_{2\sigma(2)} \widetilde{t}_{3\sigma(3)}, \tag{2.19}$$

where $\sigma \in S_3$ is a permutation of the indices $(1, 2, 3)$. Thus the number of terms is growing fairly rapidly with order.[9]

## 2.4 Decomposing $\mathcal{K}_{\mathrm{df},3}$

In order to use $\mathcal{K}_{\mathrm{df},3}$ in the quantization condition, we need to decompose it into the variables used in its matrix form. This is the analog of the partial wave decomposition of $\mathcal{K}_2$, described in section 2.1 above.

The steps in this decomposition were presented in ref. [1] and we recall them here. The total four-momentum $P^\mu$ is fixed, in our case to $(E, \vec{0})$. One each of the initial and final particles is designated as the spectator, with three-momenta denoted $\vec{k}$ and $\vec{p}$, respectively. Since $\mathcal{K}_{\mathrm{df},3}$ is symmetric separately under initial and final particle interchange, it does not matter which particles are chosen as the spectators, and we take $\vec{k} = \vec{p}_3$ and $\vec{p} = \vec{p}_3'$. The remaining two particles form the (initial and final) dimers. The total momenta of both dimers are fixed, e.g. to $P - p_3$ in the initial state. For each dimer, we can boost to its CM frame, and the only remaining degree of freedom is the direction of one of the particles in the dimer in this frame. We take this particle to be $p_1$ in the initial state, and denote its direction in the dimer CM frame by $\hat{a}^*$. Similarly, the direction of $p_1'$ in the final-state-dimer CM frame is called $\hat{a}'^*$. Using these variables we can write[10]

$$\mathcal{K}_{\mathrm{df},3} = \mathcal{K}_{\mathrm{df},3}(\vec{p}, \hat{a}'^*; \vec{k}, \hat{a}^*).\qquad(2.20)$$

The next step is to set each spectator momentum to one of the allowed finite-volume values, e.g. $\vec{k} = \vec{n}(2\pi/L)$, with $\vec{n}$ a vector of integers. The final step is then to decompose the dependence on $\hat{a}^*$ and $\hat{a}'^*$ into spherical harmonics

$$\mathcal{K}_{\mathrm{df},3}(\vec{p}, \hat{a}'^*; \vec{k}, \hat{a}^*) = 4\pi Y_{\ell'm'}^*(\hat{a}'^*)\mathcal{K}_{\mathrm{df},3;p\ell'm';k\ell m}Y_{\ell m}(\hat{a}^*),\qquad(2.21)$$

where there is an implicit sum over all angular-momentum indices. This defines the entries in the matrix form of $\mathcal{K}_{\mathrm{df},3}$.[11] In practice, we use the real version of spherical harmonics, so the complex conjugation in eq. (2.21) has no impact.

The simplest example of this decomposition is for the isotropic terms in $\mathcal{K}_{\mathrm{df},3}$, namely $\mathcal{K}^{\mathrm{iso}}$ in eq. (2.14). Recalling that $E$, and thus $\Delta$, is fixed, $\mathcal{K}^{\mathrm{iso}}$ is simply a constant. This implies that the matrix form of $\mathcal{K}^{\mathrm{iso}}$ vanishes unless $\ell' = \ell = 0$, and is independent of $\vec{p}, \vec{k}$:

$$\mathcal{K}_{\mathrm{df},3;p\ell'm';k\ell m}^{\mathrm{iso}} = \mathcal{K}^{\mathrm{iso}}\delta_{\ell'0}\delta_{m'0}\delta_{\ell 0}\delta_{m0}.\qquad(2.22)$$

The approximation $\mathcal{K}_{\mathrm{df},3} = \mathcal{K}^{\mathrm{iso}}$ is studied in ref. [13].

---

[9]We do not think that there is any significance to the fact that the number of terms depending on angular variables through cubic order, i.e. $2 + 5 = 7$, equals the number of independent angles in three-particle scattering. The dependence on these angles can be arbitrarily complicated, so there is not a one-to-one correspondence between variables and functions.

[10]As above, the $2 \cdot (3+2) = 10$ momentum components are reduced to seven independent angular variables by rotation invariance.

[11]Note that we follow ref. [1] and drop the vector symbol on the momenta in the matrix indices, in order not to overly clutter the notation.

We next work out the decomposition of $\Delta_A^{(2)}$, eq. (2.15), which is conveniently written as

$$\Delta_A^{(2)} = \left[\Delta_3^2 + \Delta_3'^2 - \Delta^2\right] + \left[\Delta_1^2 + \Delta_2^2\right] + \left[\Delta_1'^2 + \Delta_2'^2\right] . \tag{2.23}$$

The first term depends on $\vec{k}^2$ and $\vec{p}^2$, but not on $\hat{a}^*$ or $\hat{a}'^*$. This can be seen from

$$9m^2\Delta_3 = (P - p_3)^2 - 4m^2 = E^2 - 2E\omega_k - 3m^2 , \tag{2.24}$$

with $\omega_k = \sqrt{\vec{k}^2 + m^2}$, and the corresponding result for $\Delta_3'$. Thus the first term in eq. (2.23) leads to a purely $s$-wave ($\ell' = \ell = 0$) contribution to $\mathcal{K}_{\mathrm{df},3}$, although now with nontrivial dependence on $\vec{k}$ and $\vec{p}$, so this differs from an isotropic contribution.

The second term in eq. (2.23) can be rewritten using

$$\frac{9m^2}{2}\left[\Delta_1^2 + \Delta_2^2\right] = (p_+ \cdot p_3 - 2m^2)^2 + (p_- \cdot p_3)^2 = (E\omega_k - 3m^2)^2 + \frac{4E^2}{E_{2,k}^{*2}}(\vec{a}^* \cdot \vec{k})^2 , \tag{2.25}$$

where $p_\pm = p_1 \pm p_2$, and $E_{2,k}^{*2} = (P - p_3)^2$. To obtain the second form one must explicitly boost to the dimer CM frame, in which $\vec{p}_-$ equals $2\vec{a}^*$, with $a^{*2} = 9m^2\Delta_3/4$. The first term on the right-hand side of eq. (2.25) is independent of $\hat{a}^*$, and thus again contributes only an $s$-wave component. The second term, however, depends quadratically on $\hat{a}^*$, and thus, through the addition theorem for spherical harmonics,[12]

$$(\hat{a} \cdot \hat{k})^2 = \frac{1}{3} + \frac{8\pi}{15}\sum_m Y_{2m}^*(\hat{a})Y_{2m}(\hat{k}) , \tag{2.26}$$

leads to both $s$- and $d$-wave contributions. In other words, both $\mathcal{K}_{\mathrm{df},3;p00;k00}$ and $\mathcal{K}_{\mathrm{df},3;p00;k2m}$ are nonvanishing. These contributions are straightforward to work out from the above equations, and we do not display them explicitly.

The final term in eq. (2.23) differs from the second term only by changing unprimed quantities to their primed correspondents. Thus one finds contributions both to $\mathcal{K}_{\mathrm{df},3;p00;k00}$ and $\mathcal{K}_{\mathrm{df},3;p2m';k00}$. Overall, we conclude that the angular dependence in $\Delta_A^{(2)}$ leads to both $s$- and $d$-wave dimer interactions, although there are no terms with both $\ell = 2$ and $\ell' = 2$. The latter result arises from the fact that there are no terms in $\Delta_A^{(2)}$ that depend on both incoming and outgoing momenta.

Finally, we consider $\Delta_B^{(2)}$, given in eq. (2.16). This is more complicated to decompose because $\widetilde{t}_{ij}$ contains both incoming and outgoing momenta, but this same property leads to contributions with $\ell = \ell' = 2$. We provide only a sketch of the decomposition, as the details are tedious, lengthy, and straightforward to automate. Expanding $\Delta_B^{(2)}$, one finds terms that are similar to those dealt with in $\Delta_A^{(2)}$, which lead to additional contributions to $\mathcal{K}_{\mathrm{df},3;p00;k00}$, $\mathcal{K}_{\mathrm{df},3;p00;k2m}$, and $\mathcal{K}_{\mathrm{df},3;p2m';k00}$, and a term proportional to

$$(p_- \cdot p_-')^2 = a_i^* a_j^* S_{ij,rs} a_r'^* a_s'^* , \tag{2.27}$$

where $p_\pm' = p_1' \pm p_2'$, $i$, $j$, $r$, and $s$ are now spatial vector indices, and $S$ is a tensor that depends on $\vec{k}$ and $\vec{p}$ and is symmetric separately under $i \leftrightarrow j$ and $r \leftrightarrow s$. By decomposing

---

[12]Again, in practice, we use real spherical harmonics, so the complex conjugation is not needed.

$S$ into the spherical tensor basis one finds contributions to the $\ell = \ell' = 2$ part of $\mathcal{K}_{\mathrm{df},3}$, $\mathcal{K}_{\mathrm{df},3;p2m';k2m}$, as well as to the other three components.

In summary, because the terms of $\mathcal{O}(\Delta^2)$ in $\mathcal{K}_{\mathrm{df},3}$ are at most quadratic in $\vec{a}^*$ and/or $\vec{a}'^*$, they give rise to dimer interactions that are either $s$- or $d$-wave. This is the analog of the result derived in section 2.1 that, at the same order, only $\mathcal{K}_2^{(0)}$ and $\mathcal{K}_2^{(2)}$ are present.

The generalization to higher order is straightforward. Terms of $\mathcal{O}(\Delta^3)$, can, in principle be cubic in $\vec{a}^*$ and/or $\vec{a}'^*$, but Bose symmetry forbids odd powers. Thus $\mathcal{O}(\Delta^3)$ terms lead only to $s$- and $d$-wave contributions to $\mathcal{K}_{\mathrm{df},3}$, as we have checked explicitly. In order to obtain contributions with $\ell = 4$ or $\ell' = 4$ one must work at $\mathcal{O}(\Delta^4)$ in the threshold expansion. The pattern continues similarly at higher order.

## 3  Implementing the quantization condition

In this section we describe how we numerically implement the quantization condition, eq. (2.1), when working to quadratic order in the threshold expansion. The expression for $F_3$ is[13]

$$F_3 = \frac{1}{L^3}\left[\frac{\widetilde{F}}{3} - \widetilde{F}H^{-1}\widetilde{F}\right], \tag{3.1}$$

$$H = \frac{1}{2\omega\mathcal{K}_2} + \widetilde{F} + \widetilde{G}, \tag{3.2}$$

where all quantities are matrices with indices $\{k, \ell, m\}$. $\mathcal{K}_2$ is a diagonal matrix

$$\left[\frac{1}{2\omega\mathcal{K}_2}\right]_{p\ell'm';k\ell m} = \delta_{pk}\delta_{\ell'\ell}\delta_{m'm}\frac{1}{2\omega_k\mathcal{K}_{2;k}^{(\ell)}}, \tag{3.3}$$

where the only nonzero elements are the $s$- and $d$-wave terms

$$\frac{1}{\mathcal{K}_{2;k}^{(0)}} = \frac{1}{16\pi E_{2,k}^*}\left\{-\frac{1}{a_0} + r_0\frac{q_{2,k}^{*2}}{2} + P_0(r_0)^3 q_{2,k}^{*4} + |q_{2,k}^*|[1 - H(\vec{k})]\right\}, \tag{3.4}$$

$$\frac{1}{\mathcal{K}_{2;k}^{(2)}} = \frac{1}{16\pi E_{2,k}^*}\frac{1}{q_{2,k}^{*4}}\left\{-\frac{1}{a_2^5} + |q_{2,k}^{*5}|[1 - H(\vec{k})]\right\}. \tag{3.5}$$

Here $E_{2,k}^{*2} = (P - k)^2$ is the invariant mass of the dimer, while $q_k^* = \sqrt{E_{2,k}^{*2}/4 - m^2}$ is the momentum of each particle composing the dimer in its CM frame.[14] The expression (3.4) is the standard form for the effective range expansion through quadratic order, with $a_0$ the $s$-wave scattering length, $r_0$ the effective range, and $P_0$ the shape parameter. Expanding the overall factor of $E_{2,k}^*$ about threshold, and for now ignoring the $1 - H(\vec{k})$ term, one recovers the form given in eq. (2.6). Similarly, aside from the $1 - H$ term, the expression

---

[13]This is the form given in appendix C of ref. [1], with $\widetilde{F} = F/(2\omega)$ and $\widetilde{G} = G/(2\omega)$. The matrix $H$ should not be confused with the cutoff function $H(\vec{k})$, which is always shown with an argument.

[14]These quantities were denoted $s_2$ and $q_2^*$, respectively, in section 2.1, but here we need to make explicit that they depend on $\vec{k}$. The notation here is the same as in ref. [1].

– 9 –

for $\mathcal{K}_{2;k}^{(2)}$, eq. (3.5), is equivalent to the earlier result, eq. (2.7). Here the leading order term is parametrized in terms of the $d$-wave scattering length $a_2$.[15]

The $1 - H$ terms in the expressions (3.4) and (3.5) arise from the need to introduce a smooth cutoff function $H(\vec{k})$ that vanishes for $E_{2,k}^{*2} \leq 0$. We refer the reader to refs. [1, 23] for further explanation of both the need for this cutoff and the manner in which it enters these expressions. It is sufficient to note here that the $1 - H$ term turns on smoothly only well below the dimer threshold at $E_{2,k}^* = 2m$. The explicit form of $H(\vec{k})$ that we use is given in appendix A.

As noted above, the quantization condition holds only if there are no poles in $\mathcal{K}_2$ in the kinematic regime under study. The kinematic range of $q_{2,k}^*$ is given by $-m^2 < q_{2,k}^{*2} < 3m^2$ (corresponding to $0 < E_{2,k}^{*2} < 16m^2$). The parameters in eqs. (3.4) and (3.5) are thus constrained so that neither right-hand side vanishes for this range of $q_{2,k}^{*2}$. In our numerical investigations, we always use values of the scattering parameters that satisfy these constraints. For $a_2$ the constraint is that $ma_2 < 1$, with arbitrarily negative values allowed.

The other two quantities appearing in $F_3$ are the finite-volume kinematic functions $\widetilde{F}$ and $\widetilde{G}$. The former is essentially a two-particle quantity, and thus is diagonal in spectator momenta, though not in the angular-momentum indices:[16]

$$\widetilde{F}_{p\ell'm';k\ell m} \equiv \delta_{pk} H(\vec{k}) \widetilde{F}_{\ell'm';\ell m}(\vec{k}) \,. \tag{3.6}$$

$\widetilde{G}$ is a kinematic function that arises from one-particle exchange between dimers, and is thus a quantity that involves all three particles. In particular, it is not diagonal in any of the indices. We give the explicit forms of $\widetilde{F}$ and $\widetilde{G}$ in appendix A, and provide some details of their numerical evaluation of $\widetilde{F}$ in appendix B.

An important property is that $\widetilde{G}_{p\ell'm';k\ell m}$ is proportional to $H(\vec{p})H(\vec{k})$, and is thus truncated to the finite number of values of spectator momenta for which $H(\vec{k}) \neq 0$. We call this number $N_{\text{spect}}(E, L)$. The same truncation applies to $\widetilde{F}$, due to the factor of $H(\vec{k})$ in eq. (3.6). Both matrices are, however, infinite-dimensional in angular-momentum space. This is to be contrasted to $\mathcal{K}_2$ and $\mathcal{K}_{\text{df},3}$, which are (by approximation) truncated in angular momenta but not in spectator-momentum space. In angular momentum space the dimension is $1 + 5 = 6$ when keeping both $s$ and $d$ waves.

Nevertheless, it turns out that these two truncations are sufficient to reduce the quantization condition, eq. (2.1), to a determinant of a $6N_{\text{spect}}$-dimensional matrix. To show this, we first write the quantization condition as

$$\det\left[F_3^{-1}\right] \det\left[1 + F_3 \mathcal{K}_{\text{df},3}\right] = 0 \,. \tag{3.7}$$

It appears from this rewriting that there will be solutions to the quantization condition when $\det[F_3] \to \infty$, i.e., when $F_3$ has a diverging eigenvalue. However, in that case, the second determinant will, for a general $\mathcal{K}_{\text{df},3}$, also diverge, leading to a finite product. Thus

---

[15]This expansion is often written with a different definition of $a_2$, in which $a_2^5$ is replaced by $a_2$. We prefer the present form since then $a_2$ has dimensions of length.

[16]We are abusing notation here, but the two versions of $\widetilde{F}$ will always be distinguishable by the presence or absence of the argument $\vec{k}$.

we expect that the only solutions of the quantization condition (2.1) for general $\mathcal{K}_{\text{df},3}$ will be those that also satisfy

$$\det[1 + F_3 \mathcal{K}_{\text{df},3}] = 0 \,. \tag{3.8}$$

This also makes sense intuitively, since we expect all finite-volume energies to depend upon the three-particle interaction. The advantage of the form (3.8) is that it has been shown in ref. [1] that it effectively truncates all matrices that appear (i.e., $\widetilde{F}$, $\widetilde{G}$, $\mathcal{K}_2$ and $\mathcal{K}_{\text{df},3}$) to $N_{\text{spect}}$ entries in spectator-momentum space and to $s$ and $d$ waves in angular-momentum space. By "effectively" we mean that elements of the matrices that lie outside the truncated space do not contribute to the determinant.

In the following, we also consider at times the further truncation to only $s$-wave dimer interactions. This is effected by setting to zero all entries in the matrices having $\ell = 2$, so that their dimension becomes $N_{\text{spect}}$.

We have now explained how all the matrices contained in the quantization condition eq. (2.1) are constructed, for given values of $E$ and $L$. We combine these matrices to form $F_3^{-1} + \mathcal{K}_{\text{df},3}$, and calculate its eigenvalues. For a given choice of $L$, the finite-volume spectrum is then given by those values of $E$ for which an eigenvalue vanishes.

The practical calculation of this spectrum is facilitated by decomposing into irreducible representations (irreps) of the symmetry group of finite-volume scattering. For a cubic box with $\vec{P} = 0$, this is the cubic group, $O_h$. For the case of pure $s$-wave dimers, this decomposition has been worked out for the NREFT and FVU quantization conditions in ref. [14]. It has also been used implicitly in the numerical study of the isotropic approximation to the RFT quantization condition in ref. [13], since the isotropic approximation automatically involves a projection onto the trivial ($A_1^+$) irrep.[17] The new result that we now present is the generalization of the decomposition to the case in which one has both $s$- and $d$-wave dimers.

### 3.1 Projecting onto cubic group irreps

We begin by recalling some useful properties of the cubic group, $O_h$. It has dimension $[O_h] = 48$, and ten irreps. Its character table can be found, e.g. in ref. [24]. The labels for, and dimensions of, the irreps can be seen in table 1 below. Each finite-volume momentum, $\vec{k} = (2\pi/L)\vec{n}_k$, lies in a "shell" (also known as an orbit) composed of all momenta related to $\vec{k}$ by cubic group transformations. We refer to this shell as $o_k$. There are seven types of shell, differing by the symmetry properties of the individual elements. We label these by the form of $\vec{n}_k$: $(000)$, $(00a)$, $(aa0)$, $(aaa)$, $(ab0)$, $(aab)$ and $(abc)$, where $a$, $b$ and $c$ are all different, nonzero components. They have dimensions $N_o = 1, 6, 12, 8, 24, 24$ and $48$, respectively. For example, $\vec{n}_k = \hat{x}$ lies in the $(001)$ shell of type $(00a)$, and $\vec{n}_k = \hat{x} + 2\hat{z}$ lies in the $(120)$ shell of type $(ab0)$. Each element in a shell is invariant under rotations in a subgroup of $O_h$, called its little group, $L_k$. The little groups for all elements in a shell are isomorphic, with dimension $[L_k] = [O_h]/N_o$.

The four matrices that enter the quantization condition eq. (2.1), namely $2\omega\mathcal{K}_2$, $\mathcal{K}_{\text{df},3}$, $\widetilde{F}$ and $\widetilde{G}$, are all invariant under a set of orthogonal transformations $U(R)$, where $R \in O_h$.

---

[17]For a more detailed discussion of the isotropic approximation, see appendix F.

Specifically, if $M$ is one of these matrices, then

$$M = U(R)MU(R)^T, \qquad U(R)U(R)^T = \mathbf{1}, \tag{3.9}$$

$$U(R) = S(R) \otimes \mathcal{D}(R)^T, \tag{3.10}$$

$$U(R)_{p\ell'm';k\ell m} = \delta_{o_p o_k} S_{pk}^{(o_p)}(R) \delta_{\ell'\ell} \mathcal{D}_{mm'}^{(\ell)}(R). \tag{3.11}$$

Here the Wigner D-matrix is defined in eq. (A.7), while $S(R)$ permutes the spectator momenta within shells:

$$S(R)_{pk} = \delta_{o_p o_k} S_{pk}^{(o_p)}(R) = \delta_{pRk} \equiv \begin{cases} 1, & R\vec{p} = \vec{k} \\ 0, & \text{otherwise}. \end{cases} \tag{3.12}$$

For $2\omega\mathcal{K}_2$ and $\mathcal{K}_{\mathrm{df},3}$ this result follows because they are invariant under rotations, while for $\widetilde{F}$ and $\widetilde{G}$ it follows from the fact that they are form-invariant under cubic-group rotations if the quantization axis that defines the spherical harmonics is rotated along with the spectator momenta.

The matrices $\{U(R)^T\}_{R \in O_h}$ furnish a representation of $O_h$:

$$U(R_2 R_1)^T = U(R_2)^T U(R_1)^T, \quad \forall R_1, R_2 \in O_h, \text{ and } U(\mathbf{1}_3)^T = \mathbf{1}_{k\ell m}. \tag{3.13}$$

One may decompose this reducible representation into irreps $I$ of the cubic group using projection matrices (see, e.g., ref. [25])

$$P_I = \frac{d_I}{[O_h]} \sum_{R \in O_h} \chi_I(R) U(R)^T, \tag{3.14}$$

where $d_I$ is the dimension of $I$ and $\chi_I(R)$ its character.[18] An important simplifying property of $U(R)$, which carries over to $P_I$, is that it is block-diagonal. For the spectator-momentum indices, this follows because

$$U(R)_{pk}^T = S(R)_{kp} \otimes \mathcal{D}(R) = \delta_{k_R p} \mathcal{D}(R) = \begin{cases} \mathcal{D}(R), & R\vec{k} = \vec{p} \\ 0, & \text{otherwise}, \end{cases} \tag{3.15}$$

which implies that each $U(R)$ is block diagonal in shells, $o$. We label the resulting "shell blocks" of $P_I$ as $P_{I,o}$. These shell blocks inherit from $\mathcal{D}(R)$ the property of being block diagonal in $\ell$, and we label the corresponding sub-blocks as $P_{I,o(\ell)}$, with $\ell = 0$ or 2. The result is that we can write $P_I$ in the form

$$P_I = \mathrm{diag}(P_{I,o_1}, \ P_{I,o_2}, \dots), \qquad P_{I,o} = \mathrm{diag}(P_{I,o(0)}, \ P_{I,o(2)}). \tag{3.16}$$

This simplified structure allows for more efficient computation of the $P_I$ matrices, as explained in appendix C.1.

---

[18]Normally one would write $\chi_I(R)^*$ in eq. (3.14), but since $O_h$ only involves real orthogonal transformations, all characters are real and the conjugation is trivial.

Using these projectors, we can decompose the quantization condition into separate conditions for each irrep. From eq. (3.9) we know that $[P_I, M] = 0$, for each of the four matrices $M$, from which it follows that

$$[P_I, F_3^{-1} + \mathcal{K}_{df,3}] = 0 \qquad (\forall I) \,. \tag{3.17}$$

Using $\sum_I P_I = \mathbf{1}$, and the orthogonality of the projectors onto different irreps, one can then show that the determinant factorizes into that for each irrep

$$\det[F_3^{-1} + \mathcal{K}_{df,3}] = \prod_I \det_{\mathrm{sub},I}[P_I(F_3^{-1} + \mathcal{K}_{df,3})P_I] \,, \tag{3.18}$$

where the subscript indicates that the determinant is taken only over the subspace onto which $P_I$ projects. Thus the quantization condition for irrep $I$ becomes

$$\det_{\mathrm{sub},I}[P_I(F_3^{-1} + \mathcal{K}_{df,3})P_I] = 0 \,. \tag{3.19}$$

If desired, one can also apply the projectors to all the matrices contained in $F_3$, eq. (3.1), so that the entire evaluation of the quantization condition involves matrices of reduced dimensionality.

The number of eigenvalues in a given irrep is given by the dimension of the projected subspaces, $d(P_I)$. This is obtained by summing the dimensions of the sub-blocks,

$$d(P_I) = \sum_o \sum_{\ell=0,2} d(P_{I,o(\ell)}) \,, \tag{3.20}$$

where the sum over $o$ runs over all shells that are "active", i.e., that lie below the cutoff. We explain how the $d(P_{I,o(\ell)})$ are calculated in appendix C.2, and list the results in table 1. From this we learn, for example, that the $\vec{k} = \vec{0}$ shell contains one $A_1^+$ irrep for $\ell = 0$, and one each of the $E^+$ and $T_2^+$ irreps for $\ell = 2$. Note that shells can contain multiple versions of a given irrep, e.g., the $(00a)$ shell-type with $\ell = 2$ contains two versions each of the $E^+$, $T_2^+$, $T_1^-$ and $T_2^-$ irreps.

At this stage it is useful to give an example of how shells become active as $E$ and $L$ are increased. With our cutoff, described in appendix A, the maximum value of $|\vec{n}_k|$, $n_{k,\max}$, is determined by the vanishing of $E_{2,k}^{*2}$:

$$E_{2,k}^{*2} = 0 \quad \Rightarrow \quad n_{k,\max} = \frac{L}{2\pi}\left(\frac{E^2 - m^2}{2E}\right) \,. \tag{3.21}$$

This can be easily converted into the number of active shells, an example being shown in figure 1. The first fifteen shells are $(000)$, $(001)$, $(110)$, $(111)$, $(002)$, $(120)$, $(112)$, $(220)$, $(221)$, $(003)$, $(130)$, $(113)$, $(222)$, $(230)$ and $(123)$, at which point examples of all seven types have appeared.

Although each $P_I$ is block diagonal in $o$ and $\ell$, $F_3^{-1} + \mathcal{K}_{df,3}$ is generally not. Thus even though each eigenvector of $F_3^{-1} + \mathcal{K}_{df,3}$ lies in a single irrep, it will generally be a nontrivial linear combination of vectors lying in the subspaces projected onto by $P_{I,o(\ell)}$. However, we can still use table 1 to determine how many eigenvalues will be present in a given irrep for

| irrep | dim | shell types | | | | | | |
|-------|-----|-------|-------|-------|-------|-------|-------|-------|
|       |     | $(000)$ | $(00a)$ | $(aa0)$ | $(aaa)$ | $(ab0)$ | $(aab)$ | $(abc)$ |
| $A_1^+$ | 1 | $(1,0)$ | $(1,1)$ | $(1,2)$ | $(1,1)$ | $(1,3)$ | $(1,3)$ | $(1,5)$ |
| $A_2^+$ | 1 | $(0,0)$ | $(0,1)$ | $(0,1)$ | $(0,0)$ | $(1,3)$ | $(0,2)$ | $(1,5)$ |
| $E^+$ | 2 | $(0,2)$ | $(2,4)$ | $(2,6)$ | $(0,4)$ | $(4,12)$ | $(2,10)$ | $(4,20)$ |
| $T_1^+$ | 3 | $(0,0)$ | $(0,3)$ | $(0,9)$ | $(0,6)$ | $(3,21)$ | $(3,21)$ | $(9,45)$ |
| $T_2^+$ | 3 | $(0,3)$ | $(0,6)$ | $(3,12)$ | $(3,9)$ | $(3,21)$ | $(6,24)$ | $(9,45)$ |
| $A_1^-$ | 1 | $(0,0)$ | $(0,0)$ | $(0,1)$ | $(0,0)$ | $(0,2)$ | $(0,2)$ | $(1,5)$ |
| $A_2^-$ | 1 | $(0,0)$ | $(0,1)$ | $(0,1)$ | $(1,1)$ | $(0,2)$ | $(1,3)$ | $(1,5)$ |
| $E^-$ | 2 | $(0,0)$ | $(0,2)$ | $(0,4)$ | $(0,4)$ | $(0,8)$ | $(2,10)$ | $(4,20)$ |
| $T_1^-$ | 3 | $(0,0)$ | $(3,6)$ | $(3,12)$ | $(3,9)$ | $(6,24)$ | $(6,24)$ | $(9,45)$ |
| $T_2^-$ | 3 | $(0,0)$ | $(0,6)$ | $(3,12)$ | $(0,6)$ | $(6,24)$ | $(3,21)$ | $(9,45)$ |

**Table 1.** Dimension of irrep projection sub-blocks for each shell-type and angular momentum, $(d(P_{I,o(0)}), d(P_{I,o(2)}))$. Each row corresponds to an irrep of the cubic group $O_h$, whose dimension is also listed for completeness.

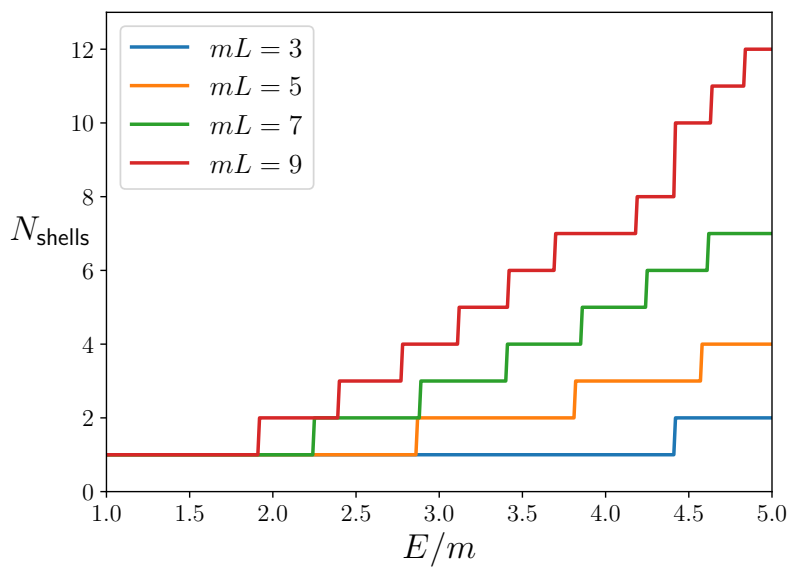

**Figure 1.** Number of active momentum shells for fixed $mL$ as a function of $E$.

a given choice of $E$ and $L$. For example, suppose we have both $s$- and $d$-wave interactions turned on and we are in the $E, L$ regime where only the first two momentum shells, $(000)$ and $(001)$, are active, so that $N_{\text{spect}} = 1 + 6 = 7$. Then the table tells us that $F_3^{-1} + \mathcal{K}_{\text{df},3}$ has 3 eigenvalues in $A_1^+$ since

$$
\begin{aligned}
d(P_{A_1^+}) &= d(P_{A_1^+,000(0)}) + d(P_{A_1^+,000(2)}) + d(P_{A_1^+,001(0)}) + d(P_{A_1^+,001(2)}) \\
&= 1 + 0 + 1 + 1 = 3 .
\end{aligned}
\tag{3.22}
$$

| level | $(\vec{n}_1^2, \vec{n}_2^2, \vec{n}_3^3)$ | degen. | irreps |
|-------|------------|--------|--------|
| 0 | $(0,0,0)$ | 1 | $A_1^+$ |
| 1 | $(1,1,0)$ | 3 | $A_1^+ + E^+$ |
| 2 | $(2,2,0)$ | 6 | $A_1^+ + E^+ + T_2^+$ |
| 3 | $(2,1,1)$ | 12 | $A_1^+ + E^+ + T_2^+ + T_1^- + T_2^-$ |
| 4 | $(3,3,0)$ | 4 | $A_1^+ + T_2^+$ |

**Table 2.** Irreps appearing in the lowest energy levels of three identical noninteracting particles. The first column gives the level number (for values of $mL \sim 5$), starting at zero. The states are labeled by the squares of the three vectors $\vec{n}_i$ that determine the momenta of the particles — see eq. (3.23) — and these are given in the second column. The third column gives the degeneracy, and the final column the irreps that appear.

Looking at the other irreps, we see that in this regime there is 1 eigenvalue in $A_2^+$, 8 in $E^+$, 3 in $T_1^+$, 9 in $T_2^+$, 0 in $A_1^-$, 1 in $A_2^-$, 2 in $E^-$, 9 in $T_1^-$, and 6 in $T_2^-$ giving the correct total of $6N_{\text{spect}} = 42$ eigenvalues. We stress that eigenvalues lying in a given irrep always come in degenerate multiplets corresponding to the dimension of the irrep. Thus, for example, the eight eigenvalues in the $E^+$ irrep in the two-shell regime consist of four two-fold-degenerate pairs.

A point that may lead to confusion when we present results in the following section is that the number of eigenvalues of $F_3^{-1} + \mathcal{K}_{\text{df},3}$ bears no direct relation to the number of solutions to the quantization condition. For there to be a solution, an eigenvalue must vanish, and this occurs only for a subset of the eigenvalues in the energy range of interest. This point can be seen explicitly if the interactions $\mathcal{K}_2$ and $\mathcal{K}_{\text{df},3}$ are weak, for then we expect the number of states to be the same as for noninteracting particles. We quote in table 2 the irreps that appear in the first few three-particle levels for noninteracting particles. These states have energies

$$E^{\text{free}}(\vec{n}_1, \vec{n}_2) = \sum_{i=1}^{3} \sqrt{m^2 + (2\pi/L)^2 \vec{n}_i^2}, \qquad \vec{n}_3 = -\vec{n}_1 - \vec{n}_2, \qquad (3.23)$$

where $\vec{n}_i$ are integer vectors. As an example of the difference between the dimensions of $F_3^{-1} + \mathcal{K}_{\text{df},3}$ and the number of solutions, we consider $mL = 5$ and the $A_1^+$ irrep, and focus on the energy range $E/m = 3$–$5$. From figure 1 we see that the number of active momentum shells begins at 2 for $E = 3m$, increases to 3 at some point, and then reaches 4 below $E = 5m$. From table 1 we deduce that the corresponding number of eigenvectors in the $A_1^+$ irrep are 3, 6 and 8. By contrast, the free levels in this irrep occur at $E = 3m$, $E = 4.21m$, $E = 5.08m$, .... For weak interactions, we expect solutions to the quantization condition only near these three values, and thus we find that, in all cases, the number of eigenvalues of $F_3^{-1} + \mathcal{K}_{\text{df},3}$ significantly exceeds the number of solutions at, or below, the given energy.

We close this section by noting that the components of $\mathcal{K}_{\text{df},3}$, given in eq. (2.13), can themselves be decomposed into different irreps. While it is clear that $\mathcal{K}_{\text{df},3}^{\text{iso}}$, eq. (2.14),

lies purely in the $A_1^+$ irrep, we also find that the same is true for the $\mathcal{K}_{\mathrm{df},3}^{(2,A)}$ term. The $\mathcal{K}_{\mathrm{df},3}^{(2,B)}$ term, however, has components that lie in the $A_1^+$, $E^+$, $T_2^+$ and $T_1^-$ irreps. For components lying in the remaining irreps one must go to cubic or higher order in the threshold expansion.

## 4 Results

The goal of this section is to illustrate the impact of including $d$-wave interactions in the quantization condition. In particular, we aim to determine which energy levels and which irreps are particularly sensitive to such interactions. We begin, however, with a case where the impact of $d$-wave interactions is small, namely the ground state energy with a weak two-particle interaction. This allows us to test of our implementation of the quantization condition in a regime where we can make an analytic prediction. We then consider the impact of a strong $d$-wave interaction, $m|a_2| \sim 1$, comparing its effect on the ground and excited states, and for different irreps. Next we study the sensitivity of the finite-volume spectrum of the physical $3\pi^+$ state, with $\mathcal{K}_2$ taken from experiment, to the various terms in $\mathcal{K}_{\mathrm{df},3}$. And, finally, we discuss the different types of unphysical solutions to the quantization condition that appear.

### 4.1 Threshold expansion including $a_2$

In this section we consider the energy of the lightest two- and three-particle states in the case of weak two-particle interactions, and with the three-particle interaction $\mathcal{K}_{\mathrm{df},3}$ set to zero. The energy of these states (called $E_2^{(0)}$ and $E_3^{(0)}$, respectively) lie close to their noninteracting values, and we define the differences as

$$\Delta E_n = E_n^{(0)} - n \times m\,. \tag{4.1}$$

These can be expanded in powers of $1/L$ (up to logarithms), the results being called the threshold expansions. These expansions have been worked out in a relativistic theory to $\mathcal{O}(L^{-6})$ in refs. [16, 21, 23]:[19]

$$\Delta E_2 = \frac{4\pi a_0}{mL^3}\left\{1 + c_1\left(\frac{a_0}{\pi L}\right) + c_2\left(\frac{a_0}{\pi L}\right)^2 + c_3\left(\frac{a_0}{\pi L}\right)^3 + \frac{2\pi r_0(a_0)^2}{L^3} - \frac{\pi a_0}{m^2 L^3}\right\} + \mathcal{O}(L^{-7})\,, \tag{4.2}$$

$$\begin{aligned} \Delta E_3 = \frac{12\pi a_0}{mL^3}\Bigg\{ &1 + d_1\left(\frac{a_0}{\pi L}\right) + d_2\left(\frac{a_0}{\pi L}\right)^2 + \frac{64\pi^2(a_0)^2\mathcal{C}_3}{mL^3} + \frac{3\pi a_0}{m^2 L^3} + \frac{6\pi r_0(a_0)^2}{L^3} \\ &+ \left(\frac{a_0}{\pi L}\right)^3\left(d_3 + c_L \log\frac{mL}{2\pi}\right)\Bigg\} - \frac{\mathcal{M}_{3,\mathrm{thr}}}{48m^3 L^6} + \mathcal{O}(L^{-7})\,. \end{aligned} \tag{4.3}$$

Here $c_L$, $\mathcal{C}_3$, and the $c_i$ and $d_i$, are numerical constant available in the aforementioned references, and $\mathcal{M}_{3,\mathrm{thr}}$ is a subtracted three-particle threshold scattering amplitude, whose definition will be discussed in appendix D.

What we observe from these results is that they depend, through $\mathcal{O}(L^{-5})$, only on the $s$-wave scattering length, $a_0$, with the effective range $r_0$ first entering at $\mathcal{O}(L^{-6})$.

---

[19]The terms up to $\mathcal{O}(L^{-5})$ agree with those obtained previously using nonrelativistic QM [26, 27].

There is no explicit dependence on the $d$-wave scattering amplitude at this order. We can understand this pattern qualitatively as follows.[20] The typical relative momentum, $q$, satisfies $\Delta E \sim q^2/m$, and thus, since $\Delta E \sim a_0/L^3$, we learn that $q^2 \sim a_0/L^3$. Using the effective range expansion, eq. (3.4), we then expect that the relative contribution from the $r_0$ term will be $r_0 a_0 q^2 \sim r_0 a_0^2/L^3$, and this is indeed what is seen in eqs. (4.2) and (4.3). By the same argument, we expect the $q^4$ terms proportional to both $P_0$ and $a_2^5$ to appear first at relative order $\mathcal{O}(L^{-6})$, and thus contribute to $\Delta E_n$ at $\mathcal{O}(L^{-9})$. If this were the case, it would be very challenging to see the dependence of the threshold energies on $a_2$.

However, it turns out that there is an additional contribution of $\mathcal{O}(L^{-6})$ to $\Delta E_3$ that depends on $a_2$, and indeed on all higher partial waves, hidden in $\mathcal{M}_{3,\mathrm{thr}}$. In appendix D we calculate the leading dependence on $a_2$ in a perturbative expansion in the scattering amplitudes, finding

$$m^2 \mathcal{M}_{3,\mathrm{thr}} \supset d_{\mathrm{thr}}(ma_0)^2(ma_2)^5 \left[1 + \mathcal{O}(a_0) + \mathcal{O}(a_2^5)\right], \quad d_{\mathrm{thr}} = -14110. \tag{4.4}$$

The appearance of $a_2^5$, rather than $a_2$, follows from our parametrization of the $d$-wave K matrix, eq. (3.5). In order to isolate the $a_2$ dependence of $\Delta E_3$, we consider the difference

$$\delta E^d(L, a_0, a_2) = \Delta E_3(L, a_0, a_2) - \Delta E_3(L, a_0, a_2 = 0). \tag{4.5}$$

Substituting eq. (4.4) into the expression for $\Delta E_3$, eq. (4.3), we obtain the theoretical prediction

$$\frac{\delta E^d}{m} = -\frac{d_{\mathrm{thr}}}{48} \frac{(ma_0)^2 (ma_2)^5}{(mL)^6} \left[1 + \mathcal{O}(a_0) + \mathcal{O}(a_2^5)\right] + \mathcal{O}\left(L^{-7}\right). \tag{4.6}$$

We have checked that the results from numerically solving the quantization condition are consistent with eq. (4.6). In particular, we have verified that the leading dependence on $a_0$, $a_2$ and $1/L$ is as predicted. An example of the comparison, showing the $L$ dependence, is given in figure 2. Agreement at the 10% level holds over many orders of magnitude. Based on our tests, we find that the major source of this small discrepancy arises from terms of higher order in $a_0$.

This comparison provides a strong cross-check of our numerical implementation. However, for weakly interacting system, such as mesons in QCD, one cannot achieve, using lattice calculations, results for the spectrum with the precision shown in the figure, nor can one work at such large values of $mL$. We now turn to situations in which $a_2$ has a numerically more significant effect.

## 4.2 Effects of $a_2$ on the three-particle spectrum

We begin by studying the strongly interacting regime, where $m|a_2| \sim 1$. This regime, although hardly conceivable in particle physics, represents an interesting academic problem that is relevant in the physics of cold atoms [29, 30].

In figure 3, we show the three particle spectrum for $E < 4m$ in two irreps, $A_1^+$ and $E^+$, as a function of negative $ma_2$. Here we have fixed the volume to $mL = 8.1$, and chosen

---

[20]See also appendix C in ref. [28].

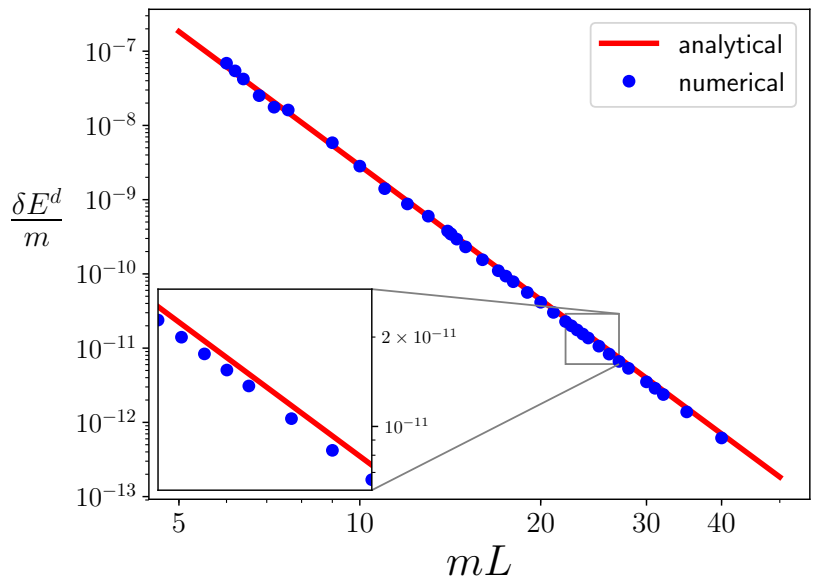

**Figure 2.** Comparison of the analytical prediction (which is absolutely normalized) with the results from a numerical solution of the quantization condition. The parameters are $ma_0 = 0.1$, $ma_2 = 0.25$, and $r_0 = P_0 = \mathcal{K}_{\mathrm{df},3} = 0$. The lack of linearity for smaller values of $mL$ is related to the opening up of new momentum shells.

a weakly attractive $s$-wave interaction, $ma_0 = -0.1$, with other scattering parameters set to zero. We choose negative values for $ma_2$ in order to avoid the possibility of a pole in $\mathcal{K}_2^{(2)}$, eq. (3.5), for which our formalism breaks down. Note that negative $a_2$ corresponds, at least for small magnitudes, to an attractive interaction, as seen from the result for $\delta E^d$, eq. (4.6). Since we use a small value of $m|a_0|$, the energy levels at the right-hand edges of both plots (where $a_2 = 0$) lie close to the energies of three noninteracting particles (which are $E/m = 3$, $3.53$, $3.97$, $4.02, \dots$ for $mL = 8.1$). As $m|a_2|$ increases, the energies are almost flat, until at a value $m|a_2| \sim 1$, the levels shift rapidly downwards. This shift begins at smaller values of $m|a_2|$ for excited states. As the magnitude of $a_2$ increases, the excited states approach lower-lying states until an avoided level crossing occurs. We also observe that states in the $E^+$ irrep are more sensitive to $d$-wave interactions, which seems to be a general feature, as will be seen in the following section.

Another interesting observation from figure 3 is the presence of a deep subthreshold state for $m|a_2| > 1$. This resembles the Efimov effect, which describes a three-particle bound state arising from an attractive two-particle interaction $m|a_0| \gg 1$ [31]. The Efimov bound state has been reproduced numerically with only $s$-wave interactions present, both in the NREFT approach [4, 14] and in the isotropic approximation of the RFT formalism [13]. Moreover, there is some theoretical work regarding the existence of this generalized Efimov scenario in the presence of $d$-wave interactions [30], although there is no result concerning the asymptotic volume dependence, unlike in the $s$-wave case [32]. We have been able to solve the quantization condition numerically up to $mL = 37.5$ and the bound state energy barely changes, which strongly suggests that it is indeed an infinite volume bound state.

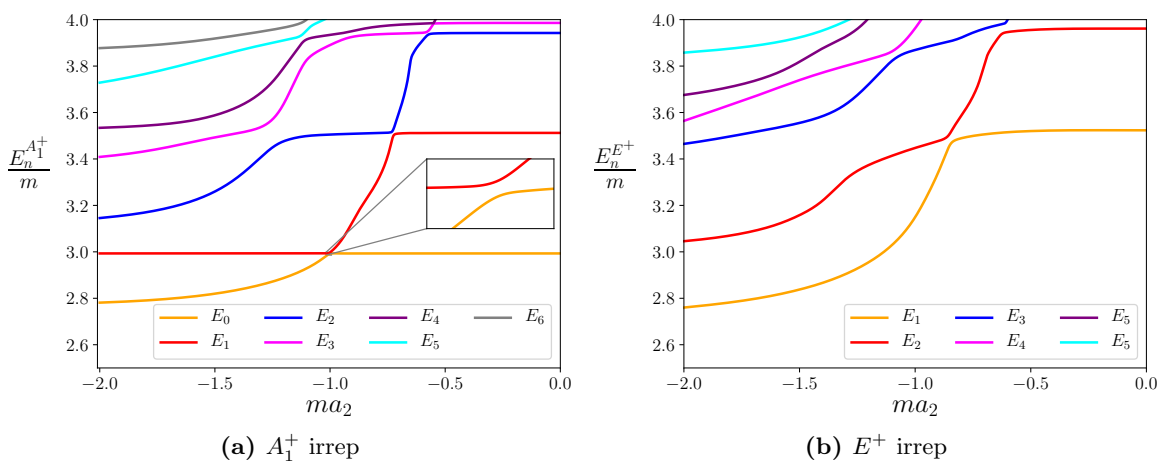

**(a)** $A_1^+$ irrep

**(b)** $E^+$ irrep

**Figure 3.** Energy levels as a function of $ma_2$ in the region $E < 4m$ with $mL = 8.1$ and $ma_0 = -0.1$, $r_0 = P_0 = \mathcal{K}_{\mathrm{df},3} = 0$ in the $A_1^+$ irrep (left) and the $E^+$ irrep (right).

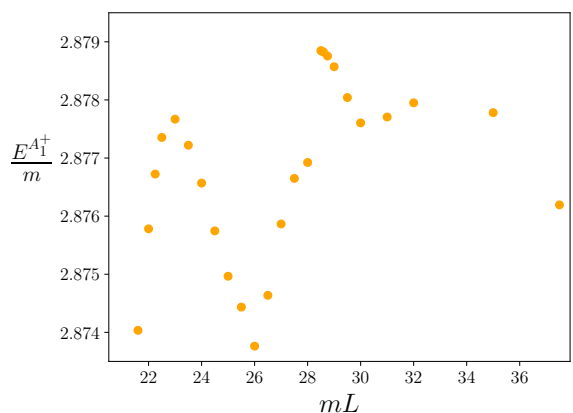

**Figure 4.** Energy of the subthreshold state in the $A_1^+$ irrep as a function of $mL$. The parameters are $ma_0 = -0.1$, $ma_2 = -1.3$ and $r_0 = P_0 = \mathcal{K}_{\mathrm{df},3} = 0$. Note the highly compressed vertical scale.

Results for $ma_2 = -1.3$ are shown in figure 4. The volume dependence of the energy is dominated by effects of the UV cut-off, which manifest themselves as small oscillations when a new shells become active. These are similar to oscillations observed in several quantities in ref. [13].

We close by commenting on the impact of using a relativistic formalism, as opposed to a NR approach, on the results of this section. We expect that the qualitative features of the results will be unchanged, but that the quantitative energy levels will be changed once they differ significantly from $3m$. Thus, for example, we expect that the energy of the subthreshold state will be only slightly changed, since it lies at the border of the NR regime.

### 4.3 Application: spectrum of $3\pi^+$ on the lattice

The simplest application in QCD for the three-particle quantization condition is the $3\pi^+$ system, not only from the theoretical point of view — no resonant subchannels — but also from the technical side — no quark-disconnected diagrams and a good signal/noise

ratio. Here we use our formalism to predict the $3\pi^+$ spectrum, using values for the two-body scattering parameters determined from experiment, and a range of choices for the parameters in $\mathcal{K}_{\mathrm{df},3}$.[21] Our focus will be on how to differentiate effects arising from the different components of $\mathcal{K}_{\mathrm{df},3}$, listed in eq. (2.13).

An important point in the following is that there is no natural size for the parameters in $\mathcal{K}_{\mathrm{df},3}$: the magnitudes of the dimensionless coefficients $\mathcal{K}_{\mathrm{df},3}^{\mathrm{iso}}$, $\mathcal{K}_{\mathrm{df},3}^{\mathrm{iso},1}$, $\mathcal{K}_{\mathrm{df},3}^{\mathrm{iso},2}$, $\mathcal{K}_{\mathrm{df},3}^{(2,A)}$, and $\mathcal{K}_{\mathrm{df},3}^{(2,B)}$ are not constrained. Strictly speaking, we know this only for $\mathcal{K}_{\mathrm{df},3}^{\mathrm{iso}}$, because, in the nonrelativistic limit, it is related to the three-particle contact interaction in NREFT (a relation given explicitly in ref. [8]), and it is well known that the latter interaction varies in a log-periodic manner from $-\infty$ to $\infty$ as the cutoff varies [33]. But we see no reason why this should not also apply to the other coefficients. In particular, we note that the physical three-particle scattering amplitude, $\mathcal{M}_3$, does not diverge when $\mathcal{K}_{\mathrm{df},3}$ does [2, 13].

We take the parameters describing isospin-2 $\pi\pi$ scattering from ref. [34]:

$$m_\pi a_0 = 0.0422, \quad m_\pi r_0 = 56.21, \quad P_0 = -3.08 \cdot 10^{-4}, \quad m_\pi a_2 = -0.1867. \quad (4.7)$$

In a lattice simulation, these parameters would be extracted from the two-pion spectrum, using the two-particle quantization condition. Indeed, there is considerable recent work on the $2\pi^+$ system using lattice methods, in some cases incorporating $d$-wave interactions [10, 35–39]. We emphasize that one must determine these parameters with high precision in order to disentangle the two- and three-body effects in the three-particle spectrum.

For the relatively weak two-particle interactions of eq. (4.7), the energy levels lie close to the noninteracting energies of eq. (3.23). For the regime of box sizes available in current lattice simulations, $4 \lesssim m_\pi L \lesssim 6$, there are at most three such levels below the five-particle threshold, $E = 5m_\pi$ (above which the quantization condition breaks down). For these levels, the solutions lie in three irreps: $\Gamma = A_1^+, E^+, T_2^+$ (see table 2). We denote the difference between the actual energy and its noninteracting value as

$$\Delta E_n^\Gamma = E_n^\Gamma - E_n^{\mathrm{free}} \quad (4.8)$$

where $n = 0, 1, \ldots$ labels the levels following the numbering scheme of table 2. It is known that, asymptotically, [40]

$$\Delta E_n^\Gamma \propto \frac{a_0}{mL^3} + \mathcal{O}(L^{-4}). \quad (4.9)$$

We stress, however, that the asymptotic result is not numerically accurate for the range of $mL$ that we consider.

Let us start from the ground state, which lies in the $A_1^+$ irrep. Here our expectations are guided by the threshold expansion, eq. (4.3). In addition to explicit dependence on $a_0$ and $r_0$, and the implicit dependence on $a_2$ worked out in section 4.1, the energy depends on $\mathcal{K}_{\mathrm{df},3}$ through the $\mathcal{M}_{3,\mathrm{thr}}/L^6$ term. Following the arguments given in section 4.1, we expect that only $\mathcal{K}_{\mathrm{df},3}^{\mathrm{iso}}$ will enter at this order, with dependence on $\mathcal{K}_{\mathrm{df},3}^{\mathrm{iso},1}$ suppressed by $1/L^3$ and that on $\mathcal{K}_{\mathrm{df},3}^{\mathrm{iso},2}$, $\mathcal{K}_{\mathrm{df},3}^{(2,A)}$ and $\mathcal{K}_{\mathrm{df},3}^{(2,B)}$ by $1/L^6$. This is borne out by our numerical results, shown

---

[21]We ignore QED effects, which are numerically small, and, in any case, cannot be incorporated into the present formalism.

– 20 –

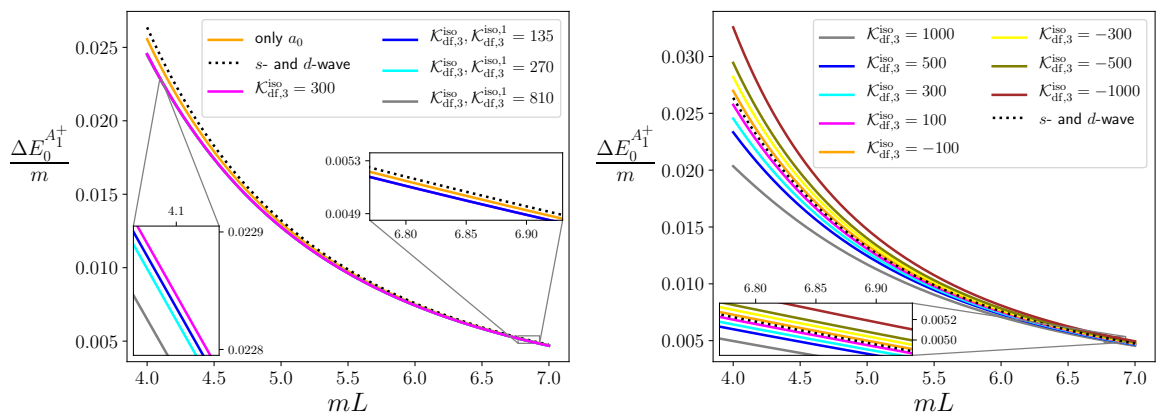

**Figure 5.** Energy shift for the ground state in the $A_1^+$ irrep, for which $E_0^{\text{free}} = 3m$. The two-particle scattering parameters are those in eq. (4.7), aside from the orange curve in the left panel, where only $a_0$ is nonzero. The three particle scattering parameters are as indicated in the legend, and explained further in the text. We use the convention that a parameter value not given explicitly is set to the value given earlier. For example, the blue line in the left panel has the parameters set to $\mathcal{K}_{\text{df},3}^{\text{iso}} = 300$ and $\mathcal{K}_{\text{df},3}^{\text{iso},1} = 135$, while $\mathcal{K}_{\text{df},3}^{\text{iso},2} = \mathcal{K}_{\text{df},3}^{(2,A)} = \mathcal{K}_{\text{df},3}^{(2,B)} = 0$.

in figure 5. The left panel compares results with several choices of parameters: (i) those of eq. (4.7) plus $\mathcal{K}_{\text{df},3} = 0$ (labeled "$s$- and $d$-wave" — black, dotted line); (ii) the same as (i) but with $\mathcal{K}_{\text{df},3}^{\text{iso}} = 300$ and all other parameters in $\mathcal{K}_{\text{df},3}$ vanishing (magenta); (iii) the same as (ii) but with $\mathcal{K}_{\text{df},3}^{\text{iso},1}$ also turned on, taking the three values 135 (blue), 270 (cyan) and 810 (grey); and (iv) the isotropic approximation, i.e., with only $s$-wave interactions, and $a_0$ the only nonzero scattering parameter (orange). We see that adding $d$-wave two-particle interactions has a similar impact to adding $\mathcal{K}_{\text{df},3}^{\text{iso}} = 300$, but that adding $\mathcal{K}_{\text{df},3}^{\text{iso},1}$ with a similar magnitude has almost no impact.

The right panel shows the dependence on $\mathcal{K}_{\text{df},3}^{\text{iso}}$, with other parameters fixed at the values in eq. (4.7). The range we consider is $\mathcal{K}_{\text{df},3}^{\text{iso}} = [-1000, +1000]$. In order to have sensitivity to $\mathcal{K}_{\text{df},3}^{\text{iso}}$ in this range, a determination of $\Delta E_0/m$ with an error of $\approx 0.01$ is needed. Such an error can be achieved with present methods. Thus, as noted in ref. [13], if one has a sufficiently accurate knowledge of the two-particle scattering parameters, one can use the ground state energy to determine the leading three-particle parameter $\mathcal{K}_{\text{df},3}^{\text{iso}}$. Indeed, this approach has been carried out successfully in refs. [11, 41].

In figure 6, we investigate the sensitivity of the energy of the first excited state to the various two-particle scattering parameters, comparing the two irreps that are present. The magnitude of the energy shifts are comparable to those for the ground state, but the dependence on the scattering parameters differs markedly. This can be understood because the relative momenta between the particles is nonvanishing for the excited state. Denoting generically the relative momenta by $q$, this satisfies $q/m \approx 2\pi/(mL) \sim \mathcal{O}(1)$. Because of this we expect that the higher-order terms in the effective range expansion, i.e. $r_0$ and $P_0$, should play a much more significant role. This is borne out by the results in the figure, particularly for the $A_1^+$ irrep. We observe that the effect of these additional terms is opposite in the two irreps, which is consistent with the prediction of the threshold

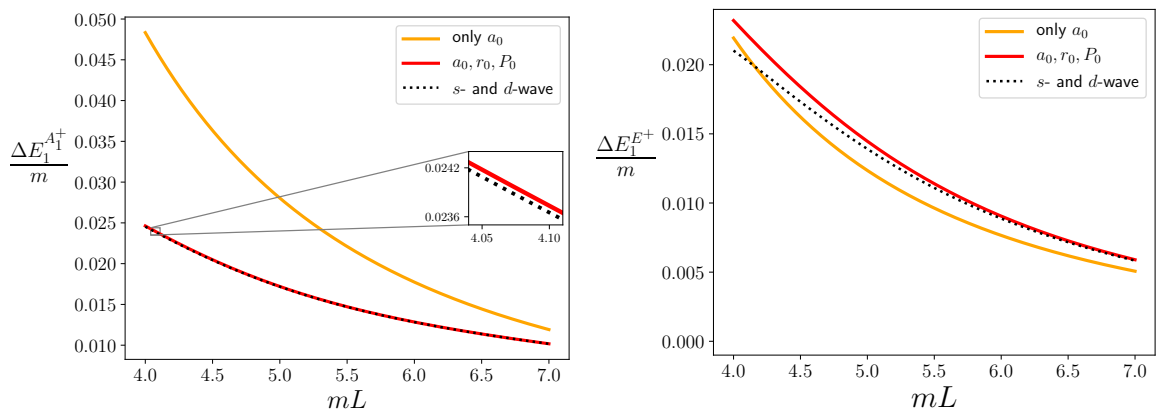

**Figure 6.** Energy shift of the first excited state in the $A_1^+$ irrep (left) and $E^+$ irrep (right). In the range of $mL$ shown, $E_1^{\text{free}}/m = 4.7$–$3.9$. The quantization condition is solved with only two-particle scattering parameters being nonzero, while $\mathcal{K}_{\text{df},3} = 0$. When a parameter is nonzero, its value is given by eq. (4.7). The solid orange and red curves include only $s$-wave dimers, the former having only $a_0$ turned on ("only $a_0$"), with the latter having all three $s$-wave parameters in $\mathcal{K}_2$ nonzero ("$a_0, r_0, P_0$"). The dotted black line shows the impact of adding $d$-wave dimers, with $a_2$ nonzero ("$s$- and $d$-wave").

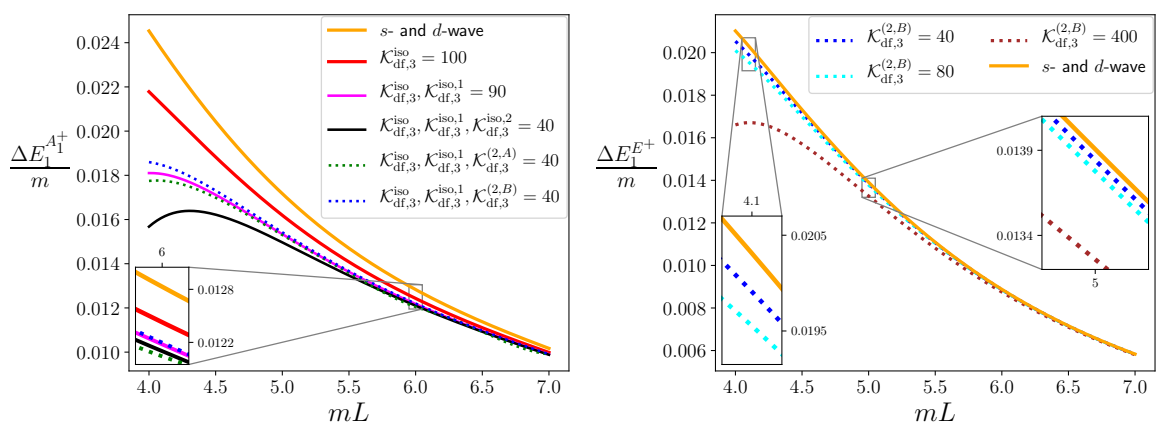

**Figure 7.** Energy shift of the first excited state in the $A_1^+$ irrep (left) and $E^+$ irrep (right) with various choices of the parameters in $\mathcal{K}_{\text{df},3}$. The two-particle scattering parameters are given by eq. (4.7) for all curves. The choices of $\mathcal{K}_{\text{df},3}$ parameters is explained by the legend, with the convention that a parameter value not given explicitly is set to the value given earlier. For example, the black line has the parameters set to $\mathcal{K}_{\text{df},3}^{\text{iso}} = 100$, $\mathcal{K}_{\text{df},3}^{\text{iso},1} = 90$, and $\mathcal{K}_{\text{df},3}^{\text{iso},2} = 40$, while $\mathcal{K}_{\text{df},3}^{(2,A)} = \mathcal{K}_{\text{df},3}^{(2,B)} = 0$.

expansion generalized to excited states [40]. We also see that adding $d$-wave dimers has almost no impact on the $A_1^+$ irrep (indeed, the effect is smaller than for the ground state) while the impact is comparable to that of $r_0$ and $P_0$ for the $E^+$ irrep. Qualitatively, this is as expected, since the averaging over orientations in the $A_1^+$ irrep suppresses the overlap with $d$-wave dimers.

In figure 7 we illustrate the dependence of the same two excited states on the five parameters in $\mathcal{K}_{\text{df},3}$, eq. (2.13). Because $q/m \sim \mathcal{O}(1)$ we expect that, unlike for the ground state, the energy should be sensitive to all five parameters, and not just to $\mathcal{K}_{\text{df},3}^{\text{iso}}$.

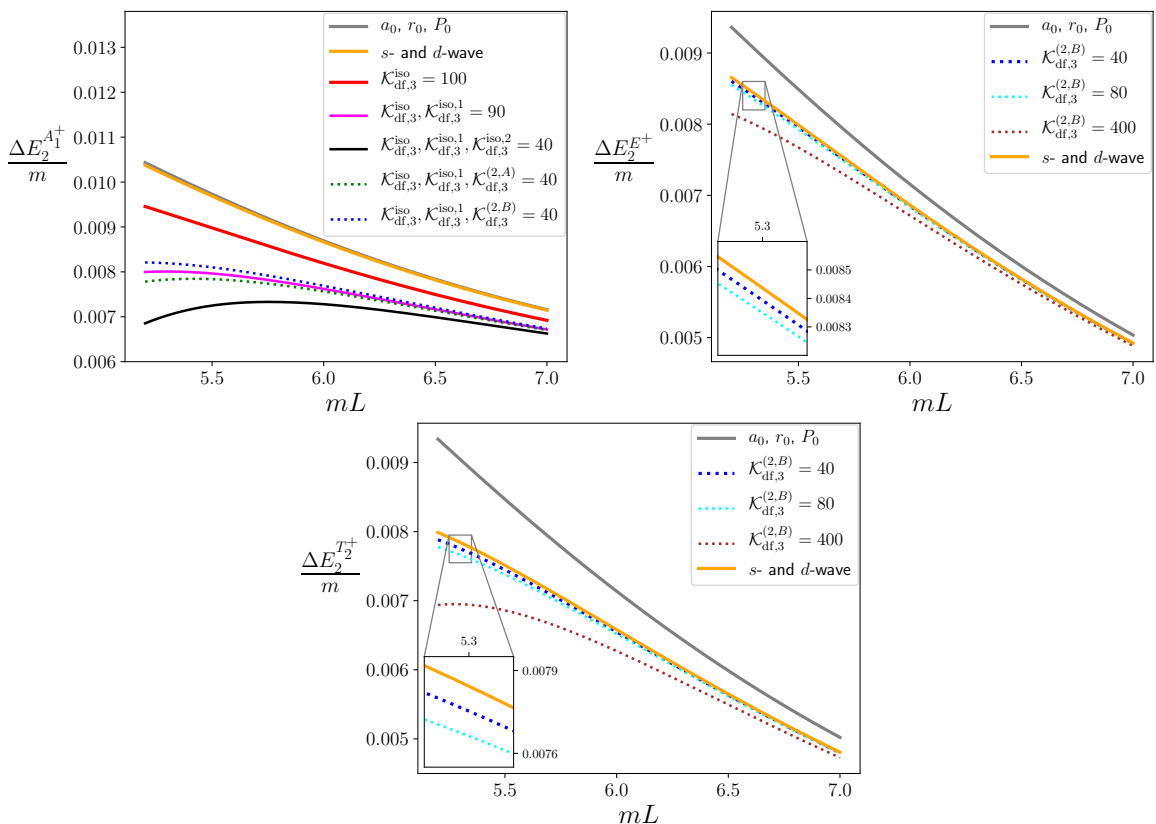

**Figure 8.** Energy shift of the second excited states in the $A_1^+$ irrep (top left), the $E^+$ irrep (top right) and $T_2^+$ irrep (bottom). The meaning of the legend is as in previous figures.

This is borne out for the $A_1^+$ irrep, where there is strong sensitivity to all three isotropic parameters, and a somewhat weaker dependence on $\mathcal{K}_{\mathrm{df},3}^{(2,A)}$ and $\mathcal{K}_{\mathrm{df},3}^{(2,B)}$. As noted above, only $\mathcal{K}_{\mathrm{df},3}^{(2,B)}$ affects the $E^+$ irrep, and figure 7 illustrates this dependence.

The energy shift for the second excited states are shown in figure 8. We show results only for those volumes for which the states lie below the five-particle threshold, which requires $mL \gtrsim 5.2$. The $A_1^+$ energy-shift depends on all parameters in $\mathcal{K}_{\mathrm{df},3}$, while the $E^+$ and $T_2^+$ irreps depend only on $\mathcal{K}_{\mathrm{df},3}^{(2,B)}$. The results show a similar dependence on parameters as for the first excited states. We also find that the $E^+$ and $T_2^+$ irreps show the greatest sensitivity to $a_2$ of all the states considered.

To sum up, a possible program for determining the coefficients in $\mathcal{K}_{\mathrm{df},3}$ up to quadratic order in the threshold expansion is as follows:

1. Determine $a_0$, $r_0$, $P_0$, and $a_2$ from the two-body sector using standard two-particle methods.

2. Extract $\mathcal{K}_{\mathrm{df},3}^{\mathrm{iso}}$ from the threshold state.

3. Use states in the $E^+$ and $T_2^+$ irreps to calculate $\mathcal{K}_{\mathrm{df},3}^{(2,B)}$.

4. Use the excited states in the $A_1^+$ irrep to obtain the rest of the parameters. The most difficult parameter to determine would be $\mathcal{K}_{\mathrm{df},3}^{(2,A)}$, because its contribution to the energy is smaller.

Further information could be obtained using moving frames, as has been done very successfully in the two-particle case. The formalism of ref. [1] is still valid, but the detailed implementation along the lines of this paper has yet to be worked out.

We close by commenting on the importance of using a relativistic formalism for the results that we have presented in this section. We note that the excited states whose energies we consider lie in the relativistic regime. For example, at $mL = 5.5$, the relativistic noninteracting energy of the second excited state is $E_2^{\text{free}} = 4.80m$, to be compared to the nonrelativistic energy $3m + 2m(2\pi/(mL))^2 = 5.61m$. Nevertheless, it may be that the energy splittings $\Delta E_n^\Gamma$ are much less sensitive to relativistic effects, and it would be interesting to implement the NREFT approach including $d$ waves in order to study this. We do expect, however, that the parametrization of the three-particle interaction will require additional terms once the constraints of relativistic invariance are removed.

### 4.4 Unphysical solutions

In this section we describe solutions to the quantization condition that are, for various reasons, unphysical. These fall roughly into two classes (although there is some overlap): solutions that occur at the energies of three noninteracting particles (which we refer to as "free solutions", occurring at "free energies"), and solutions that correspond to poles in the finite-volume correlator that have the wrong sign of the residue. The latter were first observed in ref. [13] within the isotropic approximation. In the following, we begin with a general discussion of the properties of physical solutions, and then discuss the two classes of unphysical solutions in turn.

#### 4.4.1 General properties of physical solutions

We recall here the properties that physical solutions to the quantization condition, eq. (2.1), must obey. This extends the analysis presented in ref. [13] for the isotropic approximation.

The key quantity is the two-point correlation function in Euclidean time,

$$\widetilde{C}_L(\tau) = \langle 0|\mathcal{O}(\tau)\mathcal{O}^\dagger(0)|0\rangle\,, \tag{4.10}$$

where the operator $\mathcal{O}^\dagger$ has the correct quantum numbers to create three particles (and here also has $\vec{P} = 0$). We stress that its hermitian conjugate is used to destroy the states. Inserting a complete set of finite-volume states with appropriate quantum numbers, we find the standard result

$$\widetilde{C}_L(\tau) = \sum_j \frac{c_j}{2E_j} \exp(-E_j|\tau|)\,, \tag{4.11}$$

where $E_j > 0$ are the energies relative to the vacuum, and the $c_j$ are real and positive. Fourier transforming to Euclidean energy and Wick rotating yields

$$C_L(E) = \sum_j c_j \frac{i}{E^2 - E_j^2} = \sum_j \frac{ic_j}{(E + E_j)(E - E_j)}\,, \tag{4.12}$$

where $E$ is the Minkowski energy that appears in the quantization condition. Thus $C_L(E)$ is composed of single poles whose residues, for $E > 0$, are given by $i$ times real, positive coefficients.

Next we recall from the analysis of ref. [1] that the correlator can also be written as

$$C_L(E) = A^\dagger \frac{i}{F_3^{-1} + \mathcal{K}_{\mathrm{df},3}} A = \sum_j |A^\dagger \cdot v_j(E)|^2 \frac{i}{\lambda_j(E)}, \quad (4.13)$$

where $A$ is a column vector, and to obtain the second form we have decomposed $F_3^{-1} + \mathcal{K}_{\mathrm{df},3}$ in terms of its eigenvalues $\lambda_j(E)$ and eigenvectors $v_j(E)$.[22] Since $F_3^{-1} + \mathcal{K}_{\mathrm{df},3}$ is real and symmetric, the eigenvalues are real.

It follows from comparing eqs. (4.12) and (4.13) that

(a) $\lambda_j(E)$ cannot have double zeros. This is because, in the vicinity of a double zero at $E_j$, $C_L(E)$ would have a double pole, $C_L(E) \propto 1/(E - E_j)^2$. The same prohibition applies to higher-order zeros.

(b) Eigenvalues of $F_3^{-1} + \mathcal{K}_{\mathrm{df},3}$ that pass through zero (and thus lead to solutions to the quantization condition) must do so from below as $E$ increases. To understand this, note that, if $\lambda_j(E)$ has a single zero at $E = E_j$, then

$$C_L(E) = |A^\dagger \cdot v_j(E_j)|^2 \frac{i}{\lambda_j'(E_j)(E - E_j)} + \text{non-pole}. \quad (4.14)$$

Comparing to eq. (4.12) we learn that

$$\lambda_j'(E_j) \equiv \frac{d\lambda_j(E)}{dE}\bigg|_{E=E_j} > 0. \quad (4.15)$$

This is the generalization of a condition found in ref. [13] for the isotropic approximation (where there is only a single relevant eigenvalue).

Any solutions to the quantization condition that do not satisfy both of these conditions we refer to as unphysical.

We are aware of only three possible sources for unphysical solutions. First, they can arise from the truncation of the quantization condition to a finite-number of partial waves. Second, they could be the result of an unphysical parametrization of $\mathcal{K}_2$ and $\mathcal{K}_{\mathrm{df},3}$; for example, the truncation of the threshold expansion for $\mathcal{K}_{\mathrm{df},3}$ could be unphysical. And, finally, the exponentially-suppressed terms that we have dropped could be large in some regions of parameter space, particularly for small $mL$. We now present examples of unphysical solutions that we have found in our numerical investigation.

### 4.4.2 Solutions with the wrong residue

In this section we give examples of unphysical solutions to the quantization condition that do not satisfy eq. (4.15), i.e. which lead to single poles whose residues have the wrong sign. These were observed in the isotropic approximation in ref. [13], where it was found that they occurred only when $|\mathcal{K}_{\mathrm{df},3}^{\mathrm{iso}}|$ was very large. Here we investigate how this result generalizes in the presence of $d$-wave dimers.

---

[22]For the sake of brevity, we do not show explicitly that the quantities also depend on $L$.

We first investigate whether unphysical solutions can be induced by adding $d$-wave interactions alone, with $\mathcal{K}_{\mathrm{df},3} = 0$. We do not find such solutions for large negative values of $ma_2$ — the results obtained in section 4.2 all correspond to zero crossings in the correct direction. However, as $ma_2$ approaches unity (which, as we saw in section 3, is the upper bound allowed for the formalism), we do find examples of unphysical solutions. Since we have seen in sections 4.2 and 4.3 that the impact of $d$-wave interactions is greater for irreps other than $A_1^+$, we focus on the $E^+$ irrep, and work in the vicinity of the energy of the first noninteracting excited state, $E_1^{\mathrm{free}}$. In figure 9, we plot the smallest eigenvalue in magnitude of $F_3^{-1} + \mathcal{K}_{\mathrm{df},3} = F_3^{-1}$ in the $E^+$ irrep as a function of energy, for two different values of $mL$ and a range of positive values of $ma_2$ approaching unity. The only other nonvanishing scattering parameter is $ma_0 = -0.1$. Consider first the left panel, with $mL = 8.1$. When $a_2 = 0$, there is a solution at $E \approx E_1^{\mathrm{free}} = 3.53m$, as shown by the lowest level in figure 3b. As $a_2$ is increased, the energy shifts upwards, as expected since positive $a_2$ corresponds to a repulsive interaction. When $ma_2 = 0.9$, the level is at $E_1 \approx 3.6m$, and moves to yet higher energies as $ma_2$ increases. These solutions are physical, as shown in the bottom-left inset. For $ma_2 = 0.9$ and $0.91$, however, there is also a single unphysical solution near $E = 3.85m$, which displays the additional unphysical behavior of having a decreasing energy with increasingly repulsive $a_2$. Furthermore, for $ma_2 = 0.92$, there is a triplet of solutions — two unphysical and one physical. Since they are clearly related, we consider all three to be unphysical. For even larger $ma_2$, there are no solutions in the energy range shown.

The right panel, figure 9b, displays a similar pattern, with an additional twist. Here $mL = 10$, so that $E_1^{\mathrm{free}} = 3.36m$. The energy of the physical solution lies above this, and increases with increasing $ma_2$. There is also an unphysical solution at higher energy, whose energy decreases with increasing $ma_2$. The new feature is the presence of a double zero at $E_1^{\mathrm{free}}$. As discussed above, this is manifestly unphysical since it leads to a double pole in $C_L(E)$. It is also unexpected, as its energy lies at that of noninteracting particles. We discuss such solutions in detail in the following section.

Another example of unphysical solutions in shown in figure 10, this time induced by a large, negative value of $\mathcal{K}_{\mathrm{df},3}^{(2,B)}$. Recall that, out of the parameters in $\mathcal{K}_{\mathrm{df},3}$, the $E^+$ irrep is only sensitive to $\mathcal{K}_{\mathrm{df},3}^{(2,B)}$. Again, there are physical solutions that have the expected behavior of increasing energy with increasingly negative $\mathcal{K}_{\mathrm{df},3}^{(2,B)}$ (which corresponds to a repulsive interaction), but there are also unphysical solutions at higher energy with opposite dependence on $\mathcal{K}_{\mathrm{df},3}^{(2,B)}$. Eventually, for large enough $|\mathcal{K}_{\mathrm{df},3}^{(2,B)}|$ both solutions disappear.

We do not yet understand the source of these unphysical solutions, i.e. which of the three possible sources mentioned at the end of the previous section are most important. This is a topic for future study. Our attitude is that, if a physical solution is well separated from an unphysical one, and its behavior as interactions are made more attractive or repulsive is reasonable, then we accept the physical solution and reject the unphysical one. The examples we have shown occur when the interactions are strong and repulsive, in which limit the two solutions come close together, and at some point become unreliable. For attractive interactions, the two solutions are far apart, often with the unphysical one lying outside the range in which the quantization condition is valid. In this regime, which includes that discussed in section 4.2, we trust the physical solutions.

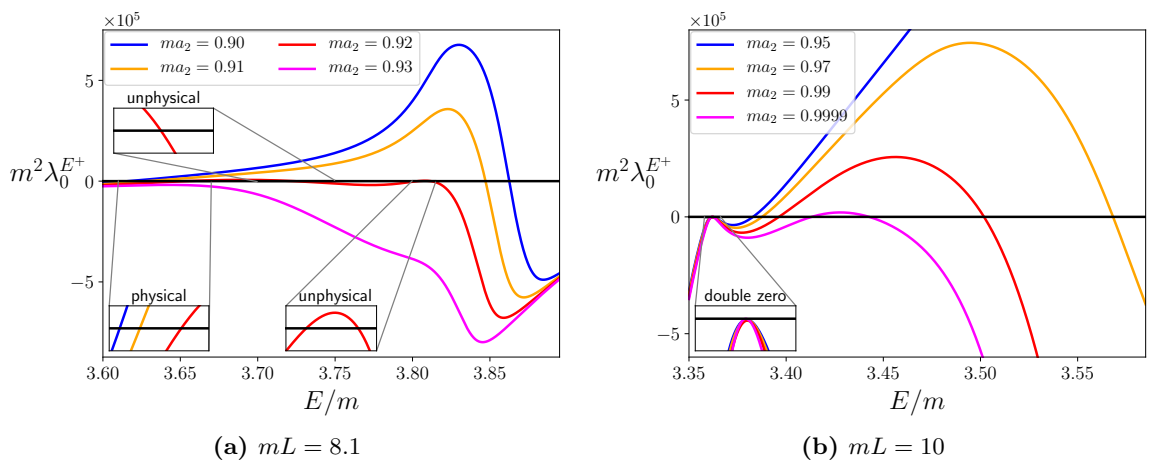

**(a)** $mL = 8.1$                    **(b)** $mL = 10$

**Figure 9.** Smallest eigenvalue in magnitude of $F_3^{-1}$ in the $E^+$ irrep as a function of the energy for two different values of $mL$. The parameters are $ma_0 = -0.1$ and $r_0 = P_0 = \mathcal{K}_{\mathrm{df},3} = 0$. Physical and unphysical solutions as well as a double pole at the free energy (to be discussed in section 4.4.3) are indicated.

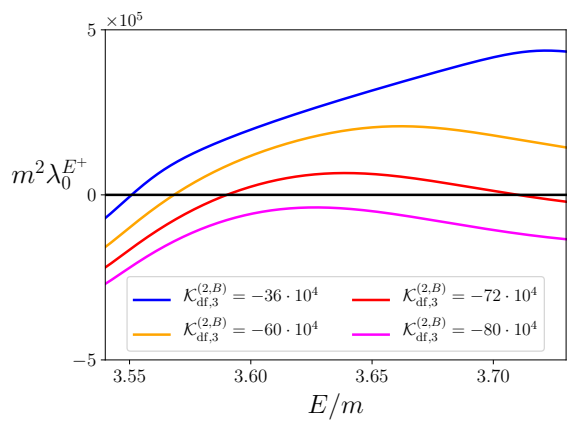

**Figure 10.** Eigenvalue of $F_3^{-1} + \mathcal{K}_{\mathrm{df},3}$ with smallest magnitude in the $E^+$ irrep as a function of the energy. The parameters are $mL = 8.1$, $ma_0 = ma_2 = 0.1$, $r_0 = P_0 = 0$, and $\mathcal{K}_{\mathrm{df},3} = 0$ for all terms except $\mathcal{K}_{\mathrm{df},3}^{(2,B)}$.

We conclude by stressing that, in the case of three pions in QCD, the interactions are relatively weak, and we do not expect unphysical solutions to be relevant.

### 4.4.3 Solutions at free particle energies

This section concerns "free solutions": solutions to the quantization condition that, even in the presence of interactions, lie at one of the energies given in eq. (3.23). We expect that, in general, there will be no such solutions. Exceptions can occur only if the symmetry of the finite-volume three-particle state is such that the chosen interactions do not couple to it. An example in the two-particle sector is that, if $\vec{P} = 0$, a finite-volume state lying in the $E^+$ irrep would not be shifted from its noninteracting value if only $s$- and $p$-wave

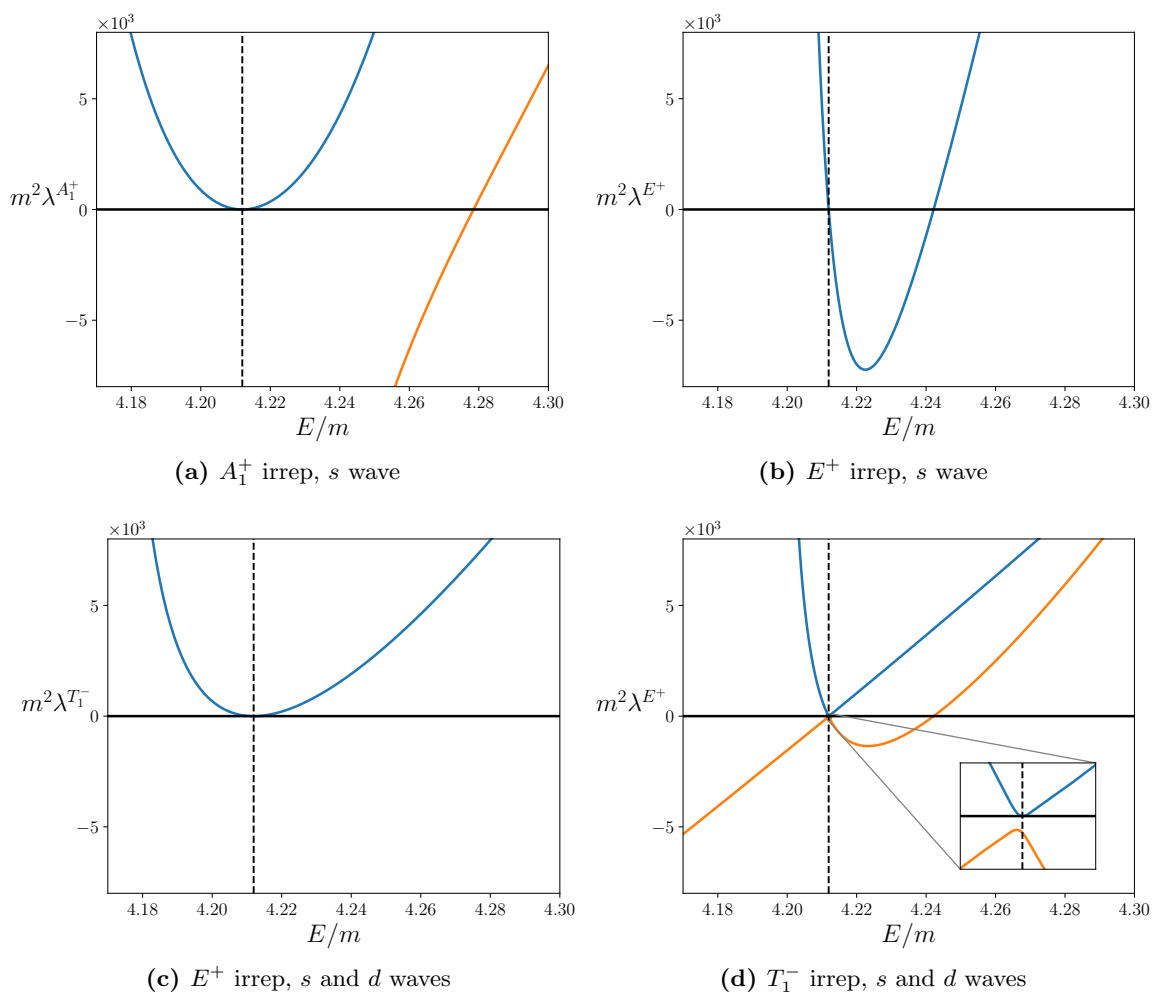

**(a)** $A_1^+$ irrep, $s$ wave

**(b)** $E^+$ irrep, $s$ wave

**(c)** $E^+$ irrep, $s$ and $d$ waves

**(d)** $T_1^-$ irrep, $s$ and $d$ waves

**Figure 11.** Examples of solutions to the quantization condition for $\mathcal{K}_{\mathrm{df},3} = 0$ occurring at the free energy $E_1^{\mathrm{free}}$ (shown in all plots as the vertical dashed line). Plots show eigenvalues of $F_3^{-1}$ as a function of $E/m$, with $ma_0 = 0.1$, $r_0 = P_0 = 0$ and $mL = 5$. Solutions to the quantization occur when an eigenvalue crosses zero. (a) $A_1^+$ irrep with only $\ell = 0$ channels; (b) $E^+$ irrep, with only $\ell = 0$ channels; (c) $T_1^-$ irrep, with both $\ell = 0$ and 2, and $ma_2 = 0.1$; (d) $E^+$ irrep, with both $\ell = 0$ and 2, and $ma_2 = 0.1$. For the $E^+/T_1^-$ irreps, all eigenvalues are doubly/triply degenerate. In (d), both apparent crossings are in fact avoided, as illustrated by the inset.

interactions were included, since the lowest wave contributing to $E^+$ has $\ell = 2$. One question we address here is where such examples occur in the three-particle sector.

We were prompted to study this issue by finding examples of free solutions in our numerical study. One example has already been seen above, in figure 9b, and further examples are shown in figure 11. The first two plots show solutions with only $s$-wave channels included. In figure 11a, which shows results for the $A_1^+$ irrep, we see a double zero at the first excited free energy, $E_1^{\mathrm{free}}$, as well as a solution shifted to slightly higher energies. The latter is expected, since the repulsive interactions should raise the energy of the free state. In the $E^+$ irrep, by contrast, there is a single zero at $E_1^{\mathrm{free}}$, with the unphysical sign

– 28 –

| Level | $\ell$ | Irreps with zeros | Zeros removed by |
|-------|--------|-------------------|------------------|
| $E_1^{\text{free}}$ | 0 | $A_1^+; T_1^-; E^+(1)$ | $(\mathcal{K}_{\text{df},3}^{(2,A)} \text{ or } \mathcal{K}_{\text{df},3}^{(2,B)}); \mathcal{K}_{\text{df},3}^{(2,B)}; \mathcal{K}_{\text{df},3}^{(2,B)}$ |
| $E_1^{\text{free}}$ | 0 & 2 | $A_1^+; T_1^-; E^+$ | $\geq$ quartic for each |
| $E_2^{\text{free}}$ | 0 | $A_1^+; T_1^-; T_2^-$ | $(\mathcal{K}_{\text{df},3}^{(2,A)} \text{ or } \mathcal{K}_{\text{df},3}^{(2,B)}); \mathcal{K}_{\text{df},3}^{(2,B)}; (\mathcal{K}_{\text{df},3}^{(3,B)} \text{ or } \mathcal{K}_{\text{df},3}^{(3,E)})$ |
| $E_2^{\text{free}}$ | 0 & 2 | $A_1^+; E^+; T_2^+; T_1^-; T_2^-$ | $\geq$ quartic for each |

**Table 3.** Irreps in which free zeros appear for the first two excited levels when $\mathcal{K}_{\text{df},3} = 0$. The "(1)" in the first row denotes that the $E_1^{\text{free}}$, $\ell = 0$ free zeros in the $E^+$ irrep are single roots with unphysical residue; all other free zeros in the table are (unphysical) double roots. Also noted are the lowest-order terms in the threshold expansion of $\mathcal{K}_{\text{df},3}$ that remove the free zeros. The notation "$\geq$ quartic" indicates that a term of at least quartic order is needed. Note that cubic-order terms are needed to remove the $E_2^{\text{free}}$, $\ell = 0$ free zeros in the $T_2^-$ irrep, as neither of the quadratic terms $\mathcal{K}_{\text{df},3}^{(2,A)}$ and $\mathcal{K}_{\text{df},3}^{(2,B)}$ has nonzero eigenvalues in this irrep.

for the residue, as well as an interacting solution at higher energy. The other two plots show examples of free zeros when $s$- and $d$-wave channels are included. Both the $T_1^-$ irrep, shown in figure 11c, and the $E^+$ irrep, shown in figure 11d, have a double-zero at $E_1^{\text{free}}$.

We find similar results for higher excited free energy levels, in which case they appear in an increasing number of irreps. We list these irreps for the first two excited free energies in table 3. There are, however, no free solutions for the lowest free energy $E_0^{\text{free}} = 3m$.[23]

In all the examples we have found, the free solutions are also unphysical — they are either double zeros or single zeros with the wrong residue. We do not know if this is a general result. Also, although the examples shown above are for $\mathcal{K}_{\text{df},3} = 0$, free solutions also occur when some components of $\mathcal{K}_{\text{df},3}$ are turned on. Indeed, one of the questions we address in the following is which components of $\mathcal{K}_{\text{df},3}$ are required to either remove the free solutions or move them away from $E_n^{\text{free}}$. Our first task, however, is to understand in more detail when and why free solutions occur. All such solutions originate from the fact that $\widetilde{F}$ and $\widetilde{G}$ have single poles at all the free energies. These can lead to poles in $F_3$ and thus zeros in $F_3^{-1}$. We analyze in detail only the lowest two free energies, i.e. those with level number $n = 0$ and 1 in the notation of table 2, and then draw some general conclusions.

For $E \approx E_0^{\text{free}} = 3m$, the only elements of $\widetilde{F}$ and $\widetilde{G}$ that have poles at $E_0^{\text{free}}$ have vanishing spectator momenta and $\ell = 0$,[24] specifically

$$\widetilde{F}_{000;000} \sim \frac{1}{2}\widetilde{G}_{000;000} \sim p_0 \equiv \frac{1}{16m^3 L^3(E - 3m)}\,. \tag{4.16}$$

Here we are using the symbol $\sim$ to indicate "up to nonpole parts". All other elements of these matrices, and of $\mathcal{K}_2$, either vanish or are of $\mathcal{O}(1)$. From table 1 it now follows that poles in $\widetilde{F}$ and $\widetilde{G}$ only appear in the $A_1^+$ irrep, and the issue is whether these lead to a pole in $F_3$.

---

[23]Strictly speaking, this is only true when one uses the improved form of the quantization condition given in eq. (A.13), and described in appendix A, which removes spurious solutions to eq. (2.1).

[24]Pole contributions with $\ell = 2$ and/or $\ell' = 2$ vanish because, at the pole, $\vec{a}^* = \vec{a}'^* = 0$.

– 29 –

To address this we consider the simplest case in which the volume is chosen such that only the lowest two momentum shells are active, which is the case for $mL \approx 5$. From table 1 we then see that in the $A_1^+$ irrep the matrices are three dimensional, with indices

$$([\text{shell } 1, \ell = 0], \; [\text{shell } 2, \ell = 0], \; [\text{shell } 2, \ell = 2]) \,. \tag{4.17}$$

We will use a $1 + 2$ block notation for the matrices, since this conveys all the necessary information. Close to $E_0^{\text{free}}$ the matrices have the form[25]

$$\widetilde{F} = \begin{pmatrix} p_0 + \mathcal{O}(1) & 0 \\ 0 & \mathcal{O}(1) \end{pmatrix}, \quad \widetilde{G} = \begin{pmatrix} 2p_0 + \mathcal{O}(1) & \mathcal{O}(1) \\ \mathcal{O}(1) & \mathcal{O}(1) \end{pmatrix}, \tag{4.18}$$

where $\mathcal{O}(1)$ elements are constrained only by the fact that $\widetilde{F}$ and $\widetilde{G}$ are symmetric. $\mathcal{K}_2$ is a diagonal matrix with $\mathcal{O}(1)$ elements. From this it follows that

$$H = \widetilde{F} + \widetilde{G} + (2\omega\mathcal{K}_2)^{-1} = \begin{pmatrix} 3p_0 + \mathcal{O}(1) & \mathcal{O}(1) \\ \mathcal{O}(1) & \mathcal{O}(1) \end{pmatrix} \;\Rightarrow\; H^{-1} = \begin{pmatrix} \frac{1}{3p_0} + \mathcal{O}(1/p_0^2) & \mathcal{O}(1/p_0) \\ \mathcal{O}(1/p_0) & \mathcal{O}(1) \end{pmatrix} \tag{4.19}$$

and thus in turn that

$$\widetilde{F} H^{-1} \widetilde{F} = \begin{pmatrix} p_0/3 + \mathcal{O}(1) & \mathcal{O}(1) \\ \mathcal{O}(1) & \mathcal{O}(1) \end{pmatrix} \;\Rightarrow\; F_3 = \mathcal{O}(1) \,. \tag{4.20}$$

We thus find that free poles at $E_0^{\text{free}}$ cancel in $F_3$. This argument generalizes to any number of active shells, since there are no additional poles, and the only change is that the second block in the above analysis is enlarged. The result agrees with our numerical finding that there are no free poles at $E_0^{\text{free}}$.

Next we consider poles at the second free energy, $E_1^{\text{free}}$. For $mL \approx 4\text{–}6$ there are then three active shells, so the matrices to consider become larger, e.g. six-dimensional in the $A_1^+$ irrep, and the analysis correspondingly more complicated. We work out the case of the $A_1^+$ irrep in appendix E, both with $\ell = 0$ channels only and with $\ell = 0$ and 2 channels included. In both cases we find that $F_3^{-1}$ has a double zero at $E = E_1^{\text{free}}$. This lies in a one-dimensional subspace of the full matrix space, and what differs between the two cases is this subspace. For $\ell = 0$ only, the matrix indices are

$$([\text{shell } 1, \ell = 0], \; [\text{shell } 2, \ell = 0], \; [\text{shell } 3, \ell = 0], \dots) \,, \tag{4.21}$$

with the dimension depending on the choice of $L$. The double zero of $F_3^{-1}$ lies, in this case, in the space spanned by

$$\langle x_1' | = \sqrt{\frac{1}{7}} \left( \sqrt{6}, -1, 0, \dots \right) \,. \tag{4.22}$$

For $\ell = 0$ and 2, the matrix indices are

$$([\text{shell } 1, \ell = 0], \; [\text{shell } 2, \ell = 0], \; [\text{shell } 2, \ell = 2], [\text{shell } 3, \ell = 0], \; \dots) \,, \tag{4.23}$$

---

[25]There are also potential poles in the $\ell = 2$ components arising from the vanishing of $q_{2,k}^*$ and $q_{2,p}^*$ in $\widetilde{G}$ and $\widetilde{F}$, eqs. (A.3) and (A.9). However, as discussed at the end of appendix A, the quantization condition can be formulated such that these purely kinematical poles are canceled, and it is legitimate to ignore them.

and the space of the double zero of $F_3^{-1}$ is spanned by

$$\langle x_1 | = \sqrt{\frac{1}{12}} \left( \sqrt{6}, -1, -\sqrt{5}, 0, \dots \right) . \tag{4.24}$$

The factors in eqs. (4.22) and (4.24) result from the form of the spherical harmonics and the size of the first two shells. They are thus kinematical.

These analytic results confirm what we find numerically. For example, the double zero at $E_1^{\text{free}}$ shown in figure 11a exactly matches that expected from the analysis of appendix E, and we have checked numerically that it lies in the predicted subspace.

We now discuss how the single zeros at free energies arise. There is a particularly simple case in which we can easily understand these analytically: the $E^+$ irrep when we keep only $s$-wave channels *and* choose $mL$ such that only the first two shells are active. We must also choose $mL$ such that $E_1^{\text{free}} < 5m$ (so that the formalism applies); one example is $mL = 3.8$, for which $E_1^{\text{free}} = 4.86m$. In fact, as shown in table 1, the first shell has no $E^+$ component for $\ell = 0$, so this simple case actually involves only the second shell, for which the $E^+$ irrep appears once. Although the $E^+$ irrep is two-dimensional, within this space all matrices are proportional to the identity. Thus the matrices are effectively one-dimensional.

The second shell consists of six elements, which we label by the direction of the spectator momentum $\vec{k}$ in the following order

$$\vec{k} \in o_{001} = (2\pi/L)\{-\hat{z}, -\hat{y}, -\hat{x}, \hat{x}, \hat{y}, \hat{z}\} . \tag{4.25}$$

In this basis, the $E^+$ eigenvectors can be chosen as

$$\frac{1}{2}(1, 0, -1, -1, 0, 1) \quad \text{and} \quad \sqrt{\frac{1}{12}}(-1, 2, -1, -1, 2, -1) . \tag{4.26}$$

It is then simple to calculate the pole terms to be

$$\widetilde{F} = \mathbf{1} \left[ p_1 + \mathcal{O}(1) \right] \quad \text{and} \quad \widetilde{G} = \mathbf{1} \left[ p_1 + \mathcal{O}(1) \right] , \tag{4.27}$$

where

$$p_1 \equiv \frac{1}{8m\omega_1^2 L^3 (E - E_1^{\text{free}})} . \tag{4.28}$$

It immediately follows that

$$F_3 = \frac{1}{L^3} \left[ \frac{\widetilde{F}}{3} - \widetilde{F} H^{-1} \widetilde{F} \right] = -\frac{p_1}{6L^3} \mathbf{1} \left[ 1 + \mathcal{O}(1/p_1) \right] . \tag{4.29}$$

Thus $F_3$ indeed has a single pole at $E = E_1^{\text{free}}$, and $F_3^{-1}$ a single (doubly degenerate) zero. Increasing $L$ so that there are more active shells does not change the pole structure or the presence of the single zero. We also see that the zero in $F_3^{-1}$ has a negative coefficient, implying that it decreases through zero, consistent with the behavior seen in figure 11b.

Thus we have understood in a few simple cases why the free zeros listed in table 3 appear. It is interesting to contrast this to the results of ref. [13], where the quantization

– 31 –

condition was studied numerically in the isotropic approximation. In that work no free zeros in $F_3^{-1}$ were found. At first this may seem puzzling, because the isotropic approximation is a subset of our analysis when we restrict to $\ell = 0$ channels. The resolution is that the additional isotropic projection that is used is orthogonal to the subspace in which the zeros live. This is demonstrated in appendix F, along with a derivation of the precise relation between the isotropic approximation and the analysis carried out here.

The final stage of our analysis is to study whether the inclusion of components of $\mathcal{K}_{\mathrm{df},3}$ removes the free zeros. Here by "remove" we mean that there is no longer a solution to the quantization condition at a free energy. This can be accomplished either by removing the solution altogether (which is possible for a double zero, which only touches the axis) or by moving it away from the free energy (the likely solution for a single zero). We expect that if $\mathcal{K}_{\mathrm{df},3}$ were not truncated then there would be no free zeros, since there would be some overlap between the state and the three-particle interaction. This is indeed consistent with what we find. What turns out to be surprising, however, is which components of $\mathcal{K}_{\mathrm{df},3}$ that are needed to remove the free zeros.

We first consider the $\ell = 0$, $A_1^+$ case. To remove the double zero, it must be that the projection of $\mathcal{K}_{\mathrm{df},3}$ into the space of zeros is nonvanishing:

$$[\mathcal{K}_{\mathrm{df},3}(E_1^{\mathrm{free}})]|x_1'\rangle \neq 0 \,, \tag{4.30}$$

where $|x_1'\rangle$ is defined in eq. (4.22). Here the square brackets indicate the matrix that results when $\mathcal{K}_{\mathrm{df},3}$ is decomposed into the $k\ell m$ basis and projected into an irrep. Note that this equation need only hold for $E = E_1^{\mathrm{free}}$, i.e. at the energy of the free zero.

The isotropic parts of $\mathcal{K}_{\mathrm{df},3}$, eq. (2.14), do not solve the problem. These terms have the matrix form

$$[\mathcal{K}^{\mathrm{iso}}] \propto |1_K\rangle\langle 1_K| \,, \tag{4.31}$$

where

$$\langle 1_K| = \left(1, \sqrt{6}, \sqrt{12}, \dots \right) \,. \tag{4.32}$$

Since this vector is orthogonal to $|x_1'\rangle$, it follows that, for all energies,

$$[\mathcal{K}^{\mathrm{iso}}]|x_1'\rangle = 0 \,, \tag{4.33}$$

so that eq. (4.30) is not satisfied. The form of $|1_K\rangle$ follows from the fact that $\mathcal{K}^{\mathrm{iso}}$ is independent of the spectator momentum, so that the $A_1^+$ projection simply gives factors of the square root of the multiplicity of the shells. We thus expect that the inclusion of *any* dependence on the spectator momentum will lead to a $[\mathcal{K}_{\mathrm{df},3}]$ satisfying eq. (4.30). This is what we find in practice with both of the quadratic terms, i.e. those with coefficients $\mathcal{K}_{\mathrm{df},3}^{(2,A)}$ and $\mathcal{K}_{\mathrm{df},3}^{(2,B)}$ [see eqs. (2.15) and (2.16)].

This result is an example of a general pattern: the part of $\mathcal{K}_{\mathrm{df},3}$ that "removes" the free zeros comes from terms that involve higher values of $\ell$ than those being included in $F_3^{-1}$. Here, we need quadratic terms, which have both $\ell = 0$ and 2 components, in order to remove the free zeros from the $\ell = 0$ part of $F_3^{-1}$. To be clear, the $\ell = 2$ components of the quadratic terms play no role; it is simply that by going to higher order one obtains

a more complicated form of the $\ell = 0$ parts, and this is sufficient to remove the unwanted free zeros. Further examples of this are shown in the last column of table 3, where we list, for all irreps that enter in a given free momentum shell, the terms in $\mathcal{K}_{\mathrm{df},3}$ that remove the free zero.

The second example we consider is the combined $\ell = 0$ and 2 part of $F_3^{-1}$ in the $A_1^+$ irrep. In this case, we need

$$[\mathcal{K}_{\mathrm{df},3}(E_1^{\mathrm{free}})]|x_1\rangle \neq 0 \tag{4.34}$$

[with $|x_1\rangle$ given in eq. (4.24)] in order to remove the free zeros. We find numerically that this equation is not satisfied by any of the quadratic or cubic terms contributing to $\mathcal{K}_{\mathrm{df},3}$, but that quartic terms do satisfy it.[26] This exemplifies the general pattern discussed above: quadratic and cubic terms contain only $\ell = 0$ and 2, while quartic terms include also $\ell = 4$ parts. We were initially surprised by this result, because $\mathcal{K}_{\mathrm{df},3}$ is an infinite-volume quantity, while $|x_1\rangle$ arises from finite-volume considerations. However, we show analytically in appendix G that orthogonality follows solely from the rotation invariance and particle-interchange symmetry of $\mathcal{K}_{\mathrm{df},3}$, together with the fact that quadratic and cubic terms contain only $\ell = 0$ and 2 parts. Thus it is an example of the phenomenon described at the beginning of this section, in which symmetries make the finite-volume state transparent to certain interactions. It is also clear from the arguments in appendix G that all that is required for eq. (4.34) to be satisfied is to use contributions to $\mathcal{K}_{\mathrm{df},3}$ that involve $\ell \geq 4$, i.e. terms of quartic or higher order in the threshold expansion.

Finally, we consider the case of the single zero in the $E^+$ irrep for $\ell = 0$ channels only, shown in figure 11b. Here we aim to shift the zero away from the free energy. This is accomplished by including a contribution from $\mathcal{K}_{\mathrm{df},3}$ that lives in the $E^+$ irrep. As noted in the final paragraph of section 3, the lowest-order term in the threshold expansion for which this is the case is the $\mathcal{K}_{\mathrm{df},3}^{(2,B)}$ term. Thus, once again, we have to use a term in $\mathcal{K}_{\mathrm{df},3}$ that contains higher values of $\ell$ (here $\ell = 2$) than are included in $F_3$.

These theoretical arguments are supported by our numerical results. We show two examples in figure 12. These correspond to the two cases shown in figures 11a and 11b, except that we have turned on $\mathcal{K}_{\mathrm{df},3}^{(2,A)}$ and $\mathcal{K}_{\mathrm{df},3}^{(2,B)}$, respectively. We expect the double-zero in the former case ($A_1^+$ irrep) to removed by the addition of any quadratic term in $\mathcal{K}_{\mathrm{df},3}$, and the figure shows that $\mathcal{K}_{\mathrm{df},3}^{(2,A)}$ does the job. In figure 12b, corresponding to the $E^+$ irrep, we need to use the $\mathcal{K}_{\mathrm{df},3}^{(2,B)}$ term, since $\mathcal{K}_{\mathrm{df},3}^{(2,A)}$ does not contain an $E^+$ component. Since this is a single zero, it is not removed, but is rather shifted to a non-free energy. Note, however, that it remains unphysical because it decreases through zero. In fact, for higher values of $\mathcal{K}_{\mathrm{df},3}^{(2,B)}$, the zeros coalesce and then disappear.

We close this section with two general comments on the nature of the resolution that we have presented to the problem of unwanted free solutions. The first concerns the result that we need higher-order terms in the threshold expansion of $\mathcal{K}_{\mathrm{df},3}$ in order to remove the free zeros of a given order in $F_3^{-1}$. On its face, this invalidates the threshold expansion, for we are evaluating distinct terms in the quantization condition at different orders. We do not think this is the case, however, because we know that, above threshold, *all terms* in

---

[26]In this case it is crucial to set the energy to $E_1^{\mathrm{free}}$; for other energies eq. (4.34) is satisfied.

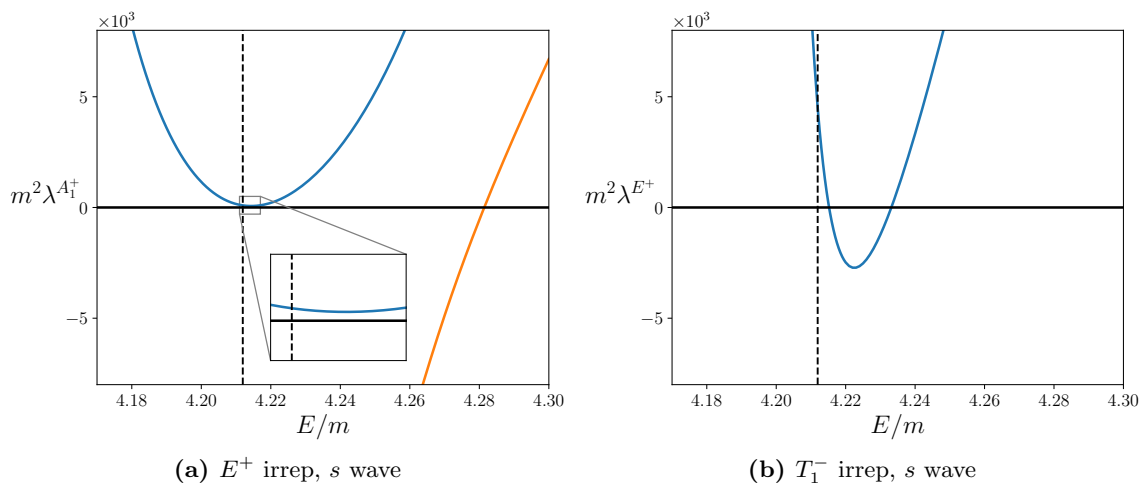

**(a)** $E^+$ irrep, $s$ wave

**(b)** $T_1^-$ irrep, $s$ wave

**Figure 12.** Effect of turning on $\mathcal{K}_{\mathrm{df},3}$ on the free solutions shown in figures 11a and 11b, with all other parameters unchanged. Eigenvalues are now those of $F_3^{-1} + \mathcal{K}_{\mathrm{df},3}$. (a) $A_1^+$ irrep with $\mathcal{K}_{\mathrm{df},3}^{(2,A)} = 8000$; (b) $E^+$ irrep with $\mathcal{K}_{\mathrm{df},3}^{(2,B)} = 8000$.

the expansion of $\mathcal{K}_{\mathrm{df},3}$ are present at some level, and it only takes an infinitesimal value for the coefficient of the requisite higher-order term to remove the unwanted solution. Thus we conclude that we can proceed, in practice, by truncating the expansion of all quantities at the same order in the threshold expansion, and simply ignore the free solutions.

The second comment concerns the fact that our resolution fails if the coefficient of the required parts of $\mathcal{K}_{\mathrm{df},3}$ vanish. In fact, this would require the simultaneous vanishing of an infinite number of terms in the threshold expansion, since higher-order terms in the correct irrep can remove the free solutions. Thus it would require an enormous fine-tuning, which seems highly implausible, especially because there is no enhancement of the symmetry of $\mathcal{K}_{\mathrm{df},3}$ at the tuned point.

## 5 Conclusions

The work presented in this paper is the first step towards the systematic inclusion of higher partial waves in the three-particle quantization condition. We have used the generic relativistic field theory (RFT) approach, formulated so that the three-particle scattering quantity, $\mathcal{K}_{\mathrm{df},3}$, is Lorentz invariant. This invariance proves very important in simplifying the threshold expansion of $\mathcal{K}_{\mathrm{df},3}$. Indeed, we find that, at quadratic order and for identical particles, only five parameters control the contribution from the three-particle sector, of which only two describe dependence on angular degrees of freedom. This provides a simple starting point for studying the impact of $\mathcal{K}_{\mathrm{df},3}$. Working at quadratic order implies keeping both $s$- and $d$-wave two-particle channels (dimers). We have numerically implemented the quantization condition at this order, and obtained several new results that we now highlight.

The first of these is to determine the projection onto irreps of the cubic group including higher partial waves. This has previously been done only for the case of $s$-wave dimers [14]. The generalization is nontrivial, since both the spectator momentum and the parameters of the dimer transform. While we have worked this out explicitly only for coupled $s$- and $d$-wave dimers, the formalism holds for dimers with any angular momentum.

Second, we have understood how the two-particle scattering amplitudes in higher partial waves enter in the $1/L$ expansion of the energy of the three-particle ground state. We find that all even partial waves enter at $\mathcal{O}(1/L^6)$, and have calculated analytically the dependence on the $d$-wave amplitude in the weak-coupling limit and for $\mathcal{K}_{\mathrm{df},3} = 0$. Although this contribution itself is likely too small to be seen in present simulations of three-particle systems, we have used it as a nontrivial check of our implementation.

Third, we have shown that $d$-wave interactions, if they are moderately strong, can have a sizable effect on the finite-volume three-particle spectrum. For example, we have presented evidence for a generalized Efimov-like three-particle bound state induced by a strongly attractive $d$-wave two-particle interaction.

Fourth, we have shown how the five parameters describing $\mathcal{K}_{\mathrm{df},3}$ lead to distinguishable effects on the spectrum of the $3\pi^+$ system, suggesting that they can be separately determined in a dedicated lattice study. Indeed, this is the system within QCD to which our truncated formalism is most applicable.

Finally, we have characterized solutions to the quantization condition that are unphysical. These presumably arise because of the truncation to a small number of partial waves, and the fact that we have dropped terms that are exponentially suppressed in $mL$. One class of solutions generally appears when either the two- or the three-particle interactions are strong and repulsive. Our approach is to use parameters such that there are no unphysical solutions near to the physical solutions of interest. The second class of solutions are those that occur at the energies of three noninteracting particles. We have presented numerical evidence and analytical arguments that these are removed if sufficiently high-order terms in $\mathcal{K}_{\mathrm{df},3}$ are included. We expect that other approaches to the three-particle quantization condition will face similar issues, for which our observations may be relevant.

There remain many directions for future study. In order to make our implementation more useful, it is important to generalize it to moving frames. The underlying formalism of ref. [1] applies in all finite-volume frames, but the projectors onto irreps will need to be generalized to account for the reduced symmetry. Another important generalization is to include subchannel resonances, i.e., dynamical poles in $\mathcal{K}_2$. For this one must implement the formalism of ref. [7], and go beyond the threshold expansion. Finally, we recall that $\mathcal{K}_{\mathrm{df},3}$ is an intermediate quantity, related to the physical three-particle scattering amplitude, $\mathcal{M}_3$, by integral equations. Since it is only by looking for complex poles in $\mathcal{M}_3$ that one can study three-particle resonances, it is crucial to develop methods to solve the necessary integral equations.

To conclude, we would like to restate that, as it is a relativistic approach, our implementation can simultaneously be useful to both the lattice QCD community and the field of cold atom physics.

## Acknowledgments

## A Definitions

Here we collect definitions of quantities appearing in $F_3$, eq. (3.1), that are not given in the main text.

We begin with the cutoff function:

$$H(\vec{k}) = J(z), \qquad z = \frac{E_{2,k}^{*2} - (1 + \alpha_H)m^2}{(3 - \alpha_H)m^2},$$ (A.1)

$$J(z) = \begin{cases} 0, & z \leq 0 \\ \exp\left(-\frac{1}{z}\exp\left[-\frac{1}{1-z}\right]\right), & 0 < z < 1 \\ 1, & 1 \leq z \end{cases}$$ (A.2)

with $\alpha_H \in [-1, 3)$ a constant. We choose $\alpha_H = -1$, corresponding to the highest cutoff, in all our numerical investigations.

For $\widetilde{G}$ we use the relativistic form suggested in ref. [5],

$$\widetilde{G}_{p\ell'm';k\ell m} \equiv \frac{1}{L^3}\frac{1}{2\omega_p}\frac{H(\vec{p})H(\vec{k})}{b^2 - m^2}\frac{4\pi \mathcal{Y}_{\ell'm'}(\vec{k}^*)\mathcal{Y}_{\ell m}(\vec{p}^*)}{q_{2,p}^{*\ell'}q_{2,k}^{*\ell}}\frac{1}{2\omega_k},$$ (A.3)

where $b = P - p - k$ is the momentum of the exchanged particle, $\vec{p}^*$ is the result of boosting $p$ to the CM frame of the dimer for which $k$ is the spectator momentum, and *vice versa*. Explicitly, we have

$$\vec{p}^* = (\gamma_k - 1)(\hat{p}\cdot\hat{k})\hat{k} + \omega_p\gamma_k\beta_k\hat{k} + \vec{p}, \quad \beta_k = \frac{|\vec{k}|}{E - \omega_k}, \quad \gamma_k = (1 - \beta_k^2)^{-1/2},$$ (A.4)

with $\vec{k}^*$ given by $\vec{p} \leftrightarrow \vec{k}$. Finally, $\mathcal{Y}_{\ell m}(\vec{k})$ are harmonic polynomials,

$$\mathcal{Y}_{\ell m}(\vec{k}) \equiv k^\ell Y_{\ell m}(\hat{k}),$$ (A.5)

where $Y_{\ell m}$ are the *real* spherical harmonics. The elements of $\widetilde{G}$ are clearly straightforward to evaluate numerically.

For completeness, we quote the real $d$-wave harmonic polynomials

$$\sqrt{4\pi}\mathcal{Y}_{2-2}(\vec{k}) = \sqrt{15}k_1 k_2\,,$$
$$\sqrt{4\pi}\mathcal{Y}_{2-1}(\vec{k}) = \sqrt{15}k_2 k_3\,,$$
$$\sqrt{4\pi}\mathcal{Y}_{20}(\vec{k}) = \sqrt{5}(2k_3^2 - k_1^2 - k_2^2)/2\,, \tag{A.6}$$
$$\sqrt{4\pi}\mathcal{Y}_{21}(\vec{k}) = \sqrt{15}k_1 k_3\,,$$
$$\sqrt{4\pi}\mathcal{Y}_{22}(\vec{k}) = \sqrt{15}(k_1^2 - k_2^2)/2\,.$$

The associated Wigner D-matrices are

$$\mathcal{D}_{\ell'm',\ell m}(R) = \int d\Omega_{\hat{r}} Y_{\ell'm'}(R\hat{r}) Y_{\ell m}(\hat{r}) = \delta_{\ell\ell'}\mathcal{D}_{m'm}^{(\ell)}(R)\,, \tag{A.7}$$

where $R$ is a rotation matrix. They are orthogonal matrices, and implement rotations of the spherical harmonics:

$$Y_{\ell m}(R\hat{r}) = \sum_{m'=-\ell}^{\ell} \mathcal{D}_{mm'}^{(\ell)}(R) Y_{\ell m'}(\hat{r})\,. \tag{A.8}$$

Finally, $\widetilde{F}(\vec{k})$ is a sum-integral difference that is proportional to the zeta functions that appear in the two-particle quantization condition [16, 17]. It requires ultraviolet (UV) regularization, and can be written in various forms that are equivalent up to exponentially-suppressed corrections. The form that follows from that presented in ref. [1] is

$$\widetilde{F}(\vec{k})_{\ell'm';\ell m} = \left[\frac{1}{L^3}\sum_{\vec{a}} - \text{PV}\int \frac{d^3 a}{(2\pi)^3}\right] \frac{1}{(q_{2,k}^*)^{\ell'+\ell}} \frac{H(\vec{a})H(\vec{b})4\pi\mathcal{Y}_{\ell'm'}(\vec{a}^*)\mathcal{Y}_{\ell m}(\vec{a}^*)}{16\omega_k\omega_a\omega_b(E-\omega_k-\omega_a-\omega_b)}\,, \tag{A.9}$$

where $b = P - k - a$ here, and $\vec{a}^*$ is the result of boosting $a$ to the dimer rest frame, with $k$ the spectator. Here the UV regularization is provided by the product of $H$ functions, and the integral over the pole is defined by the principle value prescription (leading to a real result). Instead, we use a different form that is simpler to evaluate numerically. Following the steps similar to those used in ref. [42], we change variables and introduce a new regularization, finding that, up to exponentially-suppressed corrections, $\widetilde{F}$ can be rewritten as

$$\widetilde{F}(\vec{k})_{\ell'm';\ell m} = \frac{1}{32\pi^2 L\omega_k(E-\omega_k)}\left[\sum_{\vec{n}_a} - \text{PV}\int d^3 n_a\right] \frac{e^{\alpha(x^2-r^2)}}{x^2-r^2}\frac{4\pi\mathcal{Y}_{\ell'm'}(\vec{r})\mathcal{Y}_{\ell m}(\vec{r})}{x^{\ell'+\ell}}\,, \tag{A.10}$$

where $\vec{a} = \vec{n}_a(2\pi/L)$, $x = q_{2,k}^* L/(2\pi)$, and

$$\vec{r}(\vec{n}_k, \vec{n}_a) = \vec{n}_a + \vec{n}_k\left[\frac{\vec{n}_a \cdot \vec{n}_k}{n_k^2}\left(\frac{1}{\gamma_k}-1\right) + \frac{1}{2\gamma_k}\right]\,, \tag{A.11}$$

with $\vec{k} = \vec{n}_k(2\pi/L)$. The UV regularization is now provided by the exponential in the integrand, and is parametrized by $\alpha > 0$. What is shown in ref. [42] is that the $\alpha$ dependence

is exponentially suppressed in $L$, and that, in practice, one should choose a value that is small enough that the dependence on $\alpha$ lies below the accuracy required. We find that $\alpha \approx 0.5$ is usually small enough.

An important technical point is that, as seen from eq. (3.6), in the full matrix form $\widetilde{F}_{p\ell'm';k\ell m}$, $\widetilde{F}(\vec{k})$ is always multiplied by $H(\vec{k})$, from which it follows that $\gamma_k$ is always finite and real whenever $\widetilde{F}_{p\ell'm';k\ell m}$ is nonvanishing.

We close this appendix by commenting on the factors of $q^*$ (which we use generically for $q^*_{2,k}$ or $q^*_{2,p}$) in the denominators of $\widetilde{G}$ and $\widetilde{F}$. These lead to poles for particular kinematic configurations, which in turn can lead to solutions to the quantization condition. These solutions appear to be similar to free solutions discussed in section 4.4.3, but are in fact spurious. To understand this we need an argument given in appendix A of ref. [1], which shows that the factors of $q^*$ in the denominators are always canceled by corresponding factors in the numerators of $\mathcal{K}_2$, $\mathcal{K}_{\mathrm{df},3}$, and the end cap factors $A^\dagger$ and $A$ in the finite-volume correlation function $C_L(E)$ [see eq. (4.13)]. The presence of the necessary factors of $q^*$ in $\mathcal{K}_2$ can be seen from eq. (3.5), while those in $\mathcal{K}_{\mathrm{df},3}$ arise from the quadratic dependence on $\vec{a}^*$ and $\vec{a}'^*$ described in section 3.1. Indeed, one can derive a version of the quantization condition in which all such factors are absent. To do so, we define the matrix

$$Q_{p\ell'm';k\ell m} = \delta_{pk}\delta_{\ell'\ell}\delta_{m'm}q^{*\ell}_{2,k}\,. \tag{A.12}$$

Then, from the arguments of ref. [1] we know that we can write the end caps as $A = Q\widetilde{A}$ and $A^\dagger = \widetilde{A}^\dagger Q$, with $\widetilde{A}$ and $\widetilde{A}^\dagger$ nonsingular. Thus an alternative, improved form of the quantization condition is

$$\det[(QF_3Q)^{-1} + Q^{-1}\mathcal{K}_{\mathrm{df},3}Q^{-1}] = 0\,. \tag{A.13}$$

Now we observe that, by simple algebraic manipulations, we can rewrite this form of the quantization condition in terms of $QF Q$, $Q\widetilde{G}Q$, $Q^{-1}\mathcal{K}_2Q^{-1}$ and $Q^{-1}\mathcal{K}_{\mathrm{df},3}Q^{-1}$, in all of which the factors of $q^*$ cancel. Since the difference between the two quantization conditions is a factor of $\det(Q^2)$, it follows that the solutions to the new form, eq. (A.13), are the same as those to eq. (2.1), except that spurious solutions to the latter, arising from the factors of $q^*$, are removed. In conclusion, we can use the original form of the quantization condition, eq. (2.1), as long as we ignore the spurious solutions.

# B Numerical evaluation of $\widetilde{F}$

In this appendix we describe some technical details concerning the evaluation of $\widetilde{F}(\vec{k})$.

## B.1 Evaluating the integrals

An advantage of the form eq. (A.10) is that the integrals can be evaluated analytically. Dropping overall factors, the integral that is needed is

$$I^F_{\ell'm';\ell m} = \mathrm{PV}\int d^3n_a \frac{e^{\alpha(x^2-r^2)}}{x^2-r^2} 4\pi \mathcal{Y}_{\ell'm'}(\vec{r})\mathcal{Y}_{\ell m}(\vec{r})\,. \tag{B.1}$$

– 38 –

Changing variables to $\vec{r}$, we find

$$I^F_{\ell'm';\ell m} = \gamma \text{PV} \int d^3 r \frac{e^{\alpha(x^2-r^2)}}{x^2-r^2} 4\pi \mathcal{Y}_{\ell'm'}(\vec{r}) \mathcal{Y}_{\ell m}(\vec{r}) = \delta_{\ell'\ell}\delta_{m'm} I^F_\ell , \tag{B.2}$$

$$I^F_\ell = 4\pi\gamma \text{PV} \int r^2 dr \frac{e^{\alpha(x^2-r^2)}}{x^2-r^2} r^{2\ell} . \tag{B.3}$$

The remaining integral can be evaluated analytically for all $\ell$. The explicit result for $\ell = 0$ was worked out in ref. [13], and we have extended this to the $\ell = 2$ case. For convenience, we quote both results

$$I^F_0 = 4\pi\gamma \left[ -\sqrt{\frac{\pi}{\alpha}} \frac{1}{2} e^{\alpha x^2} + \frac{\pi x}{2} \text{Erfi}\left(\sqrt{\alpha x^2}\right) \right] \tag{B.4}$$

$$I^F_2 = 4\pi\gamma \left[ -\sqrt{\frac{\pi}{\alpha^5}} \frac{3 + 2\alpha x^2 + 4\alpha^2 x^4}{8} e^{\alpha x^2} + \frac{\pi x^5}{2} \text{Erfi}\left(\sqrt{\alpha x^2}\right) \right] . \tag{B.5}$$

## B.2 Cutting off the sum

The sum in eq. (A.10) is convergent, but in practice we must introduce a cutoff in order to evaluate it numerically. We use a spherical cutoff, $|\vec{n}_a| < n_{\max}$, and in this section explain how we choose $n_{\max}$.

The basic idea is to split the sum $S$ as

$$S = S_< + S_> , \tag{B.6}$$

where $S_<$ is the contribution from below the cutoff, and $S_>$ the remainder. Assuming that the pole in the summand lies well below the cutoff, then $S_>$ can be well-approximated by a remainder integral, $R_>$. We evaluate this integral, and then choose $n_{\max}$ such that $R_>$ lies below our desired accuracy. The resulting $n_{\max}$ depends on $E$, $L$ and the orbit of $\vec{k}$.

Dropping overall factors, and changing the overall sign, the sum of interest from eq. (A.10) is

$$S = H(\vec{k}) \sum_{\vec{n}_a} \frac{e^{\alpha(x^2-r^2)}}{r^2-x^2} r^{\ell'+\ell} 4\pi Y_{\ell'm'}(\hat{r}) Y_{\ell m}(\hat{r}) . \tag{B.7}$$

Here we have included the cutoff function $H(\vec{k})$ that enters in the expression for $\widetilde{F}_{p\ell'm';k\ell m}$, eq. (3.6). Although this is an overall factor, it will play an important role in the determination of $n_{\max}$.

The integral $R_>$ that results when replacing the sum over $\vec{n}_a$ with an integral is more easily evaluated by changing variables to $\vec{r}$. The relation between $\vec{n}_a$ and $\vec{r}$, eq. (A.11), can be rewritten as

$$\gamma_k r_\parallel = n_{a,\parallel} - \frac{n_k}{2} , \qquad r_\perp = n_{a,\perp} , \tag{B.8}$$

with $\parallel$ and $\perp$ defined relative to $\vec{k}$. The cutoff is chosen such that $n_{\max} \gg n_k$, implying that the $n_k/2$ term in the expression for $r_\parallel$ is subleading. Dropping this term, we find

that a spherical cutoff on $\vec{n}_a$ corresponds to an ellipsoidal cutoff on $\vec{r}$. This makes the integral difficult to evaluate, so we replace this with a spherical cutoff, $|\vec{r}| < \Lambda$, choosing $\Lambda = n_{\max}/\gamma_k$. We call the resulting integral $R_\Lambda$. The resulting spherical region is a superset of the original ellipsoidal region, so that we overestimate the remainder, $R_\Lambda > R_>$, since the integrand is positive.

To evaluate $R_\Lambda$ we make two further approximations. First, we drop the $x^2$ term in the denominator, which is subleading since $r^2 \gg x^2$ within the region of integration. Second, we make the replacement $4\pi Y_{\ell'm'}(\hat{r}) Y_{\ell m}(\hat{r}) \to 1$, which leads to an overestimate of the integral. Then we find

$$R_\Lambda \approx \overline{R}_\Lambda \equiv \gamma_k H(\vec{k}) 4\pi \int_\Lambda^\infty dr \; e^{\alpha(x^2 - r^2)} r^{\ell' + \ell} \tag{B.9}$$

$$= \begin{cases} \gamma_k H(\vec{k}) e^{\alpha x^2} 2\pi \sqrt{\frac{\pi}{\alpha}} \mathrm{Erfc}\left[\sqrt{\alpha}\Lambda\right], & \ell' = \ell = 0 \\ \gamma_k H(\vec{k}) e^{\alpha x^2} \frac{\pi}{\alpha} \{2\Lambda e^{-\alpha\Lambda^2} + \sqrt{\frac{\pi}{\alpha}} \mathrm{Erfc}\left[\sqrt{\alpha}\Lambda\right]\}, & \ell' + \ell = 2 \\ \gamma_k H(\vec{k}) e^{\alpha x^2} \frac{\pi}{\alpha^2} \{(3\Lambda + 2\alpha\Lambda^3) e^{-\alpha\Lambda^2} + \frac{3}{2} \sqrt{\frac{\pi}{\alpha}} \mathrm{Erfc}\left[\sqrt{\alpha}\Lambda\right]\}, & \ell' + \ell = 4. \end{cases} \tag{B.10}$$

The overall factor of $\gamma_k$ is the Jacobian from changing the integration variable from $\vec{n}_a$ to $\vec{r}$. We choose the $\Lambda$ by specifying a tolerance $\epsilon$ (we use $\epsilon = 10^{-9}$) and numerically solving $\overline{R}_\Lambda = \epsilon$.[27] Given $\Lambda$, we then obtain the cutoff for the sum using $n_{\max} = \gamma_k \Lambda$.

We can now explain why we include the factor of $H(\vec{k})$ in $S$. As $|\vec{k}|$ approaches the value where $H(\vec{k})$ vanishes, $\gamma_k$ diverges. This leads to an increase in $n_{\max}$, both from the factor of $\gamma_k$ in $\overline{R}_\Lambda$, and because $n_{\max}/\Lambda = \gamma_k$. However, this increase is more than compensated by the very rapid drop in $H(\vec{k})$ near the end point, so that $n_{\max}$ is always finite.

### B.3 Using cubic symmetries

Symmetries can be exploited to optimize the computation of $\widetilde{F}$. It follows from eq. (3.6) that $\widetilde{F}(R\vec{k})$ can be obtained from $\widetilde{F}(\vec{k})$ via an orthogonal transformation for any cubic-group transformation $R \in O_h$,

$$\widetilde{F}(R\vec{k}) = \mathcal{D}(R)\widetilde{F}(\vec{k})\mathcal{D}(R)^T. \tag{B.11}$$

Here $\mathcal{D}(R)$ is the Wigner D-matrix defined in eq. (A.7). Thus once one has computed $\widetilde{F}(\vec{k})$ for some finite-volume momentum $\vec{k}$, one can use eq. (B.11) to obtain $\widetilde{F}(\vec{k}')$ for all $\vec{k}'$ in the same momentum shell. Furthermore, for each initial $\widetilde{F}(\vec{k})$ that one computes directly, any symmetries of $\vec{k}$ can be used to simplify the construction of $\widetilde{F}(\vec{k})$. In particular, if $R$ is in the little group of $\vec{k}$ (so that $R\vec{k} = \vec{k}$), then eq. (B.11) says that $\widetilde{F}(\vec{k})$ is invariant under the transformation. This often leads to linear relationships between several matrix elements $\widetilde{F}_{\ell'm',\ell m}(\vec{k})$, in which case one need only compute the linearly-independent elements in order to construct the full matrix.

---

[27]In practice we use the $\ell' = \ell = 0$ result for $\overline{R}_\Lambda$ in all cases, which is a further approximation, but one that we find makes a small numerical impact.

## C   Further details of the projection onto cubic group irreps

We collect here some results that we have found useful in the computation of the projection matrices and the determination of their properties.

### C.1   Computing $P_I$ efficiently

The projector $P_I$ is defined in eq. (3.14). As explained in the main text, it is block diagonal in momentum shells and in angular momentum, with blocks $P_{I,o(\ell)}$. Here we explain how to simplify the computation of $P_{I,o(\ell)}$ by reducing the sum in eq. (3.14), which runs over all 48 elements of $O_h$, to a sum over the elements of the little group of an element of the shell under consideration.

Let $\vec{k}'$ and $\vec{k}''$ be two elements of the orbit. Then, from eqs. (3.14) and (3.15), we have

$$\left[ P_{I,o(\ell)} \right]_{k''k'} = \frac{d_I}{[O_h]} \sum_{R \in O_h} \chi_I(R) \delta_{k'_R k''} \mathcal{D}^{(\ell)}(R) \,, \tag{C.1}$$

where $\delta_{k'_R k''}$ is unity if $R\vec{k}' = \vec{k}''$ and zero otherwise. Thus the sum is restricted to those elements of $O_h$ that rotate $\vec{k}'$ into $\vec{k}''$. A convenient representation of these elements makes use of an (arbitrarily chosen) canonical element of the orbit, denoted $\vec{k}$. Let $R_{L_k}$ be an element of the little group $L_k$ of $\vec{k}$. Then all the elements of $O_h$ that rotate $\vec{k}'$ to $\vec{k}''$ can be written as $R_{k''k}R_{L_k}R_{kk'}$, where $R_{kk'}$ is any choice of transformation from $\vec{k}'$ to $\vec{k}$, and $R_{k''k}$ is any choice of transformation from $\vec{k}$ to $\vec{k}''$. Thus the number of elements contributing to the sum in eq. (C.1) is $[L_k]$, the dimension of $L_k$. This allows us to rewrite the projector as

$$\left[ P_{I,o(\ell)} \right]_{k''k'} = \frac{d_I}{[O_h]} \sum_{R \in L_k} \chi_I(R_{k''k}RR_{kk'}) \mathcal{D}^{(\ell)}(R_{k''k}RR_{kk'}) \tag{C.2}$$

$$= \frac{d_I}{N_o} \mathcal{D}^{(\ell)}(R_{k''k}) \left[ \frac{1}{[L_k]} \sum_{R \in L_k} \chi_I(R_{k''k}RR_{kk'}) \mathcal{D}^{(\ell)}(R) \right] \mathcal{D}^{(\ell)}(R_{kk'}) \,, \tag{C.3}$$

where $N_o = [O_h]/[L_k]$ is the number of elements in the orbit.

Once we have constructed the block projectors, we combine them into $P_I$ using eq. (3.16). In practice, we want to reduce our original matrices ($M = \widetilde{F}$ etc.) down to the part that lives in the projected subspace, which has dimension $d(P_I)$. To do so, we evaluate the eigenvalues and eigenvectors of $P_I$. Since $P_I$ is a projector, its eigenvalues $\lambda_i$ are either zero or unity. We keep only the eigenvectors with unit eigenvalues, for these span the projection subspace. We orthonormalize the eigenvectors, and label them $\{\vec{v}_i\}_{i=1}^{d(P_I)}$. The reduced matrix is then given by

$$M_{ij}^{\text{red}} = \vec{v}_i^{\text{Tr}} \cdot M \cdot \vec{v}_j \qquad (i,j \in 1 - d(P_I)) \,. \tag{C.4}$$

### C.2   Dimensions of irrep projection subspaces

As explained in the main text, in order to determine the number of eigenvalues of $M$ that fall into a given irrep we need to compute the dimensions of the sub-block projectors,

$$d(P_{I,o(\ell)}) = \text{Tr } P_{I,o(\ell)} \,. \tag{C.5}$$

Using the result for the projector, eq. (C.3), we find

$$d(P_{I,o(\ell)}) = \sum_{\vec{k}' \in o} \mathrm{Tr}\left[P_{I,o(\ell)}\right]_{k'k'},$$ (C.6)

$$= \frac{d_I}{[L_k]} \sum_{R \in L_k} \chi_I(R)\,\mathrm{Tr}\,\mathcal{D}^{(\ell)}(R),$$ (C.7)

where the trace is only over the angular momentum indices, $m$, and to obtain the second line we have used the cyclicity of the trace, the fact that $R_{k'k} = R_{kk'}^{-1}$, and the standard group-theoretic result $\chi(R'RR'^{-1}) = \chi(R)$. The resulting dimensions are collected in table 1.

# D $a_2$ dependence of $\mathcal{M}_{3,\mathrm{thr}}$

In section 2, we show that to determine the $a_2$ dependence of the three-particle threshold energy, we need to calculate the corresponding dependence of $\mathcal{M}_{3,\mathrm{thr}}$. The calculation is described in this appendix.

We begin by recalling from ref. [23] that $\mathcal{M}_{3,\mathrm{thr}}$ is defined by doing the minimal subtractions necessary to have a finite quantity at threshold,

$$\mathcal{M}_{3,\mathrm{thr}} = \lim_{\delta \to 0}\left[\mathcal{M}_3(0,\hat{a}'^*;0,\hat{a}^*)\right.$$
$$\left. - I_{0,\delta}(0,\hat{a}'^*;0,\hat{a}^*) - \int_\delta \frac{d^3k_1}{(2\pi)^3}\Xi_1(\vec{k}_1) - \int_\delta \frac{d^3k_1}{(2\pi)^3}\frac{d^3k_2}{(2\pi)^3}\Xi_2(\vec{k}_1,\vec{k}_2)\right].$$ (D.1)

Here $\mathcal{M}_3$ is the three-particle scattering amplitude, expressed in terms of the same variables used for $\mathcal{K}_{\mathrm{df},3}$ in eq. (2.20). The infrared (IR) divergence of $\mathcal{M}_3$ at threshold is regularized using the $\delta$-scheme of ref. [23], and three subtractions are needed in order to obtain a finite result. The explicit expressions for $I_0$, $\Xi_1$ and $\Xi_2$ are given in section D of ref. [23], but will not be needed. All we need to know here is that the subtractions depend on $a_0$, but not on $a_2$. Thus dependence on $a_2$ can only enter through $\mathcal{M}_3$ itself.

To determine this dependence it is useful to recall the definition of the divergence-free scattering amplitude from ref. [2],

$$\mathcal{M}_{\mathrm{df},3}(\vec{p},\hat{a}'^*;\vec{k},\hat{a}^*) = \mathcal{M}_3(\vec{p},\hat{a}'^*;\vec{k},\hat{a}^*) - \mathcal{D}(\vec{p},\hat{a}'^*;\vec{k},\hat{a}^*).$$ (D.2)

Here $\mathcal{D}$ is a quantity that depends only on the two-particle scattering amplitude $\mathcal{M}_2$, whose expression will be given below. It is chosen so as to subtract IR divergences from $\mathcal{M}_3$ not only at threshold, but also above. Reordering eq. (D.2) as $\mathcal{M}_3 = \mathcal{M}_{\mathrm{df},3} + \mathcal{D}$, we note that, in general, both contributions to $\mathcal{M}_3$ depend on $a_2$. However, we also know from ref. [2] that $\mathcal{M}_{\mathrm{df},3}$ vanishes when $\mathcal{K}_{\mathrm{df},3} = 0$. So, in this limit, which is the case we consider numerically, $\mathcal{M}_3 = \mathcal{D}$. This allows us to calculate the $a_2$ dependence of $\mathcal{M}_3$. We know that this dependence is finite at threshold because no $a_2$-dependent subtraction was needed in eq. (D.1).

Before calculating the $a_2$ dependence of $\mathcal{M}_3$, it is instructive to relate the two subtracted versions of $\mathcal{M}_3$,

$$\mathcal{M}_{3,\text{thr}} = \mathcal{M}_{\text{df},3}(0, \hat{a}'^*; 0, \hat{a}^*)\Big|_{E=3m} + \text{IR finite terms}. \tag{D.3}$$

Since, as already noted, $\mathcal{M}_{\text{df},3}$ vanishes when $\mathcal{K}_{\text{df},3} = 0$, we see that it is the IR finite terms that must contain the contribution to $\mathcal{M}_{3,\text{thr}}$ from higher partial waves.

What we have learned so far is that, for $\mathcal{K}_{\text{df},3} = 0$, the $a_2$ dependence of $\mathcal{M}_{3,\text{thr}}$ is given by that of $\mathcal{D}$ evaluated at threshold. Here we are interested in determining the leading dependence, which, as discussed in the main text, is proportional to $a_2^5$. This is given by

$$\mathcal{M}_{3,\text{thr}} \supset a_2^5 \frac{d\mathcal{D}_{\text{thr}}}{d(a_2^5)}\Big|_{a_2=0}. \tag{D.4}$$

Here $\mathcal{D}_{\text{thr}}$ is $\mathcal{D}(\vec{p}, \hat{a}'^*; \vec{k}, \hat{a}^*)$ evaluated at $E = 3m$ and $\vec{p} = \vec{k} = 0$, so that there is no dependence on $\hat{a}^*$ and $\hat{a}'^*$. In fact, $\mathcal{D}$ itself diverges in this limit, but the derivative in eq. (D.4) does not.

To proceed, we need the explicit expression for $\mathcal{D}$, given in ref. [2]. It is obtained by symmetrizing over initial and final momenta the quantity $\mathcal{D}^{(u,u)}$, which is given by

$$\mathcal{D}^{(u,u)}(\vec{p}, \vec{k}) = -\mathcal{M}_2(\vec{p}) G^\infty(\vec{p}, \vec{k}) \mathcal{M}_2(\vec{p}) + \int_s \frac{1}{2\omega_s} \mathcal{M}_2(\vec{p}) G^\infty(\vec{p}, \vec{s}) \mathcal{M}_2(\vec{s}) G^\infty(\vec{s}, \vec{k}) \mathcal{M}_2(\vec{k}) + \dots. \tag{D.5}$$

Here $\int_s \equiv \int d^3s/(2\pi)^3$, and the $\hat{a}^*$ and $\hat{a}'^*$ dependence has been decomposed into partial waves, so that all quantities are implicitly matrices in angular momentum space. The spectator-momentum dependence is, however, kept explicit. $\mathcal{M}_2(\vec{p})$ is the two-particle scattering amplitude for the dimer when the spectator-momentum is $\vec{p}$. As for $\mathcal{K}_2$ [see eq. (3.3)], it is diagonal in angular momentum

$$\mathcal{M}_2(\vec{p})_{\ell'm';\ell m} = \delta_{\ell'\ell} \delta_{m'm} \mathcal{M}_2^{(\ell)}. \tag{D.6}$$

It contains all (even) partial waves, including, in particular, the $d$-wave amplitude. Finally, $G^\infty$ is given by

$$G^\infty_{\ell'm';\ell m}(\vec{p}, \vec{k}) = \frac{H(\vec{p})H(\vec{k})}{b^2 - m^2} \frac{4\pi \mathcal{Y}_{\ell'm'}(\vec{k}^*) \mathcal{Y}_{\ell m}(\vec{p}^*)}{q_{2,p}^{*\ell'} q_{2,k}^{*\ell}}, \tag{D.7}$$

where the kinematic quantities are the same as those appearing in eq. (A.3). Equation (D.7) is the relativistically-invariant version of the definition given in eq. (81) of ref. [2].

At threshold, only the $s$-wave part of $\mathcal{D}^{(u,u)}$ is nonzero, and symmetrization simply leads to an overall factor of 9:

$$\frac{d\mathcal{D}_{\text{thr}}}{d(a_2^5)}\Big|_{a_2=0} = 9 \frac{d\mathcal{D}^{(u,u)}_{00;00}(\vec{0}, \vec{0})}{d(a_2^5)}\Big|_{a_2=0}. \tag{D.8}$$

Looking at eq. (D.5), we see that the $s$-wave projection implies that the factors of $\mathcal{M}_2$ on both ends are pure $s$-wave, so the first appearance of $d$-wave scattering occurs in the

second term. This gives the leading $a_2$-dependent part of $\mathcal{D}^{(u,u)}$:

$$\mathcal{D}^{(u,u)}_{00;00}(\vec{0},\vec{0}) \supset I_d = \int_s \sum_{m=-2}^{2} \frac{1}{2\omega_s} \mathcal{M}_2^{(0)}(\vec{0}) G^\infty_{00;2m}(\vec{0},\vec{s}) \mathcal{M}_2^{(2)}(\vec{s}) G^\infty_{2m;00}(\vec{s},\vec{0}) \mathcal{M}_2^{(0)}(\vec{0}) \, . \quad \text{(D.9)}$$

At leading order in perturbation theory in $a_0$ and $a_2$, $\mathcal{M}_2^{(\ell)} = \mathcal{K}_2^{(\ell)}$, with $\mathcal{K}_2^{(\ell)}$ given by the leading terms in eqs. (3.4) and (3.5). Inserting these results, we find that $I_d$ is IR and UV convergent, so we do not need to actually take the derivative in eq. (D.8). By numerical evaluation we find

$$\mathcal{M}_{3,\text{thr}} \supset 9I_d = -\frac{14109.6}{m^2} \times (ma_0)^2(ma_2)^5[1 + \mathcal{O}(a_0) + \mathcal{O}(a_2^5)] \, . \quad \text{(D.10)}$$

This gives the leading term in the result (4.4) quoted in the main text. The corrections in (D.10) arise from the subleading terms in the expressions for $\mathcal{K}_2^{(\ell)}$.

We close with two further observations. First, a similar calculation with $\mathcal{M}_2^{(2)}$ in $I_d$ replaced by any (even) higher-order amplitude leads to a nonzero contribution to $\mathcal{M}_{3,\text{thr}}$. Thus all higher partial waves contribute to $\Delta E_3$ at $\mathcal{O}(L^{-6})$. Second, higher-order terms in $\mathcal{D}^{(u,u)}$ will also contribute to $\mathcal{M}_{3,\text{thr}}$, although suppressed by powers of $a_\ell$. For example, the first term not shown in eq. (D.5), which has four factors of $\mathcal{M}_2$, leads to contributions to $\Delta E_3$ proportional to $a_0^3 a_2^5/L^6$ and $a_0^2 a_2^{10}/L^6$. These are of the same order as the corrections in eq. (D.10).

## E  Free solutions at the first excited energy

In this appendix we analyze free solutions to the quantization condition in the $A_1^+$ irrep at the energy of the first excited noninteracting state, $E_1^{\text{free}} = m + 2\omega_1$ (with $\omega_1 = \sqrt{m^2 + k_L^2}$ and $k_L = 2\pi/L$). Our aim is to understand when $F_3^{-1}$ has zeros at this energy, and to determine their properties. We work with box lengths $4 \lesssim mL \lesssim 6$ such that there are three active shells, although the final result generalizes straightforwardly to any number of shells.

### E.1  $A_1^+$ irrep with $s$ and $d$ waves

We first consider the case in which both $\ell = 0$ and $\ell = 2$ channels are included. The matrices that enter into the quantization condition are then six dimensional: the first three indices as in eq. (4.17), and the remaining three from the third shell (one with $\ell = 0$, and two with $\ell = 2$; see table 1). The free poles enter only in the first two shells, and are proportional to

$$p = \frac{3}{8L^3 m \omega_1^2 \left(E - E_1^{\text{free}}\right)} \, . \quad \text{(E.1)}$$

It will be useful to introduce the vectors

$$\langle v_1| = (1, 0, 0, 0, 0, 0) \, , \quad \langle v_2| = \left(0, \sqrt{\frac{1}{6}}, \sqrt{\frac{5}{6}}, 0, 0, 0\right) \, , \quad \text{(E.2)}$$

– 44 –

in terms of which the pole parts are given by [using eqs. (A.9) and (A.3)]

$$\widetilde{F} = p\left(|v_1\rangle\langle v_1| + 2|v_2\rangle\langle v_2|\right) + \mathcal{O}(1)\,, \tag{E.3}$$

$$\widetilde{G} = 2p\left(|v_1\rangle\langle v_2| + |v_2\rangle\langle v_1| + |v_2\rangle\langle v_2|\right) + \mathcal{O}(1)\,. \tag{E.4}$$

As in the example discussed in section 4.4.3, all we need to know about the $\mathcal{O}(1)$ contributions are that they are real and symmetric. The relative factor of $\sqrt{5}$ between the two terms in $\langle v_2|$ arises from $Y_{20}(\hat{z}) = \sqrt{5}Y_{00}(\hat{z})$. Combining the results for $\widetilde{F}$ and $\widetilde{G}$ we find

$$H = 5p|w_1\rangle\langle w_1| + \mathcal{O}(1)\,, \quad |w_1\rangle = \sqrt{\frac{1}{5}}\left(|v_1\rangle + 2|v_2\rangle\right)\,. \tag{E.5}$$

Thus, while the pole parts of $\widetilde{F}$ and $\widetilde{G}$ are both of rank 2, that of $H$ is of rank 1, due to a partial cancelation.

In the following, we determine the pole structure of $F_3$, aiming to find a basis in which this structure is simple. We begin by changing to a more convenient basis, namely $|w_1\rangle$ combined with

$$|w_2\rangle = \sqrt{\frac{1}{5}}\left(2|v_1\rangle - |v_2\rangle\right)\,, \tag{E.6}$$

and any choice of four other vectors filling out the orthonormal set. We use a $1 + 1 + 4$ block notation, in which

$$H = p\begin{pmatrix} 5 & 0 & 0 \\ 0 & 0 & 0 \\ 0 & 0 & 0 \end{pmatrix} + \mathcal{O}(1) \quad \text{and} \quad \widetilde{F} = p\begin{pmatrix} 9/5 & -2/5 & 0 \\ -2/5 & 6/5 & 0 \\ 0 & 0 & 0 \end{pmatrix} + \mathcal{O}(1)\,. \tag{E.7}$$

The inverse of $H$ has the form

$$H^{-1} = \begin{pmatrix} 1/(5p) + \mathcal{O}(1/p^2) & \alpha_{12}/p + \mathcal{O}(1/p^2) & \vec{\alpha}_{13}/p + \mathcal{O}(1/p^2) \\ \alpha_{12}/p + \mathcal{O}(1/p^2) & \alpha_{22} + \beta_{22}/p + \mathcal{O}(1/p^2) & \vec{\alpha}_{23} + \mathcal{O}(1/p) \\ \vec{\alpha}_{13}^{\mathrm{Tr}}/p + \mathcal{O}(1/p^2) & \vec{\alpha}_{23}^{\mathrm{Tr}} + \mathcal{O}(1/p) & \overset{\leftrightarrow}{\alpha}_{33} + \mathcal{O}(1/p) \end{pmatrix}\,, \tag{E.8}$$

where the quantities $\alpha_{12}$, $\alpha_{22}$, $\beta_{22}$ etc. are given in terms of the $\mathcal{O}(1)$ parts of $H$ in a way that is not pertinent. At this stage we can see that $\widetilde{F}H^{-1}\widetilde{F}$ will contain a double pole proportional to $\alpha_{22}$ that will have the form of an outer product, as well as a complicated single-pole term. Performing the algebra we find

$$L^3 F_3 = p^2\frac{4\alpha_{22}}{25}\begin{pmatrix} -1 & 3 & 0 \\ 3 & -9 & 0 \\ 0 & 0 & 0 \end{pmatrix} + p\begin{pmatrix} a & b & -\vec{z} \\ b & -9a - 6b & 3\vec{z} \\ -\vec{z}^{\mathrm{Tr}} & 3\vec{z}^{\mathrm{Tr}} & 0 \end{pmatrix} + \mathcal{O}(1)\,, \tag{E.9}$$

where $a$, $b$ and $\vec{z}$ are given in terms of the $\alpha_{ij}$ and $\beta_{22}$.

Thus we have learned that $F_3$ contains a free double pole that can be written

$$-p^2\frac{8\alpha_{22}}{5L^3}|x_1\rangle\langle x_1|\,, \qquad \langle x_1| = \sqrt{\frac{1}{10}}\left(-1, 3, 0\right) = \sqrt{\frac{1}{2}}\left(\langle v_1| - \langle v_2|\right)\,. \tag{E.10}$$

– 45 –

The form of $|x_1\rangle$ is determined entirely by the pole structure of $\widetilde{F}$ and $H$, although the overall coefficient is determined by the $\mathcal{O}(1)$ parts. Qualitatively we can say that although $F$ contains two independent poles in this irrep, the $H^{-1}$ factor cancels one of them, leading to a left-over double pole.

We conclude by discussing the impact of the single pole contribution to $F_3$. First we note that the coefficient of $p$ can be written as

$$- (8a + 6b)|x_1\rangle\langle x_1| - \frac{1}{\sqrt{10}N_2}\left(|x_1\rangle\langle x_2| + |x_2\rangle\langle x_1|\right) \tag{E.11}$$

where the new normalized basis vector is

$$\langle x_2| = N_2\left(9a + 3b, 3a + b, -10\vec{z}\right) . \tag{E.12}$$

Thus in the basis consisting of $|x_1\rangle$, $|x_2\rangle$ and four other orthonormal vectors, $F_3$ has the $1 + 1 + 4$ block form

$$F_3 = \begin{pmatrix} fp^2 + gp & hp & 0 \\ hp & 0 & 0 \\ 0 & 0 & 0 \end{pmatrix} + \mathcal{O}(1) , \tag{E.13}$$

where $f$, $g$ and $h$ are known constants. This matrix can be diagonalized using a final, fourth basis. All we need to know here is that, close to the pole, when $|p| \gg 1$, the shift in the eigenvalues due to the off-diagonal $hp$ term is $\pm(hp)^2/(fp^2 + gp) \sim \mathcal{O}(1)$. Thus in the final basis we have

$$F_3 = \text{diag}\left[fp^2 + gp + \mathcal{O}(1), \mathcal{O}(1), \mathcal{O}(1), \mathcal{O}(1), \mathcal{O}(1), \mathcal{O}(1)\right] , \tag{E.14}$$

and thus

$$F_3^{-1} = \text{diag}\left[1/(fp^2) + \mathcal{O}(1/p^3), \mathcal{O}(1), \mathcal{O}(1), \mathcal{O}(1), \mathcal{O}(1), \mathcal{O}(1)\right] . \tag{E.15}$$

Note that the size of the change to this final basis is proportional to $1/p$, and thus vanishes at the zero of $F_3^{-1}$, so that the double zero lies in the subspace spanned by $|x_1\rangle$.

In summary, we find that the single pole in $F_3$ is hidden beneath the double pole, such that in the inverse there is simply a double zero. As $L$ is increased, there are more active shells, but the only change to the result of this section is that the number of vanishing components of $|x_1\rangle$ increases [see eq. (E.10)]. The nonvanishing components are unchanged.

## E.2  $A_1^+$ irrep with only $s$ waves

We have repeated the previous analysis for the case of only $\ell = 0$ contributions and three active shells.[28] The matrices are now three dimensional, with one entry per shell. We do not present the details, except to note that we follow the same steps as in the previous subsection, and find very similar conclusions aside from some changes in factors. In particular, $F_3^{-1}$ still has a double zero, but this now lives in the space spanned by the vector

$$\langle x_1'| = \left(\sqrt{\frac{6}{7}}, -\sqrt{\frac{1}{7}}, 0\right) , \tag{E.16}$$

where entries are ordered as in eq. (4.21).

---

[28]This builds upon, and corrects, the analysis given in appendix C of ref. [1].

## F  Properties of the isotropic approximation

This appendix recalls the definition of the isotropic approximation, describes its relation to the work of this paper, and explains why the free solutions discussed in section 4.4.3 are absent in this approximation.

The isotropic approximation was introduced in ref. [1] and used in the numerical investigation of ref. [13]. It involves three components: (1) Only $\ell = 0$ dimer channels are included in $\widetilde{F}$, $\widetilde{G}$, $\mathcal{K}_2$ and $\mathcal{K}_{\mathrm{df},3}$; (2) The resulting $\mathcal{K}_{\mathrm{df},3}$ is taken to be independent of the spectator momentum, although dependence on the total energy $E$ is allowed; (3) $F_3$ is projected onto the isotropic vector $|1_K\rangle$, which has a unit entry for every available choice of spectator momentum. Note that the third step automatically picks out solutions in the $A_1^+$ irrep.

The isotropic approximation is thus a subset of an approach we use several times in this paper, namely restricting dimers to $\ell = 0$, keeping only the isotropic part of $\mathcal{K}_{\mathrm{df},3}$ in the expansion about threshold, and projecting onto the $A_1^+$ irrep. We refer to this as the "low-energy $A_1^+$ approximation". The major difference is the absence of the third step — we do not project onto $|1_K\rangle$. A minor difference is that, for $\mathcal{K}_{\mathrm{df},3}$ to be purely isotropic, we must work only at linear order in the threshold expansion. Thus we can have at most a linear dependence of $\mathcal{K}_{\mathrm{df},3}$ on $E^2$, as opposed to the arbitrary dependence allowed in the isotropic approximation.

To explain the relationship between the two approximations, we begin in the low-energy $A_1^+$ approximation. All matrices, including $F_3$, are labeled by an index denoting the shell of the spectator momentum, as shown in eq. (4.21). All matrices have the same finite dimension given by the number of shells lying below our cutoff. Since $\mathcal{K}_{\mathrm{df},3}$ is isotropic, the quantization condition is[29]

$$\det\left([F_3]^{-1} + |1_K\rangle\mathcal{K}^{\mathrm{iso}}\langle 1_K|\right) = 0\,, \tag{F.1}$$

where the square braces indicate the $A_1^+$, $\ell = 0$ matrix, and

$$\langle 1_K| = \left(1, \sqrt{6}, \sqrt{12}, \dots\right) \tag{F.2}$$

in this basis. The entries here are the square roots of the sizes of the shells. We can rewrite the determinant in the quantization condition as

$$\det\left([F_3]^{-1}\right)\det\left(1 + [F_3]|1_K\rangle\mathcal{K}^{\mathrm{iso}}\langle 1_K|\right) = \frac{1 + \langle 1_K|[F_3]|1_K\rangle\mathcal{K}^{\mathrm{iso}}}{\det[F_3]}\,, \tag{F.3}$$

where we have used $\det(1+M) = \exp\operatorname{tr}\ln(1+M)$, expanded in $M$, used the cyclicity of the trace, and resummed. The isotropic approximation consists of keeping only the solutions arising from the numerator on the right-hand side of eq. (F.3), i.e. those satisfying

$$F_3^{\mathrm{iso}} \equiv \langle 1_K|[F_3]|1_K\rangle = -1/\mathcal{K}^{\mathrm{iso}}\,. \tag{F.4}$$

---

[29]Note that $[F_3]^{-1} = [F_3^{-1}]$ because of the cubic symmetry of the components of $F_3$.

It follows from eq. (F.3) that any solution in the isotropic approximation is also a solution in the low-energy $A_1^+$ approximation, barring an accidental, and unexpected, juxtaposition with a zero of $\det([F_3])$.[30] Thus, aside from this caveat, which appears to be irrelevant in practice, all solutions to the low-energy $A_1^+$ approximation that require a nonzero $\mathcal{K}^{\text{iso}}$ are also obtained in the isotropic approximation.

What are lost in the isotropic approximation are solutions to the quantization condition (F.1) that arise when an eigenvector of $F_3$ diverges (so that $\det([F_3]) \to \infty$) while $F_3^{\text{iso}}$ remains finite. This requires that the corresponding eigenvector of $F_3$ is orthogonal to $|1_K\rangle$. In our experience, this only happens for solutions that occur at free energies (which, we recall, means one of the energies of three noninteracting particles in the given volume), although we do not know of a fundamental reason why this should be so. Furthermore, it was found numerically in ref. [13] that there are no free solutions in the isotropic approximation. Taken together, these observations suggest that the isotropic approximation picks out all the non-free solutions to the quantization condition obtained in the low-energy $A_1^+$ approximation.

In the remainder of this appendix we explain analytically the result found numerically in ref. [13], namely that there are no free solutions in the isotropic approximation. As discussed in section 4.4.3, such solutions occur first at $E = E_1^{\text{free}}$, and there yield a double pole in $\det(F_3)$ lying in the space spanned by $|x_1'\rangle$, eq. (4.22). This pole is, however, absent in the isotropic approximation because $\langle 1_K | x_1' \rangle = 0$, so the pole is removed from $F_3^{\text{iso}}$.

Our aim is to generalize this argument to any excited free energy. We will do so for $\vec{P} = 0$, and for an excited state in which the three momenta, labeled $\vec{k}$, $\vec{p}$ and $\vec{b} = -\vec{k} - \vec{a}$, lie in different shells, e.g. $\vec{k} = k_L(0,0,1)$, $\vec{p} = k_L(1,1,0)$ and $\vec{b} = k_L(-1,-1,-1)$, with $k_L = 2\pi/L$. We denote the degeneracies of these shells by $N_1$, $N_2$, and $N_3$, respectively (6, 12 and 8 in our example). For each choice of $\vec{k}$ from shell 1, we define $N_{12}$ as the number of choices of $\vec{p}$ from shell 2 that can lead to a free solution, and define $N_{13}$ analogously. By cubic symmetry $N_{12}$ and $N_{13}$ do not depend on the choice of $\vec{k}$ from shell 1. Clearly we have $N_{12} = N_{13}$, since each solution contains both a $\vec{p}$ and $\vec{b}$. We define $N_{23} = N_{21}$ and $N_{31} = N_{32}$ analogously. The total degeneracy of free-particle solutions is then

$$N_{\text{sol}} = N_1 N_{12} = N_2 N_{23} = N_3 N_{31}\,. \tag{F.5}$$

As above, we denote the $\ell = 0$, $A_1^+$ parts of $\widetilde{F}$ and $\widetilde{G}$ by $[\widetilde{F}]$ and $[\widetilde{G}]$, which are indexed by the shell number. The poles in these matrices occur only when both indices lie in one of the three shells discussed above, and thus we can focus on this three-dimensional subspace. The matrices in this subspace have the form

$$[\widetilde{F}] = p \begin{pmatrix} N_{12} & 0 & 0 \\ 0 & N_{23} & 0 \\ 0 & 0 & N_{31} \end{pmatrix} + \mathcal{O}(1) \;\; \text{and} \;\; [\widetilde{G}] = p \begin{pmatrix} 0 & \sqrt{N_{12}N_{23}} & \sqrt{N_{12}N_{31}} \\ \sqrt{N_{23}N_{12}} & 0 & \sqrt{N_{23}N_{31}} \\ \sqrt{N_{31}N_{12}} & \sqrt{N_{31}N_{23}} & 0 \end{pmatrix} + \mathcal{O}(1)\,, \tag{F.6}$$

---

[30]This holds also when $\mathcal{K}^{\text{iso}} \to 0$, for then a solution to eq. (F.4) implies that $[F_3]$ has a diverging eigenvalue, and thus that $\det([F_3^{-1}]) \to 0$.

where

$$p = \frac{1}{8L^3\omega_k\omega_p\omega_b(E - \omega_k - \omega_p - \omega_b)}\,. \tag{F.7}$$

The coefficients in $[\widetilde{F}]$ count the number of choices of $\vec{a}$ in eq. (A.9) that lead to the pole. For example, for the $(1,1)$ element, there are $N_{12} + N_{13} = 2N_{12}$ choices, which combines with the overall factor of $1/2$ in $\widetilde{F}$ to give the quoted result $N_{12}$. To understand the form of $[\widetilde{G}]$ consider the $(1,2)$ element of the pole part. This arises from each of the $N_{\rm sol}$ solutions, multiplied by the normalization factors for the $A_1^+$ projections, $1/\sqrt{N_1 N_2}$. Then we use

$$\frac{N_{\rm sol}}{\sqrt{N_1 N_2}} = \sqrt{\frac{N_1 N_{12} N_2 N_{23}}{N_1 N_2}} = \sqrt{N_{12} N_{23}} \tag{F.8}$$

to obtain the quoted result.

Combining, we find that the pole part of $H$ lives in a one-dimensional subspace,

$$[H] = [\widetilde{F}] + [\widetilde{G}] + [1/(2\omega\mathcal{K}_2)] = |W_1\rangle\lambda p\langle W_1| + \mathcal{O}(1)\,, \tag{F.9}$$

$$\langle W_1| = \left(\sqrt{\frac{N_{12}}{\lambda}}, \sqrt{\frac{N_{23}}{\lambda}}, \sqrt{\frac{N_{31}}{\lambda}}\right)\,, \quad \lambda = N_{12} + N_{23} + N_{31}\,. \tag{F.10}$$

Here we are assuming that $\mathcal{K}_2$ does not have a zero at $E = E_1^{\rm free}$. It follows from eq. (F.9) that $[H]^{-1}$ has the form (see, e.g., eq. (C14) of ref. [1]):

$$[H]^{-1} = |W_1\rangle\frac{1 + \mathcal{O}(1/p)}{\lambda p}\langle W_1| + \mathcal{O}(1/p)\sum_{i\neq 1}(|W_1\rangle\langle W_i| + |W_i\rangle\langle W_1|) + \sum_{i,j\neq 1}|W_i\rangle\mathcal{O}(1)\langle W_j|\,. \tag{F.11}$$

Here $|W_2\rangle$ and $|W_3\rangle$ are any choice for the other two members of an orthonormal basis of which $|W_1\rangle$ is a member. Note that only the coefficient of the first term is known; for all other terms only the power of $p$ is known.

We can now calculate the pole part of $F_3^{\rm iso}$, which requires projection with $\langle 1_K|$. Within our subspace

$$\langle 1_K| \longrightarrow \left(\sqrt{N_1}, \sqrt{N_2}, \sqrt{N_3}\right)\,, \tag{F.12}$$

from which it follows that

$$\langle 1_K|[\widetilde{F}] = p\sqrt{\lambda N_{\rm sol}}\langle W_1| + \mathcal{O}(1)\,, \tag{F.13}$$

$$\langle 1_K|[\widetilde{F}]|1_K\rangle = 3pN_{sol} + \mathcal{O}(1)\,, \tag{F.14}$$

$$\langle 1_K|[\widetilde{F}][H]^{-1}[\widetilde{F}]|1_K\rangle = pN_{\rm sol} + \mathcal{O}(1)\,, \tag{F.15}$$

and thus that

$$F_3^{\rm iso} = \frac{1}{L^3}\langle 1_K|\left(\frac{[\widetilde{F}]}{3} - [\widetilde{F}][H]^{-1}[\widetilde{F}]\right)|1_K\rangle = \mathcal{O}(1)\,. \tag{F.16}$$

As claimed, all poles have canceled from $F_3^{\rm iso}$.

It is straightforward to generalize this result to the case that two or more shells are the same, and also to moving frames, i.e. $\vec{P} \neq 0$, although we do not present the details here.

## G Failure of eq. (4.34) for quadratic and cubic terms in the threshold expansion

As noted in the main text, we find numerically that the following results hold,

$$[\mathcal{K}_{\mathrm{df},3}^{(2)}(E_1^{\mathrm{free}})]|x_1\rangle = [\mathcal{K}_{\mathrm{df},3}^{(3)}(E_1^{\mathrm{free}})]|x_1\rangle = 0\,, \tag{G.1}$$

where the superscript on $\mathcal{K}_{\mathrm{df},3}$ indicates the order in the threshold expansion of $\mathcal{K}_{\mathrm{df},3}$. The vector $|x_1\rangle$ is given in eq. (4.24), and the square brackets indicate the $A_1^+$ projection of $\mathcal{K}_{\mathrm{df},3}$ expressed in the $k\ell m$ basis. Our aim here is to give an analytic explanation for these results.

We can rewrite eq. (G.1), using the symmetry of $\mathcal{K}_{\mathrm{df},3}$ and the form of $|x_1\rangle$, as

$$[\mathcal{K}_{\mathrm{df},3}^{(2,3)}]_{1i} = \sqrt{\frac{1}{6}}[\mathcal{K}_{\mathrm{df},3}^{(2,3)}]_{2i} + \sqrt{\frac{5}{6}}[\mathcal{K}_{\mathrm{df},3}^{(2,3)}]_{3i} \quad \text{at} \quad E = E_1^{\mathrm{free}}, \quad \forall i\,. \tag{G.2}$$

The ordering of the indices is given in eq. (4.23). We recall that the $\sqrt{6}$ here arises because the first shell has 6 elements, while the $\sqrt{5}$ arises because $Y_{20}(\hat{z}) = \sqrt{5}Y_{00}$. The superscript on $\mathcal{K}_{\mathrm{df},3}$ indicates that the equation should hold for both the quadratic and cubic terms in the threshold expansion.

We wish to demonstrate eq. (G.2) for any choice of $i$. To do so we first change notation, recalling from section 2.4 that the $\vec{k}, \ell, m$ indices can be replaced by dependence on $\vec{k}, \hat{a}^*$. Here we are also replacing the spectator-momentum index $k$ with $\vec{k}$, both in order to be more explicit, and because $\mathcal{K}_{\mathrm{df},3}$ is an infinite-volume quantity that is defined for all $\vec{k}$. At first, we make this change only for the initial-state indices, leading to the hybrid notation $\mathcal{K}_{\mathrm{df},3}(E; \vec{p}, \ell', m'; \vec{k}, \hat{a}^*)$.[31] In terms of this new quantity, we claim that eq. (G.2) holds for any choice of the index $i$ if

$$\begin{aligned}
\mathcal{K}_{\mathrm{df},3}^{(2,3)}(E_1^{\mathrm{free}}; \vec{0}, 0, 0; \vec{k}, \hat{a}^*) &+ c\,\mathcal{K}_{\mathrm{df},3}^{(2,3)}(E_1^{\mathrm{free}}; \vec{0}, 2, 0; \vec{k}, \hat{a}^*) \\
&= \mathcal{K}_{\mathrm{df},3}^{(2,3)}(E_1^{\mathrm{free}}; k_L\hat{z}, 0, 0; \vec{k}, \hat{a}^*) + \sqrt{5}\mathcal{K}_{\mathrm{df},3}^{(2,3)}(E_1^{\mathrm{free}}; k_L\hat{z}, 2, 0; \vec{k}, \hat{a}^*)\,,
\end{aligned} \tag{G.3}$$

is valid for all $\vec{k}$ and $\hat{a}^*$, and for one choice of $c$. To understand this, first note that (G.3) applies for an arbitrary initial state, and this subsumes all possible values of the finite-volume index $i$. As for the final state, to obtain eq. (G.2) we need to project onto the $A_1^+$ irrep. Doing so, the second term on the left-hand side of eq. (G.3) vanishes, as can be seen from the absence of an $\ell = 2$ entry in the $A_1^+$ row of the (000) shell column in table 1. This is why it is sufficient if eq. (G.3) holds for one value of $c$. The $A_1^+$ projections of the remaining three terms in eq. (G.3) leads to the three terms in eq. (G.2). The averaging over the first shell leads to the factors of $\sqrt{6}$ in the latter result. Note that to perform this averaging one must also use the rotation invariance of $\mathcal{K}_{\mathrm{df},3}$. It is also important that $m' = 0$ in the last term in eq. (G.3), since this is the component that lives in the $A_1^+$ irrep when the spectator momentum lies in the $\hat{z}$ direction.

---

[31]We are abusing notation by using the same name, $\mathcal{K}_{\mathrm{df},3}$, for the function expressed in terms of different variables, but the number of indices uniquely determines which choice of basis we are using.

In the following, we demonstrate that eq. (G.3) holds if $c = \sqrt{5}$. There are three inputs needed for this demonstration. The first is the observation that the same configuration of final-state particles can contribute to both sides of eq. (G.3). To explain this we need to write both initial and final states in the form used prior to their decomposition into harmonics, so that we have $\mathcal{K}_{\mathrm{df},3}(E; \vec{p}, \hat{a}'^*; \vec{k}, \hat{a}^*)$. Then one can show, using permutation symmetry alone, that

$$\mathcal{K}_{\mathrm{df},3}(E_1^{\mathrm{free}}; \vec{0}, \hat{z}; \vec{k}, \hat{a}^*) = \mathcal{K}_{\mathrm{df},3}(E_1^{\mathrm{free}}; k_L \hat{z}, \hat{z}; \vec{k}, \hat{a}^*) \,. \tag{G.4}$$

This result holds for any term in the threshold expansion of $\mathcal{K}_{\mathrm{df},3}$ (or, indeed, for the entire quantity), and thus we do not include a superscript. To understand eq. (G.4), note that the three particles in the final state have momenta $\vec{0}$, $k_L \hat{z}$ and $-k_L \hat{z}$. Calling $\vec{0}$ the spectator momentum yields the left-hand side of eq. (G.4), while calling $k_L \hat{z}$ the spectator yields the right-hand side. Since both choices describe the same momentum configuration, they must be equivalent due to the permutation symmetry of $\mathcal{K}_{\mathrm{df},3}$.

The second input is that $\mathcal{K}_{\mathrm{df},3}^{(2,3)}$ is either independent of, or quadratic in, $\hat{a}'^*$. This is explained in section 2.4, and is in one-to-one correspondence with the fact that only $s$- and $d$-waves contribute.

The final key input concerns angular averaging of a quadratic form:

$$(\hat{n}_i \hat{n}_j V_{ij})\big|_{\ell=0} + \sqrt{5}\,(\hat{n}_i \hat{n}_j V_{ij})\big|_{\ell=2,m=0} = \frac{1}{3} V_{ii} + \frac{1}{3}\,(2V_{33} - V_{11} - V_{22}) = V_{33} \,, \tag{G.5}$$

where $V_{ij}$ is an arbitrary tensor. In other words, the combination appearing on the left-hand side can be evaluated by setting $\hat{n} = \hat{z}$. The same is trivially true for a quantity that is independent of $\hat{n}$.

Combining the second and third key inputs, we deduce that

$$\mathcal{K}_{\mathrm{df},3}^{(2,3)}(E; \vec{p}, 0, 0; \vec{k}, \hat{a}^*) + \sqrt{5}\mathcal{K}_{\mathrm{df},3}^{(2,3)}(E; \vec{p}, 2, 0; \vec{k}, \hat{a}^*) = \mathcal{K}_{\mathrm{df},3}^{(2,3)}(E; \vec{p}, \hat{a}'^* = \hat{z}; \vec{k}, \hat{a}^*) \tag{G.6}$$

holds for any choice of $E$ and $\vec{p}$. Applying this to both sides of eq. (G.3), with $E = E_1^{\mathrm{free}}$, and $\vec{p} = \vec{0}$ for the left-hand side and $\vec{p} = k_1 \hat{z}$ for the right-hand side, we find that eq. (G.3) with $c = \sqrt{5}$ is equivalent to the first key identity eq. (G.4). This establishes the desired result.

This derivation will fail for terms of quartic and higher order in $\mathcal{K}_{\mathrm{df},3}$, since the combination of $\ell' = 0$ and 2 parts that appears in eq. (G.6) will no longer allow the replacement of $\hat{a}'^*$ with $\hat{z}$, implying that eq. (G.4) cannot be used. For example, considering one of the terms that arises in quartic terms, we find

$$(\hat{a}'^* \cdot \hat{n})^4\big|_{\ell'=0} + \sqrt{5}(\hat{a}'^* \cdot \hat{n})^4\big|_{\ell'=2,m'=0} \neq \hat{n}_z^4 \,. \tag{G.7}$$

We have checked this numerically by decomposing the simplest of the quartic terms and finding that eq. (G.1) does not hold.

# References

[1] M.T. Hansen and S.R. Sharpe, *Relativistic, model-independent, three-particle quantization condition*, *Phys. Rev.* **D 90** (2014) 116003 [arXiv:1408.5933] [inSPIRE].

[2] M.T. Hansen and S.R. Sharpe, *Expressing the three-particle finite-volume spectrum in terms of the three-to-three scattering amplitude*, *Phys. Rev.* **D 92** (2015) 114509 [arXiv:1504.04248] [inSPIRE].

[3] H.-W. Hammer, J.-Y. Pang and A. Rusetsky, *Three-particle quantization condition in a finite volume: 1. The role of the three-particle force*, *JHEP* **09** (2017) 109 [arXiv:1706.07700] [inSPIRE].

[4] H.W. Hammer, J.Y. Pang and A. Rusetsky, *Three particle quantization condition in a finite volume: 2. general formalism and the analysis of data*, *JHEP* **10** (2017) 115 [arXiv:1707.02176] [inSPIRE].

[5] R.A. Briceño, M.T. Hansen and S.R. Sharpe, *Relating the finite-volume spectrum and the two-and-three-particle S matrix for relativistic systems of identical scalar particles*, *Phys. Rev.* **D 95** (2017) 074510 [arXiv:1701.07465] [inSPIRE].

[6] M. Mai and M. Döring, *Three-body Unitarity in the Finite Volume*, *Eur. Phys. J.* **A 53** (2017) 240 [arXiv:1709.08222] [inSPIRE].

[7] R.A. Briceño, M.T. Hansen and S.R. Sharpe, *Three-particle systems with resonant subprocesses in a finite volume*, *Phys. Rev.* **D 99** (2019) 014516 [arXiv:1810.01429] [inSPIRE].

[8] M.T. Hansen and S.R. Sharpe, *Lattice QCD and Three-particle Decays of Resonances*, arXiv:1901.00483 [inSPIRE].

[9] Hadron Spectrum collaboration, *Toward the excited isoscalar meson spectrum from lattice QCD*, *Phys. Rev.* **D 88** (2013) 094505 [arXiv:1309.2608] [inSPIRE].

[10] J. Bulava, B. Fahy, B. Hörz, K.J. Juge, C. Morningstar and C.H. Wong, $I = 1$ *and* $I = 2$ $\pi - \pi$ *scattering phase shifts from* $N_f = 2 + 1$ *lattice QCD*, *Nucl. Phys.* **B 910** (2016) 842 [arXiv:1604.05593] [inSPIRE].

[11] F. Romero-López, A. Rusetsky and C. Urbach, *Two- and three-body interactions in* $\varphi^4$ *theory from lattice simulations*, *Eur. Phys. J.* **C 78** (2018) 846 [arXiv:1806.02367] [inSPIRE].

[12] R.A. Briceno, J.J. Dudek and R.D. Young, *Scattering processes and resonances from lattice QCD*, *Rev. Mod. Phys.* **90** (2018) 025001 [arXiv:1706.06223] [inSPIRE].

[13] R.A. Briceño, M.T. Hansen and S.R. Sharpe, *Numerical study of the relativistic three-body quantization condition in the isotropic approximation*, *Phys. Rev.* **D 98** (2018) 014506 [arXiv:1803.04169] [inSPIRE].

[14] M. Döring, H.W. Hammer, M. Mai, J.Y. Pang, A. Rusetsky and J. Wu, *Three-body spectrum in a finite volume: the role of cubic symmetry*, *Phys. Rev.* **D 97** (2018) 114508 [arXiv:1802.03362] [inSPIRE].

[15] M. Mai and M. Döring, *Finite-Volume Spectrum of* $\pi^+\pi^+$ *and* $\pi^+\pi^+\pi^+$ *Systems*, *Phys. Rev. Lett.* **122** (2019) 062503 [arXiv:1807.04746] [inSPIRE].

[16] M. Lüscher, *Volume Dependence of the Energy Spectrum in Massive Quantum Field Theories. 2. Scattering States*, *Commun. Math. Phys.* **105** (1986) 153 [inSPIRE].

[17] M. Lüscher, *Two particle states on a torus and their relation to the scattering matrix*, *Nucl. Phys.* **B 354** (1991) 531 [INSPIRE].

[18] C.W. Andersen, J. Bulava, B. Hörz and C. Morningstar, *Elastic I = 3/2 p-wave nucleon-pion scattering amplitude and the Δ(1232) resonance from $N_f = 2 + 1$ lattice QCD*, *Phys. Rev.* **D 97** (2018) 014506 [arXiv:1710.01557] [INSPIRE].

[19] A. Woss, C.E. Thomas, J.J. Dudek, R.G. Edwards and D.J. Wilson, *Dynamically-coupled partial-waves in ρπ isospin-2 scattering from lattice QCD*, *JHEP* **07** (2018) 043 [arXiv:1802.05580] [INSPIRE].

[20] T.D. Blanton, R.A. Briceño, M.T. Hansen, F. Romero-López and S.R. Sharpe, *Progress report on the relativistic three-particle quantization condition*, in *36th International Symposium on Lattice Field Theory (Lattice 2018)*, East Lansing, MI, United States, July 22–28, 2018 (2018) [arXiv:1810.06634] [INSPIRE].

[21] M.T. Hansen and S.R. Sharpe, *Perturbative results for two and three particle threshold energies in finite volume*, *Phys. Rev.* **D 93** (2016) 014506 [arXiv:1509.07929] [INSPIRE].

[22] S.R. Sharpe, *Testing the threshold expansion for three-particle energies at fourth order in $\phi^4$ theory*, *Phys. Rev.* **D 96** (2017) 054515 [*Erratum ibid.* **D 98** (2018) 099901] [arXiv:1707.04279] [INSPIRE].

[23] M.T. Hansen and S.R. Sharpe, *Threshold expansion of the three-particle quantization condition*, *Phys. Rev.* **D 93** (2016) 096006 [*Erratum ibid.* **D 96** (2017) 039901] [arXiv:1602.00324] [INSPIRE].

[24] P.W. Atkins, M.S. Child and C.S.G. Phillips, *Tables for group theory*, vol. 6, Oxford University Press, Oxford (1970).

[25] H. Georgi, *Lie Algebras In Particle Physics. From Isospin To Unified Theories*, *Front. Phys.* **54** (1982) 1 [INSPIRE].

[26] S.R. Beane, W. Detmold and M.J. Savage, *n-Boson Energies at Finite Volume and Three-Boson Interactions*, *Phys. Rev.* **D 76** (2007) 074507 [arXiv:0707.1670] [INSPIRE].

[27] S. Tan, *Three-boson problem at low energy and implications for dilute Bose-Einstein condensates*, *Phys. Rev.* **A 78** (2008) 013636 [arXiv:0709.2530] [INSPIRE].

[28] T. Luu and M.J. Savage, *Extracting Scattering Phase-Shifts in Higher Partial-Waves from Lattice QCD Calculations*, *Phys. Rev.* **D 83** (2011) 114508 [arXiv:1101.3347] [INSPIRE].

[29] P.M.A. Mestrom, J. Wang, C.H. Greene and J.P. D'Incao, *Efimov-van der Waals universality for ultracold atoms with positive scattering lengths*, *Phys. Rev.* **A 95** (2017) 032707.

[30] J. Wang, J.P. D'Incao, Y. Wang and C.H. Greene, *Universal three-body recombination via resonant d-wave interactions*, *Phys. Rev.* **A 86** (2012) 062511.

[31] V. Efimov, *Energy levels arising form the resonant two-body forces in a three-body system*, *Phys. Lett.* **33B** (1970) 563 [INSPIRE].

[32] U.-G. Meissner, G. Ríos and A. Rusetsky, *Spectrum of three-body bound states in a finite volume*, *Phys. Rev. Lett.* **114** (2015) 091602 [*Erratum ibid.* **117** (2016) 069902] [arXiv:1412.4969] [INSPIRE].

[33] P.F. Bedaque, H.W. Hammer and U. van Kolck, *Renormalization of the three-body system with short range interactions*, *Phys. Rev. Lett.* **82** (1999) 463 [nucl-th/9809025] [INSPIRE].

[34] F.J. Yndurain, *Low-energy pion physics*, hep-ph/0212282 [INSPIRE].

[35] NPLQCD collaboration, *The I = 2ππ S-wave Scattering Phase Shift from Lattice QCD*, *Phys. Rev.* **D 85** (2012) 034505 [`arXiv:1107.5023`] [ɪɴSPIRE].

[36] J.J. Dudek, R.G. Edwards and C.E. Thomas, *S and D-wave phase shifts in isospin-2 ππ scattering from lattice QCD*, *Phys. Rev.* **D 86** (2012) 034031 [`arXiv:1203.6041`] [ɪɴSPIRE].

[37] Z. Fu, *Lattice QCD study of the s-wave ππ scattering lengths in the I = 0 and 2 channels*, *Phys. Rev.* **D 87** (2013) 074501 [`arXiv:1303.0517`] [ɪɴSPIRE].

[38] T. Kurth, N. Ishii, T. Doi, S. Aoki and T. Hatsuda, *Phase shifts in I = 2ππ-scattering from two lattice approaches*, *JHEP* **12** (2013) 015 [`arXiv:1305.4462`] [ɪɴSPIRE].

[39] ETM collaboration, *Hadron-hadron interactions from $N_f = 2 + 1 + 1$ lattice QCD: isospin-2 ππ scattering length*, *JHEP* **09** (2015) 109 [`arXiv:1506.00408`] [ɪɴSPIRE].

[40] J.-Y. Pang, J.-J. Wu, H.W. Hammer, U.-G. Meissner and A. Rusetsky, *Energy shift of the three-particle system in a finite volume*, `arXiv:1902.01111` [ɪɴSPIRE].

[41] W. Detmold et al., *Multi-Pion States in Lattice QCD and the Charged-Pion Condensate*, *Phys. Rev.* **D 78** (2008) 014507 [`arXiv:0803.2728`] [ɪɴSPIRE].

[42] C.h. Kim, C.T. Sachrajda and S.R. Sharpe, *Finite-volume effects for two-hadron states in moving frames*, *Nucl. Phys.* **B 727** (2005) 218 [`hep-lat/0507006`] [ɪɴSPIRE].

**THE EUROPEAN
PHYSICAL JOURNAL C**

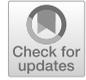

# Large $N_c$ scaling of meson masses and decay constants

P. Hernández[1], C. Pena[2], F. Romero-López[1,a]

[1] IFIC (CSIC-UVEG), Edificio Institutos Investigación, Apt. 22085, 46071 Valencia, Spain
[2] Departamento de Física Teórica and Instituto de Física Teórica UAM-CSIC, Universidad Autónoma de Madrid, 28049 Madrid, Spain

**Abstract** We perform an ab initio calculation of the $N_c$ scaling of the low-energy couplings of the chiral Lagrangian of low-energy strong interactions, extracted from the mass dependence of meson masses and decay constants. We compute these observables on the lattice with four degenerate fermions, $N_f = 4$, and varying number of colours, $N_c = 3$–6, at a lattice spacing of $a \simeq 0.075$ fm. We find good agreement with the expected $N_c$ scaling and measure the coefficients of the leading and subleading terms in the large $N_c$ expansion. From the subleading $N_c$ corrections, we can also infer the $N_f$ dependence, that we use to extract the value of the low-energy couplings for different values of $N_f$. We find agreement with previous determinations at $N_c = 3$ and $N_f = 2, 3$ and also, our results support a strong paramagnetic suppression of the chiral condensate in moving from $N_f = 2$ to $N_f = 3$.

## 1 Introduction

The 't Hooft limit of QCD [1] is well known to capture correctly most of its non-perturbative features, such as confinement and chiral symmetry breaking. Large $N_c$ inspired approximations are often employed in phenomenological approaches to hadron physics [2–11], but systematic errors from subleading $N_c$ corrections are only naively estimated.

Lattice Field Theory offers the possibility of ab initio explorations of the large $N_c$ limit of QCD, by simulating at different values of $N_c$ [12,13]. Several studies have already been performed. In Ref. [13] a thorough study of mesonic two-point functions was carried out in the quenched approximation, a limit that captures correctly the leading order terms in $N_c$, but modifies subleading corrections in an uncontrolled way. Furthermore, in Ref. [14] a similar study was performed for $N_c = 2$–5 using $N_f = 2$ dynamical fermions at rather high pion masses.

In addition to the standard approach, the study of QCD in the large $N_c$ limit can also be achieved using reduced models (see [15] for a review). In this context, there has been significant progress regarding the properties of mesons [16–20].

Besides, lattice simulations have been used to perform studies of various observables in theories with different number of colours, flavours or fermion representations in the context of Beyond-the-Standard-Model theories. Some recent results can be found in [21–26] and for recent reviews see [27,28].

In this work, we use previously generated lattice configurations with $N_c = 3$–6 and four dynamical fermions. Our particular choice of $N_f$ has also advantages for weak matrix elements [29]. On these ensembles, we compute meson masses and decay constants as a function of the quark mass at the different values of $N_c$. We fit these to chiral perturbation theory (ChPT) in order to extract the leading order and next-to-leading order low-energy chiral couplings (LECs). We then study their $N_c$ scaling and extract the first two terms in the 't Hooft series. Our study builds on previous lattice determinations of the LECs for $N_c = 3$ [30–44], whose main results are summarized in [45].

Interestingly, within the large $N_c$ expansion, the $1/N_c$ corrections have a well-defined linear dependence on $N_f$, while the 't Hooft limit is independent on $N_f$. Using this fact, we can predict the low-energy couplings at different values of $N_f$ up to higher orders in $N_c$. This allows us to compare with previous determinations, and check the prediction of *paramagnetic* suppression at large $N_f$ of Refs. [46,47].

This paper is organized as follows. First, we describe chiral perturbation theory predictions and the relation to the large $N_c$ limit in Sect. 2. In Sect. 3, we present the lattice setup that involves a mixed-action formulation. Next, we explain our scale setting procedure at different $N_c$ consistent with 't Hooft scaling in Sect. 4. In Sect. 5 we present the results of our chiral fits to the meson mass and decay constant, first at fixed $N_c$ and then combined with the large $N_c$ expansion.

a e-mail: fernando.romero@uv.es

We also present results for theories with different values of $N_f$, compare with previous literature and discuss systematic uncertainties. We conclude in Sect. 6.

## 2 Chiral perturbation Theory predictions

The light spectrum of QCD is the result of the pattern of spontaneous chiral symmetry breaking, $SU(N_f)_L \times SU(N_f)_R \rightarrow SU(N_f)_{L+R}$. ChPT represents accurately the dynamics of the expected pseudo-Nambu-Goldstone bosons (pNGB), i.e., the lightest non-singlet multiplet of pseudoscalar mesons (the octet for $N_f = 3$), at sufficiently small quark masses. The increase in the number of colours while keeping the 't Hooft coupling constant, $\lambda = g^2 N_c$, is not expected to modify these features. On the other hand, in the large $N_c$ limit, QCD reduces to a theory of narrow and non-interacting resonances and, as a result, the interactions of pNGB within the effective theory decrease with $N_c$, improving the convergence of the perturbative series. One complication of the large $N_c$ expansion is the role of the singlet pseudoscalar meson, i.e., the $\eta'$. Its mass originates in the explicit $U(1)_A$ breaking by the anomaly. In QCD this contribution to the mass is at the cutoff scale of the chiral effective theory and it is therefore integrated out. However, the anomalous contribution to the singlet mass decreases with $N_c$ and in the large $N_c$ limit the $\eta'$ becomes degenerate with the remaining pNGBs. The effective theory should consequently include an additional singlet pseudoscalar meson in the spectrum. The corresponding effective theory has been studied long ago [48–53]. A new power-counting is needed which involves a simultaneous expansion in $1/N_c$ and the usual chiral expansion in the quark mass and momenta. A consistent power counting was implemented in Refs. [52,53]:

$$\mathcal{O}(\delta) \sim \mathcal{O}(p^2) \sim \mathcal{O}(m_q) \sim \mathcal{O}(m_\pi^2) \sim \mathcal{O}(N_c^{-1}). \quad (1)$$

In the following we will concentrate on the non-singlet multiplet masses and decay constants. We now compare the usual $SU(N_f)$ ChPT to the $U(N_f)$ ChPT for these observables.

### 2.1 $SU(N_f)$ effective theory

At Leading Order (LO) in the standard $SU(N_f)$ chiral expansion there are only two couplings for any number of degenerate flavours, related to the chiral condensate and the meson decay constant. At Next-to-Leading Order (NLO), and for an arbitrary number of degenerate flavours ($N_f > 3$), 13 more LECs are needed, but only two combinations enter in the observables of interest. For $N_f$ degenerate flavours, ChPT predicts at NLO [54–56]:

$$F_\pi = F \left[ 1 - \frac{N_f}{2} \frac{M_\pi^2}{(4\pi F_\pi)^2} \log \frac{M_\pi^2}{\mu^2} \right. $$
$$\left. + 4 \frac{M_\pi^2}{F_\pi^2} \left( L_5^r + N_f L_4^r \right) \right], \quad (2)$$

and

$$M_\pi^2 = 2Bm \left[ 1 + \frac{1}{N_f} \frac{M_\pi^2}{(4\pi F_\pi)^2} \log \frac{M_\pi^2}{\mu^2} \right. $$
$$\left. + 8 \frac{M_\pi^2}{F_\pi^2} \left( 2L_8^r - L_5^r + N_f(2L_6^r - L_4^r) \right) \right], \quad (3)$$

in terms of the LO couplings, $B$, $F$, and the NLO Gasser–Leutwyler coefficients, $L_{4,5,6,8}^r(\mu)$, defined at the renormalization scale $\mu$.

Equations (2–3) are valid for an arbitrary number of colours, but the LECs scale with $N_c$ as (for a review see [57]):

$$O(N_c): F^2, L_5, L_8; \ O(1): B, L_4, L_6. \quad (4)$$

Loop corrections are suppressed in $1/F_\pi^2 = O(1/N_c)$, and hence the loop expansion is expected to converge better at larger $N_c$.

Keeping only leading and subleading dependence on $N_c$ a convenient parametrization is

$$F = \sqrt{N_c} \left( F_0 + \frac{F_1}{N_c} \right), \qquad B = B_0 + \frac{B_1}{N_c}, \quad (5)$$

and

$$L_5 + N_f L_4 \equiv L_F = N_c L_F^{(0)} + L_F^{(1)}, \quad (6)$$

$$2L_8 - L_5 + N_f(2L_6 - L_4) \equiv L_M = N_c L_M^{(0)} + L_M^{(1)}. \quad (7)$$

Note that according to the scaling of Eq. (4) and the definition of Eq. (7):

$$L_F^{(0)} = \frac{L_5}{N_c} + \mathcal{O}\left(\frac{1}{N_c}\right),$$
$$L_M^{(0)} = \frac{2L_8 - L_5}{N_c} + \mathcal{O}\left(\frac{1}{N_c}\right). \quad (8)$$

The NNLO Lagrangian of the $SU(N_f)$ theory is also known [54–56]. At this order we will instead only use the $U(N_f)$, to which we now turn.

### 2.2 $U(N_f)$ effective theory

In the $U(N_f)$ ChPT at NLO, i.e., $\mathcal{O}(\delta^1)$, the result can be read from Eqs. (2) and (3) and the different $N_c$ scalings of the LECs in Eqs. (5) and (7):

$$F_\pi = \sqrt{N_c} \left( F_0 + \frac{F_1}{N_c} \right) \left[ 1 + 4 \frac{M_\pi^2}{F_\pi^2} N_c L_F^{(0)} + \mathcal{O}(\delta^2) \right], \quad (9)$$

and

$$M_\pi^2 = 2\left(B_0 + \frac{B_1}{N_c}\right) m \left[1 + 8\frac{M_\pi^2}{F_\pi^2} N_c L_M^{(0)} + \mathcal{O}(\delta^2)\right]. \tag{10}$$

The NLO corrections are not enough to explain the data in this case, therefore going to NNLO is essential. At NNLO new features appear, because the singlet contributes to the mass loop corrections. The necessary results can be found in Ref. [58]. For degenerate flavours, they simplify to:

$$F_\pi = \sqrt{N_c}\left(F_0 + \frac{F_1}{N_c} + \frac{F_2}{N_c^2}\right)\left[1 - \frac{N_f}{2}\frac{M_\pi^2}{(4\pi F_\pi)^2}\log\frac{M_\pi^2}{\mu^2}\right.$$
$$+ 4\frac{M_\pi^2}{F_\pi^2}\left(N_c L_F^{(0)} + L_F^{(1)}\right) + N_c^2 K_F^{(0)}\left(\frac{M_\pi^2}{F_\pi^2}\right)^2$$
$$\left. + \mathcal{O}(\delta^3)\right], \tag{11}$$

and

$$M_\pi^2 = 2m\left(B_0 + \frac{B_1}{N_c} + \frac{B_2}{N_c^2}\right)$$
$$\times\left[1 + \frac{1}{N_f}\frac{M_\pi^2}{(4\pi F_\pi)^2}\log\frac{M_\pi^2}{\mu^2} - \frac{1}{N_f}\frac{M_{\eta'}^2}{(4\pi F_\pi)^2}\log\frac{M_{\eta'}^2}{\mu^2}\right.$$
$$\left. + 8\frac{M_\pi^2}{F_\pi^2}\left(N_c L_M^{(0)} + L_M^{(1)}\right) + N_c^2 K_M^{(0)}\left(\frac{M_\pi^2}{F_\pi^2}\right)^2 + \mathcal{O}(\delta^3)\right], \tag{12}$$

where $K_{F,M}^{(0)}$ are combinations of $L_F^{(0)}$, $L_M^{(0)}$ and new LECs that appear in the $U(N_f)$ case. For details see [58]. Note that for degenerate quarks, there is no $\eta$-$\eta'$ mixing.

The $\eta'$ mass in this expression can be taken in the large $N_c$ limit, where it is given by the Witten–Veneziano formula:

$$M_{\eta'}^2 = M_\pi^2 + \frac{2N_f}{F^2}\chi_t \equiv M_\pi^2 + M_0^2, \tag{13}$$

where $\chi_t$ is the topological susceptibility in pure Yang–Mills, recently computed in the large $N_c$ limit in Ref. [59].

Note that even though we use the same notation for the LECs in both chiral expansions, they are different: in the $SU(N_f)$ ChPT the LECs encode the effects of integrating out the $\eta'$. The matching of the two theories starts at NNLO [53,60] and only affects the coupling $B$ and $L_M^{(1)}$ of the above [53,60]:

$$[B]_{SU(N_f)} = [B]_{U(N_f)}\left(1 - \frac{1}{N_f}\frac{M_0^2}{(4\pi F_\pi)^2}\lambda_0\right),$$
$$\left[L_M^{(1)}\right]_{SU(N_f)} = \left[L_M^{(1)}\right]_{U(N_f)} - \frac{1}{8N_f(4\pi)^2}(\lambda_0 + 1), \tag{14}$$

with $\lambda_0 = \log\frac{M_0^2}{\mu^2}$.

### 2.3 $N_f$ versus $N_c$ dependence

A diagrammatic analysis of fermion bilinear two point functions shows that within the large $N_c$ expansion, the leading order $N_c \to \infty$ limit is $N_f$ independent and the NLO is $\mathcal{O}(N_f/N_c)$. We should confirm this expectation also in ChPT formulae above, in particular given the explicit dependence on $N_f$. It turns out that within the $U(N_f)$ expansion, the large $N_c$ expansion yields the expected behaviour: the terms in $1/N_f$ exactly cancel when the large $N_c$ expansion is taken at fixed $M_\pi$. We expect therefore that the LECs should also satisfy this same scaling.

On the other hand within the $SU(N_f)$ expansion or in the $U(N_f)$ when $M_\pi \ll M_{\eta'}$, that is when the chiral limit is taken first, anomalous $1/N_f$ terms appear coming from an expansion in $M_\pi/M_{\eta'}$. In the $U(N_f)$ expansion such dependence is explicit, but in the $SU(N_f)$ it permeates to the LECs which can no longer be assumed to have the expected $\mathcal{O}(N_f/N_c)$ dependence, as can be explicitly seen in the matching of $L_M^{(1)}$ in Eq. (14).

This way, at the order we are working, we can assume the expected scaling in $N_f$ of the $U(N_f)$ and $SU(N_f)$ couplings except in the case of $[L_M^{(1)}]_{SU(N_f)}$.

## 3 Lattice setup

We have generated ensembles for $SU(N_c)$ gauge theory with $N_f = 4$ degenerate dynamical fermions, varying $N_c = 3\text{-}6$, using the HiRep code [61]. Some of them were already presented in Ref. [62]. We have chosen the Iwasaki gauge action (following previous experience with 2+1+1 simulations [63,64]) and $O(a)$-improved[1] Wilson fermions for the sea quarks. Our simulations use the standard Hybrid Monte-carlo (HMC) algorithm with Hasenbusch acceleration. We include five layers in each of the fermionic monomials. Interestingly, we observe that the tuning of the integrator at $N_c = 3$ yields similar results at other values of $N_c$ (at similar pion mass) for the acceptance rate, which we keep at 80–90%. The computational cost of each step in Montecarlo time scales as $\sim N_c^2$, with the advantage of a more efficient parallelization at large $N_c$.

In order to achieve automatic $O(a)$ improvement and avoid the need of a non-perturbative determination of normalization factors, we employ maximally twisted valence

---

[1] For $N_c = 3$, we take the perturbative value of $c_{sw} = 1 + c_{sw}^{(1)} g^2$ from Ref. [65], where we use the plaquette-boosted coupling $g^2 = 2N_c/(\beta P) = O(1/N_c)$. For other values of $N_c$, we use the fact that the one loop coefficient is dominated by the tadpole contribution, which is of order $N_c$ (see Eq. 58 in Ref. [65]). This way, $c_{sw}$ is constant up to subleading corrections in $N_c$, which have an effect of $O(a^2/N_c)$ in physical observables. The full result cannot be easily reconstructed from Ref. [65].

quarks, i.e., the mixed-action setup [66] previously used in Refs. [67–69]. Maximal twist is ensured by tuning the untwisted bare valence mass, $m^v$ to the critical value for which the valence PCAC mass is zero:

$$\lim_{m^v \to m_{cr}} m^v_{pcac} \equiv \lim_{m^v \to m_{cr}} \frac{\partial_0 \langle A^0(x) P(y) \rangle}{2 \langle P(x) P(y) \rangle} = 0, \qquad (15)$$

where $A^\mu(x) = \bar\Psi(x)\gamma^\mu\gamma^5\Psi(x)$ and $P(x) = \bar\Psi(x)\gamma^5\Psi(x)$. The bare twisted mass parameter $\mu_0$ is tuned such that the pion mass in the sea and valence sectors coincide, $M^v_\pi = M^s_\pi$. The normalized meson decay constant $F_\pi$ can then be obtained from the bare combination [70]:

$$F_\pi = \frac{2\mu_0 \langle 0|P|\pi\rangle_{bare}}{M^2_\pi}. \qquad (16)$$

The results for the meson masses and decay constant in the mixed-action setup can be seen in Table 3. We have achieved a good tuning of $m_{pcac}$ and the pseudoscalar masses are compatible within one or two sigma with their pure Wilson value (see Table 1). When the tuning to maximal twist is not perfect, we correct the bare quark mass (and thus $F_\pi$) as follows (see also [70]):

$$a\mu_0 \to a\mu_0 \sqrt{1 + \left(\frac{Z_A am_{pcac}}{a\mu_0}\right)^2}, \qquad (17)$$

$$aF_\pi \to aF_\pi \sqrt{1 + \left(\frac{Z_A am_{pcac}}{a\mu_0}\right)^2}. \qquad (18)$$

where the axial normalization constant, $Z_A$, can be obtained non-perturbatively by matching the valence bare twisted mass with the PCAC mass measured in the sea sector:

$$\mu_0 = Z_A m^s_{pcac}, \text{ for } M^v_\pi = M^s_\pi. \qquad (19)$$

## 4 Scale setting at large $N_c$

The scale setting for different values of $N_c$ is performed using the gradient flow scale $\sqrt{8t_0}$, via the determination of $t_0/a^2$. In QCD, with $N_c = 3$, the standard definition of $t_0$ is:

$$\langle t^2 E(t) \rangle \Big|_{t=t_0} = c = 0.3. \qquad (20)$$

The leading dependence in $N_c$ is known [71] in perturbation theory:

$$\langle t^2 E(t) \rangle = \frac{3}{128\pi^2} \frac{N^2_c - 1}{N_c} \lambda_{GF}(q), \qquad (21)$$

where $\lambda_{GF}(q)$ is the gradient flow 't Hooft coupling at the scale $q = 1/\sqrt{8t}$. Hence, as in Ref. [59], we will generalize $t_0$ to an arbitrary $N_c$ as:

$$\langle t^2 E(t) \rangle \Big|_{t=t_0} = c(N_c) = \frac{3}{8} \frac{N^2_c - 1}{N_c} c(3). \qquad (22)$$

Notice that the choice here is not unique. In particular, one could choose another coupling in a different scheme (such as $\overline{MS}$), and this would induce corrections at order $O(N_f/N_c)$ in dimensionless quantities.

We also need the value of $t_0$ in physical units. This is known from lattice simulations for $N_f = 2$ [72,73] and $N_f = 3$ [74] degenerate quarks and at a reference pion mass $M_{ref} = 420$ MeV:

$$\sqrt{t_0}\Big|^{N_f=2}_{M_{ref}} = 0.1470(14) \text{ fm},$$
$$\sqrt{t_0}\Big|^{N_f=3}_{M_{ref}} = 0.1460(19) \text{ fm} \qquad (23)$$

We can use these to perform a linear extrapolation to $N_f = 4$, motivated by the weak $N_f$ dependence:

$$\sqrt{t_0}\Big|^{N_f=4}_{M_{ref}} = 0.1450(39) \text{ fm}. \qquad (24)$$

Our scale setting condition involves therefore the dimensionless quantity

$$(M_\pi \sqrt{t_0})\Big|_{M_{ref}} = 0.3091(83). \qquad (25)$$

In order to reduce discretization errors we have performed a tree level improvement of $t_0$. In Ref. [75], lattice perturbation theory is used to improve $\langle t^2 E(t) \rangle$ and thus, $t_0$. The prescription is:

$$\langle t^2 E(t) \rangle_a = \langle t^2 E(t) \rangle_{imp} \left[ 1 + \sum_n C_{2n} \left(\frac{a^2}{t}\right)^n \right], \qquad (26)$$

where the coefficients $C_{2n}$ depend on the gauge action, the flow action and the definition of $E(t)$ (clover or plaquette). The coefficients for the Iwasaki gauge action, the plaquette action for the flow and the clover definition of $E(t)$ are:

$$C_2 = -0.262333, \quad C_4 = 0.0936935,$$
$$C_6 = -0.048002, \quad C_8 = 0.0320211. \qquad (27)$$

The numerical results after the improvement, $t^{imp}_0/a^2$, are shown in Table 1.

Finally, Eq. (25) requires $t_0$ at $M_{ref}$. The mass dependence of $t_0$ has been studied in chiral perturbation theory in Ref. [77]. For degenerate flavours it is given by

$$t_0 = t^\chi_0 \left(1 + k M^2\right) + \mathcal{O}(M^4), \qquad (28)$$

where $k \propto 1/(F_\pi)^2 = O(1/N_c)$ and so the chiral dependence is suppressed in $N_c$. We have performed accordingly a linear fit in $M^2$ to extract the reference value. The mass dependence of $t^{imp}_0$ for the different values of $N_c$ can be seen in Fig. 1. As expected, the slope is suppressed with $N_c$. The

**Table 1** Summary of our ensembles: $\beta$, sea quark bare mass parameter, $m^s$, and sea pion mass $M_\pi^s$. We keep $c_{sw} = 1.69$ throughout

| Ensemble | $L^3 \times T$ | $\beta$ | $am^s$ | $aM_\pi^s$ | $t_0^{\mathrm{imp}}/a^2$ |
|---|---|---|---|---|---|
| 3A10 | $20^3 \times 36$ | 1.778 | $-0.4040$ | 0.2204 (21) | 3.263 (50) |
| 3A20 | $24^3 \times 48$ |  | $-0.4060$ | 0.1845 (14) | 3.491 (32) |
| 3A30 | $24^3 \times 48$ |  | $-0.4070$ | 0.1613 (16) | 3.740 (39) |
| 3A40 | $32^3 \times 60$ |  | $-0.4080$ | 0.1429 (12) | 3.855 (27) |
| 4A10 | $20^3 \times 36$ | 3.570 | $-0.3725$ | 0.2035 (14) | 3.494 (45) |
| 4A20 | $24^3 \times 48$ |  | $-0.3752$ | 0.1805 (7) | 3.565 (26) |
| 4A30 | $24^3 \times 48$ |  | $-0.3760$ | 0.1714 (8) | 3.593 (29) |
| 4A40 | $32^3 \times 60$ |  | $-0.3780$ | 0.1397 (8) | 3.723 (23) |
| 5A10 | $20^3 \times 36$ | 5.969 | $-0.3458$ | 0.2128 (17) | 3.532 (17) |
| 5A20 | $24^3 \times 48$ |  | $-0.3490$ | 0.1802 (6) | 3.614 (18) |
| 5A30 | $24^3 \times 48$ |  | $-0.3500$ | 0.1712 (6) | 3.664 (24) |
| 5A40 | $32^3 \times 60$ |  | $-0.3530$ | 0.1331 (7) | 3.776 (19) |
| 6A10 | $20^3 \times 36$ | 8.974 | $-0.3260$ | 0.2150 (17) | 3.619 (17) |
| 6A20 | $24^3 \times 48$ |  | $-0.3300$ | 0.1801 (5) | 3.696 (17) |
| 6A30 | $24^3 \times 48$ |  | $-0.3311$ | 0.1689 (7) | 3.721 (15) |
| 6A40 | $32^3 \times 60$ |  | $-0.3340$ | 0.1351 (6) | 3.820 (17) |

**Table 2** Results for the $t_0/a^2\big|_{M_{\mathrm{ref}}}$ and the lattice spacing as a function of $N_c$. The first error is statistical, the second comes from the uncertainty in $t_0$ in physical units, the third stems from the difference in the definitions of $E(t)$ after improvement, and the fourth are finite volume effects estimated from Ref. [76]

| $N_c$ | $t_0/a^2\big|_{M_{\mathrm{ref}}}$ | $a$ ($\times 10^{-2}$ fm) |
|---|---|---|
| 3 | $3.71(4)(7)_{t_0}(12)_a(3)_L$ | $7.53(4)(19)_{t_0}(12)_a(3)_L$ |
| 4 | $3.64(1)(3)_{t_0}(12)_a(3)_L$ | $7.60(1)(20)_{t_0}(12)_a(3)_L$ |
| 5 | $3.69(2)(3)_{t_0}(12)_a(3)_L$ | $7.54(2)(20)_{t_0}(12)_a(3)_L$ |
| 6 | $3.76(1)(2)_{t_0}(12)_a(3)_L$ | $7.48(1)(20)_{t_0}(12)_a(3)_L$ |

results of the scale setting can be seen in Table 2, where we also include the systematic uncertainties. The leading uncertainty comes from the error on the value of $t_0$ in physical units, the discretization error is estimated from the difference in two definitions of $E(t)$ after improvement, and the finite volume systematic error is estimated from Ref. [76]. As it can be seen, the scale setting yields a uniform lattice spacing for all the values of $N_c$. From now on, we will quote our results in terms of the lattice spacing $a = 0.0754$ fm, corresponding to $N_c = 5$.

## 5 Chiral perturbation theory fits

The results for $M_\pi$ and $F_\pi$ in the mixed-action setup are presented in Table 3. We want to compare these results to the expectations in ChPT described in Sec. 2 in order to the extract the LECs and study their $N_c$ scaling.

Before addressing the fits, we need to explain some technical issues regarding the finite volume effects, the renormal-

ization scale and the fitting strategy. We then perform fits at a fixed value of $N_c$ to test the ansätze for the $N_c$ scaling of the LECs in Eqs. 5 and 7. After that, we perform simultaneous chiral and $N_c$ fits. We present a selection of relevant results for the latter, and conclude the section with a discussion on systematic errors.

### 5.1 Finite volume effects

Our ensembles have $M_\pi L > 3.8$ in all cases so we expect finite volume effects to be small and suppressed as $1/N_c$. Still, we find that for the decay constant they can be of $O(1\%)$ and thus we correct them as [78,79]:

$$M_\pi(L) = M_\pi \left[ 1 + \frac{1}{2N_f} \xi \; \bar{g}_1(M_\pi L) + O(\xi^2) \right], \qquad (29)$$

$$F_\pi(L) = F_\pi \left[ 1 - \frac{N_f}{2} \xi \; \bar{g}_1(M_\pi L) + O(\xi^2) \right], \qquad (30)$$

with $\xi \equiv \frac{M_\pi^2}{(4\pi F_\pi)^2}$, while $\bar{g}_1(x)$ is given by

$$\bar{g}_1(x) \xrightarrow{x \gg 1} \frac{24}{x} K_1(x) \sim \frac{24\sqrt{2}}{\sqrt{\pi}} \frac{e^{-x}}{x^{3/2}}. \qquad (31)$$

We will use the corrected results for the analysis.

### 5.2 Renormalization scale

The NLO couplings are usually defined at $\mu = 4\pi F$ or at the $\rho$ mass, $\mu = M_\rho$. Still, in the context of the large $N_c$ expansion these are two very different choices, since the former scales with $\sqrt{N_c}$, deviating from the physical cutoff of

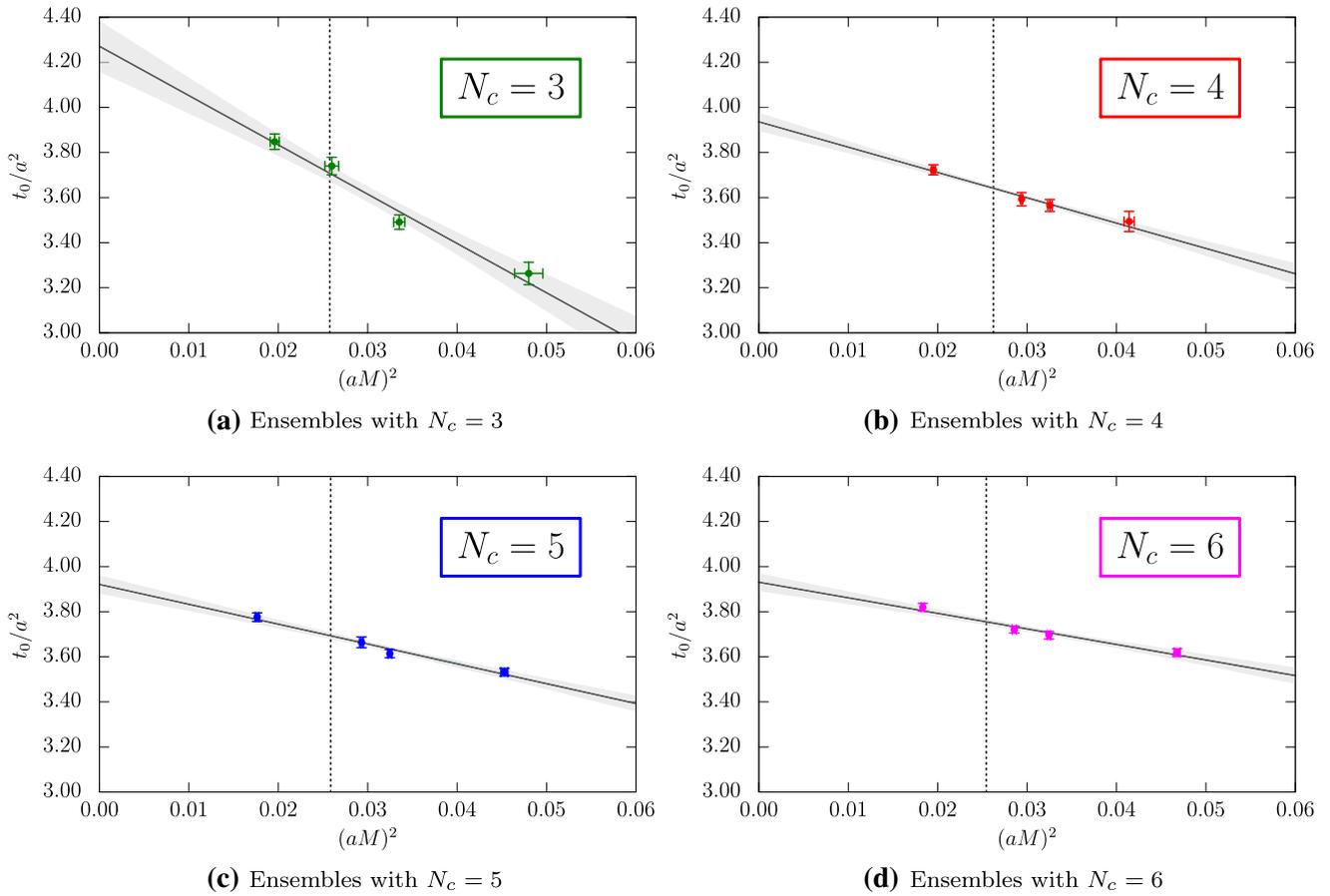

**(a)** Ensembles with $N_c = 3$

**(b)** Ensembles with $N_c = 4$

**(c)** Ensembles with $N_c = 5$

**(d)** Ensembles with $N_c = 6$

**Fig. 1** Mass dependence of $t_0^{\mathrm{imp}}/a^2$. The vertical line corresponds to the value $M^2 = M_{\mathrm{ref}}^2$

**Table 3** Results obtained in the mixed action setup, with Wilson fermions on the sea and twisted mass in the valence sector. We use $c_{sw} = 1.69$, as in the sea sector

| Ensemble | $am_{cr}$ | $a\mu_0$ | $aM_\pi^v$ | $|am_{pcac}^v|$ | $aF_\pi$ |
|---|---|---|---|---|---|
| 3A10 | $-0.4214$ | 0.01107 | 0.2216 (20) | 0.0000 (3) | 0.04405 (41) |
| 3A20 | $-0.4196$ | 0.00781 | 0.1834 (6) | 0.0001 (2) | 0.04023 (24) |
| 3A30 | $-0.4187$ | 0.00632 | 0.1613 (11) | 0.0008 (2) | 0.03678 (33) |
| 3A40 | $-0.4163$ | 0.00513 | 0.1423 (7) | 0.0006 (3) | 0.03554 (15) |
| 4A10 | $-0.3875$ | 0.01030 | 0.2037 (11) | 0.0001 (2) | 0.05131 (37) |
| 4A20 | $-0.3865$ | 0.00844 | 0.1803 (9) | 0.0000 (4) | 0.05037 (26) |
| 4A30 | $-0.3865$ | 0.00778 | 0.1717 (9) | 0.0001 (4) | 0.04913 (31) |
| 4A40 | $-0.3851$ | 0.00546 | 0.1416 (5) | 0.0001 (2) | 0.04608 (15) |
| 5A10 | $-0.3611$ | 0.01225 | 0.2114 (13) | 0.0003 (4) | 0.06125 (32) |
| 5A20 | $-0.3611$ | 0.00906 | 0.1799 (10) | 0.0001 (4) | 0.05767 (30) |
| 5A30 | $-0.3607$ | 0.00824 | 0.1706 (13) | 0.0000 (4) | 0.05647 (40) |
| 5A40 | $-0.3596$ | 0.00509 | 0.1328 (5) | 0.0002 (2) | 0.05278 (18) |
| 6A10 | $-0.3415$ | 0.01298 | 0.2142 (6) | 0.0003 (2) | 0.06813 (21) |
| 6A20 | $-0.3414$ | 0.00956 | 0.1801 (4) | 0.0002 (2) | 0.06435 (25) |
| 6A30 | $-0.3414$ | 0.00803 | 0.1668 (5) | 0.0002 (2) | 0.06278 (24) |
| 6A40 | $-0.3409$ | 0.00542 | 0.1342 (4) | 0.0000 (1) | 0.05929 (14) |

the chiral effective theory, which is expected to be set by the lighter resonances, such as the $\rho$. The scale $\mu = 4\pi F$ is instead the scale at which ChPT breaks down, which for large enough $N_c$ is much higher than the scale at which new resonances appear. In the context of large $N_c$, it is therefore sensible to choose a renormalization scale more closely related to the physical cutoff that does not scale with $N_c$. Keeping the scale related to $4\pi F$, however, has some advantages for fitting, so we choose:

$$\mu^2 = \frac{3}{N_c}(4\pi F)^2, \tag{32}$$

which has no leading dependence on $N_c$. Using this scale, the NLO predictions can be conveniently written as:

$$F_\pi = F\left[1 - 2\xi \log\left(\frac{N_c}{3}\xi\right) + 64\pi^2 \xi L_F(\mu)\right], \tag{33}$$

$$\frac{M_\pi^2}{m} = 2B\left[1 + \frac{1}{4}\xi \log\left(\frac{N_c}{3}\xi\right) + 128\pi^2 \xi L_M(\mu)\right], \tag{34}$$

where $m = \mu_0$, the bare twisted mass. Note that in this expression $B$ is bare, since the quark mass is also bare. The value of the non-singlet pseudoscalar normalization constant, $Z_P$, is thus needed.

### 5.3 Fitting strategy

Some care is needed to perform the fits in Eqs. (33) and (34). The complication comes from the fact that both coordinates, $(x, y) = (\xi, F_\pi)$ or $(x, y) = (\xi, M_\pi^2/\mu_0)$ have correlated errors. In particular the Ordinary Least Square (OLS) method is not appropriate, since it assumes no errors in $x$ coordinate. An alternative approach is the York Regression (YR) [80], in which the $\chi^2$ function is:

$$\chi^2 = \sum_i \min_{\delta_i}\left[\mathbf{R}_i^T V^{-1} \mathbf{R}_i\right], \tag{35}$$

where we have defined the two-dimensional vectors:

$$\mathbf{R}_i(\delta_i) \equiv (f(x_i + \delta_i) - y_i, \delta_i), \tag{36}$$

where $f$ is the fitting function, and $V$ is the $x$, $y$-covariance matrix, estimated using bootstrap samples. In order to account for autocorrelations, we vary the block-size of the bootstrap samples. We find that blocks of $\sim 20$ units of Montecarlo are sufficient, and we do not observe a clear $N_c$ dependence. We also estimate all the errors of the fit parameters via bootstrap resampling.

### 5.4 Fit results at fixed $N_c$

First we consider each $N_c$ separately and perform a fit of the data points to extract $F$, $L_F(\mu)$ and $B$, $L_M(\mu)$. The NLO fit results for these quantities are shown respectively in Tables 4 and 5. The $N_c$ dependence of the LECs is shown in Figs. 2

**Table 4** NLO Fits for $F_\pi$ for separate values of $N_c$

| $N_c$ | $aF/\sqrt{N_c}$ | $L_F/N_c$ | $\chi^2/dof$ |
|---|---|---|---|
| 3 | 0.0088 (9) | 0.0046 (14) | 0.7/2 |
| 4 | 0.0155 (6) | 0.0013 (3) | 3.9/2 |
| 5 | 0.0175 (4) | 0.0011 (2) | 2.2/2 |
| 6 | 0.0188 (2) | 0.0011 (1) | 0.4/2 |

**Table 5** Fits for $M_\pi$ for separate values of $N_c$

| $N_c$ | $aB$ | $L_M/N_c$ | $\chi^2/dof$ |
|---|---|---|---|
| 3 | 1.564 (55) | 0.00086 (10) | 10.2/2 |
| 4 | 1.560 (37) | 0.00064 (7) | 1.4/2 |
| 5 | 1.648 (30) | 0.00031 (6) | 0.1/2 |
| 6 | 1.610 (20) | 0.00031 (4) | 9.5/2 |

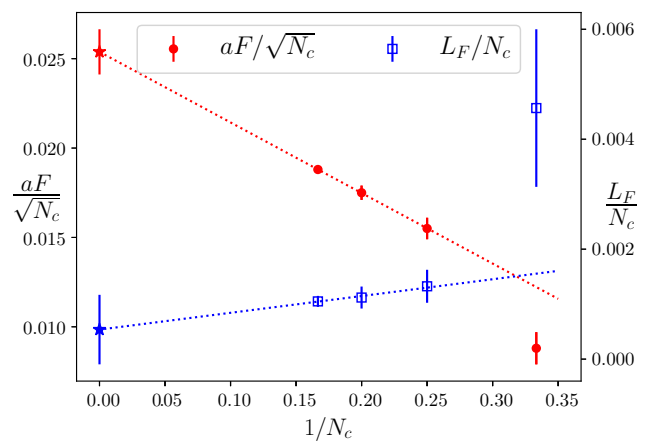

**Fig. 2** $N_c$ dependence of $F/\sqrt{N_c}$ (red) and $L_F$ (blue). The dotted lines are the best fits to Eqs. (5) and (7) excluding the data points at $N_c = 3$

and 3. It can be seen that the scaling is well described by leading and subleading $N_c$ corrections for $N_c = 4$–6, while there seems to be significant $1/N_c^2$ corrections for $N_c = 3$ in the case of $F$ and $L_F$. In the case of $B$ and $L_M$ errors are larger and there is no sign of $1/N_c^2$. Interestingly, the data suggest that the large $N_c$ limit of $L_M \sim 0$.

### 5.5 Simultaneous chiral and $N_c$ fits

We now consider a global fit including several data points at different values of $N_c$. We first perform a $SU(N_f)$-NLO fit to the subset $N_c = 4$–6, including leading and subleading $N_c$ corrections for all the LO and NLO LECs, as parametrized in Eqs. (5) and (7). We linearize the fit by considering the following parametrization

$$F_\pi = \sqrt{N_c}\left(F_0 + \frac{F_1}{N_c}\right)\left[1 - 2\xi \log\left(\frac{N_c}{3}\xi\right)\right]$$
$$+ 64\pi^2 \xi \sqrt{N_c}\left(N_c(FL_F)^{(0)} + (FL_F)^{(1)}\right), \tag{37}$$

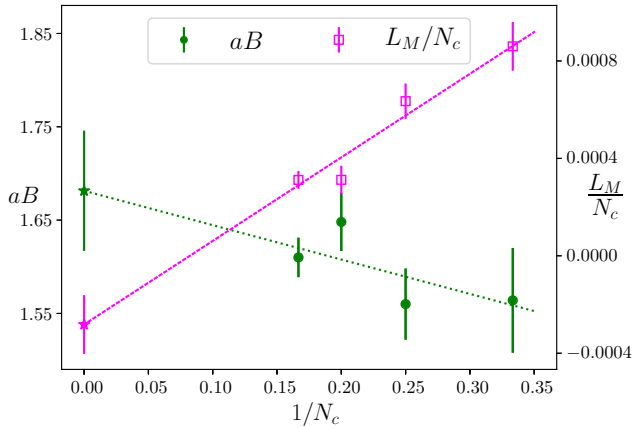

**Fig. 3** $N_c$ dependence of $B$ and $L_M$. The dotted lines are the best fits to Eqs. (5) and (7) including all points

$$
\frac{M_\pi^2}{m} = 2\left(B_0 + \frac{B_1}{N_c}\right)\left[1 + \frac{1}{4}\xi \log\left(\frac{N_c}{3}\xi\right)\right] \\
+ 256\pi^2\xi\left(N_c(BL_M)^{(0)} + (BL_M)^{(1)}\right),
\tag{38}
$$

where $(FL_F)^{(0)} \equiv F_0 L_F^{(0)}$, while $(FL_F)^{(1)} \equiv F_1 L_F^{(0)} + L_F^{(1)} F_0$, and $(BL_M)^{(0)} \equiv B_0 L_M^{(0)}$ and $(BL_M)^{(1)} \equiv B_1 L_M^{(0)} + B_0 L_M^{(1)}$.

Secondly, we consider the $U(N_f)$-NNLO expansion, since we have checked that the $U(N_f)$-NLO expressions fit the data very poorly. We also linearize the fit by considering the following fitting functions:

$$
F_\pi = \sqrt{N_c}\left(F_0 + \frac{F_1}{N_c} + \frac{F_2}{N_c^2}\right)\left[1 - 2\xi \log\left(\frac{N_c}{3}\xi\right)\right] \\
+ 64\pi^2\xi\sqrt{N_c}\left(N_c(FL_F)^{(0)} + (FL_F)^{(1)}\right) \\
+ N_c^2\sqrt{N_c}\left(16\pi^2\xi\right)^2 K_F^{(0)},
\tag{39}
$$

$$
\frac{M_\pi^2}{m} = 2\left(B_0 + \frac{B_1}{N_c} + \frac{B_2}{N_c^2}\right)\left[1 + \frac{1}{4}\xi \log\left(\frac{N_c}{3}\xi\right)\right. \\
\left. + -\frac{1}{4}\left(\xi + \frac{a_0}{N_c^2}\right)\log\left(\frac{N_c}{3}\left(\xi + \frac{a_0}{N_c^2}\right)\right)\right] \\
+ 256\pi^2\xi\left(N_c(BL_M)^0 + (BL_M)^{(1)}\right) \\
- 64 N_c^2\left(16\pi^2\xi\right)^2 K_M^{(0)},
\tag{40}
$$

where

$$
a_0 \equiv N_c^2 \frac{M_0^2}{(4\pi F)^2},
\tag{41}
$$

and $M_0^2$ is given by the Witten–Veneziano formula for the $\eta'$ mass valid in the large $N_c$ limit (see Eq. 13). We use the result for the topological susceptibility from Ref. [59],

$$
t_0^2\chi_t = 7.03(13) \cdot 10^{-4}.
\tag{42}
$$

We convert to lattice units using the value of $t_0/a^2$ in the previous section and substitute $F \to \sqrt{N_c}F_0$, as extracted from the global $F_\pi$ fit. We find $a_0 \sim 6.5$, a value we fix in the fit.

In summary we compare the following fits:

(i) Fit 1: $SU(N_f)$-NLO fit to Eqs. (37) and (38) including the data subset $N_c = 4$-6.
(ii) Fit 2: $U(N_f)$-NNLO expansion fit to Eqs. (39) and (40) including the full data set.

The results for the fitted parameters in the global fits are shown in Tables 6 and 7, and the quality of the fits is shown in Fig. 4a, b. We also quote in Table 8 the results for the NLO LECs from these fits. Errors are large, but there are significant correlations between the parameters as can be seen in Fig. 5.

### 5.6 Selected results

We will now quote some results that can be inferred from our fits. We first focus on the decay constant in the chiral limit. Using $a = 0.0754(23)$ fm, we get from our fits at fixed $N_f = 4$:

$$
\text{Fit 1}: \frac{F}{\sqrt{N_c}} = \left(67(3) - 26(4)\frac{N_f}{N_c}\right)(3\%)^a\text{MeV},
$$
$$
\text{Fit 2}: \frac{F}{\sqrt{N_c}} = \left(70(2) - 22(5)\frac{N_f}{N_c} - \frac{86(37)}{N_c^2}\right)(3\%)^a\text{MeV},
\tag{43}
$$

where the $N_f$ dependence assumed is the expected one as discussed in sec. 2. Note that no $N_f$ dependence is assumed in the $1/N_c^2$ terms. The first error is just the one obtained from the fits in Table 6 and the second error of 3% is the one corresponding to the lattice spacing determination. For two- and three-flavour QCD we get:

$$
\text{Fit 1}: \quad F^{N_c=3,N_f=2} = 86(3) \text{ MeV}, \\
F^{N_c=3,N_f=3} = 71(3) \text{ MeV},
\tag{44}
$$

$$
\text{Fit 2}: \quad F^{N_c=3,N_f=2} = 81(7) \text{ MeV}, \\
F^{N_c=3,N_f=3} = 68(7) \text{ MeV},
\tag{45}
$$

where we have taken into account the correlations between the different terms in Eq. (43), and we have assumed no $N_f$ dependence on the last term of the Fit 2. These results are in perfect agreement with phenomenological determinations:

$$
F^{N_f=2} = 86.2(5) \text{ MeV in Ref. [81]}, \\
F^{N_f=3} \simeq 71.1 \text{ MeV in Ref. [82]},
\tag{46}
$$

and also lattice results (see Ref. [45]). In addition, we can compare to previous results in the large $N_c$ limit in the quenched approximation:

**Table 6** Different fits for the decay constant as described in the text

| Fit | $F_0$ | $F_1$ | $F_2$ | $(FL_F)^{(0)}$ | $(FL_F)^{(1)}$ | $K_F^{(0)}$ | $\chi^2/dof$ |
|-----|-------|-------|-------|----------------|----------------|-------------|--------------|
| 1 | 0.0255 (12) | $-0.040$ (6) | – | $4.7$ (9.5) $\cdot 10^{-6}$ | $4.8$ (5.1) $\cdot 10^{-5}$ | – | 0.79 |
| 2 | 0.0266 (9) | $-0.034$ (8) | $-0.033$ (14) | $-8$ (10) $\cdot 10^{-6}$ | $5.6$ (4.4) $\cdot 10^{-5}$ | $7.6$ (6.4) $\cdot 10^{-7}$ | 0.9 |

**Table 7** Different fits for the meson mass as described in the text

| Fit | $B_0$ | $B_1$ | $B_2$ | $(BL_M)^{(0)}$ | $(BL_M)^{(1)}$ | $K_M^{(0)}$ | $\chi^2/dof$ |
|-----|-------|-------|-------|----------------|----------------|-------------|--------------|
| 1 | 1.70 (11) | $-0.5$ (5) | – | $-0.00046$ (29) | $0.0056$ (15) | – | 2.0 |
| 2 | 1.72 (7) | $-1.8$ (5) | $1.8$ (1.5) | $-0.00017$ (25) | $0.0066$ (10) | $1.3(9) \cdot 10^{-6}$ | 2.4 |

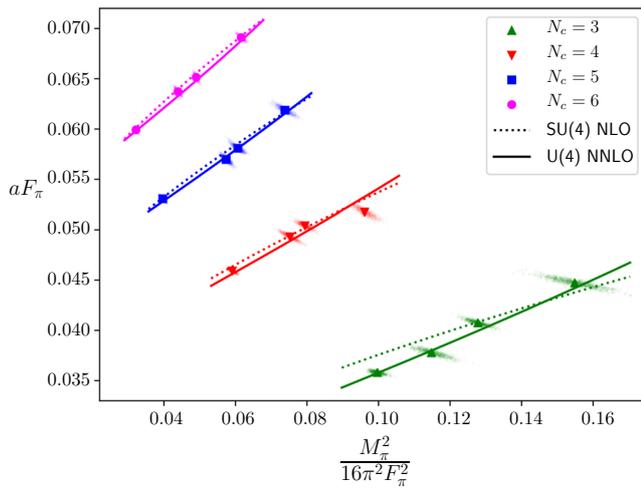

**(a)** Chiral fits for the decay constant.

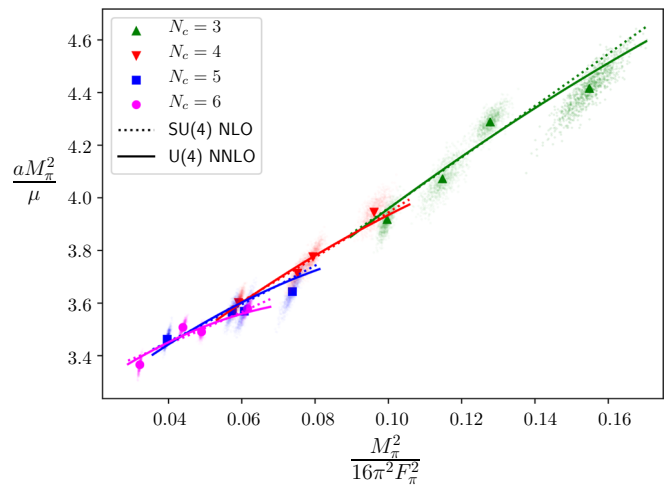

**(b)** Chiral fits for the meson mass.

**Fig. 4** Data and NLO/NNLO fits for the decay constant and meson mass. The central value is shown together with the bootstrap samples used for fitting. The results include finite-volume corrections as in Eq. (30)

$$\left. \frac{F}{\sqrt{N_c}} \right|_{N_c \to \infty} = 56(5) \text{ MeV, Ref. [13]}. \tag{47}$$

This value is $2\sigma$ away from the results in Eq. (43). This discrepancy may be explained however with the lack of non-perturbative normalization constant and discretization effects, which in their case are of $O(a)$.

Regarding the coupling, $B \equiv \Sigma/F^2$, we do not have a non-perturbative value of $Z_P$, up to this factor we get:

$$\begin{aligned} \text{Fit1}: \frac{\Sigma}{F^2} &= Z_P \left( 1.70(11) - 0.12(12)\frac{N_f}{N_c} \right), \\ \text{Fit2}: \frac{\Sigma}{F^2} &= Z_P \left( 1.72(7) - 0.45(37)\frac{N_f}{N_c} - \frac{1.8(1.5)}{N_c^2} \right). \end{aligned} \tag{48}$$

From Ref. [83], we can obtain the 1-loop perturbative result for the normalization constant:

$$Z_P(N_c = 3) = 0.555, \tag{49}$$

which at the order we are working is independent of $N_f$. With this, we obtain for $N_c = 3$:

$$\text{Fit 1}: N_f = 4 \quad \longrightarrow \frac{\Sigma}{F^2} = 2.26(11)(7)^a \text{ GeV}, \tag{50}$$

$$N_f = 3 \quad \longrightarrow \frac{\Sigma}{F^2} = 2.31(5)(7)^a \text{ GeV}, \tag{51}$$

$$N_f = 2 \quad \longrightarrow \frac{\Sigma}{F^2} = 2.35(3)(7)^a \text{ GeV}, \tag{52}$$

where the first error is systematic, the second comes from the scale setting, and we omit any systematic errors regarding the normalization constant. Combining these results with the ones in Eqs. (44) and (45), we obtain:

$$\Sigma^{1/3}(N_f = 2) = 257(2)(9)^a \text{ MeV}, \tag{53}$$

$$\Sigma^{1/3}(N_f = 3) = 223(4)(8)^a \text{ MeV}, \tag{54}$$

which is compatible within $1\sigma$ with the numbers quoted in Ref. [45]. We can also consider the ratio of condensates for

**Table 8** Values for the LECs from the fits in Tables 6 and 7

| Fit | $L_F^{(0)}$ | $L_F^{(1)}$ | $L_M^{(0)}$ | $L_M^{(1)}$ |
|---|---|---|---|---|
| 1 | $1\,(4) \cdot 10^{-4}$ | $23\,(13) \cdot 10^{-4}$ | $-20\,(15) \cdot 10^{-5}$ | $29\,(6) \cdot 10^{-4}$ |
| 2 | $-3\,(4) \cdot 10^{-4}$ | $17\,(18) \cdot 10^{-4}$ | $-1\,(1) \cdot 10^{-4}$ | $37\,(7) \cdot 10^{-4}$ |

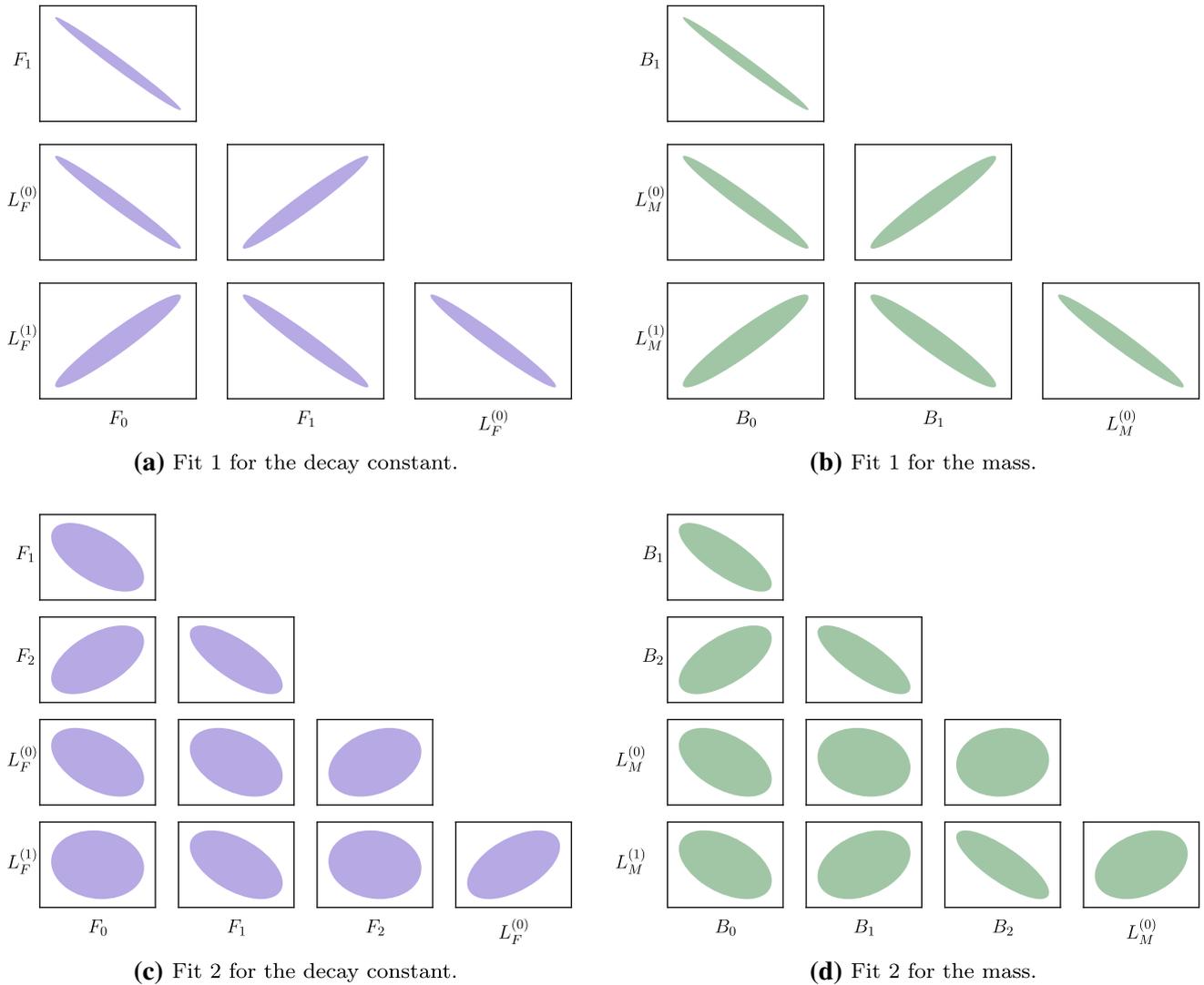

**(a)** Fit 1 for the decay constant.

**(b)** Fit 1 for the mass.

**(c)** Fit 2 for the decay constant.

**(d)** Fit 2 for the mass.

**Fig. 5** Correlations between fitted parameters

$N_f = 2$ and $N_f = 3$, where the $Z_P$ factor drops (up to subleading $N_f$ dependence):

$$\frac{\Sigma(N_f = 2)}{\Sigma(N_f = 3)} = 1.49(10), \qquad (55)$$

which shows good agreement with the prediction

$$\frac{\Sigma(N_f = 2)}{\Sigma(N_f = 3)} = 1.51(11) \text{ in Ref. [47].} \qquad (56)$$

Regarding the NLO LEC for the decay constant, we get from Fit 1:

$$\frac{L_F(\mu)}{N_c} \cdot 10^3 = 0.1(4) + 0.6(3)\frac{N_f}{N_c} + O(N_c^{-2}), \qquad (57)$$

while for the NLO LEC for the mass, we can only give the $N_c$ scaling at $N_f = 4$:

$$\frac{L_M^{N_f=4}(\mu)}{N_c} \cdot 10^3 = -0.2(2) + \frac{2.9(6)}{N_c} + O(N_c^{-2}). \qquad (58)$$

In the case of Fit 2, we can provide both results together:

$$\frac{L_F(\mu)}{N_c} \cdot 10^3 = -0.3(4) + 0.4(4)\frac{N_f}{N_c} + O(N_c^{-2}),$$
$$\frac{L_M(\mu)}{N_c} \cdot 10^3 = -0.1(1) + 0.9(2)\frac{N_f}{N_c} + O(N_c^{-2}). \tag{59}$$

From Eqs. 57 and 59, we can infer the $N_c = 3$, $N_f = 3$ results:

$$\text{Fit1}: L_F(\mu) = 2.1(3) \cdot 10^{-3},$$
$$\text{Fit2}: L_F(\mu) = 0.4(2.1) \cdot 10^{-3}, \tag{60}$$
$$L_M(\mu) = 2.4(8) \cdot 10^{-3}.$$

For $N_c = 3$, $N_f = 2$, it is more common to quote $\bar{\ell}_3$ and $\bar{\ell}_4$:

$$\bar{\ell}_3 = 2 \log\left(\frac{4\pi F_\pi}{M_\pi^{\text{phys}}}\right) - 16(4\pi)^2 L_M^{N_f=2, N_c=3},$$
$$\bar{\ell}_4 = 2 \log\left(\frac{4\pi F_\pi}{M_\pi^{\text{phys}}}\right) + 4(4\pi)^2 L_F^{N_f=2, N_c=3}. \tag{61}$$

This way, we obtain:

$$\text{Fit 1}: \bar{\ell}_4 = 5.1(3),$$
$$\text{Fit 2}: \bar{\ell}_3 = 0.4(1.6), \quad \bar{\ell}_4 = 4.1(1.1). \tag{62}$$

We stress that $U(N_f)$ $\bar{\ell}_3$ in fit 2 is not the same as the standard $\bar{\ell}_3$ in $SU(N_f)$. $\bar{\ell}_4$ agrees instead at 1–2$\sigma$ with the results quoted in Ref. [45].

### 5.7 Comments on systematics

The most important systematic uncertainty comes from the finite lattice spacing. Even though a continuum extrapolation would be needed to quantify this error properly, we can get an estimate by comparing the pion mass made of different combination of sea and valence quarks. In particular, Chiral Perturbation Theory in the mixed-action setup predicts that the chiral logs for $F_\pi$ depend upon the mixed pion mass [84]:

$$M_\pi^{\text{mixed}} \xrightarrow{\text{LO ChPT}} 2B(m^v + m^s), \tag{63}$$

where $m^v$ is the renormalized quark mass in the valence sector and $m^s$ in the sea action. We have measured this mixed pion in one ensemble:

$$\text{Ensemble 3A10} \rightarrow aM_\pi^{\text{mixed}} = 0.2201(26), \tag{64}$$

obtaining a result which is compatible within errors with both, the sea and valence quark pions.

A different estimate comes from the dependence on $c_{sw}$ in the valence sector. We have recomputed the decay constant for $c_{sw} = 0$ in the 3A10 ensemble, obtaining $[F_\pi]_{c_{sw}=0} =$ 0.04303(40), within 2% of the value at the nominal $c_{sw}$. The effects of a change in $c_{sw}$ are in principle $O(a^2)$, which can be estimated at $\sim 2\%$ for this observable. This concerns however only the charged meson sector, since the neutral pion is known to have higher discretization effects with twisted mass. That issue is out of the scope of this work, and it will be addressed in future publications. In $N_c$ we end this section with a last word on the chiral fits. We find that our data is well described by ChPT at the order we worked. Still, we cannot exclude that higher order corrections might be relevant in the range of masses we are considering. A robust study on the convergence of ChPT would require simulations at lighter quark masses and a proper continuum extrapolation.

## 6 Conclusion and outlook

In this work we presented the first lattice determination using dynamical fermions of the $N_c$ scaling of the couplings in the chiral Lagrangian that contribute to the meson masses and decay constants (see Eqs. (43), (48) and Table 8). We have been able to disentangle the leading and subleading terms and we found that the subleading contributions are typically non negligible. In fact, we find that the value for $L_M$ at $N_c = 3$ seems to be dominated by the subleading corrections, and the fit result suggests an accidental cancellation of $2L_8 - L_5$ in the large $N_c$ limit.

From our chiral fits and theoretical expectations, we have been able to infer the values of the couplings for theories with different numbers of flavours, $N_f = 2$ and $N_f = 3$ at $N_c = 3$. We find that our results nicely agree with those in the literature regarding $L_F$, $L_M$ and $F$ (see for example Ref. [45] for a summary of results). For $B$ we need to improve our determination, including a non-perturbatively determined renormalization factor. On the other hand, as long as this factor has a small $N_f$ dependence, we can estimate the ratio of B and the chiral condensate for $N_f = 2$ and $N_f = 3$. We find excellent agreement with the prediction of paramagnetic suppressions of Refs. [46,47].

We would like to stress that the results presented in this paper are complementary to similar studies that can be performed in reduced models [16–20] or the quenched approximations at large $N_c$ [13], since both of these approaches must yield the leading order result as $N_c \rightarrow \infty$. Given the strong correlations presents in our results (see Fig. 5), a precise determination of the dominant $N_c$ term would significantly improve the determination of the subleading $N_c$ corrections, and hence the determination of the physical values at $N_c = 3$. We are willing to provide the bootstrap samples if requested.

As for the future, we would like to mention that our ensembles have a big potential to study other physical observables. We plan to use them to analyse the scaling of other quantities, such as the $K \rightarrow \pi$ matrix elements (see [29,62] for

previous results). We also believe that the study of scattering amplitudes is a relevant quantity of study at large $N_c$: on one hand quantities such as the $I = 2\,\pi\pi$ scattering length give access to LECs of the chiral Lagrangian; on the other hand the study of the behaviour resonances at large $N_c$ is interesting, as it may shed light about their nature [10,11,85,86].

**Data Availability Statement** This manuscript has no associated data or the data will not be deposited. [Authors' comment: We will provide configurations and results from fits upon request.]

## References

1. G. 't Hooft, Nucl. Phys. B **72**, 461 (1974)
2. W.A. Bardeen, A.J. Buras, J.M. Gerard, Phys. Lett. B **180**, 133 (1986)
3. W.A. Bardeen, A.J. Buras, J.M. Gerard, Nucl. Phys. B **293**, 787 (1987)
4. R.S. Chivukula, J.M. Flynn, H. Georgi, Phys. Lett. B **171**, 453 (1986)
5. S.R. Sharpe, Phys. Lett. B **194**, 551 (1987)
6. A. Pich, E. de Rafael, Phys. Lett. B **374**, 186 (1996). arXiv:hep-ph/9511465
7. T. Hambye, S. Peris, E. de Rafael, JHEP **05**, 027 (2003). arXiv:hep-ph/0305104
8. A.J. Buras, J.-M. Gérard, W.A. Bardeen, Eur. Phys. J. C **74**(2871), 1401.1385 (2014)
9. H. Gisbert, A. Pich, in *21st High-Energy Physics International Conference in Quantum Chromodynamics (QCD 18) Montpellier, July 2–6, 2018*. arXiv:1810.04904 (2018)
10. J.R. Peláez, J. Nebreda, G. Ríos, Prog. Theor. Phys. Suppl. **186**(113), 1007.3461 (2010)
11. J. Nebreda, J.R. Pelaez, G. Ríos, Phys. Rev. D **84**(074003), 1107.4200 (2011)
12. B. Lucini, M. Teper, JHEP **06**, 050 (2001). arXiv:hep-lat/0103027
13. G.S. Bali, F. Bursa, L. Castagnini, S. Collins, L. Del Debbio, B. Lucini, M. Panero, JHEP **06**, 071 (2013). arXiv:1304.4437
14. T. DeGrand, Y. Liu, Phys. Rev. D **94**, 034506 (2016) [Erratum: Phys. Rev. D **95**(1), 019902 (2017)]. arXiv:1606.01277
15. B. Lucini, M. Panero, Phys. Rep. **526**(93), 1210.4997 (2013)
16. M. García Pérez, A. González-Arroyo, M. Koren, M. Okawa, JHEP **07**(169), 18070.3481 (2018)
17. M. García Pérez, A. González-Arroyo, L. Keegan, M. Okawa, PoS **LATTICE2016**, 337 (2016). arXiv:1612.07380
18. A. González-Arroyo, M. Okawa, Phys. Lett. B **755**, 132 (2016). arXiv:1510.05428
19. A. Hietanen, R. Narayanan, R. Patel, C. Prays, Phys. Lett. B **674**, 80 (2009). arXiv:0901.3752
20. R. Narayanan, H. Neuberger, Phys. Lett. B **616**, 76 (2005). arXiv:hep-lat/0503033
21. T. Appelquist et al. (Lattice Strong Dynamics (LSD)), Phys. Rev. D **89**, 094508 (2014). arXiv:1402.6656
22. T. DeGrand, Y. Liu, E.T. Neil, Y. Shamir, B. Svetitsky, Phys. Rev. D **91**, 114502 (2015). arXiv:1501.05665
23. Y. Aoki et al., (LatKMI), Phys. Rev. D **96**, 014508 (2017). arXiv:1610.07011
24. T. Appelquist et al., Phys. Rev. D **93**, 114514 (2016). arXiv:1601.04027
25. M. Hansen, T. Janowski, C. Pica, A. Toniato, EPJ Web Conf. **175**, 08010 (2018). arXiv:1710.10831
26. D. Nogradi, L. Szikszai (2019), arXiv:1905.01909
27. R.C. Brower, A. Hasenfratz, E.T. Neil, S. Catterall, G. Fleming, J. Giedt, E. Rinaldi, D. Schaich, E. Weinberg, O. Witzel (USQCD) (2019), arXiv:1904.09964
28. L. Del Debbio, PoS **ALPS2018**, 022 (2018)
29. A. Donini, P. Hernández, C. Pena, F. Romero-López, Phys. Rev. D **94**, 114511 (2016). arXiv:1607.03262
30. R.J. Dowdall, C.T.H. Davies, G.P. Lepage, C. McNeile, Phys. Rev. D **88**, 074504 (2013). arXiv:1303.1670
31. R. Frezzotti, V. Lubicz, S. Simula, (ETM), Phys. Rev. D **79**, 074506 (2009). arXiv:0812.4042
32. R. Baron et al., (ETM), JHEP **08**, 097 (2010). arXiv:0911.5061
33. B.B. Brandt, A. Jüttner, H. Wittig, JHEP **11**, 034 (2013). arXiv:1306.2916
34. V. Gülpers, G. von Hippel, H. Wittig, Eur. Phys. J. A **51**, 158 (2015). arXiv:1507.01749
35. A. Bazavov et al., (MILC), PoS **LATTICE2010**, 074 (2010). arXiv:1012.0868
36. S.R. Beane, W. Detmold, P.M. Junnarkar, T.C. Luu, K. Orginos, A. Parreno, M.J. Savage, A. Torok, A. Walker-Loud, Phys. Rev. D **86**, 094509 (2012). arXiv:1108.1380
37. S. Borsanyi, S. Durr, Z. Fodor, S. Krieg, A. Schafer, E.E. Scholz, K.K. Szabo, Phys. Rev. D **88**, 014513 (2013). arXiv:1205.0788
38. S. Dürr et al., (Budapest-Marseille-Wuppertal), Phys. Rev. D **90**, 114504 (2014). arXiv:1310.3626
39. P.A. Boyle et al., Phys. Rev. D **93**, 054502 (2016). arXiv:1511.01950
40. R. Baron et al., (ETM), PoS **LATTICE2010**, 123 (2010). arXiv:1101.0518
41. S. Aoki et al., (PACS-CS), Phys. Rev. D **79**, 034503 (2009). arXiv:0807.1661
42. J. Noaki et al., (JLQCD, TWQCD), Phys. Rev. Lett. **101**, 202004 (2008). arXiv:0806.0894
43. R. Baron et al., JHEP **06**, 111 (2010c). arXiv:1004.5284
44. V. Drach, T. Janowski, C. Pica, EPJ Web Conf. **175**, 08020 (2018). arXiv:1710.07218
45. S. Aoki et al. (Flavour Lattice Averaging Group) (2019), arXiv:1902.08191

46. S. Descotes-Genon, L. Girlanda, J. Stern, JHEP **01**, 041 (2000). arXiv:hep-ph/9910537
47. V. Bernard, S. Descotes-Genon, G. Toucas, JHEP **06**, 051 (2012). arXiv:1203.0508
48. P. Di Vecchia, G. Veneziano, Nucl. Phys. B **171**, 253 (1980)
49. C. Rosenzweig, J. Schechter, C.G. Trahern, Phys. Rev. D **21**, 3388 (1980)
50. E. Witten, Ann. Phys. **128**, 363 (1980)
51. K. Kawarabayashi, N. Ohta, Nucl. Phys. B **175**, 477 (1980)
52. P. Herrera-Siklody, J.I. Latorre, P. Pascual, J. Taron, Nucl. Phys. B **497**, 345 (1997). arXiv:hep-ph/9610549
53. R. Kaiser, H. Leutwyler, Eur. Phys. J. C **17**, 623 (2000). arXiv:hep-ph/0007101
54. J. Bijnens, J. Lu, JHEP **11**, 116 (2009). arXiv:0910.5424
55. J. Bijnens, J. Lu, JHEP **03**, 028 (2011). arXiv:1102.0172
56. J. Bijnens, K. Kampf, S. Lanz, Nucl. Phys. B **873**, 137 (2013). arXiv:1303.3125
57. A.V. Manohar, in *Probing the standard model of particle interactions. Proceedings, Summer School in Theoretical Physics, NATO Advanced Study Institute, 68th session, Les Houches, July 28–September 5, 1997. Pt. 1, 2* (1998), pp. 1091–1169. arXiv:hep-ph/9802419
58. X.-K. Guo, Z.-H. Guo, J.A. Oller, J.J. Sanz-Cillero, JHEP **06**, 175 (2015). arXiv:1503.02248
59. M. Cè, M. García Vera, L. Giusti, S. Schaefer, Phys. Lett. B **762**, 232 (2016). arXiv:1607.05939
60. P. Herrera-Siklody, Phys. Lett. B **442**, 359 (1998). arXiv:hep-ph/9808218
61. L. Del Debbio, A. Patella, C. Pica, Phys. Rev. D **81**, 094503 (2010). arXiv:0805.2058
62. F. Romero-López, A. Donini, P. Hernández, C. Pena, in *36th International Symposium on Lattice Field Theory (Lattice 2018) East Lansing, July 22–28, 2018* (2018). arXiv:1810.06285
63. J. Finkenrath, C. Alexandrou, S. Bacchio, P. Charalambous, P. Dimopoulos, R. Frezzotti, K. Jansen, B. Kostrzewa, G. Rossi, C. Urbach, EPJ Web Conf. **175**, 02003 (2018). arXiv:1712.09579
64. C. Alexandrou et al., Phys. Rev. D **98**, 054518 (2018). arXiv:1807.00495
65. S. Aoki, Y. Kuramashi, Phys. Rev. D **68**, 094019 (2003). arXiv:hep-lat/0306015
66. O. Bar, G. Rupak, N. Shoresh, Phys. Rev. D **67**, 114505 (2003). arXiv:hep-lat/0210050
67. G. Herdoíza, C. Pena, D. Preti, J.A. Romero, J. Ugarrio, EPJ Web Conf. **175**, 13018 (2018). arXiv:1711.06017
68. A. Bussone, S. Chaves, G. Herdoíza, C. Pena, D. Preti, J. A. Romero, J. Ugarrio, in *36th International Symposium on Lattice Field Theory (Lattice 2018) East Lansing, July 22–28, 2018* (2018). arXiv:1812.01474
69. A. Bussone, G. Herdoíza, C. Pena, D. Preti, J.A. Romero, J. Ugarrio (2018), arXiv:1812.05458
70. A. Shindler, Phys. Rep. **461**, 37 (2008). arXiv:0707.4093
71. M. Lüscher, JHEP **08**, 071 (2010) [Erratum: JHEP **03**, 092 (2014)]. arXiv:1006.4518
72. M. Bruno, R. Sommer, (ALPHA), PoS **LATTICE2013**, 321 (2014). arXiv:1311.5585
73. R. Sommer, PoS **LATTICE2013**, 015 (2014). arXiv:1401.3270
74. M. Bruno, T. Korzec, S. Schaefer, Phys. Rev. D **95**, 074504 (2017). arXiv:1608.08900
75. Z. Fodor, K. Holland, J. Kuti, S. Mondal, D. Nogradi, C.H. Wong, JHEP **09**, 018 (2014). arXiv:1406.0827
76. Z. Fodor, K. Holland, J. Kuti, D. Nogradi, C.H. Wong, JHEP **11**, 007 (2012). arXiv:1208.1051
77. O. Bar, M. Golterman, Phys. Rev. D **89**, 034505 (2014) [Erratum: Phys. Rev. D **89**(9), 099905 (2014)]. arXiv:1312.4999
78. J. Gasser, H. Leutwyler, Phys. Lett. B **184**, 83 (1987)
79. G. Colangelo, S. Durr, C. Haefeli, Nucl. Phys. B **721**, 136 (2005). arXiv:hep-lat/0503014
80. D. York, N.M. Evensen, M.L. Martínez, J. De Basabe Delgado, Am. J. Phys. **72**, 367 (2004). https://doi.org/10.1119/1.1632486
81. G. Colangelo, S. Durr, Eur. Phys. J. C **33**, 543 (2004). arXiv:hep-lat/0311023
82. B. Ananthanarayan, J. Bijnens, S. Ghosh, Eur. Phys. J. C **77**, 497 (2017). arXiv:1703.00141
83. C. Alexandrou, M. Constantinou, T. Korzec, H. Panagopoulos, F. Stylianou, Phys. Rev. D **86**, 014505 (2012). arXiv:1201.5025
84. J.-W. Chen, D. O'Connell, A. Walker-Loud, Phys. Rev. D **75**, 054501 (2007). arXiv:hep-lat/0611003
85. V. Bernard, M. Lage, U.G. Meißner, A. Rusetsky, JHEP **01**, 019 (2011). arXiv:1010.6018
86. J. Ruiz de Elvira, U.G. Meißner, A. Rusetsky, G. Schierholz, Eur. Phys. J. C **77**, 659 (2017). arXiv:1706.09015

# $I = 3$ Three-Pion Scattering Amplitude from Lattice QCD

Tyler D. Blanton [1,*] Fernando Romero-López [2,†] and Stephen R. Sharpe [1,‡]

[1] *Physics Department, University of Washington, Seattle, Washington 98195-1560, USA*
[2] *Instituto de Física Corpuscular, Universitat de València and CSIC, 46980 Paterna, Spain*

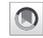

We analyze the spectrum of two- and three-pion states of maximal isospin obtained recently for isosymmetric QCD with pion mass $M \approx 200$ MeV in Hörz and Hanlon, [Phys. Rev. Lett. **123**, 142002 (2019)]. Using the relativistic three-particle quantization condition, we find $\sim 2\sigma$ evidence for a nonzero value for the contact part of the $3\pi^+$ ($I = 3$) scattering amplitude. We also compare our results to leading-order chiral perturbation theory. We find good agreement at threshold and some tension in the energy dependent part of the $3\pi^+$ scattering amplitude. We also find that the $2\pi^+$ ($I = 2$) spectrum is fit well by an $s$-wave phase shift that incorporates the expected Adler zero.

*Introduction.*—Lattice QCD (LQCD) provides a powerful (if indirect) tool for *ab initio* calculations of strong-interaction scattering amplitudes. The formalism for determining two-particle amplitudes is well understood [1–12], and there has been enormous progress in its implementation in recent years [13–32] (see Ref. [33] for a review). The present frontier is the determination of three-particle scattering amplitudes and related decay amplitudes. LQCD calculations promise access to three-particle scattering processes that are difficult or impossible to access experimentally. Examples of important applications are understanding properties of resonances with significant three-particle branching ratios (including the Roper resonance [34], and many of the $X$, $Y$, and $Z$ resonances [35]), determining the three-nucleon interaction (important for large nuclei and neutron star properties), predicting weak decays to three particles (e.g., $K \to 3\pi$), and calculating the $3\pi$ contribution to the hadronic-vacuum polarization that enters into the prediction of muonic $g - 2$ [36].

Three-particle amplitudes are determined using LQCD by calculating the energies of two- and three-particle states in a finite volume [37,38]. The challenges to carrying this out are twofold. On the one hand, the calculation of spectral levels becomes more challenging as the number of particles increases. On the other, one must develop a theoretical formalism relating the spectrum to scattering amplitudes. Significant progress has recently been achieved in both directions, with energies well above the three-particle threshold being successfully measured, and a formalism

for three identical (pseudo)scalar particles available. The formalism has been developed and implemented following three approaches: generic relativistic effective field theory (RFT) [39–45], nonrelativistic effective field theory [46–49], and (relativistic) finite volume unitarity (FVU) [50,51] (see, also, Refs. [52,53] and Ref. [54] for a review). To date, only the RFT formalism has been explicitly worked out including higher partial waves. The application to LQCD results has, so far, been restricted to the energy of the three-particle ground state, either using the threshold expansion [55–57], or, more recently, the FVU approach for $3\pi^+$ [51].

Recently, precise results were presented for the spectrum of $2\pi^+$ and $3\pi^+$ states in $O(a)$-improved isosymmetric QCD with pions having close to physical mass, $M \approx 200$ MeV [58]. These were obtained in a cubic box of length $L$ with $ML \approx 4.2$, for several values of the total momentum $\vec{P} = (2\pi/L)\vec{d}$ with $\vec{d} \in \mathbb{Z}^3$, and for several irreducible representations (irreps) of the corresponding symmetry groups. Isospin symmetry ensures that $G$ parity is exactly conserved and, thus, that the $2\pi^+$ and $3\pi^+$ sectors are decoupled. In total, sixteen $2\pi^+$ levels and eleven $3\pi^+$ levels were obtained below the respective inelastic thresholds at $E_2^* = 4M$ and $E^* = 5M$, Here, $E_2^*$ and $E^* = \sqrt{E^2 - \vec{P}^2}$ are the corresponding center-of-mass energies, with $E$ the total three-particle energy.

The purpose of this Letter is to perform a global analysis of the spectra of Ref. [58] using the RFT formalism and determine the underlying $3\pi^+$ interaction. This breaks new ground for an analysis of the three-particle spectrum in several ways: we use multiple excited states, in both trivial and nontrivial irreps, including results from moving frames. Therefore, this analysis serves as a testing ground for the utility of the three-particle formalism in an almost physical example. An additional appealing feature is that the size of

---

the $3\pi^+$ interaction can be calculated using chiral perturbation theory ($\chi$PT). We present the leading order (LO) prediction here.

After this Letter was made public, an independent study of the results of Ref. [58], using the FVU approach, appeared [59].

*Formalism and implementation.*—All approaches to determining three-particle scattering amplitudes using LQCD proceed in two steps, which we outline here. In the first step, one uses a quantization condition (QC), which predicts the finite-volume spectrum in terms of an intermediate infinite-volume three-particle scattering quantity. In the RFT approach, the QC for identical, spinless particles with a $G$-parity-like $Z_2$ symmetry takes the form (up to corrections of $\mathcal{O}(1\%)$ that are exponentially suppressed in $ML$) [39]

$$\det\left[F_3(E,\vec{P},L)^{-1} + \mathcal{K}_{\mathrm{df},3}(E^*)\right] = 0. \qquad (1)$$

Here, $F_3$ and $\mathcal{K}_{\mathrm{df},3}$ are matrices in a space describing three on-shell particles in finite volume. They have indices of angular momentum of the interacting pair, $\ell$, $m$, and finite-volume momentum of the spectator particle, $k$. $F_3$ depends on the two-particle scattering amplitude and on known geometric factors, while $\mathcal{K}_{\mathrm{df},3}$ is the three-particle scattering quantity referred to above. It is quasilocal, real, and free of singularities related to three-particle threshold (and so "divergence-free", i.e., df), thus, playing a similar role to the two-particle K matrix $\mathcal{K}_2$ in two-particle scattering. It is, however, unphysical, as it depends on an ultraviolet (UV) cutoff. Given prior knowledge of $\mathcal{K}_2$, and a parametrization of $\mathcal{K}_{\mathrm{df},3}$, the energies of finite-volume states are determined by the vanishing of the determinant in Eq. (1). The parameters in $\mathcal{K}_{\mathrm{df},3}$ are then adjusted to fit to the numerically determined spectrum. Examples on how to numerically solve Eq. (1) have been presented in Refs. [42,44,45].

The second step requires solving infinite-volume integral equations in order to relate $\mathcal{K}_{\mathrm{df},3}$ to the three-particle scattering amplitude $\mathcal{M}_3$. In fact, as explained below, it is a divergence-free version of the latter, denoted $\mathcal{M}_{\mathrm{df},3}$, that is most useful. The equations relating $\mathcal{K}_{\mathrm{df},3}$ to $\mathcal{M}_{\mathrm{df},3}$ were derived in Ref. [40], and solved in Ref. [42].

The parametrizations we use for $\mathcal{K}_2$ and $\mathcal{K}_{\mathrm{df},3}$ are based on an expansion about two- and three-particle thresholds. For $\mathcal{K}_2$, this leads to the standard effective range expansion (ERE), recalled below. At linear order in this expansion only $s$-wave interactions are nonvanishing, with $d$-wave interactions first entering at quadratic order ($p$-wave interactions are forbidden by Bose symmetry). For $\mathcal{K}_{\mathrm{df},3}$, the expansion is in powers of $\Delta = (E^{*2} - 9M^2)/(9M^2)$, and was developed in Refs. [42,44] based on the Lorentz and particle-interchange invariance of $\mathcal{K}_{\mathrm{df},3}$. Through linear order in $\Delta$, $\mathcal{K}_{\mathrm{df},3}$ is given by

$$\mathcal{K}_{\mathrm{df},3} = \mathcal{K}_{\mathrm{df},3}^{\mathrm{iso}} = \mathcal{K}_{\mathrm{df},3}^{\mathrm{iso},0} + \mathcal{K}_{\mathrm{df},3}^{\mathrm{iso},1}\Delta, \qquad (2)$$

where $\mathcal{K}_{\mathrm{df},3}^{\mathrm{iso},0}$ and $\mathcal{K}_{\mathrm{df},3}^{\mathrm{iso},1}$ are constants. There is no dependence on the momenta of the three particles at this order; this corresponds to a contact interaction, and leads to the designation "isotropic" (iso). Momentum dependence first enters at $\mathcal{O}(\Delta^2)$.

In our main analysis, we keep only the $s$-wave two-particle interaction and the isotropic terms in Eq. (2). With these approximations, the QC of Eq. (1) reduces to a finite matrix equation that can be solved by straightforward numerical methods. Previous implementations have considered only the three-particle rest frame, $\vec{P} = 0$ [42,44,45] (see, also, Ref. [48,51]). Here, we have extended the implementation to moving frames, so that we can use all the results obtained by Ref. [58].

In the Supplemental Material [60], we provide further details of the implementation for a general frame, as well as additional details concerning the fits and error estimates described in the remainder of this Letter.

$\chi$PT *prediction for $\mathcal{K}_{\mathrm{df},3}$ and $\mathcal{M}_{\mathrm{df},3}$.*—$\mathcal{M}_{\mathrm{df},3}$ and $\mathcal{K}_{\mathrm{df},3}$ have not previously been calculated in $\chi$PT, so here, we present the LO result. The LO Lagrangian in the isosymmetric two-flavor theory is [61,62]

$$\mathcal{L}_\chi = \frac{F^2}{4}\mathrm{tr}(\partial_\mu U \partial^\mu U^\dagger) + \frac{M^2 F^2}{4}\mathrm{tr}(U + U^\dagger), \quad \text{with}$$

$$U = e^{i\phi/F} \quad \text{and} \quad \phi = \begin{pmatrix} \pi^0 & \sqrt{2}\pi^+ \\ \sqrt{2}\pi^- & -\pi^0 \end{pmatrix}. \qquad (3)$$

Here, $F$ is the decay constant in the chiral limit, normalized such that $F_\pi = 92.4$ MeV. We note that, at this order, $F = F_\pi$. Expanding in powers of the pion fields, $\mathcal{L} = \mathcal{L}_{2\pi} + \mathcal{L}_{4\pi} + \mathcal{L}_{6\pi} + \cdots$, we need only the $4\pi$ and $6\pi$ vertices.

From $\mathcal{L}_{4\pi}$, we obtain the standard LO result for the $2\pi^+$ scattering amplitude [63],

$$\mathcal{M}_2 = \frac{2M^2 - E_2^{*2}}{F^2}, \qquad (4)$$

which displays the well-known Adler zero below threshold at $E_2^{*2} = 2M^2$ [64]. Given the ERE parametrization of the $s$-wave phase shift,

$$q \cot \delta_0(q) = -\frac{1}{a_0} + \frac{r q^2}{2} + P r^3 q^4 + \cdots, \qquad (5)$$

where $q^2 = E_2^{*2}/4 - M^2$, one can infer from Eq. (4) the LO results for the scattering length and effective range

$$M a_0 = \frac{M^2}{16\pi F^2} \quad \text{and} \quad M^2 r a_0 = 3. \qquad (6)$$

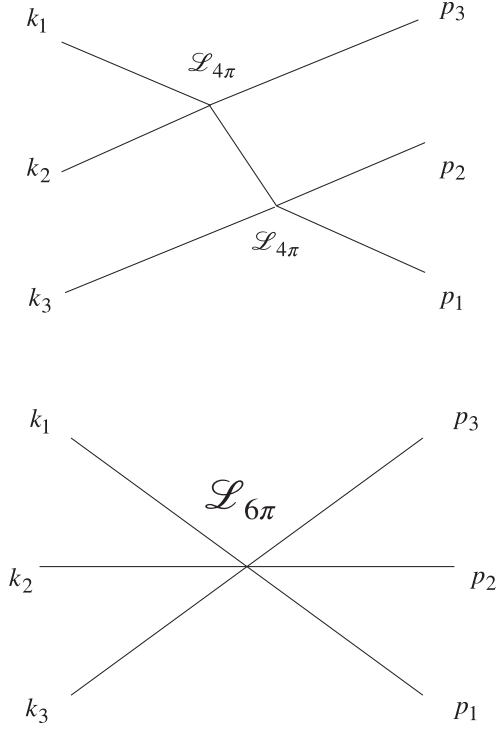

FIG. 1. LO contributions to the three-particle scattering amplitude $\mathcal{M}_3$. Momentum assignments must be symmetrized.

The $3\pi^+$ amplitude $\mathcal{M}_3$ is given at LO by the diagrams of Fig. 1. As is well known, $\mathcal{M}_3$ diverges for certain external momenta, as the propagator in Fig. 1(a) can go on shell. This motivated the introduction of a divergence-free amplitude in Ref. [39]

$$\mathcal{M}_{\mathrm{df},3} \equiv \mathcal{M}_3 - \mathcal{D}, \qquad (7)$$

$$\mathcal{D} = \mathcal{S}\left\{ -\mathcal{M}_2(s_{12}) \frac{1}{b^2 - M^2} \mathcal{M}_2(s'_{12}) \right\} + O(\mathcal{M}_2^3), \qquad (8)$$

where $s_{12} = (p_1 + p_2)^2$, $s'_{12} = (k_1 + k_2)^2$, $b = p_1 + p_2 - k_3$, and $\mathcal{S}$ indicates symmetrization over momentum assignments. $\mathcal{D}$ is defined to have the same divergences as $\mathcal{M}_3$, so that their difference is finite. At LO in $\chi$PT, only the LO term in $\mathcal{D}$ contributes, and we find

$$M^2 \mathcal{M}_{\mathrm{df},3} = \frac{M^4}{F^4}(18 + 27\Delta)$$
$$= (16\pi Ma_0)^2(18 + 27\Delta), \qquad (9)$$

a result that is real and isotropic. As a side result, we have also calculated the related threshold amplitude that enters into the $1/L$ expansion of the three-particle energy [65], finding $\mathcal{M}_{3,\mathrm{th}} = 27M^2/F^4$.

The last step is to relate $\mathcal{M}_{\mathrm{df},3}$ to $\mathcal{K}_{\mathrm{df},3}$. We find these quantities to be equal at LO. We find

$$\mathcal{K}_{\mathrm{df},3} = \mathcal{M}_{\mathrm{df},3}[1 + \mathcal{O}(M^2/F^2)], \qquad (10)$$

so that $\mathcal{K}_{\mathrm{df},3}$ is also given by Eq. (9). This implies that $\mathcal{K}_{\mathrm{df},3}$ is scheme independent at LO in $\chi$PT. We can also quantify the expected size of the corrections, finding them to range between 10% and 50%, with the larger error applying to the term linear in $\Delta$.

*Fitting the two-particle spectrum.*—Determining the two-particle phase shift is an essential step, as it enters into the three-particle QC. In particular, we need a parametrization valid below threshold, as the two-particle momentum in the three-particle QC takes values in the range $q^2/M^2 \in [-1, 3]$. We extract information on the $s$-wave phase shift using a form of the two-particle QC that holds in all frames for those irreps that couple to $J = 0$. We use the bootstrap samples provided in Ref. [58] to determine statistical errors, so that correlations are accounted for properly.

We use a parametrization of the phase shift (adapted from that of Ref. [66]; see, also, Ref. [67]) that includes the Adler zero predicted by $\chi$PT, as well as the kinematical factor $E_2^*$

$$\frac{q}{M} \cot \delta_0(q) = \frac{E_2^* M}{E_2^{*2} - 2z_2^2} \left( B_0 + B_1 \frac{q^2}{M^2} + B_2 \frac{q^4}{M^4} + \cdots \right). \qquad (11)$$

We either set $z_2^2 = M^2$, the LO value, or leave it as a free parameter. $B_0$ and $B_1$ are related in a simple way to $a_0$ and $r$. Previous lattice studies have used the ERE, Eq. (5) (see, e.g., Refs. [68–70]), but this has the disadvantage, due to the Adler zero, of having a radius of convergence of $|q^2| = |M^2 - z_2^2/2| \approx M^2/2$. In particular, the ERE gives results for $-1 < q^2/M^2 < 0$ that are substantially different

TABLE I. Fits of the two-particle spectrum to the Adler-zero form of $q \cot \delta_0$, Eq. (11).

| Fit | $B_0$ | $B_1$ | $B_2$ | $z_2^2/M^2$ | $\chi^2/\mathrm{d.o.f.}$ | $Ma_0$ | $M^2 ra_0$ |
|-----|-------|-------|-------|-------------|--------------------------|--------|------------|
| 1 | −11.2(7) | −2.1(3) | ⋯ | 1 (fixed) | 12.13/(11-2) | 0.089(6) | 2.63(8) |
| 2 | −10.4(9) | −3.7(1.0) | 0.5(3) | 1 (fixed) | 9.75/(11-3) | 0.096(8) | 2.3(3) |
| 3 | −11.7(1.8) | −2.0(4) | ⋯ | 0.94(22) | 12.06/(11-3) | 0.091(9) | 2.4(9) |

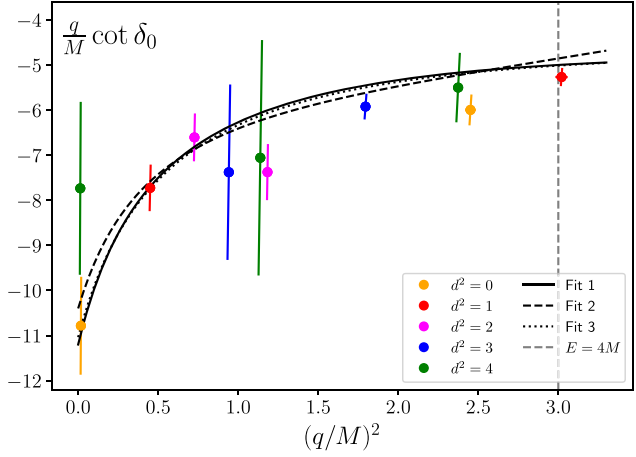

FIG. 2. Values of $q \cot \delta_0$ obtained from the two-particle spectrum of Ref. [58] using the two-particle QC, together with various fits.

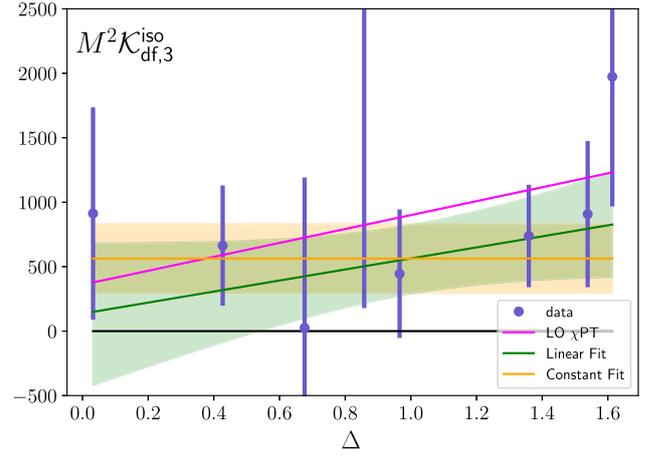

FIG. 3. Results for $M^2 \mathcal{K}_{\mathrm{df},3}^{\mathrm{iso}}$ from individual three-particle levels, using method 1, together with constant and linear fits, and the LO prediction of $\chi$PT.

from the Adler-zero form. This is related to the fact that in (11), $B_1$ and $B_2$ are both of next-to-leading order (NLO) in $\chi$PT, in contrast to the ERE form where $r$ and $P$ are both nonzero at LO, as can be seen from the explicit $\chi$PT expressions given in Ref. [68]. The formal radius of convergence of our expression (11) is $|q^2| = M^2$, due to the left-hand cut, but following common practice, we ignore this and use it up to $q^2/M^2 = 3$. We find that fitting with the restriction $|q^2|/M^2 < 1$ has only a small impact on the resulting parameters. We have also checked that fits using the ERE form provide a worse description of the data.

The results of several fits are listed in Table I and shown in Fig. 2. All fits give reasonable values of $\chi^2$ divided by the number of degrees of freedom, $\chi^2/\mathrm{d.o.f.}$, and yield values for $M^2 r a_0$ close to the predicted LO value of 3. Using the value of $F$ obtained from the same lattice configurations in Ref. [71,72], the LO chiral prediction from Eq. (6) is $M a_0 = 0.0938(12)$, and this is also in good agreement with the results of the fits. Overall, we conclude that the spectrum from Ref. [58] confirms the expectations from $\chi$PT. We choose the minimal fit 1 as our standard choice since $B_2$ is poorly determined (fit 2) and the Adler-zero position is consistent with the LO result if allowed to float (fit 3).

We have performed a similar fit to the five energy levels from Ref. [58] which are sensitive only to the $d$-wave amplitude. Despite very small shifts from the free energies, we find a $3\sigma$ signal for the $d$-wave scattering length, $(M a_2)^5 = 0.0006(2)$. The smallness of this result is

qualitatively consistent with the fact that this is a NLO effect in $\chi$PT, and justifies our neglect of $d$ waves in the three-particle analysis.

*Fitting the three-particle spectrum.*—Now, we use the three-particle spectrum to determine $\mathcal{K}_{\mathrm{df},3}^{\mathrm{iso}}$. Eight levels are sensitive to $\mathcal{K}_{\mathrm{df},3}^{\mathrm{iso}}$, while three are in irreps only sensitive to two-particle interactions. Since all levels are correlated, a global fit to two- and three-particle spectra is needed to properly estimate errors.

Before presenting the global fits, however, we use an approach ("method 1") that allows a separate determination of $\mathcal{K}_{\mathrm{df},3}^{\mathrm{iso}}$ for each of the eight levels sensitive to this parameter. Within each bootstrap sample, we fit the two-particle levels to the fit 1 Adler-zero form described above, and then adjust $\mathcal{K}_{\mathrm{df},3}^{\mathrm{iso}}$ so that the three-particle QC reproduces the energy of the level under consideration. The results are shown in Fig. 3. The values of $\mathcal{K}_{\mathrm{df},3}^{\mathrm{iso}}$ are all positive, and a constant fit yields $M^2 \mathcal{K}_{\mathrm{df},3}^{\mathrm{iso}} = 560(270)$ with $\chi^2/\mathrm{d.o.f.} = 8.5/7$. The LO $\chi$PT result (given by $M^2 \mathcal{K}_{\mathrm{df},3}^{\mathrm{iso}} = 360 + 540\Delta$, taking $M a_0$ from fit 1) is reasonably consistent with the linear fit, as shown. This indicates that a significant result for $\mathcal{K}_{\mathrm{df},3}^{\mathrm{iso}}$ of the expected size may be obtainable.

This fit does not include three-particle energy levels in irreps sensitive only to $\delta_0$. These, however, can be used as a consistency check. We find good agreement between the data and the energies predicted by the QC.

TABLE II. Global fits to the two- and three-particle spectrum using the two- and three-particle QCs.

| Fit | $B_0$ | $B_1$ | $z_2^2/M^2$ | $M^2 \mathcal{K}_{\mathrm{df},3}^{\mathrm{iso},0}$ | $M^2 \mathcal{K}_{\mathrm{df},3}^{\mathrm{iso},1}$ | $\chi^2/\mathrm{d.o.f.}$ | $M a_0$ | $M^2 r a_0$ |
|---|---|---|---|---|---|---|---|---|
| 4 | $-11.1(7)$ | $-2.3(3)$ | 1 (fixed) | 270(160) | $\cdots$ | 27.06/(22-3) | 0.090(6) | 2.59(8) |
| 5 | $-11.1(7)$ | $-2.4(3)$ | 1 (fixed) | 550(330) | $-280(290)$ | 26.04/(22-4) | 0.090(5) | 2.57(8) |

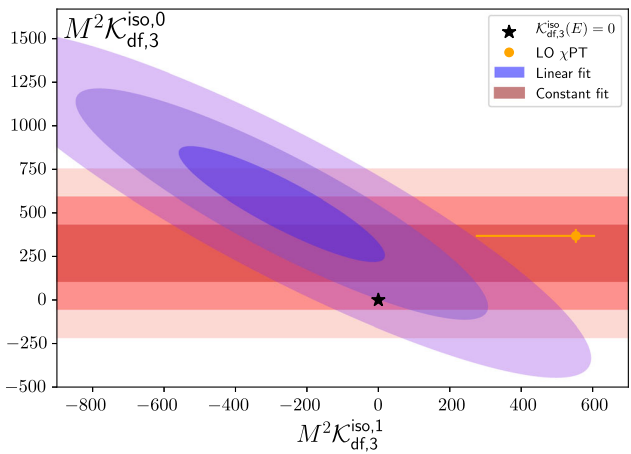

FIG. 4. One, two, and three-sigma confidence intervals for $M^2 \mathcal{K}_{\mathrm{df},3}^{\mathrm{iso}}$ for the two different global fits (4 and 5).

To establish the true significance of the results for $\mathcal{K}_{\mathrm{df},3}^{\mathrm{iso}}$ we perform global fits to the eleven two-particle and eleven three-particle levels that depend on $\delta_0$ and/or $\mathcal{K}_{\mathrm{df},3}^{\mathrm{iso}}$. We do so both for constant and linear $\mathcal{K}_{\mathrm{df},3}^{\mathrm{iso}}$. The results are collected in Table II. Fit 4 finds a value for $\mathcal{K}_{\mathrm{df},3}^{\mathrm{iso}}$ that has around $1.8\sigma$ statistical significance and also gives values for $B_0$ and $B_1$ that are consistent with those from fits 1–3 above and with the LO $\chi$PT predictions. The $p$ value of the fit is $p = 0.103$.

In fit 5, we try a linear ansatz for $\mathcal{K}_{\mathrm{df},3}^{\mathrm{iso}}$, and find that the current dataset of Ref. [58] is insufficient for a separate extraction of both constant and linear terms. We note, however, that, even in this fit, the scenario $\mathcal{K}_{\mathrm{df},3}^{\mathrm{iso}} = 0$ is excluded at $\sim 2\sigma$.

In Fig. 4, we present a summary of the errors resulting from the global fits. We also include the value from LO $\chi$PT, along with an estimate of the NLO corrections. As can be seen, the constant term agrees well with the prediction, whereas the larger disagreement for the linear term is only of marginal significance given the large uncertainty in the $\chi$PT prediction.

One concern with our global fits is that we are using the forms for $\mathcal{K}_2$ and $\mathcal{K}_{\mathrm{df},3}^{\mathrm{iso}}$ beyond their radii of convergence. For $\mathcal{K}_{\mathrm{df},3}^{\mathrm{iso}}$, we do not know the radius of convergence, but a reasonable estimate is that one should use levels only with $|\Delta| < 1$. To check the importance of this issue, we have repeated the global fits imposing $q^2/M^2 < 1$ and $\Delta < 1$, so that the fit includes only five $2\pi^+$ and five $3\pi^+$ levels. We find fit parameters that are consistent with those in Table II, but with much larger errors. For example, the result from the equivalent of fit 4 gives $M^2 \mathcal{K}_{\mathrm{df},3}^{\mathrm{iso},0} = 610(350)$.

We close by commenting on sources of systematic errors. The results of Ref. [58] are subject to discretization errors, but these are of $O(a^2)$, and likely small compared to the statistical errors from [58]. The quantization condition itself neglects exponentially suppressed corrections, but these are numerically small ($e^{-ML} \sim 1\%$) compared to our final statistical error. Errors from truncation of the threshold expansion for $\mathcal{K}_2$ and $\mathcal{K}_{\mathrm{df},3}$ are also present but harder to estimate.

*Conclusions.*—We have presented statistical evidence for a nonzero $3\pi^+$ contact interaction, obtained by analyzing the spectrum of three pion states in isosymmetric QCD with $M \approx 200$ MeV obtained in Ref. [58]. This illustrates the utility of the three-particle quantization condition. It also emphasizes the need for a relativistic formalism, since most of the spectral levels used here are in the relativistic regime. It gives an example where lattice methods can provide results for scattering quantities that are not directly accessible to experiment.

We expect that forthcoming generalizations to the formalism (to incorporate nondegenerate particles with spin, etc.), combined with advances in the methods of lattice QCD (to allow the accurate determination of the spectrum in an increasing array of systems), will allow generalization of the present results to resonant three-particle systems in the next few years.

---

*blanton1@uw.edu
†fernando.romero@uv.es
‡srsharpe@uw.edu

[1] M. Lüscher, Commun. Math. Phys. 105, 153 (1986).
[2] M. Lüscher, Nucl. Phys. B354, 531 (1991).
[3] K. Rummukainen and S. A. Gottlieb, Nucl. Phys. B450, 397 (1995).
[4] C. h. Kim, C. T. Sachrajda, and S. R. Sharpe, Nucl. Phys. B727, 218 (2005).
[5] S. He, X. Feng, and C. Liu, J. High Energy Phys. 07 (2005) 011.
[6] V. Bernard, M. Lage, and U.-G. Meißner, and A. Rusetsky, J. High Energy Phys. 01 (2011) 019.
[7] M. T. Hansen and S. R. Sharpe, Phys. Rev. D 86, 016007 (2012).
[8] R. A. Briceño and Z. Davoudi, Phys. Rev. D 88, 094507 (2013).
[9] R. A. Briceño, Phys. Rev. D 89, 074507 (2014).

[10] F. Romero-López, A. Rusetsky, and C. Urbach, Phys. Rev. D 98, 014503 (2018).

[11] T. Luu and M. J. Savage, Phys. Rev. D 83, 114508 (2011).

[12] M. Göckeler, R. Horsley, M. Lage, U. G. Meißner, P. E. L. Rakow, A. Rusetsky, G. Schierholz, and J. M. Zanotti, Phys. Rev. D 86, 094513 (2012).

[13] X. Feng, K. Jansen, and D. B. Renner, Phys. Lett. B 684, 268 (2010).

[14] M. Lage, U.-G. Meissner, and A. Rusetsky, Phys. Lett. B 681, 439 (2009).

[15] D. J. Wilson, R. A. Briceño, J. J. Dudek, R. G. Edwards, and C. E. Thomas, Phys. Rev. D 92, 094502 (2015).

[16] R. A. Briceño, J. J. Dudek, R. G. Edwards, and D. J. Wilson, Phys. Rev. Lett. 118, 022002 (2017).

[17] R. Brett, J. Bulava, J. Fallica, A. Hanlon, B. Höz, and C. Morningstar, Nucl. Phys. B932, 29 (2018).

[18] C. W. Andersen, J. Bulava, B. Hörz, and C. Morningstar, Phys. Rev. D 97, 014506 (2018).

[19] D. Guo, A. Alexandru, R. Molina, M. Mai, and M. Döring, Phys. Rev. D 98, 014507 (2018).

[20] C. Andersen, J. Bulava, B. Hörz, and C. Morningstar, Nucl. Phys. B939, 145 (2019).

[21] J. J. Dudek, R. G. Edwards, C. E. Thomas, and D. J. Wilson (Hadron Spectrum Collaboration), Phys. Rev. Lett. 113, 182001 (2014).

[22] J. J. Dudek, R. G. Edwards, and D. J. Wilson (Hadron Spectrum Collaboration), Phys. Rev. D 93, 094506 (2016).

[23] A. Woss, C. E. Thomas, J. J. Dudek, R. G. Edwards, and D. J. Wilson, J. High Energy Phys. 07 (2018) 043.

[24] A. J. Woss, C. E. Thomas, J. J. Dudek, R. G. Edwards, and D. J. Wilson, Phys. Rev. D 100, 054506 (2019).

[25] C. Helmes, C. Jost, B. Knippschild, B. Kostrzewa, L. Liu, F. Pittler, C. Urbach, and M. Werner (ETM Collaboration), Phys. Rev. D 98, 114511 (2018).

[26] L. Liu et al., Phys. Rev. D 96, 054516 (2017).

[27] C. Helmes, C. Jost, B. Knippschild, B. Kostrzewa, L. Liu, C. Urbach, and M. Werner, Phys. Rev. D 96, 034510 (2017).

[28] C. Helmes, C. Jost, B. Knippschild, C. Liu, J. Liu, L. Liu, C. Urbach, M. Ueding, Z. Wang, and M. Werner (ETM Collaboration), J. High Energy Phys. 09 (2015) 109.

[29] M. Werner et al., arXiv:1907.01237.

[30] C. Culver, M. Mai, A. Alexandru, M. Döring, and F. X. Lee, Phys. Rev. D 100, 034509 (2019).

[31] M. Mai, C. Culver, A. Alexandru, M. Döring, and F. X. Lee, Phys. Rev. D 100, 114514 (2019).

[32] M. Doring, U. G. Meissner, E. Oset, and A. Rusetsky, Eur. Phys. J. A 48, 114 (2012).

[33] R. A. Briceño, J. J. Dudek, and R. D. Young, Rev. Mod. Phys. 90, 025001 (2018).

[34] L. D. Roper, Phys. Rev. Lett. 12, 340 (1964).

[35] R. F. Lebed, R. E. Mitchell, and E. S. Swanson, Prog. Part. Nucl. Phys. 93, 143 (2017).

[36] M. Hoferichter, B.-L. Hoid, and B. Kubis, J. High Energy Phys. 08 (2019) 137.

[37] K. Polejaeva and A. Rusetsky, Eur. Phys. J. A 48, 67 (2012).

[38] S. Kreuzer and H. W. Grießhammer, Eur. Phys. J. A 48, 93 (2012).

[39] M. T. Hansen and S. R. Sharpe, Phys. Rev. D 90, 116003 (2014).

[40] M. T. Hansen and S. R. Sharpe, Phys. Rev. D 92, 114509 (2015).

[41] R. A. Briceño, M. T. Hansen, and S. R. Sharpe, Phys. Rev. D 95, 074510 (2017).

[42] R. A. Briceño, M. T. Hansen, and S. R. Sharpe, Phys. Rev. D 98, 014506 (2018).

[43] R. A. Briceño, M. T. Hansen, and S. R. Sharpe, Phys. Rev. D 99, 014516 (2019).

[44] T. D. Blanton, F. Romero-López, and S. R. Sharpe, J. High Energy Phys. 03 (2019) 106.

[45] F. Romero-López, S. R. Sharpe, T. D. Blanton, R. A. Briceño, and M. T. Hansen, J. High Energy Phys. 10 (2019) 007.

[46] H.-W. Hammer, J.-Y. Pang, and A. Rusetsky, J. High Energy Phys. 09 (2017) 109.

[47] H. W. Hammer, J. Y. Pang, and A. Rusetsky, J. High Energy Phys. 10 (2017) 115.

[48] M. Döring, H. W. Hammer, M. Mai, J. Y. Pang, A. Rusetsky, and J. Wu, Phys. Rev. D 97, 114508 (2018).

[49] J.-Y. Pang, J.-J. Wu, H. W. Hammer, Ulf.-G. Meißner, and A. Rusetsky, Phys. Rev. D 99, 074513 (2019).

[50] M. Mai and M. Döring, Eur. Phys. J. A 53, 240 (2017).

[51] M. Mai and M. Döring, Phys. Rev. Lett. 122, 062503 (2019).

[52] P. Klos, S. König, H. W. Hammer, J. E. Lynn, and A. Schwenk, Phys. Rev. C 98, 034004 (2018).

[53] P. Guo and V. Gasparian, Phys. Lett. B 774, 441 (2017).

[54] M. T. Hansen and S. R. Sharpe, Annu. Rev. Nucl. Part. Sci. 69, 65 (2019).

[55] S. R. Beane, W. Detmold, and M. J. Savage, Phys. Rev. D 76, 074507 (2007).

[56] W. Detmold, M. J. Savage, A. Torok, S. R. Beane, T. C. Luu, K. Orginos, and A. Parreno, Phys. Rev. D 78, 014507 (2008).

[57] F. Romero-López, A. Rusetsky, and C. Urbach, Eur. Phys. J. C 78, 846 (2018).

[58] B. Hörz and A. Hanlon, Phys. Rev. Lett. 123, 142002 (2019).

[59] M. Mai, M. Döring, C. Culver, and A. Alexandru, arXiv: 1909.05749.

[60] See Supplemental Material at http://link.aps.org/supplemental/10.1103/PhysRevLett.124.032001 for further details of the implementation, and additional details concerning the fits and error estimates.

[61] S. Weinberg, Physica (Amsterdam) 96A, 327 (1979).

[62] J. Gasser and H. Leutwyler, Ann. Phys. (N.Y.) 158, 142 (1984).

[63] S. Weinberg, Phys. Rev. Lett. 17, 616 (1966).

[64] S. L. Adler, Phys. Rev. 137, B1022 (1965); 137, 140 (1964).

[65] M. T. Hansen and S. R. Sharpe, Phys. Rev. D 93, 096006 (2016); 96, 039901(E) (2017).

[66] F. J. Yndurain, arXiv:hep-ph/0212282.

[67] J. R. Pelaez, A. Rodas, and J. Ruiz de Elvira, Eur. Phys. J. C 79, 1008 (2019).

[68] S. R. Beane, E. Chang, W. Detmold, H. W. Lin, T. C. Luu, K. Orginos, A. Parreno, M. J. Savage, A. Torok, and A. Walker-Loud (NPLQCD Collaboration), Phys. Rev. D **85**, 034505 (2012).

[69] J. J. Dudek, R. G. Edwards, and C. E. Thomas, Phys. Rev. D **86**, 034031 (2012).

[70] J. Bulava, B. Fahy, B. Hörz, K. J. Juge, C. Morningstar, and C. H. Wong, Nucl. Phys. **B910**, 842 (2016).

[71] M. Bruno, T. Korzec, and S. Schaefer, Phys. Rev. D **95**, 074504 (2017).

[72] M. Bruno *et al.*, J. High Energy Phys. 02 (2015) 043.

THE EUROPEAN
PHYSICAL JOURNAL C

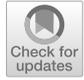

# Dissecting the $\Delta I = 1/2$ rule at large $N_c$

Andrea Donini[1], Pilar Hernández[1], Carlos Pena[2], Fernando Romero-López[1,a]

[1] IFIC (CSIC-UVEG), Edificio Institutos Investigación, Apt. 22085, 46071 Valencia, Spain
[2] Departamento de Física Teórica and Instituto de Física Teórica UAM-CSIC, Universidad Autónoma de Madrid, 28049 Madrid, Spain

**Abstract** We study the scaling of kaon decay amplitudes with the number of colours, $N_c$, in a theory with four degenerate flavours, $N_f = 4$. In this scenario, two current-current operators, $Q^\pm$, mediate $\Delta S = 1$ transitions, such as the two isospin amplitudes of non-leptonic kaon decays for $K \rightarrow (\pi\pi)_{I=0,2}$, $A_0$ and $A_2$. In particular, we concentrate on the simpler $K \rightarrow \pi$ amplitudes, $A^\pm$, mediated by these two operators. A diagrammatic analysis of the large-$N_c$ scaling of these observables is presented, which demonstrates the anticorrelation of the leading $\mathcal{O}(1/N_c)$ and $\mathcal{O}(N_f/N_c^2)$ corrections in both amplitudes. Using our new $N_f = 4$ and previous quenched data, we confirm this expectation and show that these corrections are *naturally* large and may be at the origin of the $\Delta I = 1/2$ rule. The evidence for the latter is indirect, based on the matching of the amplitudes to their prediction in Chiral Perturbation Theory, from which the LO low-energy couplings of the chiral weak Hamiltonian, $g^\pm$, can be determined. A NLO estimate of the $K \rightarrow (\pi\pi)_{I=0,2}$ isospin amplitudes can then be derived, which is in good agreement with the experimental value.

## 1 Introduction

Significant progress has been achieved recently in the lattice determination of $K \rightarrow (\pi\pi)_{I=0,2}$ and the CP violating observable $\epsilon'/\epsilon$ [1–3]. In particular, a large enhancement of the $I = 0$ amplitude over the $I = 2$ one has been reported, albeit with too large uncertainty to be considered a satisfactory first-principles determination of the $\Delta I = 1/2$ rule.[1]

In Ref. [5] an analysis of the different contributions was made and it was suggested that the main source of the enhancement lies in a strong cancellation of the isospin-two

---

[1] While this paper was under revision, a significantly improved result at the physical point was made public [4].

a e-mail: fernando.romero@uv.es (corresponding author)

amplitude, as a result of a negative relative sign between the colour-connected and colour-disconnected contractions, with the two contributions adding up in the isospin-zero channel. In Refs. [6–8] we proposed to study the $N_c$ dependence of the amplitudes, because the two contributions scale differently in large $N_c$ and therefore can be rigorously disentangled in this limit. The enhancement, if explained in this fashion, seems to require unnaturally large-$N_c$ corrections with the appropriate sign.

Interestingly, the large-$N_c$ limit of QCD [9,10] has also inspired several phenomenological determinations of these and related observables [11–19] (for a recent discussion see [20–22]). It is well known, however, that the leading-order large-$N_c$ prediction for the ratio of the amplitudes, $\lim_{N_c \to \infty} A_0/A_2 = \sqrt{2}$, i.e., no $\Delta I = 1/2$ rule whatsoever. The subleading $N_c$ corrections should therefore be very large, which could be consistent with the previous hypothesis, but casts doubts on the phenomenological approaches that make use of large-$N_c$ inspired approximations: if we know that there must be significant large-$N_c$ corrections to explain the $\Delta I = 1/2$, why should we trust approximations that neglect subleading $N_c$ terms?

The $N_c$ dependence can be studied from first-principles in lattice QCD by simply simulating at different number of colours [23–27]. In our previous work [6–8] we explored the related weak amplitudes $K \rightarrow \pi$ and $K \rightarrow \bar{K}$ in the quenched approximation, and found no unnaturally large subleading $N_c$ corrections, although we confirmed the exact anticorrelation of these corrections in the two isospin channels. The quenched approximation introduces however an uncontrollable systematic error, which in practice is often found to be relatively small in most quantities. Since we are interested in subleading $N_c$ corrections, quenching effects are expected to enter at this order of the $N_c$ expansion and therefore need to be included. The main goal of this paper is to extend our previous study beyond the quenched approximation, which will allow us to determine from first-principles the subleading $N_c$ corrections to the $\Delta I = 1/2$

rule, in a simplified setting with four degenerate flavours, $m_u = m_d = m_s = m_c$.

This paper is organized as follows: in Sect. 2 we discuss our strategy for the lattice study of $K \to \pi$ transitions; in Sect. 3 we discuss the $N_c$ scaling of the amplitudes; Sect. 4 deals with the necessary results in Chiral Perturbation Theory to connect to $K \to \pi\pi$; Sect. 5 describes the setup of our lattice computations; in Sect. 6 we discuss our physics results; and we conclude in Sect. 7.

## 2 Strategy

The Operator Product Expansion allows to represent CP-conserving $\Delta S = 1$ transitions by an effective Hamiltonian of four-fermion operators. At the electroweak scale, $\mu \simeq M_W$, we can neglect all quark masses, and the weak Hamiltonian takes the simple form:

$$H_{\mathrm{w}}^{\Delta S=1} = \int d^4x \, \frac{g_{\mathrm{w}}^2}{4M_W^2} V_{us}^* V_{ud} \sum_{\sigma=\pm} k^\sigma(\mu) \, \bar{Q}^\sigma(x,\mu), \quad (1)$$

where $g_{\mathrm{w}}^2 = 4\sqrt{2} G_{\mathrm{F}} M_W^2$. Only two four-quark operators of dimension six can appear with the correct symmetry properties under the flavour symmetry group $SU(4)_{\mathrm{L}} \times SU(4)_{\mathrm{R}}$, namely

$$\begin{aligned}\bar{Q}^\pm(x,\mu) = Z_Q^\pm(\mu) \big( & J_\mu^{su}(x) J_\mu^{ud}(x) \pm J_\mu^{sd}(x) J_\mu^{uu}(x) \\ & - [u \leftrightarrow c] \big),\end{aligned} \quad (2)$$

where $J_\mu$ is the left-handed current $J_\mu^{ij} = (\bar{\psi}^i \gamma_\mu P_- \psi^j)$; $i, j$ are quark flavour indices; $P_\pm = \frac{1}{2}(\mathbf{1} \pm \gamma_5)$; and parentheses around quark bilinears indicate that they are traced[2] over spin and colour. $Z_Q^\pm(\mu)$ is the renormalization constant of the bare operator $Q^\pm(x)$ computed in some regularization scheme as, for example, the lattice. There are other operators that could mix with those above: however, they vanish in the limit of equal up and charm masses, that we refer to as the GIM limit [28]. From the lattice point of view the GIM limit is very advantageous, not only for the simpler operator mixing, but also because no closed quark propagator contributes to the amplitudes. Even though the presence of a heavy charm was argued long ago to be at the origin of the $\Delta I = 1/2$ rule via the mixing with penguin operators [29], the relevance of penguin contributions has been found to be small in non-perturbative studies [1,30].[3] If we want to test the primary mechanism of the $\Delta I = 1/2$ enhancement proposed in [5], the GIM limit may be good enough.

---

[2] This basis can be related to the more traditional one by means of Fierz identities.

[3] The dominance of current-current operators over penguin contributions was also pointed out in the Dual QCD approach [11].

The operators $\bar{Q}^\sigma(\mu)$ are renormalized at a scale $\mu$ in some renormalization scheme, being their $\mu$ dependence exactly cancelled by that of the Wilson coefficients $k^\sigma(\mu)$. It is also possible to define renormalization group invariant (RGI) operators, which are defined by cancelling their $\mu$ dependence, as derived from the Callan–Symanzik equations,

$$\hat{Q}^\sigma \equiv \hat{c}^\sigma(\mu) \bar{Q}^\sigma(\mu), \quad (3)$$

with

$$\begin{aligned}\hat{c}^\sigma(\mu) \equiv & \left( \frac{N_c}{3} \frac{\bar{g}^2(\mu)}{4\pi} \right)^{\frac{\gamma_0^\sigma}{2b_0}} \\ & \times \exp\left\{ -\int_0^{\bar{g}(\mu)} dg \left[ \frac{\gamma^\sigma(g)}{\beta(g)} - \frac{\gamma_0^\sigma}{b_0 g} \right] \right\}, \quad (4)\end{aligned}$$

where $\bar{g}(\mu)$ is the running coupling and $\beta(g) = -g^3 \sum_n b_n g^{2n}$, $\gamma^\sigma(g) = -g^2 \sum_n \gamma_n^\sigma g^{2n}$ are the $\beta$-function and the four-fermion operator anomalous dimension, respectively. The one- and two-loop coefficients of the $\beta$-function, and the one-loop coefficient of the anomalous dimensions, are renormalization scheme-independent. Their values for the theory with $N_f$ flavours are [31–36]

$$b_0 = \frac{1}{(4\pi)^2} \left[ \frac{11}{3} N_c - \frac{2}{3} N_f \right], \quad (5)$$

$$b_1 = \frac{1}{(4\pi)^4} \left[ \frac{34}{3} N_c^2 - \left( \frac{13}{3} N_c - \frac{1}{N_c} \right) N_f \right], \quad (6)$$

and for the operators $Q^\pm$ [37,38]

$$\gamma_0^\pm = \frac{1}{(4\pi)^2} \left[ \pm 6 - \frac{6}{N_c} \right]. \quad (7)$$

The normalization of $\hat{c}^\sigma(\mu)$ coincides with the most popular one for $N_c = 3$, whilst using the 't Hooft coupling $\lambda = N_c \bar{g}^2(\mu)$ in the first factor instead of the usual coupling, so that the large-$N_c$ limit is well-defined.

Defining similarly an RGI Wilson coefficient

$$\hat{k}^\sigma \equiv \frac{k^\sigma(\mu)}{\hat{c}^\sigma(\mu)}, \quad (8)$$

we can rewrite the Hamiltonian in terms of RGI quantities, which no longer depend on the scale, so that

$$\begin{aligned}\hat{k}^\sigma \hat{Q}^\sigma &= \left[ \frac{k^\sigma(M_W)}{\hat{c}^\sigma(M_W)} \right] \left[ \hat{c}^\sigma(\mu) \bar{Q}^\sigma(\mu) \right] \\ &= k^\sigma(M_W) U^\sigma(\mu, M_W) \bar{Q}^\sigma(\mu), \quad (9)\end{aligned}$$

where $\mu$ is a convenient renormalization scale for the non-perturbative computation of matrix elements of $Q^\pm$, which will be later set to the inverse lattice scale $a^{-1}$. The factor $U^\sigma(\mu, M_W) = \hat{c}^\sigma(\mu)/\hat{c}^\sigma(M_W)$, therefore, measures the running of the renormalized operator between the scales $\mu$ and $M_W$. Ideally one would like to evaluate this factor non-perturbatively, as has been done for $N_c = 3$ [39,40], but

such a challenging endeavour is beyond the scope of this paper. We will instead use the perturbative results at two loops in the RI scheme [41,42] to evaluate the Wilson coefficients $k^\sigma (M_W)$, the running factors $U^\sigma (\mu, M_W)$, and $\hat{c}(\mu)$. This implies relying on perturbation theory at scales above $\mu \geq a^{-1} \sim 2.6$ GeV. Similarly we will also use lattice perturbation theory to estimate the renormalization factors $Z_Q^\pm$, that are known to one loop[4] [43,44].

We are interested in considering $K \rightarrow \pi$ amplitudes in the two isospin channels, that we can extract from ratios of three-point correlators

$$\mathcal{C}_3^\pm (y, z, x)$$
$$\equiv \langle P^{du}(y)[O^{suud}(z) \pm O^{sduu}(z)]P^{us}(x)\rangle, \quad (10)$$

where

$$P^{ij}(x) \equiv \bar{\psi}^i(x)\gamma_5\psi^j(x), \quad O^{ijkl} \equiv \bar{\psi}^i\gamma_\mu\psi^j\bar{\psi}^k\gamma_\mu\psi^l, \quad (11)$$

and the two-point correlators

$$\mathcal{C}_2^{ij}(y, z) \equiv \langle P^{ij}(y)A_0^{ji}(z)\rangle, \quad (12)$$

with $A_0^{ij}(x) \equiv \bar{\psi}^i(x)\gamma_0\gamma_5\psi^j(x)$.

From these correlators we define the bare lattice ratios:

$$R^\pm = \lim_{\substack{z_0 - x_0 \to \infty \\ y_0 - z_0 \to \infty}} \frac{\sum_{x,y} \mathcal{C}_3^\pm (y, z, x)}{\sum_{x,y} \mathcal{C}_2^{du}(y, z)\mathcal{C}_2^{us}(x, z)}, \quad (13)$$

which are proportional to the $K \rightarrow \pi$ matrix elements with a convenient normalization. The renormalization factors for these ratios, $Z^\pm$, are obtained from the ratio of the renormalization factors of the four fermion operators, and the current normalization factors that appear in the denominator.

From the renormalized ratios

$$\bar{R}^\sigma = Z^\sigma R^\sigma, \quad (14)$$

we can obtain the RGI normalized ratios

$$\hat{R}^\sigma = \hat{c}\left(a^{-1}\right) Z^\sigma R^\sigma, \quad (15)$$

and the normalized[5] $K \rightarrow \pi$ amplitudes, written either in terms of the RGI or the renormalized ratios, as

$$A^\sigma = \hat{k}^\sigma \hat{R}^\sigma = k^\sigma (M_W)U^\sigma (a^{-1}, M_W)\bar{R}^\sigma. \quad (16)$$

---

[4] The NLO running of the coupling and four-quark operators have been performed fully in the $N_f = 4$ theory, using the value of $\Lambda_{\overline{MS}}(N_f = 4)$ by the ALPHA Collaboration in Ref. [42]. We have checked that the effect of running from $N_f = 5$ from $M_W$ to the $b$ quark mass, and then with $N_f = 4$ down to the lattice matching scale amounts to few per mille effects on the running factors. This is completely negligible within the uncertainty of our final results.

[5] Note that our normalization in Eq. (13) cancels two powers of the decay constant in the physical amplitudes.

**Table 1** Perturbative renormalization constants and RG running factors for the ensembles with $N_f = 4$. $Z^\sigma (a^{-1})$ have been computed at one loop in tadpole-improved perturbation theory using the results in [43,44], whereas $U^\sigma$ and $k^\sigma$ are computed using the two-loop $\overline{MS}$ coupling. The star labels the simulation points with finer lattice spacing, $a \sim 0.065$ fm. In the evaluation of $\hat{c}^\sigma (a^{-1})$ we have used $\Lambda_{\overline{MS}}(N_f = 4) = 298$ MeV from Ref. [45]

| $N_c$ | $k^+(M_W)$ | $U^+(a^{-1}, M_W)$ | $Z^+(a^{-1})$ | $\hat{c}^+(a^{-1})$ |
|---|---|---|---|---|
| 3 | 1.041 | 0.843 | 0.841 | 1.456 |
| 3* | 1.041 | 0.852 | 0.844 | 1.471 |
| 4 | 1.032 | 0.877 | 0.884 | 1.367 |
| 5 | 1.026 | 0.899 | 0.909 | 1.302 |
| 6 | 1.022 | 0.914 | 0.926 | 1.255 |

| $N_c$ | $k^-(M_W)$ | $U^-(a^{-1}, M_W)$ | $Z^-(a^{-1})$ | $\hat{c}^-(a^{-1})$ |
|---|---|---|---|---|
| 3 | 0.918 | 1.433 | 1.320 | 0.488 |
| 3* | 0.918 | 1.400 | 1.314 | 0.476 |
| 4 | 0.947 | 1.254 | 1.195 | 0.602 |
| 5 | 0.961 | 1.179 | 1.137 | 0.679 |
| 6 | 0.970 | 1.137 | 1.104 | 0.731 |

All the required factors to reconstruct the physical amplitudes are summarized in Table 1 for $N_f = 4$ (this work), and in Table 2 for the quenched case [6,7].

## 3 large-$N_c$ scaling of $K \rightarrow \pi$ amplitudes

### 3.1 Diagrammatic expansion of $A^\pm$

A simple diagrammatic analysis of the three and two point correlators of Eqs. (10, 12) shows a clear pattern of the large-$N_c$ scaling, and demonstrates the expected anticorrelation of the leading large-$N_c$ corrections of the $A^\pm$ amplitudes.

After integration over fermion fields, the correlators are obtained from the gauge averages of the colour-disconnected and colour-connected contractions of Fig. 1, corresponding to the operator insertion $O^{suud}$ and $O^{sduu}$, respectively.

In Figs. 2 and 3 we show the scaling with $N_c$ of the lowest-order diagrams contributing to these correlators. The leading $N_c$ dependence of both the renormalized and bare correlators are therefore of the form:

$$\langle P^{ij}J_\mu^{ji}\rangle = N_c\left(a + b\frac{N_f}{N_c}\right) + \dots,$$

$$\langle P^{du}O^{suud}P^{us}\rangle = \langle P^{du}J_\mu^{ud}\rangle\langle P^{su}J_\mu^{us}\rangle + c + d\frac{N_f}{N_c} + \dots,$$

$$\langle P^{du}O^{sduu}P^{us}\rangle = N_c\left(e + f\frac{N_f}{N_c}\right) + \dots, \quad (17)$$

where all the coefficients $a - f$ in these expressions (each of them related to one or more diagrams in Figs. 2 and 3)

**Table 2** Perturbative renormalization constants and RG running factors for the runs with $N_f = 0$ of Refs. [6,7]. $Z^\sigma(a^{-1})$ have been computed at one loop in tadpole-improved perturbation theory using the results in [43,44], whereas $U^\sigma$ and $k^\sigma$ are computed using the two-loop $\overline{\text{MS}}$ coupling. Note that the values of $Z^\sigma(a^{-1})$ differ from those in Refs. [6,7], where bare lattice perturbation theory was used. Furthermore, the values of $k^\sigma$ and $U^\sigma$ also supersede the ones in Refs. [6,7]. In the evaluation of $\hat{c}^\sigma(a^{-1})$ we have used $\Lambda_{\overline{\text{MS}}}$ as described in Ref. [6]

| $N_c$ | $k^+(M_W)$ | $U^+(a^{-1}, M_W)$ | $Z^+(a^{-1})$ | $\hat{c}^+(a^{-1})$ |
|---|---|---|---|---|
| 3 | 1.029 | 0.877 | 0.956 | 1.412 |
| 4 | 1.025 | 0.897 | 0.963 | 1.340 |
| 5 | 1.021 | 0.911 | 0.969 | 1.285 |
| 6 | 1.018 | 0.923 | 0.973 | 1.243 |
| 7 | 1.016 | 0.932 | 0.976 | 1.212 |
| 8 | 1.014 | 0.939 | 0.979 | 1.187 |
| 17 | 1.007 | 0.969 | 0.989 | 1.091 |

| $N_c$ | $k^-(M_W)$ | $U^-(a^{-1}, M_W)$ | $Z^-(a^{-1})$ | $\hat{c}^-(a^{-1})$ |
|---|---|---|---|---|
| 3 | 0.942 | 1.312 | 1.087 | 0.511 |
| 4 | 0.959 | 1.206 | 1.061 | 0.619 |
| 5 | 0.969 | 1.153 | 1.047 | 0.690 |
| 6 | 0.975 | 1.121 | 1.038 | 0.740 |
| 7 | 0.979 | 1.101 | 1.032 | 0.776 |
| 8 | 0.982 | 1.086 | 1.027 | 0.803 |
| 17 | 0.992 | 1.037 | 1.012 | 0.907 |

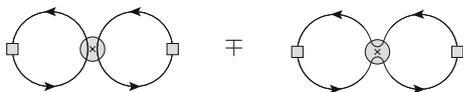

**Fig. 1** Left diagram: $O^{suud}(x)$ insertion or colour-disconnected contribution to $\mathcal{C}_3^\pm$ in Eq. (10). Right diagram: $O^{sduu}(x)$ insertion or colour-connected contribution to $\mathcal{C}_3^\pm$ in Eq. (10)

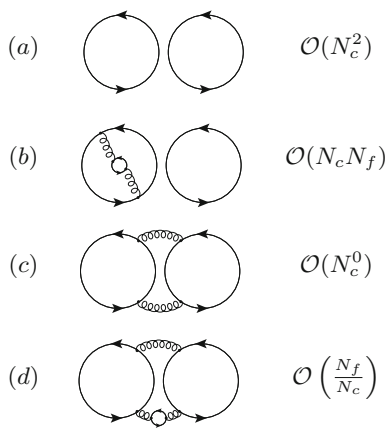

*(a)*     $\mathcal{O}(N_c^2)$

*(b)*     $\mathcal{O}(N_c N_f)$

*(c)*     $\mathcal{O}(N_c^0)$

*(d)*     $\mathcal{O}\left(\frac{N_f}{N_c}\right)$

**Fig. 2** $N_c$, $N_f$ scaling of various contributions to the colour-disconnected contraction, corresponding to the $O^{suud}(x)$ insertion

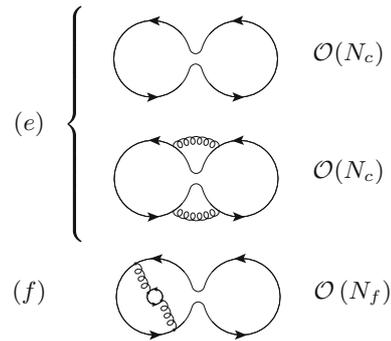

*(e)*     $\mathcal{O}(N_c)$

    $\mathcal{O}(N_c)$

*(f)*     $\mathcal{O}(N_f)$

**Fig. 3** $N_c$, $N_f$ scaling of various contributions to the colour-connected contraction, corresponding to the $O^{sduu}(x)$ insertion

are independent of $N_c$ and $N_f$. These relations imply that the leading $N_c$ corrections in the $\pm$ correlation functions of Eq. (10) are of $\mathcal{O}(N_c^2, N_f N_c)$, but factorizable. On the other hand, the leading non-factorizable corrections are of $\mathcal{O}(N_c)$ and $\mathcal{O}(N_f)$, and cancel in the sum of the $\pm$ correlators:

$$\mathcal{C}_3^+ + \mathcal{C}_3^- = \text{disconnected} + \mathcal{O}(N_c^0) + \mathcal{O}\left(\frac{N_f}{N_c}\right) + \cdots,$$
$$\mathcal{C}_3^+ - \mathcal{C}_3^- = \mathcal{O}(N_c) + \mathcal{O}(N_f) + \cdots \quad (18)$$

They are therefore fully anticorrelated in the $\pm$ correlators. Importantly, the anticorrelated terms include the leading fermion loop corrections, $\mathcal{O}(N_f)$. These relations also imply the following scaling of the renormalization factors:

$$\frac{Z_Q^+ + Z_Q^-}{2} = 1 + \mathcal{O}\left(\frac{1}{N_c^2}\right) + \mathcal{O}\left(\frac{N_f}{N_c^3}\right) + \cdots$$
$$\frac{Z_Q^+ - Z_Q^-}{2} = \mathcal{O}\left(\frac{1}{N_c}\right) + \mathcal{O}\left(\frac{N_f}{N_c^2}\right) + \cdots, \quad (19)$$

and a similar one for the Wilson coefficients, $k^\sigma$. This dependence can be explicitly checked in the perturbative coefficients known up to two loops in the $\overline{\text{MS}}$ scheme [41,42].

These results imply the following scaling of the amplitudes:

$$A^\pm = 1 \pm \tilde{a}\frac{1}{N_c} \pm \tilde{b}\frac{N_f}{N_c^2} + \tilde{c}\frac{1}{N_c^2} + \tilde{d}\frac{N_f}{N_c^3} + \cdots, \quad (20)$$

where the coefficients $\tilde{a} - \tilde{d}$ are combinations of the coefficients $a - f$ in Eq. (17), and are also independent of $N_c$ and $N_f$, and a *natural* expectation is that they are $\mathcal{O}(1)$.

Not only the leading corrections $N_c^{-1}$ are, therefore, fully anticorrelated in the ratios, but also the leading effects of dynamical quarks, $\mathcal{O}(N_f)$. Note that this analysis does not predict the sign of the different terms, i.e., the sign of the $\tilde{a} - \tilde{d}$ coefficients, only the (anti-)correlation between the two isospin channels. This way, a negative sign of $\tilde{a}$ and $\tilde{b}$ results into an enhancement of the ratio $A^-/A^+$.

## 3.2 't Hooft vs. Veneziano scaling

As we will see the number of active flavours, $N_f$, plays a relevant role in the $1/N_c$ expansion of the $K \to \pi$ amplitudes. The scaling in $N_f$ is in fact the difference between the 't Hooft and Veneziano limits of QCD. While the former keeps $N_f$ constant when taking $N_c \to \infty$, the latter keeps the ratio $N_f/N_c$ constant. From Eq. (20), it is then clear that $\tilde{a}$ and $\tilde{b}$ have the same scaling in the Veneziano limit (the same holds for $\tilde{c}$ and $\tilde{d}$). In our simulations, we will be studying the 't Hooft limit, since we keep $N_f$ fixed, but the quantity $N_f/N_c$ is large (ranging from 4/3 to 2/3, depending on $N_c$), so its contribution may be very significant even for *naturally* large $\tilde{a} - \tilde{d}$ coefficients.

## 4 $\Delta S = 1$ amplitudes in Chiral Perturbation Theory

### 4.1 Chiral Dependence of the $K \to \pi$ amplitudes

The chiral dependence of the ratios in Eq.(13) can be studied within the framework of Chiral Perturbation Theory (ChPT) with $N_f = 4$ active flavours. An extensive discussion of this framework can be found in Refs. [28,46]. Here we just summarize the required formulæ, and refer to those references for details.

The weak Hamiltonian in Eq. (1) can be translated to an effective weak Hamiltonian in terms of meson fields preserving the flavour symmetries. Since the operators $\tilde{Q}^+$ and $\tilde{Q}^-$ transform under representations of $SU(4)_L$ of dimension **84** and **20**, their ChPT counterparts must be constructed accordingly. At leading order, there are only two terms, with couplings $g^\pm$, that need to be determined non-perturbatively:

$$\mathcal{H}_W^{ChPT} = g^+ \mathcal{O}^+ + g^- \mathcal{O}^-, \qquad (21)$$

with

$$\mathcal{O}^\sigma = \sum_{ijkl} c_{ijkl}^\sigma F^4 (U \partial_\mu U^\dagger)_{ij} (U \partial^\mu U^\dagger)_{kl}, \qquad (22)$$

where $U$ is the chiral meson field, $i$, $j$, $k$, $l$ are flavour indices, and $c_{ijkl}^\sigma$ are Clebsch-Gordan coefficients (see Appendix A in Ref. [28]).

By means of the chiral weak Hamiltonian in Eq. (21) and the standard NLO ChPT Lagrangian, the chiral predictions for the normalized amplitudes in Eq. (16) are found to be:

$$A^\pm = g^\pm \left[ 1 \mp 3 \left( \frac{M_\pi}{4\pi F_\pi} \right)^2 \left( \log \frac{M_\pi^2}{\mu^2} + L_\pm^r(\mu) \right) \right], \qquad (23)$$

where $L_\pm^r$ are the NLO counterterms[6]. The NLO corrections in Eq. (23) are fully anticorrelated. Extrapolating the ratios in Eq. (13) to zero pion mass, one can determine the leading low-energy couplings (LECs) of the chiral weak Hamiltonian:

$$g^\pm = \lim_{M_\pi \to 0} A^\pm. \qquad (24)$$

The extracted values of $g^\pm$ can then be used to make predictions of other observables, such as the $K \to \pi\pi$ decay amplitudes.

We now turn to the analysis of the combined chiral and $N_c$ dependence. First, we note that Eq. (20) should hold at any pion mass, and therefore we expect:

$$g^\pm = 1 \pm a_\chi \frac{1}{N_c} \pm b_\chi \frac{N_f}{N_c^2} + c_\chi \frac{1}{N_c^2} + d_\chi \frac{N_f}{N_c^3} + \cdots \qquad (25)$$

Furthermore, by comparing the chiral dependence in Eq. (23) with the $N_c$ scaling in Eq. (20) we can see that both $L_+^r$ and $L_-^r$ must be $O(N_c^0)$, and identical at this order. The next term in the $1/N_c$ expansion for $L_\pm^r$ could in principle differ:

$$L_\pm^r = L^{(0)} + \frac{1}{N_c} L_\pm^{(1)} + \cdots. \qquad (26)$$

Hence, the combination of Eq. (23) with Eqs. (25,26) can be used to do global fits including different meson masses and values of $N_c$.

It will be convenient to also study the chiral and $N_c$ dependence of the product of $A^+ A^-$. The reason is that the leading chiral and $N_c$ corrections cancel out, which leads to a more robust chiral extrapolation. The chiral corrections for this quantity are

$$A^+ A^- = g^+ g^- \left[ 1 + 3 \left( \frac{M_\pi}{4\pi F_\pi} \right)^2 (L_-^r - L_+^r) \right], \qquad (27)$$

with

$$g^+ g^- = 1 + \alpha \frac{1}{N_c^2} + \beta \frac{1}{N_c^3} + \dots, \qquad (28)$$

$$L_-^r - L_+^r = \frac{L_-^{(1)} - L_+^{(1)}}{N_c} + \dots, \qquad (29)$$

where $\alpha$ and $\beta$ depend on the coefficients $a_\chi - d_\chi$.

### 4.2 Relation to $K \to \pi\pi$ amplitudes

Once the effective couplings $g^\pm$ have been extracted from the chiral extrapolations of the ratios $A^\pm$, they can be used to compute the $K \to \pi\pi$ weak decay amplitudes. The two

---

[6] $L_\pm^r$ are a combination of standard QCD NLO LECs with those associated to higher order operators in the chiral weak Hamiltonean. See Refs. [47] and [46] for explicit expressions.

pions in the final state can be in a state with total isospin $I = 0$ or 2:

$$i A_I e^{i\delta_I} = \langle (\pi\pi)_I | \mathcal{H}_W^{ChPT} | K^0 \rangle, \tag{30}$$

where $\delta_I$ is the two-pion scattering phase. The ratio of the two amplitudes can be calculated at leading order in ChPT using the Hamiltonian in Eq. (21) [28,48]:

$$\frac{A_0}{A_2} = \frac{1}{2\sqrt{2}} \left( 1 + 3 \frac{g^-}{g^+} \right). \tag{31}$$

The measured hierarchy of $\sim 22$ between $A_0$ and $A_2$ must then be translated into a large ratio of the couplings $g^\pm$. Note that for $g^+ = g^- = 1$, the expected large-$N_c$ result is recovered, $A_0/A_2 = \sqrt{2}$. Large $1/N_c$ corrections in the $g^-/g^+$ ratio could therefore be the origin of the $\Delta I = 1/2$ rule.

We have also derived the ChPT NLO result for the non-degenerate case in which we send the pion mass to zero, while keeping the kaon mass at its physical value[7]. As we are forced to work in the exact GIM limit, we must also send the charm quark mass to zero with the up quark mass. The calculation for $m_s > m_u = m_d = m_c = 0$ yields:

$$\begin{aligned}
\text{Re} \left. \frac{A_0}{A_2} \right|_{M_\pi, M_D \to 0, M_K^{\text{phys}}} &= \frac{1}{2\sqrt{2}} \left( 1 + 3 \frac{g^-}{g^+} \right) \\
&+ \frac{17}{12\sqrt{2}} \left( 1 + \frac{1}{17} \frac{g^-}{g^+} \right) \frac{M_K^2}{(4\pi F_K)^2} \log \frac{\Lambda_{\text{eff}}^2}{M_K^2},
\end{aligned} \tag{32}$$

where $\Lambda_{\text{eff}}$ is an unknown scale that contains information of the NLO LECs of the effective Chiral Lagrangian and the effective weak Hamiltonian. We note that the NLO effect tends to enhance (reduce) the ratio for $\Lambda_{\text{eff}} > M_K$ ($\Lambda_{\text{eff}} < M_K$).

## 5 Lattice setup

### 5.1 Simulation and matching of sea and valence sectors

Our lattice setup is the same as the one presented in Ref. [27], and we refer to it for details on the simulations and scale setting. We use ensembles with $N_f = 4$ dynamical fermions for an $SU(N_c)$ gauge theory, with $N_c = 3-6$. They have been generated using the HiRep code [50,51]. We have chosen the Iwasaki gauge action (following previous experience with 2+1+1 simulations [52]) and clover Wilson fermions for the sea quarks, with the plaquette-boosted one-loop value of $c_{sw}$. The simulation parameters are shown in Table 3. We find that a separation of $\geq 10$ units of Montecarlo time produces no autocorrelation in the ratios. The lattice spacing is found to be $a \sim 0.075$ fm for all values of $N_c$ (see also Ref. [27]). In

**Table 3** Summary of the simulation parameters of the various ensembles used in this work

| Ensemble | $N_c$ | $\beta$ | $c_{sw}$ | $T \times L$ | $am_0^s$ | # configs |
|---|---|---|---|---|---|---|
| 3A10 | 3 | 1.778 | 1.69 | $36 \times 20$ | $-0.4040$ | 195 |
| 3A11 | | | | $48 \times 24$ | $-0.4040$ | 81 |
| 3A20 | | | | $48 \times 24$ | $-0.4060$ | 155 |
| 3A30 | | | | $48 \times 24$ | $-0.4070$ | 149 |
| 3A40 | | | | $60 \times 32$ | $-0.4080$ | 94 |
| 3B10 | 3 | 1.820 | 1.66 | $48 \times 24$ | $-0.3915$ | 182 |
| 3B20 | | | | $60 \times 32$ | $-0.3946$ | 164 |
| 4A10 | 4 | 3.570 | 1.69 | $36 \times 20$ | $-0.3725$ | 82 |
| 4A30 | | | | $48 \times 24$ | $-0.3760$ | 153 |
| 4A40 | | | | $60 \times 32$ | $-0.3780$ | 55 |
| 5A10 | 5 | 5.969 | 1.69 | $36 \times 20$ | $-0.3458$ | 52 |
| 5A30 | | | | $48 \times 24$ | $-0.3500$ | 39 |
| 5A40 | | | | $60 \times 32$ | $-0.3530$ | 36 |
| 6A10 | 6 | 8.974 | 1.69 | $36 \times 20$ | $-0.3260$ | 35 |
| 6A30 | | | | $48 \times 24$ | $-0.3311$ | 30 |
| 6A40 | | | | $60 \times 32$ | $-0.3340$ | 40 |

addition, we have produced two ensembles with a finer lattice spacing, $a \sim 0.065$ fm, to estimate discretization effects.

In order to achieve automatic $O(a)$ improvement[8] [55] and avoid the mixing of different-chirality operators for weak decays, we employ maximally twisted valence quarks [56], i.e., the mixed-action setup [57] previously used in Refs. [53,54]. Working in twisted quark field variables, maximal twist is ensured by tuning the untwisted bare valence mass $m^v$ to the critical value for which the valence PCAC mass is zero:

$$\lim_{m^v \to m_{cr}} m_{\text{pcac}}^v \equiv \lim_{m^v \to m_{cr}} \frac{\partial_0 \langle A_0^{ij}(x) P^{ji}(y) \rangle}{2 \langle P^{ij}(x) P^{ji}(y) \rangle} = 0. \tag{33}$$

The bare twisted mass parameter $\mu_0$ is tuned such that the pion mass in the sea and valence sectors coincide, $M_\pi^v = M_\pi^s$.

Since twisted mass already provides $O(a)$ improvement, the clover improvement parameter $c_{sw}$ can be chosen to be an arbitrary value in the valence sector. We choose $c_{sw} = 0$ in the valence sector[9] for this work, our main motivation being that this minimizes the isospin breaking effects coming from the twisted-mass action. In addition, this will allow for a partial crosscheck of the systematics due to the use of perturbative renormalization constants, by comparing the

---

[7] See Ref. [49] for similar calculation in $N_f = 3$ ChPT.

[8] As discussed in [53,54], there are residual $O(a)$ cutoff effects from virtual sea quarks, which are proportional to $am^s$ and carry coefficients that are $O(\alpha_s^2)$ in perturbation theory. These effects are expected to be numerically very small and thus irrelevant for the discussion below. It is also worth stressing that using the one-loop value of $c_{sw}$ will also lead to residual effects of $O(a\,\alpha_s^2)$.

[9] This differs from Ref. [27], where we picked $c_{sw} = 1.69$. This value matches the one in the sea sector.

**Table 4** Summary of results for our ensembles with Iwasaki gauge action and $O(a)$-improved Wilson fermions with $c_{sw} = 0$ in the valence sector throughout. The value of the lattice spacing is $a \simeq 0.075$ fm for the "A" ensembles (see Ref. [27]), whereas it is $a \simeq 0.065$ fm for "B" ensembles. We provide the pion mass in the valence sector, $aM_\pi^v$, and the PCAC mass, $am_{pcac}^v$. We also include the results for the ratios in Eq. (13), and in the last column, the chiral parameter $\xi \equiv M_\pi^2/(4\pi F_\pi)^2$. Moreover, $\xi_L$ labels $\xi$ corrected by finite-volume effects as explained in the main text

| Ensemble | $N_c$ | $aM_\pi^s$ | $am_0^{tm}$ | $a\mu_0$ | $aM_\pi^v$ | $|am_{pcac}^v|$ | $R^+$ | $R^-$ | $\xi$ | $\xi_L$ |
|---|---|---|---|---|---|---|---|---|---|---|
| 3A10 | 3 | 0.2204(21) | $-0.9353$ | 0.01150 | 0.2220(19) | 0.0004(4) | 0.611(17) | 1.418(20) | 0.1685(56) | 0.1626(56) |
| 3A11 |   | 0.2147(18) | $-0.9353$ | 0.01150 | 0.2184(13) | 0.0004(4) | 0.627(16) | 1.389(18) | 0.1520(35) | 0.1504(35) |
| 3A20 |   | 0.1845(14) | $-0.9324$ | 0.00815 | 0.1833(12) | 0.0002(5) | 0.582(29) | 1.450(33) | 0.1352(39) | 0.1311(39) |
| 3A30 |   | 0.1613(16) | $-0.9311$ | 0.00660 | 0.1607(15) | 0.0002(3) | 0.511(44) | 1.531(50) | 0.1240(35) | 0.1165(35) |
| 3A40 |   | 0.1429(12) | $-0.9285$ | 0.00534 | 0.1413(12) | 0.0002(5) | 0.554(33) | 1.480(34) | 0.1033(19) | 0.1013(19) |
| 3B10 | 3 | 0.1755(15) | $-0.8962$ | 0.00849 | 0.1761(11) | 0.0001(3) | 0.589(16) | 1.464(19) | 0.1564(40) | 0.1495(40) |
| 3B20 |   | 0.1191(9) | $-0.8919$ | 0.00440 | 0.1206(13) | 0.0005(3) | 0.489(23) | 1.533(24) | 0.1017(30) | 0.0958(31) |
| 4A10 | 4 | 0.2035(14) | $-0.9058$ | 0.01055 | 0.2043(28) | 0.0010(7) | 0.766(14) | 1.262(17) | 0.1007(36) | 0.0978(36) |
| 4A30 |   | 0.1714(8) | $-0.9040$ | 0.00797 | 0.1736(12) | 0.0004(3) | 0.699(20) | 1.358(30) | 0.0803(18) | 0.0783(18) |
| 4A40 |   | 0.1397(8) | $-0.9030$ | 0.00551 | 0.1418(7) | 0.0003(2) | 0.699(18) | 1.379(34) | 0.0612(10) | 0.0605(10) |
| 5A10 | 5 | 0.2128(9) | $-0.8783$ | 0.01191 | 0.2112(12) | 0.0005(6) | 0.824(8) | 1.201(14) | 0.0735(20) | 0.0720(20) |
| 5A30 |   | 0.1712(6) | $-0.8768$ | 0.00810 | 0.1706(10) | 0.0001(4) | 0.761(17) | 1.274(27) | 0.0585(11) | 0.0573(11) |
| 5A40 |   | 0.1331(7) | $-0.8753$ | 0.00517 | 0.1338(10) | 0.0001(3) | 0.760(22) | 1.302(27) | 0.0407(10) | 0.0403(10) |
| 6A10 | 6 | 0.2150(7) | $-0.8562$ | 0.01280 | 0.2136(9) | 0.0001(3) | 0.842(9) | 1.170(9) | 0.0611(9) | 0.0601(9) |
| 6A30 |   | 0.1689(7) | $-0.8548$ | 0.00803 | 0.1669(7) | 0.0004(3) | 0.821(12) | 1.185(18) | 0.0455(7) | 0.0447(7) |
| 6A40 |   | 0.1351(6) | $-0.8548$ | 0.00542 | 0.1352(3) | 0.0000(2) | 0.805(9) | 1.219(8) | 0.0328(3) | 0.0325(3) |

latter to the non-perturbative determination in Ref. [58] for $N_c = 3$ (see below). Finally, we also observe that $c_{sw} = 0$ leads to smaller statistical errors.

In Table 4 we present our measurements for the ensembles used in this work. We have achieved good tuning to maximal twist, with the PCAC mass being zero within 1 or $2\sigma$. In addition, the valence and sea pion masses are matched also within 1 or $2\sigma$. The bare results for the ratios are also presented in the same table, together with the chiral parameter $\xi = M_\pi^2/(4\pi F_\pi)^2$, that will be used for the chiral extrapolations.

We conclude the discussion of the simulation setup by mentioning that we will compare the new results with dynamical fermions to the ones in Refs. [6,7]. Those results used quenched simulations, with plaquette gauge action and twisted mass fermions. The lattice spacing was $a \sim 0.093$ fm and the the pion mass was fixed at around $M_\pi = 550$–590 MeV for $N_c = 3 - 8$ and 17. In this work, we perform a reanalysis of these quenched data.

### 5.2 Comments on systematics

We conclude this section by discussing the systematic errors that can affect our results.

We start with finite-volume effects. Our ensembles have $M_\pi L > 3.8$ in all cases so we expect finite-volume effects to be small, and suppressed as $1/N_c$. Still, we find that for

the observable $\xi$ they can be of $O(1\%)$ and thus we correct for them, as explained in Ref. [27], following Refs. [59,60].

Since $B_K$ and $\bar{R}^+$ differ by a volume-independent proportionality factor, we can use the results in Ref. [61], where the finite-volume effects of $B_K$ have been calculated. In addition, it is known that the finite-volume and chiral corrections of $\bar{R}^+$ and $\bar{R}^-$ are fully anticorrelated [46]. Thus, we find:

$$\bar{R}^\pm(L) = \bar{R}^\pm \left[ 1 \pm 6\sqrt{2\pi}\xi \frac{e^{-M_\pi L}}{(M_\pi L)^{3/2}} (M_\pi L - 4) \right]. \quad (34)$$

The correction for these quantities is numerically negligible for our ensembles. While additional finite-volume effects could be present (see Ref. [60]) we observe that a factor of two increase or decrease of these finite-volume corrections alters our results well within the statistical precision.

Concerning discretization effects, we have included the results from two ensembles with a finer lattice spacing at $N_c = 3$. Assuming $O(a)$ improvement, we expect that the finer lattice spacing should reduce by $\sim 30\%$ the $O(a^2)$ discretization effects. We observe no significant difference for these data points in Fig. 6, so we see no sign of sizeable discretization errors within our statistical uncertainty. We stress however that a more extensive study is needed for a robust estimate of the discretization error.

The largest systematic error that we have found is related to the renormalization constants, which we have estimated by one-loop perturbation theory. We have first compared the non-perturbative renormalization constants of Ref. [58] to the

one-loop perturbation theory results in their setup (they used $c_{sw} = 0$). The difference is roughly $\sim 5\%$ for $N_c = 3$. On the other hand, we have computed the ratios using $c_{sw} = 1.69$ in the valence sector for the 3A10 ensemble. Using the perturbative renormalization constants for this new value of $c_{sw}$ we get a result that differs from our $c_{sw} = 0$ result by roughly 20% in the ratio. Since it is unlikely that this effect can be accounted for by discretization effects, given the tests in a finer lattice mentioned above, we conclude that there must be significant non-perturbative effects on renormalization constants for the larger $c_{sw}$ (the perturbative one-loop corrections are also significantly larger for the larger value of $c_{sw}$). This is a large error, and probably a conservative estimate, but it is comparable to the statistical error we achieve, as it will be seen later.

# 6 Results

## 6.1 $N_c$ scaling of $K \to \pi$ amplitudes

The physical amplitudes $A^{\pm}$ can be obtained, as explained in Eq. (16), from the bare ratios in Table 4, and the renormalization coefficients in Tables 1 and 2. As explained above, a rigorous way to isolate the (anti-)correlated contributions to the ratios consists in taking the half-sum and half-difference of the ratios. By doing so, the two contributions can be fitted independently since:

$$
\begin{aligned}
\frac{A^- + A^+}{2} &= 1 + \tilde{c}\,\frac{1}{N_c^2} + \tilde{d}\,\frac{N_f}{N_c^3} + \dots, \\
\frac{A^- - A^+}{2} &= -\tilde{a}\,\frac{1}{N_c} - \tilde{b}\,\frac{N_f}{N_c^2} + \dots.
\end{aligned}
\tag{35}
$$

In the following, we compare the results of the fits to Eq. (35) in three different scenarios:

1. Quenched results ($N_f = 0$) at a heavy pion mass $\sim 570$ MeV.
2. Dynamical results ($N_f = 4$) at a heavy pion mass $\sim 560$ MeV (ensembles A10).
3. Dynamical results ($N_f = 4$) at a lighter pion mass $\sim 360$ MeV (ensembles A40).

The results for the coefficients $\tilde{a} - \tilde{d}$ for the three scenarios are presented in Table 5 and Fig. 4. The coefficients are all of $\mathcal{O}(1)$ and therefore of natural size. Importantly the sign of the $\tilde{a}$ and $\tilde{b}$ coefficients is the same and negative. This implies both terms contribute to reduce the $A^+$ amplitude and enlarge, in a correlated way, the amplitude $A^-$. The fact that $\tilde{b}, \tilde{d} \sim \mathcal{O}(1)$ implies a very large unquenching effect in the large-$N_c$ scaling, and the ratio $A^-/A^+$, which is however compatible with the expansion in Eq. (35). Specifically, it is

**Table 5** Summary of results for the $1/N_c$ fits to the half-sum and half-difference of the amplitudes $A^{\pm}$. Errors are only statistical

| Case | $M_\pi$ | $\tilde{a}$ | $\tilde{b}$ | $\chi^2$/d.o.f. |
|---|---|---|---|---|
| Half-difference | | | | |
| $N_f = 0$ | 570 MeV | $-1.55(2)$ | — | 8.8/6 |
| $N_f = 4$ | 560 MeV | $-1.03(13)$ | $-1.44(13)$ | 6.6/2 |
| $N_f = 4$ | 360 MeV | $-1.49(15)$ | $-1.32(18)$ | 0.3/2 |
| Half-sum | | | | |
| $N_f = 0$ | 570 MeV | $2.1(1)$ | — | 3.5/6 |
| $N_f = 4$ | 560 MeV | $1.2(3)$ | $2.2(3)$ | 1.3/2 |
| $N_f = 4$ | 360 MeV | $2.4(4)$ | $1.6(4)$ | 3.2/2 |

due to $\tilde{b}$ and $\tilde{d}$ being absent for $N_f = 0$. The other two coefficients, $\tilde{a}$ and $\tilde{c}$, are comparable in size in the quenched and dynamical theories. We note however that uncertainties only include statistical errors, and relative discretization errors and the systematics of the perturbative renormalization constants may be significant. Finally, we observe that the mass dependence for the $N_f = 4$ results seems to affect mostly the coefficient $\tilde{a}$, which is consistent with the chiral dependence in Eq. (23), and goes also in the direction of enhancing the ratio $A^-/A^+$ towards the chiral limit.

## 6.2 Kaon B-parameter ($B_K$)

The kaon $B$-parameter, $B_K$, is defined from the matrix element of the $\Delta S = 2$ operator that mediates neutral kaon oscillations at physical kinematics:

$$
\langle \bar{K}^0 | O^{\Delta S=2}(\mu) | K^0 \rangle = \frac{8}{3} f_K^2 M_K^2\, \bar{B}_K(\mu).
\tag{36}
$$

It is customary to quote the renormalization group independent (RGI) version, labelled as $\hat{B}_K$. Its value at the physical point has been computed accurately in $N_f = 2, 2+1$, and $2+1+1$ simulations [58,62–66] (see Ref. [67] for a review).

In our setup, $\hat{B}_K$ coincides with the renormalized ratio $\bar{R}^+$ up to a normalization. Specifically, we have

$$
\hat{B}_K = \frac{3}{4}\hat{c}^+(a^{-1})\bar{R}^+
\tag{37}
$$

where $\hat{c}^+$ can be read off Table 1. There are two essential differences in our setup: all meson masses are degenerate, in particular $M_K = M_\pi$, and we have an active light charm quark. Both can significantly affect the value of $\hat{B}_K$.

We show our results in Fig. 5. We observe a very significant $N_c$ dependence of $\hat{B}_K$ for $N_f = 4$, and a much milder one for $N_f = 0$. For $N_c = 3$, the quenched result agrees with the standard value of $\hat{B}_K$, while the $N_f = 4$ result is about 25% smaller. We have included as bands the Buras-Bardeen-Gerard (BBG) Dual QCD prediction from Ref. [20], using inputs on meson masses from our own simulations in both

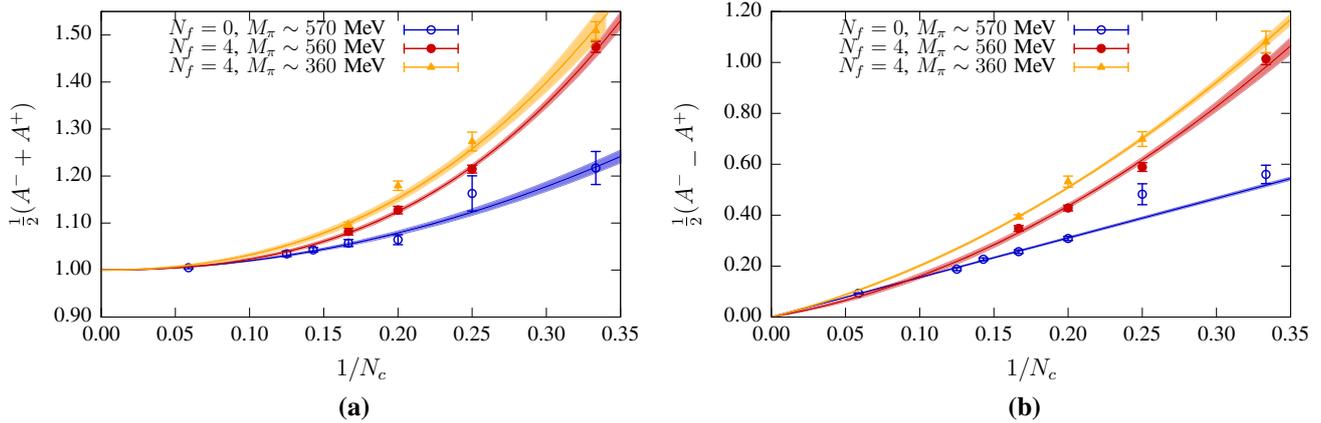

**Fig. 4** Half-sum and half-difference of the amplitudes $A^\pm$ as a function of $N_c^{-1}$ for three different cases: (i) quenched results from Ref. [6] in blue, (ii) new dynamical results at a pion similar to the quenched case (red), and (iii) dynamical results at a lighter pion mass (orange). The fit results are shown in Table 5. Error bars include only statistical errors

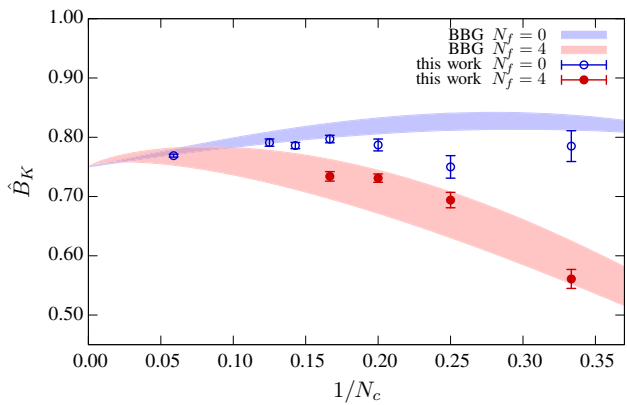

**Fig. 5** Lattice results for $\hat{B}_K$, defined in Eq. (37), in the case of $N_f = 0$ (see Refs. [6,7]), and $N_f = 4$ (this work). Error bars are only statistical errors. We also include the predictions from Ref. [20], where the band indicates the values obtained when varying the involved matching scale $M$ from 600 to 1000 MeV

cases — quenched and dynamical. We find that our results are reasonably compatible with the BBG prediction, in particular regarding the suppression of $\hat{B}_K$ in the presence of a light charm.

To conclude this subsection, we can use the scaling in $N_c$ to infer a value of $\hat{B}_K$ with three active flavours and quasi-physical kinematics. For this, we use the coefficients $\tilde{a} - \tilde{d}$ in Table 5 for the case of $N_f = 4$ and $M_\pi = 560$ MeV, and so predict the value of $A^+$ with $N_c = 3$ and $N_f = 3$ at the same value of the pion mass, degenerate with the kaon. We can the get the RGI value $\hat{B}_K$ as in Eq. (37), extracting $\tilde{R}_+$ and using the $\hat{c}^+(a^{-1})$ for three-flavour QCD [10]. We find

$$\hat{B}_K\big|_{M_K=M_\pi} = 0.67(2)_{\text{stat}}(6)_{Z}+(3)_{\text{fit}}, \tag{38}$$

_____

[10] The required parameters for $N_c = 3$, $N_f = 3$ are $k^+(M_W) = 1.038$, $U^+(a^{-1}, M_W) = 0.851$, and $\hat{c}^+(a^{-1}) = 0.841$. In the evaluation of $\hat{c}^\sigma(a^{-1})$ we have used $\Lambda_{\overline{\text{MS}}} = 341$ MeV from Ref. [45].

including statistical error, and a $\sim 10\%$ error due to the systematics of the renormalization constants. We also quote a "fit" error that we estimate by using the $N_c$ scaling derived from a direct fit of the half-sum and difference of $\tilde{R}^\pm$ instead of $A^\pm$.

We have not found results in the literature for the degenerate case that we can compare to. On the other hand, ChPT relates the value of $\hat{B}_K$ in the degenerate case, to the quasi-physical (QP) situation with $M_\pi = 0$ and $M_K$ at its physical value:

$$\hat{B}_K^{QP} = \hat{B}_K\big|_{M_K=M_\pi}\left[1 + \frac{2}{3}\left(\frac{M_K}{4\pi F_K}\right)^2\log\frac{\Lambda_{\text{eff}}^{B_K}}{M_K}\right], \tag{39}$$

where $\Lambda_{\text{eff}}^{B_K}$ labels an unknown scale that parametrizes the effect of the unknown LECs. For $\Lambda_{\text{eff}}^{B_K} > M_K$, $\hat{B}_K^{QP}$ is larger than $\hat{B}_K$ and could be compatible with the existing results at the physical point from $N_f = 2 + 1$, $N_c = 3$ simulations [58,62–66].

### 6.3 Extraction of the effective couplings $g^\pm$

The main goal of this work is to compute the ratio $g^-/g^+$ by extrapolating $A^\pm$ to the chiral limit. For the required chiral extrapolation, we follow the same strategy as in Ref. [48]. We extract $g^+$ from a chiral fit to $A^+$, and the product $g^+g^-$ from that of the product $A^+A^-$. The ratio can then be evaluated as

$$\frac{g^-}{g^+} \equiv (g^-g^+) \times \frac{1}{(g^+)^2}. \tag{40}$$

This approach results in a milder chiral extrapolation, that will hopefully introduce a smaller systematic error.

We have performed two kinds of fits. In Fit 1, we use all data points with $N_c = 3 - 6$ in a simultaneous chiral and $N_c$ fit using Eqs. (23) and (27), incorporating the $1/N_c$

**Table 6** Results for Fit 1: the simultaneous chiral and $N_c$ fits for $A^+$ and $A^+A^-$. Errors are only statistical

| Fit 1 for $A^+$ | | | | | |
|---|---|---|---|---|---|
| $a_\chi$ | $N_f b_\chi + c_\chi$ | $N_f d_\chi$ | $L^{(0)}$ | $L_+^{(1)}$ | $\chi^2$/d.o.f. |
| $-2.2(6)$ | $-3(4)$ | $7(7)$ | $2.4(8)$ | $-11(4)$ | $12.0/11$ |
| Fit 1 for $A^+A^-$ | | | | | |
| $\alpha$ | $\beta$ | | $L_-^{(1)} - L_+^{(1)}$ | | $\chi^2$/d.o.f. |
| $1.6(4)$ | $-7.2(9)$ | | $1.4(4)$ | | $26.7/13$ |

**Table 7** Results for Fit 2: the chiral fit at $N_c = 3$ for $A^+$ and $A^+A^-$. Errors are only statistical

| Fit 2 for $A^+$ | | |
|---|---|---|
| $g^+$ | $L_+^r$ | $\chi^2$/d.o.f. |
| $0.190(27)$ | $-1.1(7)$ | $4.9/5$ |
| Fit 2 for $A^+A^-$ | | |
| $g^+g^-$ | $L_-^r - L_+^r$ | $\chi^2$/d.o.f. |
| $0.80(6)$ | $0.8(2)$ | $6.2/5$ |

expansion of the couplings as in Eqs. (25,26,29). In Fit 2, we fit using only the data with $N_c = 3$, and extract the effective couplings for this theory. This way, for $N_c = 3$ we find:

$$\text{Fit 1: } g^+ = 0.187(21), \quad g^+g^- = 0.91(4),$$
$$\text{Fit 2: } g^+ = 0.190(27), \quad g^+g^- = 0.80(6). \quad (41)$$

The complete results of these fits are shown in Tables 6, and 7, and also in Fig. 6.

From these results, we obtain for the ratio of couplings at $N_c = 3$:

$$\left.\frac{g^-}{g^+}\right|_{\text{fit 1}} = 26(6), \quad \left.\frac{g^-}{g^+}\right|_{\text{fit 2}} = 22(5), \quad (42)$$

where errors are only statistical, but correlations are taken into account.

### 6.4 $K \to \pi\pi$ amplitudes in ChPT

Using the result for the ratio of couplings in Eq. (42), and the NLO ChPT prediction in Eq. (32), we can obtain an indirect result for the ratio of isospin amplitudes in the $K \to \pi\pi$ decay for $N_c = 3$. In Fig. 7, we show this prediction as a function of an unknown effective scale $\Lambda_{\text{eff}}$. This prediction, valid for $M_\pi = M_D = 0$ and physical $M_K$, shows small NLO effects in a wide range of values of the effective scale.

We are now in the position to quote a final result for the ratio of isospin amplitudes:

$$\text{Re} \left.\frac{A_0}{A_2}\right|_{N_f=4} = 24(5)_{\text{stat}}(4)_{\text{fit}}(5)_{Z^\pm}(3)_{\text{NLO}}, \quad (43)$$

where the central value comes from the fit 2 result in Eq. (42). In the previous equation, the various error sources originate as follows : (i) statistical error, (ii) systematic error from the difference between fit 1 and 2 in Eq. (42), (iii) a 20% error from the renormalization constants — see Sect. 5.2 —, and (iv) a 10% error from the NLO effects — see Fig. 7. Combining all error sources in quadrature results in a $\sim$ 30%

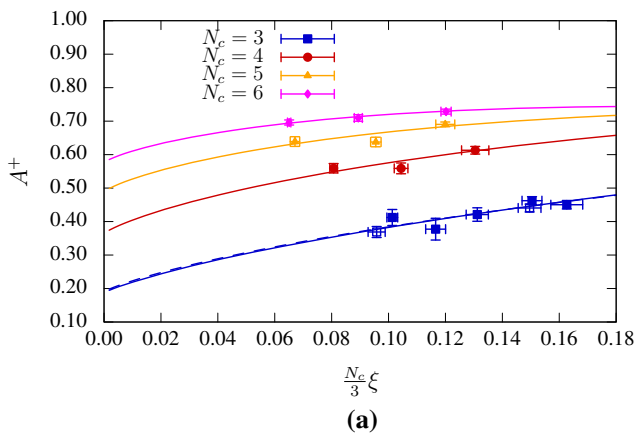
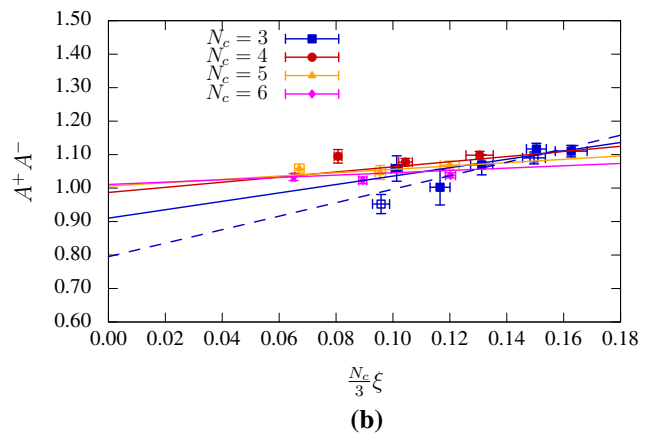

**(a)**  **(b)**

**Fig. 6** Chiral extrapolation of $A^+$ and the product $A^+A^-$. The data points are also shown in Table 4. Empty squares for $N_c = 3$ indicate a finer lattice spacing. Solid lines indicate a simultaneous chiral and $N_c$ fit as in Eq. (23). Dashed lines represent the chiral extrapolation of the data points for $N_c = 3$ following Eqs. (23) and (27). Errors are only statistical

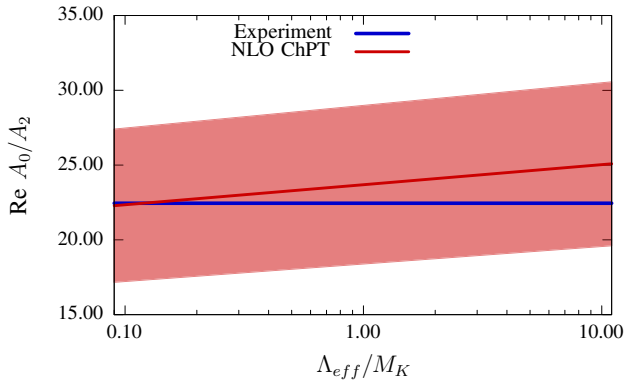

**Fig. 7** NLO ChPT prediction (in red) for the ratio of $K \to \pi\pi$ isospin amplitudes as a function of the NLO LEC, $\Lambda_{eff}$. We use the input of Fit 2 in Eq. (42). This prediction is valid for $M_\pi = M_D = 0$, and $M_K$ at its physical value. The shaded area represents the statistical error associated to the ratio of couplings — see Eq. (42). As a guideline, we also show the experimental value for the ratio of amplitudes (in blue)

uncertainty on the total result, which is dominated by systematics. We also stress that this is a result in the theory with a light charm quark. Interestingly, this indirect computation yields a value compatible with the experimental result for the $\Delta I = 1/2$ enhancement.

## 7 Conclusions

We have presented the first non-perturbative study of the scaling of $\Delta S = 1$ weak amplitudes with the number of colours, $N_c = 3 - 6$, in a theory with four degenerate light flavours $N_f = 4$. These results have been obtained from dynamical simulations with clover Wilson fermions, at $a \simeq 0.075$ fm and $a \simeq 0.065$ fm and pion masses in the range $360 - 570$ MeV. We have analysed the $K \to \pi$ amplitudes $A^\pm$, mediated by the two current-current operators $Q_\pm$ of the $\Delta S = 1$ weak Hamiltonian in Eq. (1).

The diagrammatic analysis of the large-$N_c$ scaling of these observables presented in Sect. 3 allows to classify the subleading $N_c$ corrections, and demonstrates the anticorrelation of the leading $\mathcal{O}(1/N_c)$ and $\mathcal{O}(N_f/N_c^2)$ contributions in the $A^\pm$ amplitudes. Our numerical results confirm this expectation and show that these corrections are naturally large in the Veneziano scaling limit, i.e., the coefficients of both corrections are $\mathcal{O}(1)$. They can nevertheless explain the large enhancement of the ratio $A^-/A^+$ for $N_c = 3$ with respect to the $N_c \to \infty$ limit. This involves an unprecedentedly large unquenching effect in this ratio, that is nevertheless compatible with natural size $\mathcal{O}(N_f/N_c^2)$ corrections.

The amplitudes $A^\pm$ in the chiral limit can be matched to their ChPT counterparts, which depend on the leading low-

energy couplings, $g^\pm$, of the chiral effective weak Hamiltonian. From a chiral extrapolation of the combinations $A^+$ and $A^+A^-$, we have then extracted the couplings $g^\pm$, which are finally used to predict in ChPT the ratio of $K \to (\pi\pi)_{I=0,2}$ amplitudes. In particular, we have obtained an indirect prediction of the ratio of isospin amplitudes, $A_0/A_2$, by this procedure which seems to largely account for the elusive "$\Delta I = 1/2$ rule". Our estimate for this ratio in the theory with a light charm is

$$\text{Re} \left. \frac{A_0}{A_2} \right|_{N_f=4} = 24(5)_{\text{stat}}(7)_{\text{sys}}, \tag{44}$$

which suggests that the enhancement may indeed be largely dominated by intrinsic QCD effects.

## References

1. Z. Bai et al., RBC, UKQCD. Phys. Rev. Lett. **115**, 212001 (2015). arXiv:1505.07863
2. T. Blum et al., Phys. Rev. D **91**, 074502 (2015). arXiv:1502.00263

3. N. Ishizuka, K.I. Ishikawa, A. Ukawa, T. Yoshié, Phys. Rev. D **98**, 114512 (2018). arXiv:1809.03893
4. R. Abbott et al. (RBC, UKQCD) (2020). arXiv:2004.09440
5. P.A. Boyle et al., RBC, UKQCD. Phys. Rev. Lett. **110**, 152001 (2013). arXiv:1212.1474
6. A. Donini, P. Hernández, C. Pena, F. Romero-López, Phys. Rev. D **94**, 114511 (2016). arXiv:1607.03262
7. A. Donini, P. Hernández, C. Pena, F. Romero-López, EPJ Web Conf. **175**, 13015 (2018). arXiv:1711.10248
8. F. Romero-López, A. Donini, P. Hernández, C. Pena, in *36th International Symposium on Lattice Field Theory (Lattice 2018) East Lansing, MI, United States, July 22–28, 2018* (2018). arXiv:1810.06285
9. G. Hooft, Nucl. Phys. B **72**, 461 (1974)
10. G. Hooft, Nucl. Phys. B **72**, 337 (1973)
11. W.A. Bardeen, A. Buras, J. Gerard, Phys. Lett. B **192**, 138 (1987a)
12. W.A. Bardeen, A.J. Buras, J.M. Gérard, Phys. Lett. B **180**, 133 (1986)
13. W.A. Bardeen, A.J. Buras, J.M. Gerard, Nucl. Phys. B **293**, 787 (1987b)
14. R.S. Chivukula, J.M. Flynn, H. Georgi, Phys. Lett. B **171**, 453 (1986)
15. S.R. Sharpe, Phys. Lett. B **194**, 551 (1987)
16. A. Pich, E. de Rafael, Phys. Lett. B **374**, 186 (1996). arXiv:hep-ph/9511465
17. T. Hambye, S. Peris, E. de Rafael, JHEP **05**, 027 (2003). arXiv:hep-ph/0305104
18. J. Aebischer, A.J. Buras, J.M. Garard, JHEP **02**, 021 (2019). arXiv:1807.01709
19. A.J. Buras, J.-M. Garard, Acta Phys. Polon. B **50**, 121 (2019). arXiv:1804.02401
20. A.J. Buras, J.-M. Gérard, W.A. Bardeen, Eur. Phys. J. C **74**, 2871 (2014). arXiv:1401.1385
21. H. Gisbert, A. Pich, in *21st High-Energy Physics International Conference in Quantum Chromodynamics (QCD 18) Montpellier, France, July 2–6, 2018* (2018). arXiv:1810.04904
22. V. Cirigliano, H. Gisbert, A. Pich, A. Rodríguez-Sánchez, JHEP **02**, 032 (2020). arXiv:1911.01359
23. B. Lucini, M. Teper, JHEP **06**, 050 (2001). arXiv:hep-lat/0103027
24. G.S. Bali, F. Bursa, L. Castagnini, S. Collins, L. Del Debbio, B. Lucini, M. Panero, JHEP **06**, 071 (2013). arXiv:1304.4437
25. T. DeGrand, Y. Liu, Phys. Rev. D **94**, 034506 (2016). [Erratum: Phys. Rev.D95,no.1,019902(2017)]. arXiv:1606.01277
26. F. Romero-López, A. Donini, P. Hernández, C. Pena, in *37th International Symposium on Lattice Field Theory (Lattice 2019) Wuhan, Hubei, China, June 16–22, 2019* (2019). arXiv:1910.10418
27. P. Hernández, C. Pena, F. Romero-López, Eur. Phys. J. C **79**, 865 (2019). arXiv:1907.11511
28. L. Giusti, P. Hernández, M. Laine, P. Weisz, H. Wittig, JHEP **11**, 016 (2004). arXiv:hep-lat/0407007
29. M.A. Shifman, A.I. Vainshtein, V.I. Zakharov, Nucl. Phys. B **120**, 316 (1977)
30. E. Endress, C. Pena, Phys. Rev. D **90**, 094504 (2014). arXiv:1402.0827
31. D.J. Gross, F. Wilczek, Phys. Rev. Lett. **30**, 1343 (1973)
32. D.J. Gross, F. Wilczek, Phys. Rev. Lett. **30**, 271 (1973)
33. W.E. Caswell, Phys. Rev. Lett. **33**, 244 (1974)
34. D.R.T. Jones, Nucl. Phys. B **75**, 531 (1974)
35. E. Egorian, O.V. Tarasov, Teor. Mat. Fiz. **41**, 26 (1979)
36. E. Egorian, O.V. Tarasov, Theor. Math. Phys. **41**, 863 (1979)
37. M.K. Gaillard, B.W. Lee, Phys. Rev. Lett. **33**, 108 (1974)
38. G. Altarelli, L. Maiani, Phys. Lett. B **52**, 351 (1974)
39. P. Dimopoulos, G. Herdoíza, F. Palombi, M. Papinutto, C. Pena, A. Vladikas, H. Wittig (ALPHA), JHEP **05**, 065 (2008). arXiv:0712.2429
40. M. Guagnelli, J. Heitger, C. Pena, S. Sint, A. Vladikas (ALPHA), JHEP **03**, 088 (2006). arXiv:hep-lat/0505002
41. M. Ciuchini, E. Franco, V. Lubicz, G. Martinelli, I. Scimemi, L. Silvestrini, Nucl. Phys. B **523**, 501 (1998). arXiv:hep-ph/9711402
42. A.J. Buras, M. Misiak, J. Urban, Nucl. Phys. B **586**, 397 (2000). arXiv:hep-ph/0005183
43. C. Alexandrou, M. Constantinou, T. Korzec, H. Panagopoulos, F. Stylianou, Phys. Rev. D **83**, 014503 (2011). arXiv:1006.1920
44. C. Alexandrou, M. Constantinou, T. Korzec, H. Panagopoulos, F. Stylianou, Phys. Rev. D **86**, 014505 (2012). arXiv:1201.5025
45. M. Bruno, M. Dalla Brida, P. Fritzsch, T. Korzec, A. Ramos, S. Schaefer, H. Simma, S. Sint, R. Sommer (ALPHA), Phys. Rev. Lett. **119**, 102001 (2017). arXiv:1706.03821
46. P. Hernández, M. Laine, JHEP **10**, 069 (2006). arXiv:hep-lat/0607027
47. J. Kambor, J.H. Missimer, D. Wyler, Nucl. Phys. B **346**, 17 (1990)
48. L. Giusti, P. Hernández, M. Laine, C. Pena, J. Wennekers, H. Wittig, Phys. Rev. Lett. **98**, 082003 (2007). arXiv:hep-ph/0607220
49. M.F.L. Golterman, K.C. Leung, Phys. Rev. D **56**, 2950 (1997). arXiv:hep-lat/9702015
50. L. Del Debbio, A. Patella, C. Pica, Phys. Rev. D **81**, 094503 (2010). arXiv:0805.2058
51. A. Patella, L. Del Debbio, B. Lucini, C. Pica, A. Rago, PoS **LATTICE2010**, 068 (2010). arXiv:1011.0864
52. C. Alexandrou et al., Phys. Rev. D **98**, 054518 (2018). arXiv:1807.00495
53. J. Ugarrio, A. Bussone, G. Herdoíza, C. Pena, D. Preti, J. Á. Romero (Alpha), in *Proceedings, 36th International Symposium on Lattice Field Theory (Lattice 2018): East Lansing, MI, United States, July 22–28, 2018* (SISSA, 2018), vol. LATTICE2018, p. 271. arXiv:1812.05458
54. A. Bussone, G. Herdoíza, C. Pena, D. Preti, J. Á. Romero, J. Ugarrio (ALPHA), in *booktitleProceedings, 36th International Symposium on Lattice Field Theory (Lattice 2018): East Lansing, MI, United States, July 22–28, 2018* (SISSA, 2019), vol. LATTICE2018, p. 318. arXiv:1903.00286
55. R. Frezzotti, G.C. Rossi, JHEP **08**, 007 (2004). arXiv:hep-lat/0306014
56. R. Frezzotti, P.A. Grassi, S. Sint, P. Weisz (Alpha), JHEP **08**, 058 (2001). arXiv:hep-lat/0101001
57. O. Bär, G. Rupak, N. Shoresh, Phys. Rev. D **67**, 114505 (2003). arXiv:hep-lat/0210050
58. N. Carrasco, P. Dimopoulos, R. Frezzotti, V. Lubicz, G. C. Rossi, S. Simula, C. Tarantino (ETM), Phys. Rev. D **92**, 034516 (2015). arXiv:1505.06639
59. J. Gasser, H. Leutwyler, Phys. Lett. B **184**, 83 (1987)
60. G. Colangelo, S. Dürr, C. Haefeli, Nucl. Phys. B **721**, 136 (2005). arXiv:hep-lat/0503014
61. D. Bečirević, G. Villadoro, Phys. Rev. D **69**, 054010 (2004). arXiv:hep-lat/0311028
62. S. Dürr et al., Phys. Lett. B **705**, 477 (2011). arXiv:1106.3230
63. J. Laiho, R. S. Van de Water, PoS **LATTICE2011**, 293 (2011). arXiv:1112.4861
64. T. Blum et al., RBC, UKQCD, Phys. Rev. D **93**, 074505 (2016). arXiv:1411.7017
65. B. J. Choi et al. (SWME), Phys. Rev. D **93**, 014511 (2016). arXiv:1509.00592
66. V. Bertone et al. (ETM), JHEP **03**, 089 (2013). [Erratum: JHEP07,143(2013)]. arXiv:1207.1287
67. S. Aoki et al. (Flavour Lattice Averaging Group), Eur. Phys. J. C **80**, 113 (2020). arXiv:1902.08191

# Generalizing the relativistic quantization condition to include all three-pion isospin channels

Maxwell T. Hansen,[a] Fernando Romero-López[b] and Stephen R. Sharpe[c]

[a] *Theoretical Physics Department, CERN,*
  *Geneva 23 1211, Switzerland*
[b] *IFIC, CSIC-Universitat de València,*
  *Paterna 46980, Spain*
[c] *Physics Department, University of Washington,*
  *Seattle, WA 98195-1560, U.S.A.*

  *E-mail:* maxwell.hansen@cern.ch, fernando.romero@uv.es, srsharpe@uw.edu

ABSTRACT: We present a generalization of the relativistic, finite-volume, three-particle quantization condition for non-identical pions in isosymmetric QCD. The resulting formalism allows one to use discrete finite-volume energies, determined using lattice QCD, to constrain scattering amplitudes for all possible values of two- and three-pion isospin. As for the case of identical pions considered previously, the result splits into two steps: the first defines a non-perturbative function with roots equal to the allowed energies, $E_n(L)$, in a given cubic volume with side-length $L$. This function depends on an intermediate three-body quantity, denoted $\mathcal{K}_{\mathrm{df},3}$, which can thus be constrained from lattice QCD input. The second step is a set of integral equations relating $\mathcal{K}_{\mathrm{df},3}$ to the physical scattering amplitude, $\mathcal{M}_3$. Both of the key relations, $E_n(L) \leftrightarrow \mathcal{K}_{\mathrm{df},3}$ and $\mathcal{K}_{\mathrm{df},3} \leftrightarrow \mathcal{M}_3$, are shown to be block-diagonal in the basis of definite three-pion isospin, $I_{\pi\pi\pi}$, so that one in fact recovers four independent relations, corresponding to $I_{\pi\pi\pi} = 0, 1, 2, 3$. We also provide the generalized threshold expansion of $\mathcal{K}_{\mathrm{df},3}$ for all channels, as well as parameterizations for all three-pion resonances present for $I_{\pi\pi\pi} = 0$ and $I_{\pi\pi\pi} = 1$. As an example of the utility of the generalized formalism, we present a toy implementation of the quantization condition for $I_{\pi\pi\pi} = 0$, focusing on the quantum numbers of the $\omega$ and $h_1$ resonances.

# Contents

## 1  Introduction

The computation of scattering amplitudes using lattice quantum chromodynamics (LQCD) has seen enormous progress in the last few years. The majority of calculations are based on the finite-volume formalism of Lüscher [1], which relates discrete finite-volume energies in a cubic, periodic, spatial volume of side-length $L$, to the scattering amplitude of two identical spin-zero particles. This relation is exact up to corrections scaling as $e^{-mL}$, with $m$ the pion mass, but holds only for energies in the regime of elastic scattering, i.e. below

the lowest-lying three- or four-particle threshold. The formalism has since been extended to generic two-particle systems [2–11], for which, however, the same restrictions apply. At unphysically heavy pion masses, many resonances satisfy this restriction, leading to a recent explosion of LQCD resonant studies as reviewed, for example, in ref. [12]. However, for physical masses, many experimentally observed resonances have significant branching fractions to modes containing three (or more) particles. Thus, the development of a multi-particle formalism is essential in order to gain insight into the nature of these states.

In the last few years, significant theoretical effort has been devoted to extensions and alternatives to the two-particle Lüscher formalism for more-than-two-particle systems. In particular, a three-particle quantization condition for identical (pseudo)scalars has been derived following three different approaches:[1] (i) generic relativistic effective field theory (RFT) [17–24], (ii) nonrelativistic effective field theory (NREFT) [25–28], and (iii) (relativistic) finite volume unitarity (FVU) [29–31]. (See ref. [32] for a review of the three approaches.) At this stage, only the RFT formalism has been explicitly worked out including higher partial waves.

These theoretical developments have been accompanied by significant progress in lattice calculations. In previous work, the three-particle coupling was extracted using the ground state energy in QCD [30, 33, 34], and also in $\varphi^4$ theory [35]. Going beyond this, the determination of complete spectra with quantum numbers of three pions has been achieved by multiple groups in the last two years [36–38]. In fact, very recently, a large number of three-$\pi^+$ levels (including those in moving frames) has been combined with the RFT formalism to constrain the three-particle scattering amplitude from first principles QCD [24].

As the present quantization conditions are only valid for identical particles, their use is limited to three pions (or kaons or heavy mesons) at maximal isospin, and thus only for weakly interacting channels with no resonances. Motivated by this, in the present paper we provide the generalization of the RFT approach to include nonidentical, mass-degenerate (pseudo)scalar particles. Specifically, we focus on a general three-pion state in QCD with exact isospin symmetry (and thus exact G parity, preventing two-to-three transitions).

A feature of all three-particle approaches is that the extraction of scattering amplitudes proceeds via an intermediate three-particle scattering quantity, denoted in the RFT approach by $\mathcal{K}_{\mathrm{df},3}$. In particular, the RFT quantization condition provides, for each finite-volume three-particle energy, $E_n(L)$, a combined constraint on $\mathcal{K}_{\mathrm{df},3}$ and the two-particle scattering amplitude, $\mathcal{M}_2$. Additional constraints on $\mathcal{M}_2$ are provided by the two-particle spectrum using the Lüscher formalism. Then, in a second step, infinite-volume integral equations are used to relate $\mathcal{K}_{\mathrm{df},3}$ to the physical scattering amplitude, $\mathcal{M}_3$. To implement these steps in practice, one requires a physically motivated parametrization of $\mathcal{K}_{\mathrm{df},3}$ that includes, for example, a truncation in the angular momentum of two-particle subsystems.

Our work generalizes all aspects of this work flow to three-pion scattering for all allowed values of two- and three-pion isospin. In section 2 we derive the generalized formalism. We first review the results of refs. [17, 18] for identical particles [section 2.1], before pro-

---

viding the extensions to non-identical pions, first of the relation between $E_n(L)$ to $\mathcal{K}_{\mathrm{df},3}$ [section 2.2] and then of the integral equations relating $\mathcal{K}_{\mathrm{df},3}$ to $\mathcal{M}_3$ [section 2.3]. These are presented for states with definite individual pion flavors. The change of basis to definite total isospin is given in sections 2.4 and 2.5. An important consequence of projecting to total isospin is that the results block diagonalize into four separate relations, one for each of the allowed values of the total three-pion isospin: $I_{\pi\pi\pi} = 0, 1, 2, 3$.

With the formalism in hand, in section 3 we describe strategies to parametrize $\mathcal{K}_{\mathrm{df},3}$. We determine the form of the threshold expansion for all choices of $I_{\pi\pi\pi}$, and provide expressions for $\mathcal{K}_{\mathrm{df},3}$ that produce three-particle resonant behavior for each of the choices of $I_{\pi\pi\pi}$ and $J^P$ for which such behavior is experimentally observed.

To illustrate the utility of the generalized formalism, we present a numerical implementation for the $I_{\pi\pi\pi} = 0$ channel in section 4. We do so using forms of $\mathcal{K}_{\mathrm{df},3}$ that lead to both vector and axial-vector resonances, mimicking the experimentally observed $\omega$ and $h_1$. The finite-volume energies exhibit avoided level crossings associated with the allowed cascading resonant decays, e.g. $h_1 \to \rho\pi \to \pi\pi\pi$.

This completes the main text, following which section 5 gives a brief summary of the work and a discussion of the future outlook. We include four appendices to address various technical details. First, in appendix A, we provide further discussion of the derivation of the generalized quantization condition. Second, in appendix B, we collect the definitions of the building blocks entering the quantization condition. Third, appendix C describes the different bases we use for three-pion states. Finally, appendix D summarizes some group theoretical results that are relevant to the implementation of the quantization condition.

## 2 Derivation

In this section we derive the quantization condition for general three-pion states. Following the approach of refs. [4, 17], we first introduce a matrix of correlation functions

$$C_{L;jk}(P) \equiv \int dx^0 \int_{L^3} d^3x \; e^{-i\boldsymbol{P}\cdot\boldsymbol{x}+iEt} \; \langle \mathrm{T}\mathcal{O}_j(x)\mathcal{O}_k^\dagger(0)\rangle_L \,. \tag{2.1}$$

Here $\mathcal{O}_k^\dagger$ are $\mathcal{O}_j$ are operators that, respectively, create and destroy three-pion states, with quantum numbers and additional information specified by the indices $j, k$. In the following paragraphs we give a concrete choice for these operators that is particularly convenient for the present derivation. The correlator is defined in the context of a generic, isospin-symmetric effective theory of pions. The underlying fields are denoted by $\pi_+(x)$, $\pi_-(x)$ and $\pi_0(x)$, and are normalized such that

$$\langle 0|\pi_q(x)|\pi, q, \boldsymbol{p}\rangle = e^{-ip\cdot x} \,, \tag{2.2}$$

where $|\pi, q, \boldsymbol{p}\rangle$ is a state with mass $m$ and charge $q$, and $p^0 = \omega_p = \sqrt{\boldsymbol{p}^2 + m^2}$. We use Minkowski four-vectors, adopting the metric convention $p \cdot x = p^0 x^0 - \boldsymbol{p} \cdot \boldsymbol{x}$. The finite volume is implemented by requiring that all fields satisfy periodic boundary conditions in each of the spatial directions., $\pi(x) = \pi(x + L\boldsymbol{e}_i)$.

In the derivation of refs. [17, 18], the analysis was simplified by assuming that the interactions of the identical scalar particles satisfied a $Z_2$ symmetry that led to particle number conservation modulo two.[2] This implied, for example, that there were no intermediate four-pion states in the correlator $C_L$. This simplification carries over to the present analysis because we are assuming exact isospin symmetry, so that G parity is exactly conserved, and serves as the $Z_2$ symmetry.

For a given choice of total momentum $\boldsymbol{P}$, which is constrained by the boundary conditions to take one of the values $2\pi\boldsymbol{n}/L$, with $\boldsymbol{n}$ a vector of integers, the correlator $C_{L,ij}(E, \boldsymbol{P})$ has poles in $E$ at the positions of the finite-volume eigenstates. Our aim is to derive a quantization condition whose solutions give the energies of these eigenstates.

There are 27 distinct combinations of three-pion fields, assuming that we distinguish identical fields with position labels, $x_1, x_2, x_3$. It is useful to understand this multiplicity from the viewpoint of combining three objects with isospin 1. This leads to seven irreducible representations (irreps)

$$\mathbf{1} \otimes \mathbf{1} \otimes \mathbf{1} = (\mathbf{0} \oplus \mathbf{1} \oplus \mathbf{2}) \otimes \mathbf{1} = (\mathbf{1}) \oplus (\mathbf{0} \oplus \mathbf{1} \oplus \mathbf{2}) \oplus (\mathbf{1} \oplus \mathbf{2} \oplus \mathbf{3}) \,. \tag{2.3}$$

We see that the total three-pion isospin can have values $I_{\pi\pi\pi} = 0, 1, 2, 3$, with respective multiplicities $1, 3, 2, 1$. The multiplicities correspond to the number of possible values of the two-pion isospin, $I_{\pi\pi}$, that can appear: all three values for $I_{\pi\pi\pi} = 1$, two values, $I_{\pi\pi} = 1, 2$, for $I_{\pi\pi\pi} = 2$, and only one value each for $I_{\pi\pi\pi} = 0$ and 3, namely $I_{\pi\pi} = 1$ and 2, respectively. The situation is summarized in figure 1.

Since we are treating isospin as an exact symmetry, we need only consider one choice of $I_z$ (or, equivalently, one choice of electric charge) from each of the seven irreps. A convenient choice is to use the combination with vanishing electric charge, since this appears once in each irrep. Thus, henceforth we focus on the space of the seven neutral operators:

$$\widetilde{\mathcal{O}}(a, b, k) \equiv \begin{pmatrix} \widetilde{\pi}_-(a)\, \widetilde{\pi}_0(b)\, \widetilde{\pi}_+(k) \\ \widetilde{\pi}_0(a)\, \widetilde{\pi}_-(b)\, \widetilde{\pi}_+(k) \\ \widetilde{\pi}_-(a)\, \widetilde{\pi}_+(b)\, \widetilde{\pi}_0(k) \\ \widetilde{\pi}_0(a)\, \widetilde{\pi}_0(b)\, \widetilde{\pi}_0(k) \\ \widetilde{\pi}_+(a)\, \widetilde{\pi}_-(b)\, \widetilde{\pi}_0(k) \\ \widetilde{\pi}_0(a)\, \widetilde{\pi}_+(b)\, \widetilde{\pi}_-(k) \\ \widetilde{\pi}_+(a)\, \widetilde{\pi}_0(b)\, \widetilde{\pi}_-(k) \end{pmatrix} \,. \tag{2.4}$$

Here we have written the fields in momentum space as this will prove convenient below. These operators are related to $\mathcal{O}_j(x)$ via

$$\mathcal{O}_j(x) \equiv \int_{a,b,k} f(a, b, k)\, e^{-i(a+b+k)\cdot x}\, \widetilde{\mathcal{O}}_j(a, b, k) \,, \tag{2.5}$$

where $\int_k \equiv \int dk^0/(2\pi) \sum_{\boldsymbol{k}}$, with the sum over $\boldsymbol{k}$ being over the finite-volume set introduced above for $\boldsymbol{P}$. $f(a, b, k)$ is a smooth function that specifies the detailed form of $\mathcal{O}_j$. It is

---

[2]This is not a fundamental limitation on the derivation; the generalization without a $Z_2$ symmetry is derived in ref. [19].

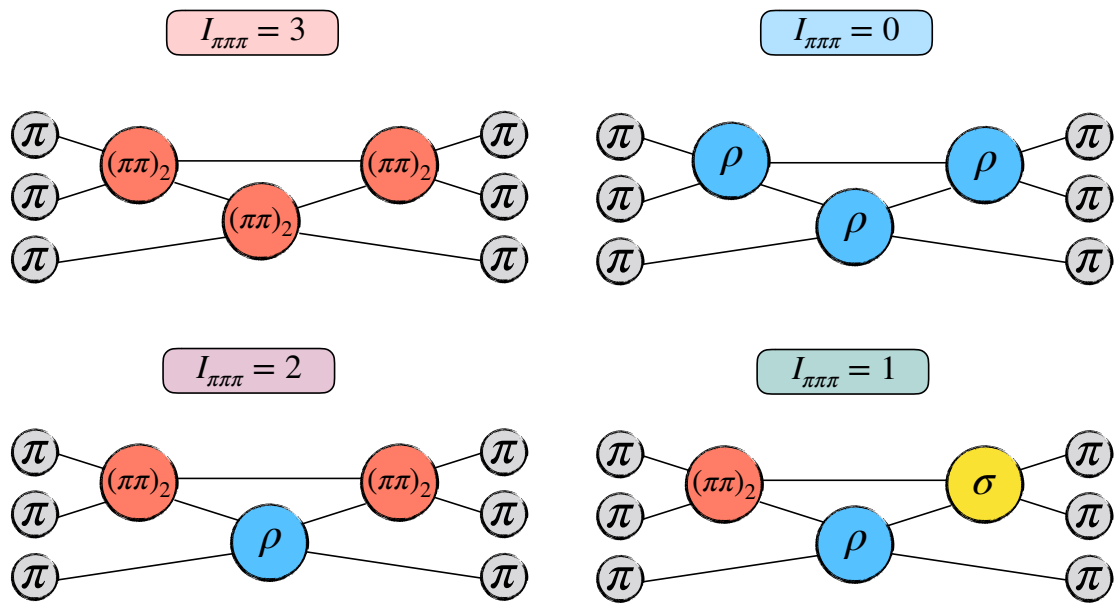

**Figure 1**. Sketch of subchannels for pairwise interactions present in each three-pion system with fixed overall isospin, $I_{\pi\pi\pi}$. For $I_{\pi\pi\pi} = 0$ and 3, only one subchannel is present, having $I_{\pi\pi} = 1$ and $I_{\pi\pi} = 2$, respectively. For $I_{\pi\pi\pi} = 2$, two subchannels are present, with $I_{\pi\pi} = 1$ and 2, implying that the three-particle quantization condition lives in a two-dimensional flavor space. For $I_{\pi\pi\pi} = 1$, all three two-pion subchannels contribute ($I_{\pi\pi} = 0, 1$, and 2), leading to a three-dimensional flavor space. For convenience, we use the shorthand notation ($I_{\pi\pi} = 0$) ≡ "$\sigma$", ($I_{\pi\pi} = 1$) ≡ "$\rho$", and ($I_{\pi\pi} = 2$) ≡ "$(\pi\pi)_2$", in which we label (when possible) the two-pion subchannels by the renonances present in them.

convenient for the subsequent derivation to choose $f(a, b, k)$ to be invariant under exchanges or permutations of its arguments.[3]

At this point, the reader may wonder why, in eq. (2.4), we have distinguished between the six different channels with charge composition $\pi_+, \pi_0, \pi_-$, by using different momentum labels, and then multiplied them by a symmetric function in eq. (2.5) so as to apparently remove the distinction between the channels. The motivation for this construction is to create a single formalism that can simultaneously treat identical and nonidentical particles. How this works will become clear below.

Having defined the column of operators, $\mathcal{O}_j$, we are now in position to derive a skeleton expansion for $C_{L;ij}$, exactly as was done in ref. [17]. The only distinction compared to the earlier work is that the endcaps, appearing on the far left and far right of every diagram, now represent a column (on the left) and row (on the right), so that each Feynman diagram encodes a $7 \times 7$ matrix, defining a contribution to the matrix of correlators, $C_{L;ij}$. As we discuss in the following, this matrix structure naturally propagates through all steps of the derivation so that the final result appears identical to that of ref. [17], but with the additional flavor channel assigned to each of the building blocks. The final step is to

---

[3]One could also, in principle choose different weight functions for the different entries of the column but this has no effect on the results derived, and leads to more complicated intermediate expressions.

perform a change of basis into states with definite two- and three-pion isospin. This block diagonalizes $C_{L;ij}$, as expected, and one recovers four distinct quantization conditions, one each for $I_{\pi\pi\pi} = 0, 1, 2, 3$. While the $I_{\pi\pi\pi} = 0$ and 3 conditions are one-dimensional in the flavor index, $I_{\pi\pi\pi} = 1$ and 2 are 3 and 2 dimensional, respectively, encoding the coupled-channel scattering of the various allowed $I_{\pi\pi}$ subchannels.

## 2.1 Formalism for identical (pseudo-)scalars

In this subsection we review the results of refs. [17, 18] for the case of three identical particles, which apply here for the $I_{\pi\pi\pi} = 3$ channel. These results will serve as stepping stones to the generalization for other values of $I_{\pi\pi\pi}$. In ref. [17], it was shown that the finite-volume correlator for three identical (pseudo-)scalars can be written

$$C_L(P) = C_\infty(P) + iA_3' F_3 \frac{1}{1 + \mathcal{K}_{\mathrm{df},3} F_3} A_3 \,, \tag{2.6}$$

where

$$2\omega L^3 \times F_3 \equiv \frac{F}{3} - F \frac{1}{1 + \mathcal{M}_{2,L} G} \mathcal{M}_{2,L} F \,, \qquad \mathcal{M}_{2,L} \equiv \frac{1}{\mathcal{K}_2^{-1} + F} \,. \tag{2.7}$$

This result holds for $m^2 < E^2 - \boldsymbol{P}^2 < (5m)^2$ and neglects $L$ dependence of the form $e^{-mL}$, while keeping all power-like scaling. The intuitive picture behind its derivation is that only three-pion states can go on shell for the kinematics considered, and only these on-shell states can propagate large distances to feel the periodicity and induce $1/L^n$ corrections. The quantities $\omega, F, G, \mathcal{K}_2, \mathcal{K}_{\mathrm{df},3}, A_3', A_3$ and $C_\infty$ are each defined in detail in ref. [17], as is the matrix space on which all quantities act.[4] Here we only give a brief summary of the most important details, with some additional definitions provided in appendix B. All objects besides $C_L$ and $C_\infty$ are defined on an index space denoted by $k, \ell, m$ where $k$ represents the three-momentum for the spectator particle, i.e. is shorthand for a finite-volume momentum $\boldsymbol{k}$, and $\ell, m$ give the angular-momentum of the non-spectator pair. A cutoff on the $k$ index is built into all matrices so that this index space is always finite. To intuitively understand the appearance of the cutoff function, note that, for fixed total energy $E$ and momentum $\boldsymbol{P}$, if the spectator carries $k^\mu = (\omega_k, \boldsymbol{k})$ then the squared invariant mass of the non-spectator pair is

$$E_{2,k}^{\star 2} \equiv (E - \omega_k)^2 - (\boldsymbol{P} - \boldsymbol{k})^2 \,. \tag{2.8}$$

This becomes negative for sufficiently large $\boldsymbol{k}^2$ implying that the state cannot go on the mass shell and therefore does not induce power-like $L$ dependence. Thus it is possible to absorb the deep subthreshold behavior into the definitions of $\mathcal{K}_2, \mathcal{K}_{\mathrm{df},3}, A_3', A_3$ and $C_\infty$ and to cut off the matrix space.

The objects $\omega$, $F$, $G$, $\mathcal{K}_2$ and $\mathcal{K}_{\mathrm{df},3}$ are all matrices on the $k, \ell, m$ space, e.g. $F = F_{k'\ell'm',k\ell m}$, whereas $A_3'$ and $A_3$ are row and column vectors respectively, e.g. $A_3 = A_{3;k\ell m}$. In this way all indices in eqs. (2.6) and (2.7) are fully contracted, with adjacent factors multiplied according to usual matrix multiplication. The $L$-dependence in these results

---

[4]The quantities we call $A_3$ and $A_3'$ here are denoted $A$ and $A'$ in refs. [17, 18].

enters both through the index space, $k$, and through explicit dependence inside of $F$ and $G$, which are defined in eqs. (B.7) and (B.3), respectively. The simplest object entering eq. (2.7) is the diagonal kinematic matrix

$$\omega_{k'\ell'm',k\ell m} \equiv \delta_{k'k}\delta_{\ell'\ell}\delta_{m'm}\sqrt{\boldsymbol{k}^2 + m^2}\,. \tag{2.9}$$

This leaves only two quantities to define: the two- and three-particle K matrices, $\mathcal{K}_2$ and $\mathcal{K}_{\mathrm{df},3}$, respectively. The former is given in eq. (B.9). It depends on the two-to-two scattering phase shift, $\delta_\ell$, in each angular momentum channel, for two-particle energies ranging from 0 (well below the threshold at $2m$) up to $E^\star - m$. Here we have introduced the notation $E^\star = \sqrt{E^2 - \boldsymbol{P}^2}$, for the three-particle center-of-momentum frame (CMF) energy. In practice, one must choose a value $\ell_{\max}$ above which the phase shift is assumed negligible, in order to render $\mathcal{K}_2$ finite-dimensional. Then it can be determined using the two-particle quantization condition, together with finite-volume energies from a numerical lattice calculation.

The remaining object, $\mathcal{K}_{\mathrm{df},3}$, encodes the short-distance part of the three-particle amplitude. We close this subsection by explaining, first, how this quantity can be constrained from finite-volume three-particle energies and, second, how it is related to the physical observable, the three-particle scattering amplitude.

The utility of eq. (2.6) is that it allows one to identify the poles in $C_L(P)$ as a function of $E$, corresponding to the three-body finite-volume spectrum for fixed values of $L$ and $\boldsymbol{P}$. These pole locations, denoted $E_n(L)$ for $n = 0, 1, 2, \ldots$, occur at energies for which

$$\det_{k,\ell,m}\left[1 + \mathcal{K}_{\mathrm{df},3}(E^\star)F_3(E, \boldsymbol{P}, L)\right] = 0\,, \tag{2.10}$$

where we have made the kinematic dependence explicit. Thus, given many values of $E_n(L)$, ideally for different $\boldsymbol{P}$ and $L$, one can identify parameterizations of $\mathcal{K}_{\mathrm{df},3}(E^\star)$ that describe the system and fix the values of the parameters. As with $\mathcal{K}_2$, also here a value of $\ell_{\max}$ must be set to render $\mathcal{K}_{\mathrm{df},3}(E^\star)$ finite-dimensional. Indeed, the angular momentum cutoffs in the two- and three-particle sectors must be performed in a self consistent way, as is described in ref. [22].

Now, taking $\mathcal{K}_{\mathrm{df},3}(E^\star)$ as known, we present its relation to the three-particle scattering amplitude, $\mathcal{M}_3$, first derived in ref. [18]. As is explained in that work, one can relate $C_L(P)$ to a new finite-volume correlator, $\mathcal{M}_{3,L}(P)$, in a two-step procedure. First we take only the second term of eq. (2.6), multiply by $i$, and amputate $A_3'F[2\omega L^3]^{-1}$ on the left and $[2\omega L^3]^{-1}FA_3$ on the right to reach

$$C_L'(P) \equiv -\left[\frac{F}{2\omega L^3}\right]^{-1}F_3\frac{1}{1 + \mathcal{K}_{\mathrm{df},3}F_3}\left[\frac{F}{2\omega L^3}\right]^{-1}\,, \tag{2.11}$$

$$= \mathcal{D}_{\mathrm{disc}}^{(u,u)} + \mathcal{D}^{(u,u)} + \mathcal{L}_L^{(u)}\frac{1}{1 + \mathcal{K}_{\mathrm{df},3}F_3}\mathcal{K}_{\mathrm{df},3}\mathcal{R}_L^{(u)}\,, \tag{2.12}$$

where in the second step we have introduced

$$\mathcal{D}_{\mathrm{disc}}^{(u,u)} \equiv -\left[\frac{F}{2\omega L^3}\right]^{-1}\left[\frac{F}{6\omega L^3} - \frac{F\mathcal{M}_{2,L}F}{2\omega L^3}\right]\left[\frac{F}{2\omega L^3}\right]^{-1}\,, \tag{2.13}$$

$$\mathcal{D}^{(u,u)} \equiv -\left[\frac{F}{2\omega L^3}\right]^{-1}F_3\left[\frac{F}{2\omega L^3}\right]^{-1} - \mathcal{D}_{\mathrm{disc}}^{(u,u)}\,, \tag{2.14}$$

$$\mathcal{L}_L^{(u)} \equiv \left[\frac{F}{2\omega L^3}\right]^{-1} F_3 \,, \tag{2.15}$$

$$\mathcal{R}_L^{(u)} \equiv F_3 \left[\frac{F}{2\omega L^3}\right]^{-1} \,. \tag{2.16}$$

Note that $\mathcal{D}^{(u,u)}$, $\mathcal{L}_L^{(u)}$ and $\mathcal{R}_L^{(u)}$ are closely related to $F_3$, differing only by the amputation factors and, in the case of $\mathcal{D}^{(u,u)}$, by the subtraction of $\mathcal{D}_{\text{disc}}^{(u,u)}$. The latter is labeled with the subscript "disc" for disconnected, referring to the fact that these terms arise from diagrams in which one of the three-particles does not interact with the other two. The second step towards defining $\mathcal{M}_{3,L}(P)$ is to drop $\mathcal{D}_{\text{disc}}^{(u,u)}$ and to symmetrize the resulting function with respect to the exchange of pion momenta. The result is

$$\mathcal{M}_{3,L}(P) \equiv \mathcal{S}\left[\mathcal{M}_{3,L}^{(u,u)}(P)\right], \tag{2.17}$$

$$\mathcal{M}_{3,L}^{(u,u)}(P) \equiv \mathcal{D}^{(u,u)} + \mathcal{L}_L^{(u)} \frac{1}{1 + \mathcal{K}_{\text{df},3} F_3} \mathcal{K}_{\text{df},3} \mathcal{R}_L^{(u)} \,, \tag{2.18}$$

where $\mathcal{S}$ indicates the symmetrization.[5] This is explained in detail in section 2.3 below, in the context of the generic isospin system.

The motivation for these seemingly *ad hoc* redefinitions is that the new correlator, $\mathcal{M}_{3,L}(P)$, is closely related to the physical, fully connected three-to-three scattering amplitude. Substituting $P = (E, \boldsymbol{P})$, the connection is given by

$$\mathcal{M}_3(E, \boldsymbol{P}) = \lim_{\epsilon \to 0^+} \lim_{L \to \infty} \mathcal{M}_{3,L}(E + i\epsilon, \boldsymbol{P}) \,. \tag{2.19}$$

This ordered double limit can be evaluated analytically to produce an integral equation relating $\mathcal{K}_{\text{df},3}$ to the $\mathcal{M}_3$. This completes the complicated mapping from the finite-volume spectrum to infinite-volume amplitudes. Again, we point the reader to ref. [18] for a full derivation and for the explicit forms of the integral equations.

## 2.2 Generalized quantization condition

In this subsection we generalize the derivation of the quantization condition [eq. (2.10)] to the system of three pions with any allowed total isospin. The relation of the generalized $\mathcal{K}_{\text{df},3}$ to the corresponding generalized scattering amplitude is discussed in the next subsection.

As explained above, the finite-volume correlator, $C_{L,ij}$, becomes a $7 \times 7$ matrix on the space of all possible neutral three-pion configurations. We find that, to generalize the quantization condition, we also need to extend all the objects in the correlator decomposition [eq. (2.6)], the quantization condition [eq. (2.10)] and the relation to $\mathcal{M}_3$ [eqs. (2.17) and (2.19)] to be matrices on the seven-dimensional flavor space. We stress that all objects,

---

[5]The quantity $\mathcal{M}_{3,L}^{(u,u)}$ given here is actually slightly different from the object with the same name defined in ref. [18]. The distinction is that the $\mathcal{M}_{3,L}^{(u,u)}$ is this work has been partially symmetrized, leading to small differences in $\mathcal{L}^{(u)}$ and $\mathcal{R}^{(u)}$. However, these differences have no impact on the fully symmetrized quantity, $\mathcal{M}_{3,L}$, which is identical to that in ref. [18].

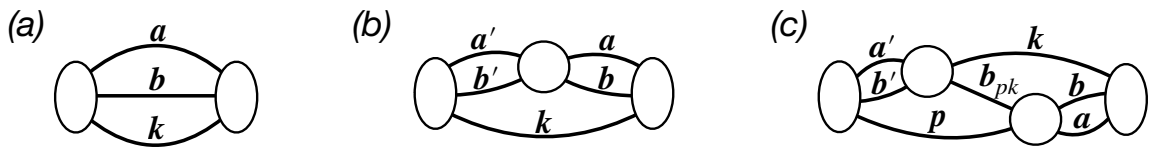

**Figure 2**. Three Feynman diagram topologies required to illustrate the extension to generic isospin.

including $C_\infty$, $A_3$ and $A_3'$ become flavor matrices, even though the latter are defined as either scalars or vectors in the $k\ell m$ indices.

In the original derivation of ref. [17], the first step was to identify a skeleton expansion that expressed $C_L$ in terms of generalized Bethe-Salpeter kernels and fully dressed propagators. Cutting rules were then applied to write each diagram as a sum of various contributions, and summing over all diagrams lead to eq. (2.6). A key feature that will simplify the present generalization is that the new matrix space can be completely implemented already at the level of Bethe-Salpeter kernels and fully dressed propagators, i.e. before the steps of decomposition and summation. These final steps, which lead to the main complications in the earlier work, can then be copied over with the new index space passing in a straightforward way into $F$, $G$, $\mathcal{K}_2$ and the other matrices entering the final results.

To illustrate this we carefully consider the three diagrams of figure 2. We give expressions for each of these in turn, first for the case of identical particles and then for the general isospin extensions. In this way, all building blocks are defined for the new quantization condition, which is then given in eq. (2.44) below.

Beginning with figure 2a, the expression in the case of three identical particles is

$$C_L^{[2a]}(P) = \frac{1}{6} \sum_{\boldsymbol{k},\boldsymbol{a}} \int \frac{da^0}{2\pi} \int \frac{dk^0}{2\pi} i\sigma(k,a)\,\Delta(a)\Delta(b)\Delta(k)\,i\sigma^\dagger(k,a)\,, \qquad (2.20)$$

where $\sigma(k,a)$ and $\sigma^\dagger(k,a)$ are endcap factors encoding the coupling of the operator to a three-particle state and $\Delta(a)$ is a fully dressed propagator. As explained in ref. [17], this can be rewritten as

$$C_L^{[2a]}(P) = C_\infty^{[2a]}(P) + i\sigma\,\frac{iF}{6\omega L^3}i\sigma^\dagger\,, \qquad (2.21)$$

where the first term on the right-hand side is the contribution from the diagram of figure 2a to the infinite-volume correlation function. In the second term we have introduced $\sigma$ and $\sigma^\dagger$ as row and column vectors, respectively, on the $k\ell m$ space. These are ultimately combined with other terms to define $A_3'$ and $A_3$, respectively.

In the extension to general three-pion isospin, eq. (2.20) is replaced with

$$C_{L,ji}^{[2a]}(P) = \frac{1}{6} \sum_{n,n'} \sum_{\boldsymbol{k},\boldsymbol{a}} \int \frac{da^0}{2\pi} \int \frac{dk^0}{2\pi} i\sigma_{jn}(k,a)\,[\Delta(a)\Delta(b)\Delta(k)]_{nn'}\,i\sigma_{n'i}^\dagger(k,a)\,, \qquad (2.22)$$

where $b = P - k - a$. Here $[\Delta(a)\Delta(b)\Delta(k)]_{nn'}$ is a diagonal matrix of propagator triplets, in which each entry is built from charged and neutral pion propagators according to eq. (2.4).

We repeat the pion content of each entry here for convenience:

$$[\Delta(a)\Delta(b)\Delta(k)] = \mathrm{diag}\Big([-][\,0\,][+],\ [\,0\,][-][+],\ [-][+][\,0\,],\ [\,0\,][\,0\,][\,0\,],$$
$$[+][-][\,0\,],\ [\,0\,][+][-],\ [+][\,0\,][-]\Big),\quad (2.23)$$

where $[-][\,0\,][+] = \Delta_-(a)\Delta_0(b)\Delta_+(k)$, etc., the subscript indicating the pion field at the sink of the two-point function defining the fully-dressed propagator. In fact, in the iso-symmetric theory, the propagators are all equal as functions, $\Delta_-(a) = \Delta_0(a) = \Delta_+(a)$. Nonetheless, it is useful to treat these objects as distinct, in order to better identify the patterns arising in our matrix representation of the Feynman rules.

The endcap matrices, $\sigma_{jl}(k,a)$ and $\sigma_{jl}^\dagger(k,a)$, are built from the function $f(a,b,k)$, introduced in eq. (2.5), that encodes how the fundamental fields, $\pi_0$, $\pi_+$ and $\pi_-$, are used to build up the annihilation operators $\mathcal{O}_j(x)$. The exact relation is $\sigma_{jl}(k,a) = M_{jl}f(a,b,k)$, where

$$M = \begin{pmatrix} \blacksquare & \blacksquare & \blacksquare & \square & \blacksquare & \blacksquare & \blacksquare \\ \blacksquare & \blacksquare & \blacksquare & \square & \blacksquare & \blacksquare & \blacksquare \\ \blacksquare & \blacksquare & \blacksquare & \square & \blacksquare & \blacksquare & \blacksquare \\ \square & \square & \square & \blacksquare & \square & \square & \square \\ \blacksquare & \blacksquare & \blacksquare & \square & \blacksquare & \blacksquare & \blacksquare \\ \blacksquare & \blacksquare & \blacksquare & \square & \blacksquare & \blacksquare & \blacksquare \\ \blacksquare & \blacksquare & \blacksquare & \square & \blacksquare & \blacksquare & \blacksquare \end{pmatrix}, \qquad \square = 0,\ \ \blacksquare = 1\,. \quad (2.24)$$

(Here and below we use empty and filled squares to present matrices of 0s and 1s as we find this form more readable.)

This complicated matrix structure in the case of the non-interacting diagram, figure 2a, may seem surprising. The structure arises simply because six of the seven entries in the column $\mathcal{O}_j(x)$ (all entries besides $j=4$) are built from $\pi^-$, $\pi^0$ and $\pi^+$, distinguished only by the momentum assignments as shown in eqs. (2.4) and (2.5). Thus, even when all interactions are turned off, $C_{L,jl}$ is still nonzero for any combination of $j,l \neq 4$.

In more detail, the definition of $M$ ensures that eq. (2.22) gives the correct expression for $C_{L,jl}^{[2a]}$, for all choices of $j$ and $l$. Here one must consider three distinct cases. First for $j = 4, l \neq 4$, as well as $j \neq 4, l = 4$, the correlator vanishes, as expected for the non-interacting contribution connecting a $[-][\,0\,][+]$ channel with a $[\,0\,][\,0\,][\,0\,]$. Second, if both $j,l \neq 4$ then one recovers a non-zero contribution with a factor of $\sum_k M_{jk}M_{kl} = 6$ arising from the contracted matrix indices. This compensates the $1/6$ pre-factor, leading to the correct expression for a diagram with three distinguishable particles. Finally, $j = l = 4$ yields the diagram with three neutral particles and in this case the $1/6$ survives and correctly gives the symmetry factor for identical particles.

Having demonstrated that eq. (2.22) gives the correct generalization of eq. (2.20), it is now very straightforward to generalize the decomposition, eq. (2.21). We find

$$\mathbf{C}_L^{[2a]}(P) = \mathbf{C}_\infty^{[2a]}(P) + \frac{1}{3}\boldsymbol{\sigma}\,\mathbf{F}\,\boldsymbol{\sigma}^\dagger\,, \quad (2.25)$$

$$[\mathbf{F}]_{jl} \equiv \frac{iF}{2\omega L^3}\delta_{jl}\,, \quad (2.26)$$

where $\delta_{jl}$ is the identity matrix on the seven-dimensional flavor space. Here we find it convenient to absorb various factors of $i$, $\omega$ and $L$ into the boldface definitions. Specifically, we use

$$[\boldsymbol{\sigma}]_{jl} = i\sigma_{jl}, \quad [\boldsymbol{\sigma}^\dagger]_{jl} = i[\sigma^\dagger]_{jl}, \quad \text{and} \quad [\mathbf{C}_L(P)]_{jl} = C_{L,jl}(P), \tag{2.27}$$

In the following we generally follow the convention of using bold-faced symbols whenever flavor-space indices are suppressed.

We turn now to the diagram shown in figure 2b. In the case of a single channel of identical particles the corresponding expression is

$$C_L^{[2b]}(P) = \frac{1}{4} \sum_{\boldsymbol{k},\boldsymbol{a},\boldsymbol{a'}} \int \frac{da^0}{2\pi} \int \frac{da'^0}{2\pi} \int \frac{dk^0}{2\pi} i\sigma(k,a')\,\Delta(a')\Delta(b')$$
$$\times i\mathcal{B}(a',b';a,b)\,\Delta(a)\Delta(b)\,\Delta(k)\,i\sigma^\dagger(k,a)\,, \tag{2.28}$$

where $\mathcal{B}$ is the infinite-volume Bethe-Salpeter kernel. As we demonstrate in ref. [17] this leads to a contribution of the form

$$C_L^{[2b]}(P) = i\sigma\,\frac{iF}{2\omega L^2}\,i\mathcal{K}_2\,iF\,i\sigma^\dagger + \cdots\,, \tag{2.29}$$

where $\mathcal{K}_2$ is the two-particle K matrix, up to some subtleties in the sub-threshold definition, as discussed in refs. [17, 18]. The ellipsis in eq. (2.29) indicates that additional terms arise containing less than two factors of $F$. Indeed, many of the complications in ref. [17] arise in the demonstration that these terms can be reabsorbed into redefinitions of $C_\infty$, $\sigma$ and $\sigma^\dagger$, in a consistent way that generalizes to all orders. It is this pattern of absorbing higher-order terms that leads to the conversion of $\mathcal{B}$ into the K matrix.

Following the pattern established above, our next step is to give the isospin generalization of eq. (2.28)

$$C_{L,jl}^{[2b]}(P) = \frac{1}{4} \sum_{\boldsymbol{k},\boldsymbol{a},\boldsymbol{a'}} \int \frac{da^0}{2\pi} \int \frac{da'^0}{2\pi} \int \frac{dk^0}{2\pi} i\sigma_{jn}(k,a')\,[\Delta(a')\Delta(b')\Delta(k)]_{nn'}$$
$$\times [\Delta(k)^{-1}i\mathcal{B}(a',b';a,b)]_{n'n''}\,[\Delta(a)\Delta(b)\Delta(k)]_{n''n'''}\,i\sigma_{n'''l}^\dagger(k,a)\,. \tag{2.30}$$

All quantities here have been defined, with the exception of $[\Delta(k)^{-1}i\mathcal{B}(a',b';a,b)]_{n'n''}$. This object is a matrix on the flavor space, with non-zero entries only when the third particles of the $n'$ and $n''$ states coincide, see again eq. (2.4). In the case where $n'$ and $n''$ do have a common spectator, the entry is defined by setting $\Delta(k)^{-1}$ to the spectator species and taking $\mathcal{B}$ as the Bethe-Salpeter kernel for the scattering of the $n'$ and $n''$ non-spectator pairs. We give a concrete expression of this matrix structure (in the context of $\mathcal{K}_2$) in eqs. (2.32)–(2.35) below.

As with eq. (2.22), it is straightforward to show that (2.30) gives the correct result for the correlator for all choices of $j$ and $l$. For example, if $j = 4$ and $l \neq 4$, then the left-hand loop (containing momenta $a'$ and $b'$) consists of three $\pi_0$s, and the expression then forces the spectator in the right-hand loop (that with momenta $a$ and $b$) to also be a $\pi_0$. There are then two options for $n'' = n'''$ available, namely $n'' = 3$ and 5 ($n'' = 4$ being

disallowed since $l \neq 4$). These two options correspond to the scattering process in the Bethe-Salpeter kernel being $\pi_0(a')\pi_0(b') \leftarrow \pi_+(a)\pi_-(b)$ and $\pi_0(a')\pi_0(b') \leftarrow \pi_-(a)\pi_+(b)$, respectively. These give equal contributions because in the loop sums/integrals we can freely interchange the dummy labels $a$ and $b$. This redundancy cancels the prefactor of $1/2$ for right-hand loop, while leaving it for the left-hand loop, as required for a diagram with only one exchange-symmetric two-particle loop.

We are now ready to present the isospin generalization of eq. (2.29),

$$\mathbf{C}_L^{[2\mathrm{b}]}(P) = \boldsymbol{\sigma}\,\mathbf{F}\,\mathbf{K}_2\,\mathbf{F}\,\boldsymbol{\sigma}^\dagger + \cdots, \tag{2.31}$$

where all objects have been defined above besides

$$\mathbf{K}_2 \equiv i[2\omega L^3] \begin{pmatrix} \mathcal{K}_+ & & \\ & \mathcal{K}_0 & \\ & & \mathcal{K}_- \end{pmatrix}. \tag{2.32}$$

Here our notation indicates a block-diagonal matrix, in which the subscript on each block denotes the charge of the spectator. The blocks are given explicitly by

$$\mathcal{K}_+ \equiv \begin{pmatrix} [\pi_-\pi_0 \leftarrow \pi_-\pi_0] & [\pi_-\pi_0 \leftarrow \pi_0\,\pi_-] \\ [\pi_0\,\pi_- \leftarrow \pi_-\pi_0] & [\pi_0\,\pi_- \leftarrow \pi_0\,\pi_-] \end{pmatrix}, \tag{2.33}$$

$$\mathcal{K}_0 \equiv \begin{pmatrix} [\pi_-\pi_+ \leftarrow \pi_-\pi_+] & [\pi_-\pi_+ \leftarrow \pi_0\,\pi_0] & [\pi_-\pi_+ \leftarrow \pi_+\pi_-] \\ [\pi_0\,\pi_0 \leftarrow \pi_-\pi_+] & [\pi_0\,\pi_0 \leftarrow \pi_0\,\pi_0] & [\pi_0\,\pi_0 \leftarrow \pi_+\pi_-] \\ [\pi_+\pi_- \leftarrow \pi_-\pi_+] & [\pi_+\pi_- \leftarrow \pi_0\,\pi_0] & [\pi_+\pi_- \leftarrow \pi_+\pi_-] \end{pmatrix}, \tag{2.34}$$

$$\mathcal{K}_- \equiv \begin{pmatrix} [\pi_0\,\pi_+ \leftarrow \pi_0\,\pi_+] & [\pi_0\,\pi_+ \leftarrow \pi_+\pi_0] \\ [\pi_+\pi_0 \leftarrow \pi_0\,\pi_+] & [\pi_+\pi_0 \leftarrow \pi_+\pi_0] \end{pmatrix}, \tag{2.35}$$

where each scattering process in square brackets indicates the corresponding two-particle K matrix. We stress that many entries in these K matrices are trivially related, e.g.

$$[\pi_-(a')\pi_+(b') \leftarrow \pi_-(a)\pi_+(b)] = [\pi_-(a')\pi_+(b') \leftarrow \pi_+(b)\pi_-(a)]. \tag{2.36}$$

This completes the discussion of figure 2b.

To conclude the extension of the quantization condition, it remains only to consider figure 2c. Here we immediately give the isospin-generalized expression

$$\begin{aligned} C_{L,jl}^{[2\mathrm{c}]}(P) = \frac{1}{4} \sum_{\boldsymbol{k},\boldsymbol{a},\boldsymbol{a}'} \int \frac{da^0}{2\pi} \int \frac{da'^0}{2\pi} \int \frac{dk^0}{2\pi}\, i\sigma_{jn}(k,a')\,[\Delta(a')\Delta(b')\Delta(k)]_{nn'} \\ \times [\Delta(k)^{-1}i\mathcal{B}(a',b';p,b_{pk})]_{n'n''}\,[\Delta(p)\Delta(b_{pk})\Delta(k)]_{n''n'''}^{\mathbf{G}} \\ \times [\Delta(p)^{-1}i\mathcal{B}(b_{pk},k;a,b)]_{n'''m''}[\Delta(a)\Delta(b)\Delta(p)]_{m''m'}\,i\sigma_{m'l}^\dagger(p,a)\,, \end{aligned} \tag{2.37}$$

– 12 –

where $b_{pk} = P - p - k$. All quantities are defined above except for the propagator triplet with the **G** superscript, which represents the contribution of the central cut in figure 2c. To give an explicit expression, we introduce the matrix

$$T_G = \begin{pmatrix} \square & \square & \square & \square & \square & \square & \blacksquare \\ \square & \square & \square & \square & \blacksquare & \square & \square \\ \square & \square & \square & \square & \square & \blacksquare & \square \\ \square & \square & \square & \blacksquare & \square & \square & \square \\ \square & \blacksquare & \square & \square & \square & \square & \square \\ \square & \square & \blacksquare & \square & \square & \square & \square \\ \blacksquare & \square & \square & \square & \square & \square & \square \end{pmatrix}, \qquad \square = 0, \ \blacksquare = 1. \tag{2.38}$$

This corresponds to interchanging the first and last particles in each channel, which is what is required by the "switching" of the spectator particle in figure 2c. Note that $T_G$ is a reducible representation of the element (13) of the permutation group $\mathcal{S}_3$ in the notation of appendix C. Using this matrix we then have

$$[\Delta(p)\Delta(b_{pk})\Delta(k)]^{\mathbf{G}}_{nn''} = [\Delta(p)\Delta(b_{pk})\Delta(k)]_{nn'}[T_G]_{n'n''}. \tag{2.39}$$

In ref. [17] we demonstrated that such exchange propagators gave rise to a new kind of finite-volume cut involving $G$. We find that the isospin-generalized result is

$$\mathbf{C}_L^{[2c]}(P) = \boldsymbol{\sigma}\,\mathbf{F}\,\mathbf{K}_2\,\mathbf{G}\,\mathbf{K}_2\,\mathbf{F}\,\boldsymbol{\sigma}^\dagger + \cdots, \tag{2.40}$$

where

$$\mathbf{G} = i\frac{1}{2\omega L^3}G\,T_G. \tag{2.41}$$

We stress that, in contrast to $\mathcal{K}_2$ and $F$, the matrix $G$ does not commute with $1/[2\omega L^3]$ on the $k,\ell,m$ index space. For this reason we have been careful to show the order of the product defining $\mathbf{G}$.

At this point we have introduced the key quantities entering the generalized quantization condition: $\mathbf{F}$, $\mathbf{K}_2$ and $\mathbf{G}$. With these objects defined, every step in the decompositions of refs. [17, 18] naturally generalizes to flavor space, with each equation carrying over essentially verbatim, but with extra flavor indices. The only significant difference is that certain steps, related to symmetrization, require additional justification when flavor is included. This is discussed in appendix A, where the additional arguments are given. In the end, one reaches a decomposition of the finite-volume correlator that is exactly analogous to eq. (2.6) above:

$$\mathbf{C}_L(P) = \mathbf{C}_\infty(P) - \mathbf{A}_3'\mathbf{F}_3\frac{1}{1 - \mathbf{K}_{\mathrm{df},3}\mathbf{F}_3}\mathbf{A}_3, \tag{2.42}$$

where

$$\mathbf{F}_3 \equiv \frac{\mathbf{F}}{3} + \mathbf{F}\frac{1}{1 - \mathbf{M}_{2,L}\mathbf{G}}\mathbf{M}_{2,L}\mathbf{F}, \qquad \mathbf{M}_{2,L} \equiv \frac{1}{\mathbf{K}_2^{-1} - \mathbf{F}}. \tag{2.43}$$

– 13 –

The sign changes in eqs. (2.42) and (2.43) as compared to eqs. (2.6) and (2.7) are due to the factors of $i$ that are absorbed into the bold-faced quantities.[6]

The endcap factors, $\mathbf{A}'_3$ and $\mathbf{A}_3$, are matrices on the seven-dimensional flavor space, describing the coupling of each of the seven operators [see eq. (2.4)] to each of the seven interacting asymptotic states. The exact definitions are unimportant for this work and it suffices to know that these quantities, like $\mathbf{C}_\infty(P)$, have only exponentially suppressed dependence on $L$, and do not contain the finite-volume poles that we are after. Thus, just as in the single channel case, the finite-volume spectrum is given by all divergences of the matrix appearing between $\mathbf{A}'_3$ and $\mathbf{A}_3$, equivalently by all solutions to the quantization condition

$$\det_{k,\ell,m,\mathbf{f}}\left[1 - \mathbf{K}_{\mathrm{df},3}(E^\star)\,\mathbf{F}_3(E,\boldsymbol{P},L)\right] = 0\,, \tag{2.44}$$

where the subscript $\mathbf{f}$ indicates that the determinant additionally runs over flavor space. Note that this expression will give the spectra of all three-particle quantum numbers simultaneously and is therefore not useful in practice. In the section 2.4 below we discuss how to project this result into the various sectors of definite total isospin.

## 2.3 Generalized relation to the three-particle scattering amplitude

First, however, we present the isospin generalizations of eqs. (2.13)–(2.19) above, thus providing the relation between $\mathbf{K}_{\mathrm{df},3}$ and the physical scattering amplitude. One first defines the modified finite-volume correlator:

$$\mathbf{M}_{3,L}(P) \equiv \mathcal{S}\left[\mathbf{M}^{(u,u)}_{3,L}(P)\right]\,, \tag{2.45}$$

$$\mathbf{M}^{(u,u)}_{3,L}(P) \equiv \mathbf{D}^{(u,u)} + \mathbf{L}^{(u)}_L \frac{1}{1 - \mathbf{K}_{\mathrm{df},3}\mathbf{F}_3}\mathbf{K}_{\mathrm{df},3}\mathbf{R}^{(u)}_L\,, \tag{2.46}$$

where

$$\mathbf{D}^{(u,u)} \equiv \mathbf{F}^{-1}\mathbf{F}_3\mathbf{F}^{-1} - \mathbf{D}^{(u,u)}_{\mathrm{disc}}\,, \qquad \mathbf{D}^{(u,u)}_{\mathrm{disc}} \equiv \mathbf{F}^{-1}\left[\frac{\mathbf{F}}{3} + \mathbf{F}\mathbf{M}_{2,L}\mathbf{F}\right]\mathbf{F}^{-1}\,, \tag{2.47}$$

$$\mathbf{L}^{(u)}_L \equiv \mathbf{F}^{-1}\mathbf{F}_3\,, \qquad\qquad \mathbf{R}^{(u)}_L \equiv \mathbf{F}_3\mathbf{F}^{-1}\,. \tag{2.48}$$

$\mathcal{S}$ now denotes a symmetrization procedure in the multi-flavor system, an extension that introduces some additional complications as we discuss in the following paragraphs. As in the case of a single channel, an ordered double limit of $\mathbf{M}_{3,L}$ gives a set of integral equations relating $\mathbf{K}_{\mathrm{df},3}$ to the physical scattering amplitude, denoted $\mathbf{M}_3$,

$$\mathbf{M}_3(E,\boldsymbol{P}) = \lim_{\epsilon\to 0^+}\lim_{L\to\infty}\mathbf{M}_{3,L}(E + i\epsilon,\boldsymbol{P})\,. \tag{2.49}$$

It is straightforward to write out the resulting integral equations explicitly, as done for identical particles in ref. [18], but they are not enlightening and we do not do so here.

---

[6]For completeness, we note that $\mathbf{A}_3$ and $\mathbf{A}'_3$ include factors of $i$: they are the flavor generalizations of $iA_3$ and $iA'_3$, respectively. They are the generalized all-orders endcaps, whose leading terms are $\boldsymbol{\sigma}^\dagger$ and $\boldsymbol{\sigma}$, respectively. Similarly $\mathbf{K}_{\mathrm{df},3}$ is the flavor generalization of $i\mathcal{K}_{\mathrm{df},3}$.

This concludes the path from finite-volume spectrum, through $\mathbf{K}_{\mathrm{df},3}$, to the scattering amplitude $\mathcal{M}_3$.

As in the single-channel case, implicit in this procedure is a conversion from the $k, \ell, m$ index space to a function of the incoming and outgoing three-momenta. This conversion is performed simultaneously with a symmetrization procedure. We stress that symmetrization is needed even for non-identical particles, to ensure that all diagrams are included, i.e. that the proper definition of the infinite-volume amplitude is recovered.

At this point, it remains only to specify the symmetrization procedure, encoded in the operator $\mathcal{S}$, for the case of general pion flavors. To do so, we begin by defining

$$X^{(u,u)}(\boldsymbol{k}', \boldsymbol{a}'; \boldsymbol{k}, \boldsymbol{a}) \equiv 4\pi Y_{\ell' m'}^{*}(\hat{\boldsymbol{a}}_{2,k'}'^{\star}) \, X^{(u,u)}_{k'\ell'm',k\ell m} \, Y_{\ell m}(\hat{\boldsymbol{a}}_{2,k}^{\star}) \,, \tag{2.50}$$

where $X^{(u,u)}_{k'\ell'm',k\ell m}$ stands for a generic, unsymmetrized quantity, e.g. $\mathcal{M}^{(u,u)}_{3,L}$ in the identical-particle case or an entry of $\mathbf{M}^{(u,u)}_{3,L}$ in flavor space. Here $\hat{\boldsymbol{a}}_{2,k}^{\star}$ is the spatial direction of $(\omega_a^{\star}, \boldsymbol{a}_{2,k}^{\star})$, the four-vector reached by boosting $(\omega_a, \boldsymbol{a})$ with velocity $\boldsymbol{\beta} = -(\boldsymbol{P} - \boldsymbol{k})/(E - \omega_k)$. In other words $\hat{\boldsymbol{a}}_2^{\star}$ gives the direction of back-to-back momenta of the non-spectator pair, which have momenta $\boldsymbol{a}$ and $\boldsymbol{P} - \boldsymbol{k} - \boldsymbol{a}$ in their two-particle CMF. The same holds for $\hat{\boldsymbol{a}}_{2,k'}'^{\star}$ with $\boldsymbol{a} \to \boldsymbol{a}'$ and $\boldsymbol{k} \to \boldsymbol{k}'$. Contracting the spherical harmonic indices, as shown on the right-hand side of eq. (2.50), leads to a function of momenta whose argument can be take as $\boldsymbol{k}, \hat{\boldsymbol{a}}_{2,k}^{\star}$ or, equally well, as $\boldsymbol{k}, \boldsymbol{a}$. Here we choose the latter convention, i.e. specifying all momenta in the finite-volume frame, as this makes the symmetrization procedure more transparent.

We begin with the case of a single channel of identical particles, where the symmetrization procedure, first introduced in ref. [18], is given by

$$X(\boldsymbol{k}', \boldsymbol{a}', \boldsymbol{b}'; \boldsymbol{k}, \boldsymbol{a}, \boldsymbol{b}) \equiv \mathcal{S}[X^{(u,u)}_{k'\ell'm',k\ell m}] \equiv \sum_{\{\boldsymbol{p}_3', \boldsymbol{p}_1'\} \in \mathcal{P}_3'} \sum_{\{\boldsymbol{p}_3, \boldsymbol{p}_1\} \in \mathcal{P}_3} X^{(u,u)}(\boldsymbol{p}_3', \boldsymbol{p}_1'; \boldsymbol{p}_3, \boldsymbol{p}_1) \,. \tag{2.51}$$

The sums here run over the sets

$$\mathcal{P}_3 = \big\{ \{\boldsymbol{k}, \boldsymbol{a}\}, \{\boldsymbol{a}, \boldsymbol{b}\}, \{\boldsymbol{b}, \boldsymbol{k}\} \big\} \quad \text{and} \quad \mathcal{P}_3' = \big\{ \{\boldsymbol{k}', \boldsymbol{a}'\}, \{\boldsymbol{a}', \boldsymbol{b}'\}, \{\boldsymbol{b}', \boldsymbol{k}'\} \big\} \,, \tag{2.52}$$

with $\boldsymbol{b} \equiv \boldsymbol{P} - \boldsymbol{a} - \boldsymbol{k}$ and $\boldsymbol{b}' \equiv \boldsymbol{P} - \boldsymbol{a}' - \boldsymbol{k}'$. As discussed in ref. [18], this step is necessary to reach the correct definition of $\mathcal{M}_3$, a quantity that is invariant under the exchange of any two incoming or outgoing momenta. The essential point is that the sum runs over all assignments of the spectator momentum for both incoming and outgoing particles in $X^{(u,u)}$.

To generalize this to non-trivial flavors, we first note that the identical-particle prescription, i.e. simply summing $\mathbf{M}^{(u,u)}_{3,L}$ over all permutations of the momenta, is clearly incorrect. The issue is that, for example, the $\pi_0 \pi_+ \pi_- \to \pi_0 \pi_+ \pi_-$ scattering amplitude is not, in general, invariant under permutations of either the incoming or the outgoing momenta. Instead, the required property is that amplitudes must be invariant under the simultaneous exchange of flavor and momentum labels. Summing over such exchanges ensures that the all choices of the spectator pion flavor are included, as illustrated in figure 3.

To express this we introduce matrices that rearrange flavors in accordance with a given momentum permutation. For example, the second element in the set $\mathcal{P}_3$ corresponds to

$\boldsymbol{k} \to \boldsymbol{a}$, $\boldsymbol{a} \to \boldsymbol{b}$, $\boldsymbol{b} \to \boldsymbol{k}$, and should be matched with the following flavor rearrangement:

$$\mathbf{R}_{\boldsymbol{k}\to\boldsymbol{a}} \equiv \begin{pmatrix} \square & \square & \square & \square & \blacksquare & \square & \square \\ \square & \square & \square & \square & \square & \square & \blacksquare \\ \square & \blacksquare & \square & \square & \square & \square & \square \\ \square & \square & \square & \blacksquare & \square & \square & \square \\ \square & \square & \square & \square & \square & \blacksquare & \square \\ \blacksquare & \square & \square & \square & \square & \square & \square \\ \square & \square & \blacksquare & \square & \square & \square & \square \end{pmatrix}, \qquad \square = 0, \quad \blacksquare = 1, \qquad (2.53)$$

We additionally define $\mathbf{R}_{\boldsymbol{k}\to\boldsymbol{k}} \equiv \mathbb{I}$ (the identity) and $\mathbf{R}_{\boldsymbol{k}\to\boldsymbol{b}} \equiv \mathbf{R}_{\boldsymbol{k}\to\boldsymbol{a}}^2$. The matrices $\mathbf{R}_{\boldsymbol{k}\to\boldsymbol{b}}$, $\mathbf{R}_{\boldsymbol{k}\to\boldsymbol{a}}$, and $\mathbf{R}_{\boldsymbol{k}\to\boldsymbol{k}}$ are reducible representations of elements (231), (321), and (1) of $\mathcal{S}_3$ [see again appendix C]. This then allows us to succinctly express the generalization of eq. (2.51) to the space of all possible three-pion flavors

$$\mathbf{X}_{\mathbf{f}',\mathbf{f}}(\boldsymbol{k}', \boldsymbol{a}', \boldsymbol{b}'; \boldsymbol{k}, \boldsymbol{a}, \boldsymbol{b}) \equiv \mathcal{S}[\mathbf{X}_{\mathbf{f}'k'\ell'm',\mathbf{f}k\ell m}^{(u,u)}], \qquad (2.54)$$

$$\equiv \sum_{\{\boldsymbol{p}_3',\boldsymbol{p}_1'\}\in\mathcal{P}_3'} \sum_{\{\boldsymbol{p}_3,\boldsymbol{p}_1\}\in\mathcal{P}_3} \mathbf{R}_{\boldsymbol{k}'\to\boldsymbol{p}_3'}^T \cdot \mathbf{X}^{(u,u)}(\boldsymbol{p}_3', \boldsymbol{p}_1'; \boldsymbol{p}_3, \boldsymbol{p}_1) \cdot \mathbf{R}_{\boldsymbol{k}\to\boldsymbol{p}_3}. \qquad (2.55)$$

Note that the symmetrization also converts us from the index space to the momentum coordinates $(\boldsymbol{k}', \boldsymbol{a}', \boldsymbol{b}'; \boldsymbol{k}, \boldsymbol{a}, \boldsymbol{b})$, and thus leads to the proper dependence for the three-body scattering amplitude. In fact, the scattering amplitude does not depend on this full set of vectors, but rather on the subset built from the eight possible Poincaré invariants that can be built from six on-shell four-vectors. This statement holds regardless of whether or not the particles are identical.

We conclude this subsection by commenting that, as for the quantization condition in eq. (2.44), the relation (2.49) is in the basis of three-pion states labeled by individual pion flavors. The conversion to definite three-pion isospin, and the resulting block diagonalization, will be addressed in section 2.5.

### 2.4 Block diagonalization in isospin: quantization condition

We now project the above expressions onto definite two- and three-pion isospin. To achieve this we require a matrix $\mathcal{C}$ such that

$$\begin{pmatrix} {}_3\langle(\pi\pi)_2\pi| \\ {}_2\langle(\pi\pi)_2\pi| \\ {}_2\langle\rho\pi| \\ {}_1\langle(\pi\pi)_2\pi| \\ {}_1\langle\rho\pi| \\ {}_1\langle\sigma\pi| \\ {}_0\langle\rho\pi| \end{pmatrix} = \mathcal{C} \cdot \begin{pmatrix} \langle\pi_-\,,\,\pi_0\,,\,\pi_+| \\ \langle\pi_0\,,\,\pi_-\,,\,\pi_+| \\ \langle\pi_-\,,\,\pi_+\,,\,\pi_0| \\ \langle\pi_0\,,\,\pi_0\,,\,\pi_0| \\ \langle\pi_+\,,\,\pi_-\,,\,\pi_0| \\ \langle\pi_0\,,\,\pi_+\,,\,\pi_-| \\ \langle\pi_+\,,\,\pi_0\,,\,\pi_-| \end{pmatrix}, \qquad (2.56)$$

where the subscripts on the bras on the left-hand side indicate the total isospin, $I_{\pi\pi\pi}$, and we have indicated the isospin of the first two pions with the shorthand $(\pi\pi)_2$ for $I_{\pi\pi} = 2$, $\rho$

**Figure 3**. Representation of the symmetrization procedure applied to the outgoing particles. Colors indicate different flavors.

for $I_{\pi\pi} = 1$ and $\sigma$ for $I_{\pi\pi} = 0$. This notation and some related results are discussed further in appendix C. A simple exercise using Clebsch-Gordon coefficients shows that the result is given by the orthogonal matrix

$$
\mathcal{C} = \begin{pmatrix}
\frac{1}{\sqrt{10}} & \frac{1}{\sqrt{10}} & \frac{1}{\sqrt{10}} & \sqrt{\frac{2}{5}} & \frac{1}{\sqrt{10}} & \frac{1}{\sqrt{10}} & \frac{1}{\sqrt{10}} \\
-\frac{1}{2} & -\frac{1}{2} & 0 & 0 & 0 & \frac{1}{2} & \frac{1}{2} \\
-\frac{1}{2\sqrt{3}} & \frac{1}{2\sqrt{3}} & -\frac{1}{\sqrt{3}} & 0 & \frac{1}{\sqrt{3}} & -\frac{1}{2\sqrt{3}} & \frac{1}{2\sqrt{3}} \\
\frac{\sqrt{\frac{3}{5}}}{2} & \frac{\sqrt{\frac{3}{5}}}{2} & -\frac{1}{\sqrt{15}} & -\frac{2}{\sqrt{15}} & -\frac{1}{\sqrt{15}} & \frac{\sqrt{\frac{3}{5}}}{2} & \frac{\sqrt{\frac{3}{5}}}{2} \\
\frac{1}{2} & -\frac{1}{2} & 0 & 0 & 0 & -\frac{1}{2} & \frac{1}{2} \\
0 & 0 & \frac{1}{\sqrt{3}} & -\frac{1}{\sqrt{3}} & \frac{1}{\sqrt{3}} & 0 & 0 \\
-\frac{1}{\sqrt{6}} & \frac{1}{\sqrt{6}} & \frac{1}{\sqrt{6}} & 0 & -\frac{1}{\sqrt{6}} & -\frac{1}{\sqrt{6}} & \frac{1}{\sqrt{6}}
\end{pmatrix} .
\tag{2.57}
$$

The block-diagonalized finite-volume correlator is then given by

$$
\mathcal{C} \cdot \mathbf{C}_L(P) \cdot \mathcal{C}^T = \mathcal{C} \cdot \left[ \mathbf{C}_\infty(P) - \mathbf{A}'_3 \mathbf{F}_3 \frac{1}{1 - \mathbf{K}_{\mathrm{df},3} \mathbf{F}_3} \mathbf{A}_3 \right] \cdot \mathcal{C}^T .
\tag{2.58}
$$

To further reduce these expressions one can insert $\mathcal{C}^T \cdot \mathcal{C} = 1$ between all adjacent factors, so that every matrix is replaced according to $\mathbf{X} \to \mathcal{C} \cdot \mathbf{X} \cdot \mathcal{C}^T$. One can explicitly check that this transformation block diagonalizes $\mathbf{F}$, $\mathbf{K}_2$, $\mathbf{G}$ and $\mathbf{K}_{\mathrm{df},3}$ so that the final quantization condition factorizes into four results, one each for the four possibilities of total three-pion isospin, $I_{\pi\pi\pi} = 0, 1, 2, 3$. For example, starting with eq. (2.41) above, one finds (with blank entries vanishing)

$$
\mathcal{C} \cdot \mathbf{G} \cdot \mathcal{C}^T = i \frac{1}{2\omega L^3} G \begin{pmatrix}
1 & & & & & & \\
& -\frac{1}{2} & -\frac{\sqrt{3}}{2} & & & & \\
& -\frac{\sqrt{3}}{2} & \frac{1}{2} & & & & \\
& & & \frac{1}{6} & \frac{\sqrt{15}}{6} & \frac{\sqrt{5}}{3} & \\
& & & \frac{\sqrt{15}}{6} & \frac{1}{2} & -\frac{1}{\sqrt{3}} & \\
& & & \frac{\sqrt{5}}{3} & -\frac{1}{\sqrt{3}} & \frac{1}{3} & \\
& & & & & & -1
\end{pmatrix} .
\tag{2.59}
$$

$$\det\left[1 - \mathbf{K}_{\mathrm{df},3}^{[I]}(E^\star)\,\mathbf{F}_3^{[I]}(E,\boldsymbol{P},L)\right] = 0$$

$$\mathbf{F}_3^{[I]} \equiv \frac{\mathbf{F}^{[I]}}{3} + \mathbf{F}^{[I]}\frac{1}{1 - \mathbf{M}_{2,L}^{[I]}\mathbf{G}^{[I]}}\mathbf{M}_{2,L}^{[I]}\mathbf{F}^{[I]} \qquad \mathbf{M}_{2,L}^{[I]} \equiv \frac{1}{\mathbf{K}_2^{[I]-1} - \mathbf{F}^{[I]}}$$

| $I$ | $\mathbf{F}^{[I]}$ | $\mathbf{K}_2^{[I]}$ | $\mathbf{G}^{[I]}$ |
|---|---|---|---|
| 3 | $\frac{iF}{2\omega L^3}$ | $i[2\omega L^3]\mathcal{K}_{(\pi\pi)_2}$ | $i\frac{1}{2\omega L^3}G$ |
| 2 | $\frac{iF}{2\omega L^3}\begin{pmatrix}1 & 0\\0 & 1\end{pmatrix}$ | $i[2\omega L^3]\begin{pmatrix}\mathcal{K}_{(\pi\pi)_2} & 0\\0 & \mathcal{K}_\rho\end{pmatrix}$ | $i\frac{1}{2\omega L^3}G\begin{pmatrix}-\frac{1}{2} & -\frac{\sqrt{3}}{2}\\-\frac{\sqrt{3}}{2} & \frac{1}{2}\end{pmatrix}$ |
| 1 | $\frac{iF}{2\omega L^3}\begin{pmatrix}1 & 0 & 0\\0 & 1 & 0\\0 & 0 & 1\end{pmatrix}$ | $i[2\omega L^3]\begin{pmatrix}\mathcal{K}_{(\pi\pi)_2} & 0 & 0\\0 & \mathcal{K}_\rho & 0\\0 & 0 & \mathcal{K}_\sigma\end{pmatrix}$ | $i\frac{1}{2\omega L^3}G\begin{pmatrix}\frac{1}{6} & \frac{\sqrt{15}}{6} & \frac{\sqrt{5}}{3}\\\frac{\sqrt{15}}{6} & \frac{1}{2} & -\frac{1}{\sqrt{3}}\\\frac{\sqrt{5}}{3} & -\frac{1}{\sqrt{3}} & \frac{1}{3}\end{pmatrix}$ |
| 0 | $\frac{iF}{2\omega L^3}$ | $i[2\omega L^3]\mathcal{K}_\rho$ | $-i\frac{1}{2\omega L^3}G$ |

**Table 1**. Summary of quantization conditions for all allowed values of the total isospin $I = I_{\pi\pi\pi}$.

We introduce the shorthand $\mathbf{G}^{[I]}$ to indicate the block within $\mathcal{C}\cdot\mathbf{G}\cdot\mathcal{C}^T$ corresponding to a given total isospin. See table 1 for the explicit definitions. It is interesting to note that $\mathbf{G}^{[3]}$, $\mathbf{G}^{[0]}$, and $\mathbf{G}^{[2]}$ each correspond to the element (13), as it is defined, respectively, in the trivial, sign and standard irreps of $\mathcal{S}_3$. In addition $\mathbf{G}^{[1]}$ is this same element in a reducible representation, the direct sum of the trivial and the standard irreps.

For the two-particle K matrix, $\mathbf{K}_2$, the change of basis gives an exact diagonalization, with each total-isospin block populated by the possible two-pion subprocesses, as illustrated in figure 1. The quantity $\mathbf{F}$ is trivial under the change of basis, since it is proportional to the identity matrix. Finally, the exchange properties of the pions within $\mathbf{K}_{\mathrm{df},3}$ (which are the same as those of $\mathbf{M}_{3,L}$ and $\mathbf{M}_3$) are enough to show that it too block diagonalizes, but now with all elements non-zero in a given total-isospin sector. We conclude that the quantization condition divides into four separate relations, compactly represented by adding superscripts $[I]$ to all quantities. The resulting forms of $\mathbf{K}_2^{[I]}$ and $\mathbf{F}^{[I]}$ as well as the corresponding quantization conditions, are summarized in table 1. One noteworthy result is the change in the sign of the $G$ term for $I_{\pi\pi\pi} = 0$ compared to that for $I_{\pi\pi\pi} = 3$, which is a consequence of the antisymmetry of the isospin wavefunction in the former case.

### 2.5 Block diagonalization in isospin: relation to $\mathbf{M}_3$

To conclude our construction of the general isospin formalism, it remains only to express the relations between $\mathbf{K}_{\mathrm{df},3}$ and the scattering amplitude, $\mathbf{M}_3$, described in section 2.3, in

the definite-isospin basis. Exactly as with the quantization condition, the approach is to left- and right-multiply the finite-volume correlator, $\mathbf{M}_{3,L}(P)$, by $\mathcal{C}$ and $\mathcal{C}^T$ respectively

$$\mathcal{C} \cdot \mathbf{M}_{3,L}(P) \cdot \mathcal{C}^T = \mathcal{C} \cdot \mathcal{S}\left[\mathbf{M}_{3,L}^{(u,u)}(P)\right] \cdot \mathcal{C}^T = \sum_{\{\boldsymbol{p}_3',\boldsymbol{p}_1'\}\in\mathcal{P}_3'}\sum_{\{\boldsymbol{p}_3,\boldsymbol{p}_1\}\in\mathcal{P}_3}$$
$$\times\, \mathcal{C}\cdot\mathbf{R}_{\boldsymbol{k}'\to\boldsymbol{p}_3'}^T\cdot\mathcal{C}^T\cdot\mathcal{C}\cdot\mathbf{M}_L^{(u,u)}(\boldsymbol{p}_3',\boldsymbol{p}_1';\boldsymbol{p}_3,\boldsymbol{p}_1)\cdot\mathcal{C}^T\cdot\mathcal{C}\cdot\mathbf{R}_{\boldsymbol{k}\to\boldsymbol{p}_3}\cdot\mathcal{C}^T\,. \tag{2.60}$$

One can then verify that the change of basis block diagonalizes the various $\mathbf{R}_{\boldsymbol{k}\to\boldsymbol{p}_3}$ as well as $\mathbf{M}_L^{(u,u)}$. In other words, the symmetrization does not mix the different total isospin so that we can write

$$\mathbf{M}_{3,L}^{[I]}(P) = \sum_{\{\boldsymbol{p}_3',\boldsymbol{p}_1'\}\in\mathcal{P}_3'}\sum_{\{\boldsymbol{p}_3,\boldsymbol{p}_1\}\in\mathcal{P}_3}\mathbf{R}_{\boldsymbol{k}'\to\boldsymbol{p}_3'}^{[I]\,T}\,\mathbf{M}_L^{[I](u,u)}(\boldsymbol{p}_3',\boldsymbol{p}_1';\boldsymbol{p}_3,\boldsymbol{p}_1)\,\mathbf{R}_{\boldsymbol{k}\to\boldsymbol{p}_3}^{[I]}\,, \tag{2.61}$$

where each object on the right-hand is reached by identifying a specific block after the change of basis. The symmetrizing matrices are defined as follows: $\mathbf{R}_{\boldsymbol{k}'\to\boldsymbol{k}'}^{[I]} = \mathbf{R}_{\boldsymbol{k}\to\boldsymbol{k}}^{[I]} = \mathbb{I}$, $\mathbf{R}_{\boldsymbol{k}\to\boldsymbol{b}}^{[I]} = \mathbf{R}_{\boldsymbol{k}'\to\boldsymbol{b}'}^{[I]} = \left(\mathbf{R}_{\boldsymbol{k}\to\boldsymbol{a}}^{[I]}\right)^2$, and $\mathbf{R}_{\boldsymbol{k}\to\boldsymbol{a}}^{[I]} = \mathbf{R}_{\boldsymbol{k}'\to\boldsymbol{a}'}^{[I]}$ are given in table 2. For $I_{\pi\pi\pi} = 0,\ 2$, and 3, $\mathbf{R}_{\boldsymbol{k}\to\boldsymbol{a}}^{[I]}$ coincides with the element (321) in the irreps of $\mathcal{S}_3$, see eqs. (C.9) and (C.10).

To conclude we only need the isospin specific definitions for the building blocks entering $\mathbf{M}_{3,L}^{[I]}(P)$. These are natural generalizations of eqs. (2.46)–(2.48) but we repeat the expressions here for convenience:

$$\mathbf{M}_{3,L}^{[I](u,u)}(P) \equiv \mathbf{D}^{[I](u,u)} + \mathbf{L}_L^{[I](u)}\frac{1}{1-\mathbf{K}_{\mathrm{df},3}^{[I]}\mathbf{F}_3^{[I]}}\mathbf{K}_{\mathrm{df},3}^{[I]}\mathbf{R}_L^{[I](u)}\,, \tag{2.62}$$

where

$$\begin{aligned}
\mathbf{D}_{\mathrm{disc}}^{[I](u,u)} &\equiv \left(\mathbf{F}^{[I]}\right)^{-1}\left[\frac{\mathbf{F}^{[I]}}{3}+\mathbf{F}^{[I]}\mathbf{M}_{2,L}\mathbf{F}^{[I]}\right]\left(\mathbf{F}^{[I]}\right)^{-1}\,,\\
\mathbf{D}^{[I](u,u)} &\equiv \left(\mathbf{F}^{[I]}\right)^{-1}\mathbf{F}_3^{[I]}\left(\mathbf{F}^{[I]}\right)^{-1}-\mathbf{D}_{\mathrm{disc}}^{[I](u,u)}\,,\\
\mathbf{L}_L^{[I](u)} &\equiv \left(\mathbf{F}^{[I]}\right)^{-1}\mathbf{F}_3^{[I]}\,,\\
\mathbf{R}_L^{[I](u)} &\equiv \mathbf{F}_3^{[I]}\left(\mathbf{F}^{[I]}\right)^{-1}\,.
\end{aligned} \tag{2.63}$$

# 3 Parametrization of $\mathbf{K}_{\mathrm{df},3}$ in the different isospin channels

In order to use the quantization condition detailed in the previous section, $\mathcal{K}_{\mathrm{df},3}$ must be parametrized in a manner that is consistent with its symmetries. In the ideal situation, only a few free parameters will be needed describe $\mathcal{K}_{\mathrm{df},3}$ in the kinematic range of interest, such that one can overconstrain the system with many finite-volume energies and thereby extract reliable predictions for the three-particle scattering amplitude. There are two regimes in which this is expected to hold: near the three-particle threshold and in the vicinity of

| $I$ | $\mathbf{R}_{\boldsymbol{k}\to\boldsymbol{a}}^{[I]}$ |
|---|---|
| 3 | $1$ |
| 2 | $\begin{pmatrix} -\frac{1}{2} & -\frac{\sqrt{3}}{2} \\ \frac{\sqrt{3}}{2} & -\frac{1}{2} \end{pmatrix}$ |
| 1 | $\begin{pmatrix} \frac{1}{6} & \frac{\sqrt{\frac{5}{3}}}{2} & \frac{\sqrt{5}}{3} \\ -\frac{\sqrt{\frac{5}{3}}}{2} & -\frac{1}{2} & \frac{1}{\sqrt{3}} \\ \frac{\sqrt{5}}{3} & -\frac{1}{\sqrt{3}} & \frac{1}{3} \end{pmatrix}$ |
| 0 | $1$ |

**Table 2**. Summary of the symmetrization matrices entering the relation between the scattering amplitude and $\mathbf{K}_{\mathrm{df},3}^{[I]}$.

a three-particle resonance. In this section we describe the parametrizations in these two regimes.

An important property of $\mathcal{K}_{\mathrm{df},3}$ that has been left implicit heretofore is that it can be chosen real.[7] This applies when $\mathcal{K}_{\mathrm{df},3}$ is expressed as a function of momenta, using eqs. (2.50) and (2.51), rather than in the $\{k\ell m\}$ basis.[8] The reality of $\mathcal{K}_{\mathrm{df},3}$ in the case of identical scalars arises in the derivation of ref. [17] from the use of the PV prescription to define integrals over poles. The same argument applies here, except that, in addition, one must choose the relative phases between different flavor channels to be real. This additional condition is relevant for the multichannel cases, $I = 1$ and 2.

### 3.1 Threshold expansion of $\mathbf{K}_{\mathrm{df},3}$

Although in the discussion above $\mathbf{K}_{\mathrm{df},3}$ appears in the finite-volume quantization condition, it is important to remember that it is an infinite-volume quantity. In addition, like the physical scattering amplitude, it is a Poincare-invariant function (equivalently a Lorentz-invariant and momentum-conserving function) of the six on-shell momenta. It also inherits from $\mathbf{M}_3$ invariance under the simultaneous exchange of particle species and momenta in

---

[7]This assumes that, as is the case for QCD, the underlying theory is invariant under T, or equivalently CP, so that coupling constants in the effective field theory can be chosen to be real.

[8]In the $\{k\ell m\}$ basis, $\mathcal{K}_{\mathrm{df},3}$ becomes complex due to the spherical harmonics in the decomposition (2.50). This applies also to $F$, $G$ and $\mathcal{K}_2$. The key point, however, is that each of these objects, and thus any symmetric product built from them, is an hermitian matrix on the $\{k\ell m\}$ space. The determinant of any such matrix, in particular the determinant defining the quantization condition, must then be a real function. Similarly, since $\mathcal{M}_{3,L}^{(u,u)}$ is hermitian, one recovers a real function upon contracting with spherical harmonics. This subtlety can be avoided by using real spherical harmonics, as we do in our numerical implementation below.

both the initial and final state, as well symmetry under charge conjugation (C), parity (P) and time-reveral (T) transformations [19].

To make this final point clear it is useful to introduce $\mathcal{K}_{\mathrm{df},3}$ (representing here a generic entry of the flavor matrix $\mathbf{K}_{\mathrm{df},3}$) as a function of six three-vectors, in direct analogy to the left-hand side of eq. (2.54). Working in the basis of definite individual pion flavors allows us to readily express the consequences of various symmetries. For example, the exchange symmetry can be written as

$$\mathcal{K}_{\mathrm{df},3;[\pi^+\pi^0\pi^- \leftarrow \pi^+\pi^0\pi^-]}(\boldsymbol{p}_1', \boldsymbol{p}_2', \boldsymbol{p}_3'; \boldsymbol{p}_1, \boldsymbol{p}_2, \boldsymbol{p}_3) = \\ \mathcal{K}_{\mathrm{df},3;[\pi^+\pi^0\pi^- \leftarrow \pi^+\pi^-\pi^0]}(\boldsymbol{p}_1', \boldsymbol{p}_2', \boldsymbol{p}_3'; \boldsymbol{p}_1, \boldsymbol{p}_3, \boldsymbol{p}_2) , \quad (3.1)$$

where we have swapped the second and third species and momenta on the in-state.[9] Using T invariance then implies the following relation,

$$\mathcal{K}_{\mathrm{df},3;[\pi^+\pi^0\pi^- \leftarrow \pi^+\pi^0\pi^-]}(\boldsymbol{p}_1', \boldsymbol{p}_2', \boldsymbol{p}_3'; \boldsymbol{p}_1, \boldsymbol{p}_2, \boldsymbol{p}_3) = \\ \mathcal{K}_{\mathrm{df},3;[\pi^+\pi^0\pi^- \leftarrow \pi^+\pi^0\pi^-]}(-\boldsymbol{p}_1, -\boldsymbol{p}_2, -\boldsymbol{p}_3; -\boldsymbol{p}_1', -\boldsymbol{p}_2', -\boldsymbol{p}_3') . \quad (3.2)$$

Combining with parity implies that $\mathcal{K}_{\mathrm{df},3}$ is unchanged when the initial- and final-state momenta triplets are swapped:

$$\mathcal{K}_{\mathrm{df},3;[\pi^+\pi^0\pi^- \leftarrow \pi^+\pi^0\pi^-]}(\boldsymbol{p}_1', \boldsymbol{p}_2', \boldsymbol{p}_3'; \boldsymbol{p}_1, \boldsymbol{p}_2, \boldsymbol{p}_3) = \\ \mathcal{K}_{\mathrm{df},3;[\pi^+\pi^0\pi^- \leftarrow \pi^+\pi^0\pi^-]}(\boldsymbol{p}_1, \boldsymbol{p}_2, \boldsymbol{p}_3; \boldsymbol{p}_1', \boldsymbol{p}_2', \boldsymbol{p}_3') . \quad (3.3)$$

This result holds for all theories that are PT invariant.

As proposed in ref. [20], and worked out in ref. [22] for three identical bosons, one can expand $\mathcal{K}_{\mathrm{df},3}$ (which in the present case is replaced with the matrix $\mathbf{K}_{\mathrm{df},3}$) about the three-particle threshold in a consistent fashion, and use the symmetries to greatly restrict the number of terms that appear. The results of ref. [22] apply to the $I_{\pi\pi\pi} = 3$ three-pion system; here we generalize them to the $I_{\pi\pi\pi} = 0$, 1 and 2 channels. The new feature is the need to include isospin indices in the particle interchange transformations.

For the parametrizations, we use the same building blocks as in ref. [22],

$$\Delta \equiv \frac{s - 9m^2}{9m^2} , \quad \Delta_i \equiv \frac{s_{jk} - 4m^2}{9m^2} , \qquad \Delta_i' \equiv \frac{s_{jk}' - 4m^2}{9m^2} , \quad \widetilde{t}_{ij} \equiv \frac{t_{ij}}{9m^2} , \qquad (3.4)$$

with generalized Mandelstam variables defined as

$$s \equiv E^2 , \quad s_{ij} \equiv (p_i + p_j)^2 = s_{ji} , \quad s_{ij}' \equiv (p_i' + p_j')^2 = s_{ji}' , \quad t_{ij} \equiv (p_i - p_j')^2 . \qquad (3.5)$$

The power counting scheme for the expansion will be

$$\Delta \sim \Delta_{ij} \sim \Delta_{ij}' \sim \widetilde{t}_{ij} . \qquad (3.6)$$

_____

[9]This property may seem obvious, but we stress that it does not hold for individual Feynman diagrams. Because the definition for $\mathcal{K}_{\mathrm{df},3}$ is built up diagrammatically, the exchange invariance does not hold for various intermediate quantities entering the original derivation and only emerges in the final definition. This point is discussed in more detail in appendix A.

As discussed in ref. [22], only eight of the sixteen quantities in eq. (3.4) are independent — the overall CMF energy, and seven angular variables. The relations between the quantities will be used to simplify the threshold expansions.

In the following, we work out the leading two or three terms in the parametrizations of $\mathbf{K}_{\mathrm{df},3}$ in each of the isospin channels. A summary of key aspects of the results is given in table 3. The presence of even or odd values of $\ell$ is determined by whether the states in the isospin decomposition are given by $|(\pi\pi)_2\pi\rangle$ and $|\sigma\pi\rangle$, leading to even angular momentum in the first two pions, or else $|\rho\pi\rangle$, leading to odd angular momenta.[10] The fact that only small values of angular momentum appear in the table ($\ell, \ell' \leq 2$) is due to our consideration of only the lowest few terms in the threshold expansion. Only a few cubic-group irreps appear for the same reason. All values of $\ell$ and $\ell'$, as well as all cubic-group irreps, will appear at some order in the expansion.

### 3.1.1 $I_{\pi\pi\pi} = 3$

This is the simplest channel, and has been analyzed previously in ref. [22], from which we simply quote the results. The $I_{\pi\pi\pi} = 3$ state is fully symmetric in isospin, so the momentum-dependent part of $\mathbf{K}_{\mathrm{df},3}^{[I=3]}$ must be symmetric under particle interchanges. In the charge neutral sector, there is only a single $I_{\pi\pi\pi} = 3$ state, and thus no isospin indices are needed. $\mathbf{K}_{\mathrm{df},3}^{[I=3]}$ is therefore a function only of the momenta, and, through quadratic order, there are only five independent terms that can appear:

$$m^2 \mathbf{K}_{\mathrm{df},3}^{[I=3]} = \mathcal{K}^{\mathrm{iso}} + \mathcal{K}_{\mathrm{df},3}^{(2,A)} \Delta_A^{(2)} + \mathcal{K}_{\mathrm{df},3}^{(2,B)} \Delta_B^{(2)} + \mathcal{O}(\Delta^3) \,, \tag{3.7}$$

$$\mathcal{K}^{\mathrm{iso}} = \mathcal{K}_{\mathrm{df},3}^{\mathrm{iso}} + \mathcal{K}_{\mathrm{df},3}^{\mathrm{iso},1} \Delta + \mathcal{K}_{\mathrm{df},3}^{\mathrm{iso},2} \Delta^2 \tag{3.8}$$

$$\Delta_A^{(2)} = \sum_{i=1}^{3} (\Delta_i^2 + \Delta_i'^2) - \Delta^2, \tag{3.9}$$

$$\Delta_B^{(2)} = \sum_{i,j=1}^{3} \widetilde{t}_{ij}^2 - \Delta^2 \,. \tag{3.10}$$

Here $\mathcal{K}_{\mathrm{df},3}^{\mathrm{iso}}, \mathcal{K}_{\mathrm{df},3}^{\mathrm{iso},1}, \mathcal{K}_{\mathrm{df},3}^{\mathrm{iso},2}, \mathcal{K}_{\mathrm{df},3}^{(2,A)}$ and $\mathcal{K}_{\mathrm{df},3}^{(2,B)}$ are numerical constants. An extensive study of how these terms affect the finite-volume spectrum has been performed in ref. [22].

### 3.1.2 $I_{\pi\pi\pi} = 0$

The three-pion state with $I_{\pi\pi\pi} = 0$ is totally antisymmetric under the permutation of isospin indices, as shown explicitly by the last row of $\mathcal{C}$ in eq. (2.56). Thus, to satisfy the exchange symmetry exemplified by eq. (3.1), the momentum-dependent part of $\mathbf{K}_{\mathrm{df},3}^{[I=0]}$ must also be totally antisymmetric under particle exchange, in order that the full three-pion state remains symmetric. Again, no explicit isospin indices are needed, as there is only one $I_{\pi\pi\pi} = 0$ state.

---

[10]We stress that the notation $|\rho\pi\rangle$ indicates only that the first two pions are combined into an isotriplet. This implies that their relative angular momentum must be odd, but does not restrict the pions to be in a $p$-wave.

| $I_{\pi\pi\pi}$ | term | $(\ell', \ell)$ | irreps |
|---|---|---|---|
| 3 | $\mathcal{K}_{\mathrm{df},3}^{\mathrm{iso}}$ | $(0,0)$ | $A_1^-$ |
| 3 | $\mathcal{K}_{\mathrm{df},3}^{(2,A)}$ | $(0,0),(0,2),(2,0)$ | $A_1^-$ |
| 3 | $\mathcal{K}_{\mathrm{df},3}^{(2,B)}$ | $(0,0),(0,2),(2,0),(2,2)$ | $A_1^-, E^-, T_2^-, T_1^+$ |
| 0 | $\mathcal{K}_{\mathrm{df},3}^{(\mathrm{AS})}$ | $(1,1)$ | $T_1^-, T_1^+$ |
| 0 | $\mathcal{K}_{\mathrm{df},3}^{(\mathrm{AS},2)}$ | $(1,1)$ | $T_1^-$ |
| 2 | $\mathcal{K}_{\mathrm{df},3}^{T}$ | $\begin{pmatrix} (0,0) & (0,1) \\ (1,0) & (1,1) \end{pmatrix}$ | $A_1^-, T_1^+$ |
| 2 | $\mathcal{K}_{\mathrm{df},3}^{T,2}$ | $\begin{pmatrix} (0,0) & (0,1) \\ (1,0) & (1,1) \end{pmatrix}$ | $A_1^-$ |
| 2 | $\mathcal{K}_{\mathrm{df},3}^{T,3}$ | $\begin{pmatrix} (0,0),(0,2),(2,0) & (0,1),(2,1) \\ (1,0),(1,2) & (1,1) \end{pmatrix}$ | $A_1^-, T_1^+$ |
| 2 | $\mathcal{K}_{\mathrm{df},3}^{T,4}$ | $\begin{pmatrix} (0,0),(0,2),(2,0),(2,2) & (0,1),(2,1) \\ (1,0),(1,2) & (1,1) \end{pmatrix}$ | $A_1^-, E^-, T_2^-, T_1^+$ |
| 1 | $\mathcal{K}_{\mathrm{df},3}^{\mathrm{SS}}$ | $\begin{pmatrix} (0,0) & - & - \\ - & - & - \\ - & - & - \end{pmatrix}$ | $A_1^-$ |
| 1 | $\mathcal{K}_{\mathrm{df},3}^{\mathrm{SD}}$ | $\begin{pmatrix} - & (0,0) & (0,1) \\ (0,0) & - & - \\ (1,0) & - & - \end{pmatrix}$ | $A_1^-$ |
| 1 | $\mathcal{K}_{\mathrm{df},3}^{\mathrm{DD}}$ | $\begin{pmatrix} - & - & - \\ - & (0,0) & (0,1) \\ - & (1,0) & (1,1) \end{pmatrix}$ | $A_1^- T_1^-$ |

**Table 3**. Properties of low-order terms in the threshold expansion of $\mathbf{K}_{\mathrm{df},3}$. The terms are specified by their coefficients in eqs. (3.7), (3.12), (3.27), and (3.30). The values of $(\ell', \ell)$ are obtained by decomposing the expressions into the $k\ell m$ basis, following the method of ref. [22]. The matrix structure corresponds to the isospin decomposition of appendix C, which is also used in the aforementioned equations. The final column lists the cubic-group irreps that are present in finite volume when one considers the rest frame, $\boldsymbol{P} = 0$. The superscript gives the parity, which includes the intrinsic negative parity of the three-pion state. The irreps are determined by first working out which $J^P$ values are present, and then subducing to the cubic group. Results for $I_{\pi\pi\pi} = 3$ are taken from ref. [22].

It is straightforward to see that the leading completely antisymmetric term that can appear in the momentum-dependent part of $\mathbf{K}_{\mathrm{df},3}^{[I=0]}$ is of quadratic order in the threshold expansion:

$$\mathbf{K}_{\mathrm{df},3}^{[I=0]} \supset \mathcal{K}_{\mathrm{df},3}^{\mathrm{AS}} \sum_{\substack{ijk \\ mnr}} \epsilon_{ijk}\epsilon_{mnr} t_{im} t_{jn} \equiv \mathcal{K}_{\mathrm{df},3}^{\mathrm{AS}} \Delta_{\mathrm{AS}}^{(2)}. \tag{3.11}$$

At next order two new structures arise and the full form can be written

$$\mathbf{K}_{\mathrm{df},3}^{[I=0]} = \left( \mathcal{K}_{\mathrm{df},3}^{\mathrm{AS}} + \mathcal{K}_{\mathrm{df},3}^{\mathrm{AS},1} \Delta \right) \Delta_{\mathrm{AS}}^{(2)} + \mathcal{K}_{\mathrm{df},3}^{\mathrm{AS},2} \Delta_{\mathrm{AS}}^{(3)} + O(\Delta^4), \tag{3.12}$$

with

$$\Delta_{\mathrm{AS}}^{(3)} \equiv \sum_{\substack{ijk \\ mnr}} \epsilon_{ijk}\epsilon_{mnr} t_{im} t_{jn} t_{kr}. \tag{3.13}$$

### 3.1.3 $I_{\pi\pi\pi} = 2$

As discussed in the previous section, and summarized in table 1, the isotensor channel involves a two-dimensional flavor space. This space can be understood in terms of the permutation group $S_3$, as described in appendix C. The two isospin basis vectors, $|\chi_1\rangle_2 = |(\pi\pi)_2\pi\rangle_2$ and $|\chi_2\rangle_2 = |\rho\pi\rangle_2$, also given in eqs. (C.12) and (C.13), transform in the standard irrep of $S_3$. To satisfy the exchange relations exemplified by eq. (3.1), the combined transformation of isospin indices and momenta must lie in the trivial irrep of $S_3$. This requires combining the isospin doublet with a momentum-space doublet also transforming in the standard irrep. At linear order, there are three momenta, and these decompose into a symmetric singlet $(p_1 + p_2 + p_3)$ and the standard-irrep doublet

$$\xi_1 = \frac{1}{\sqrt{6}}(2p_3 - p_1 - p_2) \quad \text{and} \quad \xi_2 = \frac{1}{\sqrt{2}}(p_2 - p_1). \tag{3.14}$$

There is an analogous doublet, $\xi_i'$, built from final-state momenta. The symmetric combinations are then

$$|\psi_{\mathrm{sym}}\rangle = \xi_1 |\chi_1\rangle_2 + \xi_2 |\chi_2\rangle_2 \equiv \begin{pmatrix} \xi_1 \\ \xi_2 \end{pmatrix} \equiv \vec{\xi}, \tag{3.15}$$

$$|\psi_{\mathrm{sym}}'\rangle = \xi_1' |\chi_1\rangle_2 + \xi_2' |\chi_2\rangle_2 \equiv \begin{pmatrix} \xi_1' \\ \xi_2' \end{pmatrix} \equiv \vec{\xi}', \tag{3.16}$$

where the last two forms introduce a convenient column vector notation. The leading term in $\mathbf{K}_{\mathrm{df},3}^{[I=2]}$ then becomes

$$\begin{aligned}
\mathbf{K}_{\mathrm{df},3}^{[I=2]} &\supset \mathcal{K}_{\mathrm{df},3}^{\mathrm{ST}} |\psi_{\mathrm{sym}}'\rangle \cdot \langle \psi_{\mathrm{sym}}| \equiv \mathcal{K}_{\mathrm{df},3}^{\mathrm{ST}} \begin{pmatrix} \xi_1' \cdot \xi_1 & \xi_1' \cdot \xi_2 \\ \xi_2' \cdot \xi_1 & \xi_2' \cdot \xi_2 \end{pmatrix} \equiv \mathcal{K}_{\mathrm{df},3}^{\mathrm{ST}} \vec{\xi}'^{\mu} \otimes \vec{\xi}_\mu, \\
&= \frac{\mathcal{K}_{\mathrm{df},3}^{\mathrm{ST}}}{6} \begin{pmatrix} (2p_3' - p_1' - p_2') \cdot (2p_3 - p_1 - p_2) & \sqrt{3}(2p_3' - p_1' - p_2') \cdot (p_2 - p_1) \\ \sqrt{3}(p_2' - p_1') \cdot (2p_3 - p_1 - p_2) & 3(p_2' - p_1') \cdot (p_2 - p_1) \end{pmatrix},
\end{aligned} \tag{3.17}$$

where $\mathcal{K}_{\mathrm{df},3}^{\mathrm{ST}}$ is a constant. Note that this is of linear order in $\Delta$, since the inner products $\xi_i \cdot \xi_j'$ can be written as linear combinations of the $t_{ij}$. There are no terms of $\mathcal{O}(\Delta^0)$.

At next order, there are three sources of contributions. First, one can multiply the term in eq. (3.17) by $\Delta$. Second, one can build additional basis vectors transforming as doublets, but of higher order in momentum. Third, one can form Lorentz singlets in more than one way. We discuss the latter two issues in turn.

To proceed systematically, we begin by classifying objects quadratic in momenta, of the general form $p_i^\mu p_j^\nu$. The nine such objects contain three standard-irrep doublets:

$$\xi(S)_i^{\mu\nu} = \xi_i^\mu P^\nu + \mu \leftrightarrow \nu\,, \quad \xi(A)_i^{\mu\nu} = \xi_i^\mu P^\nu - \mu \leftrightarrow \nu\,, \tag{3.18}$$

and

$$\vec{\xi}(\bar{S})^{\mu\nu} \equiv \left(\xi(\bar{S})_1^{\mu\nu}, \xi(\bar{S})_2^{\mu\nu}\right) = \left(\xi_2^\mu \xi_2^\nu - \xi_1^\mu \xi_1^\nu,\ \xi_1^\mu \xi_2^\nu + \xi_2^\mu \xi_1^\nu\right)\,. \tag{3.19}$$

The latter is the standard irrep that results from the direct product of $\vec{\xi}$ with itself. Each of these doublets can be combined with the isospin-space doublet to make fully symmetric objects out of both initial- and final-state momenta. These are then combined as in eq. (3.17) to give a contribution to $\mathcal{K}_{\mathrm{df},3}$. When Lorentz contractions are included, as discussed below, symmetric doublets ($\xi(S)$ and $\xi(\bar{S})$) must be combined with other symmetric objects, and similarly for the antisymmetric doublet $\xi(A)$. Taking into account also CPT symmetry, there are then four possible combinations, schematically given by

$$\xi(S)'\xi(S)\,, \quad \xi(S)'\xi(\bar{S}) + \xi(\bar{S})'\xi(S)\,, \quad \xi(\bar{S})'\xi(\bar{S}) \ \text{ and } \ \xi(A)'\xi(A)\,. \tag{3.20}$$

Lorentz indices can be contracted in three ways:

$$(i)\ g_{\mu\nu}g_{\mu'\nu'}\,, \quad (ii)\ g_{\mu\mu'}g_{\nu\nu'} \ \text{ and } \ (iii)\ \epsilon_{\mu\nu\mu'\nu'}\,. \tag{3.21}$$

The first two can be used only for the symmetric objects, while the last two can be used for the antisymmetric objects. We begin with the Lorentz contractions of type $(i)$. Here it turns out that all three symmetric combinations lead to the same result, namely the outer product

$$\mathbf{K}_{\mathrm{df},3}^{[I=2]} \supset \vec{\xi}'^{(2)} \otimes \vec{\xi}^{(2)}\,, \tag{3.22}$$

where

$$\vec{\xi}^{(2)} = \left(\frac{2\Delta_3 - \Delta_1 - \Delta_2}{\sqrt{6}}, \frac{\Delta_2 - \Delta_1}{\sqrt{2}}\right) \propto (\xi_1 \cdot P, \xi_2 \cdot P)\,, \tag{3.23}$$

with $P = p_1 + p_2 + p_3 = p_1' + p_2' + p_3'$. Next we consider Lorentz contractions of type $(ii)$. Here we find only two combinations lead to new structures, namely,

$$\mathbf{K}_{\mathrm{df},3}^{[I=2]} \supset \vec{\xi}(\bar{S})'^{\mu\nu} \otimes \vec{\xi}(S)_{\mu\nu} + \vec{\xi}(S)'^{\mu\nu} \otimes \vec{\xi}(\bar{S})_{\mu\nu}\,, \tag{3.24}$$

and

$$\mathbf{K}_{\mathrm{df},3}^{[I=2]} \supset \vec{\xi}(\bar{S})'^{\mu\nu} \otimes \vec{\xi}(\bar{S})_{\mu\nu}\,. \tag{3.25}$$

Finally, the contraction of type $(iii)$ leads to

$$\mathbf{K}_{\mathrm{df},3}^{[I=2]} \supset \epsilon_{\mu\nu\rho\sigma}\ \vec{\xi}(A)'^{\mu\nu} \otimes \vec{\xi}(A)^{\rho\sigma}\,, \tag{3.26}$$

which vanishes identically.

Thus, at this stage, we have found four terms of $\mathcal{O}(\Delta^2)$. A further potential source of such terms is to combine contributions linear in $\xi$ with those cubic in $\xi'$, and *vice versa*. Carrying out an analysis similar to that above, we find, however, that all such terms can be written in terms of those already obtained. Thus the final form of $\mathbf{K}_{\mathrm{df},3}^{[I=2]}$ is

$$\mathbf{K}_{\mathrm{df},3}^{[I=2]} = \left(\mathcal{K}_{\mathrm{df},3}^{\mathrm{T}} + \mathcal{K}_{\mathrm{df},3}^{\mathrm{T},1}\Delta\right)\vec{\xi}\,'^{\mu}\otimes\vec{\xi}_{\mu} + \mathcal{K}_{\mathrm{df},3}^{\mathrm{T},2}\,\vec{\xi}\,'^{(2)}\otimes\vec{\xi}\,^{(2)} +$$
$$+ \mathcal{K}_{\mathrm{df},3}^{\mathrm{T},3}\left(\vec{\xi}(\bar{S})\,'^{\mu\nu}\otimes\vec{\xi}(S)_{\mu\nu} + \vec{\xi}(S)\,'^{\mu\nu}\otimes\vec{\xi}(\bar{S})_{\mu\nu}\right) + \mathcal{K}_{\mathrm{df},3}^{\mathrm{T},4}\,\vec{\xi}(\bar{S})\,'^{\mu\nu}\otimes\vec{\xi}(\bar{S})_{\mu\nu} + \mathcal{O}(\Delta^3)\,, \quad (3.27)$$

where the superscript T refers to isotensor.

### 3.1.4 $I_{\pi\pi\pi} = 1$

Lastly, we consider the parametrization of $\mathbf{K}_{\mathrm{df},3}^{[I=1]}$. Here the isospin subspace is three-dimensional and in section 2 we used a basis with definite two-pion isospin,

$$\{|(\pi\pi)_2\pi\rangle_1\,,\ |\rho\pi\rangle_1\,,\ |\sigma\pi\rangle_1\}\,. \quad (3.28)$$

In this section we find it convenient to use a different basis, consisting of a singlet transforming in the trivial irrep of $S_3$ and a doublet in the standard irrep. The relation between bases is shown explicitly in eqs. (C.15)–(C.18) and, in the matrix notation that follows, we order the basis vectors such that the singlet comes first:

$$\{|\chi_s\rangle_1\,,\ |\chi_1\rangle_1\,,\ |\chi_2\rangle_1\}\,. \quad (3.29)$$

The presence of two irreps implies a greater number of options for building a fully symmetric object. In particular, the analysis for the symmetric singlet component is identical to that for the $I_{\pi\pi\pi} = 3$ sector, with the leading two terms being of $\mathcal{O}(\Delta^0)$ and $\mathcal{O}(\Delta)$, respectively. Combining a final-state singlet with an initial-state doublet, an overall singlet of $\mathcal{O}(\Delta)$ is obtained using the Lorentz-scalar doublet $\vec{\xi}\,^{(2)}$ of eq. (3.23). An analogous term is obtained by interchanging initial and final states. At this same order, initial- and final-state doublets can be combined as in eq. (3.17). In total, enforcing CPT invariance, we end up with

$$\mathbf{K}_{\mathrm{df},3}^{[I=1,|\chi\rangle]} = \left(\mathcal{K}_{\mathrm{df},3}^{\mathrm{SS}} + \mathcal{K}_{\mathrm{df},3}^{\mathrm{SS},1}\Delta\right)\begin{pmatrix} 1 & 0 & 0 \\ 0 & 0 & 0 \\ 0 & 0 & 0 \end{pmatrix} + \mathcal{K}_{\mathrm{df},3}^{\mathrm{SD}}\begin{pmatrix} 0 & \xi_1^{(2)} & \xi_2^{(2)} \\ \xi_1'^{(2)} & 0 & 0 \\ \xi_2'^{(2)} & 0 & 0 \end{pmatrix}$$
$$+ \mathcal{K}_{\mathrm{df},3}^{\mathrm{DD}}\begin{pmatrix} 0 & 0 & 0 \\ 0 & \xi_1'\cdot\xi_1 & \xi_1'\cdot\xi_2 \\ 0 & \xi_2'\cdot\xi_1 & \xi_2'\cdot\xi_2 \end{pmatrix} + O(\Delta^2)\,, \quad (3.30)$$

where the $|\chi\rangle$ superscript on the left-hand side emphasizes that we are using the new basis, introduced in (3.29). The SS and DD superscripts on the right-hand side refer to singlet and doublet irreps.

### 3.2 Three-particle resonances

The threshold expansion derived in the previous section plays a similar role for three-particle interactions as the effective-range expansion does for the two-particle K matrix. It provides a smooth parametrization of the interaction, valid for some range around threshold, that respects the symmetries. However, we expect that the convergence of the series is limited by the singularities in $\mathcal{K}_{\mathrm{df},3}$ closest to the three-particle threshold, just as the expansion for $\mathcal{K}_2$ is limited either by the nearest poles, possibly associated with a two-resonance, or else by the $t$-channel cut. As studying three-particle resonances is one of the major goals behind the development of the three-particle quantization condition, it is important to determine appropriate forms of $\mathcal{K}_{\mathrm{df},3}$ in the channels that contain such resonances. This is the task of the present section.

We begin by listing, in table 4, the total $J^P$ and isospin for the resonant channels observed in nature that couple to three pions [39]. We include only cases where the coupling is allowed in isosymmetric QCD. Resonances are present only for $I_{\pi\pi\pi} = 0$ and $I_{\pi\pi\pi} = 1$. We note the absence of the $J^P = 0^+$, $I_{\pi\pi\pi} = 1$, $a_0(980)$, for which no three-pion coupling is possible that is simultaneously consistent with angular momentum and parity conservation. For each resonance, we also note the corresponding subduced cubic group irreps. The cubic symmetry group including parity (also called the achiral or full octahedral group) defines the symmetry of the system provided that the total momentum is set to zero. In a lattice QCD calculation, one can project the three-pion states onto definite cubic-group irreps by choosing appropriate three-pion interpolating operators, as discussed in appendix D. Note that, for the values of $J^P$ arising in the table, a finite-volume irrep can always be identitifed that does not couple to any other listed values. The final column in the table gives the lowest three-pion orbit that couples to the irrep(s) for the corresponding state. The ordering of the orbits is described in appendix D; see in particular table 5.

In the remainder of this section we determine the forms of the entries of $\mathbf{K}_{\mathrm{df},3}$ that couple to three pions having each of the quantum numbers listed in table 4. We stress that, as in the previous section, this is an infinite-volume exercise. When using the resulting forms for $\mathbf{K}_{\mathrm{df},3}^{[I]}$ in the quantization condition, one must covert the forms given here to the $k\ell m$ index set introduced above. This is a straightforward exercise that we do not discuss further here.

By analogy with the two-particle case, we expect that a three-particle resonance can be represented by a pole in the part of $\mathbf{K}_{\mathrm{df},3}^{[I]}$ with the appropriate quantum numbers [20], i.e.

$$\mathbf{K}_{\mathrm{df},3}^{[I,|\chi\rangle]} = \mathcal{K}_{\mathrm{df},3}^X \frac{c_X}{s - M_X^2} + \mathcal{O}\big[(s - M_X^2)^0\big]\,, \tag{3.31}$$

where the superscript $|\chi\rangle$ on the left-hand side emphasizes that we work in the basis of definite symmetry states for $I_{\pi\pi\pi} = 1$ (see also appendix C). On the right-hand side, $X$ labels the quantum numbers, $M_X$ is close to the resonance mass (at least in the case of narrow resonances), the real constant $c_X$ is related to the width of the resonance, and $\mathcal{K}_{\mathrm{df},3}^X$ carries the overall quantum numbers. The precise relationship of $c_X$ and $M_X$ to the resonance parameters in $\mathcal{M}_3$ is not known analytically, since determining $\mathcal{M}_3$ requires solving the non-trivial integral equations discussed above.

| Resonance | $I_{\pi\pi\pi}$ | $J^P$ | Irrep ($\boldsymbol{P} = 0$) | $3\pi$ orbit |
|:---:|:---:|:---:|:---:|:---:|
| $\omega(782)$ | 0 | $1^-$ | $T_1^-$ | 4 |
| $h_1(1170)$ | 0 | $1^+$ | $T_1^+$ | 2 |
| $\omega_3(1670)$ | 0 | $3^-$ | $A_2^-$ | 4 |
| $\pi(1300)$ | 1 | $0^-$ | $A_1^-$ | 1 |
| $a_1(1260)$ | 1 | $1^+$ | $T_1^+$ | 2 |
| $\pi_1(1400)$ | 1 | $1^-$ | $T_1^-$ | 4 |
| $\pi_2(1670)$ | 1 | $2^-$ | $E^-$ and $T_2^-$ | 2 |
| $a_2(1320)$ | 1 | $2^+$ | $E^+$ and $T_2^+$ | 3 |
| $a_4(1970)$ | 1 | $4^+$ | $A_1^+$ | 16 |

**Table 4**. Lowest lying resonances with negative G-parity, and which couple to three pions, in the different isospin and $J^P$ channels. The fourth column shows the cubic group irreps that are subduced from the rotation group irreps, assuming that the resonance is at rest ($\boldsymbol{P} = 0$). The final column gives the lowest three-pion momentum orbit that contains the corresponding cubic group irrep, again assuming $\boldsymbol{P} = 0$.

We stress that, once a form for $\mathcal{K}_{\mathrm{df},3}^X$ is known, only one sign of $c_X$ will lead to a resonance pole with the physical sign for the residue. The correct choice can be identified by requiring that the finite-volume correlator $C_L$ has a single pole with the correct residue [20, 22]. In the limit $c_X \to 0$, one recovers an additional decoupled state in the finite-volume spectrum at energy $E = M_X$ (assuming $\boldsymbol{P} = 0$), corresponding to a stable would-be resonance. The form in eq. (3.31) was proposed in ref. [20] for the case of identical scalars (which is equivalent to the $I_{\pi\pi\pi} = 3$ channel here) for which $\mathcal{K}_{\mathrm{df},3}^X$ is a constant. As noted above, however, there are no resonances in nature in the $I_{\pi\pi\pi} = 3$ or $I_{\pi\pi\pi} = 2$ channels, so the example given in ref. [20] is for illustrative purposes only. In the following we determine forms for $\mathcal{K}_{\mathrm{df},3}^X$ that can be used for all the resonant channels listed in table 4.

We also enforce an additional requirement on $\mathcal{K}_{\mathrm{df},3}^X$, namely that it has a factorized form in isospin space. This is motivated by the fact that the residues of resonance poles in $\mathcal{M}_2$ and $\mathcal{M}_3$ do factorize, and it was argued in ref. [21] that this carries over to poles in $\mathcal{K}_2$ evaluated at off-shell momenta. Here we assume that this holds also for resonance poles in $\mathcal{K}_{\mathrm{df},3}$. We view this as plausible, but leave the proof to future work.

Before turning to the detailed parametrizations, we comment on the range of validity for the quantization condition. All the resonances in table 4 have, in principle, additional decay channels, such as $5\pi$ or $K\bar{K}$. One must consider on a case by case basis whether neglecting these is justified, based on the couplings between the resonance of interest to the neglected channels, as well as the target precision of the calculation. Another possibility is to work at unphysically heavy pion masses, such that some of the neglected channels are

kinematically forbidden. While the procedure for including additional two-particle channels should be given by a straightforward generalization of ref. [19], rigorously accommodating the $5\pi$ state would be a significant formal undertaking.

### 3.2.1 Isoscalar resonances

The symmetry requirements for the $\mathcal{K}_{\mathrm{df},3}^X$ are exactly as in the threshold expansion. For $I_{\pi\pi\pi} = 0$, this means complete antisymmetry under particle exchange. Useful building blocks are the following objects:

$$V^\alpha = P_\mu \sum_{ijk} \epsilon_{ijk}\, p_j^\mu p_k^\alpha \qquad \xrightarrow{\text{CMF}} \qquad \frac{E}{2}\left(0, -3\omega_-\boldsymbol{p}_3 - \boldsymbol{p}^-[E - 3\omega_3]\right), \qquad (3.32)$$

$$A^\alpha = \epsilon_{\alpha\beta\gamma\delta}\, p_1^\beta p_2^\gamma p_3^\delta \qquad \xrightarrow{\text{CMF}} \qquad E\left(0,\, \boldsymbol{p}_1 \times \boldsymbol{p}_2\right) = E\left(0,\, \boldsymbol{p}_2 \times \boldsymbol{p}_3\right), \qquad (3.33)$$
$$= E\left(0,\, \boldsymbol{p}_3 \times \boldsymbol{p}_1\right),$$

where $p^{-\mu} = p_1^\mu - p_2^\mu = (\omega^-, \boldsymbol{p}^-)$, $p_3^\mu = (\omega_3, \boldsymbol{p}_3)$, etc. The quantities $V^\alpha$ and $A^\alpha$ are fully antisymmetric under particle exchange, and describe a vector and axial vector, respectively, as can be seen from their forms in the CMF. In particular, the vanishing of the temporal components in this frame shows the absence of scalar and pseudoscalar contributions (with the respect to the three-dimensional rotation group).

Taking into account the negative parity of the pion, the momentum-space amplitude for the $J^P = 1^-$ $\omega(782)$ to decay to three pions must transform as an axial vector. This leads to the following form for $\mathcal{K}_{\mathrm{df},3}$,

$$\mathcal{K}_{\mathrm{df},3}^\omega = A'^\mu A_\mu, \qquad (3.34)$$

where $A'^\mu$ has the same form as $A^\mu$ but expressed in terms of final-state momenta. The expression (3.34) is manifestly Lorentz and CPT invariant. We have checked explicitly that, when reduced to the $k\ell m$ basis used in the quantization condition, this expression transforms purely as a $T_1^-$ under the cubic group. Indeed, it turns out to be proportional to the operator $\Delta_{\mathrm{AS}}^{(3)}$, given in eq. (3.13), that arises in the threshold expansion. Furthermore, we note from table 5 in appendix D that the lowest three-pion state in a cubic box that transforms in the $T_1^-$ irrep lies in the fourth orbit and has momenta $(1, 1, 0)$, $(-1, 0, 0)$ and $(0, -1, 0)$ (or a cubic rotation thereof) in units of $2\pi/L$. This can be understood from the fact that, in the CMF, $A^\mu$ vanishes if *any* of the three pion momenta vanish, as can be seen from eq. (3.33).

These results have implications for a practical study of the $\omega$ resonance. As is known from the study of two-particle resonances, to map out the resonant structure (e.g. the rapid rise in the phase shift) requires many crossings between the finite-volume resonance level and those of weakly-interacting multi-particle states. Since the lowest, non-interacting three-pion state with the quantum numbers of the $\omega$ lies in the fourth orbit, it occurs at relatively high energy. Thus for small to moderate volumes, the finite-volume level corresponding to the $\omega$ will be the lowest lying state and there will be no avoided level crossings. Only by going to larger boxes will the level-crossings needed to constrain $\mathcal{K}_{\mathrm{df},3}$ in

detail be present. For physical pion masses the constraint is not too strong — an avoided level crossing requires $mL \gtrsim 4.6$. However, if working with heavier-than-physical pions, such as in the example presented in section 4, larger values of $mL$ are needed ($mL \gtrsim 6.5$ in the toy model). These constraints apply, however, only in the overall rest frame. It is likely that moving frames, for which the constraints will be relaxed, will play an important role in any detailed investigation of the $\omega$ resonance.

For the $J^P = 1^+$ $h_1(1170)$, the momentum-space decay amplitude must transform as a vector, leading to

$$\mathcal{K}^{h_1}_{\mathrm{df},3} = V'^\mu V_\mu \,. \tag{3.35}$$

Only two momenta need to be nonzero for $V^\mu$ to be nonvanishing, and indeed the lowest momentum configuration transforming as the required $T_1^+$ lies in the second orbit and has momenta $(1,0,0)$, $(-1,0,0)$ and $(0,0,0)$ (see table 5). Applying the same estimate as above based on the non-interacting energy, the first CMF avoided-level crossing for physical pion masses is already expected for $mL \gtrsim 1.8$. Thus, for all volumes where the neglected $e^{-mL}$ is a reasonable approximation (typically requiring $mL \gtrsim 4$), we expect to recover useful constraints on the $h_1$ width in all finite-volume frames.

Finally, for the $J^P = 3^-$ $\omega_3(1670)$, the momentum-space amplitude must transform as $J^P = 3^+$. One possible form is

$$\mathcal{K}^{\omega_3}_{\mathrm{df},3} = (A_\mu A'^\mu)^3 - \frac{3}{5}(A^2)(A'^2)(A_\mu A'^\mu) \,, \tag{3.36}$$

where the second term is required to project against a $J^P = 1^+$ component. The corresponding cubic-group irrep, $A_2^-$, appears first in the same three-pion orbit as for the $\omega$, for then the axial current $A_\mu$ is nonzero.

### 3.2.2 Isovector resonances

We turn now to parameterizations of $\mathcal{K}^X_{\mathrm{df},3}$ in the three-dimensional isovector case, working always in the $\chi$-basis of (3.29) [defined explicitly in eqs. (C.15)–(C.18)].

Beginning with the $J^P = 0^-$ $\pi(1300)$, the simplest case in this sector, we note that these quantum numbers can be obtained from three pions at rest, so that no momentum dependence is required in $\mathcal{K}^\pi_{\mathrm{df},3}$. However, as we have seen in section 3.1.4, momentum-independence is possible only for the component connecting permutation-group-singlets in the initial and final states. For the other components momentum dependence is needed to obtain a form that is fully symmetric under permutations. Using results from our discussion of the threshold expansion, we find the following possible form[11]

$$\mathcal{K}^\pi_{\mathrm{df},3} = \begin{pmatrix} s_\pi \\ d_\pi\,\xi_1'^{(2)} \\ d_\pi\,\xi_2'^{(2)} \end{pmatrix} \otimes \left( s_\pi,\, d_\pi\,\xi_1^{(2)},\, d_\pi\,\xi_2^{(2)} \right) \,. \tag{3.37}$$

---

[11]We stress that we are not here doing an expansion in momenta, but rather writing a simple form that has the appropriate symmetries. More complicated expressions consistent with the desired quantum numbers are certainly possible.

Here $s_\pi$ and $d_\pi$ are real constants, corresponding to couplings to the singlet and doublet components, respectively. The outer product structure is necessary due to the factorization of the residue at the K-matrix pole. We stress that the components of the two vectors in the outer product must be Lorentz scalars in order that $\mathcal{K}^\pi_{\mathrm{df},3}$ couples to $J^P = 0^-$. Thus, for example, $\xi_1^{(2)}$ cannot be replaced by $\xi_1^\mu$. We also note that we do not expect the momentum-dependent parts of this expression to be suppressed relative to the momentum-independent ones, since we are far from threshold.

We can use the properties of the physical $\pi(1300)$ resonance to guide our expectations concerning $s_\pi$ and $d_\pi$. In particular, the resonance has been observed to have both $\sigma\pi$ and $\rho\pi$ final states [39]. Recalling from appendix C that the first two entries of the vector space are linear combinations of the states $|(\pi\pi)_2\pi\rangle_1$ and $|\sigma\pi\rangle_1$, while the third is $|\rho\pi\rangle_1$, we see that $s_\pi$ describes the coupling to the former two states, while $d_\pi$ couples to all three. Thus $d_\pi$ must be nonzero to describe the physical resonance, with its $\rho\pi$ decay, while the importance of $s_\pi$ depends on the details of the amplitude.

Next we turn to the $J^P = 1^+$ $a_1(1260)$. Taking into account the intrinsic parity of the pion, the decay amplitude must transform as a vector. A possible form is thus

$$\mathcal{K}^{a_1}_{\mathrm{df},3} = g^P_{\mu\nu} \begin{pmatrix} s_{a_1} V'^\mu_S \\ d_{a_1}\, \xi'^\mu_1 \\ d_{a_1}\, \xi'^\mu_2 \end{pmatrix} \otimes \left( s_{a_1} V^\nu_S,\; d_{a_1}\, \xi^\nu_1,\; d_{a_1}\, \xi^\nu_2 \right), \qquad (3.38)$$

where

$$V^\nu_S = \xi^\nu_1 \xi^{(2)}_1 + \xi^\nu_2 \xi^{(2)}_2, \qquad (3.39)$$

is a vector that is symmetric under permutations, and

$$g^P_{\mu\nu} = (g_{\mu\nu} - P_\mu P_\nu / P^2), \qquad (3.40)$$

is the projector that arises from the sum over polarizations of $\epsilon_\mu \epsilon^*_\nu$. It projects against $P^\mu$, and in the CM frame it picks out the spatial part, $\boldsymbol{V}_S$, which transforms as a vector, while removing the $J^P = 0^+$ quantity, $V^0_S$. We are forced to use a form for $V^\nu_S$ that is cubic in momenta because the only symmetric vector linear in momenta is simply $P^\mu$, which vanishes when contracted with $g^P$. In contrast to the form for the $\pi(1300)$, eq. (3.37), the doublet portion of the amplitude in eq. (3.38) has a simpler momentum-dependence than the singlet part. The real constants $s_{a_1}$ and $d_{a_1}$ play the same role as for the $\pi(1300)$, and again $d_{a_1}$ must be nonzero since $\rho\pi$ and $\sigma\pi$ decays are observed.

Next we turn to the $J^P = 1^-$ $\pi_1(1400)$. It is not possible to construct a fully symmetric axial vector from three momenta, and thus the decay amplitude of the symmetric component vanishes. For the doublet part, a nonzero amplitude can be obtained by combining the completely antisymmetric axial vector $A^\mu$ [eq. (3.33)] with the doublet $\vec{\xi}^{(2)}$ in the appropriate manner. This leads to

$$\mathcal{K}^{\pi_1}_{\mathrm{df},3} = A'^\mu g^P_{\mu\nu} A^\nu \begin{pmatrix} 0 \\ -\xi'^{(2)}_2 \\ \xi'^{(2)}_1 \end{pmatrix} \otimes \left( 0,\; -\xi^{(2)}_2,\; \xi^{(2)}_1 \right). \qquad (3.41)$$

To parametrize the $J^P = 2^-$ $\pi_2(1670)$ requires a tensor composed of momentum vectors, with the appropriate symmetry properties. Using the constructions from the previous section, we find the following form:

$$\mathcal{K}_{\mathrm{df},3}^{\pi_2} = \left( g_{\rho\mu}^P g_{\sigma\nu}^P - \frac{1}{3} g_{\rho\sigma}^P g_{\mu\nu}^P \right) \begin{pmatrix} s_{\pi_2} T'^{\rho\sigma} \\ d_{\pi_2}\, \xi(\bar{S})_1'^{\rho\sigma} \\ d_{\pi_2}\, \xi(\bar{S})_2'^{\rho\sigma} \end{pmatrix} \otimes \left( s_{\pi_2} T^{\mu\nu},\, d_{\pi_2}\, \xi(\bar{S})_1^{\mu\nu},\, d_{\pi_2}\, \xi(\bar{S})_2^{\mu\nu} \right) , \quad (3.42)$$

where

$$T^{\mu\nu} = \xi_1^\mu \xi_1^\nu + \xi_2^\mu \xi_2^\nu , \qquad (3.43)$$

is a Lorentz tensor that is an $S_3$ singlet. The tensor containing $g^P$ projects out the $J = 2$ part in the CM frame.

For the $J^P = 2^+$ $a_2(1320)$ we need to construct a pseudotensor from momentum vectors. The simplest form that we have found is

$$\mathcal{K}_{\mathrm{df},3}^{a_2} = \left( g_{\rho\mu}^P g_{\sigma\nu}^P - \frac{1}{3} g_{\rho\sigma}^P g_{\mu\nu}^P \right) \begin{pmatrix} s_{a_2} A'^\rho V'^\sigma \\ -d_{a_2} A'^\rho \xi_2'^\sigma \\ d_{a_2} A'^\rho \xi_1'^\sigma \end{pmatrix}_{\mathrm{sym}} \otimes \left( s_{a_2} A^\mu V^\nu,\, -d_{a_2} A^\mu \xi_2^\nu,\, d_{a_2} A^\mu \xi_1^\nu \right)_{\mathrm{sym}} ,$$

$$(3.44)$$

where the subscript "sym" indicates symmetrizing the tensors.

Finally, for the $J^P = 4^+$ $a_4(1970)$, we need to construct an $\ell = 4$ pseudotensor from momentum vectors. One possible form is

$$\mathcal{K}_{\mathrm{df},3}^{a_4} = \left( g_{\mu'\mu}^P g_{\nu'\nu}^P g_{\rho'\rho}^P g_{\sigma'\sigma}^P - \frac{6}{7} g_{\mu'\nu'}^P g_{\mu\nu}^P g_{\rho'\sigma'}^P g_{\sigma'\sigma}^P + \frac{3}{35} g_{\mu'\nu'}^P g_{\rho'\sigma'}^P g_{\mu\nu}^P g_{\rho\sigma}^P \right) T_4'^{\mu'\nu'\rho'\sigma'} \otimes T_4^{\mu\nu\rho\sigma} ,$$

$$(3.45)$$

$$T_4^{\mu\nu\rho\sigma} = \left( s_{a_4}(A^\mu A^\nu A^\rho V^\sigma),\, -d_{a_4}(A^\mu A^\nu A^\rho \xi_2^\sigma),\, d_{a_4}(A^\mu A^\nu A^\rho \xi_1^\sigma) \right)_{\mathrm{sym}} , \qquad (3.46)$$

$$T_4'^{\mu\nu\rho\sigma} = \begin{pmatrix} s_{a_4}(A'^\mu A'^\nu A'^\rho V'^\sigma) \\ -d_{a_4}(A'^\mu A'^\nu A'^\rho \xi'^{2\sigma}) \\ d_{a_4}(A'^\mu A'^\nu A'^\rho \xi'^{1\sigma}) \end{pmatrix} . \qquad (3.47)$$

An alternative form replaces two of the axial vectors with vectors (in either or both the initial and final states).

# 4 Toy model: spectrum in $I_{\pi\pi\pi} = 0$ channel

The goal of this section is to present an example of the implementation of the new quantization conditions derived in this paper. We choose the $I_{\pi\pi\pi} = 0$ channel, which is the simplest of the new results, since the quantization condition is one-dimensional in isospin space. The extension of the implementation to the other channels is, however, straightforward.

The $I_{\pi\pi\pi} = 0$ channel is of direct phenomenological relevance, due to the presence of two (relatively) light three-particle resonances, the $\omega(782)$ and the $h_1(1170)$. In particular, at physical pion masses, the $\omega$ lies only slightly above the five-pion inelastic threshold, and the isospin-violating couplings to two and four pions are weak, so that the three-particle quantization condition is likely to provide a good description. Indeed, at somewhat heavier-than-physical pion masses (e.g. $M_\pi \sim 200\,\mathrm{MeV}$), the $\omega$ should lie between the three- and five-pion thresholds. If, in addition, one has exact isospin symmetry, there will be no coupling to channels with an even number of pions. This example can thus be explored in a rigorous way using the quantization condition derived in this work, and is an excellent candidate for the first lattice QCD study of a three-particle resonance.

Another feature of interest in these examples is the presence of the $\rho$ resonance in two-particle subchannels. Although the decay $\omega \to \rho\pi$ is kinematically forbidden, we expect, given the width of the resonance, that it will have a significant impact on the energy levels in the vicinity of the $\omega$ mass. For the $h_1$, the $\rho\pi$ decay is allowed (and seen experimentally), and thus the system provides an example in which the full complication of cascading resonant decays, $h_1 \to \rho\pi \to 3\pi$, occurs. We also note that, away from the three-particle resonance energy, the dominant effect on the three-pion spectrum arises from pairwise interactions, and thus this spectrum provides an alternative source of information on the $\rho$ resonance. Indeed, the effect on the three-particle spectrum is enhanced relative to that for two pions due to the presence of three pairs.

The implementation of the isoscalar three-particle quantization condition requires only minor generalizations of the $I_{\pi\pi\pi} = 3$ case implemented previously in refs. [20, 22–24]. Specifically, appendices A and B of ref. [22] provide a summary of all necessary results. The new features here are two-fold: (i) the expression for $F_3$ contains a relative minus sign for $G$ compared to that for $I_{\pi\pi\pi} = 3$ (see table 1), which is trivial to implement; (ii) the angular momentum indices $\ell, m$ of the interacting pair contain only odd partial waves. Concerning the latter point, in our illustrative example we restrict to the lowest allowed partial wave, namely $\ell = 1$. While odd two-particle partial waves have not previously been implemented in the three-particle quantization condition, this requires only a simple generalization from the work in ref. [22], where $\ell = 0$ and 2 were considered. In particular, we follow that work in using real spherical harmonics, and in the method of projection onto different irreps of the cubic group.

We now describe how the resonances are included in our example. We stress at the outset that the parameters we choose are not intended to be close to those for the physical particles, but rather are choices that allow certain features of the resulting spectrum to be clearly seen. For the $\rho$, we use the Breit-Wigner parametrization:

$$\left(\frac{k}{M_\pi}\right)^3 \cot\delta_1 = \frac{M_\rho^2 - E^2}{E M_\pi} \frac{6\pi}{g^2} \frac{E^2}{M_\rho^2}, \tag{4.1}$$

with $g = 1$ and $M_\rho = 2.8 M_\pi$.[12] As explained in ref. [23], in order for the three-particle

---

[12]Our chosen value of $M_\rho/M_\pi$ corresponds to a theory with $M_\pi \sim 320\,\mathrm{MeV}$ (see ref. [41]). Our choice of the coupling $g$ is, however, significantly smaller than the observed value (corresponding to a narrower-than-physical decay width).

quantization condition to remain valid in the presence of two-particle resonances, we must use a modified principal value prescription. This requires the following changes to $\widetilde{F}$ and $\widetilde{\mathcal{K}}_2$:

$$[F]_{k\ell'm';p\ell m} \to [F]_{k\ell'm';p\ell m} + \delta_{kp}\delta_{\ell'\ell}\delta_{m'm}H(\boldsymbol{k})\frac{I_{\mathrm{PV}}^{(\ell)}(q_{2,k}^{\star 2})}{32\pi}\,, \tag{4.2}$$

$$\left[(\mathcal{K}_2)^{-1}\right]_{k\ell'm';p\ell m} \to \left[(\mathcal{K}_2)^{-1}\right]_{k\ell'm';p\ell m} - \delta_{kp}\delta_{\ell'\ell}\delta_{m'm}H(\boldsymbol{k})\frac{I_{\mathrm{PV}}^{(\ell)}(q_{2,k}^{\star 2})}{32\pi}\,, \tag{4.3}$$

where $\ell$ and $\ell'$ are odd, and in this case $\ell = \ell' = 1$. We find that $I_{\mathrm{PV}}^{(\ell=1)}(q) = C/q^2$, with $C \lesssim -50M_\pi^2$ is enough to accommodate any resonance in the region $M_\rho < 5M_\pi$.[13]

For the three-particle resonances, we use the general form given in eq. (3.31) for $\mathcal{K}_{\mathrm{df},3}$, with the specific momentum-dependent expressions for $\mathcal{K}_{\mathrm{df},3}^\omega$ and $\mathcal{K}_{\mathrm{df},3}^{h_1}$ given in eqs. (3.34) and (3.35), respectively. We choose $C = -100M_\pi^2$, and set $M_\omega = 4.3M_\pi$, $M_{h_1} = 4.7M_\pi$, $c_\omega = 0.02$, and $c_{h_1} = 0.42$. These choices are motivated by the hierarchy of the resonance parameters known from experiment, i.e., $M_{h_1} > M_\omega$, $\Gamma_{h_1} > \Gamma_\omega$. We stress, however, that we do not at present know how to relate the parameters $c_X$ to the physical width, and that these values are chosen only for illustrative purposes.

The resulting three-pion spectra for two different irreps, $T_1^\mp$, are shown in figure 4 as a function of $M_\pi L$. As described in section 3.2, these irreps couple to resonances with $J^P = 1^\mp$, i.e. to the $\omega$ and $h_1$ channels, respectively. For comparison, we include noninteracting energies for the finite-volume $3\pi$, $\rho\pi$, and $\omega/h_1$ states. The actual spectral lines show significant shifts from the noninteracting levels, as well as the usual pattern of avoided level crossings. For our choice of parameters of the $\omega$ and $h_1$, the avoided level crossings are quite narrow. This could be a result of the resonance being narrow, or a volume suppression of the gap in the avoided level crossings.

Moreover, the finite-volume state related to the toy $h_1$ is significantly shifted with respect to the position of the pole in $\mathcal{K}_{\mathrm{df},3}$. To further investigate this feature, in figure 5 we study the effects of varying $c_{h_1}$ [the residue of the pole in $\mathcal{K}_{\mathrm{df},3}$] as well as $C$ [parametrizing the scheme dependence in eqs. (4.2) and (4.3)]. We stress that $C$ ultimately encodes a scheme dependence of $\mathcal{K}_{\mathrm{df},3}$, in that one can vary $C$ and $\mathcal{K}_{\mathrm{df},3}$ simultaneously to keep the finite-volume spectrum and the three-particle scattering amplitude unchanged. It follows that varying $C$ at fixed $\mathcal{K}_{\mathrm{df},3}$ corresponds to a change in the physical system, so that the finite-volume energies should also shift. In short, the four panels of figure 5 correspond to four different physical systems with the common feature that $\mathcal{K}_{\mathrm{df},3}$, in some given scheme, has the $h_1$ pole position. We find that the position of interacting levels moves closer to the pole position (horizontal dashed line) when either $c_{h_1}$ or $C$ is reduced. This shows that the large shift in figure 4 is a result of the specific parameters chosen, and not a general

---

[13]A technical aspect of our numerical implication is that the matices $F$, $G$ and $\mathcal{K}_2$ are truncated slightly before $H(\boldsymbol{k}) = 0$, by already discarding entries for which $H(\boldsymbol{k}) \lesssim 10^{-8}$. This corresponds to truncating values of $E_{2,k}^{\star 2}$ slightly above zero and is required because the boost factor $\gamma_k = (E - \omega_k)/E_{2,k}^\star$ [also defined in eq. (B.4) below] can become arbitrarily enhanced for near-zero values, leading to numerical instabilities. In the present case this cut also serves to avoid the unphysical pole in $\mathcal{K}_2$ [due to the $1/E$ term in eq. (4.1)], which is present even after the $I_{\mathrm{PV}}$ shift is applied.

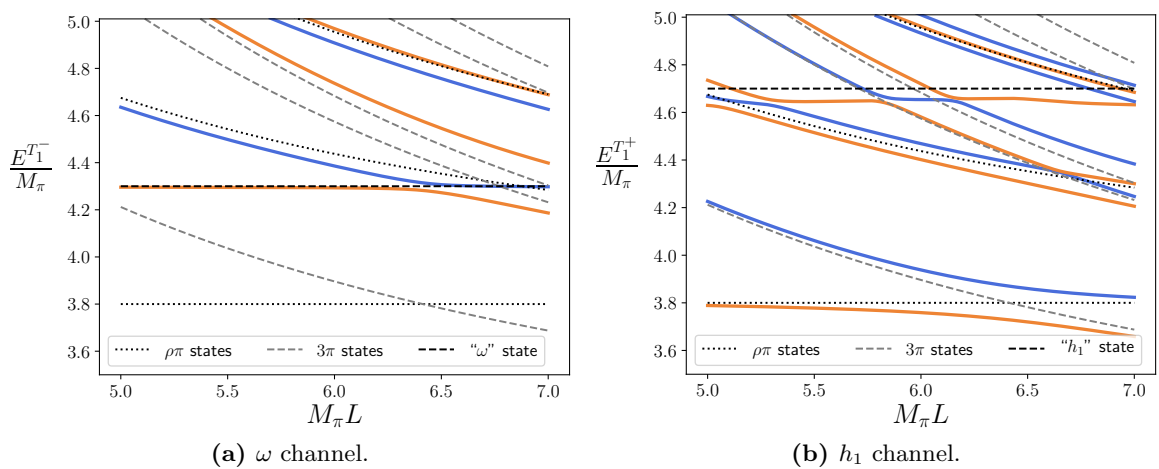

**(a)** $\omega$ channel.

**(b)** $h_1$ channel.

**Figure 4**. Illustrative finite-volume spectra for three pions with $I_{\pi\pi\pi} = 0$ and irreps (a) $T_1^-$ and (b) $T_1^+$, plotted versus $M_\pi L$. The interacting spectrum is shown by solid lines, with the alternating orange and blue colors only used to distinguish adjacent levels. Dashed and dotted grey lines show the comparison with different noninteracting levels. The parameters used for $\mathcal{K}_2$ and $\mathcal{K}_{\mathrm{df},3}$ are described in the text.

feature of the system considered. Clearly, future work is needed to fully understand the interplay of $\mathcal{K}_{\mathrm{df},3}$ with the physical resonance parameters and the finite-volume energies.

Finally, we comment that the smaller number of observed levels in the $T_1^-$ plot, as compared to the $T_1^+$, can be understood in terms of the antisymmetry of the momentum wavefunctions — as discussed in appendix D. Indeed, one can understand precisely the counting of levels in both plots, as we explain in that appendix.

## 5 Conclusion

This work constitutes the first extension of the finite-volume three-particle formalism to include nonidentical particles. We have focused on the description of a generic three-pion system in QCD with exact isospin symmetry. The main difference with the original quantization condition of refs. [17, 18] is that there are different subchannels for pairwise interactions ($I_{\pi\pi} = 0, 1, 2$) that must be taken into account. The new three-particle quantization condition, and the infinite-volume three-particle integral equations, look formally identical to those for identical particles, but live in an enlarged matrix space with additional flavor indices. The central point of this work is to give the explicit forms of all building blocks in this enlarged space, and to outline a strategy for extracting three-pion scattering amplitudes, in both weakly-interacting and resonant systems, for all possible quantum numbers.

As described in section 2, to carry out the derivation it is convenient to first generalize the quantization condition using the basis with definite individual pion flavors. The final result is then block-diagonalized by performing a standard change of basis in flavor space, with the resulting blocks labeled by the three-pion isospin $I_{\pi\pi\pi} = 0 - 3$, and the elements

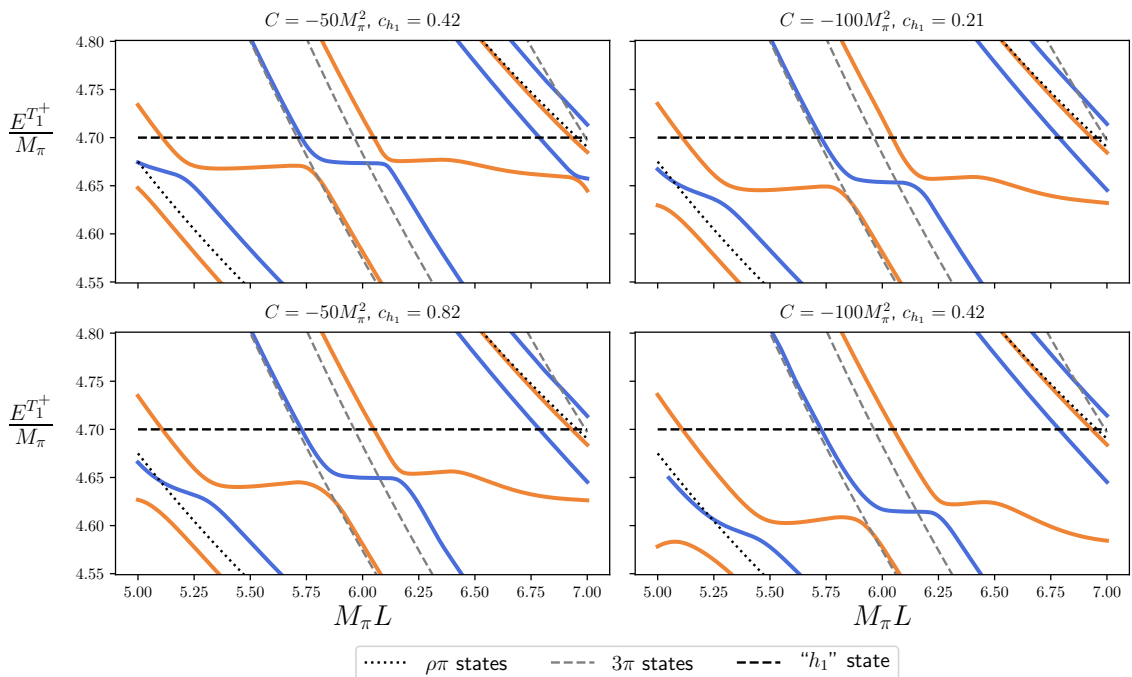

**Figure 5**. Finite-volume energies for various scattering parameters in the $T_1^+$ irrep, zoomed in to focus on energies close to the toy $h_1$ resonance. As explained in the text, changing either $C$ or $c_{h_1}$, changes the physical three-particle scattering amplitude while leaving the pole in $\mathcal{K}_{\mathrm{df},3}$ fixed. The bottom right panel corresponds to the parameters of the figure 4b.

within each block labeled by the allowed values of incoming and outgoing two-pion isospin $I_{\pi\pi}$. In this way, the three-pion quantization condition turns into a set of four independent expressions, to be applied separately to finite-volume energies with the corresponding quantum numbers. The $I_{\pi\pi\pi} = 3$ quantization condition is the same as that for three identical (pseudo-)scalars derived in refs. [17, 18], while those for $I_{\pi\pi} = 0, 1, 2$ are new. The implementation of the new quantization conditions is of similar complexity to the $I_{\pi\pi\pi} = 3$ case, where there have been extensive previous studies [20, 22–24]. They do, however, exhibit some new features, such as the presence of odd partial waves and different relative signs between the finite-volume objects involved.

In section 3, we also have addressed the parametrization of $\mathcal{K}_{\mathrm{df},3}$ in a general isospin channel, which is a crucial point for the extraction of three-particle scattering amplitudes from lattice QCD. First, we have extended the threshold expansion of $\mathcal{K}_{\mathrm{df},3}$ to all values of $I_{\pi\pi\pi}$. This is a series expansion about threshold based on symmetry properties of $\mathcal{K}_{\mathrm{df},3}$: Lorentz invariance, CPT and particle exchange. We have worked out the first few terms for all isospin channels. In addition, we propose parametrizations of $\mathcal{K}_{\mathrm{df},3}$ to describe all three-particle resonances present in the $I_{\pi\pi\pi} = 0$ and 1 channels. These generate an additional state in the spectrum, which decouples in the limit of zero coupling.

Given these results, all ingredients are now available for lattice studies of resonances with three-particle decay channels, such as the $\omega(782)$ and the $h_1(1170)$. These two $I_{\pi\pi\pi} = 0$ resonances are particularly good candidates for a first study, as they lie be-

– 36 –

low the $5M_\pi$ threshold for slightly heavier-than-physical pions. In section 4 we use the new quantization condition to determine the finite-volume spectrum for these two channels in a toy model motivated by the experimentally observed hierarchies of masses and widths. Our exploration suggests that, in practice, moving frames will be needed to gain insight in the nature of the resonances, especially in the case of the $\omega(782)$. We stress, however, not yet established how the parameters of $\mathcal{K}_{\mathrm{df},3}$ relate to the physical masses and widths of the resonances and thus more investigation is needed.

Going forward, the next steps fall into three basic categories. First, it would be instructive to study various limiting cases, in order to provide useful crosschecks and gain insights into the structure of the new quantization conditions. One concrete example would be to study the $I_{\pi\pi\pi} = 2$ expressions, continued to parameters such that the $\rho$ resonance becomes a stable particle. In this case one can restrict to the energy regime $M_\rho + M_\pi < \sqrt{s} < 3M_\pi$, and the result should coincide with the two-particle, finite-volume formalism for vector-scalar scattering [43], already used to analyze finite-volume energies in ref. [38]. Second, it is necessary to further generalize the formalism, so as to describe all possible systems of two- and three-particles with generic interactions, quantum numbers, and degrees of freedom. Specific cases, ranked from most straightforward to most difficult, include three pseudoscalar particles in $\mathrm{SU}(N_f)$-symmetric QCD, three-nucleon systems (i.e. the inclusion of spin) and, by far the most challenging, $N\pi \to N\pi\pi$ transitions in the Roper channel (requiring spin, $2 \to 3$ transitions, and non-identical and non-degenerate particles). Finally, and most importantly, the application of this formalism to three-pion resonances using lattice QCD is now well within reach. This will represent the achievement of a long-standing milestone on the way towards unlocking the exotic excitations of the strong force.

## Acknowledgments

## A    Further details of the derivation

In this appendix we provide more details of the derivation of the result for the generalized finite-volume correlator, eq. (2.42). As noted in the main text, most of the steps in the

original derivation of ref. [17] go through, with the only change being the need to generalize the core quantities $F$, $G$ and $\mathcal{K}_2$ in the presence of flavor [using the definitions of eqs. (2.26), (2.32) and (2.41)]. In other words, almost all of the equations in ref. [17] can be taken over unchanged as long as one adds flavor indices and uses the new definitions. There is, however, one step in the derivation that needs further generalization, as we now explain.

The most challenging part of the derivation of ref. [17] is to show that $\mathcal{K}_{\mathrm{df},3}$ has the appropriate symmetry. Since the symmetrization procedure must be generalized here, as described in section 2.3, a natural question is whether the derivation of the quantization condition in the presence of flavor leads to the appropriately symmetrized version of $\mathcal{K}_{\mathrm{df},3}$, denoted $\mathbf{K}_{\mathrm{df},3}$. A second aim of this appendix is to explain why this is indeed the case.

For the sake of brevity, we assume that the reader has a copy of ref. [17] in front of them and we do not repeat equations from that work. We refer to equations from ref. [17] as (HS1), (HS2), etc.[14]

The first place in ref. [17] where the discussion does not generalize in a simple way is in the discussion between (HS140) and (HS146). This concerns the introduction of quantities with a superscript $(s)$, e.g. $A'^{(1,s)}$ in (HS140). These are to be contrasted with quantities having a superscript $(u)$, such as $\mathbf{D}^{(u,u)}$ in eq. (2.47). For the latter quantities, the matrix index $k$ corresponds to the spectator momentum, while for quantities with superscript $(s)$, $k$ labels the momentum of one of the nonspectator pair. To be more precise, in the symmetrization described in eq. (2.51), the choice $\mathcal{P}_3 = \{\boldsymbol{k}, \boldsymbol{a}\}$ from eq. (2.52) corresponds to a $(u)$ quantity, while that with $\mathcal{P}_3 = \{\boldsymbol{a}, \boldsymbol{b}\}$ corresponds to an $(s)$ quantity. The third choice, $\mathcal{P}_3 = \{\boldsymbol{b}, \boldsymbol{k}\}$, leads to quantities denoted by $(\tilde{s})$ in ref. [17]. These three choices are illustrated in figure 13b of ref. [17].

We choose our flavor generalizations of $A'^{(1,u)}$ and $A'^{(1,s)}$ such that (HS140) maintains its form, becoming[15]

$$\mathbf{A}_L'^{(2,u)} = \mathbf{A}'^{(2,u)} + 2\mathbf{A}'^{(1,s)} \, \mathbf{F} \, \mathbf{K}_2 \,. \tag{A.1}$$

With this choice, the coupling of flavor and momentum labels is automatically maintained. For example, in the product $[\mathbf{A}'^{(1,s)}]_{ij}[\mathbf{F}]_{jl}$, if $j = 2$, corresponding to $\widetilde{\pi}_0(a)\widetilde{\pi}_-(b)\widetilde{\pi}_+(k)$, then the spectator attaching to the endcap has momentum $\boldsymbol{a}$ and is a neutral pion. Thus no additional permutation matrix is needed. With this choice the symmetrized endcap is simply given by[16]

$$\mathbf{A}' = \mathbf{A}'^{(u)} + \mathbf{A}'^{(s)} + \mathbf{A}'^{(\tilde{s})} \,. \tag{A.2}$$

---

[14]Some aspects of the derivation of ref. [17] were streamlined in ref. [21], which generalized the derivation to include a K-matrix pole. We do not refer to the latter work, however, since the notation therein is quite involved, as there is an additional channel needed for the K-matrix pole, which is not relevant here. In any case, our aim is not to repeat the derivation, but rather to describe how it can be taken over wholesale. The more pedestrian approach of ref. [17] is adequate for this purpose.

[15]The numerical superscripts indicate the order in an expansion in numbers of "switch states". The details, described in ref. [17], are not important for the present discussion.

[16]A potentially confusing issue is why there are only three terms in the symmetrization sums, as opposed to six, the number of permutations of the three momenta. In other words, why is it sufficient to have one contribution from each of the different choices of spectator momenta, while the order of the nonspectator momenta is irrelevant? In the case of three neutral pions ($j = 4$) this is because the amplitude is symmetric under exchange of the nonspectator pair. For other choices of the flavor index $j$, the two pions in the

Here we are considering endcaps obtained by summing to all orders in perturbation theory, and thus there is no numerical superscript. In this notation the complete endcap appearing in the main text is $\mathbf{A}'_3 = \boldsymbol{\sigma} + \mathbf{A}'$ [see, e.g., eq. (2.42)].

Now we come to the core issue of this appendix. The derivation of ref. [17] produces, in many places,[17] the combination $\mathbf{A}'^{(u)} + 2\mathbf{A}'^{(s)}$, rather than the desired symmetric quantity $\mathbf{A}'$. The key results needed to allow symmetrization generalize here to

$$\left\{\mathbf{A}'^{(u)} + 2\mathbf{A}'^{(s)}\right\}\mathbf{F}\mathbf{A}^{(u)} = \mathbf{A}'\mathbf{F}\mathbf{A}^{(u)} \qquad \Leftrightarrow \qquad \mathbf{A}'^{(s)}\mathbf{F}\mathbf{A}^{(u)} = \mathbf{A}'^{(\tilde{s})}\mathbf{F}\mathbf{A}^{(u)}, \qquad (\text{A.3})$$

$$\left\{\mathbf{A}'^{(u)} + 2\mathbf{A}'^{(s)}\right\}\mathbf{F}\mathbf{A} = \mathbf{A}'\mathbf{F}\mathbf{A} \qquad \Leftrightarrow \qquad \mathbf{A}'^{(s)}\mathbf{F}\mathbf{A} = \mathbf{A}'^{(\tilde{s})}\mathbf{F}\mathbf{A}. \qquad (\text{A.4})$$

In each line, the two forms are algebraically equivalent, and we will demonstrate the second forms. The argument for (the ungeneralized form of) these results given in ref. [17] applies only for identical particles. Here we give the generalization.

In both eqs. (A.3) and (A.4) there is an implicit sum over the flavor indices. The matrix $\mathbf{F}$ is diagonal in flavor [see eq. (2.26)], so the right-hand flavor index of the left endcap and the left-hand flavor index of the right endcap are the same, and we call this common index $j$. In the all-neutral case, $j = 4$, the arguments of ref. [17] hold and demonstrate the equalities. For other choices, the equalities hold only after summing over the pairs of values of $j$ that are related by interchanging the first two pions, i.e. $j = \{1, 2\}$, $\{3, 5\}$ and $\{6, 7\}$. For each of these pairs, we denote the two values as $j_1$ and $j_2$. The new results that are needed are

$$(\mathbf{A}^{(u)}_{j_1 i})_{k\ell m} = (-1)^\ell (\mathbf{A}^{(u)}_{j_2 i})_{k\ell m}, \qquad (\text{A.5})$$

$$(\mathbf{A}'^{(s)}_{ij_1})_{k\ell'm'} = (-1)^{\ell'} (\mathbf{A}'^{(\tilde{s})}_{ij_2})_{k\ell'm'}, \qquad (\text{A.6})$$

as well as a result derived in ref. [17],

$$(-1)^{\ell'} F_{k'\ell'm';k\ell m}(-1)^\ell = F_{k'\ell'm';k\ell m}, \qquad (\text{A.7})$$

using which it is simple to derive eqs. (A.3) and (A.4).

We discuss eqs. (A.5)–(A.7) in turn. Note that in the first two of these equations, the flavor label $i$ plays no role. What eq. (A.5) states is that, if we interchange the momenta $\boldsymbol{a}$ and $\boldsymbol{b}$, and interchange the flavors $j_1$ and $j_2$, then we obtain the same amplitude. The factor of $(-1)^\ell$ arises because we are decomposing into spherical harmonics with respect to $\widehat{\boldsymbol{a}}^\star$ on the left-hand side and $\widehat{\boldsymbol{b}}^\star$ on the right-hand side, corresponding to a parity flip in the CMF of the nonspectator pair. The same explanation holds for eq. (A.6), except here there is the additional feature that interchanging $\boldsymbol{a}$ and $\boldsymbol{b}$ also interchanges $(s)$ and $(\tilde{s})$. Finally, eq. (A.7) encodes the statement that $F$ vanishes (up to exponentially suppressed corrections) unless $\ell + \ell'$ is even.

---

nonspectator pair have different charges, and their order has no meaning in the context of a Feynman diagram, as long as we associate a given momentum label always with a given flavor, as is the case here.

[17]Strictly speaking, these quantities should have a common numerical superscript indicating the order in the expansion in switch states, but this plays no role in the present derivation, so we drop it for the sake of brevity.

The remainder of the derivation in ref. [17] generalizes step by step in the presence of flavor. Each equation holds when the original quantities are replaced by their flavored (bold faced) generalizations (taking into account the factors of $i$ and $2\omega L^3$ absorbed into the bold faced definitions). No new results are needed. For example, the key result given in (HS196)–(HS198), which is also crucial to allow symmetrization, carries over verbatim for each choice of flavor indices. Also, the complicated steps in (HS213)–(HS239), which result in a symmetrized $\mathcal{K}_{df,3}$, carry over and (using the key results given above) lead to a $\mathbf{K}_{df,3}$ with exactly the generalized symmetry properties described in section 2.3. Finally we note that the inclusion of the generalized three-particle Bethe-Salpeter kernel, $\mathbf{B}_3$, also follows the same steps as in section IV.E of ref. [17], because $\mathbf{B}_3$ has the same symmetry properties as $\boldsymbol{\sigma}$, namely those of $\mathbf{M}_3$.

## B  Building blocks of the quantization condition

This appendix provides a self-contained collection of all necessary definitions to implement the three-particle quantization condition.

First, we define the cutoff function:

$$H(\boldsymbol{k}) = J(z), \qquad z = \frac{E_{2,k}^{\star 2} - (1 + \alpha_H)m^2}{(3 - \alpha_H)m^2}, \tag{B.1}$$

$$J(z) = \begin{cases} 0, & z \le 0, \\ \exp\left(-\frac{1}{z}\exp\left[-\frac{1}{1-z}\right]\right), & 0 < z < 1, \\ 1, & 1 \le z, \end{cases} \tag{B.2}$$

where $E_{2,k}^{\star 2} = (E - \omega_k)^2 - (\boldsymbol{P} - \boldsymbol{k})^2$ and $\alpha_H \in [-1, 3)$ a constant that sets the scheme for $\mathcal{K}_{df,3}$ but does not affect the relation between finite-volume energies and the physical amplitude. We typically choose $\alpha_H = -1$, corresponding to the highest cutoff.

For $G$ we use the relativistic form described in ref. [19],

$$G_{p\ell'm';k\ell m}(E, \boldsymbol{P}, L) \equiv \frac{1}{L^3} \frac{H(\boldsymbol{p})H(\boldsymbol{k})}{b^2 - m^2} \frac{4\pi \mathcal{Y}_{\ell'm'}(\boldsymbol{k}^\star)\mathcal{Y}_{\ell m}^*(\boldsymbol{p}^\star)}{q_{2,p}^{\star \ell'} q_{2,k}^{\star \ell}} \frac{1}{2\omega_k}, \tag{B.3}$$

where $b = P - p - k$ is the momentum of the exchanged particle and $q_{2,k}^{\star 2} = E_{2,k}^{\star 2}/4 - m^2$ is the squared back-to-back momentum of the non-$b$ pair in its CMF. We have also used the two-particle CMF quantities $\boldsymbol{p}^\star$ and $\boldsymbol{k}^\star$, defined via

$$\boldsymbol{p}^\star = (\gamma_k - 1)\left(\boldsymbol{p} \cdot (\widehat{\boldsymbol{k-P}})\right)(\widehat{\boldsymbol{k-P}}) + \omega_p \gamma_k \beta_k(\widehat{\boldsymbol{k-P}}) + \boldsymbol{p},$$

$$\beta_k = \frac{|\boldsymbol{P} - \boldsymbol{k}|}{E - \omega_k}, \quad \gamma_k = (1 - \beta_k^2)^{-1/2}, \tag{B.4}$$

where $\hat{\boldsymbol{x}} = \boldsymbol{x}/|\boldsymbol{x}|$. The definition for $\boldsymbol{k}^\star$ is given by exchanging $\boldsymbol{p} \leftrightarrow \boldsymbol{k}$ everywhere. Finally, $\mathcal{Y}_{\ell m}(\boldsymbol{k})$ are harmonic polynomials,

$$\mathcal{Y}_{\ell m}(\boldsymbol{k}) \equiv |\boldsymbol{k}|^\ell Y_{\ell m}(\hat{\boldsymbol{k}}), \tag{B.5}$$

where $Y_{\ell m}$ are the spherical harmonics. In practice, it is more convenient to use the real spherical harmonics, as discussed in ref. [22].

Next,

$$F_{k'\ell'm';k\ell m}(E, \boldsymbol{P}, L) \equiv \delta_{k'k} F_{\ell'm',\ell m}(\boldsymbol{k}) \,, \tag{B.6}$$

where $F(\boldsymbol{k})$ is a sum-integral difference that is proportional to the zeta functions that appear in the two-particle quantization condition [1, 2]. This object also depends on $(E, \boldsymbol{P}, L)$ but we leave this implicit, focusing on the role of the spectator momentum. $F(\boldsymbol{k})$ requires ultraviolet (UV) regularization, and can be written in various forms that are equivalent up to exponentially-suppressed corrections. The original form, presented in ref. [17], uses a product of $H$ functions as a UV regulator. Here, we give a different form that is simpler to evaluate numerically. Following ref. [4], we write

$$F_{\ell'm';\ell m}(\boldsymbol{k}) = \frac{1}{16\pi^2 L(E-\omega_k)} \left[ \sum_{\boldsymbol{n}_a} -\mathrm{PV} \int d^3\boldsymbol{n}_a \right] \frac{e^{\alpha(x^2-r^2)}}{x^2-r^2} \frac{4\pi \mathcal{Y}_{\ell'm'}(\boldsymbol{r}) \mathcal{Y}_{\ell m}^*(\boldsymbol{r})}{x^{\ell'+\ell}} \,, \tag{B.7}$$

where $\boldsymbol{n}_a = \boldsymbol{a}L/(2\pi)$, $x = q_{2,k}^\star L/(2\pi)$, and

$$\boldsymbol{r}(\boldsymbol{n}_k, \boldsymbol{n}_a) = \boldsymbol{n}_a + \boldsymbol{n}_{kP} \left[ \frac{\boldsymbol{n}_a \cdot \boldsymbol{n}_{kP}}{n_{kP}^2} \left( \frac{1}{\gamma_k} - 1 \right) + \frac{1}{2\gamma_k} \right] \,, \tag{B.8}$$

with $\boldsymbol{k} - \boldsymbol{P} = \boldsymbol{n}_{kP}(2\pi/L)$, and $\gamma_k$ as in eq. (B.4). The UV regularization is now provided by the exponential in the integrand with $\alpha > 0$. The $\alpha$ dependence is exponentially suppressed in $L$ but can become numerically significant if $\alpha$ is taken too large. We find that $\alpha \lesssim 0.5$ is usually sufficient. In this regularization, the integral can be performed analytically, as explained in appendix B of ref. [22].

Finally, we turn to $\mathcal{K}_2$, which is a diagonal matrix:

$$\left[ \frac{1}{\mathcal{K}_2} \right]_{p\ell'm';k\ell m} = \delta_{pk} \delta_{\ell'\ell} \delta_{m'm} \frac{1}{\mathcal{K}_{2;k}^{(\ell)}} \,, \tag{B.9}$$

$$\frac{1}{\mathcal{K}_{2;k}^{(\ell)}} = \frac{1}{16\pi E_{2,k}^\star} \left\{ q_{2,k}^\star \cot \delta_\ell(q_{2,k}^\star) + |q_{2,k}^\star| [1 - H(\boldsymbol{k})] \right\} \,, \tag{B.10}$$

where $\delta_\ell(q_{2,k}^\star)$ is the two-particle phase-shift in the $\ell$th partial wave.

## C  Three-pion states

We collect in this appendix some additional details concerning the basis we use for the neutral three-pion states. The first two pions are combined into a state of definite isospin. The $I_{\pi\pi} = 2$, 1 and 0 states are denoted $(\pi\pi)_2^q$, $\rho^q$, and $\sigma$, respectively, with $q$ the charge. The two-pion state is then combined with the remaining pion to create a state of total

isospin $I_{\pi\pi\pi}$ (denoted by a subscript on the kets listed below). This leads to

$$|(\pi\pi)_2\pi\rangle_3 = \frac{1}{\sqrt{5}} \left( |(\pi\pi)_2^+\pi_-\rangle + \sqrt{3}\,|(\pi\pi)_2^0\pi_0\rangle + |(\pi\pi)_2^-\pi_+\rangle \right), \tag{C.1}$$

$$|(\pi\pi)_2\pi\rangle_2 = \frac{1}{\sqrt{2}} \left( |(\pi\pi)_2^+\pi_-\rangle - |(\pi\pi)_2^-\pi_+\rangle \right), \tag{C.2}$$

$$|\rho\pi\rangle_2 = \frac{1}{\sqrt{6}} \left( |\rho^+\pi_-\rangle + 2\,|\rho^0\pi_0\rangle + |\rho^-\pi_+\rangle \right), \tag{C.3}$$

$$|(\pi\pi)_2\pi\rangle_1 = \frac{1}{\sqrt{10}} \left( \sqrt{3}\,|(\pi\pi)_2^+\pi_-\rangle - 2\,|(\pi\pi)_2^0\pi_0\rangle + \sqrt{3}\,|(\pi\pi)_2^-\pi_+\rangle \right), \tag{C.4}$$

$$|\rho\pi\rangle_1 = \frac{1}{\sqrt{2}} \left( |\rho^+\pi_-\rangle - |\rho^-\pi_+\rangle \right), \tag{C.5}$$

$$|\sigma\pi\rangle_1 = |\sigma\pi_0\rangle, \tag{C.6}$$

$$|\rho\pi\rangle_0 = \frac{1}{\sqrt{3}} \left( |\rho^+\pi_-\rangle - |\rho^0\pi_0\rangle + |\rho^-\pi_+\rangle \right). \tag{C.7}$$

The right-hand sides can be further decomposed into the $|\pi\pi\pi\rangle$ basis used in the main text, resulting in eqs. (2.56) and (2.57).

We make extensive use of the irreducible representations (irreps) of the symmetry group $S_3$, which describes permutations of three objects. It has 6 elements, divided into three conjugacy classes as

$$\{(1)\}, \; \{(12), \, (23), \, (13)\} \; \text{and} \; \{(231), \, (321)\}. \tag{C.8}$$

The three irreps are as follows.

1. The trivial representation, with all elements being the identity. States transforming according this irrep are denoted $|\chi_s\rangle$.

2. The sign or alternating representation:

$$\begin{aligned}
(1), \, (231), \, (312) &\to +1, \\
(12), \, (23), \, (13) &\to -1.
\end{aligned} \tag{C.9}$$

   States transforming according to this irrep are denoted $|\chi_a\rangle$.

3. The standard representation, which is two dimensional. A convenient choice of basis vectors, denoted $|\chi_1\rangle$ and $|\chi_2\rangle$, leads to:

$$(1) \to \begin{pmatrix} 1 & 0 \\ 0 & 1 \end{pmatrix}, \qquad (12) \to \begin{pmatrix} 1 & 0 \\ 0 & -1 \end{pmatrix}, \qquad (13) \to \frac{1}{2} \begin{pmatrix} -1 & -\sqrt{3} \\ -\sqrt{3} & 1 \end{pmatrix},$$

$$(23) \to \frac{1}{2} \begin{pmatrix} -1 & \sqrt{3} \\ \sqrt{3} & 1 \end{pmatrix}, \quad (231) \to \frac{1}{2} \begin{pmatrix} -1 & \sqrt{3} \\ -\sqrt{3} & -1 \end{pmatrix}, \quad (312) \to \frac{1}{2} \begin{pmatrix} -1 & -\sqrt{3} \\ \sqrt{3} & -1 \end{pmatrix}. \tag{C.10}$$

The three-pion states listed above can be classified according to their transformations under permutations. The $I_{\pi\pi\pi} = 3$ state transforms in the symmetric irrep, the $I_{\pi\pi\pi} = 2$ states in the standard irrep, the $I_{\pi\pi\pi} = 1$ states in a direct sum of the symmetric and standard irreps, and the $I_{\pi\pi\pi} = 0$ state in the sign irrep. The linear combinations that lie in the permutation-group irreps are (with the subscript on the ket again denoting isospin)

$$|\chi_s\rangle_3 = |(\pi\pi)_2\pi\rangle_3\,, \tag{C.11}$$

$$|\chi_1\rangle_2 = |(\pi\pi)_2\pi\rangle_2\,, \tag{C.12}$$

$$|\chi_2\rangle_2 = |\rho\pi\rangle_2\,, \tag{C.13}$$

$$|\chi_1\rangle_1 = -\frac{\sqrt{5}}{3}|(\pi\pi)_2\pi\rangle_1 + \frac{2}{3}|\sigma\pi\rangle_1\,, \tag{C.14}$$

$$\quad = \frac{1}{\sqrt{12}}\Big(2\,|\pi_+,\pi_-,\pi_0\rangle + 2\,|\pi_-,\pi_+,\pi_0\rangle - |\pi_+,\pi_0,\pi_-\rangle$$
$$\quad\quad - |\pi_0,\pi_+,\pi_-\rangle - |\pi_0,\pi_-,\pi_+\rangle - |\pi_-,\pi_0,\pi_+\rangle\Big)\,, \tag{C.15}$$

$$|\chi_2\rangle_1 = |\rho\pi\rangle_1\,, \tag{C.16}$$

$$|\chi_s\rangle_1 = \frac{2}{3}|(\pi\pi)_2\pi\rangle_1 + \frac{\sqrt{5}}{3}|\sigma\pi\rangle_1\,, \tag{C.17}$$

$$\quad = \frac{1}{\sqrt{15}}\Big(|\pi_+,\pi_-,\pi_0\rangle + |\pi_0,\pi_+,\pi_-\rangle + |\pi_-,\pi_0,\pi_+\rangle + |\pi_-,\pi_+,\pi_0\rangle$$
$$\quad\quad + |\pi_0,\pi_-,\pi_+\rangle + |\pi_+,\pi_0,\pi_-\rangle - 3\,|\pi_0,\pi_0,\pi_0\rangle\Big)\,, \tag{C.18}$$

$$|\chi_a\rangle_0 = |\rho\pi\rangle_0\,. \tag{C.19}$$

# D  Group-theoretic results

In this appendix we collect some group-theoretic results that are relevant for the practical implementation of the quantization condition described in the main text. We restrict our considerations to the overall rest frame, i.e. we set $\boldsymbol{P} = 0$; generalizations to moving frames are straightforward but tedious.

We begin by listing the irreps that are created and annihilated by operators with $(I_{\pi\pi\pi})_z = 0$, having the form of three noninteracting pions, each with a definite momentum. Focusing on annihilation operators, we write

$$\widetilde{\pi}_i(\boldsymbol{a})\widetilde{\pi}_j(\boldsymbol{b})\widetilde{\pi}_k(\boldsymbol{c})\,, \tag{D.1}$$

with $\widetilde{\pi}$ the Fourier transform of some choice of local pion operator. The indices $i, j, k$ denote $(I_\pi)_z$, and the constraint that the total operator is neutral restricts the choices of indices to seven options, as described in appendix C. The momenta are $\boldsymbol{a} = 2\pi\boldsymbol{m}_1/L$, $\boldsymbol{b} = 2\pi\boldsymbol{m}_2/L$, and $\boldsymbol{c} = -\boldsymbol{a}-\boldsymbol{b} = 2\pi\boldsymbol{m}_3/L$. One then projects onto definite isospin using the results given in eqs. (2.56) and (2.57) and appendix C. Operators of this type are typically used as part of the variational basis in lattice QCD calculations, and the energies of the

| orb. | $(\boldsymbol{m}_1^2, \boldsymbol{m}_2^2, \boldsymbol{m}_3^2)$ | dim. | $I_{\pi\pi\pi} = 0$ | $I_{\pi\pi\pi} = 1$ | $I_{\pi\pi\pi} = 2$ | $I_{\pi\pi\pi} = 3$ |
|---|---|---|---|---|---|---|
| 1 | (0,0,0) | 2 | — | $R_2^{(o)} + R_3^{(o)}$ | — | $A_1^-$ |
| 2 | (1,1,0) | 21 | $T_1^+$ | $R_2^{(o)} + R_3^{(o)}$ | $A_1^-, E^-, T_1^+$ | $A_1^-, E^-$ |
| 3 | (2,2,0) | 42 | $T_1^+, T_2^+$ | $R_2^{(o)} + R_3^{(o)}$ | $A_1^-, E^-, T_2^-, T_1^+, T_2^+$ | $A_1^-, E^-, T_2^-$ |
| 4 | (2,1,1) | 84 | $R_0^{(4)}$ | $R_2^{(o)} + R_3^{(o)}$ | $R_2^{(4)}$ | $R_3^{(4)}$ |
| 5 | (3,3,0) | 28 | $A_2^+, T_1^+$ | $R_2^{(o)} + R_3^{(o)}$ | $A_1^-, T_2^-, A_2^+, T_1^+$ | $A_1^-, T_2^-$ |
| 6 | (4,1,1) | 24 | — | $R_2^{(o)} + R_3^{(o)}$ | $A_1^-, E^-, T_1^+$ | $A_1^-, E^-, T_1^+$ |
| 7 | (3,2,1) | 168 | $R_3^{(7)}$ | $R_2^{(o)} + R_3^{(o)}$ | $2\ R_3^{(7)}$ | $R_3^{(7)}$ |
| 16 | (5,3,2) | 336 | $R_3^{(16)}$ | $R_2^{(o)} + R_3^{(o)}$ | $2\ R_3^{(16)}$ | $R_3^{(16)}$ |

**Table 5**. Cubic-group irreps for the three-pion operators with $\boldsymbol{P} = 0$ and total charge zero for isospin $I_{\pi\pi\pi} = 0$, 2 and 3. These results include the intrinsic negative parity of the pions. The operators are those with the lowest seven noninteracting energies for a cubic box with $mL \approx 4$, together with the lowest-lying orbit having the maximal possible dimension. The first column gives the orbit number, $o$, the second specifies the orbit, as described in the text, while the third gives the dimension of the orbit. The remaining columns list the irreps appearing in the orbit, $R_I^{(o)}$. As indicated, results for $I_{\pi\pi\pi} = 1$ are given by summing the irreps in the $I_{\pi\pi\pi} = 2$ and $I_{\pi\pi\pi} = 3$ columns. Entries in the $I_{\pi\pi\pi} = 3$ column agree with those in table 2 of ref. [22] (up to intrinsic parity, which is omitted in the earlier work). The missing entries are $R_0^{(4)} = A_2^+, E^-, T_1^-, T_1^+, T_2^+$, $R_2^{(4)} = A_1^-, A_2^-, 2E^-, T_1^-, T_2^-, 2T_1^+, 2T_2^+$, $R_3^{(4)} = A_1^-, E^-, T_2^-, T_1^+, T_2^+$, $R_3^{(7)} = A_1^-, E^-, T_1^-, 2T_2^-, A_2^+, E^+, 2T_1^+, T_2^+$, and $R_3^{(16)} = A_1^-, A_2^-, 2E^-, 3T_1^-, 3T_2^-, A_1^+, A_2^+, 2E^+, 3T_1^+, 3T_2^+$.

corresponding noninteracting states provide points of comparison for the spectrum of the interacting theory (see, e.g., figure 4).

Each choice of $\boldsymbol{m}_1$ and $\boldsymbol{m}_2$ (which fixes $\boldsymbol{m}_3 = -\boldsymbol{m}_1 - \boldsymbol{m}_2$) is related to some number of other choices by cubic group transformations. We specify the resulting orbit by giving the values of $m_1^2$, $m_2^2$ and $m_3^2$, which provide a unique specification for the examples we consider (although not in general). Each orbit decomposes into irreps of the cubic group, and these are listed in table 5 for the operators coupling to the seven lowest-energy states in the absence of interactions. We recall that the irreps for the 48-element cubic group (including parity) are $A_1^\pm$, $A_2^\pm$, $E^\pm$, $T_1^\pm$ and $T_2^\pm$, with dimensions of $\{1, 1, 2, 3, 3\}$, respectively. The result from appendix C that the $I_{\pi\pi\pi} = 1$ triplets decompose into a trivial singlet and a standard irrep doublet under the permutation group $S_3$, leads to the result shown in the table that the irreps for $I_{\pi\pi\pi} = 1$ are simply the sum of those for $I_{\pi\pi\pi} = 2$ and $I_{\pi\pi\pi} = 3$.

We stress that it is always possible to choose particular linear combinations of operators that pick out each of the irreps in a given orbit. This is very useful in practice as it restricts the number of terms in $\mathbf{K}_{\mathrm{df},3}$ that contribute (see section 3.1), and allows one to consider the resonances discussed in section 3.2 one by one. We note that certain irreps do not appear until quite high orbits, e.g. $A_2^-$ and $T_1^-$ do not appear until the fourth orbit, while

| orb. | $m_\rho^2$ | $m_\pi^2$ | dim. | irreps |
|------|-----------|-----------|------|--------|
| 1 | 0 | 0 | 3 | $T_1^+$ |
| 2 | 1 | 1 | 18 | $A_1^-, E^-, T_1^-, T_2^-, 2T_1^+, T_2^+$ |
| 3 | 2 | 2 | 36 | $A_1^-, A_2^-, 2E^-, 2T_1^-, 2T_2^-, A_2^+, E^+, 3T_1^+, 2T_2^+$ |
| 4 | 3 | 3 | 24 | $A_1^-, E^-, T_1^-, 2T_2^-, A_2^+, E^+, 2T_1^+, T_2^+$ |

**Table 6**. Cubic-group irreps contained in $\rho\pi$ states. The intrinsic negative parity of the pion and the rho are included. Orbits are numbered, and specified by the squares of the momenta, with $\boldsymbol{p}_\rho = 2\pi\boldsymbol{m}_\rho/L$ and $\boldsymbol{p}_\pi = 2\pi\boldsymbol{m}_\pi/L$. The irreps shown are present for each the three allowed isospins, $I_{\rho\pi} = 0$, 1, and 2. The dimensions of the orbits apply separately for each choice of isospin.

$E^+$ and $A_2^+$ do not appear until the seventh. This leaves only $A_1^+$, which does not appear until the sixteenth orbit. This is the lowest "generic" orbit, i.e. one in which all nontrivial cubic-group transformations have vanishing characters.

In order to interpret the interacting spectra in the presence of narrow two-particle resonances, it is also useful to determine which irreps are present assuming that the resonance is a stable particle. In practice, for the energy range of interest, the most important such resonance is the $\rho$, as shown by the examples in figure 4. Thus we have determined the irreps created by $\rho\pi$ operators, treating the $\rho$ as a stable particle with $J^P = 1^-$. There are three isospin combinations with total $(I_{\rho\pi})_z = 0$, and these decompose into total isospin $I_{\rho\pi} = 0$, 1 and 2. Since the $\rho$ and $\pi$ are different particles, the cubic-group irreps that appear are the same for all choices of isospin, and the results for the lowest few momentum orbits are given in table 6. The multiplicities of the $T_1^-$ irrep agree with the results from table 3 of ref. [38].

We can use the results in tables 5 and 6 to understand the level-counting in figure 4, which shows the spectra for $I_{\pi\pi\pi} = 0$ and irreps (a) $T_1^-$ and (b) $T_1^+$. The energies of the second to the sixth noninteracting $3\pi$ orbits are shown in both panels (the first orbit, having $E/m = 3$, lies below the plotted range), as well as the first three noninteracting $\rho\pi$ levels.

For $T_1^-$ (the $\omega$ channel), we see from table 5 that, for the energy range shown in the figure, only the fourth orbit contains this irrep. From table 6, we see that the second and third $\rho\pi$ orbits contain the $T_1^-$, but not the first. In all but one case, there is only a single $T_1^-$ irrep present, the exception being the third $\rho\pi$ orbit, which contains two such irreps. These results are consistent with the interacting energies plotted in figure 4a, which can be interpreted, for $mL \lesssim 6$, as roughly corresponding to the $\omega$ resonance, second $\rho\pi$ orbit, fourth $3\pi$ orbit, and a pair of $\rho\pi$ third orbits.

The results for the $T_1^+$ irrep, displayed in figure 4b, can be interpreted in a similar manner. All the $3\pi$ and $\rho\pi$ orbits shown in the figure contain this irrep, with unit multiplicities except for the second and third $\rho\pi$ orbits, which have multiplicities 2 and 3, respectively. This counting, together with the $h_1$ state, matches that seen in the figure.

## References

[1] M. Lüscher, *Volume Dependence of the Energy Spectrum in Massive Quantum Field Theories. 2. Scattering States*, *Commun. Math. Phys.* **105** (1986) 153 [INSPIRE].

[2] M. Lüscher, *Two particle states on a torus and their relation to the scattering matrix*, *Nucl. Phys.* **B 354** (1991) 531 [INSPIRE].

[3] K. Rummukainen and S.A. Gottlieb, *Resonance scattering phase shifts on a nonrest frame lattice*, *Nucl. Phys.* **B 450** (1995) 397 [hep-lat/9503028] [INSPIRE].

[4] C. Kim, C.T. Sachrajda and S.R. Sharpe, *Finite-volume effects for two-hadron states in moving frames*, *Nucl. Phys.* **B 727** (2005) 218 [hep-lat/0507006] [INSPIRE].

[5] S. He, X. Feng and C. Liu, *Two particle states and the S-matrix elements in multi-channel scattering*, *JHEP* **07** (2005) 011 [hep-lat/0504019] [INSPIRE].

[6] V. Bernard, M. Lage, U.G. Meißner and A. Rusetsky, *Scalar mesons in a finite volume*, *JHEP* **01** (2011) 019 [arXiv:1010.6018] [INSPIRE].

[7] M.T. Hansen and S.R. Sharpe, *Multiple-channel generalization of Lellouch-Lüscher formula*, *Phys. Rev.* **D 86** (2012) 016007 [arXiv:1204.0826] [INSPIRE].

[8] R.A. Briceño and Z. Davoudi, *Moving multichannel systems in a finite volume with application to proton-proton fusion*, *Phys. Rev.* **D 88** (2013) 094507 [arXiv:1204.1110] [INSPIRE].

[9] R.A. Briceño, *Two-particle multichannel systems in a finite volume with arbitrary spin*, *Phys. Rev.* **D 89** (2014) 074507 [arXiv:1401.3312] [INSPIRE].

[10] F. Romero-López, A. Rusetsky and C. Urbach, *Vector particle scattering on the lattice*, *Phys. Rev.* **D 98** (2018) 014503 [arXiv:1802.03458] [INSPIRE].

[11] A.J. Woss, D.J. Wilson and J.J. Dudek, *Efficient solution of the multi-channel Lüscher determinant condition through eigenvalue decomposition*, arXiv:2001.08474 [INSPIRE].

[12] R.A. Briceño, J.J. Dudek and R.D. Young, *Scattering processes and resonances from lattice QCD*, *Rev. Mod. Phys.* **90** (2018) 025001 [arXiv:1706.06223] [INSPIRE].

[13] K. Polejaeva and A. Rusetsky, *Three particles in a finite volume*, *Eur. Phys. J.* **A 48** (2012) 67 [arXiv:1203.1241] [INSPIRE].

[14] P. Guo and V. Gasparian, *An solvable three-body model in finite volume*, *Phys. Lett.* **B 774** (2017) 441 [arXiv:1701.00438] [INSPIRE].

[15] P. Klos, S. König, H.W. Hammer, J.E. Lynn and A. Schwenk, *Signatures of few-body resonances in finite volume*, *Phys. Rev.* **C 98** (2018) 034004 [arXiv:1805.02029] [INSPIRE].

[16] P. Guo, M. Döring and A.P. Szczepaniak, *Variational approach to N-body interactions in finite volume*, *Phys. Rev.* **D 98** (2018) 094502 [arXiv:1810.01261] [INSPIRE].

[17] M.T. Hansen and S.R. Sharpe, *Relativistic, model-independent, three-particle quantization condition*, *Phys. Rev.* **D 90** (2014) 116003 [arXiv:1408.5933] [INSPIRE].

[18] M.T. Hansen and S.R. Sharpe, *Expressing the three-particle finite-volume spectrum in terms of the three-to-three scattering amplitude*, *Phys. Rev.* **D 92** (2015) 114509 [arXiv:1504.04248] [INSPIRE].

[19] R.A. Briceño, M.T. Hansen and S.R. Sharpe, *Relating the finite-volume spectrum and the two-and-three-particle S matrix for relativistic systems of identical scalar particles*, *Phys. Rev.* **D 95** (2017) 074510 [arXiv:1701.07465] [INSPIRE].

[20] R.A. Briceño, M.T. Hansen and S.R. Sharpe, *Numerical study of the relativistic three-body quantization condition in the isotropic approximation*, *Phys. Rev.* **D 98** (2018) 014506 [arXiv:1803.04169] [INSPIRE].

[21] R.A. Briceño, M.T. Hansen and S.R. Sharpe, *Three-particle systems with resonant subprocesses in a finite volume*, *Phys. Rev.* **D 99** (2019) 014516 [arXiv:1810.01429] [INSPIRE].

[22] T.D. Blanton, F. Romero-López and S.R. Sharpe, *Implementing the three-particle quantization condition including higher partial waves*, *JHEP* **03** (2019) 106 [arXiv:1901.07095] [INSPIRE].

[23] F. Romero-López, S.R. Sharpe, T.D. Blanton, R.A. Briceño and M.T. Hansen, *Numerical exploration of three relativistic particles in a finite volume including two-particle resonances and bound states*, *JHEP* **10** (2019) 007 [arXiv:1908.02411] [INSPIRE].

[24] T.D. Blanton, F. Romero-López and S.R. Sharpe, *I = 3 Three-Pion Scattering Amplitude from Lattice QCD*, *Phys. Rev. Lett.* **124** (2020) 032001 [arXiv:1909.02973] [INSPIRE].

[25] H.-W. Hammer, J.-Y. Pang and A. Rusetsky, *Three-particle quantization condition in a finite volume: 1. The role of the three-particle force*, *JHEP* **09** (2017) 109 [arXiv:1706.07700] [INSPIRE].

[26] H.W. Hammer, J.Y. Pang and A. Rusetsky, *Three particle quantization condition in a finite volume: 2. General formalism and the analysis of data*, *JHEP* **10** (2017) 115 [arXiv:1707.02176] [INSPIRE].

[27] M. Döring, H.W. Hammer, M. Mai, J.Y. Pang, A. Rusetsky and J. Wu, *Three-body spectrum in a finite volume: the role of cubic symmetry*, *Phys. Rev.* **D 97** (2018) 114508 [arXiv:1802.03362] [INSPIRE].

[28] J.-Y. Pang, J.-J. Wu, H.W. Hammer, U.-G. Meißner and A. Rusetsky, *Energy shift of the three-particle system in a finite volume*, *Phys. Rev.* **D 99** (2019) 074513 [arXiv:1902.01111] [INSPIRE].

[29] M. Mai and M. Döring, *Three-body Unitarity in the Finite Volume*, *Eur. Phys. J.* **A 53** (2017) 240 [arXiv:1709.08222] [INSPIRE].

[30] M. Mai and M. Döring, *Finite-Volume Spectrum of $\pi^+\pi^+$ and $\pi^+\pi^+\pi^+$ Systems*, *Phys. Rev. Lett.* **122** (2019) 062503 [arXiv:1807.04746] [INSPIRE].

[31] M. Mai, M. Döring, C. Culver and A. Alexandru, *Three-body unitarity versus finite-volume $\pi^+\pi^+\pi^+$ spectrum from lattice QCD*, *Phys. Rev.* **D 101** (2020) 054510 [arXiv:1909.05749] [INSPIRE].

[32] M.T. Hansen and S.R. Sharpe, *Lattice QCD and Three-particle Decays of Resonances*, *Ann. Rev. Nucl. Part. Sci.* **69** (2019) 65 [arXiv:1901.00483] [INSPIRE].

[33] S.R. Beane, W. Detmold, T.C. Luu, K. Orginos, M.J. Savage and A. Torok, *Multi-Pion Systems in Lattice QCD and the Three-Pion Interaction*, *Phys. Rev. Lett.* **100** (2008) 082004 [arXiv:0710.1827] [INSPIRE].

[34] W. Detmold et al., *Multi-Pion States in Lattice QCD and the Charged-Pion Condensate*, *Phys. Rev.* **D 78** (2008) 014507 [arXiv:0803.2728] [INSPIRE].

[35] F. Romero-López, A. Rusetsky and C. Urbach, *Two- and three-body interactions in $\varphi^4$ theory from lattice simulations*, *Eur. Phys. J.* **C 78** (2018) 846 [arXiv:1806.02367] [INSPIRE].

[36] B. Hörz and A. Hanlon, *Two- and three-pion finite-volume spectra at maximal isospin from lattice QCD*, *Phys. Rev. Lett.* **123** (2019) 142002 [arXiv:1905.04277] [INSPIRE].

[37] C. Culver, M. Mai, R. Brett, A. Alexandru and M. Döring, *Three body spectrum from lattice QCD*, arXiv:1911.09047 [INSPIRE].

[38] A. Woss, C.E. Thomas, J.J. Dudek, R.G. Edwards and D.J. Wilson, *Dynamically-coupled partial-waves in $\rho\pi$ isospin-2 scattering from lattice QCD*, *JHEP* **07** (2018) 043 [arXiv:1802.05580] [INSPIRE].

[39] PARTICLE DATA GROUP, *Review of Particle Physics*, *Phys. Rev.* **D 98** (2018) 030001 [INSPIRE].

[40] R.A. Briceño, M.T. Hansen, S.R. Sharpe and A.P. Szczepaniak, *Unitarity of the infinite-volume three-particle scattering amplitude arising from a finite-volume formalism*, *Phys. Rev.* **D 100** (2019) 054508 [arXiv:1905.11188] [INSPIRE].

[41] EXTENDED TWISTED MASS collaboration, *Hadron-Hadron Interactions from $N_f = 2 + 1 + 1$ Lattice QCD: The $\rho$-resonance*, *Eur. Phys. J.* **A 56** (2020) 61 [arXiv:1907.01237] [INSPIRE].

[42] M.T. Hansen and S.R. Sharpe, *Threshold expansion of the three-particle quantization condition*, *Phys. Rev.* **D 93** (2016) 096006 [*Erratum ibid.* **D 96** (2017) 039901] [arXiv:1602.00324] [INSPIRE].

[43] R.A. Briceño and M.T. Hansen, *Multichannel $0 \to 2$ and $1 \to 2$ transition amplitudes for arbitrary spin particles in a finite volume*, *Phys. Rev.* **D 92** (2015) 074509 [arXiv:1502.04314] [INSPIRE].

# Erratum: Generalizing the relativistic quantization condition to include all three-pion isospin channels

Maxwell T. Hansen,[a] Fernando Romero-López[b] and Stephen R. Sharpe[c]

[a] *Theoretical Physics Department, CERN,*
 *Geneva 23 1211, Switzerland*
[b] *IFIC, CSIC-Universitat de València,*
 *Paterna 46980, Spain*
[c] *Physics Department, University of Washington,*
 *Seattle, WA 98195-1560, U.S.A.*

*E-mail:* maxwell.hansen@cern.ch, fernando.romero@uv.es, srsharpe@uw.edu

We have found an error in a statement following eq. (2.5) of our paper, concerning the function $f(a, b, k)$ that first appears in that equation. The error does not affect the main results of the published manuscript but leads to various erroneous statements about auxiliary quantities used in the derivation, specifically the matrices $\sigma(k, a)$, $\sigma^\dagger(k, a)$ as well as the endcap factors $A_3$ and $A'_3$.

The issue arises in the statement that it is convenient to take the function $f(a, b, k)$ to be exchange symmetric with respect to its three arguments. This has the unwanted consequence of making six of the seven operators in the column vector of eq. (2.4) identically equal. This, in turn, implies that many operators are identically zero in the definite isospin basis, considered in section 2.4. To repair this, the last sentence of the paragraph containing eq. (2.4), starting "It is convenient for the subsequent...", should be removed, as should footnote 3 and the next paragraph, beginning with "At this point, the reader may wonder why...". The deleted text should be replaced with the following:

> As we discuss further below, it is crucial that $f(a, b, k)$ is *not* symmetric with respect to permutations of its arguments. More precisely, $f(a, b, k)$ is defined such that all seven operators defining $\mathcal{O}_j(x)$ are in fact distinct, which is necessary to ensure that all definite-isospin operators are non-zero.

The next modification begins with the sentence "The exact relation is $\sigma_{jl}(k,a) = M_{jl}f(a,b,k)$, where" appearing immediately before eq. (2.24). The text and also eq. (2.24) should be removed and replaced with the following:

The exact relation is

$$
\sigma(k,a) = \begin{pmatrix}
f(a,b,k) & f(b,a,k) & f(a,k,b) & 0 & f(b,k,a) & f(k,a,b) & f(k,b,a) \\
f(b,a,k) & f(a,b,k) & f(k,a,b) & 0 & f(k,b,a) & f(a,k,b) & f(b,k,a) \\
f(a,k,b) & f(b,k,a) & f(a,b,k) & 0 & f(b,a,k) & f(k,b,a) & f(k,a,b) \\
0 & 0 & 0 & g(a,b,k) & 0 & 0 & 0 \\
f(k,a,b) & f(k,b,a) & f(b,a,k) & 0 & f(a,b,k) & f(b,k,a) & f(a,k,b) \\
f(b,k,a) & f(a,k,b) & f(k,b,a) & 0 & f(k,a,b) & f(a,b,k) & f(b,a,k) \\
f(k,b,a) & f(k,a,b) & f(b,k,a) & 0 & f(a,k,b) & f(b,a,k) & f(a,b,k)
\end{pmatrix},
$$
(2.24)

where

$$
g(a,b,k) \equiv f(a,b,k) + f(b,a,k) + f(a,k,b) + f(b,k,a) + f(k,a,b) + f(k,b,a)
$$

is the symmetrized version of $f(a,b,k)$. Here the $(i,j)$ entry of the matrix can be understood as the non-interacting overlap of the operator $\mathcal{O}_i(0)$ with the $j$th state. The latter is defined with the convention of eq. (2.4). So, for example, the $(1,3)$ entry follows from

$$
\langle 0|\mathcal{O}_1(0)|\pi\pi\pi, j=3\rangle = \int_{a',b',k'} \langle 0|f(a',b',k') \widetilde{\pi}_-(a') \widetilde{\pi}_0(b') \widetilde{\pi}_+(k')|\pi_-(a)\pi_+(b)\pi_0(k)\rangle,
$$
$$
= f(a,k,b),
$$

where $|\pi\pi\pi, j=3\rangle$ represents the non-interacting state with momentum assignment given by the index. The 6 different terms in the $(4,4)$ entry arise from the 3! contractions of the neutral operator with the neutral state.

In these adjustments the quantity $M$ is no longer needed and is replaced everywhere with $\sigma_{ij}(k,a)$. In addition the parenthetical remark: "(Here and below we use empty and filled squares to present matrices of 0s and 1s as we find this form more readable.)" is now first relevant after eq. (2.38) and should be moved to this location.

Finally, in the paragraph preceding eq. (2.25), the discussion of the factor of 6 should be modified. Starting with the sentence "Second, if both $j, l \neq 4$ then one recovers a non-zero contribution with a factor of. . . ", the remainder of the paragraph should be replaced with the following:

Second, if both $j, l \neq 4$ then one recovers a non-zero contribution with a factor of 6 arising from the contracted matrix indices. For example for $j = 1, l = 2$ one finds

$$
\sum_k \sigma_{1k}(k,a)\sigma_{k2}^\dagger(k,a) = f(a,b,k)f^*(b,a,k) + f(b,a,k)f^*(a,b,k) + f(a,k,b)f^*(k,a,b)
$$
$$
+ f(b,k,a)f^*(k,b,a) + f(k,a,b)f^*(a,k,b) + f(k,b,a)f^*(b,k,a)
$$
$$
\rightarrow 6f(a,b,k)f^*(b,a,k),
$$

where the arrow represents a replacement in the integral that is justified as all other factors are exchange symmetric with respect to $a$, $b$, and $k$. This compensates the $1/6$ pre-factor, leading to the correct expression for a diagram with three distinguishable particles. Finally, $j = l = 4$ yields the diagram with three neutral particles and in this case the $1/6$ survives and correctly gives the symmetry factor for identical particles.

These adjustments complete the redefinitions of $\sigma(k, a)$ and $\sigma^\dagger(k, a)$ and the adjustments to $A_3$ and $A_3'$ follow automatically. All remaining equations and, in particular, all the main results and conclusions of the paper are unchanged. The implicit assumption that the on-diagonal entries of $A_3$ and $A_3'$ are non-zero in the isospin basis, introduced in section 2.4, is now correct.

For completeness, we also note an additional minor correction. The sum following eq. (2.5) should be defined with an implicit factor of $1/L^3$, but this is never stated explicitly. To correct this, the following should be inserted after the sentence containing eq. (2.5): "We also adopt the convention here, and below, that the factor of $1/L^3$ accompanying each sum is left implicit".

**Open Access.** This article is distributed under the terms of the Creative Commons Attribution License ([CC-BY 4.0](CC-BY 4.0)), which permits any use, distribution and reproduction in any medium, provided the original author(s) and source are credited.

# Decay amplitudes to three hadrons from finite-volume matrix elements

Maxwell T. Hansen,[a] Fernando Romero-López[b] and Stephen R. Sharpe[c]

[a]*Higgs Centre for Theoretical Physics, School of Physics and Astronomy,*
 *The University of Edinburgh,*
 *Edinburgh EH9 3FD, U.K.*
[b]*IFIC, CSIC-Universitat de València,*
 *46980 Paterna, Spain*
[c]*Physics Department, University of Washington,*
 *Seattle, WA 98195-1560, U.S.A.*

 *E-mail:* maxwell.hansen@ed.ac.uk, fernando.romero@uv.es,
 srsharpe@uw.edu

ABSTRACT: We derive relations between finite-volume matrix elements and infinite-volume decay amplitudes, for processes with three spinless, degenerate and either identical or non-identical particles in the final state. This generalizes the Lellouch-Lüscher relation for two-particle decays and provides a strategy for extracting three-hadron decay amplitudes using lattice QCD. Unlike for two particles, even in the simplest approximation, one must solve integral equations to obtain the physical decay amplitude, a consequence of the nontrivial finite-state interactions. We first derive the result in a simplified theory with three identical particles, and then present the generalizations needed to study phenomenologically relevant three-pion decays. The specific processes we discuss are the CP-violating $K \to 3\pi$ weak decay, the isospin-breaking $\eta \to 3\pi$ QCD transition, and the electromagnetic $\gamma^* \to 3\pi$ amplitudes that enter the calculation of the hadronic vacuum polarization contribution to muonic $g-2$.

## Contents

## 1 Introduction

The theoretical formalism for extracting three-hadron scattering amplitudes using lattice QCD has grown apace in recent years [1–20], and applications to simple systems have been successfully undertaken [21–31]. In all such studies, the basic approach is to extract the spectrum of three-hadron states in a finite spatial volume, and to use this information, by means of general relations, to constrain the infinite-volume scattering amplitudes. In particular, the spectrum of three-pion and three-kaon states of maximal isospin has been obtained in multiple calculations with different geometries, and with many values of total momentum in the finite-volume frame. In the following we abbreviate the latter as "different frames".

A natural extension of this work is to consider electroweak transitions to three particles, e.g. the $K \to 3\pi$ decay. Although challenging, one can now conceive of undertaking a lattice calculation of finite-volume matrix elements of the form $\langle 3\pi, L | \mathcal{H}_W | K, L \rangle$, where $\mathcal{H}_W$ is the weak Hamiltonian density, and $\langle 3\pi, L |$ is a finite-volume state whose energy

and momentum are tuned to match that of the initial kaon. Here we restrict attention to a cubic, periodic spatial volume, and $L$ denotes the periodicity (i.e. the box length) in each of the three spatial dimensions. The question is then how to convert knowledge of several such matrix elements (with different volumes and frames) into information on the corresponding infinite-volume decay amplitude, including its dependence on the momenta of the three outgoing pions. In this work we answer this question, providing the formalism for a first-principles calculation of the amplitudes for $K \to 3\pi$ and related decays.

The corresponding problem for two-particle $K \to \pi\pi$ decays was solved in a seminal paper by Lellouch and Lüscher (LL) [32], where it was shown, for the case of a kaon at rest in the finite-volume frame, that the relation between the squared finite-volume matrix element and the magnitude squared of the infinite-volume decay amplitude is an overall multiplicative factor, the LL factor. This result was subsequently generalized in many ways [33–49], with the most important extension for our purposes being the work of refs. [42, 44], in which an alternative and more general formalism was developed for calculating the LL factors for arbitrary $1 \to 2$ processes mediated by an external operator. It is this approach that we use in the main text below to determine the generalization to three-particle final states.

To derive this generalization we first consider a final state consisting of three identical particles, and then move to the more phenomenologically interesting case of three pions in isosymmetric QCD. Exactly as in the two-particle case, the relation between finite-volume matrix elements and decay amplitudes follows from a quantization condition, which can be understood as a relation between finite-volume energies and hadronic scattering amplitudes. In this article we use the form of the quantization condition derived by two of us in refs. [3, 4] together with its extension to all possible three pion states, derived by all of us in ref. [18]. We refer to this approach as the relativistic field theory method.

We note, as was already stressed in refs. [42, 44], that the basic methodology of relating finite-volume matrix elements to infinite-volume amplitudes can be applied to a wide range of processes. To emphasize this in the context of three-hadron final states, in this work we also describe in some detail how the approach may be applied to the virtual photon decay $\gamma^* \to 3\pi$ as well as the isospin breaking transition $\eta \to 3\pi$. The former process is relevant for quantifying finite-volume corrections to the hadronic-vacuum-polarization contribution to $(g-2)_\mu$ arising from the isoscalar part of the photon, along the same lines that $\gamma^* \to \pi\pi$ is used for the isovector part as described in refs. [50, 51].

The remainder of the paper is organized into two parts. In the first, contained in section 2, we derive the necessary formalism for decays to states containing three identical particles. To do so, we first summarize the three-particle scattering formalism in section 2.1. Then, in section 2.2, we derive the relation between the finite-volume matrix elements and a scheme-dependent intermediate infinite-volume quantity, $A_{K3\pi}^{\mathrm{PV}}$. In section 2.3, we explain how to systematically expand $A_{K3\pi}^{\mathrm{PV}}$ about threshold based on symmetries, following which we explain how $A_{K3\pi}^{\mathrm{PV}}$ can be connected to the physical decay amplitude via integral equations (section 2.4). To conclude the discussion for identical particles, in section 2.5 we consider the isotropic approximation in which a more explicit and much simpler expression can be given, results from which we illustrate with numerical examples.

The second part of the paper, contained in section 3, concerns the case of decays to three pions in isosymmetric QCD. We begin, in section 3.1, by presenting the appropriate generalization of the formalism. We then consider the processes $\gamma^* \to 3\pi$, $\eta \to 3\pi$ and $K^+ \to 3\pi$ in sections 3.2, 3.3 and 3.4, respectively. We present our conclusions and outlook in section 4.

We included four appendices. Appendix A derives a technical result needed in the main text. Appendix B presents an alternative derivation of the relation obtained in section 2.2 using the method of Lellouch and Lüscher. Appendix C collects relevant results concerning the isospin decomposition of three-pion states. Finally, appendix D presents the generalization of the results of section 3.4 to the decays of neutral kaons.

While this work was in preparation, a formalism for determining three-particle decay amplitudes to identical scalars in non-relativistic effective field theory (NREFT) was made public [52]. The authors considered only leading-order (non-derivative) couplings for the decay and scattering vertices. The formalism presented here goes beyond that of ref. [52] in several ways: (i) it is valid for nonidentical particles, and thus for the three-pion system; (ii) no approximations concerning the couplings are made, and no truncation in angular momenta is required; (iii) it is valid for generic moving frames; (iv) it is derived in a fully relativistic formalism. We include additional brief comments on the relationship between the approaches in section 2.5.

## 2 Derivation for identical particles

We consider first a simple theory consisting of two real scalar fields, the "kaon" $K$ and "pion" $\phi$, both having an associated $\mathbb{Z}_2$ symmetry that conserves particle number modulo 2. Aside from this symmetry constraint, the interactions between these fields are arbitrary. The physical masses of the particles are $m_K$ and $m_\pi$, respectively, and satisfy

$$3m_\pi < m_K < 5m_\pi \,. \tag{2.1}$$

Both the kaon and the pion are stable particles in this theory. To induce decays, we add an interaction Hamiltonian, suggestively denoted $\mathcal{H}_W$, that violates both $\mathbb{Z}_2$ symmetries, and is chosen to couple the kaon to the odd-pion-number sector. A simple example of the required Hamiltonian density is

$$\mathcal{H}_W(x) = c_W \frac{K(x)\phi(x)^3}{3!} \,, \tag{2.2}$$

but we need not commit to a particular form; all that matters is that the interaction is local and has the correct quantum numbers. We treat $c_W$ as small, such that we need only work to first order in this parameter. Decays of the kaon to even numbers of pions, although kinematically allowed for two pions and possibly also for four pions, are forbidden by symmetries. The potential decay $K \to 5\pi$ is kinematically disallowed for the mass range in eq. (2.1).

To understand the intuition behind the following analysis, consider a diagrammatic representation of the $K \to 3\pi$ amplitude, to leading order in $c_W$ but to all orders in the

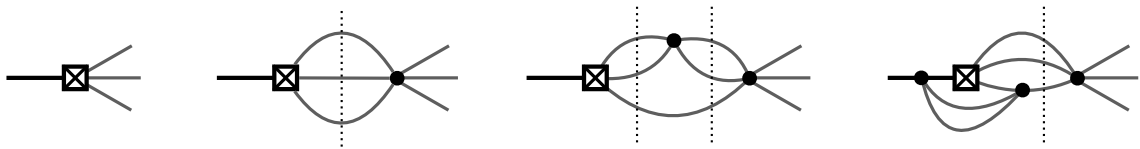

**Figure 1**. Examples of the underlying diagrams describing the $K \to 3\pi$ decay and the corresponding finite-volume matrix element. The left-most diagram represents a local one-to-three transition with only exponentially suppressed finite-volume effects. By contrast the middle two diagrams have power-like $L$ dependence due to the on-shell intermediate states, indicated by the vertical dashed line. Finally, the rightmost diagram indicates a strong-interaction induced dressing to the weak vertex. All such interactions, as well as all dressing on the incoming and outgoing vertices are included in the formalism.

$\mathbb{Z}_2$ preserving interactions. As we illustrate in figure 1, in such an expansion, the only on-shell intermediate states are those involving three pions. Arbitrary virtual interactions between the incoming (dressed) kaon and the final-state pions are allowed, but do not lead to on-shell intermediate states. One can think of such virtual loops as resulting from propagation that is localized near $\mathcal{H}_W$, and they lead to an effective renormalization of the bare coupling $c_W$. This is the physics that one expects to be captured by a calculation of the matrix element in a finite volume. On the other hand, the final-state interactions, which involve long distance, near on-shell propagation, will be mangled in finite volume, and it is these distortions that are corrected by the formalism developed in this work.

As stressed in the introduction, throughout this article we take finite volume to mean a cubic box of side $L$ with periodic boundary conditions on the fields $K$ and $\phi$. This restricts momenta to lie in the finite-volume set $\boldsymbol{p} = \boldsymbol{n}(2\pi/L)$, where $\boldsymbol{n}$ is a three-vector of integers. In our derivation, we drop volume-dependent terms that fall as $\exp(-m_\pi L)$ or faster. For typical volumes used in actual simulations, these exponentially-suppressed terms are much smaller than the power-law volume dependence that we keep. As is quite standard in these types of analyses, we take the temporal extent to be infinite. We also work in a continuum effective field theory with the assumption that the discretization effects entering a numerical lattice QCD calculation using these methods are small and included in the systematic uncertainties of the finite-volume matrix elements and energies.

## 2.1 Recap of three-particle quantization condition and related formalism

We make extensive use of the formalism developed to relate the finite-volume spectrum of three-particle states to the infinite-volume two- and three-particle scattering amplitudes. A general feature of the formalism is that it involves two steps. In the first, the finite-volume spectrum is related to an intermediate, unphysical infinite-volume three-particle K matrix ($\mathcal{K}_{\mathrm{df},3}$ in the approach of this paper), while, in the second, the K matrix is related to the scattering amplitudes by solving integral equations. This two-step procedure carries over naturally to the extension we develop here, with an intermediate, unphysical decay amplitude ($A_{K3\pi}^{\mathrm{PV}}$ below) determined from the finite-volume matrix elements, and the physical decay amplitude then obtained from $A_{K3\pi}^{\mathrm{PV}}$ via integral equations.

As noted above, we use the approach developed in refs. [3, 4], and our aim in this subsection is to recall its essential results. One important feature of this formalism for the case of identical particles is that the intermediate three-particle K matrix, $\mathcal{K}_{\mathrm{df},3}$, is symmetric under separate interchanges of initial and final momenta. This symmetry will carry over to the intermediate one-to-three amplitude, $A_{K3\pi}^{\mathrm{PV}}$, that arises here.[1]

The central result of ref. [3] concerns the following three-particle finite-volume correlator:

$$C_L^{\mathsf{M}}(E, \boldsymbol{P}) = \int_{-\infty}^{\infty} dx_0 \int_L d^3x \, e^{i(Ex^0 - \boldsymbol{P}\cdot\boldsymbol{x})} \langle 0|\mathrm{T}\sigma(x)\sigma^\dagger(0)|0\rangle \,, \tag{2.3}$$

where the superscript indicates that the underlying correlation function is evaluated in Minkowski space, and T stands for time-ordering. Here $\sigma \sim \phi^3$ couples to three pions, but is otherwise an arbitrary operator possibly containing derivatives. Assuming the $\mathbb{Z}_2$ symmetry described above, the kinematic range of interest is

$$m_\pi < E^* = \sqrt{E^2 - \boldsymbol{P}^2} < 5m_\pi \,. \tag{2.4}$$

Within this range, it is shown in ref. [3] that the difference between the finite- and infinite-volume versions of this correlator takes the form[2]

$$C_L^{\mathsf{M}}(E, \boldsymbol{P}) - C_\infty^{\mathsf{M}}(E, \boldsymbol{P}) = iA_3' \frac{1}{F_3^{-1} + \mathcal{K}_{\mathrm{df},3}} A_3 \,. \tag{2.5}$$

Here all quantities have matrix indices $\{k\ell m\}$, with $A_3'$ a row vector, $A_3$ a column vector, while $F_3$ and $\mathcal{K}_{\mathrm{df},3}$ are matrices. The index $k$ is shorthand for the momentum $\boldsymbol{k}$ of one of the three particles, referred to as the spectator. The values of this index are drawn from the finite-volume set. The indices $\ell m$ give the decomposition into spherical harmonics of the angular dependence of the nonspectator pair, when boosted to the pair center-of-momentum frame (CMF). The sum over $\boldsymbol{k}$ is cut off by a smooth function contained in $F$ and $G$, while the sum over $\ell$ is not cut off at this stage. All quantities are also implicit functions of $E$ and $\boldsymbol{P}$, with $F_3$ also depending on $L$. $F_3$ is given by

$$F_3 = \frac{1}{2\omega L^3}\left[\frac{F}{3} - F\frac{1}{1 + \mathcal{M}_{2,L}G}\mathcal{M}_{2,L}F\right] \,, \quad \mathcal{M}_{2,L}^{-1} = \mathcal{K}_2^{-1} + F \,, \tag{2.6}$$

where $\omega$, $F$, $G$, and $\mathcal{K}_2$ are matrices defined in ref. [3], and (with the exception of $\omega$) are also implicit functions of $E$, $\boldsymbol{P}$ and, in the case of $F$ and $G$, also $L$. The only detail we need to know now is that $F$, $G$ and $\mathcal{K}_2$ pick out one of the three particles as the spectator, so that these are intrinsically asymmetric quantities, an asymmetry that is inherited by $F_3$. By contrast, the endcaps $A_3'$ and $A_3$, as well as $\mathcal{K}_{\mathrm{df},3}$, are intrinsically symmetric quantities that are being expressed in terms of asymmetric variables.

---

[1] It is also possible to derive a simpler (though equivalent) version of the three-particle formalism that involves an asymmetric K matrix [16] or the asymmetric R matrix [17]. We do not use these results, however, as the resulting renormalized decay amplitude is less constrained by symmetry, leading to a more complicated parametrization.

[2] We are following the notation of ref. [18] since we use results from this work in the physical $K \to 3\pi$ case below. The notation differs slightly from that of refs. [3, 4].

The endcaps play an important role in the determination of the decay amplitude, as we will see below. The derivation of ref. [3] defines these quantities by an all-orders constructive procedure, the key feature of which is that it involves loop integrals regulated by a principal value (PV) scheme. Thus one can think of the endcaps as, roughly speaking, the sum of all vacuum to three-pion diagrams in which only the short distance contributions from loops are kept. The long distance part, which leads to final state interactions, and the associated complex phases, is removed by the use of the PV prescription. We stress, however, that this qualitative interpretation of the endcaps is not needed to carry through the derivation described below. A technical result that is important below is that, if the creation and annihilation operators in $C_L^{\mathsf{M}}$ are related by hermitian conjugation, then $A_3' = A_3^\dagger$. We prove this fact in appendix A.

From the result (2.5) for the correlator, the quantization condition is seen to be

$$\det(F_3^{-1} + \mathcal{K}_{\mathrm{df},3}) = 0\,. \tag{2.7}$$

As written here, this equation ignores the residual symmetries of the finite-volume system that can be used to block diagonalize the matrix $F_3^{-1} + \mathcal{K}_{\mathrm{df},3}$. The relevant symmetry group depends on the value of $\boldsymbol{P}$. For the purposes of this work it suffices to note that for each group one can identify a set of irreducible representations (irreps), denoted by $\Lambda$, and for each irrep a row index, denoted $\mu$. Each set of $\Lambda\mu$ then corresponds to a block so that eq. (2.7) breaks into a set of independent quantization conditions of the form

$$\det_{\Lambda\mu}\left[\mathbb{P}_{\Lambda\mu} \cdot (F_3^{-1} + \mathcal{K}_{\mathrm{df},3}) \cdot \mathbb{P}_{\Lambda\mu}\right] = 0\,, \tag{2.8}$$

where $\mathbb{P}_{\Lambda\mu}$ projects out a given irrep and row.

To give the definition of $\mathbb{P}_{\Lambda\mu}$, we introduce $\mathbb{R}$ as a unitary matrix with the property that

$$\mathbb{R}^\dagger \cdot (F_3^{-1} + \mathcal{K}_{\mathrm{df},3}) \cdot \mathbb{R}\,, \tag{2.9}$$

is block diagonal with one block corresponding to each possible value of $\Lambda\mu$. The construction of this matrix is a standard group-theoretic exercise, described, for example, in ref. [12]. We then define $\widetilde{\mathbb{P}}_{\Lambda\mu}$ as a diagonal matrix of ones and zeroes that annihilates all blocks besides that corresponding to the target irrep and row. Finally we define

$$\mathbb{P}_{\Lambda\mu} = \mathbb{R} \cdot \widetilde{\mathbb{P}}_{\Lambda\mu} \cdot \mathbb{R}^\dagger\,, \tag{2.10}$$

which projects to the target irrep while preserving the $\{k\ell m\}$ matrix space. The matrix $\mathbb{P}_{\Lambda\mu} \cdot (F_3^{-1} + \mathcal{K}_{\mathrm{df},3}) \cdot \mathbb{P}_{\Lambda\mu}$ will always have vanishing determinant, since the projection amounts to setting all eigenvalues with eigenvectors outside the $\Lambda\mu$ subspace to zero. For this reason, we include the $\Lambda\mu$ subscript on the determinant, indicating that this is evaluated only over the nontrivial subspace.

We stress that eqs. (2.7)–(2.10) are formal relations involving infinite-dimensional matrices and must be truncated in practice. This is done by assuming that the two- and three-particle interactions vanish above some value of $\ell$. For a given $\boldsymbol{P}$, $\Lambda\mu$ and $L$, this equation will be satisfied for a discrete set of values of $E$, which we label $E_n^\Lambda(\boldsymbol{P}, L)$ and often abbreviate as $E_n$.

The final result we need concerns the finite-volume three-particle scattering amplitude, $\mathcal{M}_{3,L}$, defined in ref. [4]. This is the finite-volume version of the amputated, connected infinite-volume amplitude $\mathcal{M}_3$. What will be important here is how $\mathcal{M}_{3,L}$ can be obtained from $C_L$ by an amputation procedure discussed in refs. [4, 9]. The idea is that, as we move in from the endcaps we may encounter a factor of $F$, and this sets the three particles on shell. An unsymmetrized form of the scattering amplitude, $\mathcal{M}_{3,L}^{(u,u)}$, is then obtained by keeping terms in $C_L$ that have at least two factors of $F$ — one for incoming and the other for outgoing particles — and dropping all but the contributions between the two outermost $F$s. In fact, this includes some disconnected three-particle diagrams that must also be dropped. In a final step, the resulting connected amplitude is symmetrized.

We now explain the resulting procedure in detail. We first remove the factors of $i$, $A_3'$ and $A_3$, and rewrite the result as[3]

$$\frac{1}{F_3^{-1} + \mathcal{K}_{\mathrm{df},3}} = F_3 - F_3 \frac{1}{1 + \mathcal{K}_{\mathrm{df},3}F_3} \mathcal{K}_{\mathrm{df},3}F_3 \,, \tag{2.11}$$

$$= \frac{F}{6\omega L^3} - \frac{F}{2\omega L^3} \frac{1}{1 + \mathcal{M}_{2,L}G} \mathcal{M}_{2,L} 2\omega L^3 \frac{F}{2\omega L^3} - F_3 \frac{1}{1 + \mathcal{K}_{\mathrm{df},3}F_3} \mathcal{K}_{\mathrm{df},3}F_3 \,. \tag{2.12}$$

We drop the first term on the right-hand side as it contains a single $F$, and complete the amputation by multiplying by the inverse of $iF/(2\omega L^3)$ on both ends. This leads to

$$\frac{1}{1 + \mathcal{M}_{2,L}G} \mathcal{M}_{2,L} 2\omega L^3 + \left(\frac{F}{2\omega L^3}\right)^{-1} F_3 \frac{1}{1 + \mathcal{K}_{\mathrm{df},3}F_3} \mathcal{K}_{\mathrm{df},3}F_3 \left(\frac{F}{2\omega L^3}\right)^{-1} \,. \tag{2.13}$$

Expanding out the first term in a geometric series, the leading contribution, $\mathcal{M}_{2,L} 2\omega L^3$, is disconnected and thus dropped, leading to the final result for $\mathcal{M}_{3,L}^{(u,u)}$,

$$\mathcal{M}_{3,L}^{(u,u)} = \mathcal{D}^{(u,u)} + \mathcal{L}_L^{(u)} \frac{1}{1 + \mathcal{K}_{\mathrm{df},3}F_3} \mathcal{K}_{\mathrm{df},3}\mathcal{R}_L^{(u)} \,, \tag{2.14}$$

$$\mathcal{D}^{(u,u)} = -\frac{1}{1 + \mathcal{M}_{2,L}G} \mathcal{M}_{2,L} G \mathcal{M}_{2,L} 2\omega L^3 \,, \tag{2.15}$$

$$\mathcal{L}_L^{(u)} = \left(\frac{F}{2\omega L^3}\right)^{-1} F_3 = \frac{1}{3} - \frac{1}{1 + \mathcal{M}_{2,L}G} \mathcal{M}_{2,L}F \,, \tag{2.16}$$

$$\mathcal{R}_L^{(u)} = F_3 \left(\frac{F}{2\omega L^3}\right)^{-1} = \frac{1}{3} - F\mathcal{M}_{2,L} \frac{1}{1 + G\mathcal{M}_{2,L}} \,. \tag{2.17}$$

The full amplitude is then given by

$$\mathcal{M}_{3,L} = \mathcal{S}\left\{\mathcal{M}_{3,L}^{(u,u)}\right\} \,, \tag{2.18}$$

where the symmetrization operator is defined in ref. [4], and discussed in more detail in ref. [18]. We also note that, following ref. [4], $\mathcal{M}_3$ can be obtained from $\mathcal{M}_{3,L}$ by taking the $L \to \infty$ limit in which poles in $F$ and $G$ are shifted from the real axis by the usual $i\epsilon$ prescription.

---

[3]We remove the $i$ since the result of removing $A_3'$ and $A_3$ alone is $i\mathcal{M}_{3,L}$.

## 2.2 Residue method to obtain intermediate decay matrix elements

The approach we follow is adapted from that of ref. [44], and also draws from ref. [42]. The matrix elements that can be determined in finite volume are

$$\langle E_n, \boldsymbol{P}, \Lambda\mu, L|\mathcal{H}_W(0)|K, \boldsymbol{P}, L\rangle \,. \tag{2.19}$$

Here $|K, \boldsymbol{P}, L\rangle$ is a single kaon state, with momentum $\boldsymbol{P}$ drawn from the finite-volume set, while $|E_n, \boldsymbol{P}, \Lambda\mu, L\rangle$ is a three-particle finite-volume state with the same momentum $\boldsymbol{P}$, and with energy $E_n$. It transforms in the irrep $\Lambda$ and in the row $\mu$ of that irrep. Both states are normalized to unity. The energy of the kaon state is $E_K(\boldsymbol{P}) = (\boldsymbol{P}^2 + m_K^2)^{1/2}$, with no volume dependence aside from exponentially suppressed effects. The energy of the three-particle state, by contrast, has a power-law dependence on $L$. In order to obtain a matrix element related to the infinite volume decay amplitude, $L$ should be tuned so that $E_n^\Lambda(\boldsymbol{P}, L) = E_K(\boldsymbol{P})$, implying that four-momentum is conserved.[4] There can be many such matrix elements, each corresponding to a different finite-volume level, with a different choice of $L$ needed in each case.

It is useful to sketch how the matrix elements (2.19) would be determined from a simulation of the theory, carried out necessarily in Euclidean space. We idealize the setup by assuming an infinite Euclidean time direction, and work with correlators fully transformed to momentum space. The three correlators that are needed are

$$C_{K,L}(P) = Z_K \int_{-\infty}^{\infty} dx_4 \int_L d^3x\, e^{-iPx}\langle 0|\mathsf{T}_\mathsf{E} K(x_4, \boldsymbol{x})K(0)|0\rangle \,, \tag{2.20}$$

$$C_{3\pi,L}(P) = \int_{-\infty}^{\infty} dx_4 \int_L d^3x\, e^{-iPx}\langle 0|\mathsf{T}_\mathsf{E} \mathcal{A}_{3\pi}(x_4, \boldsymbol{x})\mathcal{A}_{3\pi}^\dagger(0)|0\rangle \,, \tag{2.21}$$

$$C_{K3\pi,L}(P) = \int_{-\infty}^{\infty} dx_4 \int_L d^3x\, e^{-iPx}\langle 0|\mathsf{T}_\mathsf{E} \mathcal{A}_{3\pi}(x_4, \boldsymbol{x})\mathcal{B}_{K3\pi}(0)|0\rangle \,, \tag{2.22}$$

where $P = (\boldsymbol{P}, P_4)$ and $x = (\boldsymbol{x}, x_4)$ are Euclidean four-vectors, whose inner product is denoted $Px$, and $\mathsf{T}_\mathsf{E}$ denotes Euclidean time ordering.

The correlator $C_{K,L}$ determines the normalization constant $Z_K$. It should be chosen so that

$$\lim_{P_4 \to iE_K(\boldsymbol{P})}(P^2 + m_K^2)C_{K,L}(P) = 1 \,, \tag{2.23}$$

which implies that the renormalized kaon field satisfies

$$|\langle K, \boldsymbol{P}, L|\sqrt{Z_K}K(0)|0\rangle| = \frac{1}{\sqrt{2E_K(\boldsymbol{P})L^3}} \,. \tag{2.24}$$

The correlator $C_{3\pi,L}$ determines the coupling of the operator $\mathcal{A}_{3\pi}$ to the finite-volume states $|E_n, \boldsymbol{P}, \Lambda\mu, L\rangle$. Here, $\mathcal{A}_{3\pi}$ is an operator chosen to couple to three-pion states in a particular row of the desired finite-volume irrep. In practice, $\mathcal{A}_{3\pi}$ will involve pion fields with phase factors such that they have appropriate relative momenta, and thus will be

---

[4]If one were interested in the matrix element (2.19) in which $\mathcal{H}_W(0)$ inserted energy, then the subsequent derivation would still hold in an appropriate kinematic regime. The analysis can also be straightforwardly generalized to the case where $\mathcal{H}_W(0)$ inserts momentum.

complex. Other details of the operator are not relevant in the following. The correlator will consist of a sum of poles, and we pick out the contribution of the desired state from the residue

$$R_{3\pi}(E_n, \boldsymbol{P}, \Lambda\mu, L) \equiv \lim_{P_4 \to iE_n} (E_n + iP_4) C_{3\pi,L}(P) = L^3 \left| \langle 0|\mathcal{A}_{3\pi}(0)|E_n, \boldsymbol{P}, \Lambda\mu, L\rangle \right|^2 . \quad (2.25)$$

The final correlator, $C_{K3\pi,L}$, can then be used to determine the desired matrix element. Here, following ref. [44], we use a composite operator $\mathcal{B}_{K3\pi}$ that both creates the initial kaon (implicitly having momentum $\boldsymbol{P}$) and includes the action of the weak Hamiltonian,

$$\mathcal{B}_{K3\pi}(x) = \sqrt{Z_K} \lim_{P_4 \to iE_K(\boldsymbol{P})} \left[ P^2 + m_K^2 \right] \int d^4y\, e^{iPy} \mathcal{H}_W(x) K(x+y) , \quad (2.26)$$

where $P = (P_4, \boldsymbol{P})$. The limit picks out the incoming kaon pole, while the factor of $P^2 + m_K^2$ amputates the kaon propagator.[5] Including all factors we obtain

$$R_{K3\pi}(E_n, \boldsymbol{P}, \Lambda\mu, L) \equiv \lim_{P_4 \to iE_n} (E_n + iP_4) C_{K3\pi,L}(P) , \quad (2.27)$$

$$= L^3 \langle 0|\mathcal{A}_{3\pi}(0)|E_n, \boldsymbol{P}, \Lambda\mu, L\rangle \langle E_n, \boldsymbol{P}, \Lambda\mu, L|\mathcal{H}_W(0)|K, \boldsymbol{P}, L\rangle \sqrt{2E_K(\boldsymbol{P})L^3} . \quad (2.28)$$

Without loss of generality, we can choose the phase of the operator and state such that $\langle 0|\mathcal{A}_{3\pi}(0)|E_n, \boldsymbol{P}, \Lambda\mu, L\rangle$ is real and positive. Then, combining eqs. (2.25) and (2.28), we obtain

$$\langle E_n, \boldsymbol{P}, \Lambda\mu, L|\mathcal{H}_W(0)|K, \boldsymbol{P}, L\rangle \sqrt{2E_K(\boldsymbol{P})L^3} = \frac{R_{K3\pi}(E_n, \boldsymbol{P}, \Lambda\mu, L)}{\sqrt{L^3 R_{3\pi}(E_n, \boldsymbol{P}, \Lambda\mu, L)}} . \quad (2.29)$$

This matrix element will only be nonvanishing if $\Lambda$ and $\mu$ are chosen to match the transformation properties of $\mathcal{H}_W(0)|K, \boldsymbol{P}, L\rangle$. If not, then the correlator $C_{K3\pi,L}(P)$ and the residue $R_{K3\pi}$ will vanish. For a rotationally invariant $\mathcal{H}_W$, only the trivial irrep of the little group for momentum $\boldsymbol{P}$ will appear (or else the corresponding parity conjugate irrep), but we develop the formalism allowing for more general cases.

We now evaluate this ratio using the results from the previous subsection. To do so we first generalize the correlator $C_L$ of eq. (2.3) by replacing $\sigma$ and $\sigma^\dagger$ with general operators $\mathcal{A}$ and $\mathcal{B}$ that couple the vacuum to three-pion states, but are, in general, unrelated to each other:

$$C_{AB,L}^{\mathsf{M}}(E, \boldsymbol{P}) = \int_{-\infty}^{\infty} dx_0 \int_L d^3x\, e^{i(Ex^0 - \boldsymbol{P}\cdot\boldsymbol{x})} \langle 0|\mathsf{T}\mathcal{A}(x)\mathcal{B}(0)|0\rangle . \quad (2.30)$$

The analysis of ref. [3] remains valid for $C_{AB,L}^{\mathsf{M}}$, since it requires only that the allowed on-shell intermediate states involve three pions. Thus the expression (2.5) still holds, except that the endcaps $A_3'$ and $A_3$ are replaced by new quantities that we call, respectively, $A^{\mathrm{PV}}$

---

[5]Note that a subtlety arises here due to the fact that the operator $\mathcal{B}_{K3\pi}$ is not local in time. This is not an issue because the $P_4 \to iE_K(\boldsymbol{P})$ limit is dominated by early $y_4$ so that the $K(x+y)$ operator is ordered far to the right. Thus only one time-ordering arises, that with the intermediate finite-volume states that we analyze explicitly.

and $B^{\mathrm{PV}}$. The superscript is a reminder that loops in these quantities are defined using a PV prescription.

We next do a Wick rotation $(x_0 \to -ix_4)$ on the underlying correlation function, so that it is evaluated in Euclidean space-time. This results in

$$C^{\mathsf{M}}_{AB,L}(E, \boldsymbol{P}) = -iC_{AB,L}(P)\big|_{P_4=iE}, \tag{2.31}$$

$$C_{AB,L}(P) = \int_{-\infty}^{\infty} dx_4 \int_L d^3x \, e^{-iPx} \langle \mathsf{T}_{\mathsf{E}} \mathcal{A}(x) \mathcal{B}(0) \rangle, \tag{2.32}$$

where again $P = (\boldsymbol{P}, P_4)$. It follows that $C_{AB,L}$ can be written

$$C_{AB,L}(P) = C_{AB,\infty}(P) - A^{\mathrm{PV}} \frac{1}{F_3^{-1} + \mathcal{K}_{\mathrm{df},3}} B^{\mathrm{PV}}, \tag{2.33}$$

where now $A^{\mathrm{PV}}$, $F_3$, $\mathcal{K}_{\mathrm{df},3}$ and $B^{\mathrm{PV}}$ are written as functions of $P$ by setting $E = -iP_4$. The poles now lie on the imaginary axis, at the positions $P_4 = iE_n$, where $E_n$ is a solution of the quantization condition eq. (2.7).

The reason for these manipulations is that the two correlators that enter into the expression (2.29) for the desired matrix element, $C_{3\pi}$ and $C_{K3\pi}$, are in the class for which eq. (2.33) holds. In particular, we can use the results of ref. [3] to write these correlators as

$$C_{3\pi,L}(P) = C_{3\pi,\infty}(P) - A^{\mathrm{PV}}_{3\pi} \frac{1}{F_3^{-1} + \mathcal{K}_{\mathrm{df},3}} A^{\mathrm{PV}\dagger}_{3\pi}, \tag{2.34}$$

$$C_{K3\pi,L}(P) = C_{K3\pi,\infty}(P) - A^{\mathrm{PV}}_{3\pi} \frac{1}{F_3^{-1} + \mathcal{K}_{\mathrm{df},3}} A^{\mathrm{PV}}_{K3\pi}. \tag{2.35}$$

In eq. (2.34) we are using the result, demonstrated in appendix A, that if the source and sink operators are related by hermitian conjugation, then the same holds for the endcap factors. Note that this only holds because the latter are defined with the PV prescription.

We next evaluate the residues that enter eq. (2.29). Since the infinite-volume correlators and the endcaps are smooth, infinite-volume functions, $L$-dependent poles only arise from the zero eigenvalues in $F_3^{-1} + \mathcal{K}_{\mathrm{df},3}$. The required residues are thus

$$\mathcal{R}_{\Lambda\mu}(E_n^\Lambda, \boldsymbol{P}, L) = \lim_{P_4 \to iE_n^\Lambda} -(E_n^\Lambda + iP_4) \, \mathbb{P}_{\Lambda\mu} \cdot \frac{1}{F_3^{-1} + \mathcal{K}_{\mathrm{df},3}} \cdot \mathbb{P}_{\Lambda\mu}, \tag{2.36}$$

where the minus sign is for later convenience, and $E_n^\Lambda$ is one of the finite-volume three-pion energies for the given choice of $\boldsymbol{P}$, $\Lambda$ and $L$. $\mathcal{R}_{\Lambda\mu}$ is a matrix in the $\{k\ell m\}$ space, which can be evaluated explicitly given expressions for $\mathcal{K}_2$ (contained in $F_3$) and $\mathcal{K}_{\mathrm{df},3}$. The idea here is that these quantities have been previously determined (or, more realistically, constrained within some truncation scheme) by using the two- and three-particle quantization conditions applied to the spectrum of two- and three-particle states.

An important property of $\mathcal{R}_{\Lambda\mu}$ is that it has rank one. This is because only one of the formally infinite tower of eigenvalues of $\mathbb{P}_{\Lambda\mu} \cdot (F_3^{-1} + \mathcal{K}_{\mathrm{df},3}) \cdot \mathbb{P}_{\Lambda\mu}$ will vanish for a given finite-volume energy $E_n^\Lambda(\boldsymbol{P}, L)$. Denoting the relevant eigenvalue by $\lambda(E, \boldsymbol{P}, \Lambda\mu, L)$ and the corresponding normalized eigenvector by $\boldsymbol{e}(E, \boldsymbol{P}, \Lambda\mu, L)$, one finds

$$\mathcal{R}_{\Lambda\mu}(E_n^\Lambda, \boldsymbol{P}, L) = \left(\frac{\partial \lambda(E, \boldsymbol{P}, \Lambda\mu, L)}{\partial E}\bigg|_{E=E_n^\Lambda(\boldsymbol{P}, L)}\right)^{-1} \boldsymbol{e}(E, \boldsymbol{P}, \Lambda\mu, L) \, \boldsymbol{e}^\dagger(E, \boldsymbol{P}, \Lambda\mu, L). \tag{2.37}$$

This rank one property of $\mathcal{R}_{\Lambda\mu}$ was first described in the two-particle case in refs. [42, 44]. As is discussed, e.g. in refs. [10, 12, 38], the eigenvalue must satisfy the inequality

$$\left(\frac{\partial\lambda(E,\boldsymbol{P},\Lambda\mu,L)}{\partial E}\bigg|_{E=E_n^{\Lambda}(\boldsymbol{P},L)}\right)^{-1} > 0 \,. \tag{2.38}$$

Thus, defining

$$v(E_n^{\Lambda},\boldsymbol{P},\Lambda\mu,L) \equiv \left(\frac{\partial\lambda(E,\boldsymbol{P},\Lambda\mu,L)}{\partial E}\bigg|_{E=E_n^{\Lambda}(\boldsymbol{P},L)}\right)^{-1/2} \boldsymbol{e}(E,\boldsymbol{P},\Lambda\mu,L) \,, \tag{2.39}$$

$\mathcal{R}_{\Lambda\mu}$ can be written as a simple outer product

$$\mathcal{R}_{\Lambda\mu}(E_n^{\Lambda},\boldsymbol{P},L) = v(E_n^{\Lambda},\boldsymbol{P},\Lambda\mu,L)v^{\dagger}(E_n^{\Lambda},\boldsymbol{P},\Lambda\mu,L) \,. \tag{2.40}$$

Since $F_3^{-1} + \mathcal{K}_{\mathrm{df},3}$ is a real, symmetric matrix (assuming that we use real spherical harmonics), the elements of each $v$ are relatively real, with only the overall phase undetermined.

Using these results, we can immediately evaluate the required residues, obtaining

$$R_{3\pi}(E_n^{\Lambda},\boldsymbol{P},\Lambda\mu,L) = |A_{3\pi}^{\mathrm{PV}}v|^2 \,, \tag{2.41}$$

$$R_{K3\pi}(E_n^{\Lambda},\boldsymbol{P},\Lambda\mu,L) = (A_{3\pi}^{\mathrm{PV}}v)(v^{\dagger}A_{K3\pi}^{\mathrm{PV}}) \,, \tag{2.42}$$

where $v$ is an abbreviation for $v(E_n^{\Lambda},\boldsymbol{P},\Lambda\mu,L)$. All quantities on the right-hand side are (implicitly) evaluated at $P = (\boldsymbol{P},iE_n^{\Lambda})$, with $E_n^{\Lambda} = E_K(\boldsymbol{P})$. The overall sign in eq. (2.36) can now be justified. From eq. (2.25), we know that $R_{3\pi}$ is positive, and thus the overall sign in eq. (2.41) must be positive, as shown.[6]

Choosing the phase of $v$ such that $A_{3\pi}^{\mathrm{PV}}v$ is real and positive, and inserting these results into eq. (2.29), we obtain

$$\sqrt{2E_K(\boldsymbol{P})}L^3\langle E_n,\boldsymbol{P},\Lambda\mu,L|\mathcal{H}_W(0)|K,\boldsymbol{P},L\rangle = v^{\dagger}A_{K3\pi}^{\mathrm{PV}} \,. \tag{2.43}$$

This achieves the aim of relating the finite-volume decay matrix element (which could be determined by a numerical simulation) to a quantity in the generic relativistic field theory, namely a projection of the quantity $A_{K3\pi}^{\mathrm{PV}}$. By using multiple matrix elements, one could determine the parameters in a truncated approximation to $A_{K3\pi}^{\mathrm{PV}}$. The result (2.43) can also be derived by a generalization of the method of Lellouch and Lüscher [32], as we show in appendix B.

Before turning to parametrizations of $A_{K3\pi}^{\mathrm{PV}}$, we close this section with a few more comments on the phase conventions entering the various relations on matrix elements. We first review the requirements we have imposed above. First, we have fixed the phase of the state $\mathcal{A}_{3\pi}(0)|E_n,\boldsymbol{P},\Lambda\mu,L\rangle$ by requiring that $\langle 0|\mathcal{A}_{3\pi}(0)|E_n,\boldsymbol{P},\Lambda\mu,L\rangle$ is real and positive. Second, we have required that, while $A_{3\pi}^{\mathrm{PV}}$ and $v(E_n^{\Lambda},\boldsymbol{P},\Lambda\mu,L)$ may individually carry phases, these must cancel such that $A_{3\pi}^{\mathrm{PV}}v$ is real and positive. We have then demonstrated that, with these two convention choices, the finite-volume matrix element appearing in

---

[6]This is in fact the criterion introduced in ref. [10], and studied in refs. [12, 15], to determine whether solutions to the three-particle quantization condition are physical.

eq. (2.43) must have the same phase as the combination $v^\dagger A_{K3\pi}^{\mathrm{PV}}$. Finally, to extract the value of $A_{K3\pi}$, we must establish the phase of $v$ itself, which has been left open so far. The most natural convention is to simply require $A_{3\pi}^{\mathrm{PV}}$ and $v$ to be individually real. In this convention $v^\dagger$ is also real, so any phase in the finite-volume matrix element on the left-hand side of eq. (2.43) (resulting, for example, from a CP-violating phase in $\mathcal{H}_W$) will be inherited by $A_{K3\pi}^{\mathrm{PV}}$.

As was already discussed in refs. [42, 44], the utility in carefully tracking this phase information is that it allows one to extract relative phases between various matrix elements. For example, if the weak Hamiltonian density is decomposed into operators $\mathcal{O}_1(x)$ and $\mathcal{O}_2(x)$, it follows from eq. (2.43) that

$$\frac{\langle E_n, \boldsymbol{P}, \Lambda\mu, L | \mathcal{O}_1(0) | K, \boldsymbol{P}, L \rangle}{\langle E_n, \boldsymbol{P}, \Lambda\mu, L | \mathcal{O}_2(0) | K, \boldsymbol{P}, L \rangle} = \frac{v^\dagger A_{K3\pi[\mathcal{O}_1]}^{\mathrm{PV}}}{v^\dagger A_{K3\pi[\mathcal{O}_2]}^{\mathrm{PV}}}. \tag{2.44}$$

The overall phase in $v^\dagger$ cancels, so the phase in the ratio of PV amplitudes on the right-hand side is given by that of the ratio of the matrix elements on the left-hand side. This phase information will be passed on to the decay matrix elements by solving the integral equations described below in section 2.4.

## 2.3 Threshold expansion of $A_{K3\pi}^{\mathrm{PV}}$

Since $A_{K3\pi}^{\mathrm{PV}}$ is an unfamiliar quantity, we discuss its properties in this brief subsection. We recall that it is an infinite-volume on-shell quantity, given, crudely speaking, by calculating all $K \to 3\pi$ diagrams with PV regulation for the poles. Thus it is an analytic function of the kinematic variables, symmetric under interchange of any pair of final-state momenta.

A useful parametrization of $A_{K3\pi}^{\mathrm{PV}}$ is given by the threshold expansion, which is an expansion in powers of relativistic invariants that vanish at threshold, for instance

$$\Delta = \frac{m_K^2 - 9m_\pi^2}{9m_\pi^2}. \tag{2.45}$$

For the decays $K^+ \to \pi^+\pi^+\pi^-$ and $K^+ \to \pi^+\pi^0\pi^0$, for example, $\Delta \approx 0.39$ and $0.45$, respectively. Labelling the pion four-momenta $p_1$, $p_2$, and $p_3$, so that $P = p_K = p_1+p_2+p_3$, the three Mandelstam variables are

$$s_i = (p_j + p_k)^2 = (P - p_i)^2, \quad \sum_{i=1}^{3} s_i = m_K^2 + 3m_\pi^2, \tag{2.46}$$

where $\{i, j, k\}$ are ordered cyclically. We will expand in dimensionless quantities that vanish at threshold, namely $\Delta$ and

$$\Delta_i = \frac{s_i - 4m_\pi^2}{9m_\pi^2}, \tag{2.47}$$

which satisfy $\sum_i \Delta_i = \Delta$. Using this sum rule, and enforcing particle-interchange symmetry and smoothness, we find[7]

$$A_{K3\pi}^{\mathrm{PV}} = A^{\mathsf{iso}} + A^{(2)} \sum_i \Delta_i^2 + A^{(3)} \sum_i \Delta_i^3 + A^{(4)} \sum_i \Delta_i^4 + \mathcal{O}(\Delta^5). \tag{2.48}$$

---

[7]The presence of only a single term in each of the second, third and fourth orders is a pattern that does not continue to higher orders.

Here "iso" refers to the isotropic limit, in which the amplitude is independent of the momenta of the decay products. To obtain a strict expansion in powers of $\Delta$, one would need to expand the coefficients, e.g.

$$A^{\text{iso}} = \sum_{n=0}^{\infty} \Delta^n A^{\text{iso,n}} \,, \tag{2.49}$$

keeping only the appropriate number of terms (e.g. the first five terms if working to fourth order in $\Delta$).

To use the threshold expansion (2.48) in the result from the previous subsection, eq. (2.43), one must convert $A_{K3\pi}^{\text{PV}}$ to the $\{k\ell m\}$ basis. We recall here how this is done [3]. We first note that the on-shell three-particle phase space with fixed total four-momentum (and ignoring Lorentz invariance) is five-dimensional. We can parametrize this space in various ways, one choice being to use a set of five momentum coordinates: $p_{1,x}$, $p_{1,y}$, $p_{1,z}$, $p_{2,x}$, $p_{2,y}$. The remaining five coordinates are then set by the fixed total energy and momentum. To connect to the $\{k\ell m\}$ basis we make a different choice, labelled $\{\boldsymbol{k}, \widehat{\boldsymbol{a}}^*\}$. Here $\boldsymbol{k}$ is one of the three momenta, e.g. $\boldsymbol{k} = \boldsymbol{p}_1$, while $\widehat{\boldsymbol{a}}^*$ is the result of boosting the remaining two particles to their CMF and picking the direction of one of them, say particle 2. Here we are using the notation that a quantity with a superscript $*$ is evaluated in a boosted frame. We then decompose the amplitude into spherical harmonics in the pair CMF,

$$A_{K3\pi}^{\text{PV}}(\boldsymbol{k}, \widehat{\boldsymbol{a}}^*) = \sum_{\ell m} \sqrt{4\pi}\, Y_{\ell m}(\widehat{\boldsymbol{a}}^*) A_{K3\pi}^{\text{PV}}(\boldsymbol{k})_{\ell m} \,. \tag{2.50}$$

To use the result of the previous subsection we must restrict $\boldsymbol{k}$ to lie in the finite-volume set,

$$A_{K3\pi;k\ell m}^{\text{PV}} \equiv A_{K3\pi}^{\text{PV}}(\boldsymbol{k})_{\ell m}\Big|_{\boldsymbol{k}=2\pi\boldsymbol{n}/L} \,. \tag{2.51}$$

The decomposition of the terms in the threshold expansion into the $\{k\ell m\}$ basis is straightforward but tedious, and we do not present it here. It follows closely the corresponding decomposition of $\mathcal{K}_{\text{df},3}$ worked out in ref. [12].

## 2.4 Relating $A_{K3\pi}^{\text{PV}}$ to the physical decay amplitude

In this subsection we show how the physical $K \to 3\pi$ decay amplitude can be obtained by solving appropriate integral equations, once the endcap $A_{K3\pi}^{\text{PV}}$ has been determined using the results of the previous two subsections. This is the second step of the general procedure described in section 2.1, and involves relations between infinite-volume quantities. The method we use follows the strategy introduced in ref. [4]: we consider a finite-volume correlator whose infinite-volume limit produces the physical decay amplitude, and write this correlator in terms of $\mathcal{K}_2$, $\mathcal{K}_{\text{df},3}$, and in particular $A_{K3\pi}^{\text{PV}}$.

We begin by recalling that the infinite-volume decay matrix element can be defined by

$$T_{K3\pi} = \langle 3\pi, \text{out}|\, \mathcal{H}_W(0)\, |K, \boldsymbol{P}\rangle \,, \tag{2.52}$$

where states are defined using the standard relativistic normalization. The decay rate is then given by

$$\Gamma = \frac{1}{3!} \frac{1}{2m_K} \int d\text{LIPS}\, |T_{K3\pi}|^2 \,, \tag{2.53}$$

where 1/3! is the identical-particle symmetry factor, and dLIPS is the Lorentz-invariant phase-space measure. We will use the $\{\boldsymbol{k}, \hat{\boldsymbol{a}}^*\}$ variables introduced above, in terms of which the measure becomes

$$\text{dLIPS} = \frac{d^3\boldsymbol{k}}{2\omega_k(2\pi)^3}\frac{a^*}{4\pi\omega_{a^*}}\frac{d^2\Omega_{\hat{a}^*}}{4\pi}. \tag{2.54}$$

Here $a^{*2} = q_{2,k}^{*2}$ is the squared momentum of one of the nonspectator pair in their CMF, with

$$q_{2,k}^{*2} = (E_K(\boldsymbol{P}) - \omega_k)^2 - (\boldsymbol{P} - \boldsymbol{k})^2, \tag{2.55}$$

and $\omega_{a^*} = \sqrt{a^{*2} + m_\pi^2}$ is the corresponding energy.

In order to obtain an expression for $T_{K3\pi}$ in terms of $A_{K3\pi}^{\text{PV}}$, we consider the finite-volume decay matrix element, $T_{K3\pi,L}$. This is defined as the sum of all Feynman diagrams contributing to $T_{K3\pi}$, including appropriate amputations, but evaluated with finite-volume Feynman rules. A subtlety arises because the energies of three external on-shell pions, each with a momentum from the finite-volume set will, not, in general, sum to $E_K(\boldsymbol{P})$. To have an energy-conserving process, the external momenta in $T_{K3\pi,L}$ must be adjusted. This on-shell projection is done using the method introduced in ref. [3]. The spectator momentum, $\boldsymbol{k}$, is held fixed at a finite-volume value, while the magnitude of $\boldsymbol{a}^*$ (the momentum of one of the nonspectator pair boosted to the pair CMF) is adjusted until energy is conserved. This requires setting $\boldsymbol{a}^* = q_{2,k}^*\hat{\boldsymbol{a}}^*$, and leads to the third particle having momentum $-\boldsymbol{a}^*$ in the pair CMF. This is the on-shell projection that appears in all quantities adjacent to factors of $F$ and $G$. The projection only affects the external momenta for $T_{K3\pi,L}$ — when written as a skeleton expansion in terms of Bethe-Salpeter kernels, the internal loop momenta are all drawn from the finite-volume set. This point is discussed at length in ref. [4]. The result is the quantity $T_{K3\pi,L}(\boldsymbol{k}, \hat{\boldsymbol{a}}^*)$.

We will need a variant of this quantity in the following, namely $T_{K3\pi,L}^{(u)}(\boldsymbol{k}, \hat{\boldsymbol{a}}^*)$, which we refer to as the asymmetric decay amplitude. This is defined as the sum of the same set of amputated diagrams with two restrictions: first, if the final interaction involves a two-particle Bethe-Salpeter kernel, then $\boldsymbol{k}$ is chosen as the momentum for the spectator particle. Second, if the final interaction involves a three-particle kernel, then the diagram is multiplied by 1/3. In fact, what appears in the expressions below is $T_{K3\pi,L,\ell m}^{(u)}$, which results when we decompose the $\hat{\boldsymbol{a}}^*$ dependence into spherical harmonics as in eq. (2.50).

To obtain the desired expression for $T_{K3\pi,L,\ell m}$, we begin from the correlator $C_{K3\pi,L}(P)$, introduced in eq. (2.22), which describes a finite-volume $K \to 3\pi$ process. We consider the Minkowski version of this correlator, given by

$$C_{K3\pi,L}^{\text{M}}(E, \boldsymbol{P}) = C_{K3\pi,\infty}^{\text{M}}(E, \boldsymbol{P}) + A_{3\pi}^{\text{PV}}\frac{i}{F_3^{-1} + \mathcal{K}_{\text{df},3}}A_{3\pi}^{\text{PV}}. \tag{2.56}$$

We obtain $T_{K3\pi,L}$ by keeping contributions that have at least one factor of $F$ (since this puts the intermediate three-particle state on shell) and amputating all that lies to the left of the left-most $F$. Only the second term on the right-hand side contains $F$s, and we amputate it as described in section 2.1 by removing $A_{3\pi}^{\text{PV}}$ and multiplying by the inverse of

$iF/(2\omega L^3)$, leading to

$$T_{K3\pi,L}^{(u)} = \left(\frac{iF}{2\omega L^3}\right)^{-1} F_3 \frac{i}{1+\mathcal{K}_{\mathrm{df},3}F_3} A_{K3\pi}^{\mathrm{PV}}, \tag{2.57}$$

$$= \mathcal{L}_L^{(u)} \frac{1}{1+\mathcal{K}_{\mathrm{df},3}F_3} A_{K3\pi}^{\mathrm{PV}}, \tag{2.58}$$

where $\mathcal{L}_L^{(u)}$ is given in eq. (2.16). Note that, unlike in the construction of $\mathcal{M}_{3,L}^{(u,u)}$ described in section 2.1, here there are no disconnected terms to drop.

With the expression for $T_{K3\pi;k\ell m}^{(u)}$ in hand, we next note, following ref. [4], that the result can be extended to an arbitrary choice of $\boldsymbol{k}$, not just one in the finite-volume set. The form of eq. (2.58) remains unchanged, and the various quantities extend simply to arbitrary $\boldsymbol{k}$, as explained in ref. [4]. The result, $T_{K3\pi,L}^{(u)}(\boldsymbol{k})_{\ell m}$, is still a finite-volume quantity, since internal loops remain summed. We now insert $i\epsilon$ factors to regulate the poles in $F$ and $G$, and take the infinite-volume limit holding $\boldsymbol{k}$ fixed

$$T_{K3\pi}^{(u)}(\boldsymbol{k})_{\ell m} = \lim_{\epsilon\to 0^+}\lim_{L\to\infty} T_{K3\pi,L}^{(u)}(\boldsymbol{k})_{\ell m}\Big|_{E\to E+i\epsilon}. \tag{2.59}$$

This gives the correct asymmetric infinite-volume decay amplitude because, in the limit, all sums in Feynman diagrams that run over a pole (which are those in which three particles can go on shell) are replaced by integrals in which the pole is regulated by the standard $i\epsilon$ prescription.

The final step is to obtain the complete decay amplitude by symmetrizing, which corresponds to adding all possible attachments of the momentum labels to the Feynman diagrams. This is effected by

$$T_{K3\pi}(\boldsymbol{k},\widehat{\boldsymbol{a}}^*) \equiv \mathcal{S}\left\{T_{K3\pi}(\boldsymbol{k})_{\ell m}\right\}, \tag{2.60}$$

$$= T_{K3\pi}^{(u)}(\boldsymbol{k},\widehat{\boldsymbol{a}}^*) + T_{K3\pi}^{(u)}(\boldsymbol{a},\widehat{\boldsymbol{b}}^*) + T_{K3\pi}^{(u)}(\boldsymbol{b},\widehat{\boldsymbol{k}}^*), \tag{2.61}$$

where $T_{K3\pi}^{(u)}(\boldsymbol{k},\widehat{\boldsymbol{a}}^*)$ is obtained by combining $T_{K3\pi,L}^{(u)}(\boldsymbol{k})_{\ell m}$ with spherical harmonics as in eq. (2.50). The notation in eq. (2.61) is the natural generalization of that given above: just as $(\omega_{a^*},\boldsymbol{a}^*)$ is the result of boosting $(\omega_a,\boldsymbol{a})$ to the CMF of the $\{\boldsymbol{a},\boldsymbol{b}\}$ pair (with $\boldsymbol{b} = \boldsymbol{P}-\boldsymbol{k}-\boldsymbol{a}$), so $(\omega_{b^*},\boldsymbol{b}^*)$ is the result of boosting $(\omega_b,\boldsymbol{b})$ to the CMF of the $\{\boldsymbol{b},\boldsymbol{k}\}$ pair, while $(\omega_{k^*},\boldsymbol{k}^*)$ is the result of boosting $(\omega_k,\boldsymbol{k})$ to the CMF of the $\{\boldsymbol{k},\boldsymbol{a}\}$ pair.

Applying this procedure to the result eq. (2.58) for $T_{K3\pi,L}^{(u)}$ leads to a set of integral equations. Since the steps are very similar to those in ref. [4], we simply quote the final results. As for $T_{K3\pi}^{(u)}$, the $\{k\ell m\}$ indices used in finite volume go over in infinite-volume to a dependence on the continuous spectator momentum, $\boldsymbol{k}$, as well as an unchanged dependence on $\ell$ and $m$. Thus the matrix indices $\ell m$ remain, and will be implicit in the following equations, while the dependence on $\boldsymbol{k}$ will be explicit.

The combination $(1+\mathcal{M}_{2,L}G)^{-1}\mathcal{M}_{2,L}$, which appears in $\mathcal{L}_L^{(u)}$ and in $F_3$, goes over in infinite volume to $\mathcal{D}_{23}^{(u,u)}(\boldsymbol{p},\boldsymbol{k})_{\ell'm';\ell m}$ (using the notation of ref. [16]), which satisfies

$$\mathcal{D}_{23}^{(u,u)}(\boldsymbol{p},\boldsymbol{k}) = \bar{\delta}(\boldsymbol{p}-\boldsymbol{k})\mathcal{M}_2(\boldsymbol{k}) - \mathcal{M}_2(\boldsymbol{p})\int_{\boldsymbol{r}} G^\infty(\boldsymbol{p},\boldsymbol{r})\mathcal{D}_{23}^{(u,u)}(\boldsymbol{r},\boldsymbol{k}), \tag{2.62}$$

where $G^\infty$ is defined in eq. (81) of ref. [4], and includes an $i\epsilon$-regulated pole, while

$$\overline{\delta}(\boldsymbol{p} - \boldsymbol{k}) = 2\omega_p(2\pi)^3\delta^3(\boldsymbol{p} - \boldsymbol{k}),\tag{2.63}$$

$$\mathcal{M}_2(\boldsymbol{k})_{\ell'm';\ell m} = \delta_{\ell'\ell}\delta_{m'm}\mathcal{M}_2^{(\ell)}(q_{2,k}^*),\tag{2.64}$$

$$\int_r = \int \frac{d^3r}{2\omega_r(2\pi)^3}.\tag{2.65}$$

Here $\mathcal{M}_2^{(\ell)}$ is the $\ell$th partial wave of $\mathcal{M}_2$, evaluated for the CMF momentum of one of the scattering pair. Given a solution to the integral equation (2.62), and the relation of $F_3$ to $\mathcal{L}_L^{(u)}$, eq. (2.16), the equation satisfied by the infinite-volume limit of $X = (1 + \mathcal{K}_{\mathrm{df},3}F_3)^{-1}$ is

$$X(\boldsymbol{p}, \boldsymbol{k}) = \overline{\delta}(\boldsymbol{p} - \boldsymbol{k}) - \int_{r,s}\mathcal{K}_{\mathrm{df},3}(\boldsymbol{p}, \boldsymbol{r})\widetilde{\rho}_{\mathrm{PV}}(\boldsymbol{r})\mathcal{L}^{(u)}(\boldsymbol{r}, \boldsymbol{s})X(\boldsymbol{s}, \boldsymbol{k}).\tag{2.66}$$

In the first term there is an implicit identity matrix in $\ell m$ space. The quantity $\widetilde{\rho}_{\mathrm{PV}}$ results from the infinite-volume limit of $F$, and is

$$\widetilde{\rho}_{\mathrm{PV}}(\boldsymbol{r})_{\ell'm';\ell m} = \delta_{\ell'\ell}\delta_{m'm}\,\widetilde{\rho}_{\mathrm{PV}}^{(\ell)}(q_{2,r}^*),\tag{2.67}$$

where $\rho_{\mathrm{PV}}^{(\ell)}$ is a modified phase space factor given in eq. (B6) of ref. [16]. Finally,

$$\mathcal{L}^{(u)}(\boldsymbol{r}, \boldsymbol{s}) = \frac{1}{3}\overline{\delta}(\boldsymbol{r} - \boldsymbol{s}) - \mathcal{D}_{23}^{(u,u)}(\boldsymbol{r}, \boldsymbol{s})\widetilde{\rho}_{\mathrm{PV}}(\boldsymbol{s}),\tag{2.68}$$

which is the infinite-volume limit of $\mathcal{L}_L^{(u)}$.

With these ingredients we can write down the relationship of the asymmetric decay amplitude to $A_{K3\pi}^{\mathrm{PV}}$,

$$T_{K3\pi}^{(u)}(\boldsymbol{k}) = \int_{r,s}\mathcal{L}^{(u)}(\boldsymbol{k}, \boldsymbol{r})X(\boldsymbol{r}, \boldsymbol{s})A_{K3\pi}^{\mathrm{PV}}(\boldsymbol{s}).\tag{2.69}$$

The full amplitude is then given by symmetrization

$$T_{K3\pi}(\boldsymbol{k}, \widehat{\boldsymbol{a}}^*) = \mathcal{S}\left\{T_{K3\pi}^{(u)}(\boldsymbol{k})_{\ell m}\right\},\tag{2.70}$$

using the definition in eq. (2.60) above. This completes the procedure for determining the decay amplitude from the finite-volume decay matrix elements. The physical interpretation of the factors in eq. (2.69) is as follows. $\mathcal{L}^{(u)}$ incorporates pairwise final state interactions, through multiple factors of $\mathcal{M}_2$ alternating with switch factors $G^\infty$. $T_{K3\pi}^{(u)}$ becomes complex both because $\mathcal{M}_2$ itself is complex, and due to the $i\epsilon$ in $G^\infty$. The quantity $X$ incorporates final state interactions involving all three particles, with intermediate pairwise scattering. Since this result derives from an all-orders diagrammatic derivation, the amplitude $T_{K3\pi}$ will automatically satisfy the required unitarity constraints, and in particular those that lead to Khuri-Treiman relations describing final-state interactions [53].

### 2.5 Isotropic approximation

We close this section by giving an explicit example of how the formalism works when making the simplest approximations to the decay and scattering amplitudes. We assume that only the leading, isotropic term in the threshold expansion of the decay amplitude, $A^{\mathsf{iso}}$, is nonvanishing — see eq. (2.48). This implies that $A^{\mathrm{PV}}_{K3\pi;k\ell m}$ is only nonzero for $\ell = m = 0$, and is independent of $k$. In addition, it couples only to three-pion states in the trivial irrep of the appropriate little group, e.g., the $A_1^-$ irrep for $\boldsymbol{P} = 0$ (for pions with negative intrinsic parity). For the amplitudes $\mathcal{M}_2$ and $\mathcal{K}_{\mathrm{df},3}$, we assume that only the $s$-wave contributes (so again $\ell = m = 0$) and that $\mathcal{K}_{\mathrm{df},3}$ is independent of the spectator momentum. This is equivalent to keeping only the isotropic term in the threshold expansion of $\mathcal{K}_{\mathrm{df},3}$ [10, 12].

Given these approximations, all quantities entering the definition of $F_3$ depend only on the spectator momenta. The isotropic nature of $A^{\mathrm{PV}}_{K3\pi}$ and $\mathcal{K}_{\mathrm{df},3}$ is represented by introducing the vector $|1\rangle$ in spectator-momentum space, which equals unity for all choices of $\boldsymbol{k}$ in the finite-volume set that lie below the cutoff. Specifically,

$$A^{\mathrm{PV}}_{K3\pi} \longrightarrow |1\rangle \, A^{\mathsf{iso}} \quad \text{and} \quad \mathcal{K}_{\mathrm{df},3} \longrightarrow |1\rangle \, \mathcal{K}^{\mathsf{iso}}_{\mathrm{df},3} \, \langle 1| \, , \tag{2.71}$$

where $A^{\mathsf{iso}}$ and $\mathcal{K}^{\mathsf{iso}}_{\mathrm{df},3}$ are constants. Using eq. (2.11), one then finds that

$$\frac{1}{F_3^{-1} + \mathcal{K}_{\mathrm{df},3}} \longrightarrow F_3 - F_3 \, |1\rangle \, \frac{1}{F_3^{\mathsf{iso}} + (\mathcal{K}^{\mathsf{iso}}_{\mathrm{df},3})^{-1}} \, \langle 1| \, F_3 \, , \tag{2.72}$$

where $F_3^{\mathsf{iso}}$ is the isotropic component of $F_3$,

$$F_3^{\mathsf{iso}} \equiv \langle 1| \, F_3 \, |1\rangle \, . \tag{2.73}$$

It follows that the only poles in three-particle correlators [e.g. $C_L^{\mathsf{M}}$ of eq. (2.5)] that depend on $\mathcal{K}^{\mathsf{iso}}_{\mathrm{df},3}$ occur when the isotropic quantization condition is satisfied, i.e.

$$F_3^{\mathsf{iso}} = -(\mathcal{K}^{\mathsf{iso}}_{\mathrm{df},3})^{-1} \, . \tag{2.74}$$

There are also solutions at free energies resulting from the $F_3$ terms in eq. (2.72), but these are an artifact of the isotropic approximation, as discussed in appendix F of ref. [12]. From eq. (2.72), we can determine the residue using eq. (2.36), finding

$$\mathcal{R}_n^{\mathsf{iso}} = F_3 \, |1\rangle \, r_n^{\mathsf{iso}} \, \langle 1| \, F_3 \, , \tag{2.75}$$

where we have abbreviated the arguments of $\mathcal{R}_{\Lambda\mu}(E_n, \boldsymbol{P}, L)$, and defined

$$r_n^{\mathsf{iso}} = -\left( \frac{\partial F_3^{\mathsf{iso}}(E, \boldsymbol{P}, L)}{\partial E} + \frac{\partial[1/\mathcal{K}^{\mathsf{iso}}_{\mathrm{df},3}(E^*)]}{\partial E} \right)^{-1} \Bigg|_{E = E_n^{A_1}(\boldsymbol{P}, L)} \, . \tag{2.76}$$

Here all derivatives are evaluated at the energy $E_n^{A_1}(\boldsymbol{P}, L)$, a solution to the isotropic quantization condition. The quantity $r_n^{\mathsf{iso}}$ is real in general, and positive for a physical solution. Thus we can read off the vector $v(E_n, \boldsymbol{P}, \Lambda\mu = A_1, L)$ defined in eq. (2.40),

$$(v_n^{\mathsf{iso}})^\dagger = (r_n^{\mathsf{iso}})^{1/2} \, \langle 1| \, F_3 \, . \tag{2.77}$$

– 17 –

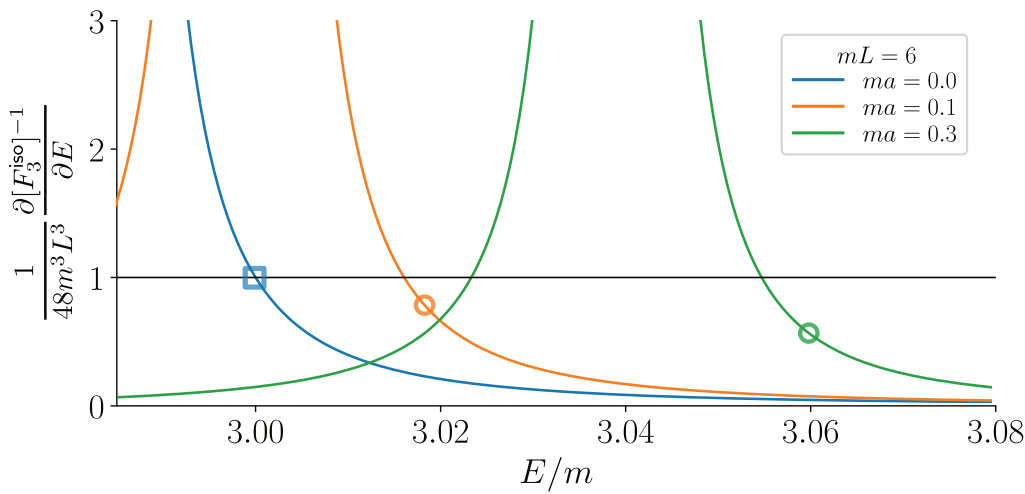

**Figure 2**. Plot of the conversion factor appearing in eq. (2.79) (rescaled as indicated by the plot label) in the vicinity of the three-particle threshold for the case of constant $\mathcal{K}_{\mathrm{df},3}^{\mathsf{iso}}$. The factor is plotted versus energy $E$ for $\boldsymbol{P} = \boldsymbol{0}$ and $mL = 6$. The two-particle K matrix, entering $F_3^{\mathsf{iso}}$, determined by keeping only the scattering length, $a$, in the effective range expansion. The three curves correspond to three values of the scattering length, as indicated by the legend, and each unfilled marker corresponds to the ground-state energy for the corresponding $ma$ value when $\mathcal{K}_{\mathrm{df},3}^{\mathsf{iso}} = 0$. In particular, the blue square corresponds to the non-interacting limit. The fact that the conversion factor is unity in the latter case indicates that the non-interacting matrix elements are equal in finite and infinite volume, up to a trivial normalization. More generally, once the scattering length is determined, these types of curves allow one to directly relate — within the isotropic approximation — any value of measured three-particle energy (horizontal axis) to a matrix element conversion factor (vertical axis).

Here we have chosen the overall phase according to the convention discussed above, so that $v_n^{\mathsf{iso}}$ is real. Using eq. (2.43) we now obtain

$$\sqrt{2E_K(\boldsymbol{P})}L^3 \langle E_n, \boldsymbol{P}, A_1, L| \, \mathcal{H}_W(0) \, |K, \boldsymbol{P}, L\rangle = (r_n^{\mathsf{iso}})^{1/2} F_3^{\mathsf{iso}} A^{\mathsf{iso}} \,. \tag{2.78}$$

This can be massaged into a simple form for determining $A^{\mathsf{iso}}$

$$
\begin{aligned}
A^{\mathsf{iso}}(E_n^*)^2 = {} & 2E_K(\boldsymbol{P})L^6 \langle E_n, \boldsymbol{P}, A_1, L| \, \mathcal{H}_W(0) \, |K, \boldsymbol{P}, L\rangle^2 \\
& \times \left( \frac{\partial F_3^{\mathsf{iso}}(E, \boldsymbol{P}, L)^{-1}}{\partial E} + \frac{\partial \mathcal{K}_{\mathrm{df},3}^{\mathsf{iso}}(E^*)}{\partial E} \right)_{E = E_n^{A_1}(\boldsymbol{P}, L)} \,.
\end{aligned} \tag{2.79}
$$

Thus, in the isotropic approximation, we need to measure the matrix element to only a single three-pion state in order to determine $A^{\mathsf{iso}}$ at that energy. In figure 2 we plot the conversion factor appearing on the second line of this equation for the case of constant $\mathcal{K}_{\mathrm{df},3}^{\mathsf{iso}}$, implying $\partial \mathcal{K}_{\mathrm{df},3}^{\mathsf{iso}}(E^*)/\partial E = 0$.

The relationship of $A^{\mathsf{iso}}$ to $T_{K3\pi}$ is also substantially simplified in the isotropic approximation. We first note that eq. (2.58) simplifies to

$$T_{K3\pi,L}^{(u),\mathsf{iso}} = \mathcal{L}_L^{(u)} |1\rangle \, \frac{1}{1 + \mathcal{K}_{\mathrm{df},3}^{\mathsf{iso}} F_3^{\mathsf{iso}}} A^{\mathsf{iso}} \,. \tag{2.80}$$

Taking the infinite volume limit as before, we obtain

$$T_{K3\pi}^{(u),\text{iso}}(\boldsymbol{k}, \widehat{\boldsymbol{a}}^*) = \mathcal{S}\left\{T_{K3\pi}^{(u),\text{iso}}(\boldsymbol{k})\right\}, \tag{2.81}$$

where

$$T_{K3\pi}^{(u),\text{iso}}(\boldsymbol{k}) = \mathcal{L}^{(u),\text{iso}}(\boldsymbol{k})\frac{A^{\text{iso}}}{1 + \mathcal{K}_{\text{df},3}^{\text{iso}}F_3^{\infty,\text{iso}}}. \tag{2.82}$$

Here the momentum dependence arises solely from the final-state interactions in

$$\mathcal{L}^{(u),\text{iso}}(\boldsymbol{k}) = \frac{1}{3} - \int_s \mathcal{D}_{23}^{(u,u)}(\boldsymbol{k}, \boldsymbol{s})\widetilde{\rho}_{\text{PV}}(\boldsymbol{s}), \tag{2.83}$$

where $\mathcal{D}_{23}^{(u,u)}(\boldsymbol{k}, \boldsymbol{s})$ still satisfies eq. (2.62), but now with all quantities restricted to $\ell = m = 0$, and

$$F_3^{\infty,\text{iso}} = \int_r \widetilde{\rho}_{\text{PV}}(\boldsymbol{r})\mathcal{L}^{(u),\text{iso}}(\boldsymbol{r}). \tag{2.84}$$

In this case, the only integral equation that has to be solved is that for $\mathcal{D}_{23}^{(u,u)}$, as has been done recently in refs. [27, 54]. We note that $F_3^{\infty,\text{iso}}$ and $\mathcal{L}^{(u),\text{iso}}$ are, in general, complex.

The expressions in the isotropic approximation are sufficiently simple that one can readily combine eqs. (2.79) and (2.82) to display the direct relation between the finite-volume matrix element and the physical amplitude. Unpacking the compact notation used above slightly, we reach

$$|T_{K3\pi}^{\text{iso}}(E^*, m_{12}^2, m_{23}^2)|^2 = 2E_K(\boldsymbol{P})L^6\Big|\langle E_n, \boldsymbol{P}, A_1, L|\,\mathcal{H}_W(0)\,|K, \boldsymbol{P}, L\rangle\Big|^2$$
$$\times \left|\mathcal{L}^{\text{iso}}(E^*, m_{12}^2, m_{23}^2)\frac{1}{1 + \mathcal{K}_{\text{df},3}^{\text{iso}}(E^*)F_3^{\infty,\text{iso}}(E^*)}\right|^2 \left(\frac{\partial F_3^{\text{iso}}(E, \boldsymbol{P}, L)^{-1}}{\partial E} + \frac{\partial \mathcal{K}_{\text{df},3}^{\text{iso}}(E^*)}{\partial E}\right), \tag{2.85}$$

where $E$ (and thus $E^*$) is fixed by the value of finite-volume energy, tuned to $E^* = M_K$ for a physical decay amplitude. We have emphasized that the right-hand side depends on the two squared invariant masses $m_{12}^2$ and $m_{23}^2$, defined by

$$m_{12}^2 = (E - \omega_k)^2 - (\boldsymbol{P} - \boldsymbol{k})^2, \tag{2.86}$$
$$m_{23}^2 = (E - \omega_a)^2 - (\boldsymbol{P} - \boldsymbol{a})^2, \tag{2.87}$$

and have also introduced the symmetrized final-state interaction factor.

$$\mathcal{L}^{\text{iso}}(E^*, m_{12}^2, m_{23}^2) \equiv \mathcal{L}^{(u),\text{iso}}(\boldsymbol{k}) + \mathcal{L}^{(u),\text{iso}}(\boldsymbol{a}) + \mathcal{L}^{(u),\text{iso}}(\boldsymbol{b}). \tag{2.88}$$

At this stage we can comment on the relationship of our result to that of ref. [52]. We expect that the isotropic limit, given in eq. (2.85), is equivalent to the result of ref. [52], aside from differences in the schemes used to define the short-distance quantities. Indeed, the equations have the same basic structure, with a contribution resulting from final state interactions (the term involving $\mathcal{L}^{\text{iso}}$) and a Lellouch-Lüscher-like correction factor. Demonstrating the precise equivalence, however, is nontrivial, since our approach based

in short-distance quantities, $\mathcal{K}_{\mathrm{df},3}$ and $A_{K3\pi}^{\mathrm{PV}}$, that are symmetric under particle exchange, whereas the approach of ref. [52] does not symmetrize until the very end. Presumably, the mapping can be determined using the relation between symmetric and asymmetric approaches explained in refs. [16, 17], but this is beyond the scope of the present work.

In closing, we note that eq. (2.85) is analogous to the original Lellouch-Lüscher relation presented in ref. [32]. In particular, the two-particle result is reached by making the replacements

$$T_{K3\pi}^{\mathsf{iso}}(E^*, m_{12}^2, m_{23}^2) \longrightarrow T_{K2\pi}(E) \,, \tag{2.89}$$

$$\mathcal{L}^{\mathsf{iso}}(E^*, m_{12}^2, m_{23}^2) \longrightarrow 1 \,, \tag{2.90}$$

$$\mathcal{K}_{\mathrm{df},3}^{\mathsf{iso}}(E^*) \longrightarrow \mathcal{K}_2(E) \,, \tag{2.91}$$

$$F_3^{\infty,\mathsf{iso}}(E^*) \longrightarrow -i\rho(E) \equiv -i\frac{q}{16\pi E} \,, \tag{2.92}$$

$$F_3^{\mathsf{iso}}(E, \boldsymbol{P}, L) \longrightarrow F(E, L) \,, \tag{2.93}$$

where we have also restricted attention to the $\boldsymbol{P} = \boldsymbol{0}$ frame. On the right-hand side we have introduced the physical $K \to \pi\pi$ amplitude $T_{K2\pi}(E)$, extended to allow for final-state energies different from the kaon mass. We have also used the two-particle K-matrix, $\mathcal{K}_2$, and the two-particle finite-volume function, $F$, both restricted to the $s$-wave. These are essentially the same quantities as appearing in eq. (2.6), in the definition of $F_3$, but without the implicit sub-threshold regulator used there and without the spectator-momentum index. We have also introduced the two-particle phase-space, $\rho(E)$, with $q = \sqrt{E^2/4 - m^2}$.

Making the indicated substitutions into eq. (2.85) yields

$$|T_{K2\pi}(E)|^2 = 2M_K L^6 \, |\langle E_n, A_1, L| \, \mathcal{H}_W(0) \, |K, L\rangle|^2$$
$$\times \left| \frac{1}{1 - i\mathcal{K}_2(E)\rho(E)} \right|^2 \left( \frac{\partial F(E, L)^{-1}}{\partial E} + \frac{\partial \mathcal{K}_2(E)}{\partial E} \right) . \tag{2.94}$$

Substituting the definitions of the scattering phase $\delta(E)$ and the $L$-dependent, so-called pseudophase $\phi(E, L)$

$$\mathcal{K}_2(E) = \frac{16\pi E \tan \delta(E)}{q} \,, \qquad F(E, L)^{-1} = \frac{16\pi E \tan \phi(E, L)}{q} \,, \tag{2.95}$$

one can easily reach eq. (4.5) of ref. [32], after some algebraic manipulations.

This completes our discussion of the formalism in the context of the simplified theory. We now turn to realistic applications of these results.

## 3 Applications to physical processes

In this section, we describe the generalization of the previous analysis to processes involving three-pion final states in isosymmetric QCD. This allows our results to be applied to several processes of phenomenological interest: (i) the electromagnetic transition $\gamma^* \to 3\pi$, which contributes to the hadronic vacuum polarization piece of the muon's magnetic momentum,

$(g-2)_\mu$; (ii) the isospin-violation strong decay $\eta \to 3\pi$; and (iii) the weak decay $K \to 3\pi$, which has both CP-conserving and violating amplitudes.

The generalization presented here requires the generic three-pion quantization condition derived in ref. [18]. We start this section by recalling some results from that work, and presenting the generalization of the formulae derived above to the three-pion system. We then describe the specific applications to the three processes listed above.

### 3.1 General considerations

In the derivation in section 2, the "kaon" and "pion" fields were taken to be real scalars with separate $\mathbb{Z}_2$ symmetries. Here we consider the physical kaon and pion fields. The former, which can be either charged or neutral, are complex fields with strangeness conservation playing the role of the $\mathbb{Z}_2$ symmetry. The pions are represented by a triplet of fields, with two complex fields in the definite charge basis ($\pi^+$ and $\pi^-$) and one real filed ($\pi^0$), with the $\mathbb{Z}_2$ symmetry being G parity. Both kaons and pions are stable particles in QCD, with masses satisfying the required inequality, eq. (2.1). The form of the weak Hamiltonian depends on the decay being considered, but its essential property, unchanged from above, is that it annihilates one of the kaons and creates three pions. The new feature is the presence of multiple three-pion intermediate states, e.g. $\pi^+\pi^0\pi^-$ and $\pi^0\pi^0\pi^0$ in the neutral sector, and it is this feature that the derivation of ref. [18] takes into account.

We stress again that, since the weak interactions are added by hand as external operators, we can choose to separately consider operators that create three and two pions, with G parity ensuring that these two sectors do not mix. We can also consider one at a time operators that create three pions in states of definite isospin. Indeed, the quantization condition of ref. [18] decomposes into separate results for each choice of total isospin. Finally, we note that, although we couch the discussion in this subsection in terms of the $K \to 3\pi$ decay, the essential aspects of the discussion apply equally well if the kaon is replaced by a $\gamma^*$ or $\eta$, and the weak operator is replaced by the electromagnetic current or the isospin-breaking Hamiltonian, respectively.

A generic three-pion state can have total isospin $I = 0, 1, 2$ and 3. It is, however, important to note that the isospin of any pair of particles is not conserved — for a given total isospin there can be several two-pion subchannels with pairwise interactions. As discussed in ref. [18], the following subchannels contribute

$$
\begin{aligned}
I &= 0: \left\{ \, |\rho\pi\rangle_0 \, \right\}, \\
I &= 1: \left\{ \, |\sigma\pi\rangle_1, |\rho\pi\rangle_1, |(\pi\pi)_2\pi\rangle_1 \, \right\}, \\
I &= 2: \left\{ \, |\rho\pi\rangle_2, |(\pi\pi)_2\pi\rangle_2 \, \right\}, \\
I &= 3: \left\{ \, |(\pi\pi)_2\pi\rangle_3 \, \right\},
\end{aligned}
\tag{3.1}
$$

where "$\sigma$","$\rho$","$(\pi\pi)_2$" label a two-pion combination with isospin 0,1, and 2, respectively, and the subscripts on the kets denotes the total isospin. Explicit expressions for these states for the charge zero ($I_3 = 0$) sector are given in appendix C of ref. [18].

The order of pion fields in each state of eq. (3.1) is a shorthand for the interplay of momentum and isospin assignment. In particular, if we consider asymptotic states with

fixed total energy and momentum $(E, \boldsymbol{P})$ then the remaining degrees of freedom, $\ell m$ and $\boldsymbol{k}$, are assigned to the leading pion pair and the third pion field, respectively. As emphasized in section 2.1, the asymmetric description is natural from the perspective of the finite-volume formalism, since many of the quantities appearing there, in particular $F$, $G$, $\mathcal{K}_2$ and $F_3$, single out a pion pair in their definition. The result is that there are additional flavor spaces with dimensions one, three, two and one, for $I = 0, 1, 2, 3$ respectively. Aside from this feature, and a minor change in notation (to be described below), the forms of the final results in ref. [18] are the same as those for identical particles reviewed in section 2.1.

The simplicity of the generalization from three identical particles to three-pion states carries over to the new quantities needed to discuss decay matrix elements. For this reason we only quote the results. We begin with the generalization of the Euclidean correlator $C_{AB,L}(P)$, defined in eq. (2.32). The operators $\mathcal{A}$ and $\mathcal{B}$ now respectively destroy and create a three-pion state of definite isospin. The expression for this correlator, previously given by eq. (2.33), now becomes

$$C_{AB,L}^{[I]} = C_{AB,\infty}^{[I]} - i\mathbf{A}^{\mathrm{PV},[I]} \frac{1}{[\mathbf{F}_3^{[I]}]^{-1} - \mathbf{K}_{\mathrm{df},3}^{[I]}} \mathbf{B}^{\mathrm{PV},[I]} \,. \tag{3.2}$$

The notation for bold-faced quantities is taken over from ref. [18]: they contain a factor of $i$ compared to those used for identical particles [which explains differences in signs and factors of $i$ compared to eq. (2.33)] and also have an additional index corresponding to the flavor space described above. For example, for $I = 1$, the endcap $\mathbf{A}^{\mathrm{PV},[I]}$ is a three-dimensional row vector in these indices (in addition to being a row vector in the $k\ell m$ indices), while $\mathbf{F}_3^{[I]}$ and $\mathbf{K}_{\mathrm{df},3}^{[I]}$ are $3 \times 3$ flavor matrices (as well as being matrices in the $k\ell m$ indices).[8] The explicit expressions for the flavor structure of $\mathbf{F}_3^{[I]}$ are given in table 1 of ref. [18] and we do not repeat them here.

With eq. (3.2) in hand, the derivation in section 2.2 goes over almost verbatim. One uses the same three correlators, eqs. (2.20)–(2.22), except for the above-described changes to the kaon field and the three-pion operators. The final result is a generalization of eq. (2.43):

$$\sqrt{2E_K(\boldsymbol{P})}L^3 \langle E_n^{\Lambda,[I]}, \boldsymbol{P}, I, I_3, \Lambda\mu, L | \mathcal{H}_W(0) | K, \boldsymbol{P}, L \rangle = \mathbf{v}^\dagger \mathbf{A}_{K3\pi}^{\mathrm{PV},[I]} \,. \tag{3.3}$$

The matrix element on the left-hand side is obtained from the lattice simulation with the kaon state having the desired quantum numbers, and $E_n^{\Lambda,[I]}$ being the energy of a three-pion state of chosen isospin and hypercubic-group irrep. We assume that the weak Hamiltonian couples the kaon to this state, for otherwise the equation is trivially satisfied as both sides vanish. On the right-hand side the column vector $\mathbf{v}$ is an abbreviation for

$$\mathbf{v}(E_n^{\Lambda,[I]}, \boldsymbol{P}, I, I_3, \Lambda\mu, L) \,, \tag{3.4}$$

---

[8] One difference compared to ref. [18] is that the endcaps in that work are matrices in flavor space, while those here are row or column vectors. This reflects the fact that creation and annihilation operators in ref. [18] were chosen to create three-pion states of all isospins, whereas here we consider single operators with definite three-pion isospin.

which is a row vector having both $\{k\ell m\}$ and flavor indices, and includes a factor of $i$ relative to the $v$ of section 2.2 in order to cancel the factor of $i$ in $\mathbf{A}_{K3\pi}^{\mathrm{PV},[I]}$. It is an eigenvector of $[\mathbf{F}_3^{[I]}]^{-1} - \mathbf{K}_{\mathrm{df},3}^{[I]}$ with vanishing eigenvalue, and is defined by the generalization of eq. (2.40):

$$\mathbf{R}_{\Lambda\mu}^{[I,I_3]}(E_n^{\Lambda,[I]}, \boldsymbol{P}, L) = \lim_{P_4 \to iE_n^{\Lambda,[I]}} -(E_n^{\Lambda,[I]} + iP_4)\mathbb{P}_{\Lambda\mu}^{[I,I_3]}\frac{(-i)}{1/\mathbf{F}_3^{[I]} - \mathbf{K}_{\mathrm{df},3}^{[I]}}\mathbb{P}_{\Lambda\mu}^{[I,I_3]} \equiv \mathbf{v}\,\mathbf{v}^\dagger. \quad (3.5)$$

We stress that we do *not* include a relative factor of $i$ between the definitions of $\mathbf{R}_{\Lambda\mu}^{[I,I_3]}$ and $\mathcal{R}_{\Lambda\mu}$ of section 2.2. The bold quantity defined here thus differs from the $\mathcal{R}_{\Lambda\mu}$ only by the addition the flavor index.

The workflow for using eq. (3.3) is as follows: first, one chooses the initial kaon quantum numbers and the form of $\mathcal{H}_W$ based on the physical process under consideration. This determines the allowed values of $I$ and $I_3$ for the three-pion final states. Second, one calculates the three-pion energy spectrum for one of the allowed values of $\{I, I_3\}$, using a range of choices of $\boldsymbol{P}$, and picking irreps/rows $\Lambda\mu$ such that the desired $K \to 3\pi$ matrix element is nonvanishing. Third, one compares this spectrum to the result from the quantization condition of ref. [18],

$$\det\left([\mathbf{F}_3^{[I]}]^{-1} - \mathbf{K}_{\mathrm{df},3}^{[I]}\right) = 0\,, \quad (3.6)$$

and uses this to determine (a parameterized form of) $\mathbf{K}_{\mathrm{df},3}^{[I]}$. Fourth, with this form in hand one uses eq. (3.5) to determine the vectors $\mathbf{v}$ for levels that have their energies matched to $E_K(\boldsymbol{P})$. Finally, one uses eq. (3.3) to provide a constraint on the row vector $\mathbf{A}_{K3\pi}^{\mathrm{PV},[I]}$. By combining several such constraints can determine a (parametrized form of) $\mathbf{A}_{K3\pi}^{\mathrm{PV},[I]}$.

The second step — connecting to the physical decay amplitude — also mirrors that for identical particles, which was described in section 2.4. One first introduces an asymmetric finite-volume amplitude that generalizes eq. (2.58),

$$\mathbf{T}_{K3\pi,L}^{[I](u)} = \left(\mathbf{F}^{[I]}\right)^{-1}\mathbf{F}_3^{[I]}\frac{1}{1 - \mathbf{K}_{\mathrm{df},3}^{[I]}\mathbf{F}_3^{[I]}}\mathbf{A}_{K3\pi}^{\mathrm{PV},[I]}\,, \quad (3.7)$$

where $\mathbf{F}^{[I]}$ is $iF/(2\omega L^3)$ tensored with the identity in the corresponding flavor space [18]. Here again the boldfaced quantity $\mathbf{T}_{K3\pi,L}^{[I](u)}$ differs from the $T_{K3\pi,L}^{(u)}$ used in section 2.4 both by the addition of flavor indices and by a factor of $i$. The physical amplitude is then obtained by taking the appropriate ordered limit and symmetrizing,

$$\mathbf{T}_{K3\pi}^{[I]} = \mathcal{S}\left\{\lim_{\epsilon\to 0^+}\lim_{L\to\infty}\mathbf{T}_{K3\pi,L}^{[I](u)}\right\}\,. \quad (3.8)$$

This limit leads to integral equations that are simple generalizations of those presented in section 2.4, and which we do not display explicitly. The only subtlety that is introduced by the flavor indices is the need to generalize the definition of symmetrization, as is explained in section 2.3 of ref. [18]. We stress that the symmetrization here acts on a column vector with a single index, rather than on a matrix as in ref. [18].

The results of these steps are the infinite-volume decay amplitudes in the isospin basis. To convert to a measurable amplitude, e.g. that for $K^+ \to \pi^+\pi^0\pi^0$, one must combine the

isospin amplitudes appropriately. The results needed to do this are collected in appendix C. In this regard there is a further subtlety concerning the amplitudes that have a multi-dimensional flavor space, i.e. those with $I = 1$ and 2. To explain this point (which is not discussed in ref. [18]) we focus on the example of $I = 1$. The result from eq. (3.8) is then *three* $K \to [3\pi]_{I=1}$ amplitudes, each expressed as a function of the three pion momenta. The issue is that, when one has the full momentum dependence, these three amplitudes are not independent. In fact, as we explain below, one needs to know only two of the three in order to completely reconstruct the $I = 1$ amplitude. Similarly, for the $I = 2$ case, only one of the two amplitudes is needed. This redundancy does not, however, lead to any simplification in the solution of the integral equations implicit in eq. (3.8).

## 3.2 The electromagnetic transition $\gamma^* \to 3\pi$

The electromagnetic process $\gamma^* \to 3\pi$ is of phenomenological interest as it contributes, via the hadronic vacuum polarization (HVP) and the hadronic light-by-light scattering (HLbL), to the anomalous magnetic moment of the muon [55–59]. Our formalism allows one to determine the infinite volume amplitude using a finite volume lattice QCD calculation. In particular, although this is not a decay, the results above are readily adapted — one simply takes advantage of the fact that one can allow the final three-particle state to take on any energy and momentum in the relations given above. This then corresponds to a timelike photon with virtuality $q^2 = E_n^\Lambda(L, \boldsymbol{P})^2 - \boldsymbol{P}^2$. The analogous two-particle process, $\gamma^* \to \pi\pi$, and its relation to finite-volume matrix elements is discussed in ref. [37].

The replacement of the kaon with a virtual photon simplifies the required lattice calculation. The composite operator $\mathcal{B}_{K3\pi}(x)$ in eq. (2.26) is replaced by the electromagnetic current $\mathcal{J}_\nu(x)$, and the kaon correlator is not required. We consider here only the part of this current that involves up and down quarks,

$$\mathcal{J}_\nu = \frac{2}{3}\bar{u}\gamma_\nu u - \frac{1}{3}\bar{d}\gamma_\nu d \,, \tag{3.9}$$

as this leads to the dominant contribution to $\gamma^* \to 3\pi$. No tuning of the volume is needed to match a given energy; instead, each finite-volume three pion state with appropriate quantum numbers leads to a result for the desired amplitude with photon virtuality given by the energy of the state.

The electromagnetic current contains both isoscalar and isovector parts. The latter has positive G parity and thus, in isosymmetric QCD, couples only to even numbers of pions, and in particular to the $\rho$ resonance. What is of interest here is the isoscalar part,

$$\mathcal{J}_\nu^0 = \frac{1}{6}\left(\bar{u}\gamma_\nu u + \bar{d}\gamma_\nu d\right) \,, \tag{3.10}$$

which has negative G parity and thus couples to three pions. The dominant contribution in the energy range of interest for muonic $g - 2$ is from the $\omega(782)$ resonance.

The desired amplitude is obtained using the two-step process explained above. Each matrix element obtained from a lattice calculation is related to the intermediate PV amplitude by

$$L^{3/2}\langle E_n^{\Lambda,[0]}, \boldsymbol{P}, I = 0, \Lambda\mu, L|\mathcal{J}_\nu^0(0)|0\rangle = \mathbf{v}^\dagger \mathbf{A}_{\gamma3\pi,\nu}^{\mathrm{PV},[0]} \,, \tag{3.11}$$

where $\mathbf{v} = \mathbf{v}^{[0]}(E_n^{\Lambda,[0]}, \boldsymbol{P}, I = 0, \Lambda\mu, L)$ is obtained from the spectrum of $I = 0$ three pion states using eq. (3.5). The irreps $\Lambda$ and rows $\mu$ that lead to nonzero matrix elements depend on the total momentum $\boldsymbol{P}$ and the Lorentz index $\nu$. Note that for $I = 0$ the flavor space is one dimensional, so $\mathbf{A}_{\gamma 3\pi,\nu}^{\mathrm{PV},[0]}$ and $\mathbf{v}$ can be viewed as vectors in $\{k\ell m\}$ space alone. We also comment that the left-hand side of eq. (3.11) differs from the corresponding results for kaon decays, eqs. (2.43) and (3.3), by the absence of a factor of $(2E_K(\boldsymbol{P})L^3)^{1/2}$. This is because, in contrast to the unit normalized finite-volume kaon state, there is no need to correct the normalization of the vacuum, which matches between the finite- and infinite-volume theories.

To implement eq. (3.11), the infinite-volume PV amplitude $\mathbf{A}_{\gamma 3\pi,\nu}^{\mathrm{PV},[0]}$ must be parametrized. This is most easily done by using eq. (2.50) to convert from $\{k\ell m\}$ space to a function of three on-shell momenta, $p_1$, $p_2$ and $p_3$. Up to the overall factor of $i$, the amplitude is a real, smooth function of momenta, antisymmetric under the interchange of any pair of momenta, and transforming as an axial vector.[9] Expanding about threshold as in section 2.3 (with $m_K^2 \to q^2$), the general form satisfying these properties is

$$\mathbf{A}_{\gamma 3\pi,\mu}^{\mathrm{PV},[0]} = i\epsilon_{\mu\nu\rho\sigma}p_1^\nu p_2^\rho p_3^\sigma \left( A_{\gamma 3\pi}^{(0)} + A_{\gamma 3\pi}^{(2)} \sum_i \Delta_i^2 + \dots \right) . \tag{3.12}$$

Here the $\Delta_i$ are the threshold expansion parameters defined in eq. (2.47), and the coefficients $A_{\gamma 3\pi}^{(n)}$ are functions of $\Delta = q^2/(9m_\pi^2) - 1$. For a consistent threshold expansion, $A_{\gamma 3\pi}^{(0)}$ should be a quadratic function of $\Delta$, while $A_{\gamma 3\pi}^{(2)}$ should be a constant. The ellipsis represents higher order terms. We observe that the threshold expansion begins at higher order than for the symmetric amplitude discussed in section 2.3.

The second step is to solve the integral equations encoded in the $I = 0$ versions of eqs. (3.7) and (3.8), which convert $\mathbf{A}_{\gamma 3\pi,\nu}^{\mathrm{PV},[0]}$ into the $\gamma^* \to [3\pi]_{I=0}$ amplitude, $\mathbf{T}_{\gamma 3\pi,\nu}^{[0]}(p_1, p_2, p_3)$. Recalling from ref. [18] that the $I = 0$ state is given by

$$\frac{1}{\sqrt{6}} \left( |\pi^+\pi^0\pi^-\rangle - |\pi^0\pi^+\pi^-\rangle + |\pi^0\pi^-\pi^+\rangle - |\pi^-\pi^0\pi^+\rangle + |\pi^-\pi^+\pi^0\rangle - |\pi^+\pi^-\pi^0\rangle \right) , \tag{3.13}$$

with the three pions in each ket having the momenta $p_1$, $p_2$ and $p_3$, respectively, and noting that only the $I = 0$ amplitude is nonzero, we obtain the physical amplitude as

$$i\mathcal{T}\left[ \gamma^* \to \pi^+(p_1)\pi^0(p_2)\pi^-(p_3) \right] = \sqrt{\frac{1}{6}}\mathbf{T}_{\gamma 3\pi,\nu}^{[0]}(p_1, p_2, p_3), \tag{3.14}$$

where the index $\nu$ refers to the polarization of the virtual photon.[10]

### 3.3 The isospin-violating strong decay $\eta \to 3\pi$

The decay $\eta \to 3\pi$ provides an example where our formalism can be used within the context of the strong interactions. The key point is that the $\eta$ is stable in isosymmetric

---

[9]If the intrinsic negative parity of the pions is included the amplitude transforms as a vector, as required to couple to the electromagnetic current.

[10]This can also be obtained from the bottom row of the matrix $\mathcal{R}$ given in eq. (C.3). Since only the $I = 0$ amplitude is nonzero, the rightmost entry in this row gives the relevant factor.

QCD, but can decay to three pions in the presence of isospin violation.[11] The decay has a very small partial width, $\Gamma(\eta \to 3\pi) \approx 0.7$ keV [60], and can be treated at leading order in an expansion in isospin breaking. Isospin violation in the Standard Model arises both from the up-down quark mass difference in QCD and from electromagnetic effects. Here, however, isospin breaking from QCD dominates, since electromagnetic effects are of second order in $\alpha$ due to the neutrality of the $\eta$. Thus this process is uniquely sensitive to the up-down quark mass difference. We refer the reader to ref. [61] for a recent review of the status of phenomenological predictions for these decays.

A natural approach for a first-principles lattice QCD calculation of these decay amplitudes is to simulate isosymmetric QCD with mass term

$$\mathcal{H}^{\Delta I=0} = \frac{m_u + m_d}{2} \left( \bar{u}u + \bar{d}d \right) , \tag{3.15}$$

but introduce isospin violation through the insertion of the mass difference operator[12]

$$\mathcal{H}^{\Delta I=1} = \frac{m_u - m_d}{2} \left( \bar{u}u - \bar{d}d \right) . \tag{3.16}$$

This brings the calculation into the same class as that for $K \to 3\pi$ decays, with the initial kaon replaced by the $\eta$ and $\mathcal{H}_W$ replaced by $\mathcal{H}^{\Delta I=1}$. We observe that, although isospin-breaking is being included only at leading order, our formalism includes all rescattering effects due to final state interactions. Thus it provides an alternative to the dispersive methods used in present analyses [64, 65].

Since the initial $\eta$ has $I = 0$, the final three pion state has $I = 1$. Thus to obtain the $\eta \to 3\pi$ amplitude we can use the results of section 3.1, by simply making the replacement $K \to \eta$, and taking $I = 1$. In this way, we can use the formalism to determine the intermediate PV amplitude $\mathbf{A}_{\eta 3\pi}^{\mathrm{PV},[1]}$ and the final, physical amplitude $\mathbf{T}_{\eta 3\pi}^{[1]}$. We note that these amplitudes have a three-dimensional flavor space. For a practical implementation one needs a parametrization of $\mathbf{A}_{\eta 3\pi}^{\mathrm{PV},[1]}$, and the relation of $\mathbf{T}_{\eta 3\pi}^{[1]}$ to the amplitudes into charged and neutral pions. We provide these results in the remainder of this subsection.

To present the parametrization of $\mathbf{A}_{\eta 3\pi}^{\mathrm{PV},[1]}$, it is convenient to use a different basis for the flavor space of three-pion states than that of eq. (3.1). The new basis, which we denote the $\chi$ basis, uses states that lie in irreps of the symmetric group $S_3$ corresponding to permutations of the three particles. It is given by [18]

$$\begin{aligned}
\{ |\chi_s\rangle_1 , |\chi_1\rangle_1 , |\chi_2\rangle_1 \} = \\
\left\{ \frac{2}{3} |(\pi\pi)_2\pi\rangle_1 + \frac{\sqrt{5}}{3} |\sigma\pi\rangle_1 , \ -\frac{\sqrt{5}}{3} |(\pi\pi)_2\pi\rangle_1 + \frac{2}{3} |\sigma\pi\rangle_1 , \ |\rho\pi\rangle_1 \right\} ,
\end{aligned} \tag{3.17}$$

where $|\chi_s\rangle$ transforms in the trivial irrep of $S_3$, while $\{|\chi_1\rangle , |\chi_2\rangle\}$ transform in the two-dimensional standard irrep. We refer to appendix C in ref. [18] for explicit expressions for the isospin states, as well as further discussion of the group properties.

---

[11]Potential decays to $2\pi$ and $4\pi^0$ that are allowed by G parity and kinematics are forbidden by parity conservation, irrespective of isospin breaking.

[12]We note that this method of calculating isospin-violating effects is similar to the perturbative method introduced in refs. [62, 63], but differs in that here we imagine inserting the operator at a single position rather than over the entire volume.

We now adapt the results obtained in ref. [18] for the parametrizations of scattering amplitudes to that of the intermediate PV amplitude. Working to quadratic order in the threshold expansion, we find

$$\mathbf{A}_{\eta 3\pi}^{\mathrm{PV},[1]} = i \left( A_{\eta 3\pi}^{\mathsf{s},0} + A_{\eta 3\pi}^{\mathsf{s},1} \Delta + A_{\eta 3\pi}^{\mathsf{s},2} \Delta^2 + A_{\eta 3\pi}^{\mathsf{s},2a} \sum_i \Delta_i^2 \right) \begin{pmatrix} 1 \\ 0 \\ 0 \end{pmatrix}$$

$$+ i \left( A_{\eta 3\pi}^{\mathsf{d},1} + A_{\eta 3\pi}^{\mathsf{d},2} \Delta \right) \begin{pmatrix} 0 \\ P \cdot \xi_1 \\ P \cdot \xi_2 \end{pmatrix} + i A_{\eta 3\pi}^{\mathsf{d},2a} \begin{pmatrix} 0 \\ (P \cdot \xi_2)^2 - (P \cdot \xi_1)^2 \\ 2 P \cdot \xi_1 P \cdot \xi_2 \end{pmatrix} + \ldots, \quad (3.18)$$

where $A_{\eta 3\pi}^{\mathsf{s},0}$, etc. are real coefficients. The notation is as in section 2.3, except for the replacement $m_K \to m_\eta$, and the use of the new quantities

$$\xi_1 = \frac{1}{\sqrt{6}} \left( 2p_3 - p_1 - p_2 \right), \quad \text{and} \quad \xi_2 = \frac{1}{\sqrt{2}} \left( p_2 - p_1 \right). \quad (3.19)$$

The superscripts $\mathsf{s}$ and $\mathsf{d}$ refer to the "singlet" symmetric and "doublet" standard irrep of $S_3$, respectively. We observe that the symmetric part of the amplitude begins at leading order in the threshold expansion, while that transforming in the doublet enters only at linear order.

Finally we describe the reconstruction of the decay amplitudes into final states composed of pions with definite charges, which are

$$\mathcal{T}_\eta^{000}(p_1, p_2, p_3) \equiv \mathcal{T}[\eta \to \pi^0(p_1)\pi^0(p_2)\pi^0(p_3)], \quad (3.20)$$

$$\mathcal{T}_\eta^{+0-}(p_1, p_2, p_3) \equiv \mathcal{T}[\eta \to \pi^+(p_1)\pi^0(p_2)\pi^-(p_3)]. \quad (3.21)$$

Our formalism yields the $I = 1$ amplitude, which, expressed in the $\chi$ basis, is

$$\mathbf{T}_{\eta 3\pi}^{[1]}(p_1, p_2, p_3) = i \begin{pmatrix} \mathcal{T}_\mathsf{s}^{[1]}(p_1, p_2, p_3) \\ \mathcal{T}_{\mathsf{d},1}^{[1]}(p_1, p_2, p_3) \\ \mathcal{T}_{\mathsf{d},2}^{[1]}(p_1, p_2, p_3) \end{pmatrix}. \quad (3.22)$$

The relation between the $\chi$ basis and that involving particles of definite charge is given in eq. (C.3). Using this result, and the fact that the amplitudes for $I = 0$, 2, and 3 vanish, we obtain

$$\mathcal{T}_\eta^{000}(p_1, p_2, p_3) = -\sqrt{\frac{3}{5}} \mathcal{T}_\mathsf{s}^{[1]}(p_1, p_2, p_3), \quad (3.23)$$

$$\mathcal{T}_\eta^{+0-}(p_1, p_2, p_3) = \frac{1}{\sqrt{15}} \mathcal{T}_\mathsf{s}^{[1]}(p_1, p_2, p_3) - \frac{1}{\sqrt{12}} \mathcal{T}_{\mathsf{d},1}^{[1]}(p_1, p_2, p_3) + \frac{1}{2} \mathcal{T}_{\mathsf{d},2}^{[1]}(p_1, p_2, p_3). \quad (3.24)$$

We note that all three $I = 1$ amplitudes are invariant under the interchange $p_1 \leftrightarrow p_3$, so that $\mathcal{T}_\eta^{+0-}(p_1, p_2, p_3) = \mathcal{T}_\eta^{+0-}(p_3, p_2, p_1)$, which is consistent with the positive charge conjugation parity of the pseudoscalar mesons.

As noted earlier, the two doublet amplitudes are not independent when one uses the freedom to permute the momenta. A convenient form of this relationship is

$$\mathcal{T}_{\mathsf{d},2}^{[1]}(p_1, p_2, p_3) = \frac{1}{\sqrt{3}} \mathcal{T}_{\mathsf{d},1}^{[1]}(p_1, p_2, p_3) + \frac{2}{\sqrt{3}} \mathcal{T}_{\mathsf{d},1}^{[1]}(p_1, p_3, p_2), \quad (3.25)$$

where we stress that the order of the momentum arguments differs in the last term. Using this result, eq. (3.24) can be rewritten as

$$\mathcal{T}_\eta^{+0-}(p_1, p_2, p_3) = \frac{1}{\sqrt{15}} \mathcal{T}_{\mathsf{s}}^{[1]}(p_1, p_2, p_3) + \frac{1}{\sqrt{3}} \mathcal{T}_{\mathsf{d},1}^{[1]}(p_1, p_3, p_2) \,. \tag{3.26}$$

## 3.4 The weak decay $K \to 3\pi$

Finally, we turn to the $K \to 3\pi$ decays that are the primary motivation for this work. We have left these processes to the end as they are the most complicated to analyze. The main reason for developing the formalism for a lattice calculation of the $K \to 3\pi$ amplitudes is to provide a method for determining the CP-violating contribution, so as to allow further tests of the Standard Model. This is analogous to the situation with $K \to 2\pi$ decays, where the well-measured CP-violating quantity $\epsilon'/\epsilon$ can now be predicted reliably in the Standard Model using lattice QCD [66–68].

In the three-particle case, the decay amplitudes are

$$\begin{aligned}
\mathcal{T}_K^{+00}(p_1, p_2, p_3) &\equiv \mathcal{T}[K^+ \to \pi^+(p_1)\pi^0(p_2)\pi^0(p_3)] \,, \\
\mathcal{T}_K^{-++}(p_1, p_2, p_3) &\equiv \mathcal{T}[K^+ \to \pi^-(p_1)\pi^+(p_2)\pi^+(p_3)] \,,
\end{aligned} \tag{3.27}$$

together with their charge conjugates, and the neutral kaon amplitudes

$$\begin{aligned}
\mathcal{T}_{K_S}^{+-0}(p_1, p_2, p_3) &\equiv \mathcal{T}[K_S \to \pi^+(p_1)\pi^0(p_2)\pi^-(p_3)] \,, \\
\mathcal{T}_{K_S}^{000}(p_1, p_2, p_3) &\equiv \mathcal{T}[K_S \to \pi^0(p_1)\pi^0(p_2)\pi^0(p_3)] \,, \\
\mathcal{T}_{K_L}^{+-0}(p_1, p_2, p_3) &\equiv \mathcal{T}[K_L \to \pi^+(p_1)\pi^0(p_2)\pi^-(p_3)] \,, \\
\mathcal{T}_{K_L}^{000}(p_1, p_2, p_3) &\equiv \mathcal{T}[K_L \to \pi^0(p_1)\pi^0(p_2)\pi^0(p_3)] \,.
\end{aligned} \tag{3.28}$$

In the absence of CP violation, all are nonzero except for $\mathcal{T}_{K_S}^{000}$. All have been measured except for those for neutral kaon decays to $3\pi^0$ [60]. The effects of CP violation that are measurable at present involve the charged kaon decays. Specifically, CP violation shows up as a difference between Dalitz plot slope parameters in $K^+$ and $K^-$ decays (see ref. [69] for a review). Experimentally, these differences are on the edge of observability [70, 71]. Phenomenological predictions for CP violating observables achieve a comparatively higher accuracy [72, 73]. In light of this situation, we focus here on the formalism for the decays of charged kaons, and specifically on the $K^+$ decay. The generalization to the $K^-$ decay is straightforward, and that for the neutral kaon decays is summarized in appendix D.

The operators needed for a lattice study of this process are those of the effective electroweak Hamiltonian, $\mathcal{H}_W$. The set of operators that are relevant after running to scales below the charm mass is given for instance in refs. [74, 75]. Since $\mathcal{H}_W$ contains only operators that change isospin by $1/2$ or $3/2$, the allowed total isospin of the $3\pi$ state is $I = 0$, 1 and 2. For charged kaons only decays to $I = 1$ and 2 amplitudes are allowed. Using the formalism described above, a lattice calculation can determine (constraints on) the intermediate amplitudes $\mathbf{A}_{K3\pi}^{\mathrm{PV},[1]}$ and $\mathbf{A}_{K3\pi}^{\mathrm{PV},[2]}$. We stress that this can be done separately for each choice of total isospin, and for the CP-conserving and CP-violating parts of each operator contained in $\mathcal{H}_W$. To carry this out in practice one needs,

as usual, parametrizations of the PV amplitudes. That for $\mathbf{A}_{K3\pi}^{\text{PV},[1]}$ is identical in form to the result given for the $\eta \to 3\pi$ amplitude in eq. (3.18), with only the labels on the coefficients changing:

$$\mathbf{A}_{K3\pi}^{\text{PV},[1]} = i\left(A_{K3\pi}^{[1],\text{s},0} + A_{K3\pi}^{[1],\text{s},1}\Delta + A_{K3\pi}^{[1],\text{s},2}\Delta^2 + A_{K3\pi}^{[1],\text{s},2a}\sum_i \Delta_i^2\right)\begin{pmatrix} 1 \\ 0 \\ 0 \end{pmatrix} \tag{3.29}$$

$$+ i\left(A_{K3\pi}^{[1],\text{d},1} + A_{K3\pi}^{[1],\text{d},2}\Delta\right)\begin{pmatrix} 0 \\ P\cdot\xi_1 \\ P\cdot\xi_2 \end{pmatrix} + iA_{K3\pi}^{[1],\text{d},2a}\begin{pmatrix} 0 \\ (P\cdot\xi_2)^2 - (P\cdot\xi_1)^2 \\ 2P\cdot\xi_1 P\cdot\xi_2 \end{pmatrix} + \dots.$$

The corresponding result for the $I = 2$ case is

$$\mathbf{A}_{K3\pi}^{\text{PV},[2]} = i\left(A_{K3\pi}^{[2],\text{d},1} + A_{K3\pi}^{[2],\text{d},2}\Delta\right)\begin{pmatrix} P\cdot\xi_1 \\ P\cdot\xi_2 \end{pmatrix} + iA_{K3\pi}^{[2],\text{d},2a}\begin{pmatrix} (P\cdot\xi_2)^2 - (P\cdot\xi_1)^2 \\ 2P\cdot\xi_1 P\cdot\xi_2 \end{pmatrix} + \dots \tag{3.30}$$

Here we are using the basis [18]

$$\{|\chi_1\rangle_2, |\chi_2\rangle_2\} = \{|(\pi\pi)_2\pi\rangle_2, |\rho\pi\rangle_2\}, \tag{3.31}$$

which is further discussed in appendix C. We have worked to quadratic order in the expansions of $\mathbf{A}_{K3\pi}^{[I]}$, since fits to experimentally measured Dalitz plots usually work only to this order.

Given a determination of $\mathbf{A}_{K3\pi}^{\text{PV},[1]}$ and $\mathbf{A}_{K3\pi}^{\text{PV},[2]}$, the second step of solving the integral equations leads to the decay amplitudes in the isospin basis. There are five amplitudes[13]

$$\mathbf{T}_{K3\pi}^{[1]}(p_1, p_2, p_3) = i\begin{pmatrix} \mathcal{T}_{\text{s}}^{[1]}(p_1, p_2, p_3) \\ \mathcal{T}_{\text{d},1}^{[1]}(p_1, p_2, p_3) \\ \mathcal{T}_{\text{d},2}^{[1]}(p_1, p_2, p_3) \end{pmatrix}, \qquad \mathbf{T}_{K3\pi}^{[2]}(p_1, p_2, p_3) = i\begin{pmatrix} \mathcal{T}_{\text{d},1}^{[2]}(p_1, p_2, p_3) \\ \mathcal{T}_{\text{d},2}^{[2]}(p_1, p_2, p_3) \end{pmatrix}, \tag{3.32}$$

although, as above, only one from each doublet is independent. The form of this redundancy is exactly as in eq. (3.25) for both $I = 1$ and 2. The relationship of the isospin-basis states to those with pions of definite charges is given in appendix C. Using these results, and simplifying using the redundancy equation (3.25), we find

$$\mathcal{T}^{+00}(p_1, p_2, p_3) = -\frac{1}{\sqrt{15}}\mathcal{T}_{\text{s}}^{[1]}(p_1, p_2, p_3) + \frac{1}{\sqrt{3}}\left[\mathcal{T}_{\text{d},1}^{[1]}(p_1, p_2, p_3) + \mathcal{T}_{\text{d},1}^{[2]}(p_1, p_2, p_3)\right]$$
$$+ \frac{1}{\sqrt{3}}\left[\mathcal{T}_{\text{d},1}^{[1]}(p_1, p_3, p_2) + \mathcal{T}_{\text{d},1}^{[2]}(p_1, p_3, p_2)\right], \tag{3.33}$$

$$\mathcal{T}^{-++}(p_1, p_2, p_3) = \frac{2}{\sqrt{15}}\mathcal{T}_{\text{s}}^{[1]}(p_1, p_2, p_3) + \frac{1}{\sqrt{3}}\left[\mathcal{T}_{\text{d},1}^{[1]}(p_1, p_2, p_3) - \mathcal{T}_{\text{d},1}^{[2]}(p_1, p_2, p_3)\right]$$
$$+ \frac{1}{\sqrt{3}}\left[\mathcal{T}_{\text{d},1}^{[1]}(p_1, p_3, p_2) - \mathcal{T}_{\text{d},1}^{[2]}(p_1, p_3, p_2)\right], \tag{3.34}$$

where we have used the vanishing of the $I = 3$ amplitude.

---

[13]There is a potential confusion with the amplitudes for $\eta$ decay that have the same names — see eq. (3.22). It should, however, be clear from the context to which process the amplitudes apply.

## 4 Conclusion

In this article we have derived the formalism that allows the study of three-particle decay processes using input from lattice QCD calculations. This generalizes the well-established formalism for two-particle decays developed by Lellouch and Lüscher [32] and its subsequent extensions. Specifically, our formalism applies for decays in which the three particles are degenerate and spinless, although they do not need to be identical. Thus, in particular, the phenomenologically important $K \to 3\pi$ decays are now accessible to lattice methods in the isospin-symmetric limit. Our formalism applies not only to $1 \to 3$ decay processes, but also $0 \to 3$ transitions in the strong interactions, such as that for $\gamma^* \to 3\pi$, which is relevant for lattice calculations of the hadronic vacuum polarization contribution to muonic $g - 2$.

We have divided the presentation into two parts. In the first, given in section 2, we give a detailed derivation in a simplified theoretical set up in which the "pions" are identical. This allows us to focus on the essential new features that are introduced when moving from two to three particles. The derivation is carried out by extending the relativistic three-particle finite-volume formalism for identical scalar particles [3, 4]. Just as in the relation between the finite-volume spectrum and scattering amplitudes, the relation we find between finite-volume decay matrix elements and physical decay amplitudes requires two steps. In the first, finite-volume matrix elements are used to constrain an infinite-volume but scheme-dependent intermediate quantity, $A_{K3\pi}^{\mathrm{PV}}$. This quantity plays a role that is analogous to that of $\mathcal{K}_{\mathrm{df},3}$ in the scattering formalism of refs. [3, 4]. The second step in the formalism is to relate $A_{K3\pi}^{\mathrm{PV}}$ to the physical decay amplitude, and is analogous to the relation between $\mathcal{K}_{\mathrm{df},3}$ and the physical scattering amplitude [4]. This relation is achieved by solving integral equations in infinite-volume that incorporate the effects of two- and three-particle final state interactions (entering through the two-particle K matrix $\mathcal{K}_2$ and $\mathcal{K}_{\mathrm{df},3}$, respectively) and leads to a decay amplitude satisfying the constraints of unitarity.

Our derivation is independent of the details of the effective theory, aside from the assumption of a $\mathbb{Z}_2$ symmetry analogous to G parity. It holds for decays of "kaons" with masses up to the first inelastic threshold, $m_K < 5m_\pi$. The approach is relativistic, implying, for one thing, that the intermediate amplitude $A_{K3\pi}^{\mathrm{PV}}$ is Lorentz invariant. We use this constraint to develop an expansion of $A_{K3\pi}^{\mathrm{PV}}$ about threshold.

It is instructive to compare the two and three-particle formalisms in more detail. The first step of our formalism is the analog of the multiplication by the LL factor that is required for two-particle decays involving only a single channel. In particular, the vector $v$ that enters the key relation, eq. (2.43), is determined by a combination of scattering amplitudes and kinematic factors, just as the LL factor is in the two-particle case. The main new feature compared to the two-particle analysis is the need for the second step. In the original LL derivation, this step is essentially replaced by the multiplication by the final-state phase required by Watson's theorem. It is the more complicated nature of three-particle final-state interactions that necessitates the solution of integral equations. Another difference from the original LL result is that, in general, each finite-volume three-particle matrix element serves only to constrain $A_{K3\pi}^{\mathrm{PV}}$, rather than provide a direct determination. This difference is, however, only due to the simplicity of the set-up considered in the

original LL work. If one considers a multiple-channel two-particle system, then each lattice matrix element again only provides a constraint on physical decay amplitudes [38, 42, 44]. Conversely, if we consider the simplest approximation for three-particle scattering and decay amplitude, then, as shown in section 2.5, only a single finite-volume matrix element is required to determine $A_{K3\pi}^{\mathrm{PV}}$.

In the second part of our presentation, given in section 3, we generalize the formalism so that it applies for decays to a general three-pion state in isosymmetric QCD. This builds upon our recent generalization of the formalism for three-particle scattering to include all three-pion isospin channels [18]. It allows us to address phenomenologically relevant processes, and we have discussed in detail three applications: the electromagnetic transition $\gamma^* \to 3\pi$, the isospin-violating decay $\eta \to 3\pi$, and the weak decay $K \to 3\pi$. While most of the features of the formalism for identical particles also hold for three-pion decays, the key difference is that all quantities have an additional isospin index. One impact of this change is that the symmetry properties of the generalization of $A^{\mathrm{PV}}$ differ from those for identical particles, and we have presented explicit expressions in a threshold expansion that should suffice for realistic calculations.

An important difference between the process $\gamma^* \to 3\pi$ on the one hand and the decays $\eta \to 3\pi$ and $K \to 3\pi$ on the other, is that the latter two have a clear physical interpretation only when the initial and final state energies match, whereas the virtual photon transition is meaningful for all final state energies. However, the formalism presented here also holds for matrix elements in which the kinematics are not perfectly matched. In practice, this freedom can be used to extract $A_{K3\pi}^{\mathrm{PV}}$ as a function of the final state energy, e.g. by fitting to multiple closely spaced states. This could be useful both for giving stronger constraints on the target amplitude and for interpreting the value, including the role of resonance enhancement in the amplitude, by considering the result for energies away from physical kinematics.

Although a controlled computation of the $K \to 2\pi$ decay amplitude using lattice QCD has only been achieved very recently [68], we are hopeful that the extension to $K \to 3\pi$ decays can be undertaken in the next few years. This will require a program of calculations of the finite-volume three-pion spectrum with all allowed total isospins, in addition to the calculation of the finite volume $K \to 3\pi$ matrix elements. We note that work on the second step of our formalism — which requires solving integral equations — can begin independently of lattice simulations, since the methods required do not depend on the functional form of the necessary input quantities ($\mathcal{K}_2$, $\mathcal{K}_{\mathrm{df},3}$ and $A_{K3\pi}^{\mathrm{PV}}$). Indeed, methods for solving the closely-related integral equations required for three-particle scattering are under active development [27, 54].

Finally, we note that further generalizations of the formalism derived here will be needed to allow lattice calculations of all three-particle decay amplitudes of interest. For example, to address isospin breaking in $K \to 3\pi$ decays requires formalism for three non-degenerate particles, as well as for multiple, nondegenerate channels. The recent extension of the three-particle quantization condition to the case of nondegenerate particles is a first step in this direction [19].

## Acknowledgments

## A Proof that $A_3' = A_3^\dagger$

In this appendix, we prove that the quantities $A_3$ and $A_3'$, introduced in eq. (2.5), are related by hermitian conjugation, provided that the same is true of the two operators entering the corresponding correlation function, eq. (2.3). This result is required to reach eq. (2.34), which is used, in turn, to derive the main result of section 2.

A constructive definition of the quantities $A_3$ and $A_3'$ is provided in ref. [3], but it is cumbersome and difficult to use in proving basic relations. Therefore, here we find it easier to pursue an indirect method. Our approach is in the spirit of ref. [4] in which $\mathcal{K}_{\mathrm{df},3}$ is related to the physical scattering amplitude via a finite-volume quantity, without making direct use of the complicated constructive definition of ref. [3].

The key idea is to use the relation between $A_3$, $A_3'$ and their corresponding finite-volume decay amplitudes. To define the latter we first introduce matrix elements defined in terms of physical, asymptotic three-particle states:

$$T'(E, \boldsymbol{k}, \widehat{\boldsymbol{a}}^\star) = \langle 0|\sigma(0)|3\pi, \mathrm{in}\rangle \,, \tag{A.1}$$

$$T(E, \boldsymbol{k}, \widehat{\boldsymbol{a}}^\star) = \langle 3\pi, \mathrm{out}|\sigma^\dagger(0)|0\rangle \,, \tag{A.2}$$

where the arguments on the left-hand side provide a description of the three incoming or outgoing pions, as described in the text following (2.61). Starting from these, one can give diagrammatic definitions of $T_L^{(u)}$ and $T_L'^{(u)}$, the asymmetric finite-volume decay amplitudes corresponding to $A_3$ and $A_3'$ respectively. For concreteness, we focus on $T_L^{(u)}$; the argument for $T_L'^{(u)}$ is analogous. The definition of $T_L^{(u)}$ is essentially the same as that for $T_{K3\pi, L}^{(u)}$ given in section 2.4, except that the initial amputated kaon propagator is absent, so that the initial kaon state in eq. (2.52) is replaced in eq. (A.2) with the vacuum. In words, $T_L^{(u)}(E, \boldsymbol{k}, \widehat{\boldsymbol{a}}^\star)$ is the asymmetric finite-volume vacuum to three pion amplitude in which, if the final interaction involves a $2 \to 2$ Bethe-Salpeter kernel, then $\boldsymbol{k}$ is the momentum assigned to the spectator, and if the final interaction involves the $3 \to 3$ kernel, the diagram is multiplied by $1/3$. The amplitude $T(E, \boldsymbol{k}, \widehat{\boldsymbol{a}}^\star)$ in eq. (A.2) is then obtained by taking

the appropriate $L \to \infty$ limit and symmetrizing, just as for $T_{K3\pi}$ in eqs. (2.59)–(2.61) of the main text.

From the analysis given in section 2.4, it then follows that

$$T_L^{(u)} = X_L A_3 \,, \qquad X_L = \mathcal{L}_L^{(u)} \frac{1}{1 + \mathcal{K}_{\mathrm{df},3} F_3} \,, \tag{A.3}$$

with $T_L^{(u)}$ a column vector in $\{k\ell m\}$ space, and $\mathcal{L}_L^{(u)}$ is given in eq. (2.16). This has exactly the same structure as eq. (2.58), with $A_{K3\pi}^{\mathrm{PV}}$ replaced here with $A_3$. A similar analysis leads to

$$T_L'^{(u)} = A_3' X_R \,, \qquad X_R = \frac{1}{1 + F_3 \mathcal{K}_{\mathrm{df},3}} \mathcal{R}_L^{(u)} \,, \tag{A.4}$$

with $T_L'^{(u)}$ a row vector in $\{k\ell m\}$ space, and $\mathcal{R}_L^{(u)}$ given in eq. (2.17). The first key observation is now that

$$X_R = X_L^\dagger \,, \tag{A.5}$$

which follows because $\mathcal{L}_L^{(u)\dagger} = \mathcal{R}_L^{(u)}$, $F_3 = F_3^\dagger$ and $\mathcal{K}_{\mathrm{df},3} = \mathcal{K}_{\mathrm{df},3}^\dagger$. These results themselves follow from the hermiticity of the building blocks $F$, $\mathcal{K}_2$ and $(2\omega L^3)^{-1} G$.

The second key relation that we need is

$$T_L'^{(u)} = (T_L^{(u)})^\dagger \,, \tag{A.6}$$

which follows directly from the diagrammatic definitions of $T_L^{(u)}$ and $T_L'^{(u)}$ [without reference to eqs. (A.3) and (A.4)], assuming T and P invariance of the effective field theory, and P invariance of the operator $\sigma$ (ignoring the intrinsic parity of the pion). To make the argument, we first note that, aside from phases arising from the operators $\sigma^\dagger$ and $\sigma$, each diagram contributing to $T_L^{(u)}$ and $T_L'^{(u)}$ is real. This is because we are working in finite volume. One way to show this result is to evaluate diagrams using time-ordered perturbation theory, in which case the only source of imaginary contributions is the $i\epsilon$ in the energy denominators. But in finite volume, the sums over spatial momenta do not require that the poles from these denominators be regulated, so that $\epsilon$ can be set to zero. Next we note that T invariance implies the relation $T_L'^{(u)}(E, \boldsymbol{k}, \hat{\boldsymbol{a}}^\star) = T_L^{(u)}(E, -\boldsymbol{k}, -\hat{\boldsymbol{a}}^\star)^*$, where complex conjugation is only needed because of possible phases arising from $\sigma^\dagger$ and $\sigma$. Now, using parity invariance, we have that $T_L^{(u)}(E, -\boldsymbol{k}, -\hat{\boldsymbol{a}}^\star) = T_L^{(u)}(E, \boldsymbol{k}, \hat{\boldsymbol{a}}^\star)$. Finally, decomposing into the $\{k\ell m\}$ basis, and taking into account that $T_L'^{(u)}$ is a row vector and $T_L^{(u)}$ a column vector, we obtain eq. (A.6).

Combining eqs. (A.3), (A.4) and (A.6) yields

$$A_3' X_R = A_3^\dagger X_R \,. \tag{A.7}$$

The final step is to note that, for any total energy $E$, $X_R$ is well-defined and invertible away from a discrete set of values of $L$ for which one of its eigenvalues vanishes or diverges. Away from these "singular" values of $L$, we can apply the inverse of $X_R$ to both sides of eq. (A.7), and conclude that $A_3' = A_3^\dagger$. This demonstrates the desired equality for all values of the spectator momentum $\boldsymbol{k}$ that lie in the finite-volume sets of the nonsingular values of $L$. Assuming that the nonsingular values of $L$ form a dense set, then, given that $A_3$ and $A_3'$ are continuous functions of the spectator momentum, we find that $A_3' = A_3^\dagger$ in general.

Here we follow the approach of ref. [32], which provides an alternative to the first step of the derivation, which is presented in the main text in section 2.2. We consider the same theory as in section 2 but now imagine determining the finite-volume spectrum in the two-pion and three-pion sectors in the presence of the weak interaction, with Hamiltonian density $\mathcal{H}_W(x)$. These sectors are still decoupled in the presence of $\mathcal{H}_W$, differing by whether the total number of particles is even or odd. The logic of the approach is that the weak interactions shift the spectrum, beginning at linear order, and these shifts can be calculated in two ways: (i) from the finite-volume matrix element; (ii) using the quantization condition, due to a shift in the infinite-volume interactions that depends on the infinite-volume decay amplitude. Comparing the two results for the shift leads to the desired relation. We stress that throughout this section we drop contributions of quadratic or higher order in $\mathcal{H}_W$ from all equations.

We begin with the two-pion sector. A key distinction here, as compared to the $K \to \pi\pi$ case of ref. [32], is that $\mathcal{H}_W$ only couples the single kaon to states with G parity minus. Thus, the lightest new intermediate state coupling to $\pi\pi$ via the weak interactions is the $K\pi$ state, which, given the constraint eq. (2.1), has a CMF energy $E_2^*$ that exceeds $4m_\pi$. It follows that the spectrum in the energy range $E_2^* < 4m_\pi$ will only be shifted by second-order weak processes involving off-shell intermediate $K\pi$ states. Since we work at linear order, these can be ignored. Thus the energy levels are unchanged, which, using the two-particle quantization condition, implies that the two-particle scattering amplitude $\mathcal{M}_2$ is also unchanged. The latter result can also be seen by studying the modifications to this amplitude directly in infinite volume.

The situation is different in the three-pion sector. Here the lightest new intermediate state consists of a single kaon, and this is kinematically allowed; see again (2.1). Levels away from the kaon energy will be shifted only at second order in perturbation theory. However, if the volume is tuned so that there is a three-pion level in the theory without weak interactions whose CMF energy matches that of a finite-volume kaon, then we must use degenerate perturbation theory at leading order.[14] We consider here only a rotationally invariant, local form of $\mathcal{H}_W(x)$ [such as that of eq. (2.2)]. In this case, only the trivial irrep of the appropriate little group will be coupled to the kaon and thus only the tuned QCD level in this irrep is relevant. The degenerate sector is thus $(|K, \boldsymbol{P}, L\rangle, |E_n, \boldsymbol{P}, A_1, L\rangle)$, and the Hamiltonian restricted to this sector is

$$\begin{pmatrix} E_K(\boldsymbol{P}) & M(\boldsymbol{P}) \\ M^*(\boldsymbol{P}) & E_K(\boldsymbol{P}) \end{pmatrix}, \quad M(\boldsymbol{P}) = L^3 \langle E_n, \boldsymbol{P}, A_1, L | \mathcal{H}_W(0) | K, \boldsymbol{P}, L \rangle, \tag{B.1}$$

where the factor of $L^3$ arises due to the difference between Hamiltonian and Hamiltonian

---

[14]The difference between finite- and infinite-volume kaon energies is exponentially suppressed in $L$ and thus neglected in this derivation. Therefore, strictly speaking, the approach described in this appendix is equally valid whether one tunes the three-pion level to the finite- or the infinite-volume kaon energy. However, in practice, the tuning should be to the finite-volume kaon as this is the quantity available in the lattice calculation.

density. Diagonalizing, we obtain the energies to first order in $\mathcal{H}_W$,

$$E_K(\boldsymbol{P}) \to E_K^\pm(\boldsymbol{P}) \equiv E_K(\boldsymbol{P}) \pm |M(\boldsymbol{P})|. \tag{B.2}$$

This is the first result for the energy shifts.

To obtain the second result for the shifts we begin by noting that, when the total CMF energy $E_3^*$ lies within $\mathcal{O}(\mathcal{H}_W)$ of $m_K$, the three-particle scattering amplitude is changed at linear order in $\mathcal{H}_W$. This is because of the nearly on-shell process $3\pi \to K \to 3\pi$, which leads to

$$i\delta\mathcal{M}_3(E_3^*) \equiv i\mathcal{M}_3^{[\mathcal{H}_W \neq 0]}(E_3^*) - i\mathcal{M}_3^{[\mathcal{H}_W = 0]}(E_3^*), \tag{B.3}$$

$$= \langle 3\pi, \text{out}| \, [-i\mathcal{H}_W(0)] \, |K, \boldsymbol{P}\rangle \, \frac{i}{E_3^{*2} - m_K^2 + i\epsilon} \, \langle K, \boldsymbol{P}| \, [-i\mathcal{H}_W(0)] \, |3\pi, \text{in}\rangle, \tag{B.4}$$

where we have used the superscripts $[\mathcal{H}_W \neq 0]$ and $[\mathcal{H}_W = 0]$ to indicate whether the $3\pi \to K \to 3\pi$ transition is present or absent. Here the dependence on the initial and final pion momenta is implicit. Although this appears to be of second order in $\mathcal{H}_W$, the denominator of the propagator is

$$E_3^{*2} - m_K^2 = E_3(\boldsymbol{P})^2 - E_K(\boldsymbol{P})^2, \tag{B.5}$$

$$= 2E_K(\boldsymbol{P})[E_3(\boldsymbol{P}) - E_K(\boldsymbol{P})] + \mathcal{O}\big[(E_3(\boldsymbol{P}) - E_K(\boldsymbol{P}))^2\big], \tag{B.6}$$

and thus of $\mathcal{O}(\mathcal{H}_W)$ for $E_3(\boldsymbol{P}) = E_K^\pm(\boldsymbol{P})$. It follows that the difference between the perturbed and unperturbed amplitudes at the shifted finite-volume energy is $\mathcal{O}(\mathcal{H}_W)$:

$$\delta_\pm\mathcal{M}_3 \equiv \delta\mathcal{M}_3\big([E_K^\pm(\boldsymbol{P})^2 - \boldsymbol{P}^2]^{1/2}\big), \tag{B.7}$$

$$= \mp \frac{\langle 3\pi, \text{out}| \, \mathcal{H}_W(0) \, |K, \boldsymbol{P}\rangle \, \langle K, \boldsymbol{P}| \, \mathcal{H}_W(0) \, |3\pi, \text{in}\rangle}{2E_K(\boldsymbol{P})|M(\boldsymbol{P})|}. \tag{B.8}$$

Our next task is to determine the shift in $\mathcal{K}_{\text{df},3}$ that corresponds to that in $\mathcal{M}_3$, for the former is the quantity that enters the quantization condition. For the sake of brevity, we write the following expressions in terms of finite-volume quantities, with the $L \to \infty$ limit implied. We use the expression for $\mathcal{M}_{3,L}^{(u,u)}$, eq. (2.14), but need keep only the second, divergence-free term, since $\mathcal{D}^{(u,u)}$ does not depend on $\mathcal{K}_{\text{df},3}$:

$$\delta\mathcal{M}_3 = \mathcal{S}\left\{\delta\mathcal{M}_{\text{df},3,L}^{(u,u)}\right\}, \tag{B.9}$$

$$\mathcal{M}_{\text{df},3,L}^{(u,u)} = \mathcal{L}_L^{(u)} \frac{1}{1 + \mathcal{K}_{\text{df},3}F_3} \mathcal{K}_{\text{df},3}\mathcal{R}_L^{(u)}, \tag{B.10}$$

$$\delta\mathcal{M}_{\text{df},3,L}^{(u,u)} = \mathcal{L}_L^{(u)} \frac{1}{1 + \mathcal{K}_{\text{df},3}F_3} \delta\mathcal{K}_{\text{df},3} \frac{1}{1 + F_3\mathcal{K}_{\text{df},3}} \mathcal{R}_L^{(u)}. \tag{B.11}$$

Next we use eq. (2.58) for the decay amplitude, and the conjugate result for the $3\pi \to K$ amplitude, to rewrite eq. (B.8) as

$$\delta_\pm\mathcal{M}_3 = \mp\mathcal{S}\left\{\mathcal{L}_L^{(u)} \frac{1}{1 + \mathcal{K}_{\text{df},3}F_3} \frac{A_{K3\pi}^{\text{PV}}A_{K3\pi}^{\text{PV}\dagger}}{2E_K|M|} \frac{1}{1 + F_3\mathcal{K}_{\text{df},3}} \mathcal{R}_L^{(u)}\right\}, \tag{B.12}$$

where we have suppressed the $\boldsymbol{P}$ dependence in $E_K$ and $M$. Matching eqs. (B.9) and (B.11) with eq. (B.12), we find

$$\delta_\pm \mathcal{K}_{\mathrm{df},3} = \mp \frac{A_{K3\pi}^{\mathrm{PV}} A_{K3\pi}^{\mathrm{PV}\,\dagger}}{2E_K |M|} \,. \tag{B.13}$$

The outer product structure reflects the factorization of the residue at the pole in $\mathcal{M}_3$.

The final step is to enforce the quantization condition with the shifted amplitude at the shifted energies. To this end we define

$$A(E) \equiv F_3(E, \boldsymbol{P}, L)^{-1} + \mathcal{K}_{\mathrm{df},3}(E^*) \,. \tag{B.14}$$

Then the unshifted quantization condition can be written as $\det[A(E_K)] = 0$, and the shifted version as

$$\det[A(E_K^\pm) + \delta_\pm \mathcal{K}_{\mathrm{df},3}] = \det[A(E_K) + \delta_\pm A] = 0 \,, \tag{B.15}$$

where we have introduced

$$\delta_\pm A = \pm |M| \frac{dA}{dE}\bigg|_{E_K} + \delta_\pm \mathcal{K}_{\mathrm{df},3} \,. \tag{B.16}$$

Recalling that $v$ is the eigenvector of $A(E_K)$ with vanishing eigenvalue, and defining $v + \delta_\pm v$ as the corresponding eigenvector for $A(E_K) + \delta_\pm A$, we have

$$(v^\dagger + \delta_\pm v^\dagger) \cdot [A(E_K) + \delta_\pm A] \cdot (v + \delta_\pm v) = 0 \,. \tag{B.17}$$

Multiplying out this result, using $A(E_K) \cdot v = 0 = v^\dagger \cdot A(E_K)$, and using the fact that the left-hand side of eq. (B.17) must vanish order by order in $\mathcal{H}_W$ (in particular at linear order) yields

$$v^\dagger \cdot \delta_\pm A \cdot v = 0 \,. \tag{B.18}$$

Substituting eqs. (B.13) and (B.16) then gives

$$|M(\boldsymbol{P})| \left[ v^\dagger \cdot \frac{dA}{dE}\bigg|_{E_K} \cdot v \right] = v^\dagger \cdot \frac{A_{K3\pi}^{\mathrm{PV}} A_{K3\pi}^{\mathrm{PV}\,\dagger}}{2E_K(\boldsymbol{P}) |M(\boldsymbol{P})|} \cdot v \,. \tag{B.19}$$

To evaluate the quantity in square brackets we use eqs. (2.36) and (2.40) of the main text, which imply

$$A(E) = (E - E_K) \frac{v v^\dagger}{|v|^4} + X(E) \,, \tag{B.20}$$

where the first term results from

$$A(E)^{-1} = \frac{v v^\dagger}{E - E_K} + \mathcal{O}\big[(E - E_K)^0\big] \,, \tag{B.21}$$

and $X(E)$ arises from the non-singular part of $A^{-1}$. Here we only require that $X(E)$ satisfies $v \cdot X(E) \cdot v^\dagger = 0$. This relies on the fact that the eigenvectors of $A(E_K)$ form a complete set that can be used for any $A(E)$. Then $X(E)$ is built from the sum over all

eigenvector pairs $\boldsymbol{e}^{(i)}\boldsymbol{e}^{(j)\dagger}$, weighted by $E$-dependent coefficients, with at least one of the two vectors $\boldsymbol{e}^{(i)}$ and $\boldsymbol{e}^{(j)}$ orthogonal to $v$. From this it immediately follows that

$$v^{\dagger} \cdot \left.\frac{dA}{dE}\right|_{E_K} \cdot v = 1 \,. \tag{B.22}$$

Finally, inserting eqs. (B.1) and (B.22) into eq. (B.19), we obtain

$$|v^{\dagger} A_{K3\pi}^{\mathrm{PV}}|^2 = 2E_K(\boldsymbol{P})L^6 \left| \langle E_n, \boldsymbol{P}, A_1, L| \, \mathcal{H}_W(0) \, |K, \boldsymbol{P}, L\rangle \right|^2 \,. \tag{B.23}$$

This agrees with eq. (2.43) in the main text.

## C    Relations between three-pion states

In ref. [18], we provided the isospin decomposition for all neutral ($I_3 = 0$) three-pion states, and described the decomposition into irreducible representations of the group $S_3$. Here we provide a result for the neutral sector not given explicitly in ref. [18], since this is needed in the discussion of the $\gamma^* \to 3\pi$ and $\eta \to 3\pi$ processes. In addition, we generalize the results to the charge 1 ($I_3 = 1$) sector, as these are needed in the discussion of $K^+$ decays.

The first result is for the matrix $\mathcal{R}$ defined by

$$
\begin{pmatrix} |-0+\rangle \\ |0-+\rangle \\ |-+0\rangle \\ |0\ 0\ 0\rangle \\ |+-0\rangle \\ |0+-\rangle \\ |+0-\rangle \end{pmatrix} = \mathcal{R} \cdot \begin{pmatrix} |(\pi\pi)_2\pi\rangle_3 \\ |(\pi\pi)_2\pi\rangle_2 = |\chi_1\rangle_2 \\ |\rho\pi\rangle_2 = |\chi_2\rangle_2 \\ |\chi_s\rangle_1 \\ |\chi_1\rangle_1 \\ |\chi_2\rangle_1 \\ |\rho\pi\rangle_0 \end{pmatrix} \,, \tag{C.1}
$$

where we are using the shorthands

$$|-0+\rangle \equiv |\pi^-(p_1)\pi^0(p_2)\pi^+(p_3)\rangle \,, \quad |+0-\rangle \equiv |\pi^+(p_1)\pi^0(p_2)\pi^-(p_3)\rangle \,, \text{ etc.} \tag{C.2}$$

We find

$$
\mathcal{R} = \begin{pmatrix} \frac{1}{\sqrt{10}} & -\frac{1}{2} & -\frac{1}{\sqrt{12}} & \frac{1}{\sqrt{15}} & -\frac{1}{\sqrt{12}} & \frac{1}{2} & -\frac{1}{\sqrt{6}} \\ \frac{1}{\sqrt{10}} & -\frac{1}{2} & \frac{1}{\sqrt{12}} & \frac{1}{\sqrt{15}} & -\frac{1}{\sqrt{12}} & -\frac{1}{2} & \frac{1}{\sqrt{6}} \\ \frac{1}{\sqrt{10}} & 0 & -\frac{2}{\sqrt{12}} & \frac{1}{\sqrt{15}} & \frac{2}{\sqrt{12}} & 0 & \frac{1}{\sqrt{6}} \\ \frac{2}{\sqrt{10}} & 0 & 0 & -\frac{3}{\sqrt{15}} & 0 & 0 & 0 \\ \frac{1}{\sqrt{10}} & 0 & \frac{2}{\sqrt{12}} & \frac{1}{\sqrt{15}} & \frac{2}{\sqrt{12}} & 0 & -\frac{1}{\sqrt{6}} \\ \frac{1}{\sqrt{10}} & \frac{1}{2} & -\frac{1}{\sqrt{12}} & \frac{1}{\sqrt{15}} & -\frac{1}{\sqrt{12}} & -\frac{1}{2} & -\frac{1}{\sqrt{6}} \\ \frac{1}{\sqrt{10}} & \frac{1}{2} & \frac{1}{\sqrt{12}} & \frac{1}{\sqrt{15}} & -\frac{1}{\sqrt{12}} & \frac{1}{2} & \frac{1}{\sqrt{6}} \end{pmatrix} \,. \tag{C.3}
$$

We use the last row of $\mathcal{R}$ in sections 3.2 and 3.3.

We now turn to the charge 1 sector of three pions, giving our conventions for the states and the relation between the isospin and definite-charge bases. In this sector, the total isospin can only be $I = 1, 2$ or $3$, with degeneracies $3, 2, 1$, respectively [18]. The $S_3$ irreps that appear are the symmetric irrep, labeled $|\chi_s\rangle_I$, and the two-dimensional standard irrep, labeled $\{|\chi_1\rangle_I, |\chi_2\rangle_I\}$.

The relation to the states in the basis with definite isospin for the first pair is

$$|\chi_s\rangle_3^+ = |(\pi\pi)_2\pi\rangle_3^+ \tag{C.4}$$

$$|\chi_1\rangle_2^+ = |(\pi\pi)_2\pi\rangle_2^+ \tag{C.5}$$

$$|\chi_2\rangle_2^+ = |\rho\pi\rangle_2^+ \,, \tag{C.6}$$

$$|\chi_s\rangle_1^+ = \frac{2}{3}|(\pi\pi)_2\pi\rangle_1^+ + \frac{\sqrt{5}}{3}|\sigma\pi\rangle_1^+ \,, \tag{C.7}$$

$$|\chi_1\rangle_1^+ = -\frac{\sqrt{5}}{3}|(\pi\pi)_2\pi\rangle_1^+ + \frac{2}{3}|\sigma\pi\rangle_1^+ \,, \tag{C.8}$$

$$|\chi_2\rangle_1^+ = |\rho\pi\rangle_1^+ \,. \tag{C.9}$$

From this, the relation to the states composed of pions of definite charges is simple to obtain. What we need in section 3.4 is this inverse of this relation,

$$
\begin{pmatrix}
|+\ 0\ 0\rangle \\
|0\ +\ 0\rangle \\
|0\ 0\ +\rangle \\
|-++\rangle \\
|+-+\rangle \\
|++-\rangle
\end{pmatrix}
= \mathcal{R}_1 \cdot
\begin{pmatrix}
|\chi_s\rangle_3^+ \\
|\chi_1\rangle_2^+ \\
|\chi_2\rangle_2^+ \\
|\chi_s\rangle_1^+ \\
|\chi_1\rangle_1^+ \\
|\chi_2\rangle_2^+
\end{pmatrix} \,, \tag{C.10}
$$

where

$$
\mathcal{R}_1 =
\begin{pmatrix}
\frac{2}{\sqrt{15}} & \frac{1}{\sqrt{12}} & \frac{1}{2} & -\frac{1}{\sqrt{15}} & \frac{1}{\sqrt{12}} & \frac{1}{2} \\
\frac{2}{\sqrt{15}} & \frac{1}{\sqrt{12}} & -\frac{1}{2} & -\frac{1}{\sqrt{15}} & \frac{1}{\sqrt{12}} & -\frac{1}{2} \\
\frac{2}{\sqrt{15}} & -\frac{2}{\sqrt{12}} & 0 & -\frac{1}{\sqrt{15}} & -\frac{2}{\sqrt{12}} & 0 \\
\frac{1}{\sqrt{15}} & -\frac{1}{\sqrt{12}} & -\frac{1}{2} & \frac{2}{\sqrt{15}} & \frac{1}{\sqrt{12}} & \frac{1}{2} \\
\frac{1}{\sqrt{15}} & -\frac{1}{\sqrt{12}} & \frac{1}{2} & \frac{2}{\sqrt{15}} & \frac{1}{\sqrt{12}} & -\frac{1}{2} \\
\frac{1}{\sqrt{15}} & \frac{2}{\sqrt{12}} & 0 & \frac{2}{\sqrt{15}} & -\frac{2}{\sqrt{12}} & 0
\end{pmatrix} \,. \tag{C.11}
$$

## D  Formalism for $K^0 \to 3\pi$ decays

For completeness, we collect here the results needed to apply the formalism to the decays of neutral kaons. We do so for the $K^0$ decay. That for $\overline{K}^0$ decay is identical in form, and by forming appropriate combinations one can determine the amplitudes for $K_S$ and $K_L$ decays.

The major change compared to $K^+$ decays is the presence of the $I = 0$ final state in addition to those with $I = 1$ and $2$. The parametrization of the intermediate PV $I = 0$ amplitude requires an antisymmetric combination of the pion momenta that is a Lorentz invariant. In terms of the parameters defined in section 2.3, we find that the leading term is of cubic order in the threshold expansion,

$$\mathbf{A}_{K3\pi,[0]}^{\mathrm{PV},[0]} = iA_{K3\pi}^a \left[ \Delta_3^2(\Delta_1 - \Delta_2) + \Delta_1^2(\Delta_2 - \Delta_3) + \Delta_2^2(\Delta_3 - \Delta_1) \right] + \dots . \tag{D.1}$$

The parametrizations of the $I = 1$ and $2$ amplitudes are as for the $K^+$ decay discussed in section 3.4.

We use the same notation for the isospin-basis amplitudes as in eq. (3.32), but now add

$$\mathbf{T}_{K3\pi}^{[0]}(p_1, p_2, p_3) \equiv i\mathcal{T}_a^{[0]}(p_1, p_2, p_3) , \tag{D.2}$$

where the subscript "$a$" denotes the antisymmetric irrep of $S_3$. Using $\mathcal{R}$ in eq. (C.3) and the redundancy result eq. (3.25) we obtain the relation between isospin amplitudes and those for pions of definite charge,

$$\mathcal{T}^{000}(p_1, p_2, p_3) = -\frac{3}{\sqrt{15}} \mathcal{T}_s^{[1]}(p_1, p_2, p_3) \tag{D.3}$$

$$\begin{aligned}
\mathcal{T}^{+-0}(p_1, p_2, p_3) = {} & \frac{1}{\sqrt{15}} \mathcal{T}_s^{[1]}(p_1, p_2, p_3) + \frac{2}{3} \mathcal{T}_{d,1}^{[2]}(p_1, p_2, p_3) + \frac{1}{3} \mathcal{T}_{d,1}^{[2]}(p_1, p_3, p_2) \\
& + \frac{1}{\sqrt{3}} \mathcal{T}_{d,1}^{[1]}(p_1, p_3, p_2) + \frac{1}{\sqrt{6}} \mathcal{T}_a^{[0]}(p_1, p_2, p_3) .
\end{aligned} \tag{D.4}$$

**Open Access.** This article is distributed under the terms of the Creative Commons Attribution License ([CC-BY 4.0](https://creativecommons.org/licenses/by/4.0/)), which permits any use, distribution and reproduction in any medium, provided the original author(s) and source are credited.

## References

[1] R.A. Briceno and Z. Davoudi, *Three-particle scattering amplitudes from a finite volume formalism*, *Phys. Rev. D* **87** (2013) 094507 [arXiv:1212.3398] [INSPIRE].

[2] K. Polejaeva and A. Rusetsky, *Three particles in a finite volume*, *Eur. Phys. J. A* **48** (2012) 67 [arXiv:1203.1241] [INSPIRE].

[3] M.T. Hansen and S.R. Sharpe, *Relativistic, model-independent, three-particle quantization condition*, *Phys. Rev. D* **90** (2014) 116003 [arXiv:1408.5933] [INSPIRE].

[4] M.T. Hansen and S.R. Sharpe, *Expressing the three-particle finite-volume spectrum in terms of the three-to-three scattering amplitude*, *Phys. Rev. D* **92** (2015) 114509 [arXiv:1504.04248] [INSPIRE].

[5] R.A. Briceño, M.T. Hansen and S.R. Sharpe, *Relating the finite-volume spectrum and the two-and-three-particle S matrix for relativistic systems of identical scalar particles*, *Phys. Rev. D* **95** (2017) 074510 [arXiv:1701.07465] [INSPIRE].

[6] H.-W. Hammer, J.-Y. Pang and A. Rusetsky, *Three-particle quantization condition in a finite volume: 1. The role of the three-particle force*, *JHEP* **09** (2017) 109 [arXiv:1706.07700] [INSPIRE].

[7] H.W. Hammer, J.Y. Pang and A. Rusetsky, *Three particle quantization condition in a finite volume: 2. General formalism and the analysis of data*, *JHEP* **10** (2017) 115 [arXiv:1707.02176] [INSPIRE].

[8] M. Mai and M. Döring, *Three-body Unitarity in the Finite Volume*, *Eur. Phys. J. A* **53** (2017) 240 [arXiv:1709.08222] [INSPIRE].

[9] R.A. Briceño, M.T. Hansen and S.R. Sharpe, *Three-particle systems with resonant subprocesses in a finite volume*, *Phys. Rev. D* **99** (2019) 014516 [arXiv:1810.01429] [INSPIRE].

[10] R.A. Briceño, M.T. Hansen and S.R. Sharpe, *Numerical study of the relativistic three-body quantization condition in the isotropic approximation*, *Phys. Rev. D* **98** (2018) 014506 [arXiv:1803.04169] [INSPIRE].

[11] A.W. Jackura et al., *Equivalence of three-particle scattering formalisms*, *Phys. Rev. D* **100** (2019) 034508 [arXiv:1905.12007] [INSPIRE].

[12] T.D. Blanton, F. Romero-López and S.R. Sharpe, *Implementing the three-particle quantization condition including higher partial waves*, *JHEP* **03** (2019) 106 [arXiv:1901.07095] [INSPIRE].

[13] R.A. Briceño, M.T. Hansen, S.R. Sharpe and A.P. Szczepaniak, *Unitarity of the infinite-volume three-particle scattering amplitude arising from a finite-volume formalism*, *Phys. Rev. D* **100** (2019) 054508 [arXiv:1905.11188] [INSPIRE].

[14] M.T. Hansen and S.R. Sharpe, *Lattice QCD and Three-particle Decays of Resonances*, *Ann. Rev. Nucl. Part. Sci.* **69** (2019) 65 [arXiv:1901.00483] [INSPIRE].

[15] F. Romero-López, S.R. Sharpe, T.D. Blanton, R.A. Briceño and M.T. Hansen, *Numerical exploration of three relativistic particles in a finite volume including two-particle resonances and bound states*, *JHEP* **10** (2019) 007 [arXiv:1908.02411] [INSPIRE].

[16] T.D. Blanton and S.R. Sharpe, *Alternative derivation of the relativistic three-particle quantization condition*, *Phys. Rev. D* **102** (2020) 054520 [arXiv:2007.16188] [INSPIRE].

[17] T.D. Blanton and S.R. Sharpe, *Equivalence of relativistic three-particle quantization conditions*, *Phys. Rev. D* **102** (2020) 054515 [arXiv:2007.16190] [INSPIRE].

[18] M.T. Hansen, F. Romero-López and S.R. Sharpe, *Generalizing the relativistic quantization condition to include all three-pion isospin channels*, *JHEP* **07** (2020) 047 [*Erratum ibid.* **02** (2021) 014] [arXiv:2003.10974] [INSPIRE].

[19] T.D. Blanton and S.R. Sharpe, *Relativistic three-particle quantization condition for nondegenerate scalars*, *Phys. Rev. D* **103** (2021) 054503 [arXiv:2011.05520] [INSPIRE].

[20] F. Müller, A. Rusetsky and T. Yu, *Finite-volume energy shift of the three-pion ground state*, *Phys. Rev. D* **103** (2021) 054506 [arXiv:2011.14178] [INSPIRE].

[21] M. Mai and M. Döring, *Finite-Volume Spectrum of $\pi^+\pi^+$ and $\pi^+\pi^+\pi^+$ Systems*, *Phys. Rev. Lett.* **122** (2019) 062503 [arXiv:1807.04746] [INSPIRE].

[22] B. Hörz and A. Hanlon, *Two- and three-pion finite-volume spectra at maximal isospin from lattice QCD*, *Phys. Rev. Lett.* **123** (2019) 142002 [arXiv:1905.04277] [INSPIRE].

[23] T.D. Blanton, F. Romero-López and S.R. Sharpe, *I = 3 Three-Pion Scattering Amplitude from Lattice QCD*, *Phys. Rev. Lett.* **124** (2020) 032001 [arXiv:1909.02973] [INSPIRE].

[24] C. Culver, M. Mai, R. Brett, A. Alexandru and M. Döring, *Three pion spectrum in the $I = 3$ channel from lattice QCD*, *Phys. Rev. D* **101** (2020) 114507 [arXiv:1911.09047] [INSPIRE].

[25] M. Mai, M. Döring, C. Culver and A. Alexandru, *Three-body unitarity versus finite-volume $\pi^+\pi^+\pi^+$ spectrum from lattice QCD*, *Phys. Rev. D* **101** (2020) 054510 [arXiv:1909.05749] [INSPIRE].

[26] M. Fischer, B. Kostrzewa, L. Liu, F. Romero-López, M. Ueding and C. Urbach, *Scattering of two and three physical pions at maximal isospin from lattice QCD*, arXiv:2008.03035 [INSPIRE].

[27] HADRON SPECTRUM collaboration, *Energy-Dependent $\pi^+\pi^+\pi^+$ Scattering Amplitude from QCD*, *Phys. Rev. Lett.* **126** (2021) 012001 [arXiv:2009.04931] [INSPIRE].

[28] A. Alexandru et al., *Finite-volume energy spectrum of the $K^-K^-K^-$ system*, *Phys. Rev. D* **102** (2020) 114523 [arXiv:2009.12358] [INSPIRE].

[29] R. Brett, C. Culver, M. Mai, A. Alexandru, M. Döring and F.X. Lee, *Three-body interactions from the finite-volume QCD spectrum*, arXiv:2101.06144 [INSPIRE].

[30] F. Romero-López, A. Rusetsky and C. Urbach, *Two- and three-body interactions in $\varphi^4$ theory from lattice simulations*, *Eur. Phys. J. C* **78** (2018) 846 [arXiv:1806.02367] [INSPIRE].

[31] F. Romero-López, A. Rusetsky, N. Schlage and C. Urbach, *Relativistic N-particle energy shift in finite volume*, *JHEP* **02** (2021) 060 [arXiv:2010.11715] [INSPIRE].

[32] L. Lellouch and M. Lüscher, *Weak transition matrix elements from finite volume correlation functions*, *Commun. Math. Phys.* **219** (2001) 31 [hep-lat/0003023] [INSPIRE].

[33] C.J.D. Lin, G. Martinelli, C.T. Sachrajda and M. Testa, *$K \to \pi\pi$ decays in a finite volume*, *Nucl. Phys. B* **619** (2001) 467 [hep-lat/0104006] [INSPIRE].

[34] W. Detmold and M.J. Savage, *Electroweak matrix elements in the two nucleon sector from lattice QCD*, *Nucl. Phys. A* **743** (2004) 170 [hep-lat/0403005] [INSPIRE].

[35] Ch. Kim, C.T. Sachrajda and S.R. Sharpe, *Finite-volume effects for two-hadron states in moving frames*, *Nucl. Phys. B* **727** (2005) 218 [hep-lat/0507006] [INSPIRE].

[36] N.H. Christ, C. Kim and T. Yamazaki, *Finite volume corrections to the two-particle decay of states with non-zero momentum*, *Phys. Rev. D* **72** (2005) 114506 [hep-lat/0507009] [INSPIRE].

[37] H.B. Meyer, *Lattice QCD and the Timelike Pion Form Factor*, *Phys. Rev. Lett.* **107** (2011) 072002 [arXiv:1105.1892] [INSPIRE].

[38] M.T. Hansen and S.R. Sharpe, *Multiple-channel generalization of Lellouch-Lüscher formula*, *Phys. Rev. D* **86** (2012) 016007 [arXiv:1204.0826] [INSPIRE].

[39] R.A. Briceno and Z. Davoudi, *Moving multichannel systems in a finite volume with application to proton-proton fusion*, *Phys. Rev. D* **88** (2013) 094507 [arXiv:1204.1110] [INSPIRE].

[40] V. Bernard, D. Hoja, U.G. Meissner and A. Rusetsky, *Matrix elements of unstable states*, *JHEP* **09** (2012) 023 [arXiv:1205.4642] [INSPIRE].

[41] A. Agadjanov, V. Bernard, U.G. Meißner and A. Rusetsky, *A framework for the calculation of the $\Delta N\gamma^*$ transition form factors on the lattice*, *Nucl. Phys. B* **886** (2014) 1199 [arXiv:1405.3476] [INSPIRE].



\end{document}